%% file: thesis.tex
\documentclass[physics,dissertation]{puthesis}

\usepackage{amsmath}
\usepackage{amssymb}

\usepackage{color}

\usepackage{multicol}
\usepackage{multirow}
\usepackage{array}
\newcolumntype{x}[1]{>{\centering\arraybackslash\hspace{0pt}}p{#1}}
\newcolumntype{y}[1]{>{\centering\arraybackslash\hspace{0pt}}m{#1}}

\usepackage{subfigure}
\usepackage{epsfig}
\usepackage{epstopdf}
\usepackage[pdftitle={Charge Multiplicity Asymmetry Correlation Study Searching for Local Parity Violation at RHIC for STAR Collaboration},pdfauthor={Quan~Wang}]{hyperref}

\hypersetup{
pdfborder= 0 0 0
}

\title{
  Charge Multiplicity Asymmetry Correlation Study Searching for Local Parity Violation at RHIC for STAR Collaboration
}

\author{Quan Wang}{Wang, Quan.}

\pudegree{Doctor of Philosophy}{Ph.D.}{May}{2012}

\majorprof{Fuqiang Wang}

\campus{West Lafayette}

\input{mydefs}

\newcommand\red[1]{{#1}}
\let\en=\ensuremath

\begin{document}

\volume

\include{front}

\include{intro}

\include{experiment}

\include{DataAna}

\include{result}

\include{summary}



\include{bibliography}

\appendix
\nochapterblankpages
\include{moreplots}


\include{vita}

\end{document}

%% file: mydefs.tex
\newcommand{\be}{\begin{equation}}
\newcommand{\ee}{\end{equation}}
\newcommand{\degree}{^\circ}

\newcount{\myi}
\newcommand{\Repeat}[2]{%
    \myi=0
    \loop
        \ifnum\myi<#2
        #1
        \advance\myi by 1
    \repeat
}

\def\gsim{\mathrel{\rlap{\lower4pt\hbox{\hskip1pt$\sim$}}
    \raise1pt\hbox{$>$}}}                

%% file: front.tex
%
%
%
%
%

\begin{dedication}
	To My Family
\end{dedication}

\begin{acknowledgments}
	First and foremost I would like to thank my advisor Prof. Fuqiang Wang, for his encouragement, support and guidance throughout my research.
	The thesis would not have been into the current stage without the help from him.
	I would like to thank Prof. Wei Xie.
	He was always ready to help me with the technical details and provided lots of inspiring discussions.
	I would also like to thank Prof. Denes Molnar for the helping of theory behind the analysis.
	My thank goes to all other members of heavy-ion group as well: Andrew Hirsch, Rolf Scharenberg and Brijish Srivastava, for their advices they have given me.
	I would also thank the graduate students whom I share office with, Jason Ulery, Terence Tarnowsky, Joshua Konzer, Michael Skoby, David Garand and Lingshan Xu, and also other graduate students in the heavy-ion group, Xin Li, Mustafa Mustafa, Kurt Jung, Cristina Moody, Deke Sun, and post-doc Daniel Kikola.
	Finally, I thank my family for the support all these years.
\end{acknowledgments}

%
%
%

\tableofcontents

\listoftables

\listoffigures


\begin{abbreviations}
  CME&  Chiral Magnetic Effect\cr
  EP&   Event Plane\cr
  LPV&  Local Parity Violation\cr
  QGP&  Quark Gluon Plasma\cr
  RHIC& Relativistic Heavy Ion Collider\cr
  RP&   Reaction Plane\cr
  STAR& Solinoid Tracker At RHIC\cr
  TPC&  Time Projection Chamber\cr
\end{abbreviations}



\begin{abstract}
The strong force is one of the four fundamental interactions in particle physics describing the interaction between partons (quarks and gluons) which make up hadrons.
The theory of the strong force is called quantum chromodynamics (QCD), which is a quantum field theory of the color charged partons.
The force between color charges does not diminish while they are separated.
This property causes the color charges to be confined in to hadrons, in ordinary matter.
Quark-Gluon Plasma (QGP) is one phase of the QCD matter at extremely high temperature and/or pressure, where the partons are asymptotically free.
Experimentally, QGP might be created in ultra relativistic heavy ion collisions \cite{Adams:2005dq,Adcox:2004mh,Arsene:2004fa,Back:2004je}.

It has been suggested that in such deconfined QCD matter, the metastable domains with non-zero topological charge $Q_{\text{w}}$ will generate charge separation along the system angular momentum direction caused by chiral magnetic effect (CME).
The charge separation direction is random as the sign of $Q_{\text{w}}$ is random from domain to domain.
The event-by-event charge separation along the system angular momentum direction violates the parity and time-reversal symmetries locally (LPV) \cite{Morley:1983wr,Kharzeev:1998kz,Kharzeev:2004ey,Kharzeev:2007tn,Fukushima:2008xe,Kharzeev:2007jp}.
In this analysis, we measure the CME/LPV in heavy ion collisions with charge multiplicity asymmetry correlations.

We separate a heavy ion collision event into up and down, or left and right hemispheres according to the reconstructed event-plane and the plane perpendicular to the event-plane.
We then calculate the multiplicity asymmetries of the positive and negative charges by taking the multiplicity difference between up and down hemispheres ($A_{\pm,UD}$), as well as left and right hemispheres ($A_{\pm,LR}$), and divide by the total multiplicities.
Since the event-plane does not distinguish between up and down nor left and right, the average of the asymmetries are consistent with zero, $\langle A_{\pm,UD} \rangle = \langle A_{\pm,LR} \rangle = 0$.
However, the correlations between the asymmetries are non-zero due to the physical correlations between particles.

We study the variances ($\langle A^2_{UD} \rangle$ and $\langle A^2_{LR} \rangle$) and covariances ($\langle A_+A_- \rangle_{UD}$ and $\langle A_+A_- \rangle_{LR}$) of the charge multiplicity asymmetries.
The asymmetries are calculated using the multiplicity from one side of the TPC tracks with respect to the event-plane reconstructed from the other side of the TPC tracks in order to avoid self-correlation.
We also apply single particle detector efficiency correction on asymmetry calculation and event-plane reconstruction.
The variance results are alway positive because they are the square of real numbers, which is the effect of statistical fluctuations.
We subtract the statistical fluctuation and the detector non-uniformity effects to obtain the dynamical variances $\delta\langle A^2_{UD} \rangle$ and $\delta\langle A^2_{LR} \rangle$.

We show the dynamical variances and covariances of Au+Au 200 GeV collisions and d+Au 200 GeV collisions.
Data show the dynamical variances are positive at peripheral collisions consistent with d+Au data.
This suggests that same-sign particle pairs are emitted preferentially in the same direction.
Both variances in $UD$ and $LR$ drop in mid-central and central collisions and turn to negative,
which suggests that the same-sign pairs are more likely to be emitted symmetrically, more back-to-back in other words, regardless of the directions.
The covariances are largely positive for both $UD$ and $LR$ directions, which suggests the opposite-sign particles are strongly correlated, and emitted with small angle correlation.

The CME/LPV expects charge separation across the event-plane ($UD$ direction), 
which gives additional correlation to the same-sign particle pairs in out-of-plane ($UD$) direction than in-plane ($LR$) direction, i.e. a wider distribution of the asymmetries in $UD$ direction.
We should expect $\Delta\langle A^2 \rangle = \delta\langle A^2_{UD} \rangle - \delta\langle A^2_{LR} \rangle > 0$.
One also expects that the positive and negative charges are anti-correlated in $UD$ direction, so that the covariances $\Delta\langle A_+A_- \rangle = \langle A_+A_- \rangle_{UD} - \langle A_+A_- \rangle_{LR}$ are negative.
We show the $UD-LR$ correlations of the dynamical variances and covariances.
Both the variance and covariance differences are positive for all centralities except the most peripheral bins.
The variance $UD-LR$ correlation is positive, which is consistent with CME/LPV expectation. 
However we also know same-sign pairs are preferentially back-to-back from mid-central to central collisions.
The covariance $UD-LR$ correlation is also positive, which is not consistent with the naive expectation of CME/LPV.

We study the transverse momentum ($p_T$) dependence of the correlations.
The CME/LPV expects the charge separation is mostly a low-$p_T$ effect.
However data show the correlations increase with $p_T$ in the mid-central collisions.

The dynamical variances and covariance as a function of event-by-event anisotropy $v_2^{obs}$ for low-$p_T$ ($p_T < 2$ GeV/$c$) and high-$p_T$ ($p_T > 2$ GeV/$c$) particles are analysed.
The $UD-LR$ variance and covariance show opposite trend of low-$p_T$ $v_2^{obs}$, but with very weak dependence of the high-$p_T$ $v_2^{obs}$.
We use four different data and cuts to verify the result: sub-events with $\eta>0$ and $\eta<0$, sub-events with large pseudo-rapidity gap $\eta>0.5$ and $\eta<-0.5$,
events with the first order ZDC-SMD event-plane and top 2\% most central data.
They all show similar $v_2^{obs}$ dependence.

There might be charge independent common background sitting between the same-sign and opposite-sign correlations.
So we define charge separation as the difference of same-sign and opposite-sign correlations, $\Delta = \Delta\langle A^2 \rangle - \Delta\langle A_+A_- \rangle$, to cancel the background, and show it as a function of the wedge size, azimuthal region of the analysed particles.
The charge separation vanishes with the decrease of the wedge size, which suggests the charge separation effect is within the vicinity of the reaction-plane.

We also show the charge separation as a function of event-by-event low-$p_T$ $v_2^{obs}$.
The CME/LPV effect does not expect event anisotropy dependence.
However, we see the charge separation is strongly and linearly depending on low-$p_T$ $v_2^{obs}$ for four different cases.
The linear dependence intercept at zero or slightly positive, when the sub-event is isotropic in low-$p_T$ particle azimuth angle, i.e. $v_2^{obs} = 0$.

Because the average $v_2^{obs}$ is positive due to elliptic flow, the charge separation is positive if we integrate over all events.
The linear $v_2^{obs}$ dependence of charge separation is more likely an intrinsic charge dependent bulk correlation of the medium.
A more precise measurement of the CME/LPV effect may lie on the events with a more isotropic shape.
We show the charge separation of events with $|v_2^{obs}|<0.04$ as a function of centrality, which is consistent with zero.
Thus we give the upper limit of the CME/LPV effect based on the linear fit of the charge separation $\Delta$ as a function of $v_2^{obs}$, that $\Delta = 4.7 \times 10^{-5}$ with 98\% CL.

\end{abstract}

%% file: intro.tex
%
%
%

\chapter{INTRODUCTION}

\section{Strong Interaction and Quantum Chromodynamics}

\begin{figure}[hbt]
	\begin{center}
		\psfig{figure=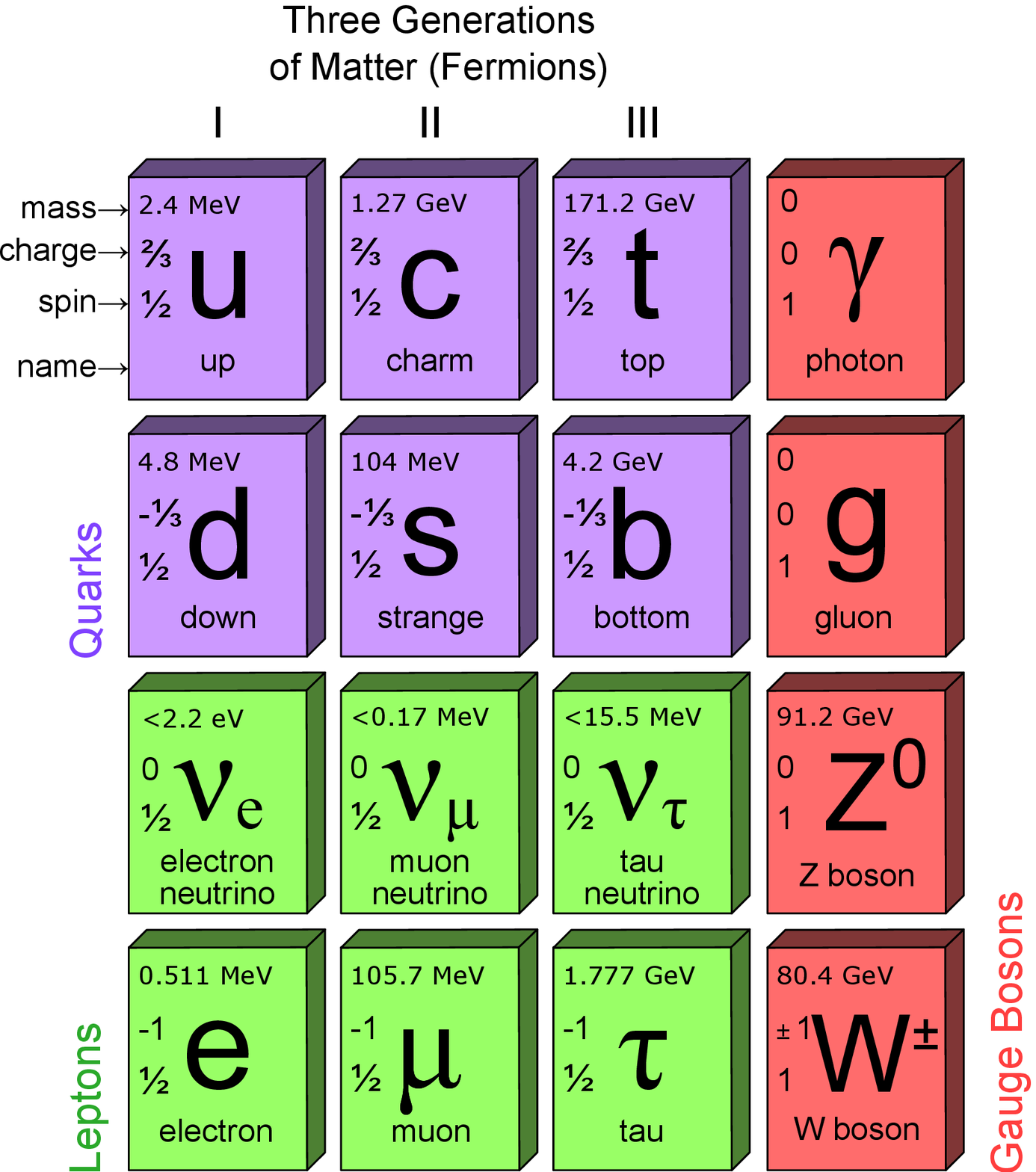,width=0.6\textwidth}
	\end{center}
	\caption[Elementary particles]{The three generations of quarks and leptons, and the gauge bosons.
	Figure is taken from wikipedia.}
	\label{fig:elepart}
\end{figure}

There are four fundamental interactive forces in particle physics, which describe the way elementary particles interact with each other.
They are electromagnetism, strong interaction, weak interaction and gravitation.
Modern physics attempts to explain all physically observed phenomena by the theories of these fundamental interactions.
Except for gravitation, theories of electromagnetism, the strong interaction and the weak interaction are well established in the standard model.
In the concept model, matter consists of elementary particles, which are spin one-half fermions 
and interact with one another according to their properties (charges).
They interact by exchanging spin one gauge bosons, also called force carriers.
A summary of the fundamental interactions with their theories and properties is shown in table~\ref{tab:intact}.
Figure \ref{fig:elepart} shows the three generations of elementary leptons and quarks, as well as the gauge bosons.

\begin{table}[thb]
	\centering
	\caption{Fundamental interactions.}
	\begin{tabular}{y{60pt} y{140pt} y{90pt} y{50pt}}
		\hline
		\hline
		Interaction& Theory & Force Carriers (gauge boson)& Relative Strength \\
		\hline
		Strong & Quantum ChromoDynamics (QCD) & gluon~($g$) & $10^{38}$ \\
		\hline
		Electromagnetic & Quantum ElectroDynamics (QED) & photons~($\gamma$) & $10^{36}$ \\
		\hline
		Weak & Electroweak Theory & W~($W^{\pm}$) and Z~($Z^0$) bosons & $10^{25}$ \\
		\hline
		Gravitation & General Relativity~(GR) & gravitons~(hypothetical) & $1$ \\
		\hline
		\hline
	\end{tabular}
	\label{tab:intact}
\end{table}

Specifically in this thesis, the focus is on the study of the strong interaction.
It is a short range interaction comparing to other three interactions that binds protons and neutrons together to form the nucleus of atoms (in the range of 1-3 fm), 
and also binds quarks and anti-quarks to form hadron particles (in the range of less than 1 fm).
The strong interaction is carried out by gluons exchanging ``color charge'', 
an analogous to the electronic charge in electromagnetism between quarks, anti-quarks and gluons.
Different from the force carrier photon ($\gamma$) in electromagnetism, gluon ($g$) can interact between themselves.
Unlike electromagnetism's electric charges (positive and negative), there are three types of color charges, resulting in different behaviors of strong interaction.
The behavior of the color charges and the interactions of quark-gluon are detailed in the theory of quantum chromodynamics (QCD),
the quantum field theory of the nuclear interaction.
As part of the standard model, mathematically, the theory is a non-Abelian gauge theory based on a local symmetry group $SU(3)$.

There are two unique properties of the QCD theory: color confinement and asymptotic freedom.
Quarks and gluons are the only elementary particles carrying color charges.
The strong force between color charges, unlike all other forces, doesn't diminish with increasing the distance of the color charges.
It takes an infinite amount of energy to separate two quarks.
Thus, before the quarks can be separated, the energy is large enough to create quark and anti-quark pairs to combine with the original quarks.
Experimentally, isolated quarks have never been observed, i.e. free color charges.
Any ordinary matter which can be observed are color neutral.
This phenomenon is called color confinement.

Asymptotic freedom is a property of the $SU(3)$ gauge theory. 
At high energy, or equivalently at very short distance ($\ll$ 1 fm), the interaction between quarks becomes weak,
while at low energy or equivalently at large distance ($\gsim$ 1 fm), the interaction becomes strong.
This phenomenon prevents the baryons and mesons from unbinding.

\section{Quark-Gluon Plasma and Chiral Symmetry}

\begin{figure}[thb]
	\begin{center}
		\psfig{figure=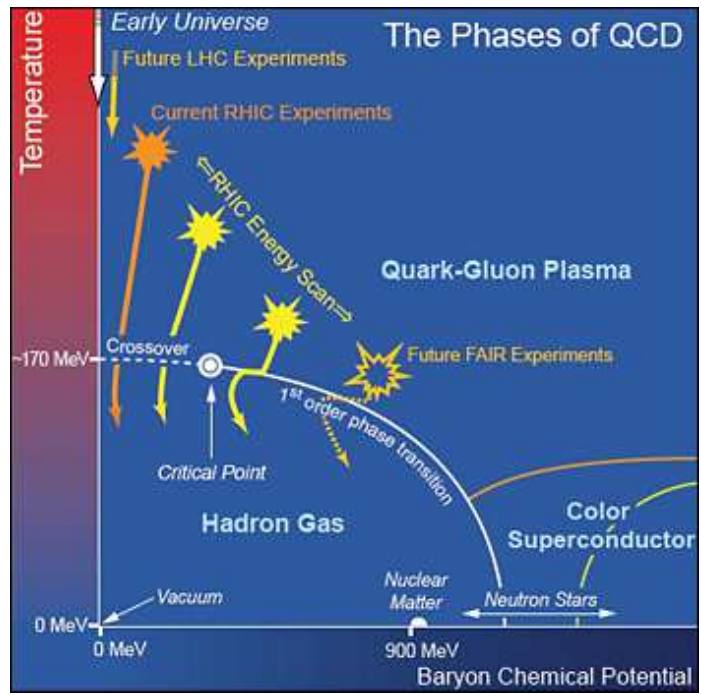,width=0.6\textwidth}
	\end{center}
	\caption[QCD Phase diagram]{
	The illustration phase diagram of quark matter.
	}
	\label{fig:phase}
\end{figure}

Quark-Gluon Plasma (QGP) is a phase of QCD matter at extreme conditions, such as very high temperature ($T$) and/or high baryon chemical potential $\mu_B$ \cite{Stephanov:2004wx,Aggarwal:2010cw}.
Figure~\ref{fig:phase} shows an illustrated phase diagram of quark matter.
QGP might be produced in heavy ion collisions at the Relativistic Heavy Ion Collider~(RHIC) \cite{Adams:2005dq,Adcox:2004mh,Arsene:2004fa,Back:2004je}, which is similar to the environment of the universe a few milliseconds after the Big Bang.
The hot dense matter created in the heavy ion collisions at very high energy is considered to be in thermal equilibrium.
Further studies show that extra degrees of freedom could be released at the quark level.
Quarks have been relatively freed from the confined nuclei.
This phenomenon is called color deconfinement, one of two fundamental properties of QGP.

The other fundamental property, also considered as a signature of the QGP, is chiral symmetry restoration.
Chirality in physics is a phenomenon that an object is not identical to its mirror image, i.e. its mirror image cannot be mapped by only rotations and translations.
In the high energy limit, chirality can be treated as helicity~(handedness).
The helicity is defined as the sign of projecting the particle spin onto its direction of motion.
The chirality is positive (right-handed) if the direction of the particle's spin is aligned in the same direction as its motion.
It is negative (left-handed) if the spin is opposite to its motion.

For massless particles, the helicity cannot be reversed by a Lorentz boost because no observer can travel faster than light.
Therefore the massless particles have their helicity fixed for all reference frames, such as photon ($\gamma$) and gluon ($g$).
Their helicity are invariant under Lorentz transformation.
On the other hand, massive particles have slower speed than light.
Then one can always boost a reference frame, so that the momentum reverses the direction.
Massive particles thus may change their helicity signs after the Lorentz boost.

In QCD theory, the Lagrangian can be written as
\begin{equation}
	\mathcal{L} = -{1 \over 4} F^{\alpha}_{\mu\nu}F^{\mu\nu}_{\alpha}-\sum_{f}\bar{\psi}_{f}\gamma^{\mu}\left[ \partial_{\mu}-i g A_{\mu}^{\alpha}t_{\alpha}\right]\psi_{f} - \sum_{f} m_{f} \bar{\psi}_{f} \psi_{f},
  \label{eq:lag}
\end{equation}
where $f$ and $g$ denote flavor index and the QCD coupling constant, 
$F^{\alpha}_{\mu\nu}$ denotes the spin-1 gluonic field strength tensor.
$A_{\mu}^{\alpha}$ is the vector potential of the color field.
$\psi_{f}$ and $t_{\alpha}$ are the quark fields and the generators of the color $SU(3)$ group.
The mass term $m_f \bar{\psi}_{f} \psi_{f}$ explicitly breaks the chiral symmetry of the QCD Lagrangian.

Quark masses come from two sources.
One is the ``naked'' quark mass, also called current mass, which is considered to originate from the Higgs mechanism in standard model.
The other source is from the gluon field induced by a valence quark (quark which determines the hadron's quantum number), where the quark is surrounded like a cloud by sea quarks called covering.
The two terms together give rise to the effective quark mass called the constituent mass.

For light quarks, i.e. $up$ and $down$, the constituent mass is much larger than the current mass, while for heavy quarks, i.e. $charm$, $bottom$ and $top$,
the constituent mass is nearly the same as the current mass.
Figure \ref{fig:elepart} shows the current masses of the quarks.
To show the constituent quark mass, we use proton as an example.
The proton is a composite of three valence quarks ``$uud$'' with the total current mass approximately 10 MeV.
However, the mass of proton is 938 MeV, which is much larger than the total current masses of the valence quarks.
The difference comes from the gluon field, the binding energy of quantum chromodynamics, while the gluons are massless.

In the QGP phase as shown in the phase diagram, the quarks can be released from the confined matters and move relatively ``freely''.
They are not really free, but relatively free.
Thus, in such quark matter, quarks loose their covering.
The light quarks will have their mass greatly reduced to nearly zero, such that the mass term in the QCD Lagrangian vanishes, 
which results in the chiral symmetry becoming restored in the quark matter.
In such chiral limits ($m_{u}=0$, $m_{d}=0$), for light quarks, all left-handed quarks remain left-handed, and all right-handed quarks remain right-handed.
Each chiral state has a chiral symmetry partner with the opposite parity and equal mass.

\section{Chiral Magnetic Effect and Local Parity Violation}
\label{CME}

Based on the well defined gauge theory, many remarkable properties of QCD matter have been discovered.
One of the properties is that the field configurations can be characterized by a topological invariant, the topological charge $Q_\text{w}$ \cite{Kharzeev:2007jp}.
It is defined as
\begin{equation}
	Q_\text{w} = {g^2 \over 32 \pi^2} \int \mathrm{d}^4 x F^{\alpha}_{\mu\nu} \tilde{F}^{\mu\nu}_{\alpha} \in \mathcal{Z},
	\label{eq:tcharge}
\end{equation}
where $g$ is the QCD coupling constant, and $F^{\alpha}_{\mu\nu}$ and $\tilde{F}^{\mu\nu}_{\alpha} = {1\over 2} {\epsilon_{\mu\nu}}^{\rho\sigma}F^{\alpha}_{\rho\sigma}$ denote the gluonic field tensor and its dual.

In QCD matter with the chiral limits (massless quarks $m_f = 0$) satisfied, chiral symmetry can be restored in the initial state with $N_{LH} = N_{RH}$, 
where $N_{LH}$ and $N_{RH}$ denote number of left-handed and right-handed quarks.
There are metastable domains with certain topological charge $Q_\text{w}$ forming in the vicinity of the deconfined QCD matter \cite{Kharzeev:1998kz}, which lead to parity ($\mathcal{P}$) and charge-conjugation and parity ($\mathcal{CP}$) violation, if $Q_\text{w}$ is non-zero.
The result is to convert right-handed~(left-handed) quarks to left-handed~(right-handed) quarks,
with $N_{LH}-N_{RH} = 2 N_{f} Q_\text{w}$ in the final state,
where $N_{f}$ is the number of flavors \cite{Kharzeev:2007jp}.

\begin{figure}[thb]
	\begin{center}
		\psfig{figure=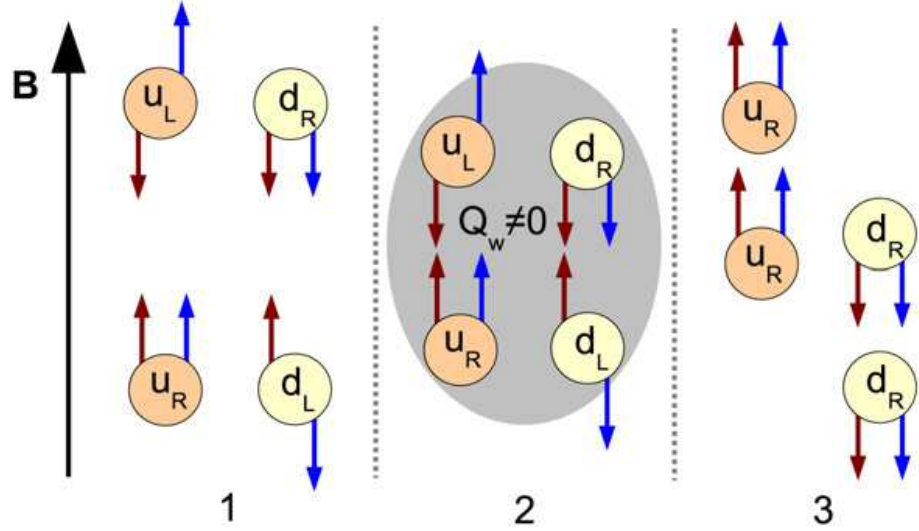,width=0.8\textwidth}
	\end{center}
	\caption[Chiral magnetic effect]{
	Illustration of the chiral magnetic effect in a very large homogeneous magnetic field.
	The red arrows denote the momentum direction, and the blue arrows denote the spin of quarks.
	See text for the details of the effect.
	Figure is taken from reference \cite{Kharzeev:2007jp}.
	}
	\label{fig:cme}
\end{figure}

Figure~\ref{fig:cme} shows an illustration of the Chiral Magnetic Effect~(CME) inside the quark matter with the presence of a large and uniform magnetic field B.
Within the quark matter, all the quarks are deconfined, and chiral symmetry is restored with chiral limit~($m_u=m_d=0$).
The red arrows denote the momentum direction, and the blue arrows denote the spin direction of the quarks.
Due to the large magnetic field, quarks will eventually occupy the lowest Landau level after equilibrium, with their magnetic moments align in the same direction as the B field.
Thus, $u$ quarks with positive charges have their spin in the same direction of the B field, and $d$ quarks with negative charges in the opposite direction of the B field as shown in part (1).
Then the gauge field with non-zero topological charge $Q_{\text{w}}$ interacts within the metastable domain, as shown in part (2), 
and breaks the chiral symmetry by, for example with negative $Q_\text{w}$, converting left-handed $u$ and $d$ quarks into right-handed quarks.
The result is to flip the momentum direction of the left-handed quarks to the opposite direction.
In the end as shown in part (3), all the $u$ quarks are right-handed and moving upwards carrying positive charges, and all the $d$ quarks are also right-handed but moving downwards carrying negative charges.
This charge separation effect is then called the Chiral Magnetic Effect~(CME), and could possibly be measured in experiment if indeed true.

\begin{figure}[thb]
	\begin{center}
		\subfigure[Parity violation]{\label{fig:chargesep-a}\includegraphics[width=0.3\textwidth]{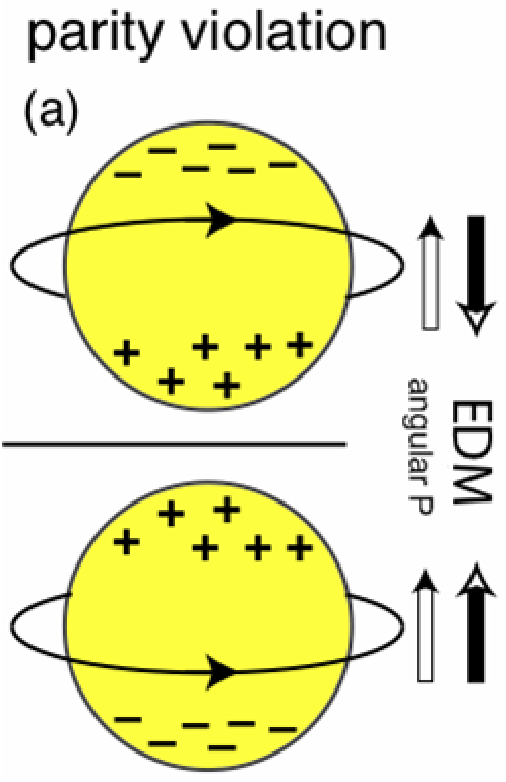}}
		\subfigure[Charge separation]{\label{fig:chargesep-b}\includegraphics[width=0.6\textwidth]{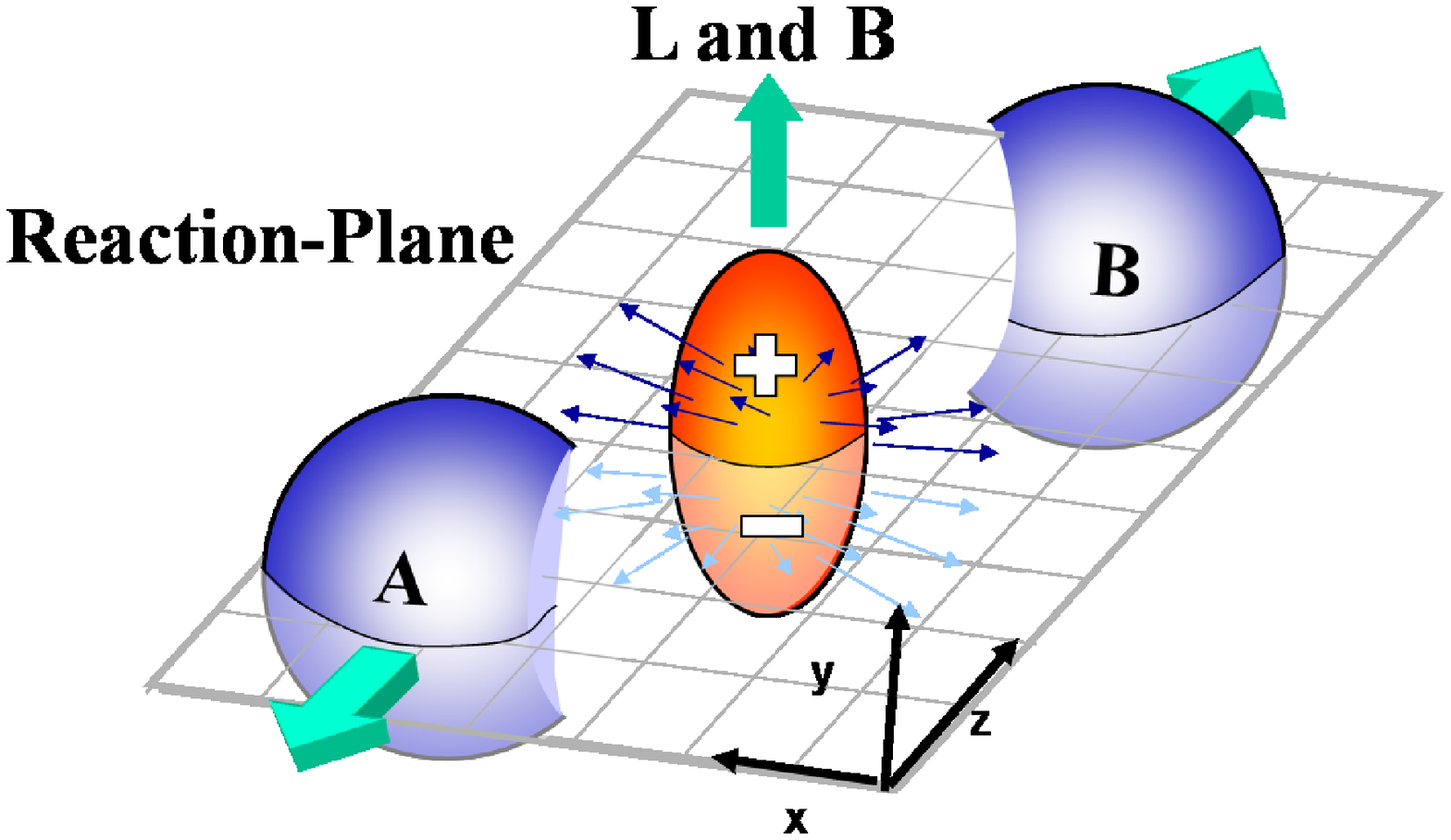}}
	\end{center}
	\caption[Charge separation and parity violation]{
	Panel (a): Illustration of parity violation with charge separation in the system angular momentum direction.
	Panel (b): Illustration of charge separation along the system angular momentum direction (L) and magnetic field direction (B) in center of mass frame.
	The $z$ direction is defined as the projectile (A) and target (B) nuclei momentum direction.
	The two heavy ion nuclei undergo a mid-central collision with the reaction area shown in orange color.
	See text for details.
	}
	\label{fig:chargesep}
\end{figure}

The schematic view of the charge separation effect in a mid-central heavy ion collision is illustrated in figure~\ref{fig:chargesep-b}.
In the center of mass frame, the beam direction ($z$ direction in the figure) and the direction connecting the centers of two colliding nuclei ($x$ direction) define the ``Reaction-Plane'' of the collision.
The overlapping reaction area, has an elliptical shape and contains the hot dense QCD matter which could have chiral symmetry restored.
The spectators, the wounded nuclei A and B in the figure, carry positive charges and create a magnetic field $B$ when passing the reaction area.
The direction of the magnetic field is the same as the QCD system's angular momentum $L$ direction.
If CME is indeed true and the effect can survive through the hot dense medium evolution to the detectors, one should observe charge separation along the magnetic field, i.e. system angular momentum direction, which is perpendicular to the reaction plane.

As shown in figure \ref{fig:chargesep-a}, charge separation gives the system an electric dipole moment~(EDM), with its direction pointing from the negative charge to the positive charge.
When the system has its angular momentum aligned (anti-aligned) with the direction of the EDM, applying a parity operation to
the system will change its parity state to anti-alignment (alignment).
Thus, charge separation in the angular momentum direction is a phenomenon of parity violation.
It is known that the parity symmetry is violated in some weak interactions,
while it is well preserved in all other three interactions including strong interaction.
If we could measure the charge separation in the system angular momentum direction in QCD matter,
it may indicate that the parity symmetry could be violated in strong interaction.
However, the topological charge $Q_\text{w}$ of the metastable domains are random within the QCD matter.
Thus, the direction of charge separation is also random.
The charge separation effect cancels out over repeating experiments, in other words, it only happens on the event-by-event basis.
So, the effect is only local, namely local parity violation~(LPV).
Globally, the parity symmetry is still conserved over a large number of events.

The estimates of the charge separation signal are proportional to the topological charge $Q_\text{w}$, 
and diluted by the event multiplicity. 
It was first calculated in reference \cite{Kharzeev:2004ey} 
that the asymmetry of $u$ quarks, for example, can be estimated as 
\begin{equation}
	A_{u} = {N_{R} - N_{L} \over N_{R} + N_{L}},
\end{equation}
when assuming $\mathcal{P}$ and $\mathcal{CP}$ are preserved in the hadronic process.
Then the asymmetry should translate to the hadron multiplicity asymmetry of charged pions
\begin{equation}
	A_{\pi^+} \simeq {Q_\text{w} \over N_{+}},
	\label{eq:asymmag}
\end{equation}
where $N_{+}$ is taken as positively charged pion multiplicity in one unit of rapidity.
This is because soft particles are usually correlated in one unit of rapidity range,
which is also the extent of the parity violation domain in the rapidity space \cite{Kharzeev:2004ey}.
At STAR experiment, the reference multiplicity $N_{RefMult}$ is the total number of charged particles recorded by the main TPC
with pseudo-rapidity range of $-0.5 < \eta < 0.5$, see section \ref{CENT}.
At RHIC energy of 200 GeV per nucleon pair in Au+Au central collision, 
$N_{RefMult}$ is typically around 300, and drops to 150 in mid-central collisions,
and 30-50 in peripheral collisions \cite{Adams:2003xp}.
Although the gold nucleon carries positive charges causing
positive charged particles to produced slightly more frequently than the negatived charge particles (about 2\%),
we still have $A_+ = A_-$ because they are normalized by the total number.
Putting those numbers together, we have the charge multiplicity asymmetries at the order of $\sim 10^{-2}$ in mid-central collisions \cite{Kharzeev:2004ey,Kharzeev:2007tn,Fukushima:2008xe},
and $\sim 10^{-4}$ to $\sim 10^{-3}$ for the asymmetry correlations.

Apparently we do not take into account some of factors in the above estimation which may vary the final result,
such as the magnetic field strength and duration time.
More accurate estimations can be found in reference \cite{Kharzeev:2007jp},
where theoretical calculations have included magnetic field strength and fluctuations.
They all give similar estimated results.
However, the in-medium interaction and final state interactions before freeze out may also play an important role.
Because the strong interaction conserves parity symmetry,
the in-medium interactions could not contribute to the parity odd signal, 
but they will destroy the charge separation signal which are generated in early stage of the collisions.
The effect is to smear out the correlations between charges.
The estimated effect is to reduce the signal for at least one order of magnitude \cite{Ma:2011uma}, with the estimate around $\sim 10^{-5}$ to $\sim 10^{-4}$ for the asymmetry correlations \cite{Kharzeev:2007jp}.
A recent estimate shows that the CME/LPV induced charge asymmetry correlation is less than $10^{-6}$ in \cite{Muller:2010jd}, and $10^{-4}$ after multiplied by the $N_{part}$.
Some even claim it is possible that the radial flow can even push the opposite-sign pairs into the same direction \cite{:2009txa}.
Thus, the sign of the opposite-sign correlation may possibly be even positive.

\section{Anisotropic Flow and Three-Particle Correlator}
\label{flow}

To study the properties of the QCD matter, 
physicists study heavy ion collisions at ultra relativistic conditions by accelerating heavy nuclei such as gold to the speed close to the light, 
and colliding them to create the new state of hot dense matter.
A massive amount of particles are created during the collision, which mimics the early time of the Big Bang of the universe.

In non-central heavy ion collisions shown in figure \ref{fig:chargesep-b}, the initial spacial anisotropy will cause a pressure gradient in the azimuthal angle.
The pressure in in-plane direction is larger than that in out-of-plane, which translates to larger transverse momentum ($p_T$) and more particles are emitted in-plane than out-of-plane in the final state hadrons.
The spacial and momentum space anisotropy of the event is driven by the initial pressure gradient, and affected by the medium strong interactions,
which can be used as a probe of the initial collision geometry and the medium properties.
The anisotropy is characterized by the Fourier expansion of the event azimuthal angle $\phi$:
\begin{equation}
	{\operatorname{d} N \over {\operatorname{d} \phi} } \propto 1+\sum_{n=1}^{\infty} 2 v_n \cos \left[n\left(\phi - \psi_{RP}\right)\right],
  \label{eq:fourier}
\end{equation}
where $\psi_{RP}$ denotes the true reaction plane as shown in figure \ref{fig:chargesep-b}.
The Fourier coefficient $v_n$ stands for the $n$-th harmonic of the event azimuthal anisotropy.
If we apply orthogonal condition of $\cos n(\phi-\psi_{RP})$ to the above equation, we can get the $n$-th harmonic coefficient $v_n$ as:
\begin{equation}
	v_n = \langle \cos[n(\phi-\psi_{RP})] \rangle,
  \label{eq:vn}
\end{equation}
where $\langle \ldots \rangle$ denotes an average over all the particles of each event.
Due to the reflection symmetry, the sine terms vanish.
Specifically, we refer the first order harmonic $v_1$ as directed flow,
and the second order harmonic $v_2$ as elliptic flow, respectively.
The directed flow $v_1$ is an odd function of the rapidity due to momentum conservation,
which is measured very small at RHIC $\sqrt{s_{NN}}=200$ GeV Au+Au collisions with typically $|v_1|<0.005$ for $|\eta|<1$ \cite{:2011eg}.
The elliptic flow $v_2$ is an even function of the rapidity,
and is measured sizable positive at 200 GeV Au+Au mid-central collisions about 6 percent for particle within $0.15 <p_T < 2.0$ GeV/$c$ \cite{:2008ed}.

Both the first and second order anisotropy are directly related to the initial condition of the collision.
Thus, the directed flow and elliptic flow provide us the experimental tools to determine the reaction-plane direction.
The methods are introduced in sections \ref{TPCEP} and \ref{ZDCEP}.

Equation \ref{eq:fourier} is parity even because cosine is an even function to the mirror reflection.
In order to study the $\mathcal{P}$-violation across the reaction-plane, we have to account for the parity odd terms, sine.
The modified Fourier expansion can be written as
\begin{equation}
	{\operatorname{d} N \over {\operatorname{d} \phi} } \propto 1+\sum_{n=1}^{\infty} 2\left( v_n \cos \left[n\left(\phi - \psi_{RP}\right)\right]
	+ a_n \sin \left[n\left(\phi - \psi_{RP}\right)\right] \right).
  \label{eq:fourier1}
\end{equation}
The coefficient $a_n$ stands for the $\mathcal{P}$-violation terms across the reaction-plane.
As we introduced in previous section, $a_n$ is due to the local parity violation with the topological charge $Q_{\text{w}}$.
The signs of $a_n$ vary with the fluctuation of $Q_\text{w}$.
If we average a large amount of events, the averages of $a_n$ vanish because the topological charge is random.
Thus, the direct measurement of $a_n$ is not possible.
However, the charge separation effect will not vanish, and can be measured through correlation methods.

STAR has published measurements of the first order charge dependent coefficient $a_1$ correlations.
A charge dependent three-particle correlator \cite{Voloshin:2004vk,:2009uh,:2009txa} is introduced as
\begin{align}
	\langle \cos (\phi_{\alpha} + \phi_{\beta} - 2 \phi_{c}) \rangle / v_{2,c} &\approx
	\langle \cos (\phi_{\alpha} + \phi_{\beta} - 2 \psi_{RP}) \rangle \\
	&= \langle \cos \Delta\phi_{\alpha} \cos \Delta\phi_{\beta}  \rangle - \langle \sin \Delta\phi_{\alpha} \sin \Delta\phi_{b} \rangle \\
	&= [\langle v_{1,\alpha} v_{1,\beta} \rangle + B_{in}] - [\langle a_{1,\alpha} a_{2,\beta} \rangle + B_{out}],
	\label{eq:3partdef}
\end{align}
where $\alpha$, $\beta$ and $c$ are particle charge labels, and $\Delta\phi = \phi - \psi_{RP}$ refers to the particle azimuthal angle relative to the reaction-plane.
Assuming firstly, directed flow term $\langle v_{1,\alpha} v_{1,\beta} \rangle$ vanishes because it is an odd function of the rapidity and its fluctuation is small.
Secondly, the average background from in-plane $B_{in}$ and out-of-plane $B_{out}$ cancels out, assuming the reaction-plane dependent background $[B_{in}-B_{out}]$ is small.
Lastly, only the first order $a_1$ dominates.
Under such assumptions, the three-particle correlators are reported as the first evidence of the CME/LPV.
We will review the assumptions and compare our observables to the three-particle correlations in section \ref{3P}.

%% file: experiment.tex
%
%
%

\chapter{EXPERIMENT}

In this chapter, we introduce the experiment of relativistic heavy ion collision.
We introduce the facility and detectors used for data taking.
We also introduce the kinematic variables measured for this analysis.

\section{Relativistic Heavy Ion Collider}

The Relativistic Heavy Ion Collider~(RHIC) is one of the high energy heavy-ion colliders located at Brookhaven National Laboratory in Upton, New York on Long Island.
By accelerating and colliding heavy ion and polarized proton beams, physicists study the matter created at extremely high temperature and density,
which is the QCD matter with strongly interacting partons (quarks and gluons).

Protons and heavy ion nuclei are accelerated in two independent pipes to nearly the speed of light and may collide at four intersecting points where the pipes cross.
So far, several particle species have been accelerated for collisions at different energies, 
including proton-on-proton (p+p), deuterium-on-gold (d+Au), copper-on-copper (Cu+Cu), gold-on-gold (Au+Au), copper-on-gold (Cu+Au) and uranium-on-uranium (U+U).
For heavy ion collisions, the center of mass energy can reach 200 GeV  per nucleon pair.
For p+p collisions, it achieved 500 GeV in 2009.

\begin{figure}[thb]
	\begin{center}
		\psfig{figure=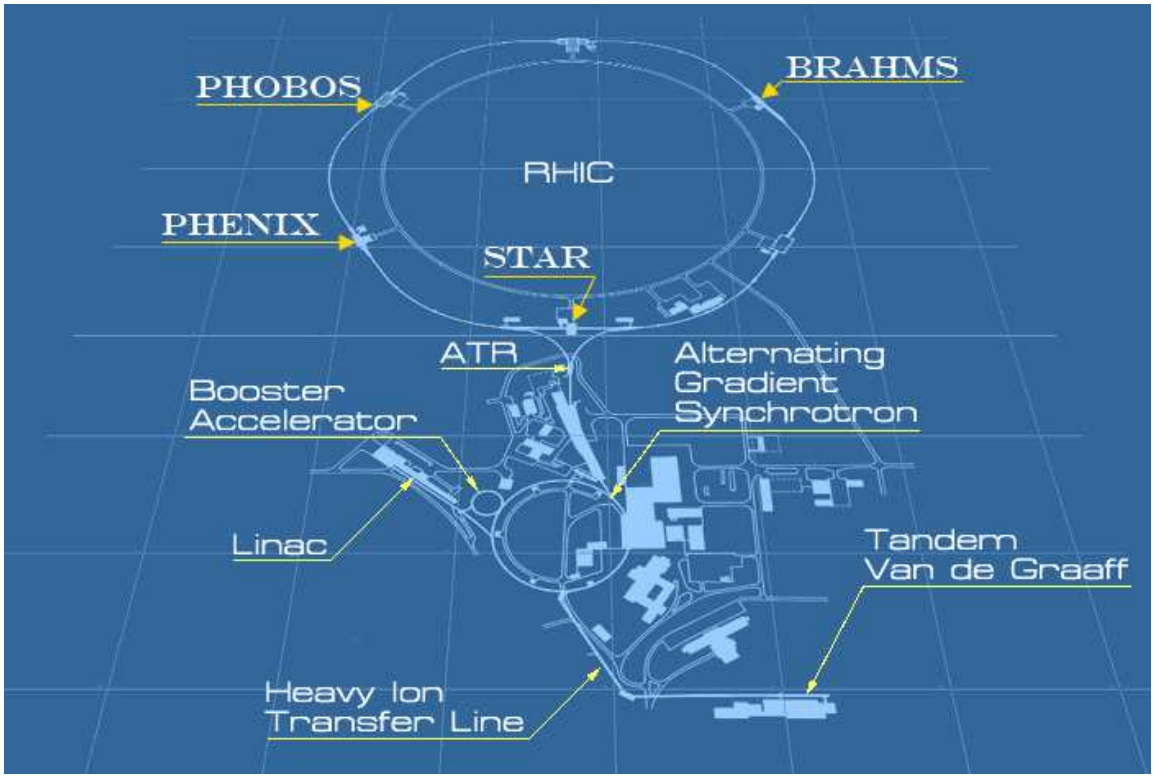,width=\textwidth}
	\end{center}
	\caption[RHIC at BNL]{
	The schematic plot of the Relativistic Heavy Ion Collider (RHIC) facility located at Brookhaven National Laboratory (BNL).
	Diagram taken from \cite{Ackermann:2002ad}.
	}
	\label{fig:rhic}
\end{figure}

As shown in figure~\ref{fig:rhic}, the RHIC accelerator ring is 3,834 m long in circumference,
and there are four experiments on RHIC collision points.
They are STAR~(6 o'clock), PHENIX~(8 o'clock), PHOBOS~(10 o'clock) and BRAHMS~(1 o'clock).
PHOBOS and BRAHMS have completed their commissioning and been shut down after 2005 and 2006,
while STAR and PHENIX are still running since 2000.
This thesis is based on the data taken by STAR experiment of Au+Au and d+Au collisions at $\sqrt{s_{NN}} = $ 200 GeV.

\section{STAR Experiment}

The Solenoidal Tracker at RHIC (STAR) detector is located at the 6 o'clock interaction region of the RHIC accelerator ring.
The main physics goal is to study the formation, evolution and characteristics of the strongly coupled Quark Gluon Plasma (sQGP) \cite{Adams:2005dq,Adcox:2004mh,Arsene:2004fa,Back:2004je}, 
a state of QCD matter which is believed to be formed at very high temperature and/or high energy density.
It is designed primarily for charged hadron production measurements with high precision tracking and momentum over a large solid angle.

\begin{figure}[thb]
	\begin{center}
		\psfig{figure=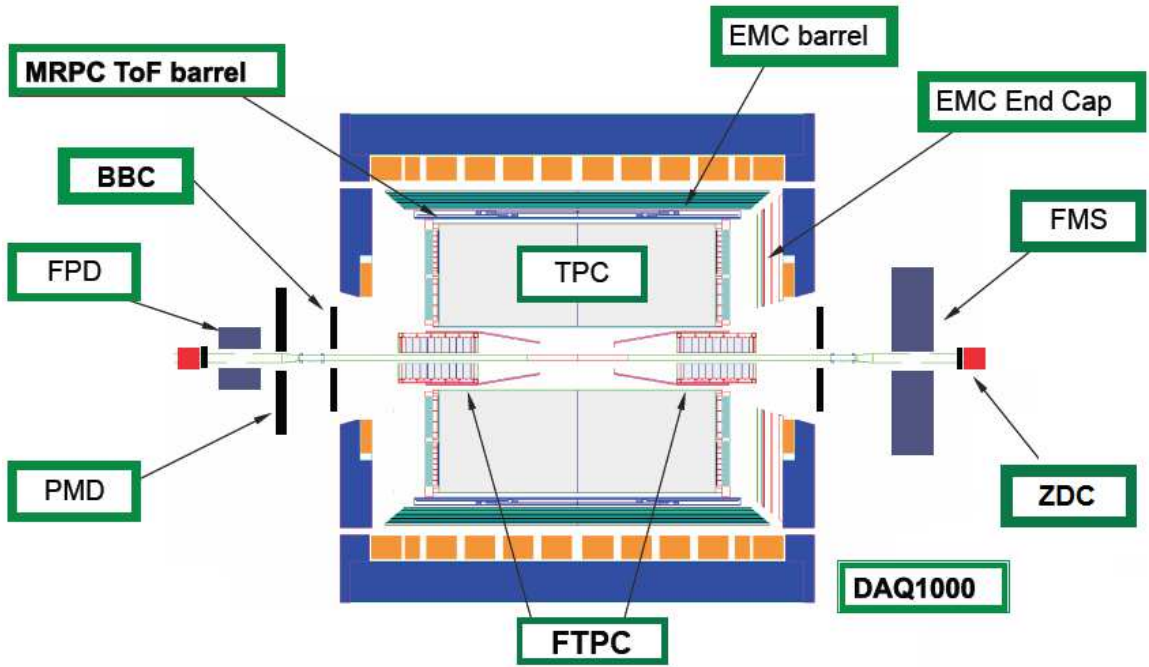,width=\textwidth}
	\end{center}
	\caption[STAR detector]{
	Schematic STAR detector layout with selected sub-systems.
	Diagram taken from \cite{Ackermann:2002ad}.
	}
	\label{fig:star}
\end{figure}

STAR is a massive detector weighing 1,200 tons and as big as a building.
It has a large relatively uniform acceptance in azimuthal angle, and also a large polar angle coverage in mid-rapidity.
It consists of several sub-systems as shown in figure~\ref{fig:star}.
System upgrades are constantly going on since the first build of the detector.
Some of the sub-detectors are removed and replaced for better measurement or more tracking capability.

For convenience, we set up the STAR coordinate system with its origin located at the STAR detector geometry center.
The $z$ direction points to the west, which is parallel to the beam pipe direction.
And the $x$ direction points to the south, and the $y$ direction points to up.
The whole detector is surrounded by the main magnetic coil 
which generates a field in the $z$ direction with the maximum of $\left|B_z\right| = 0.5$ T \cite{Bergsma:2002ac}.

\subsection{Time Projection Chamber}

The main tracking device is the Time Projection Chamber~(TPC), 
which records the tracks of particles and provides the kinematic information of each track \cite{Ackermann:1999kc,Anderson:2003ur}.
It is located within the magnetic coil, with a 4.2 m cylinder length, 0.5 m inner radii and 2 m outer radii.
With the full magnetic field ($\left| B_z\right| = 0.5$ T) turned on, the TPC can identify a broad transverse momentum range of charged particles from 0.15 GeV/$c$ to 30 GeV/$c$ depending on particles.

\begin{figure}[thb]
	\begin{center}
		\psfig{figure=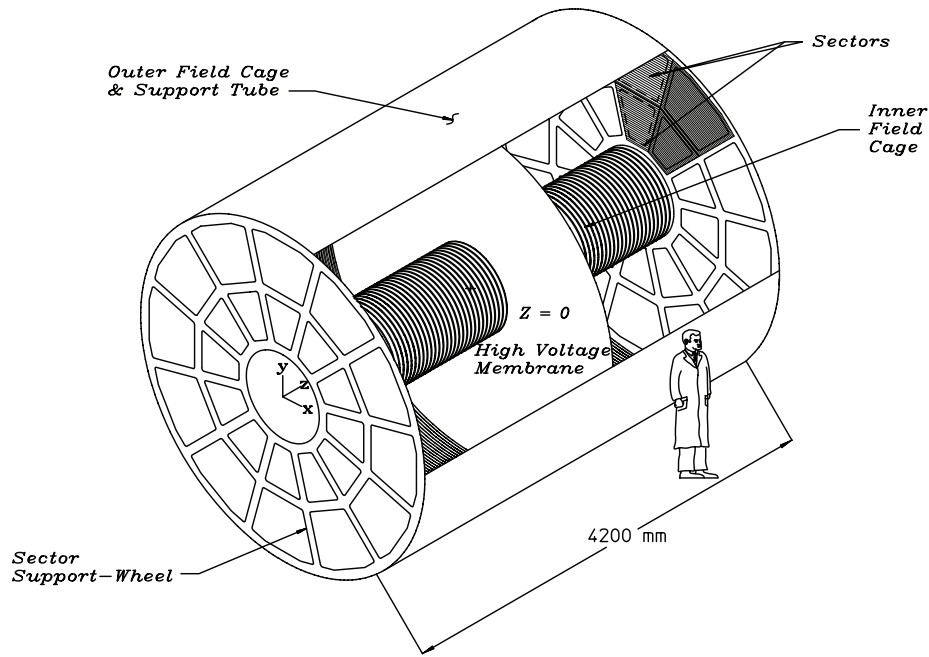,width=0.8\textwidth}
	\end{center}
	\caption[Time Projection Chamber]{
	The schematic cutaway view of STAR TPC detector.
	Figure is taken from \cite{Anderson:2003ur}.
	}
	\label{fig:tpc}
\end{figure}

A diagram of the TPC is shown in figure~\ref{fig:tpc}.
The TPC is a particle detector which consists of a gas-filled cylindrical chamber with multi-wire proportional chambers on the endcaps at each side of the cylinder.
There is a high voltage central membrane disc which divides the cylindrical chamber into two halves.
The central membrane, together with the ``Outer Field Cage'' and ``Inner Field Cage'' as shown in figure~\ref{fig:tpc}, 
provide a nearly uniform electric field along the $z$ direction parallel to the beam pipe and magnetic field, pointing from the endcaps to the center.

\begin{figure}[thb]
	\begin{center}
		\subfigure[Transverse view]{\label{fig:kin-a} \includegraphics[width=0.6\textwidth]{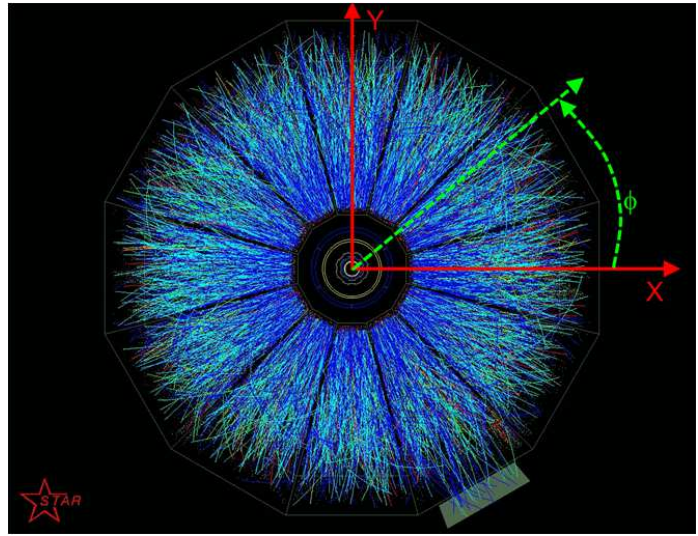}}
		\subfigure[Side view]{\label{fig:kin-b} \includegraphics[width=0.6\textwidth]{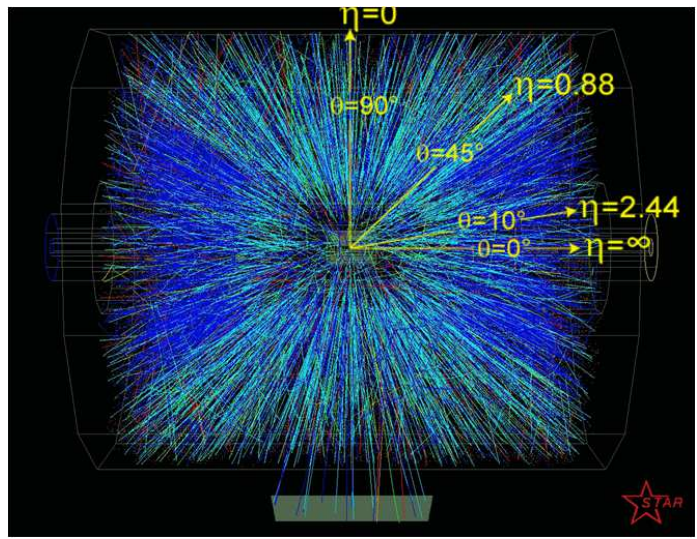}}
	\end{center}
	\caption[Event example]{
	An example event of central Au+Au collisions in transverse view~(a) and side view~(b).
	The definition of azimuthal angle $\phi$ and pseudo-rapidity $\eta$ are shown in panel (a) and (b) respectively.
	}
	\label{fig:kin}
\end{figure}

When a charged particle is generated in the collision, it traverses the TPC volume, ionizes gas atoms every few tenths of a millimeter along its trajectory, and leaves a cluster of electrons behind.
The electron clusters then drift within the electric field toward the sectors of the endcaps.
Each electron cluster will be accelerated by the electric field around the anode and cause a localized cascade of ionization,
which is collected on the high voltage wire and results in an electric current.
The position of the hit point is then recorded in the $r$ and $\phi$ dimension, 
and the amplitude of the current is proportional to the energy of the detected particle.
The endcaps are made by 24 identical sectors, 12 on each side.
Each sector covers about $\pi / 6$ in azimuthal angle, and totally full coverage of $2\pi$.
The $z$ position is obtained by measuring the drift time from the collision to the time when the electron cluster is recorded at the endcaps because the drift velocity can be precisely measured beforehand.
One can then reconstruct the tracks by fitting the 3-dimensional hit points collected by the TPC.
Track in the TPC usually has a helix shape, and the maximum hit points at the endcaps can be as many as 45 hits.
The transverse momentum $p_T$ and the charge sign of a particle can be calculated from the curvature of the trajectory with the applied magnetic field.
Figure~\ref{fig:kin} shows the side view ($xz$ plane), and front view ($xy$ plane) of an Au+Au collision event.

In this thesis, the particle information is taken mostly from the TPC.
The most frequently used kinematic variables $\phi$ and $\eta$ are shown in figure~\ref{fig:kin}.
The azimuthal angle $\phi$ (figure~\ref{fig:kin-a}) is the angle of a particle transverse momentum $p_T$ relative to $x$ axis in STAR coordinates.
Pseudo-rapidity $\eta$ describes the angle relative to the beam direction, i.e. $z$ direction.
It is defined as
\begin{align}
	\eta &= -\ln \left[ \tan \left( {\theta \over 2}\right)\right] \\
	& = {1 \over 2} \ln \left( { \left| \vec{p} \right| + p_L \over \left| \vec{p}\right| - p_L }\right),
	\label{eq:eta}
\end{align}
where $\theta$ is the angle of the particle momentum $\vec{p}$ relative to the beam axis, and $p_L$ denotes the longitudinal component of the particle momentum $\vec{p}$.
In the relativistic limit when particle's speed is close to the speed of light, 
or the particle mass is small compared to its total energy, 
the rest mass could be ignored to good approximation. 
Then pseudo-rapidity is numerically close to rapidity which is defined as
\begin{equation}
	y = {1 \over 2} \ln \left( { E+ p_L \over E- p_L}\right),
	\label{eq:rapidity}
\end{equation}
where $E$ is the energy of the particle.
The rapidity definition is used in theoretical calculations, while it requires two parameters $E$ and $\vec{p}$ of a particle to be measured at the same time, which is inconvenient experimentally.
However, the pseudo-rapidity requires only one measurement, $\vec{p}$, and it is close to rapidity for light particles (i.e. pions, electrons).
The STAR detector has a large and uniform pseudo-rapidity coverage of $-1 < \eta < 1$.

When an event is recorded, the tracks reconstructed from the TPC hits are called global tracks.
By extrapolating the tracks back to the center of the detector, the original collision vertex can then be fitted from all the trajectories very precisely.
The position of the collision vertex is noted as $\vec{v}$ in STAR coordinates.
We can then do the tracking reconstruction again with all the TPC hits plus the additional collision vertex.
The tracks reconstructed with collision vertex are called primary tracks.
The advantage of primary tracks is that particles created from the collision vertex will get a better resolution since the collision vertex has a very good resolution.
However, the tracks from secondary decay will have worse resolution because the collision vertex might be off from the track helix.
Thus, we introduce the Distance of Closest Approach (DCA).
It is defined as the closest distance from the collision vertex to a track helix.
By cutting on DCA, one can manually select the tracks preferentially from the collision vertex or the secondary decay vertex.

\subsection{Zero Degree Calorimeters}

The Zero Degree Calorimeter (ZDC) detectors are located at $\pm 18~m$ away from the geometry center of the STAR detector as shown in red in figure~\ref{fig:star} \cite{Adler:2003sp}.
They measure the energy deposition of neutrons, which are associated with the spectator matter, in the three tungsten plates on each side of the ZDC.
The ZDC has been used for beam monitoring and event triggering.

There was an upgrade in 2003 of the ZDC by adding Shower Maximum Detectors (SMD), which gives the ZDC-SMD the capability of recording the shower profile of neutron clusters in the transverse plane ($xy$ plane) of the collisions.
The SMD information contains 7-slate (vertical) by 8-slate (horizontal) readouts from the ADC photomultiplier tubes which connect to the SMD scintillators.
The raw readouts are corrected for the background by subtracting the pedestal.
The readouts also have to be adjusted for the distortion of the electronics by applying the gain correction, which is from the cosmic ray calibration.
Finally, the vertical and horizontal signals can present a well defined spectator position in the $xy$ plane, which can be used for the direct event-plane reconstruction.

\subsection{Event Triggering}
\label{trig}

The collision rate at RHIC 2004 is about 10M Hz, which means there are 10M collision events per second happening in the STAR detector.
However, the STAR TPC is a slow detector because the electrons need time to drift to the endcaps to be recorded,
and the DAQ system is also limited by the bandwidth.
At 2004, the DAQ operated at rates about 100 Hz of Au+Au 200 GeV collisions.

In order to reduce the recorded collision events, trigger detectors are needed to select 100 events out of 10M events per second, based on our interests \cite{Bieser:2002ah}.
The fast detectors are used as trigger detectors. 
They are the Zero Degree Calorimeters (ZDC), 
the Central Trigger Barrel (CTB) and the Beam-Beam Counters (BBC).
The ZDC is introduced in previous section.
The CTB detector is located between the outer cage of the TPC and the ToF detector, which measures the charge multiplicity in the same pseudo-rapidity range of $|\eta|<1$ as the TPC with full azimuthal coverage.
The BBC detector is located outside of the pole tip magnets as shown in figure \ref{fig:star}.
It measures the multiplicity in forward region and provides vertex location information of the collision.

The Au+Au minimum bias collisions, the least biased data sample, are triggered by both ZDCs (east- and west-side) and the CTB.
The event is cut on the ZDC coincidence rate above certain value from the east and west ZDC, and the primary vertex from the ZDC signals.
And there is also a cut on the CTB multiplicity to reject the non-hadronic events.
However, some of the low multiplicity hadronic events are also rejected, thus there are bias on the low multiplicity events at the most peripheral collisions below 80\% centrality.
For this reason, our analysis is focusing on the data above 80\% centrality only.

The d+Au minimum bias collisions used in this analysis are triggered on ZDC detector from the east side only, which is where the Au beam is from.

We also use the most central Au+Au collisions in the analysis which are triggered on ZDC detectors and the CTBs.
It requires the ZDC detector with a high coincidence rate, and the CTBs with a large multiplicity matching the most central minimum bias collisions, which eventually selects the most central collisions about 12\% centrality.

\subsection{Centrality Definition}
\label{CENT}

The collision initial condition is of great importance in heavy ion nucleus-nucleus collisions.
The impact parameter $b$ is defined as the distance between the geometric center of the two colliding nuclei in the transverse plane.
The nucleons are then undergo interactions with each other.
The number of participants referred to as $N_{part}$ is defined as the number of nucleons that participate in at least one inelastic nucleon-nucleon reactions.
And the number of binary collisions is defined as the number of such inelastic nucleon-nucleon reactions, usually referred to as $N_{coll}$.
However, they cannot be measured directly from the experiment, and have to be deduced from other experimental observables and combined with Monte-Carlo simulations.
The simulation is done in the geometry model with experimental measurement of nucleon-nucleon cross sections and considering the multiple scattering of nucleons in the heavy ion collisions.
Such Monte-Carlo techniques is generally referred to as Glauber model~\cite{Miller:2007ri}.

\begin{figure}[thb]
	\begin{center}
		\psfig{figure=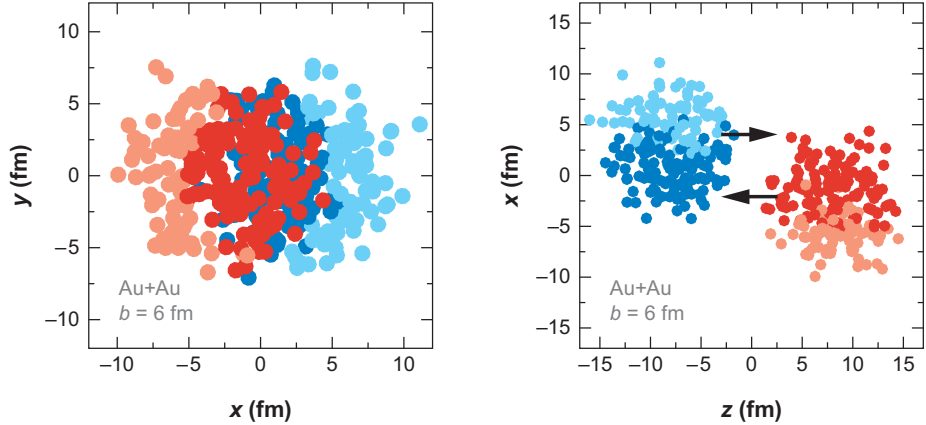,width=0.8\textwidth}
	\end{center}
	\caption[Glauber Monte-Carlo event]{
	A Glauber Monte-Carlo event (Au+Au at $\sqrt{s_{NN}}$ = 200 GeV with impact parameter $b=6$ fm) viewed in transverse plane (left) and along the beam axis (right).
	Figure is taken from \cite{Miller:2007ri}.
	}
	\label{fig:cent1}
\end{figure}

\begin{figure}[thb]
	\begin{center}
		\psfig{figure=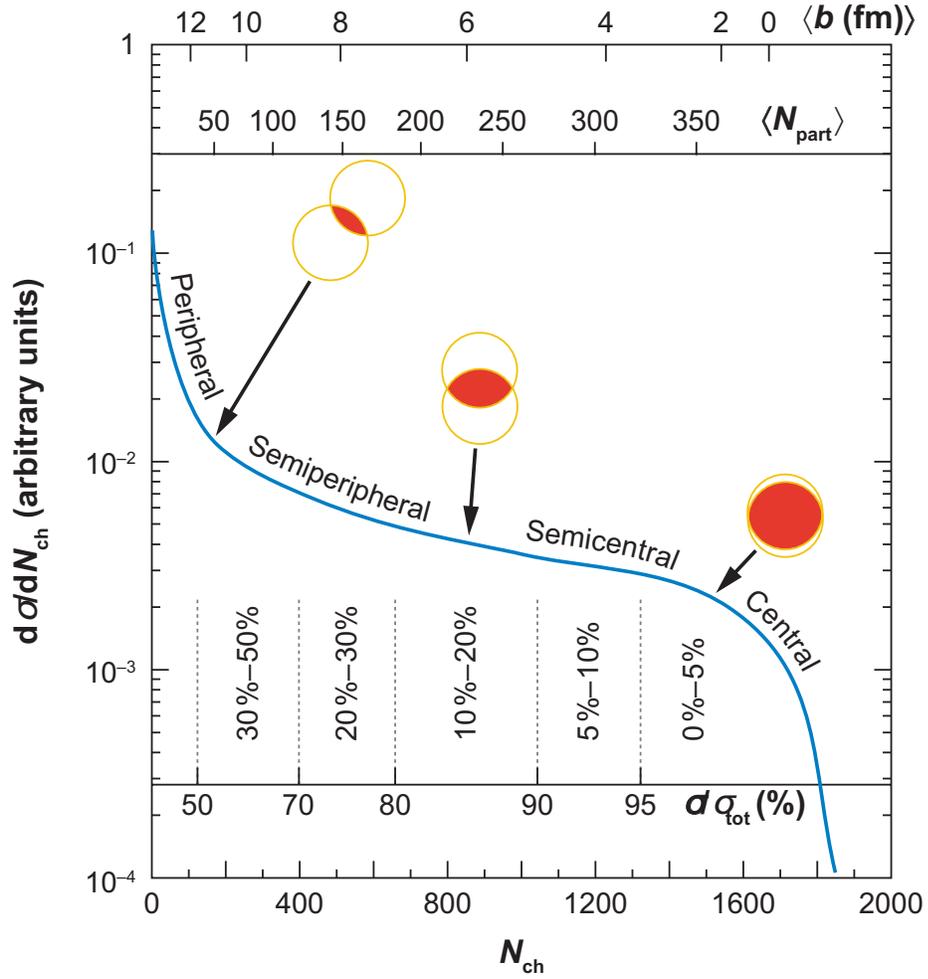,width=0.8\textwidth}
	\end{center}
	\caption[Number of participants]{
	An illustrated example of the total final state inclusive charged particle multiplicity $N_{ch}$ distribution with Glauber-calculated quantities ($b$, $N_{part}$).
	The plotted distribution and various values are illustrative and not actual measurements.
	Figure is taken from \cite{Miller:2007ri}.
	}
	\label{fig:cent2}
\end{figure}

Figure~\ref{fig:cent1} shows a Glauber Monti-Carlo Au+Au collision event at center of mass $\sqrt{s_{NN}}$ = 200 GeV.
The impact parameter $b = 6$ fm. The dark circles represent the participating nucleons.
It is obvious that the smaller the impact parameter, the more participating nucleons and binary collisions in an event,
hence more generated particles detected by the detector.
Figure~\ref{fig:cent2} shows the illustration plot of the event multiplicity within $|\eta|<1$ distribution corresponding to the impact parameter $b$ and the number of participants $N_{part}$.
Then experimentally, one can relate the final state multiplicity to the $N_{part}$ and $b$.

In the STAR experiment, the efficiency uncorrected charged particle multiplicity in one unit of the pseudo-rapidity $\mathrm{d} N_{ch}/ \mathrm{d} \eta$ is used instead of $N_{ch}$ to deduce the $N_{part}$ and $b$ parameters, which is called reference multiplicity $N_{RefMult}$ \cite{Adams:2003xp,:2008ez}.
The $N_{RefMult}$ is the number of charged particles recorded by the main TPC at mid-rapidity range of $-0.5<\eta<0.5$.
With the $N_{RefMult}$ distribution, we can cut on certain fraction of the events to correspond to the impact parameter $b$ and $N_{part}$.
Typically, we cut on 0-5\%, 5-10\%, 10-20\%, 20-30\%, 30-40\%, 40-50\%, 50-60\%, 60-70\% and 70-80\% from most central to most peripheral collisions.

%% file: DataAna.tex
%
%
%

\chapter{DATA ANALYSIS}

In this chapter, we first give the definition of our charge multiplicity asymmetry variables and their correlations.
We then introduce the data and quality cuts used in the analysis,
followed by related analysis procedures and corrections.
At the end, we check for consistency and study the systematic uncertainties.

\section{Charge Multiplicity Asymmetry Observables}

\subsection{Charge Multiplicity Asymmetries}

\begin{figure}[h]
	\begin{center}
		\psfig{figure=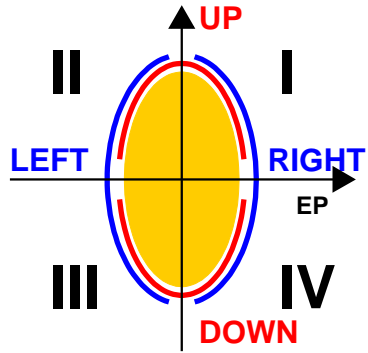,width=0.5\textwidth}
	\end{center}
	\caption[Charge multiplicity asymmetry definition]{Schematic view of the overlap region on transverse plane in a typical non-central heavy ion collision.
	The four quadrants are labeled. \en{UP=I+II, DOWN=III+IV, LEFT=II+III, RIGHT=I+IV}.
	}
	\label{fig:def}
\end{figure}

In heavy ion collisions, the overlap area of the collision can be illustrated as an elliptical shape on the transverse plane as shown in figure \ref{fig:chargesep}.
The event anisotropy can then be used to estimate the reaction-plane (RP) direction for a given event.
The estimation is defined as event-plane (EP).
Once the event-plane is determined, 
we can then separate the overlap area into hemispheres in the collision transverse plane ($xy$ plane in figure \ref{fig:chargesep}).
As shown in figure~\ref{fig:def}, UP- and DOWN-hemispheres are separated by the EP.
LEFT- and RIGHT-hemispheres are separated by the plane perpendicular to the EP.
Particle multiplicity asymmetries are then defined, on event-by-event basis, as
\begin{align}
&A_{+,UD} = (N_{+,U}-N_{+,D})/(N_{+,U}+N_{+,D}),\nonumber\\
&A_{-,UD} = (N_{-,U}-N_{-,D})/(N_{-,U}+N_{-,D}),\nonumber\\
&A_{+,LR} = (N_{+,L}-N_{+,R})/(N_{+,L}+N_{+,R}),\nonumber\\
&A_{-,LR} = (N_{-,L}-N_{-,R})/(N_{-,L}+N_{-,R}).\label{eq:asymdef}
\end{align}
Here $N_{+,U}$, $N_{+,D}$, $N_{+,L}$ and $N_{+,R}$ are positively charged particle multiplicities in the UP (quadrants I and II), DOWN (III and IV),
LEFT (II and III), and RIGHT (I and IV) hemispheres as in figure~\ref{fig:def}, respectively. 
Those of negatively charge particle multiplicities are represented by $N_{-,U}$, $N_{-,D}$, $N_{-,L}$, and $N_{-,R}$.

\subsection{Charge Multiplicity Asymmetry Correlations}

As introduced in section~\ref{CME}, the non-zero topological charges ($Q_\text{w}$), gauge configurations, 
change the quark chirality which causes the asymmetry in the number of left- and right-handed quarks,
$N_{LH}-N_{RH} = 2 N_{f} Q_\text{w}$, where $N_{f}$ is the number of light quark flavors.
Since the sign of the topological charges ($Q_\text{w}$) is random,
the chirality changes are also random from domain to domain
in a single event and from event to event.
Moreover, the event-plane reconstructed from event anisotropy does not distinguish up from down, neither left from right.
This causes the charge separation direction random from event to event.
Therefore, only the magnitude of the up-down~($UD$) multiplicity asymmetries ($A_{+,UD}$ and $A_{-,UD}$) 
gets larger due to CME/LPV, while the sign is still random.
As a result, the average asymmetries of $UD$ remain zero over all events,
but the distributions of $A_{\pm,UD}$ (we use $A_{\pm,UD}$ to collectively denote $A_{+,UD}$ and $A_{-,UD}$) would be wider comparing to those of $A_{\pm,LR}$ (i.e. $A_{+,LR}$ and $A_{-,LR}$), to which CME/LPV doesn't contribute.
In other words, the variances $\langle A_{\pm,UD}^2\rangle$ should be larger than the respective $\langle A_{\pm,LR}^2\rangle$,
where $\langle \ldots \rangle$ denotes the average over the event sample.
Hence, we study the variance of the charge multiplicity asymmetries,
which provide the dynamic informations about the particle correlations with the same charge signs.

We also study the covariances ($\langle A_+A_-\rangle_{UD}$ and $\langle A_+A_- \rangle_{LR}$) of the charge multiplicity asymmetries as the opposite-sign correlations.
The covariance measurement $\langle A_+A_-\rangle$ is a more traditional measurement of parity violation.
The positive charged particle multiplicity asymmetry $A_{+,UD}$ represents, on average, the ``parity-axis'' of the orbital angular momentum and the topological charge sign;
the covariance $\langle A_+A_- \rangle_{UD}$ is then a measurement of $A_{-,UD}$ with respect to this ``parity-axis''.
For local parity violation, the positively and negatively charged particles tend to move in opposite direction across the event-plane.
This will generate additional negative correlation between the asymmetries in up-down direction comparing to left-right direction, in other words, $\langle A_+A_- \rangle_{UD} < \langle A_+A_- \rangle_{LR}$.

The charge multiplicity asymmetry correlations themselves are, however, parity even, 
and subject to physics backgrounds similar to those in the charge correlator measurement. 
The backgrounds may be assessed by the left-right~($LR$) asymmetry correlations ($\langle A^2_{LR} \rangle$ and $\langle A_+A_- \rangle_{LR}$),
to which CME/LPV does not contribute.
This is because, CME/LPV is an effect along the system angular momentum direction, which is perpendicular to the event-plane direction.
In other words, the left-right~($LR$) observables can serve as our null-reference.
However, as we will show in the next chapter, the physics backgrounds are likely different in $UD$ and $LR$ measurements.
This introduces complications in the interpretations of the measured $UD$ and $LR$ correlation differences.
Details will be discussed in section \ref{discuss}.

\subsection{Dynamical Correlation and Charge Separation}

In this analysis, we compute the charge multiplicity asymmetries $A_{\pm,UD}$ and $A_{\pm,LR}$ on the event-by-event basis.
We obtain their variances (widths of their distributions) $\langle A_{\pm,UD}^2\rangle$ and $\langle A_{\pm,LR}^2\rangle$, and their covariances $\langle A_+A_- \rangle_{UD}$ and $\langle A_+A_- \rangle_{LR}$.
The variances are always positive because they are the squares of a real number.
They subject to statistical fluctuation and detector non-uniformity acceptance effects.
We have to subtract those effects to get the dynamical correlations, the physics we are interested in.

The dynamical variances are defined as
\begin{align}
	\delta\langle A_{\pm,UD}^2 \rangle = \langle A_{\pm,UD}^2 \rangle - \langle A_{\pm,UD,stat+det}^2 \rangle,\nonumber\\
	\delta\langle A_{\pm,LR}^2 \rangle = \langle A_{\pm,LR}^2 \rangle - \langle A_{\pm,LR,stat+det}^2 \rangle,\label{eq:dynasym1}
\end{align}
where subscript ``stat+det'' stands for statistical fluctuation plus detector effects.
As will be presented later,
the dynamical asymmetry correlations of positively charged particles ($\delta\langle A_+^2 \rangle$) and negatively charged particles ($\delta\langle A_- \rangle^2$) are consistent.
We therefore report the average dynamical variances (with prefix ``$\delta$'') as
\begin{align}
	\delta \langle A_{UD}^2 \rangle = \left(\delta \langle A_{+,UD}^2 \rangle + \delta \langle A_{-,UD}^2 \rangle\right)/2,\nonumber\\
	\delta \langle A_{LR}^2 \rangle = \left(\delta \langle A_{+,LR}^2 \rangle + \delta \langle A_{-,LR}^2 \rangle \right)/2.\label{eq:dynasym}
\end{align}

We also analyze covariances $\langle A_+A_- \rangle_{UD}$ and $\langle A_+A_- \rangle_{LR}$.
As shown later, because the positively and negatively charged particles are not statistically correlated, the statistical fluctuation and detector effects are consistent with zero for the covariances.
We thus do not subtract them from the covariances.

We report the differences between the $UD$ and $LR$ measurements, which may be directly sensitive to the CME/LPV.
Namely (with prefix ``$\Delta$''),
\begin{align}
	\Delta\langle A^2 \rangle &\equiv \delta\langle A_{\pm,UD}^2 \rangle - \delta \langle A_{\pm,LR}^2 \rangle \approx \langle A_{\pm,UD}^2 \rangle - \langle A_{\pm,LR}^2 \rangle,\label{eq:DA2}\\
	\Delta\langle A_+A_- \rangle &\equiv \langle A_+A_- \rangle_{UD} - \langle A_+A_- \rangle_{LR},\label{eq:DAA}
\end{align}
where, ideally, the statistical fluctuation, detector effects and other backgrounds that are not related to the reaction plane
cancel in the last step of subtraction in equation \ref{eq:DA2}.
However, as we will show later, the detector non-uniformity causes the statistical fluctuation and detector effects $\langle A_{\pm,UD,stat+det}^{2} \rangle$ and $\langle A_{\pm,LR,stat+det}^{2} \rangle$ not identical 
for different pseudo-rapidity regions.
Thus the background from $UD$ and $LR$ directions will not cancel completely.
To be precise, we keep track of all the statistical fluctuations and detector effects from equation \ref{eq:dynasym1} 
for each variances and subtract them accordingly to get the dynamical correlations.
And this procedure is necessary for the wedge size and wedge location analysis introduced in the following section,
because the $stat+det$ effects are largely varied with the wedge size and wedge location being studied, see next section.

\subsection{Wedge Size and Wedge Location}
\label{wedge}

\begin{figure}[thb]
	\begin{center}
		\subfigure[Configuration for the wedge size dependence]{\label{fig:wedge-a}\includegraphics[width=0.4\textwidth]{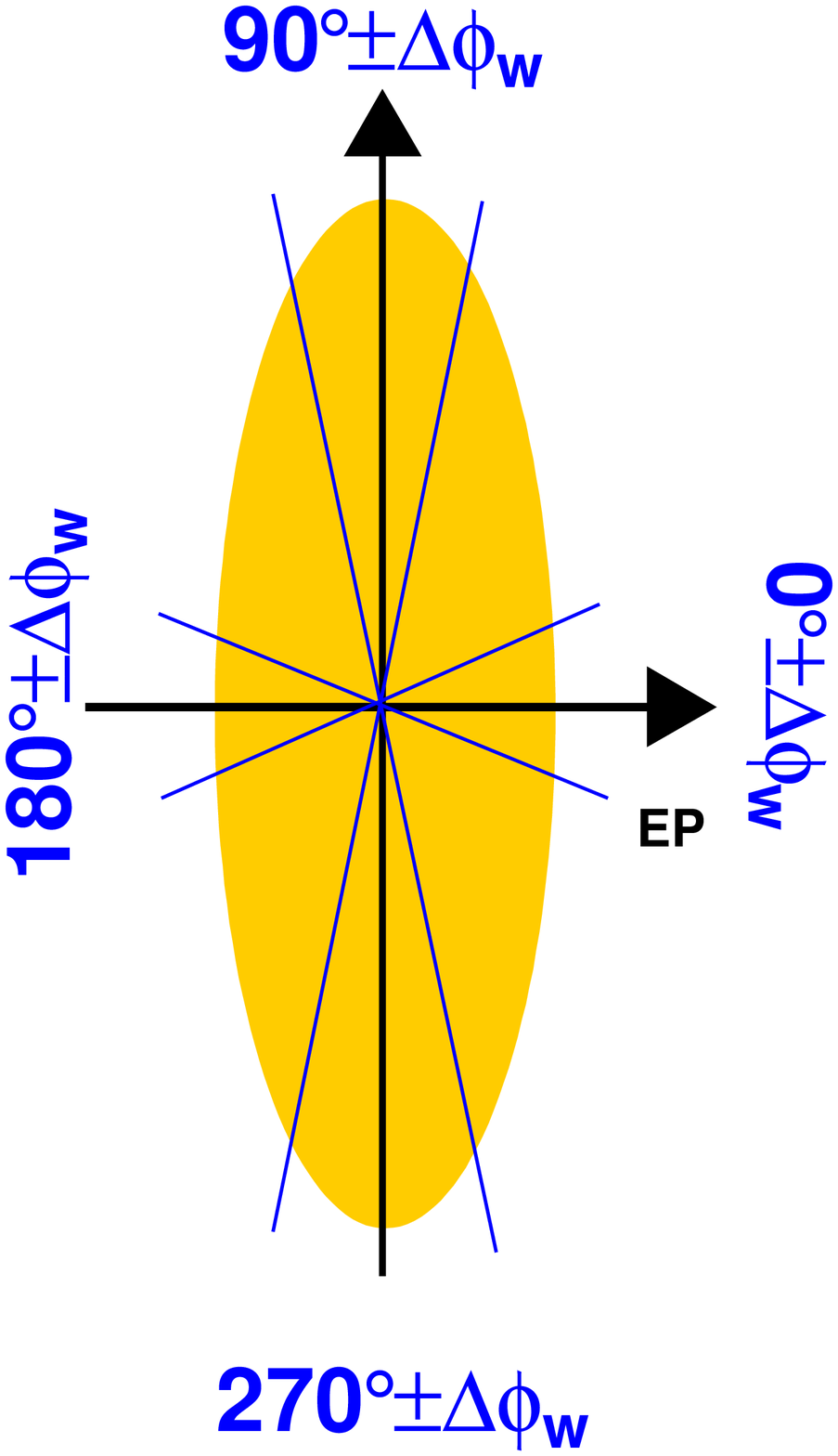}}
		\subfigure[Configuration for the wedge location dependence]{\label{fig:wedge-b}\includegraphics[width=0.4\textwidth]{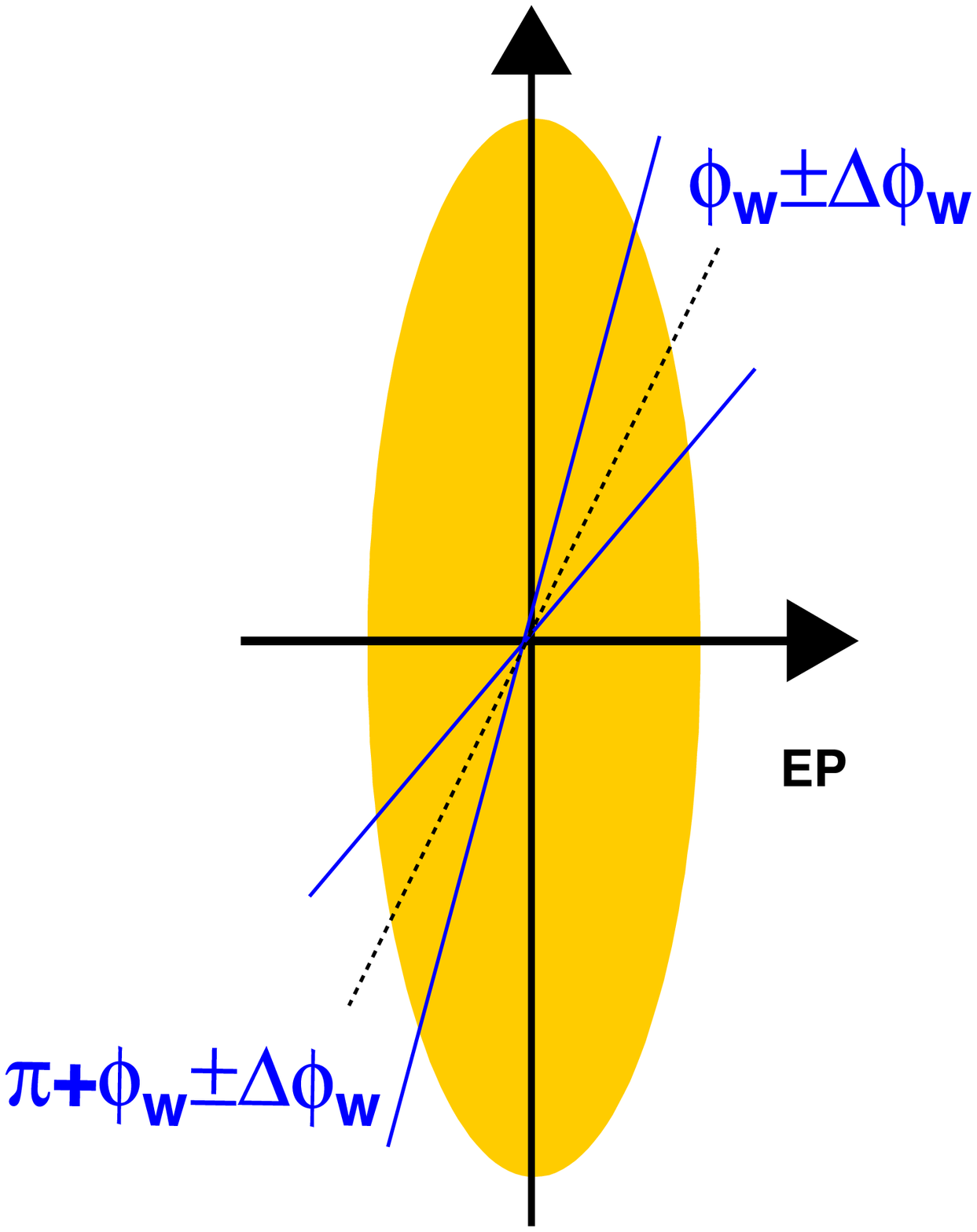}}
	\end{center}
	\caption[Charge asymmetry wedge definition]{Schematic view of collision overlap area on transverse plane, and the azimuthal back-to-back wedge regions, where the charge asymmetries can be calculated.}
	\label{fig:wedge}
\end{figure}

We can not only investigate the charge multiplicity asymmetry between hemispheres, 
but also between different wedge sizes and different wedge axis locations.
As shown in figure~\ref{fig:wedge}, we can study the charge multiplicity asymmetries with their correlations for any given axis with any given opening angle $2\Delta\phi_\text{w}$.
For example in figure~\ref{fig:wedge-a}, if we want to study the out-of-plane asymmetries, 
we count charged particle multiplicities within the azimuthal angle relative to the EP between $90^\circ-\Delta\phi_\text{w}$ and $90^\circ+\Delta\phi_\text{w}$ as $N_{\pm,U}$, and that between $270^\circ-\Delta\phi_\text{w}$ and $270^\circ+\Delta\phi_\text{w}$ as $N_{\pm,D}$.
Similarly for in-plane asymmetries, we count particle multiplicities between $0^\circ-\Delta\phi_\text{w}$ and $0^\circ+\Delta\phi_\text{w}$ as $N_{\pm,R}$, and that between $180^\circ-\Delta\phi_\text{w}$ and $180^\circ+\Delta\phi_\text{w}$ as $N_{\pm,L}$.
Then the asymmetries are calculated in the same way as those defined in equation \ref{eq:asymdef}.

Furthermore, figure~\ref{fig:wedge-b} shows the schematic configuration of the back-to-back wedges with axis centered at $\phi_\text{w}$ and wedge size of $2\Delta\phi_\text{w}$.
The asymmetries are calculated between $\phi_\text{w}\pm\Delta\phi_\text{w}$ and that between $(\pi+\phi_\text{w})\pm\Delta\phi_\text{w}$.
In such configuration, we can vary the asymmetry axis $\phi_\text{w}$ to study the progressive evolution of the asymmetries and their correlations from in-plane to out-of-plane.

We use $A_{\pm,\phi_\text{w}\pm\Delta\phi_\text{w}}$ to stand for the multiplicity asymmetries of positively and negatively charged particles
with wedge axis located at $\phi_\text{w}$ and wedge size of $2\Delta\phi_\text{w}$.
Particularly, the $A_{\pm,UD}$ and $A_{\pm,LR}$ we described earlier are equivalent to $A_{\pm,90^\circ\pm90^\circ}$ and $A_{\pm,0^\circ\pm90^\circ}$ respectively, and we refer this configuration as hemisphere asymmetry.

We have to be cautious when using non-hemispherical ranges. 
For the variances, it is obvious that the statistical fluctuation and detector effects do not cancel 
between the asymmetry correlations in $\phi_\text{w}$ ($UD$) and $\phi_{\text{w}+90^\circ}$ ($LR$) wedges 
when we take the difference between $UD$ and $LR$ as in equation \ref{eq:DA2}.
Due to the anisotropy flow, we expect more particles emitted in-plane, and fewer particles emitted out-of-plane.
Thus, the statistical fluctuation is expected to be different for non-hemisphere wedges in non-central collisions,
because it is approximately proportional to ``1/N'' where N is the multiplicity within the wedge.
\red{
On top of that, the electronic inefficiency will cause difference in the observables.
The detector inefficient sector could fall into either $UD$ or $LR$ asymmetries, 
while in the hemisphere asymmetries which covers the full azimuthal angle, 
the inefficient sector will be guaranteed to fall into both $UD$ and $LR$ asymmetries.
When calculating the $UD-LR$ difference, the detector effects introduced ``dynamical'' correlation will not cancel in the subtraction in the non-hemisphere asymmetry correlations.
These two factors combined together are non-trivial and may vary significantly for wedges with different sizes and locations.
And this is the other reason we mentioned in last section that we want to carry the statistical fluctuation and detector effects,
denoted as $\langle A_{\pm,\phi_\text{w}\pm\Delta\phi_\text{w},stat+det} \rangle$, 
all the way through the analysis and subtract them respectively according to the variances.
}
The statistical fluctuation and detector effects can be estimated in the same way we describe in section \ref{stat}.
Thus the dynamical correlation is obtained as
\begin{equation}
  \delta\langle A_{\pm,\phi_\text{w}\pm\Delta\phi_\text{w}}^2 \rangle = \langle A_{\pm,\phi_\text{w}\pm\Delta\phi_\text{w}}^2 \rangle - \langle A_{\pm,\phi_\text{w}\pm\Delta\phi_\text{w}, stat+det}^2 \rangle.
  \label{eq:wedgeA2}
\end{equation}
We report the average variance of positive and negative charged particles
\begin{equation}
	\delta\langle A_{\phi_\text{w}\pm \Delta\phi_\text{w}}^2 \rangle = \left(\delta\langle A_{+,\phi_\text{w}\pm \Delta\phi_\text{w}}^2 \rangle + \delta\langle A_{-,\phi_\text{w}\pm \Delta\phi_\text{w}}^2 \rangle \right)/2,
	\label{eq:wedgedynA2}
\end{equation}
and the difference between the correlations with respect to the \red{perpendicular} asymmetry axises.
The differences are taken between the dynamical variances after the subtracting statistical fluctuation and detector effects of the variances.

For the covariances, they are not affected by such statistical fluctuation and detector effects.
We thus report the raw correlations.

\section{Data Sets}

The data we present in this analysis were taken by STAR experiment at Brookhaven National Laboratory.
We use minimum-bias and ZDC central triggered Au+Au collision data with center of mass energy $\sqrt{s_{NN}}=200$ GeV per nucleon pair.
We focus on the data taken between year 2004 and 2005 (RUN IV) with 24 million events in total.
We also use minimum-bias triggered 200 GeV Au+Au data taken between year 2007 and 2008 (RUN VII) for ZDC-SMD event-plane study with total 56 million events.
As a reference, we use minimum-bias triggered d+Au data with center of mass energy $\sqrt{s_{NN}}=200$ GeV.
Total events for d+Au are 9 million.
The results are presented as a function of the number of participants, $N_{part}$.

The centrality definition is based on a Glauber Model simulation as introduced in section \ref{CENT}.
We summarize the $N_{RefMult}$ cut in table \ref{tab:cent} for RUN IV and RUN VII.

\begin{table}[thb]
	\centering
	\caption[Centrality definition]{Centrality definition of $\sqrt{s_{NN}}$ = 200 GeV Au+Au collisions and the number of participants $N_{part}$ \cite{:2008ez,Adams:2004cb} in RUN IV and RUN VII and the average event-by-event $v_{2,p_T < 2\text{GeV/}c}^{obs}$ of RUN IV.}
	\begin{tabular}{c c c x{64pt} x{64pt} c}
		\hline
		\hline
		\multirow{2}{*}{\#} & \multirow{2}{*}{Centrality} & \multirow{2}{*}{$N_{part}$} & \multicolumn{2}{c}{Lower $N_{RefMult}$ cut ($\ge$)} & \multirow{2}{*}{$\langle v_{2,p_T < 2 \text{GeV/}c}^{obs} \rangle$} \\
		\cline{4-5}
		& & & RUN IV & RUN VII & \\
		\hline
		9& $0-5\%$ & $352.4^{+3.4}_{-4.0}$ & 520	& 485 & 0.011\\
		\hline
		8& $5-10\%$ & $299.3^{+6.6}_{-6.7}$ & 441	& 399 & 0.020 \\
		\hline 
		7& $10-20\%$ & $234.6^{+8.3}_{-9.3}$ & 319 	& 269 & 0.032 \\
		\hline 
		6& $20-30\%$ & $166.7^{+9.0}_{-10.6}$ & 222 	& 178 & 0.043 \\
		\hline 
		5& $30-40\%$ & $115.5^{+8.7}_{-11.2}$ & 150 	& 114 & 0.047 \\
		\hline 
		4& $40-50\%$ & $76.6^{+8.5}_{-10.4}$ & 96 	& 69  & 0.044 \\
		\hline 
		3& $50-60\%$ & $47.8^{+7.6}_{-9.5}$ & 57 	& 39  & 0.036 \\
		\hline 
		2& $60-70\%$ & $27.4^{+5.5}_{-7.5}$ & 31 	& 21  & 0.025 \\
		\hline 
		1& $70-80\%$ & $14.1^{+3.6}_{-5.0}$ & 14 	& 10  & 0.017 \\
		\hline 
		\hline 
	\end{tabular}
	\label{tab:cent}
\end{table}

\section{Quality Cuts}

In order to get quality event and track information, we apply STAR standard cuts to ensure the events and tracks are with good precision,
yet least biased by the detector imperfection.
The cuts used here are described in previous chapter.

\subsection{Event Selection}

Event wise, a minimum-bias trigger (section \ref{trig}) was used at data taking.
Furthermore, events are required to have the collision vertices within 30 cm from the STAR detector center along the beam line ($|v_z| <$ 30 cm),
in order to ensure the events are not biased toward one side of the TPC.

To reject collisions from beam halo and beam pipe, or collisions from possible secondary vertices,
we cut on the radius of the vertex position to the beam pipe center with $v_r\equiv\sqrt{v_x^2+v_y^2} <$ 2 cm.

During data taking, the TPC magnet operated in two polarity configurations, full field (FF) and reverse full field (RFF),
with magnetic field strength of 0.5 T in beam line direction.
We then use different magnetic polarity to assess the detector effects which is effectively the same as
switching positive and negative charges.

The reference multiplicity of an event is also required to be less than $1000$ in order to reject pile-up events.

\subsection{Track Selection}

Track wise, tracks are required to be within $\pm 1$ in pseudo rapidity ($\lvert \eta \rvert < 1$), 
where the TPC has the best tracking performance.
Each track is required at least 20 hit points (out of 45 at most) used in track reconstruction from the TPC.
Also, the ratio of the number of hit points in track reconstruction to the most possible number of hit points is required to be larger than 51\%, 
which will eliminate partial track reconstructed from a single track.
We also require a lower transverse momentum cut with $p_T>0.15$ GeV/$c$, which is the lower limit of STAR detector.
We require each track with $\text{DCA}<$ 2 cm to ensure the particle is from the collision vertex.

We vary the event cuts and track cuts to study the systematic uncertainties,
which is discussed in section \ref{systematic}.

We summarize the cuts in the table \ref{tab:cuts} and the number of events after the cuts for each dataset in table \ref{tab:dataset}.

\begin{table}[thb]
	\centering
	\caption{The standard event and track selection cut.}
	\begin{tabular}{c c}
		\hline
		\hline
		\multicolumn{2}{c}{Event selection cut} \\
		\hline
		Vertex z position $v_{z}$ 	& $|v_z| <$ 30 cm	\\
		Vertex radius cut $v_{r} = \sqrt{v_x^2+v_y^2}$ 	& $v_r <$ 2 cm	\\
		Reference multiplicity $N_{RefMult}$ 	& $N_{RefMult} <$ 1000	\\
		\hline
		\hline
		\multicolumn{2}{c}{Track selection cut} \\
		\hline
		Pseudo-rapidity $\eta$ 	& $|\eta| <$ 1	\\
		Number of hits $nfit$ 	& $nfit >$ 20	\\
		Ratio of $nfit$ to maximum fit points	& $rfit >$ 0.51 \\
		Lower transverse momentum $p_T$		& $p_T >$ 0.15 GeV/$c$	\\
		DCA 	&	$\text{DCA} <$ 2 cm	\\
		\hline
		\hline
	\end{tabular}
	\label{tab:cuts}
\end{table}

\begin{table}[thb]
	\centering
	\caption{Dataset and statistics.}
	\begin{tabular}{c c}
		\hline
		\hline
		Dataset	&	Number of events after cuts\\
		\hline
		RUN IV Au+Au 200 GeV min-bias	&	22.5M \\
		RUN IV Au+Au 200 GeV top 2\% central	&	5.5M \\
		RUN IV d+Au 200 GeV min-bias	&	8.9M \\
		RUN VII Au+Au 200 GeV min-bias	&	56.4M \\
		\hline
		\hline
	\end{tabular}
	\label{tab:dataset}
\end{table}

\section{Detector Efficiency Correction}
\label{det}

STAR detector has full azimuthal coverage in $2\pi$.
For a perfect detector, the particle distribution accumulating over a large amount of events should be uniform with respect to the azimuthal angle.
But in reality, the TPC is made of 12 separated sectors on each side of the endcaps.
The sector boundaries and deficit sectors due to electronics failure will reduce tracking efficiency,
and create ``dynamical'' correlation on the event-by-event basis.
To reduce the ``dynamical'' correlation introduced by non perfect detector, we will have to correct the single track efficiency to flatten the azimuthal angle $\phi$ distribution.

\begin{figure}[thb]
	\begin{center}
		\subfigure[Uncorrected $\phi$ distribution]{\label{fig:acc-a}\includegraphics[width=0.6\textwidth]{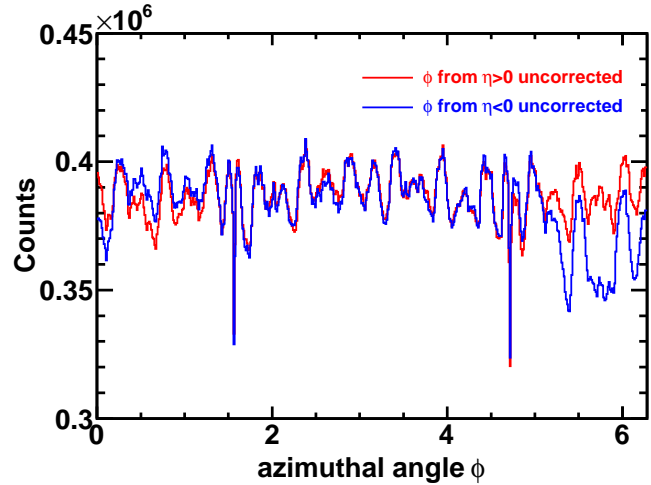}}
		\subfigure[Corrected $\phi$ distribution]{\label{fig:acc-b}\includegraphics[width=0.6\textwidth]{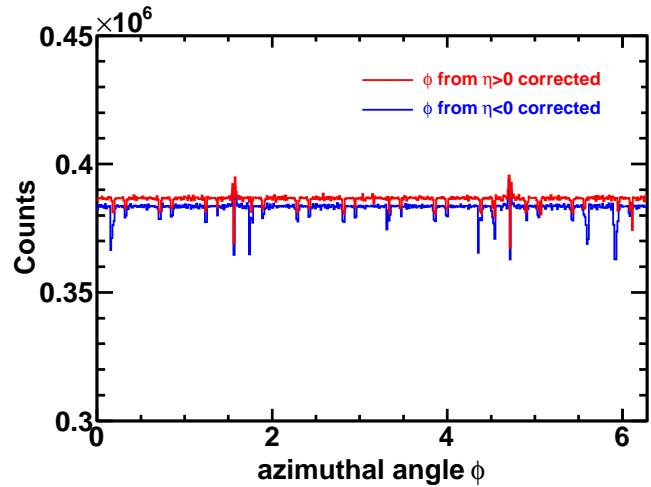}}
	\end{center}
	\caption[Single track acceptance correction]
	{Example plot of detector acceptance effect for positively charged particles in 30-40\% Au+Au 200 GeV RUN IV collisions.
	Panel (a) shows the raw (uncorrected) lab frame azimuthal angle ($\phi$) distribution.
	Blue curve is for those particles in $\eta < 0$ (east-side TPC) region,
	and red curve is for particles in $\phi > 0$ (west-side TPC) region.
	Data include both full field and reverse full field magnetic polarities, and the transverse momentum is integrated over $0.15 < p_T < 2$ GeV/$c$.
	Panel (b) shows the corrected $\phi$ distribution from the same data sample.
	}
	\label{fig:acc}
\end{figure}

Figure~\ref{fig:acc-a} shows an example of the lab frame azimuthal angle distributions of positively charged particles before the acceptance correction.
Data used in the figure are from RUN IV Au+Au 200 GeV collisions from east-side ($\eta <0$) in red and west-side TPC ($\eta >0$) in blue separately.
The centrality range is 30-40\%; \red{for all centralities and charges, refer to figure \ref{fig:appacc1}}.
The magnetic field polarities and the different charges have been summed together and the transverse momentum $p_T$ is integrated over $0.15 < p_T < 2$ GeV/$c$.
From the uncorrelated raw distributions, the repeatedly dropping pattern, \red{especially in most central collisions with large statistics (\ref{fig:appacc1}),} is clearly seen because of the sector boundaries of the TPC.
And there is additional inefficiency in east-side TPC ($\eta <0$, blue), range within $5\pi/3<\phi<2\pi$.
This is due to the inefficiency in the electronics readout system (RDO) of two out of twelve sectors in the east-side TPC.
We found the effect persists over the entire data set of RUN IV period, with no significant time variation.

We correct for the single particle $\phi$ dependent inefficiency, 
mainly due to sector boundaries and electronic dead sectors, 
by giving each particle a weight depending on the particle $\phi$ position.
We normalize the raw $\phi$ distribution in figure~\ref{fig:acc-a} to average unity.
The normalized $\phi$ distributions are defined as acceptance $\times$ efficiency, $\epsilon(\phi)$.
Those $\epsilon(\phi)$ distributions are separated according to magnetic field polarities, particle charges and centrality bins.
They are then further separated for positive and negative $\eta$ (corresponding to east- and west-side of the TPC tracks),
positive and negative vertex in $z$ direction ($v_z$),
and for different $p_T$ bins as following, $0.15-0.5$, $0.5-1.0$, $1.0-1.5$, $1.5-2.0$ GeV/$c$.
Then, we take $1/\epsilon(\phi)$ as the single particle weight as our correction factor for the detector acceptance $\times$ efficiency.
Note, the weight $1/\epsilon(\phi)$ could be either larger or smaller than unity.
Figure \ref{fig:acc-b} shows the corrected single particle $\phi$ distributions of the same set of particles in figure \ref{fig:acc-a}.
The corrected distributions are \red{almost uniform, although some jitter effects are still not completely removed due to binning and fluctuation issues,} which means we compensate the inefficiency in the single particle $\phi$ non-uniformity.
The weight $1/\epsilon(\phi)$ is then applied to all the asymmetry multiplicities and the TPC event-plane reconstruction through this analysis.
\red{For all centralities and charges, the corrected $\phi$ distributions are shown in appendix figure~\ref{fig:appacc2}.}

There are two main reasons we have to apply the single particle $\phi$ weighted correction.
The first reason is to flatten the event-plane reconstructed from the TPC tracks.
By nature, event-plane direction is random and should be flat in azimuthal direction.
As we will introduce in section \ref{TPCEP}, the second order event-plane reconstruction method is based on 
particle azimuthal distribution anisotropy.
If we do not correct the detector non-uniform $\phi$ inefficiency, the artifact anisotropy will couple with the particle distribution anisotropy making the reconstructed event-plane have preferred azimuth direction.
We do see this effect in data for about several percent magnitude.
After we applied the $\phi$ weighted correction, the event-plane distribution is flat for all centralities.
We will show that in section \ref{TPCEP}.

\begin{figure}[thb]
	\begin{center}
		\subfigure[$\langle A_{+,UD}^2\rangle$]{\label{fig:accasym-a}\includegraphics[width=0.3\textwidth]{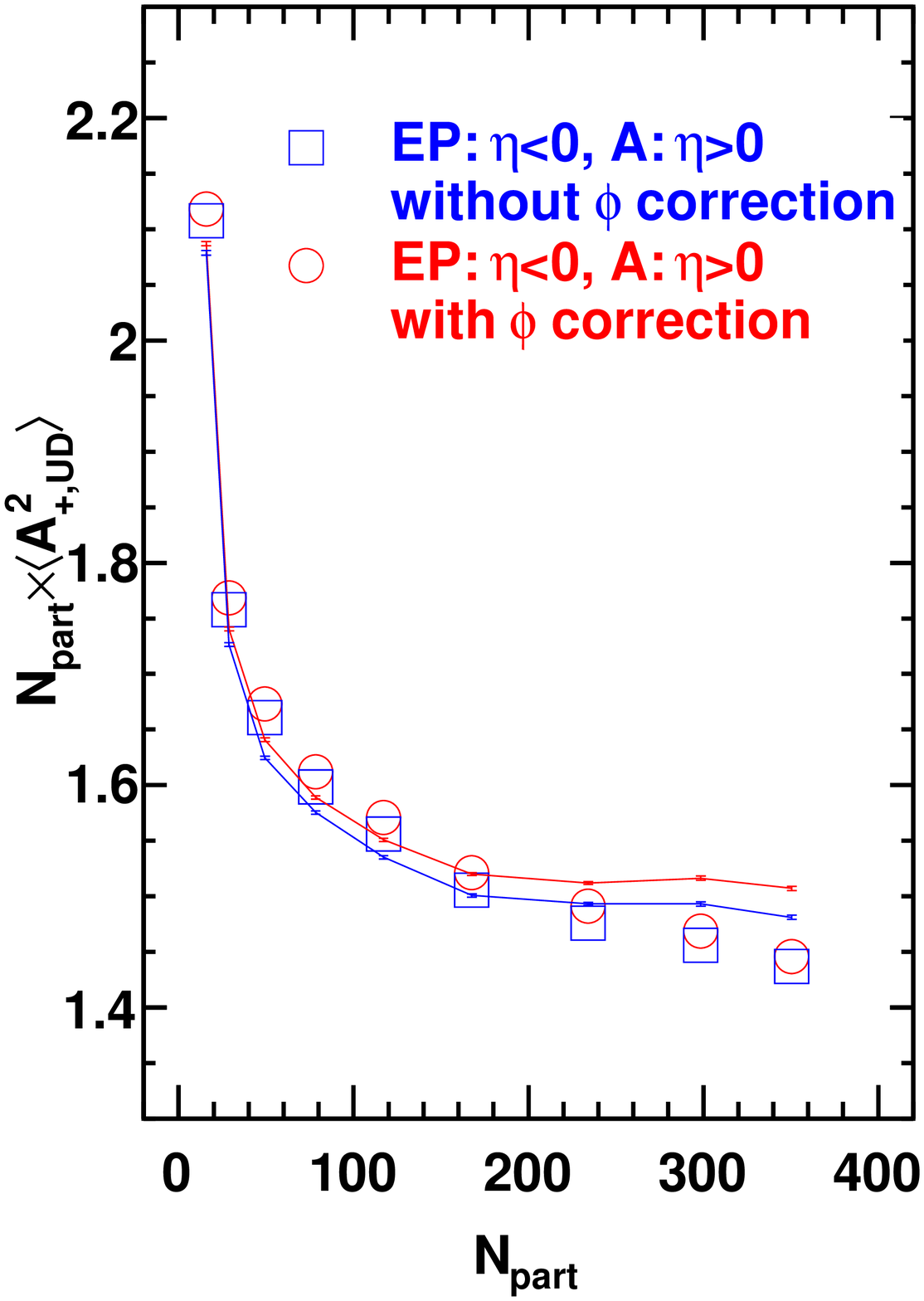}}
		\subfigure[$\langle A_{-,UD}^2\rangle$]{\label{fig:accasym-b}\includegraphics[width=0.3\textwidth]{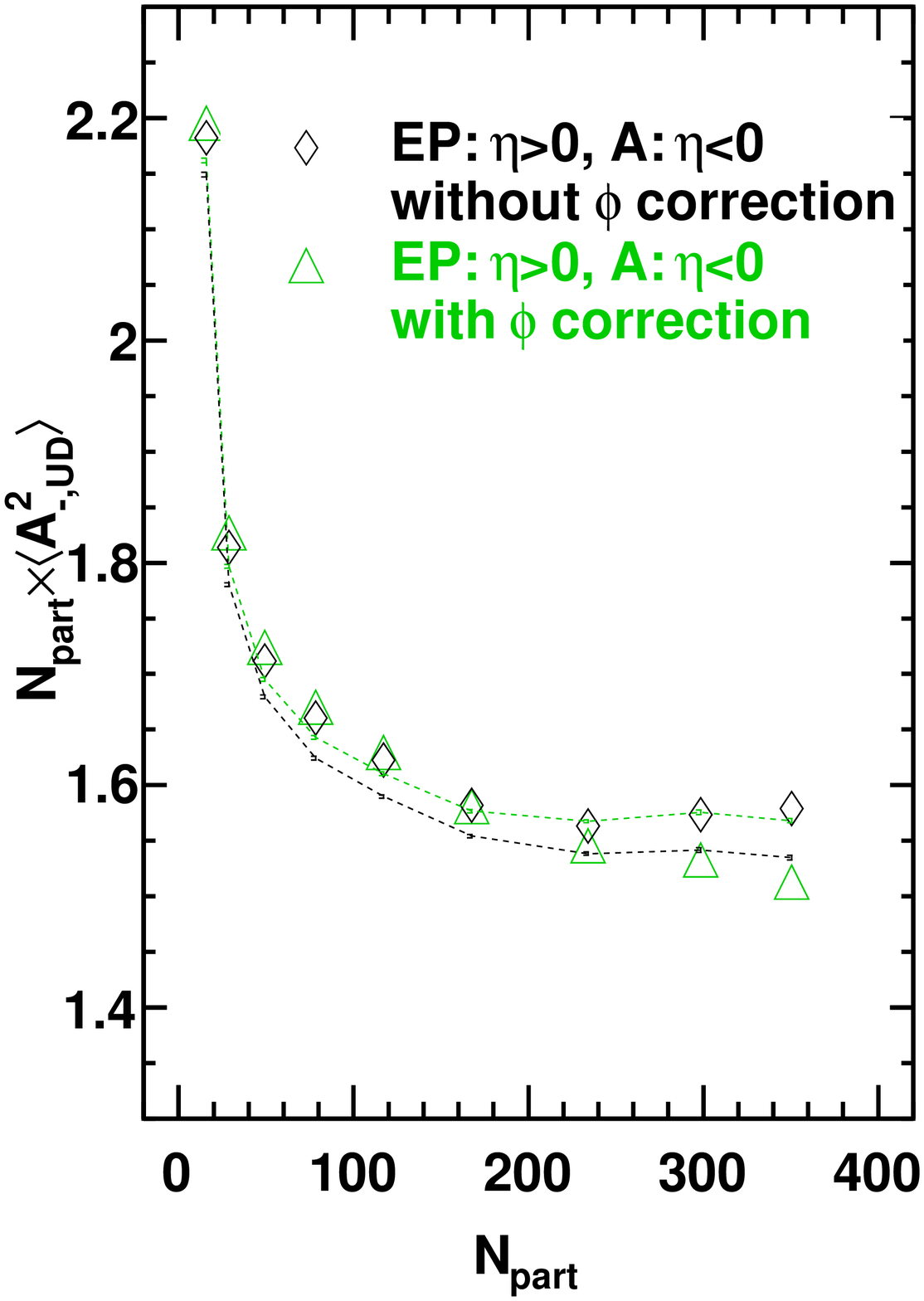}}
		\subfigure[$\langle A_+ A_- \rangle_{UD}$]{\label{fig:accasym-c}\includegraphics[width=0.3\textwidth]{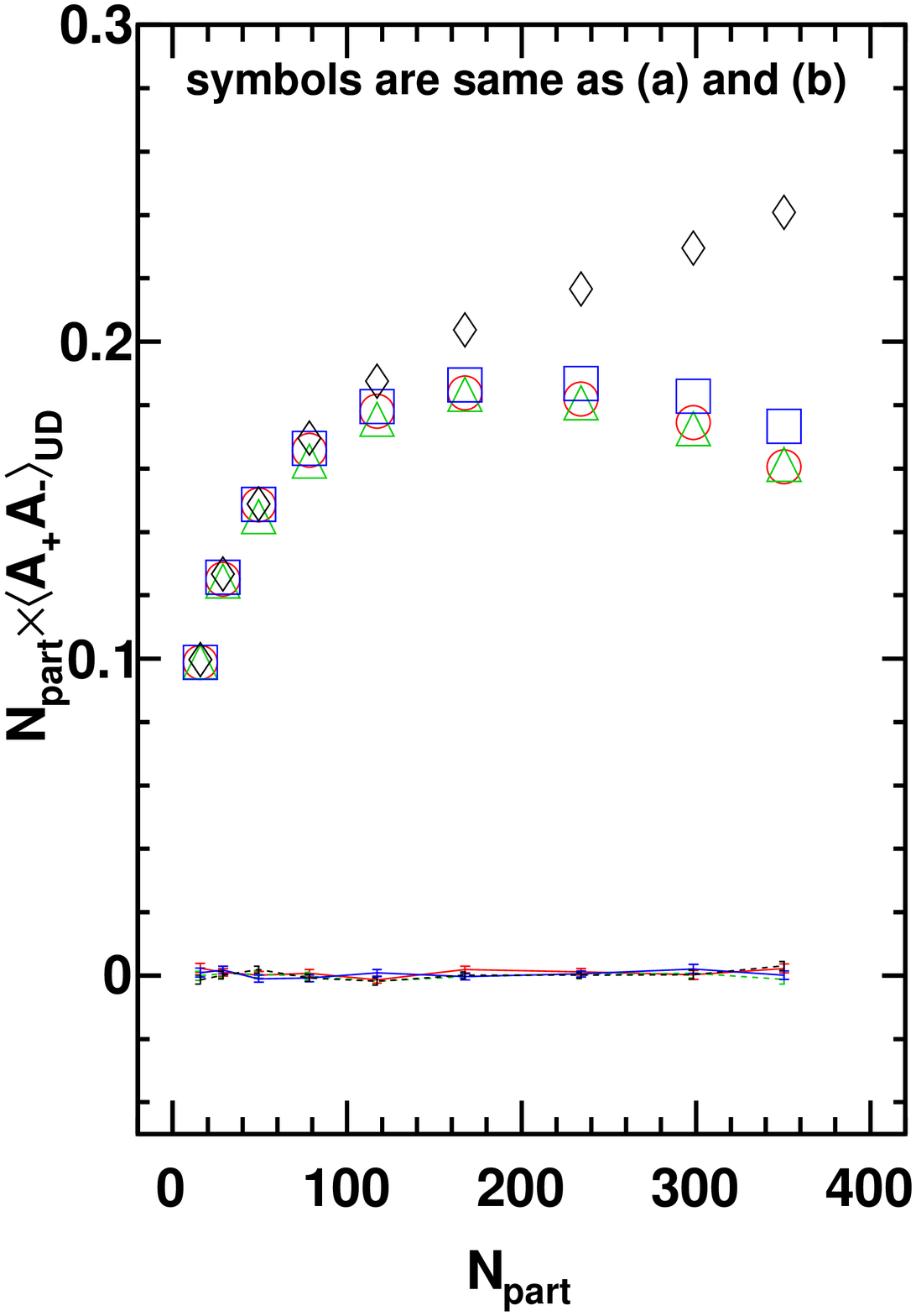}}
	\end{center}
	\caption[Charge asymmetry correlations efficiency correction in $UD$]{
	Asymmetry correlations: panel (a) $\langle A_{+,UD}^2\rangle$, (b) $\langle A_{-,UD}^2\rangle$, (c) $\langle A_+A_- \rangle_{UD}$
	(scaled by number of participants, $N_{part}$), before and after single particle corrections for the $\phi$ dependent acceptance $\times$ efficiency.
	The EP is reconstructed by charged particles with $p_T$ range of $0.15 < p_T < 2.0$ GeV/$c$ from one side of the TPC, and the asymmetry correlations are calculate in the same $p_T$ range but from the other side of the TPC.
	The lines are the statistical fluctuation and detector effect estimation.
	For clarity, only selected asymmetry correlations are shown.
	For all plots, see appendix.
	}
	\label{fig:accasym}
\end{figure}

The other reason for $\phi$ dependent efficiency correction is to compensate the bias of charge asymmetry correlations due to the detector effect.
Figure~\ref{fig:accasym} shows the selected charge asymmetry correlations in $UD$ examples:
$\langle A_{+,UD}^2\rangle$ (\ref{fig:accasym-a}), $\langle A_{-,UD}^2\rangle$ (\ref{fig:accasym-b}) 
and $\langle A_+A_- \rangle$ (\ref{fig:accasym-c}), before and after the $\phi$ dependent efficiency correction.
In the plots, all the asymmetry correlations are multiplied by the number of participants $N_{part}$ to better show the magnitude.
Note, asymmetries are calculated with tracks from half side of the TPC $\eta$ region ($\eta>0$ or $\eta<0$) with respect to
the EP reconstructed with tracks from the other half side of the TPC $\eta$ region ($\eta<0$ and $\eta>0$).
The reason for this setup is to avoid self-correlation, which is detailed in section \ref{selfcorr}.

We show only selected charge and $UD$ correlations in the figures for clarity reason.
\red{The $\phi$ efficiency correction effect for other charge asymmetry combinations and correlations in $LR$ is similar to $UD$. 
They are shown in figure~\ref{fig:appaccasymUD} and \ref{fig:appaccasymLR}.}
We see greater effect on correlations for asymmetries from $\eta < 0$ region than that from $\eta>0$ region.
This is because of the electronics RDO system inefficiency within $5\pi/3<\phi<2\pi$ for east-side of the TPC ($\eta<0$ region).
The inefficiency creates additional non-uniformity, and more ``dynamical'' correlations from the east-side than the west-side of the TPC by comparing the difference before correction in figure \ref{fig:accasym-b} to \ref{fig:accasym-a}.

The \red{correction} effect is larger in the opposite-sign correlations ($\langle A_+A_- \rangle$ in figure \ref{fig:accasym-c}) rather than the same sign correlations ($\langle A^2 \rangle$ in figure \ref{fig:accasym-b}).
As seen in figure~\ref{fig:accasym-c}, the $\langle A_+A_- \rangle_{UD}$ for the $\eta<0$ region is significantly larger
than the $\eta>0$ region before the $\phi$ efficiency correction.
After the single particle efficiency correction, the correlations are consistent between different charges, and different $\eta$ regions, which are shown in section \ref{consist}.
The lines in the plots will be discussed in section \ref{stat}.

\section{Event Plane Reconstruction}

In this section, we introduce two methods to reconstruct the event-plane.
The second order event-plane is reconstructed from the TPC tracks.
And the first order event-plane is reconstructed from the ZDC-SMD neutron energy shower profiles.
We will also discuss the event-plane resolution, 
which defines how well the estimation is for the reconstructed event-plane $\psi_{EP}$ compare to the real reaction plane $\psi_{RP}$,
even we don't know it exactly.

\subsection{Second Order Event Plane Reconstruction from TPC}
\label{TPCEP}

As we introduced in section \ref{flow}, 
the initial geometry eccentricity of a collision event will translate into the final state particle azimuth and transverse momentum distribution anisotropy, which is called flow.
The anisotropic flow can be described by a series of non-zero Fourier coefficients shown in equation \ref{eq:fourier}.
Then the reaction plane can be estimated from each term of the Fourier components.

In general, for a given event, its $n$-th order reaction-plane can be estimated in the following way \cite{Poskanzer:1998yz}:
\begin{equation}
	\psi_{n,EP} = \left(\arctan { \sum\limits_{i} w_i \sin (n \phi_i) \over \sum\limits_{i} w_i \cos (n \phi_i)}\right)/n.
  \label{eq:EPdef}
\end{equation}
The sums go through all the particles used in the reaction-plane reconstruction.
$w_i$ is the weight used to get the optimized reaction plane resolution for each particle.

In this analysis, we mainly focus on the study of asymmetry correlations with respect to the second order event-plane,
which gives the best reaction-plane $\psi_{2,RP}$ estimation,
because the second order term in the Fourier expansion is the dominant term of the event shape anisotropy.
For better event-plane resolution, the particle transverse momentum $p_T$ is used as the weight, $w_i = p_{T,i}$ because $v_{2}$ increases with $p_T$ \cite{Adams:2004bi}.
Also we have to apply the $\phi$ dependent single particle efficiency correction.
The second order event-plane $\psi_{EP}$ is then calculated as
\begin{equation}
	\psi_{EP} = \left(\arctan { \sum\limits_{i} p_{T,i} \sin (2 \phi_i) / \epsilon(\phi_i) \over \sum\limits_{i} p_{T,i} \cos (2 \phi_i) / \epsilon(\phi_i) }\right)/2.
  \label{eq:EPcalc}
\end{equation}

\begin{figure}[thb]
  \begin{center}
    \psfig{figure=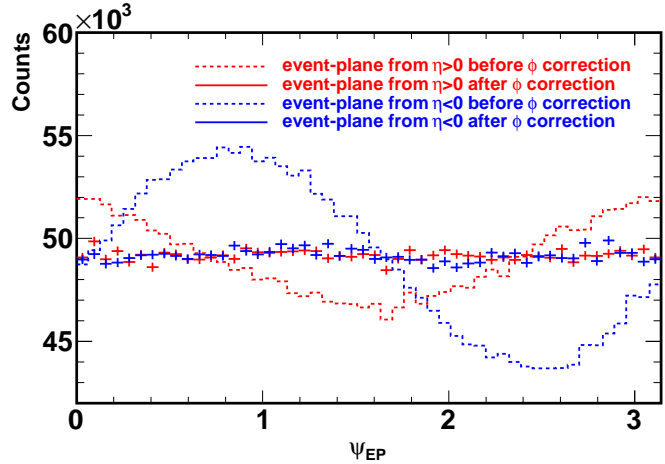,width=0.6\textwidth}
  \end{center}
  \caption[TPC second order EP distribution]
  {Reconstructed second order event-plane azimuthal distribution for RUN IV Au+Au 200 GeV collisions in 30-40\% centrality.
  The event-plane is reconstructed from charged particles within $0.15 < p_T < 2$ GeV/$c$ from $\eta<0$ (red) and $\eta>0$ (blue) separately.
  The lines are event-planes reconstructed with particles without single particle $\phi$ acceptance correction, and data points are with $\phi$ correction.
  }
  \label{fig:EP}
\end{figure}

Figure~\ref{fig:EP} shows the reconstructed event-plane azimuthal angle $\psi_{EP}$ distributions example from RUN IV 200 GeV Au+Au collision in 30-40\%.
The complete event-plane distributions of all centralities are shown in figure \ref{fig:appEP}.

\red{
Before applying the single particle $\phi$ acceptance correction, the event-plane distributions are shown as lines.
The distribution of event-plane reconstructed with the particles from east-side TPC within electronic deficit region ($\eta<0$), shown in dashed blue line, has a larger deviation from flatness compared to the other side.
This is because the EP reconstruction method based on the particle distribution anisotropy will try to find a direction with more particles emitted in the event-plane direction, and fewer particles emitted in out-of-plane direction.
And as shown in figure \ref{fig:acc-a}, the inefficient sector in $\eta<0$ region is located at $5\pi/3<\phi<2\pi$ in azimuth angle.
The inefficiency gives a preferred direction to the event-plane at approximately $\pi/3$, which is perpendicular to the center of the deficit sector around $11\pi/6$.
The peak of the blue line in figure \ref{fig:EP} is found at $\pi/4$, which is roughly consistent with our expectation.

To correct such detector effects, we apply the single particle $\phi$ dependent correction to all the tracks used in the EP reconstruction.
The results are shown as data points in figure \ref{fig:EP}, which are flat for both $\eta>0$ and $\eta<0$ regions.
So, we are confident with the reconstruction method and the necessary correction applied to remove detector non-perfection effect in the EP reconstruction.
}

\red{
Although we corrected the detector effects in event-plane reconstruction, the reconstructed EP is still an approximation of the true reaction-plane.
}
In equation \ref{eq:EPdef}, the event-plane will approach to the true reaction-plane if the event had infinite number of particles.
However, a real event has only finite number of particles recorded in TPC.
This will limit the accuracy of the estimated reaction-plane.
The inaccuracy is defined as event-plane resolution \cite{Poskanzer:1998yz} 
\begin{equation}
  \epsilon_{EP} = \langle \cos (2(\psi_{EP}-\psi_{RP}))\rangle,
  \label{eq:res2def}
\end{equation}
for second order event-plane $\psi_{EP}$ and reaction plane $\psi_{RP}$ here.
Under such definition, $\epsilon_{EP}$ is 1 if the $\psi_{EP}$ is exactly the same as $\psi_{RP}$,
and 0 if $\psi_{EP}$ is completely random relative to $\psi_{RP}$.

Note we separate an event into two sub sets, particles from east- and west-side of the TPC tracks.
By nature, the RP is a collision parameter which should be identical for both sub sets because they are from the same event.
However, the event-plane reconstructed from the two sub sets may not be the same.
And it is even possible that, due to other physics mechanism such as resonance decay or jet quenching effects, 
there are events with odd shapes making the event-planes significantly different from one another.
Although we do not know the true value of $\psi_{RP}$,
the event-planes reconstructed from the sub events are subject to the same resolution effects with respect to the true reaction-plane.
We can thus show that the event-plane resolution can be calculated from the sub events approximately as \cite{Poskanzer:1998yz},
\begin{align}
	\langle \cos 2 \left( \psi_{EP,\eta>0} - \psi_{EP,\eta<0} \right) \rangle &= 
	\langle \cos \left( 2\left( \psi_{EP,\eta>0} - \psi_{RP}\right) - 2\left( \psi_{EP,\eta<0} - \psi_{RP}\right)\right) \rangle \nonumber\\
	&\approx \langle \cos 2\left( \psi_{EP,\eta>0} - \psi_{RP}\right) \cos 2\left( \psi_{EP,\eta<0} - \psi_{RP}\right) \rangle \nonumber\\
	&\approx \epsilon_{EP}^2, \nonumber \\
	\epsilon_{EP} &= \sqrt{\langle \cos 2(\psi_{EP,\eta>0} - \psi_{EP,\eta<0})\rangle},
	\label{eq:res2}
\end{align}
where $\psi_{EP,\eta>0}$ and $\psi_{EP,\eta<0}$ are the reconstructed event-plane azimuthal angles from particles in $\eta>0$ and $\eta<0$ regions respectively.
Note, this particular event-plane resolution comes from the sub-events from the same event, which is most relevant for the studies we carry out in this thesis.

\begin{figure}[thb]
  \begin{center}
    \psfig{figure=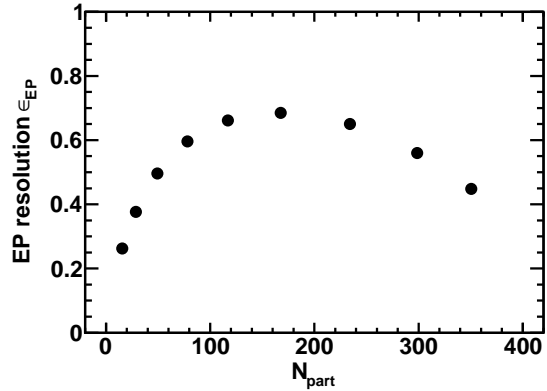,width=0.5\textwidth}
  \end{center}
  \caption[Second order EP resolution for RUN IV Au+Au 200 GeV]{Second order event-plane resolution from sub-events ($\eta < 0$ and $\eta > 0$) as a function of the number of participants in RUN IV Au+Au 200 GeV collisions.
  Statistical error bars are smaller than the symbols.}
  \label{fig:res2}
\end{figure}

Figure~\ref{fig:res2} shows the EP resolution $\epsilon_{EP}$ as a function of centrality (number of participants $N_{part}$).
The maximum event-plane resolution is found in medium central collisions. 
The resolution decreases in the peripheral and most central collisions.
This is because the event-plane reconstruction method is essentially based on the particle azimuthal anisotropy which defines the event-plane separating the particles into equal halves.
In very peripheral collisions, the multiplicity is low.
Those events are more likely being affected by non-flow and fluctuation.
For example, a pair of back-to-back di-jets may define the event-plane direction, which is unrelated to the reaction-plane.
Thus the resolution is low.
On the other hand in the most central collisions, 
the two colliding nuclei have the maximum overlapping area, less anisotropy in other words.
Thus the EP resolution is also low.
In the medium central collisions, they have more significant anisotropy, resulting in the largest EP resolution.

\begin{figure}[thb]
	\begin{center}
		\subfigure[$\langle A^{2} \rangle$ vs EP resolution]{\label{fig:EPres-a}\includegraphics[width=0.45\textwidth]{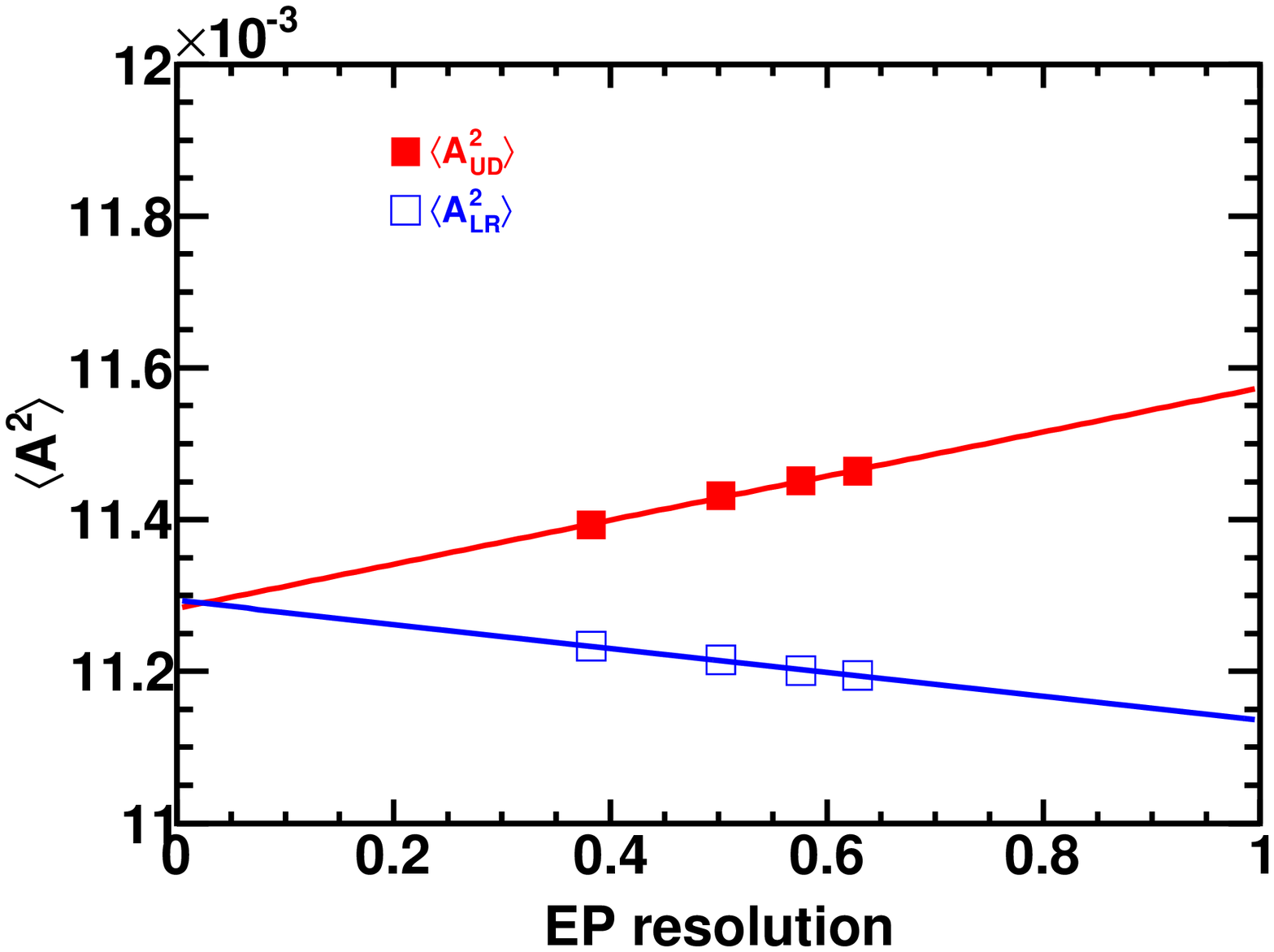}}
		\subfigure[$\langle A_+A_- \rangle$ vs EP resolution]{\label{fig:EPres-b}\includegraphics[width=0.45\textwidth]{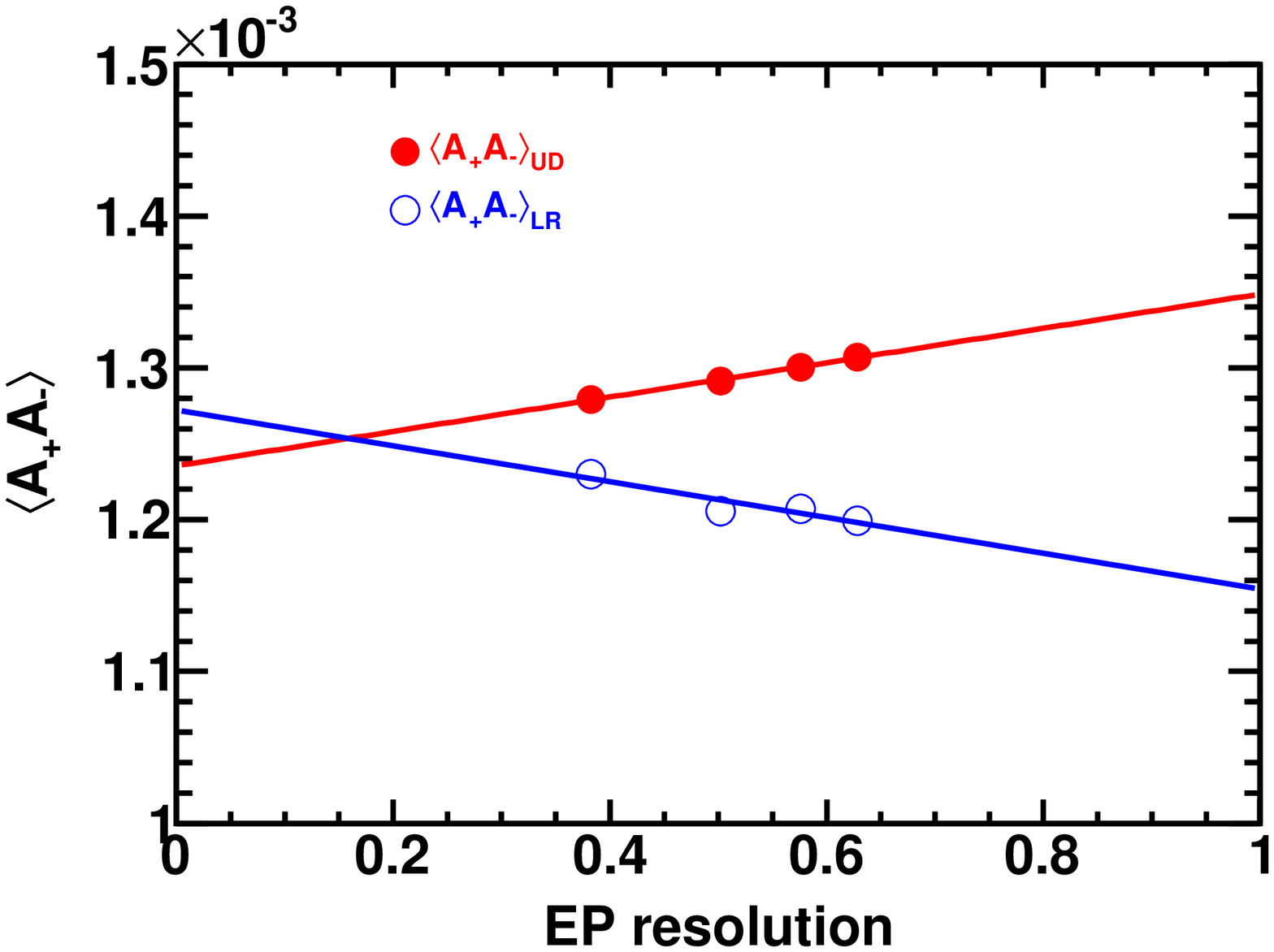}}
		\subfigure[$UD-LR$ differences vs EP resolution]{\label{fig:EPres-c}\includegraphics[width=0.45\textwidth]{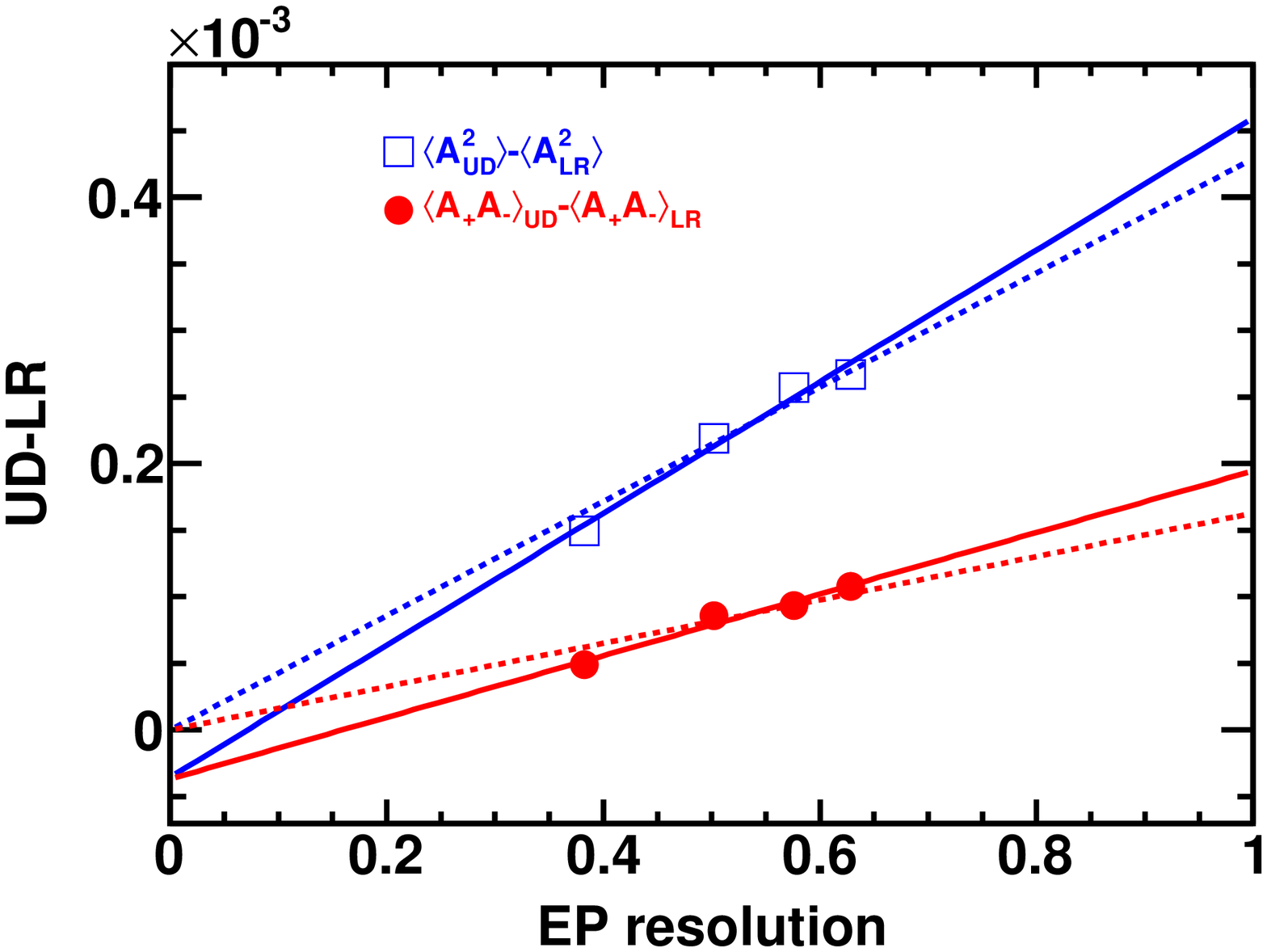}}
	\end{center}
	\caption[EP resolution dependence of mid-central asymmetry correlations]{
	Charge multiplicity asymmetry correlations: (a) $\langle A^{2} \rangle$, (b) $\langle A_+A_- \rangle$ and (c) their differences between $UD$ and $LR$, as a function of the event-plane resolution $\epsilon_{EP}$ in 20-40\% mid-central centrality.
	The solid lines are linear fits to the data.
	The dashed lines are linear fits with fixed zero intercept at $\epsilon_{EP} = 0$.
	Error bars are statistical.
	}
	\label{fig:EPres}
\end{figure}

\red{
After reconstructing the event-plane, correcting for the detector effects and having the EP resolution under control, 
we want to study how the EP resolution affects our charge asymmetry variances and covariances.
We cannot reach perfect EP resolution, but we can reduce the resolution by randomly throwing away a certain fraction of the particles during the event-plane reconstruction.
In this way, we can show and understand how the asymmetry correlations vary with different EP resolutions.

Figure \ref{fig:EPres} shows EP dependences of the 20-40\% mid-central Au+Au 200 GeV RUN IV collisions. 
The rightmost data points in each figure are asymmetry correlations with respect to the event-plane reconstructed using all charged particles from the half side of the TPC.
Then from right to the left, we artificially reduce the EP resolution by discarding 25\%, 50\%, and 75\% tracks used in the EP reconstruction.
The EP resolution is estimated by $\epsilon_{EP} = \sqrt{\langle \cos 2(\psi_{EP,\eta>0} - \psi_{EP,\eta<0})\rangle}$.
We then plot the charge asymmetry variances $\langle A^{2} \rangle$ (figure \ref{fig:EPres-a}), covariances $\langle A_+A_- \rangle$ (figure \ref{fig:EPres-b}) and their differences between $UD$ and $LR$ directions (figure \ref{fig:EPres-c}) as a function of the corresponding EP resolution.
Note that the asymmetries are calculated with all charged particles from half side of the TPC. 
We do not discard particles in the asymmetry calculation.
}

\red{
In figure \ref{fig:EPres}, we apply a linear fit of the correlation data to extrapolate the EP resolution to zero and unity.
The fits are presented as solid lines for all the variances, covariances and the $UD-LR$ differences.
When EP resolution approaches zero, i.e. the event-plane is random, we cannot distinguish between $UD$ and $LR$.
Then the asymmetry correlations of $UD$ and $LR$ should converge at the same intercept.
This is shown in figure \ref{fig:EPres-a} and \ref{fig:EPres-b}, where the linear fits of variances and covariances in $UD$ and $LR$ roughly converge at the zero EP resolution.
The differences between the $UD$ and $LR$, $\Delta\langle A^{2} \rangle$ and $\Delta\langle A_+A_- \rangle$ correlations should vanish at zero EP resolution 
because $UD$ and $LR$ are identical at zero EP resolution.
This is shown in figure \ref{fig:EPres-c}, where the linear fit in solid lines roughly converge in zero at zero EP resolution.
The dashed lines in the same figure are the fits with fixed intercept of 0 at zero EP resolution,
which are visually consistent with what we expected.

The linear fits in all three figures indicate that EP resolution only smears out the correlations between $UD$ and $LR$.
If we extrapolate the EP resolution to unity, the magnitude of the correlation differences between $UD$ and $LR$ will only be larger than what we actually measured from data.
In this analysis, we do not correct for EP resolution.
This is because, the linear extrapolation only works when the high-order harmonic terms of asymmetry correlations in equation \ref{eq:asymdiffexpand} are negligible, which is not true for the variances as we shall discuss in section \ref{3P}.
The high-order terms of the variance contribute significantly to the $UD-LR$ correlation.
It remains unknown how the high-order harmonic terms respond to the EP resolution.
Similar figure \ref{fig:appEPres} shows the most central and peripheral charge asymmetry correlations and their $UD$ and $LR$ differences as a function of the EP resolution.
And more detailed plots of each centrality are shown in figure \ref{fig:appEPresA2} for variances, figure \ref{fig:appEPresAA} for covariances and figure \ref{fig:appEPresd} for the differences of $UD$ and $LR$.
It is important to realize that our qualitative conclusions will not change if we have perfect EP resolution.
}

\red{
Note that, the second order event-plane azimuthal angle $\psi_{EP}$ range from 0  to $\pi$.
The event-plane angle $\psi_{EP}$ is equivalent to $\psi_{EP}+\pi$.
When calculating the asymmetries, we randomly flip the reconstructed event-plane to make it range from 0 to $2\pi$.
This is because, if we do not flip $\psi_{EP}$, the event-plane will have a preferred azimuthal direction in one half side of the TPC.
Thus the $UD$ and $LR$ will also have a preferred direction in azimuth.
This will introduce systematic errors due to any residual effect from the imperfect detector efficiency.
After random flipping, the preferred direction is avoided, so is the systematic uncertainty due to the preferred direction.
}

\subsection{First Order Event Plane from ZDC-SMD}
\label{ZDCEP}

The Zero Degree Calorimeters and Shower Maximum Detectors (ZDC-SMD) are located $\pm 18$~m away from the center of the STAR detector.
It records the neutron energy deposit profile from the deflected spectators,
which can be used to measure the first order event-plane determined by direct flow.
Since its pseudo-rapidity coverage ($\left| \eta \right|> 6$) is far away from TPC ($\left| \eta\right| < 1$), 
and the measured neutrons are from the fragmented gold nuclei which do not participate in the reaction,
there is little correlation between the ZDC-SMD signal to TPC tracks.
By using the event-plane reconstructed from ZDC-SMD, we can further remove possible physics correlations between the charge asymmetry correlations and the event-plane.
However, the first order event-plane resolution is not as good as the TPC event-plane as we will show below.

The ZDC-SMD detector measures the energy deposit profile with 7-slate vertical and 8-slate horizontal channels from both east-side and west-side of the STAR detectors.
The raw signals are corrected by pedestal subtraction and electronic gain corrections \red{for each channel}.
The pedestal subtraction is applied to correct the different electronic background of each readout channel.
And the gain correction is used to correct the linearity of the ADC response to the neutron energy deposit.
After the corrections, the signals give a hit profile in the transverse plane \red{in the manner of vertical and horizontal distributions}.
The vector from the beam center (averaged for each run) to the profile center gives the direct flow direction of the collision on each side of ZDC-SMD in transverse plane,
which is the measurement of the first order event-plane direction from one side of the ZDC-SMD.
Combining the two vectors from east- and west-side, we can get a better measurement of a single first order event-plane for each event \cite{JiayunChen:phd}.
Note that the two vectors are mostly back-to-back due to momentum conservation of the fragmented spectators.
Also note that, the first order event-plane ranges from 0 to $2\pi$, while second order event-plane ranges from 0 to $\pi$.

\begin{figure}[thb]
	\begin{center}
		\psfig{figure=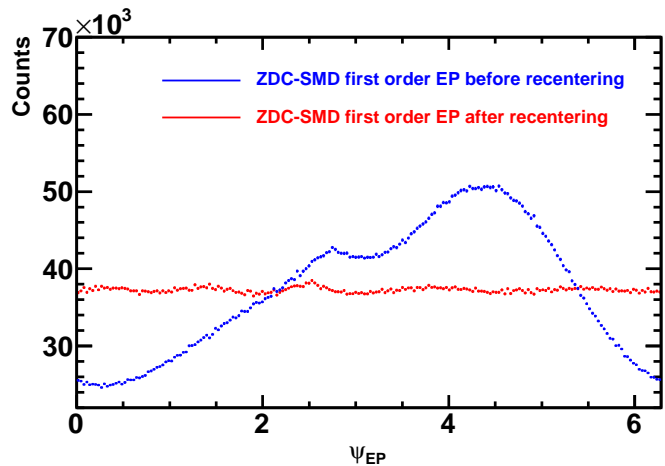,width=0.6\textwidth}
	\end{center}
	\caption[ZDC-SMD EP distribution]{Reconstructed ZDC-SMD first order event-plane distribution for RUN VII Au+Au 200 GeV 30-40\% centrality.
	The blue curve shows the uncorrected raw EP distribution. 
	The red curve shows the corrected EP distribution with the recentering method.
	The event-plane is reconstructed from combining the shower profile from east- and west-side ZDC-SMD.
	Error bars are statistical only.
	}
	\label{fig:zdcep}
\end{figure}

\red{
Figure~\ref{fig:zdcep} blue curve shows the raw ZDC-SMD first order event-plane of RUN VII 200 GeV Au+Au collision at 30-40\% centrality.
As we can see, the raw event-plane distribution is largely non-uniform in the azimuth, and has the preference direction with a fluctuation as large as about 50\%.
Thus the ZDC-SMD event-plane has to be corrected for non-uniformity in the azimuthal direction same as for the second order event-plane.
The major difference between these two is that,
the TPC event-plane reconstruction is track based, while ZDC-SMD event-plane reconstruction is profile based.
We can correct each track to flatten the TPC event-plane, but it is impossible to do the same to the energy profiles.
So another method, the so-called ``recentering'' method, has to be used for the analysis \cite{GangWang:phd}.
}

The idea is to shift the event-plane angle $\psi_{EP}$ by a correction according to its location $\delta\psi(\psi_{EP})$,
such that the new event-plane angle distribution $\psi'_{EP} = \psi_{EP}+\delta\psi(\psi_{EP})$ is flat for the whole event sample.
The raw event-plane distribution 
can be Fourier decomposed as the following:
\begin{align}
	{\mathrm{d} N(\psi_{EP}) \over \mathrm{d} \psi_{EP}} &= \frac{a_0}{2} + \sum_{n}\, [a_n \cos(n \psi_{EP}) + b_n \sin(n \psi_{EP})], \quad n \ge 1,\nonumber
\end{align}
with
\begin{align}
	a_n &= \frac{1}{\pi}\int_{0}^{2\pi} N(\psi_{EP}) \cos(n \psi_{EP})\, \mathrm{d} \psi_{EP}, \quad n \ge 0, \nonumber\\
	b_n &= \frac{1}{\pi}\int_{0}^{2\pi} N(\psi_{EP}) \sin(nx)\, \mathrm{d} \psi_{EP}, \quad n \ge 1. \nonumber
\end{align}
In order to obtain a flat event-plane distribution, the correction term must satisfy the following:
\begin{align}
	\psi'_{EP} &= \psi_{EP} + \delta\psi(\psi_{EP}), \nonumber \\
	{\mathrm{d} N(\psi'_{EP}) \over \mathrm{d} \psi'_{EP} } &= \frac{a_0}{2}, \nonumber
\end{align}
which leads to
\begin{align}
	\delta\psi(\psi_{EP}) &= \sum_{n}\, [A_n \cos(n \psi_{EP}) + B_n \sin(n \psi_{EP})], \quad n \ge 1 \nonumber . 
\end{align}
We can then easily derive that
\begin{align}
	A_n &= -{2\over n} \langle \sin (n\psi_{EP}) \rangle, \nonumber\\
	B_n &= {2\over n} \langle \cos (n\psi_{EP}) \rangle. \nonumber
\end{align}

In principle, if we apply infinite number of orders ($n$) to the correction, we definitely will get a flat first order event-plane.
In this analysis, we take the shifting up to the 4th order correction ($n = 1, 2, 3, 4$),
and the corrected event-plane distribution is shown in figure~\ref{fig:zdcep} in red data points.
With up to the 4th order correction, the final EP distribution is flat enough in the azimuthal angle.
\red{
Also in figure \ref{fig:appZDCEP}, we show all centrality first order event-plane azimuth distributions before and after recentering correction.
The data we used are RUN VII 200 GeV Au+Au collisions.
}

\begin{figure}[thb]
	\begin{center}
		\psfig{figure=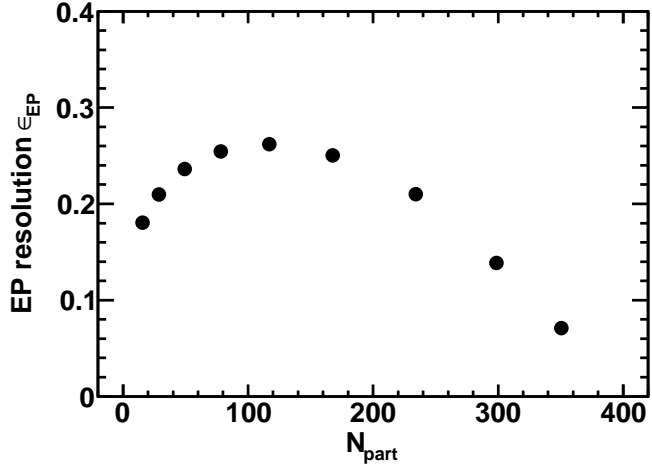,width=0.6\textwidth}
	\end{center}
	\caption[ZDC-SMD EP resolution]{The first order event-plane resolution as a function of the number of participants in RUN VII Au+Au 200 GeV collisions.
	Statistical error bars are too small to be seen.}
	\label{fig:zdcepres}
\end{figure}

To show how good the first order event-plane is compare to the reaction-plane, 
we calculate the event-plane resolution which is similar to the second order event-plane.
It is calculated as:
\begin{align}
  \epsilon_{EP} = \sqrt{2 \langle \cos (\psi_{EP,ZDCEast} - \psi_{EP,ZDCWest}) \rangle}.
\end{align}
Note that we combine the two event-planes $\psi_{EP,ZDCEast}$ and $\psi_{EP,ZDCWest}$ to form a single event-plane,
so the resolution has a $\sqrt{2}$ difference to the second event-plane.
Figure \ref{fig:zdcepres} shows the first order event-plane resolution as a function of $N_{part}$ for Au+Au 200 GeV RUN VII data.
It is lower than the second order event-plane shown in figure \ref{fig:res2}.

\section{Self-Correlation}
\label{selfcorr}

In general, when working on correlation study, one has to be cautious about self-correlation.
The problem rises when we calculate an observable from one set of data, and then correlate it with another observable calculated from the same set of data.
The two observables are intrinsically related, and the correlation between these two is automatically affected by self-correlation.

Particularly in this analysis, the reconstructed second order event-plane method utilizes the particle distribution anisotropy.
The EP reconstruction algorithm guarantees that the EP always divides the event multiplicity into more or less two equally halves
in UP and DOWN hemispheres.
For example, in one set of particles from the same event, 
if the positively charged particle multiplicity is unbalanced toward one side of the event-plane  
either due to fluctuation or underlying physics,
the negatively charged particle multiplicity is more likely unbalanced toward the other side of the event-plane.
Therefore, the asymmetries of positively and negatively charged particles are anti-correlated between $UD$ with respect to the EP reconstructed from the same set of particles.
However it does not affect the correlations between $LR$.
This results in smaller covariance $\langle A_+A_- \rangle_{UD}$ than $\langle A_+A_- \rangle_{LR}$,
and the difference is an artifact of self-correlation, which does not indicate physics dynamics.

\begin{figure}[thb]
	\begin{center}
		\subfigure[$\langle A_+A_- \rangle_{UD}$]{\label{fig:selfAAUD}\includegraphics[width=0.4\textwidth]{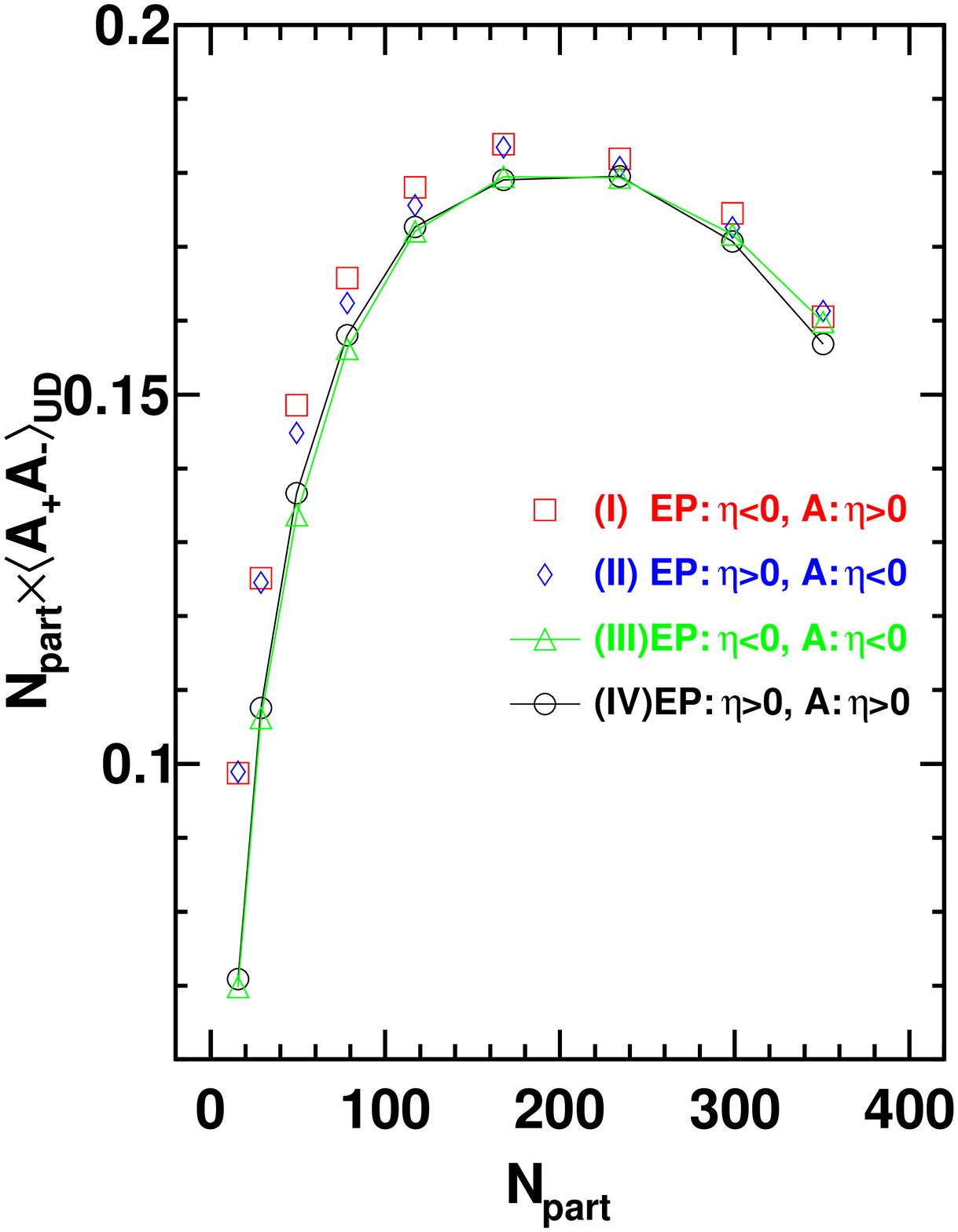}}
		\subfigure[$\langle A_+A_- \rangle_{LR}$]{\label{fig:selfAALR}\includegraphics[width=0.4\textwidth]{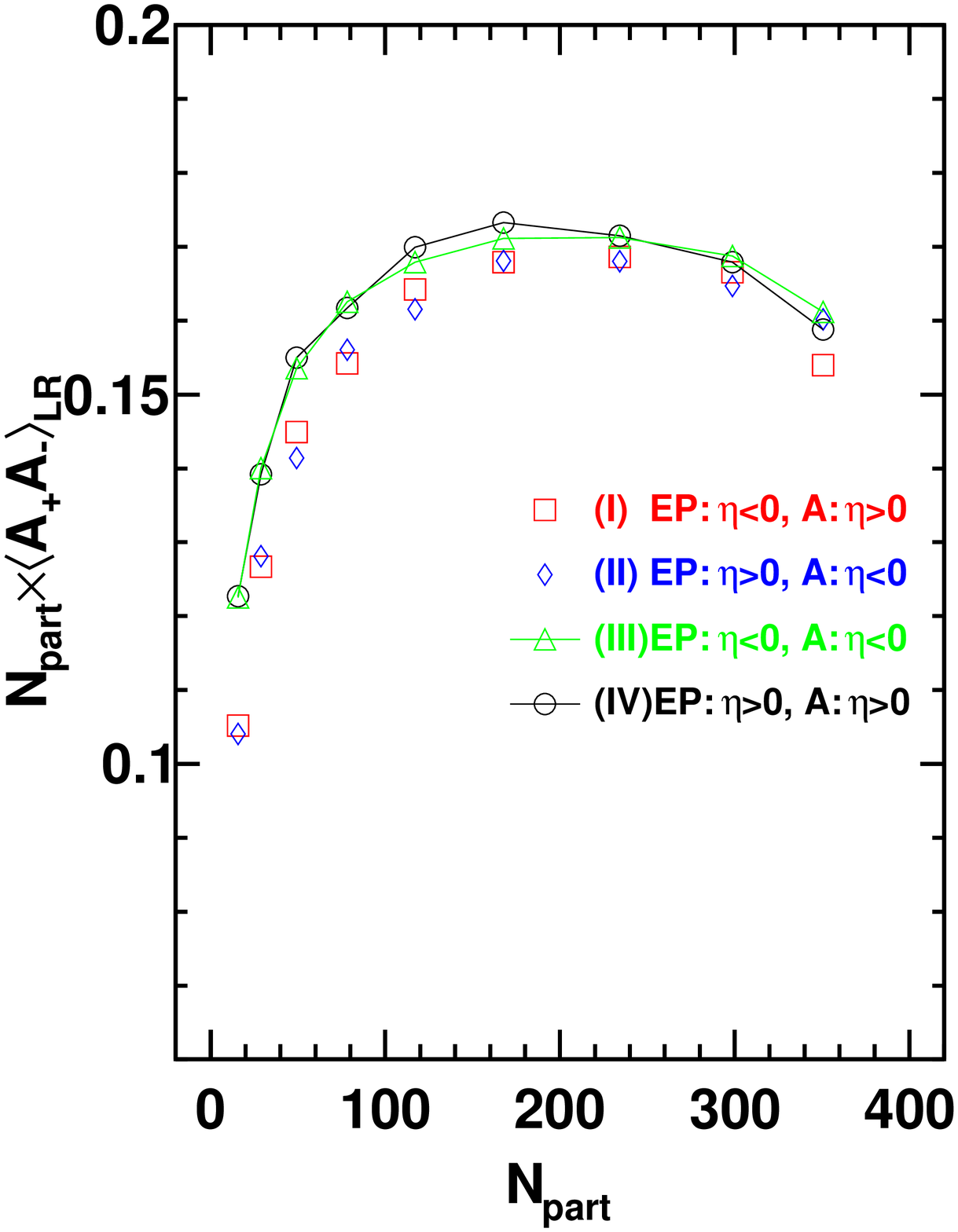}}
	\end{center}
	\caption[Self-correlation of covariances]{
	Panel (a): The covariances of the four cases in $UD$ direction.
	Panel (b): The covariances of the four cases in $LR$ direction.
	The covariances are scaled by the number of participants $N_{part}$ and plot against centrality ($N_{part}$) for
	four cases.
	The particle $p_T$ range of $0.15 < p_T < 2.0$ GeV/$c$ is used for both EP reconstruction and asymmetry calculation.
	Error bars are statistical.
	}
	\label{fig:selfAA}
\end{figure}

\begin{figure}[thb]
	\begin{center}
		\subfigure[$\langle A^2_{UD} \rangle$]{\label{fig:selfA2UD}\includegraphics[width=0.4\textwidth]{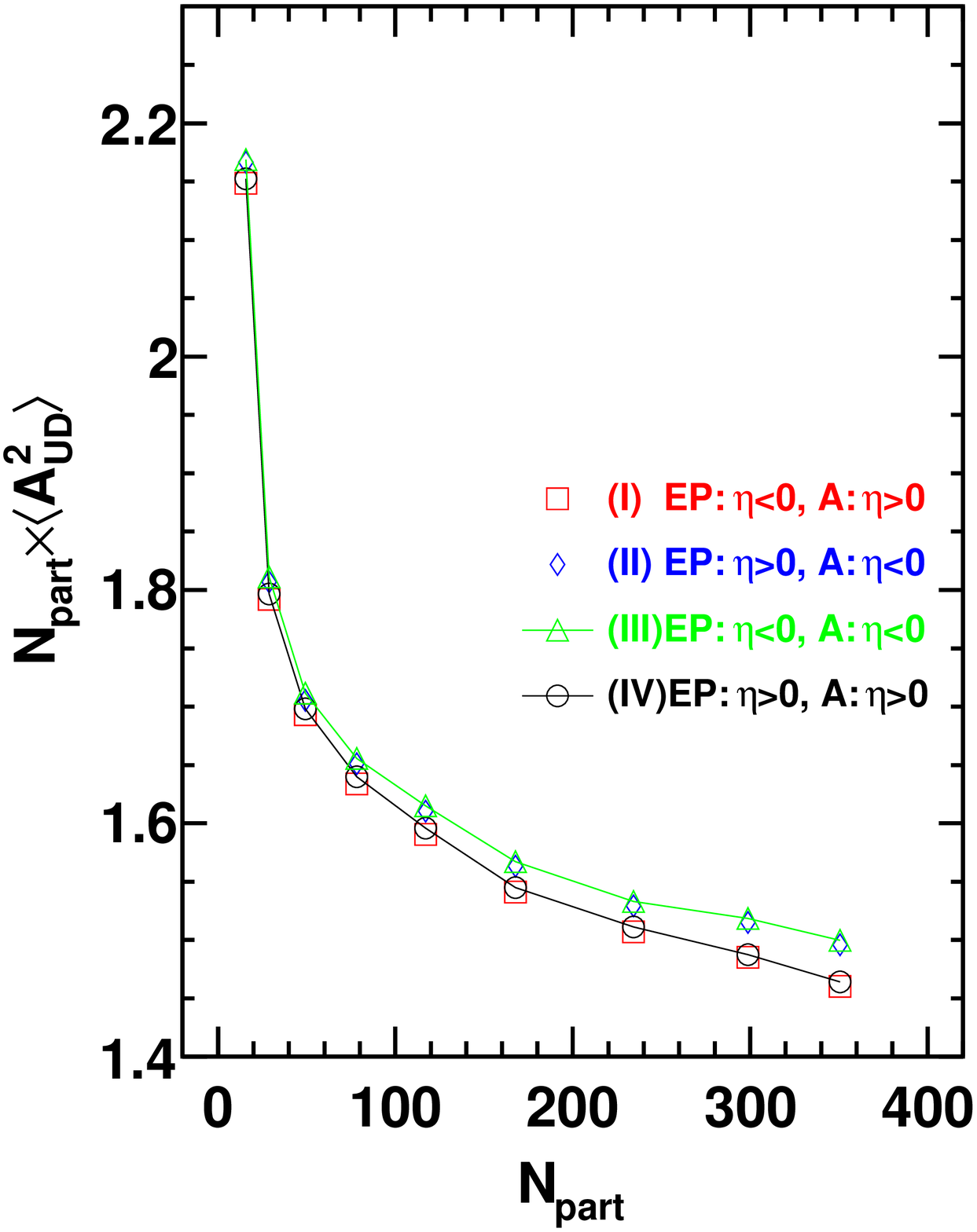}}
		\subfigure[$\langle A^2_{LR} \rangle$]{\label{fig:selfA2LR}\includegraphics[width=0.4\textwidth]{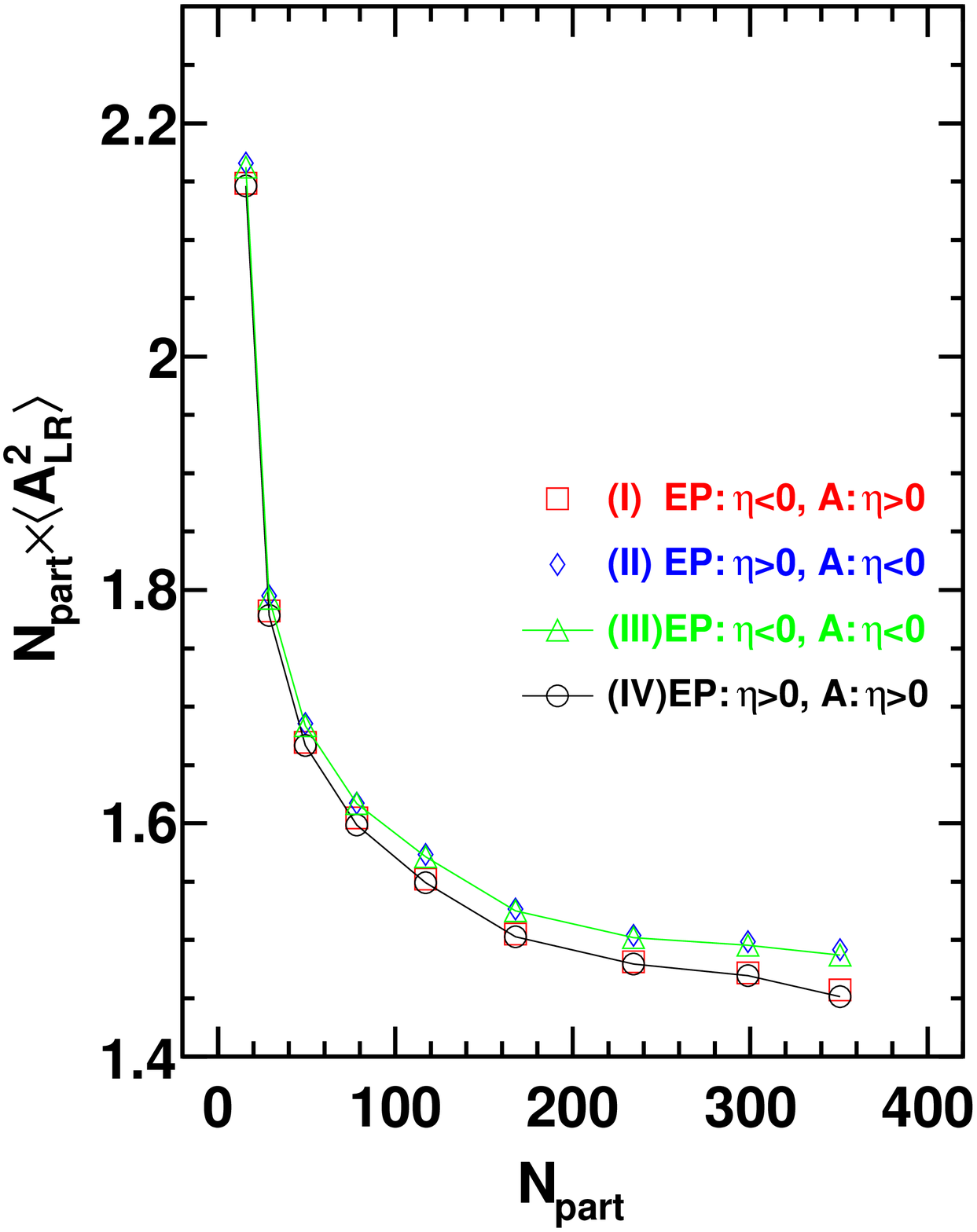}}
	\end{center}
	\caption[Self-correlation of variances]{
	Panel (a): The variances of the four cases in $UD$ direction.
	Panel (a): The variances of the four cases in $LR$ direction.
	The variances are scaled by the number of participants $N_{part}$ and plot against centrality ($N_{part}$) for
	four cases.
	The particle $p_T$ range of $0.15 < p_T < 2.0$ GeV/$c$ is used for both EP reconstruction and asymmetry calculation.
	Error bars are statistical.
	}
	\label{fig:selfA2}
\end{figure}

To show the self-correlation effect, we calculate four sets of asymmetry correlations and compare them.
They are separated according to $\eta$ regions for EP reconstruction and asymmetry calculation in different combinations.
\begin{itemize}
	\item[(I)] Using particles within $-1<\eta<0$ for EP reconstruction and within $0<\eta<1$ for asymmetry correlations;
	\item[(II)] Using particles within $0<\eta<1$ for EP reconstruction and within $-1<\eta<0$ for asymmetry correlations;
	\item[(III)] Using particles within $-1<\eta<0$ for EP reconstruction and within $-1<\eta<0$ for asymmetry correlations;
	\item[(IV)] Using particles within $0<\eta<1$ for EP reconstruction and within $0<\eta<1$ for asymmetry correlations;
\end{itemize}
\red{
The results of the four cases are shown in figure \ref{fig:selfAA} for the covariances and figure \ref{fig:selfA2} for the variances.
}

\begin{figure}[thb]
	\begin{center}
		\subfigure[$\langle A_+A_- \rangle_{UD} / \langle A_+A_- \rangle_{LR}$]{\label{fig:selfcorr-a}\includegraphics[width=0.4\textwidth]{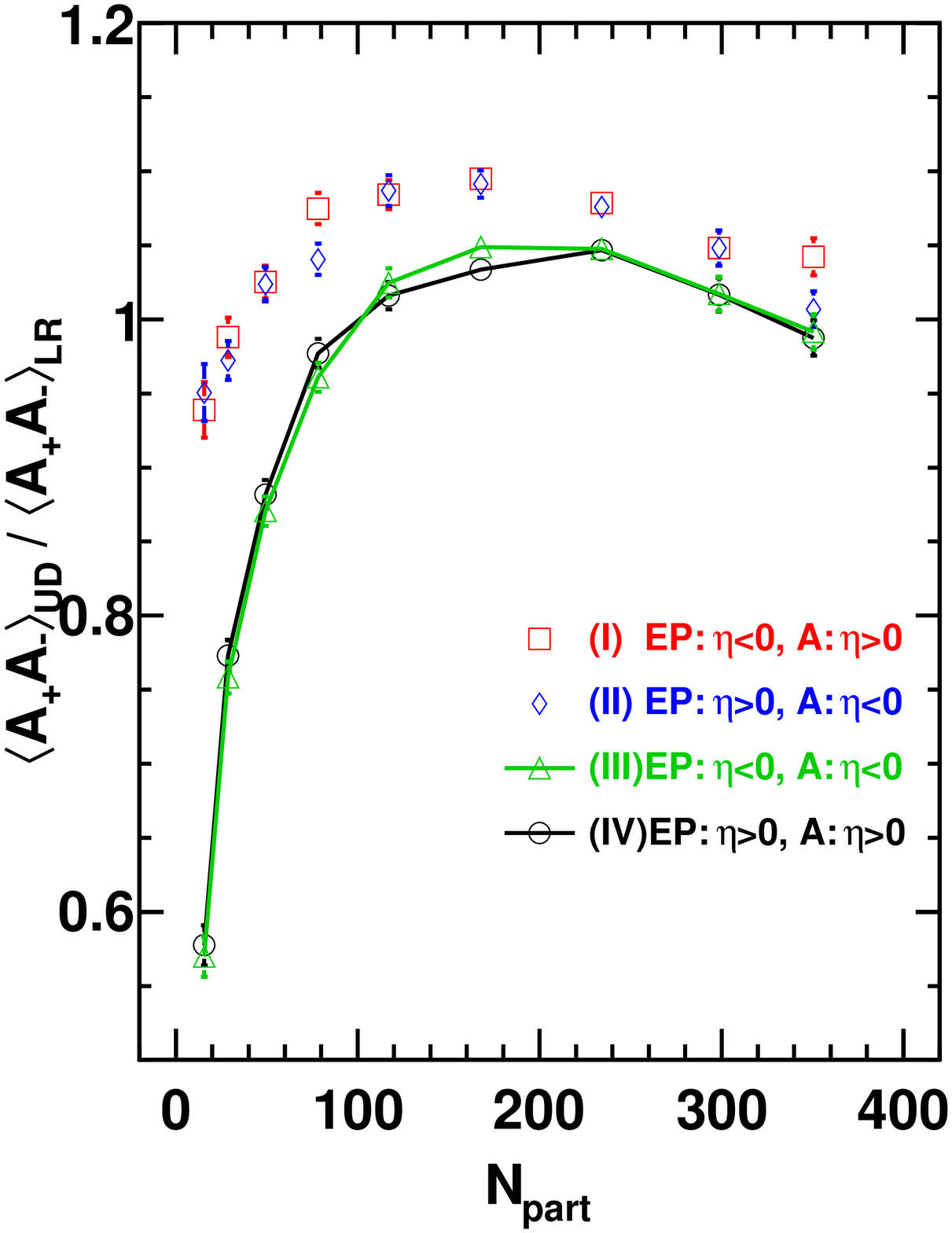}}
		\subfigure[$\langle A^2_{UD} \rangle / \langle A^2_{LR} \rangle$]{\label{fig:selfcorr-b}\includegraphics[width=0.4\textwidth]{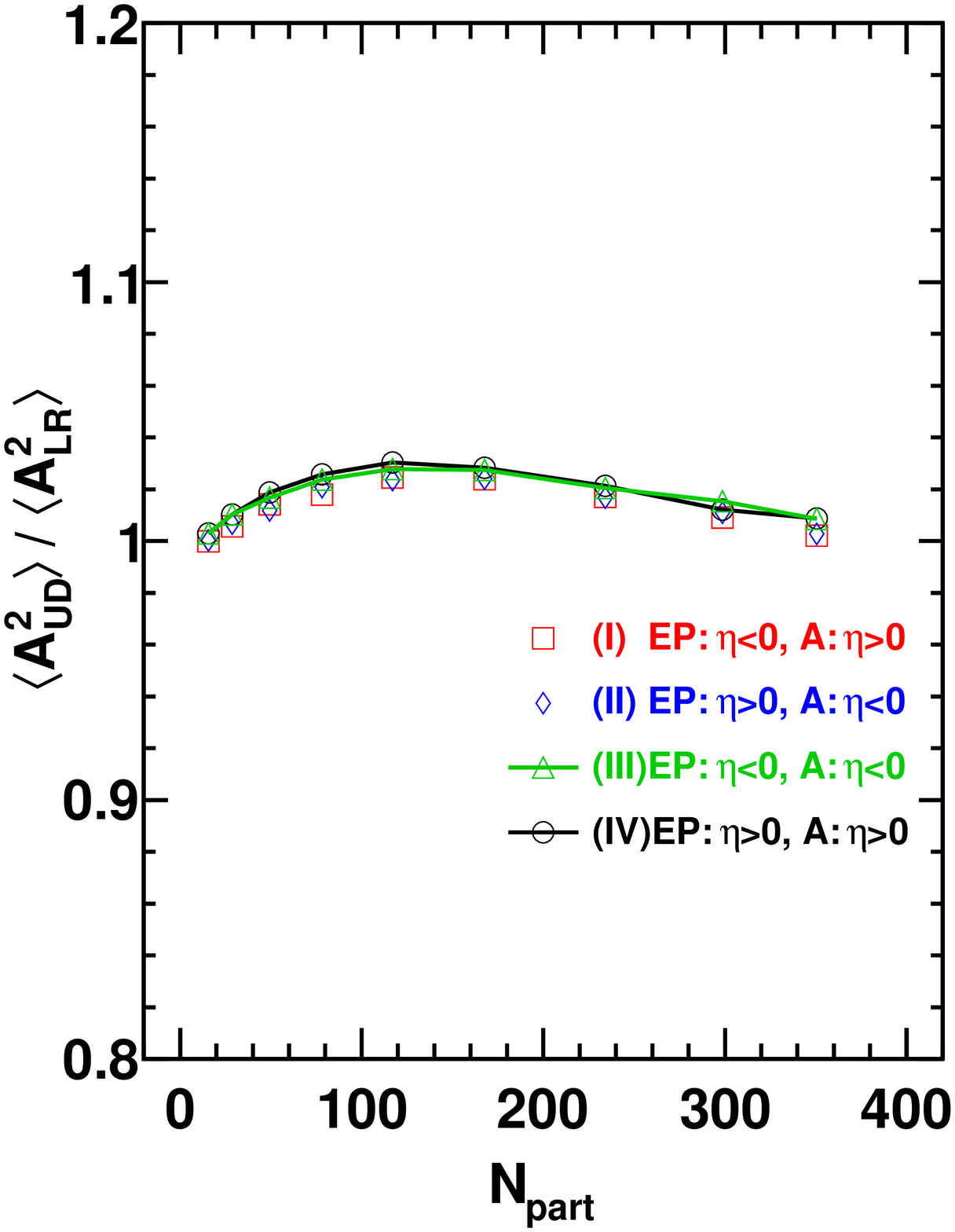}}
	\end{center}
	\caption[Self-correlation effect]{
	Relative ratios of charge multiplicity correlations as a function of the number of participants $N_{part}$ for
	four combinations of $\eta$ ranges used in EP reconstruction and asymmetry calculation.
	Panel (a): $\langle A_+A_- \rangle_{UD} / \langle A_+A_- \rangle_{LR}$,
	Panel (b): $\langle A^2_{UD} \rangle / \langle A^2_{LR} \rangle$.
	The particle $p_T$ range of $0.15 < p_T < 2.0$ GeV/$c$ is used for both EP reconstruction and asymmetry calculation.
	Error bars are statistical.
	}
	\label{fig:selfcorr}
\end{figure}

To better illustrate the differences,
figure~\ref{fig:selfcorr} shows the relative ratios between $UD$ and $LR$ for the covariances
$\langle A_+A_-\rangle_{UD} / \langle A_+A_- \rangle_{LR}$ in figure~\ref{fig:selfcorr-a} and the variances $\langle A^2_{UD} \rangle / \langle A^2_{LR} \rangle$ in figure~\ref{fig:selfcorr-b} with the above four combinations.
The significant self-correlation effect is observed in the covariances from figure \ref{fig:selfcorr-a} of cases (III) and (IV),
where the ratios of the cases with the asymmetry correlations and event-plane reconstructed from the same side of the TPC are diverged from those cases using different side of TPC tracks.
\red{
The effect is more significant for the peripheral events, where the total multiplicity is less than in the central events.
Thus, the fluctuation to multiplicity ratio due to self-correlation is larger in the peripheral events than in the more central events.
}
However, the ratios from different side of the TPC tracks are relatively stable over all centralities,
which indicates a smaller self-correlation effect.
In this analysis, we use the average of cases (I) and (II) as our result correlations to avoid the self-correlation.

On the other hand, the same-sign correlations, the variances $\langle A_{\pm,UD}^2 \rangle$ and $\langle A_{\pm,LR}^2\rangle$ show no significant self-correlation effect as shown in \ref{fig:selfcorr-b}.
This is because the particles used in variances and the EP reconstruction are not identical set of particles.
Only half of the particles (positive or negative charged particles) are used in the asymmetry calculation.
Thus the self-correlation effect is not obvious from the ratio plot.
We still use the average of cases (I) and (II) for the variances to consistent with the covariances.

In such setup, we calculate the charge asymmetries and their correlations from one side of the TPC tracks (from $\eta>0$ or $\eta<0$) 
with respect to the event-plane reconstructed from the other side of the TPC tracks (from $\eta<0$ or $\eta>0$).
For each event, we have two sets of asymmetries and their correlations, and we take the average of them to increase statistics.

\red{
To further remove the short range correlation which has a bulk correlation with the rapidity span around one unit for the soft particles,
the ZDC-SMD first order event-plane is used. 
It extends the pseudo-rapidity range to $|\eta|>6$.
The correlation between the TPC tracks and the ZDC-SMD signal are reduced to minimum.
In order to make direct comparison, we still divide an event into two sub-events according to the pseudo-rapidity range for the asymmetry and their correlation calculation with respect to the full event-plane reconstructed from ZDC-SMD.
}

\section{Statistical Fluctuation and Detector Effect}
\label{stat}

\begin{figure}[thb]
	\begin{center}
		\subfigure[$stat+det$ for $\langle A_{+}^2 \rangle_{LR}$ in $\eta<0$ region]
		{\label{fig:stat-a}\includegraphics[width=0.45\textwidth]{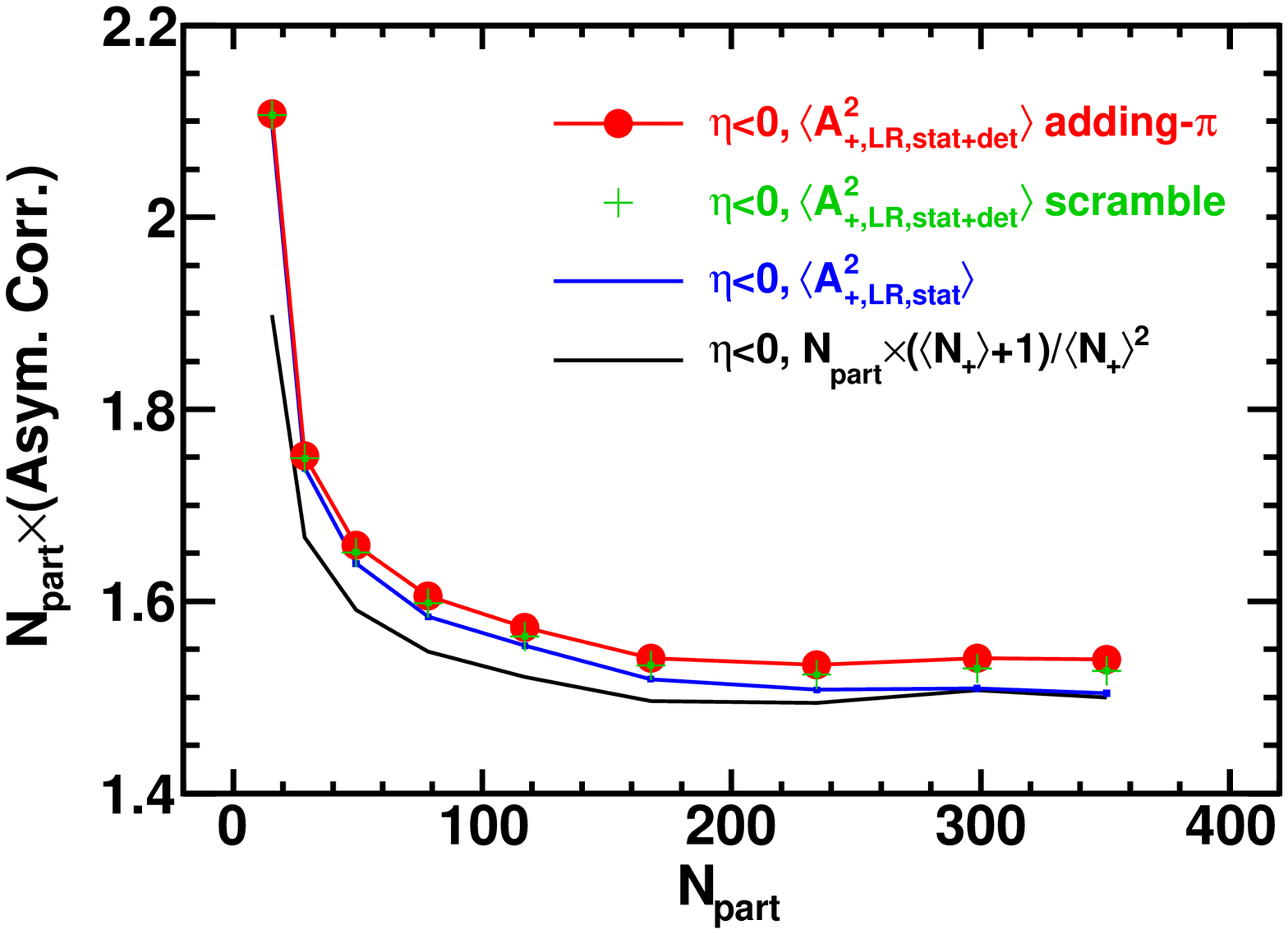}}
		\subfigure[$stat+det$ for $\langle A_{+}^2 \rangle_{LR}$ in $\eta>0$ region]
		{\label{fig:stat-a0}\includegraphics[width=0.45\textwidth]{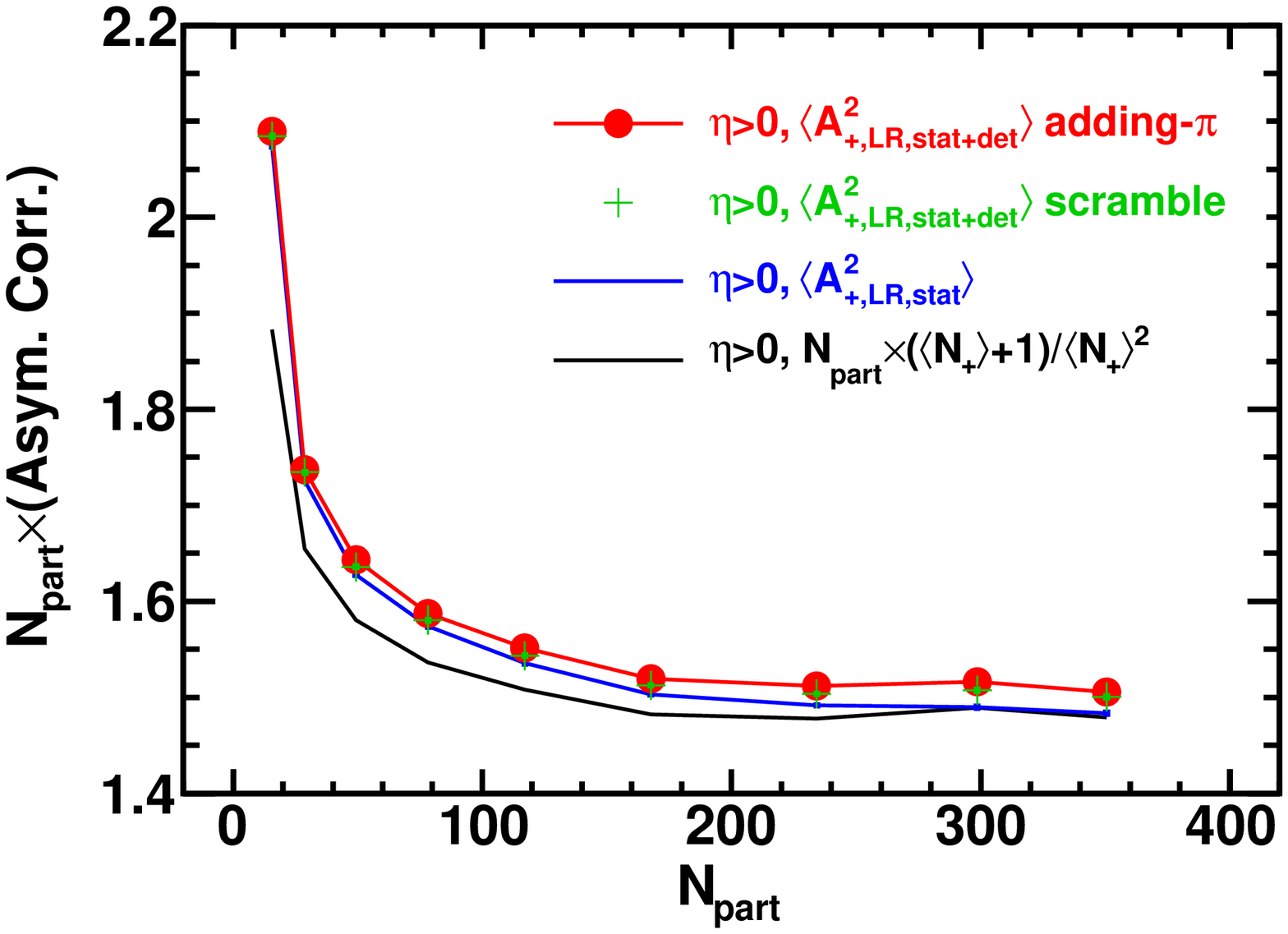}}
		\subfigure[Ratios between methods]
		{\label{fig:stat-b}\includegraphics[width=0.45\textwidth]{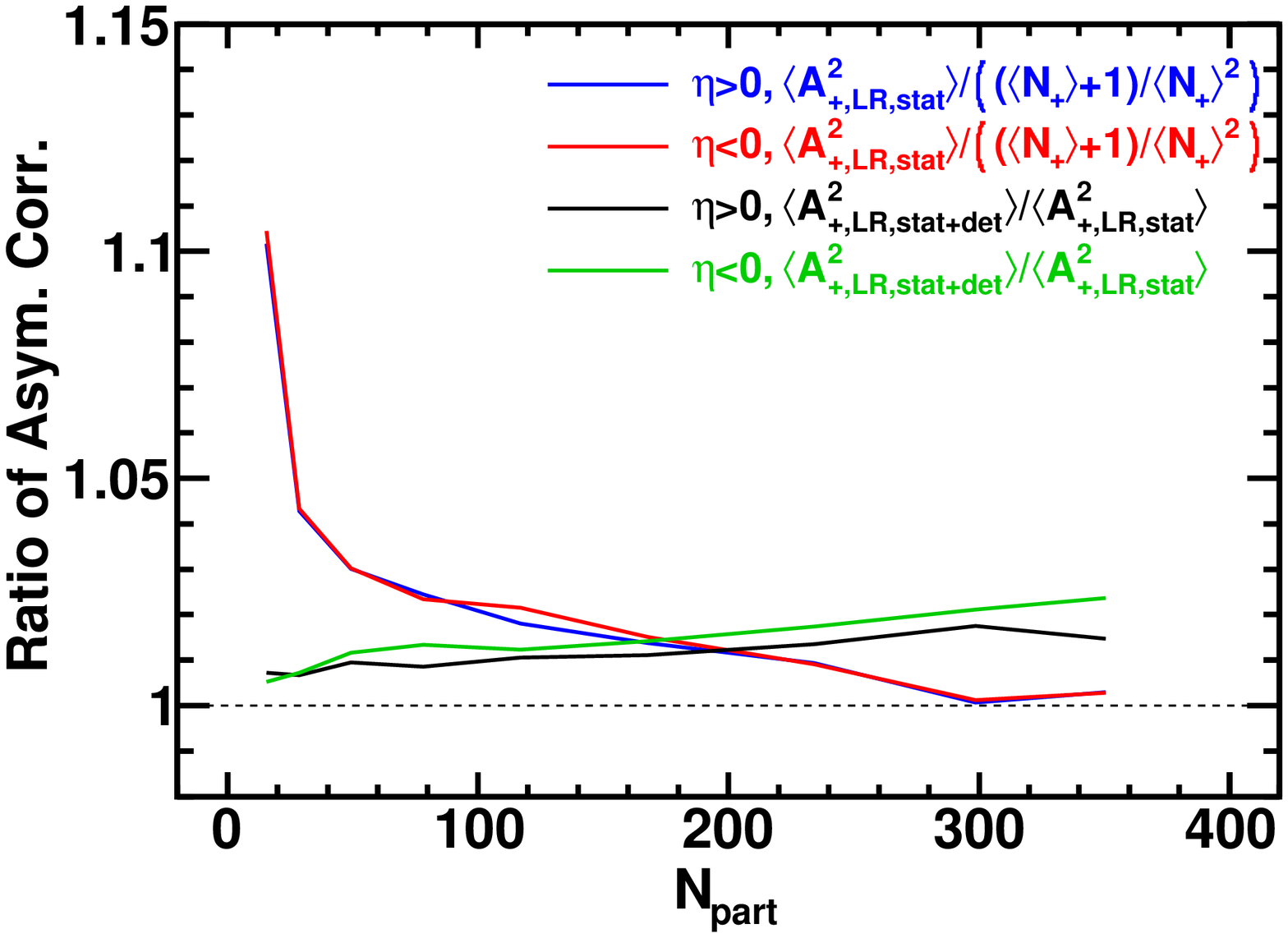}}
		\subfigure[Ratios between charges and hemispheres]
		{\label{fig:stat-c}\includegraphics[width=0.45\textwidth]{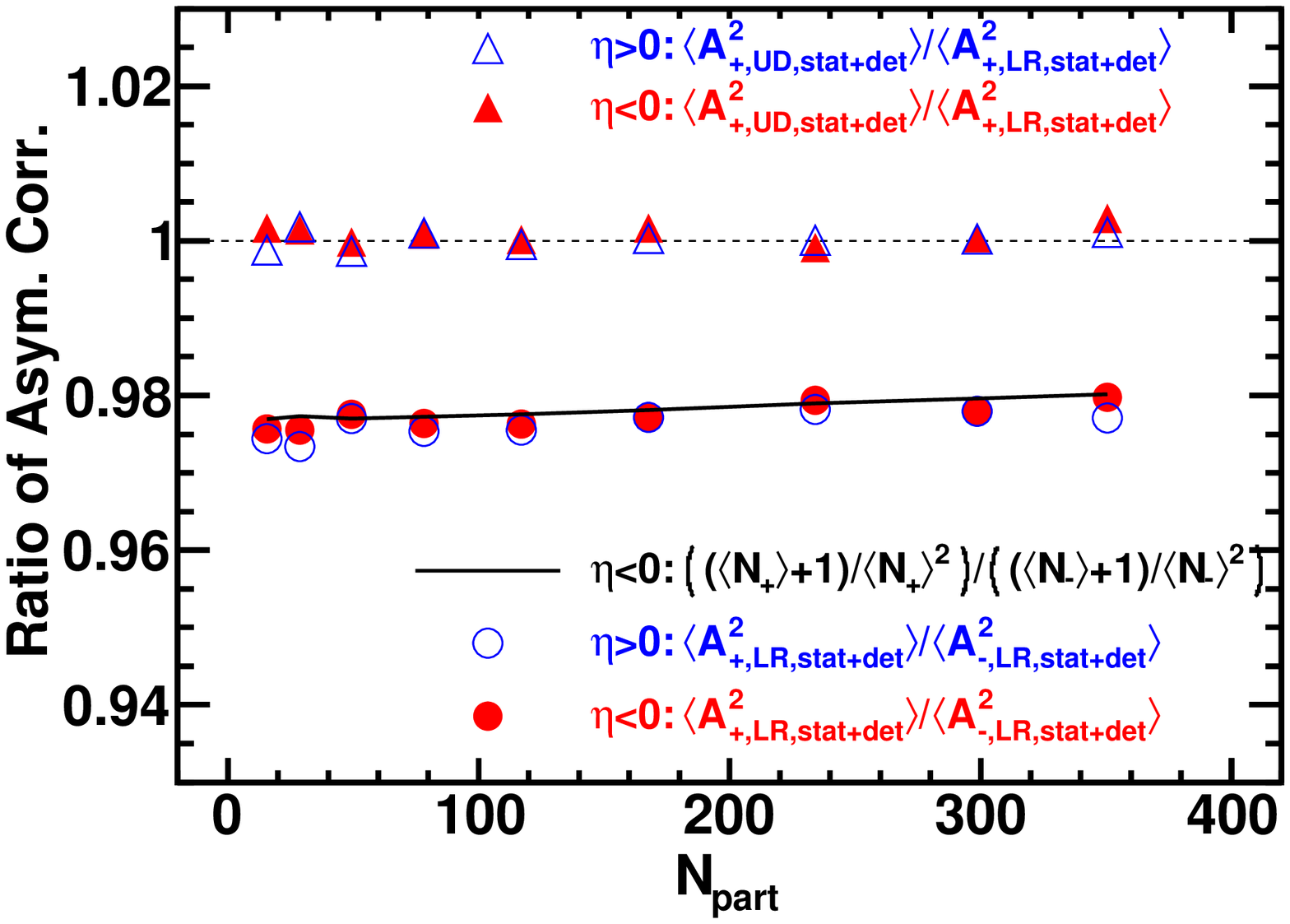}}
	\end{center}
	\caption[Statistical fluctuation and detector effect]{Panel (a): Statistical fluctuation and detector effects in charge asymmetry variances scaled by the number of participants $N_{part}$ from east-side of the TPC, $\eta<0$ region, 
	with respect to the EP reconstructed from west-side of the TPC, $\eta>0$ region.
	The black curve shows the ``$1/N$'' approximation by equation~\ref{eq:stat1}.
	The blue curve shows the pure statistical fluctuation $\langle A^2_{+,LR,stat} \rangle$ with ``50-50'' method.
	The statistical fluctuation plus detector effects $\langle A^2_{+,LR,stat+det} \rangle$ are shown in green crosses
	using scramble method, and red circles using flipping-$\pi$ method.
	Panel (b): Same as (a) but for $\eta>0$ region with EP reconstructed from $\eta<0$ region.
	Panel (c): The ratios of statistical fluctuation estimations between different methods of different $\eta$ regions.
	Panel (d): The consistency check of ratios of $UD$ to $LR$ and positive to negative charge obtained by adding-$\pi$ method.
	Data are from RUN IV Au+Au 200 GeV collisions. The particle $p_T$ range is integrated over $0.15 < p_T < 2.0$ GeV/$c$.
	}
	\label{fig:stat}
\end{figure}

It is obvious that the variances are non-zero even if there are no dynamical fluctuation presents,
because they are defined as the squares of the ratios which are real numbers.
This is a trivial effect of the statistical fluctuation of the finite multiplicity.
Take $N_{\pm,U} = \langle N_{\pm,U} \rangle +\delta N_{\pm,U}$ as the collective notation for $N_{+,U}$ and $N_{-,U}$,
and $N_{\pm,D} = \langle N_{\pm,D} \rangle +\delta N_{\pm,D}$ for $N_{+,D}$ and $N_{-,D}$.
If we assume their fluctuations are Poisson, the statistical fluctuation effect of the variances can be expanded in $\delta N_{\pm}/N_{\pm}$ and approximated by
\begin{align}
	\langle A_{\pm,UD,stat}^2\rangle &= {1 \over \langle N_{\pm}^2\rangle} \left \langle \left( {\delta N_{\pm,U}-\delta N_{\pm,D} \over 1 + \left( \delta N_{\pm,U}+\delta N_{\pm,D} \right)/\langle N_{\pm}\rangle } \right)^2 \right \rangle \nonumber \\
	&\approx {\langle N_{\pm}\rangle + 1 \over \langle N_{\pm}\rangle^{2}}, 
	\label{eq:stat1}
\end{align}
where $\langle N_{\pm} \rangle =  \langle N_{\pm,U} \rangle + \langle N_{\pm,D} \rangle$, and likewise for $\langle A_{\pm,LR,stat}^2\rangle$.
The multiplicities in equation~\ref{eq:stat1} are the measured multiplicities before any acceptance corrections.
The acceptance corrections cancel in the correlations and do not contribute to the statistical fluctuations.
Example results are shown for $\langle A^2_{+,LR} \rangle$ in $\eta<0$ region in figure~\ref{fig:stat-a} 
and $\eta>0$ region in \ref{fig:stat-a0} as black curves. 
The method  is referred as ``$1/N$'' approximation.
The data are scaled by the number of participants $N_{part}$ to better show the difference.
The approximation works when the multiplicity is large enough, and the multiplicity distribution is close to Poisson distribution.

The Poisson assumption of the multiplicities may not be true in real data.
To improve the estimate, we can count the multiplicities of each charges, and then obtain the statistical fluctuation from real data.
It is done by generating a random number with 50\% chance to flip a particle to its opposite direction in azimuthal angle, 
in other words, randomly flipping a particle to up or down hemisphere, and left or right hemisphere, 
without considering single particle efficiency correction $1 / \epsilon(\phi)$.
After the random flipping, the asymmetries and their correlations are calculated again as the statistical fluctuation and noted as $\langle A^2_{\pm,stat} \rangle$.
Under such process, the charge correlations are completely destroyed within $\langle A^2_{\pm,stat} \rangle$ because the flipping is random and does not depend on charges.
The result has been shown in figure \ref{fig:accasym-a} and \ref{fig:accasym-b} without $\phi$-correction.
We refer this method as the ``50-50'' method, which doesn't require the multiplicities to be Poisson distributions or very large numbers.
\red{
After the random flipping, the single particle distribution does not reflect the detector acceptance efficiency; in other words, it would work only when the detector is perfectly uniform,
because the flipping does not dependent on the detector performance.
}
Thus, this method is an estimation of the pure statistical fluctuation only.
The blue curves in figure~\ref{fig:stat-a} ($\eta<0$) and \ref{fig:stat-a0} ($\eta>0$) are the statistical fluctuations obtained by this method from the same sets of data used in the ``$1/N$'' method.
As shown in the figures, the ``$1/N$'' method underestimates the statistical fluctuation especially in low multiplicity events, 
because the large multiplicity and Poisson distribution assumptions cannot describe the data precisely in those events.

The ratio of the statistical fluctuations between ``50-50'' method and ``$1/N$'' method is shown as the red line 
from $\eta<0$ region and blue line from $\eta>0$ region on figure~\ref{fig:stat-b}
to quantify the difference between the two methods.
\red{
The ``$1/N$'' method underestimates the statistical fluctuation for about several percent in peripheral to mid central collisions, 
and agrees with ``50-50'' method very well in the most central collisions.
It is because the multiplicity Poisson distribution assumption breaks down in low multiplicity events, where the ``$1/N$'' method underestimates.
In high multiplicity events, the ``$1/N$'' method works well where the Poisson distribution holds.
The $\eta>0$ and $\eta<0$ regions show similar effect of the multiplicity off Poisson distributions.
}

The ``50-50'' approximation assumes a perfect detector with uniform distribution in the azimuthal direction.
However, our detector performance is subject to the sector boundaries, hardware efficiencies and other deficit electronics, etc.
As mentioned in section \ref{det}, the non-uniformity has to be corrected for the asymmetries and event-plane reconstructions 
with the average $\phi$ dependent correction factor $1/\epsilon(\phi)$ respectively.
The correction factor can correct the single particle distribution and the event-plane distribution,
however, the event-by-event ``dynamical'' fluctuations introduced by the asymmetry of detector inefficiencies cannot be corrected.
We refer detector induced event-by-event ``dynamical'' fluctuations as ``detector effects'', which have to be corrected to obtain the physics dynamics.
To assess the detector effects, we modify the ``50-50'' method to account for the detector non-uniformity.

To do so is, for each particle, we still flip its azimuthal angle by adding $\pi$ to its azimuthal angle, or do nothing.
But instead of 50\% chance of flipping as we did in ``50-50'' method, the probability for adding $\pi$ to the azimuthal angle
is determined by the relative azimuthal acceptance $\times$ efficiency $1/\epsilon(\phi)$ and $1/\epsilon(\phi+\pi)$.
For example a particle with azimuthal angle $\phi$, the chance of flipping it to its opposite direction is 
$\epsilon(\phi+\pi)/(\epsilon(\phi)+\epsilon(\phi+\pi))$.
\red{
We then calculate the asymmetries and their correlations after the random flipping and note that as $\langle A^2_{\pm,stat+det} \rangle$.
The result has been shown in figure \ref{fig:accasym-a} and figure \ref{fig:accasym-b} with $\phi$-correction.
The method is referred as adding-$\pi$ method, and the selected correlations are shown with red circles in figure \ref{fig:stat-a} for $\langle A^2_{+,LR,stat+det} \rangle$ of $\eta<0$ region and figure \ref{fig:stat-a0} of $\eta>0$ region.
The magnitude of $\langle A^2_{stat+det} \rangle$ is larger than the pure statistical fluctuation $\langle A^2_{stat} \rangle$.
We show the ratios of the two in figure \ref{fig:stat-b} with black ($\eta>0$ region) and green ($\eta<0$) lines.
In the figure, we can see that the detector effect is a few percentage larger than the pure statistical fluctuation.
The detector effect is larger in the most central collisions because the particle multiplicity is larger in more central collisions, therefore, the detector effect and tracking inefficiency are more significant.
Also note, the ratio of $\eta<0$ region in green line is slightly larger than the black link which is the $\eta>0$ region, because the detector deficit electronics effect induces the ``dynamic'' correlations in this region.
}

\red{
Using Monte-Carlo simulation is another way to assess the statistical fluctuation plus the detector effects.
For a given event, we generate the same amount of particles as the TPC recorded with random azimuthal angles according the corresponding $\phi$ distribution.
During the particle generation, we keep other parameters unchanged, such as $p_T$ and charge.
We use the generated events as mixed event to calculate the charge asymmetries and their correlations.
Since the Monte-Carlo simulation does not depend on the particle charges, the mixed event asymmetry correlations are actually the measurement of the statistical fluctuation.
Because we also use the detector acceptance to mimic the detector effect, the asymmetry correlations from the mixed events also take care of the detector effects.
We refer this method as scramble method, and the selected results are shown in figure \ref{fig:stat-a} for $\langle A^2_{+,LR,stat+det} \rangle$ in $\eta<0$ region and in figure \ref{fig:stat-a0} in $\eta>0$ region with green crosses.
The results show good consistency with the adding-$\pi$ method for all charge, $\eta$, and $UD$ and $LR$ combinations.
}

\red{
For other statistical fluctuation and detector effects of different methods comparison, please refer to figure \ref{fig:appstat} for positive charge and figure \ref{fig:appstatn} for negative charge.
They are further separated by $\eta > 0$ and $\eta<0$ regions and $UD$/$LR$ directions.
To check for consistency, we plot the ratios of the $\langle A^2_{stat+det} \rangle$ between the positive and negative charges and the $UD$ and $LR$ directions in figure \ref{fig:stat-c}, data points are obtained by adding-$\pi$ method.
Since the events are divided into hemispheres, the total multiplicity of the charged particles are same in $UD$ and $LR$ directions.
In principle, $\langle A^2_{UD,stat+det} \rangle$ and $\langle A^2_{LR,stat+det} \rangle$ are the same, which is shown with the triangle symbols for positive charge in $\eta>0$ and $\eta<0$ regions.
The ratios are consistent with unity for all centralities.
However, the ratios between positive and negative charges, shown in circles, are consistent with 98\% instead of unity for all centralities.
This is because the total multiplicity of the positively and negatively charged particle average multiplicities are different due to isospin asymmetry.
We collide gold nuclei with net charge of positive 79 from each gold ion.
In the TPC covered $-1<\eta<1$ region, the total multiplicity of recorded positively charged particles is then roughly 2\% larger than the multiplicity of negative charges.
As we mentioned, the statistical fluctuation is approximately proportional to the inverse of the total multiplicity.
Then, the ratio of between $\langle A^2_{+,LR,stat+det} \rangle$ and $\langle A^2_{-,LR,stat+det} \rangle$ is off unity by 2\%.
We also show the ratio of positive and negative charges using ``$1/N$'' approximation in black line, which is consistent with the adding-$\pi$ method.
}

\red{
For simplicity and computing efficiency reason, we use adding-$\pi$ method for our statistical fluctuation and detector effects measurement in this analysis.
We have calculated the adding-$\pi$ results in figure \ref{fig:accasym-c} to demonstrate the stat+det for covariances are consistent with zero.
}

\section{Consistency Check}
\label{consist}

\begin{figure}[thb]
	\begin{center}
		\subfigure[$\langle A^{2}_{+,UD} \rangle$ and $\langle A^{2}_{+,UD,stat+det} \rangle$]{\label{fig:cons-a}\includegraphics[width=0.4\textwidth]{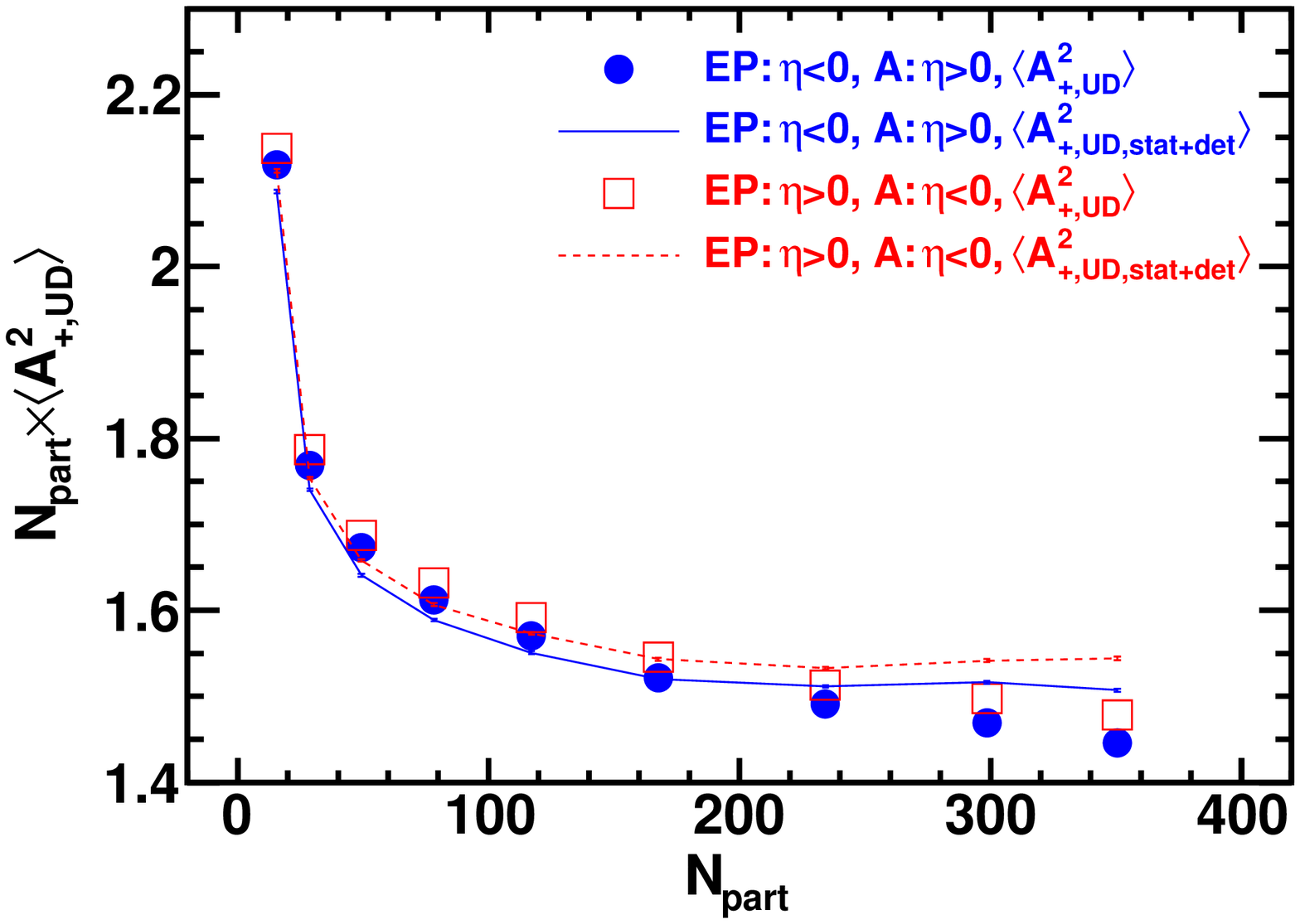}}
		\subfigure[$\langle A^{2}_{+,LR} \rangle$ and $\langle A^{2}_{+,LR,stat+det} \rangle$]{\label{fig:cons-b}\includegraphics[width=0.4\textwidth]{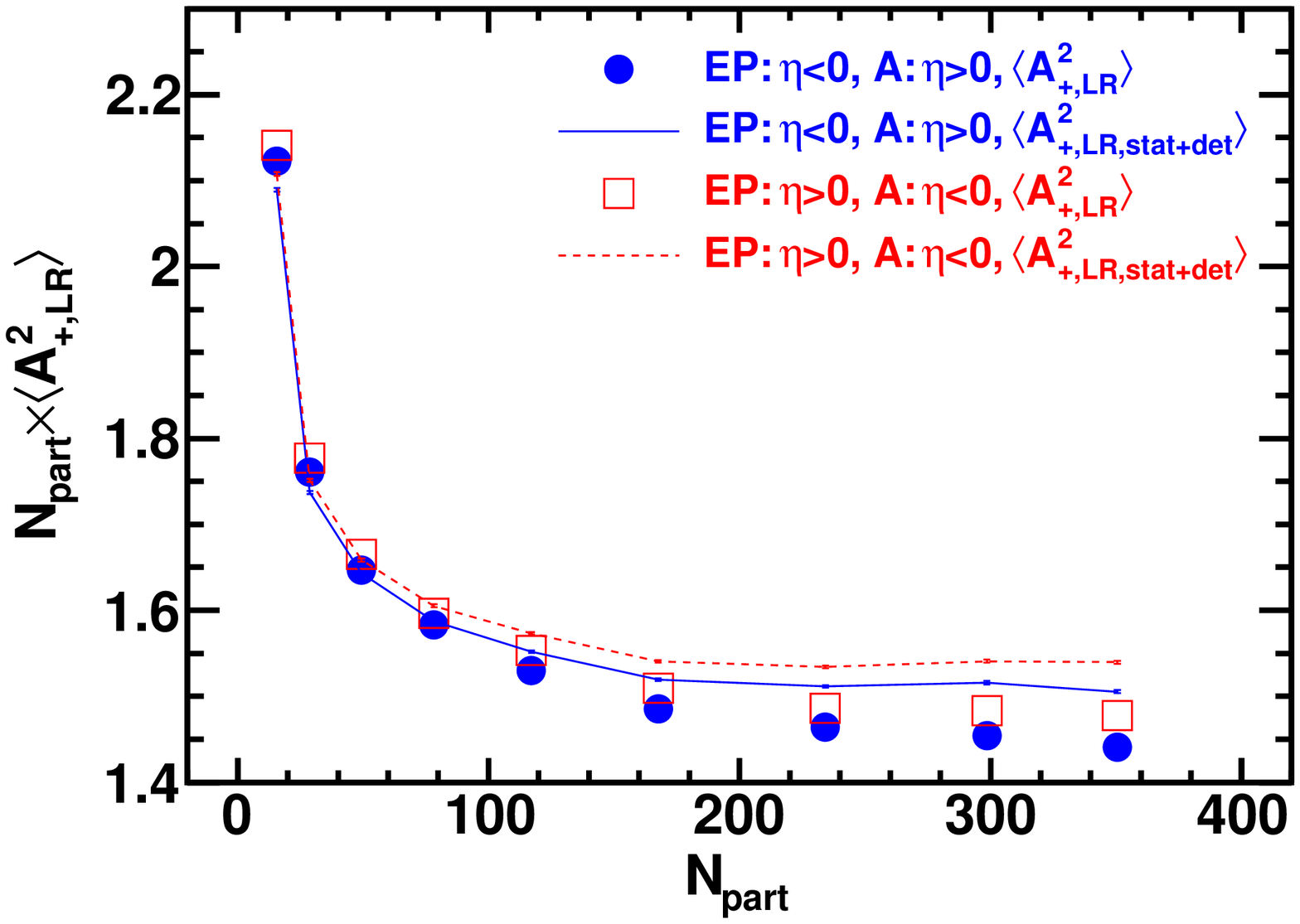}}
		\subfigure[$\langle A^{2}_{-,UD} \rangle$ and $\langle A^{2}_{-,UD,stat+det} \rangle$]{\label{fig:cons-c}\includegraphics[width=0.4\textwidth]{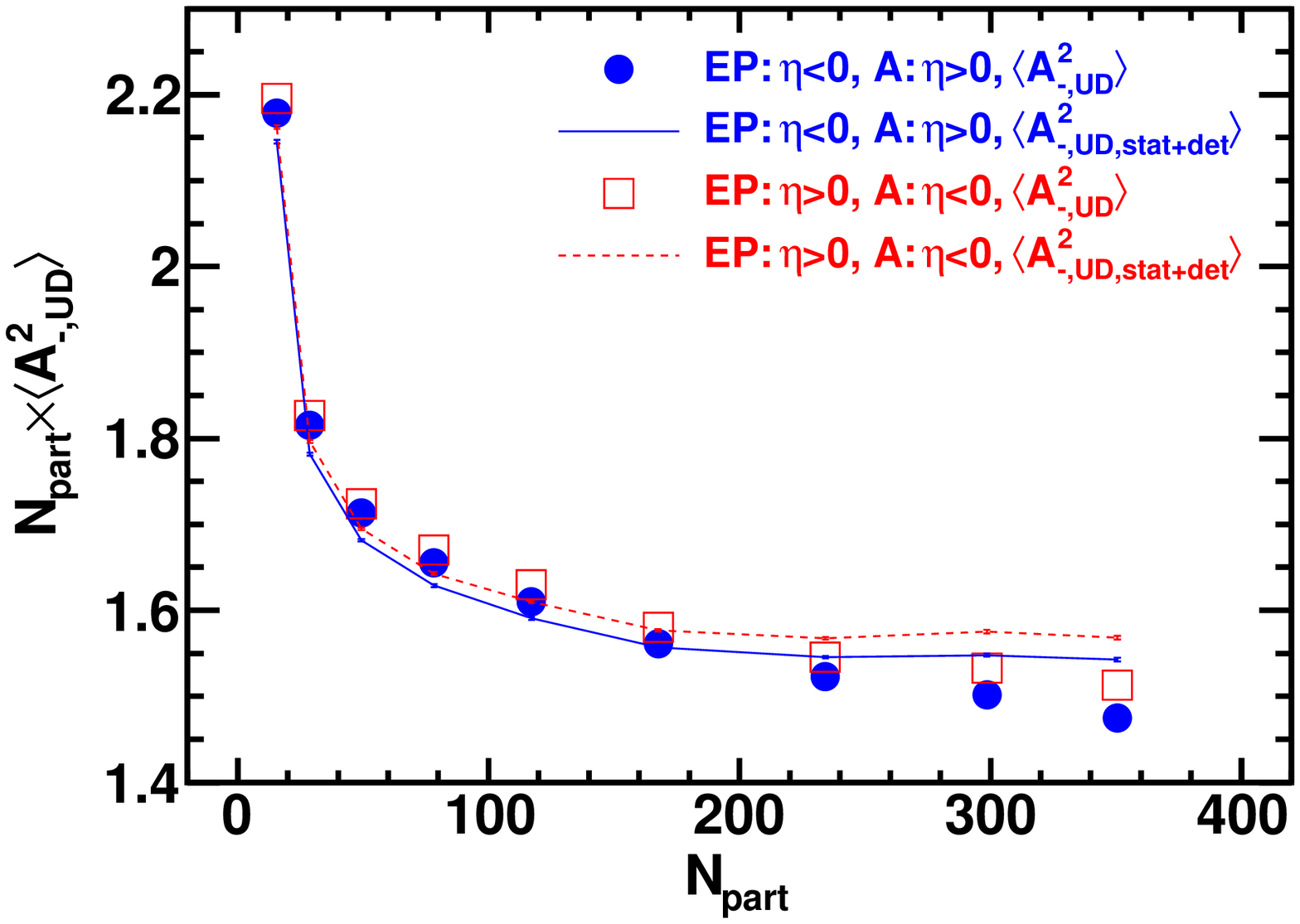}}
		\subfigure[$\langle A^{2}_{-,LR} \rangle$ and $\langle A^{2}_{-,LR,stat+det} \rangle$]{\label{fig:cons-d}\includegraphics[width=0.4\textwidth]{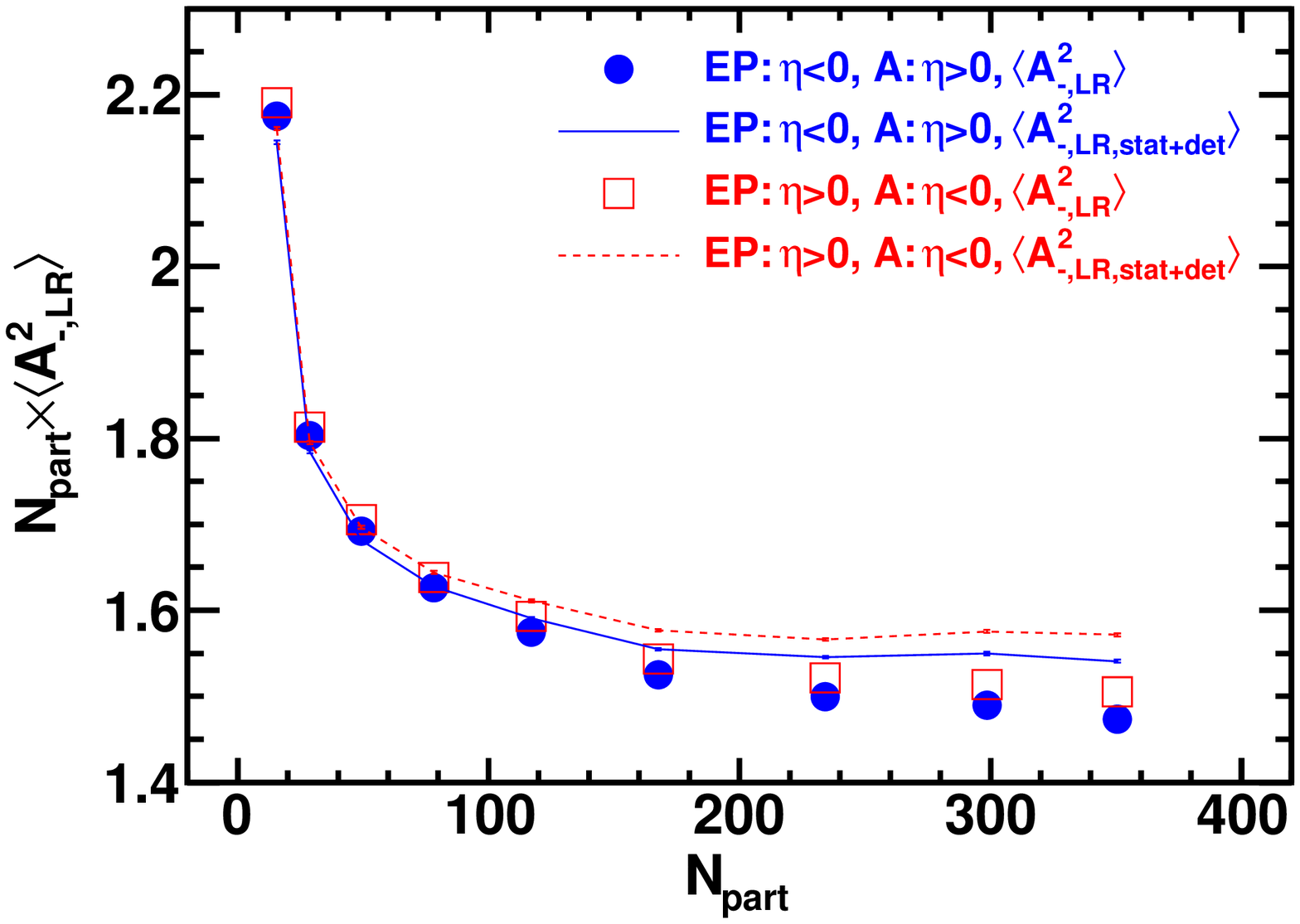}}
		\subfigure[$\delta\langle A^{2}_{UD} \rangle$]{\label{fig:cons-e}\includegraphics[width=0.4\textwidth]{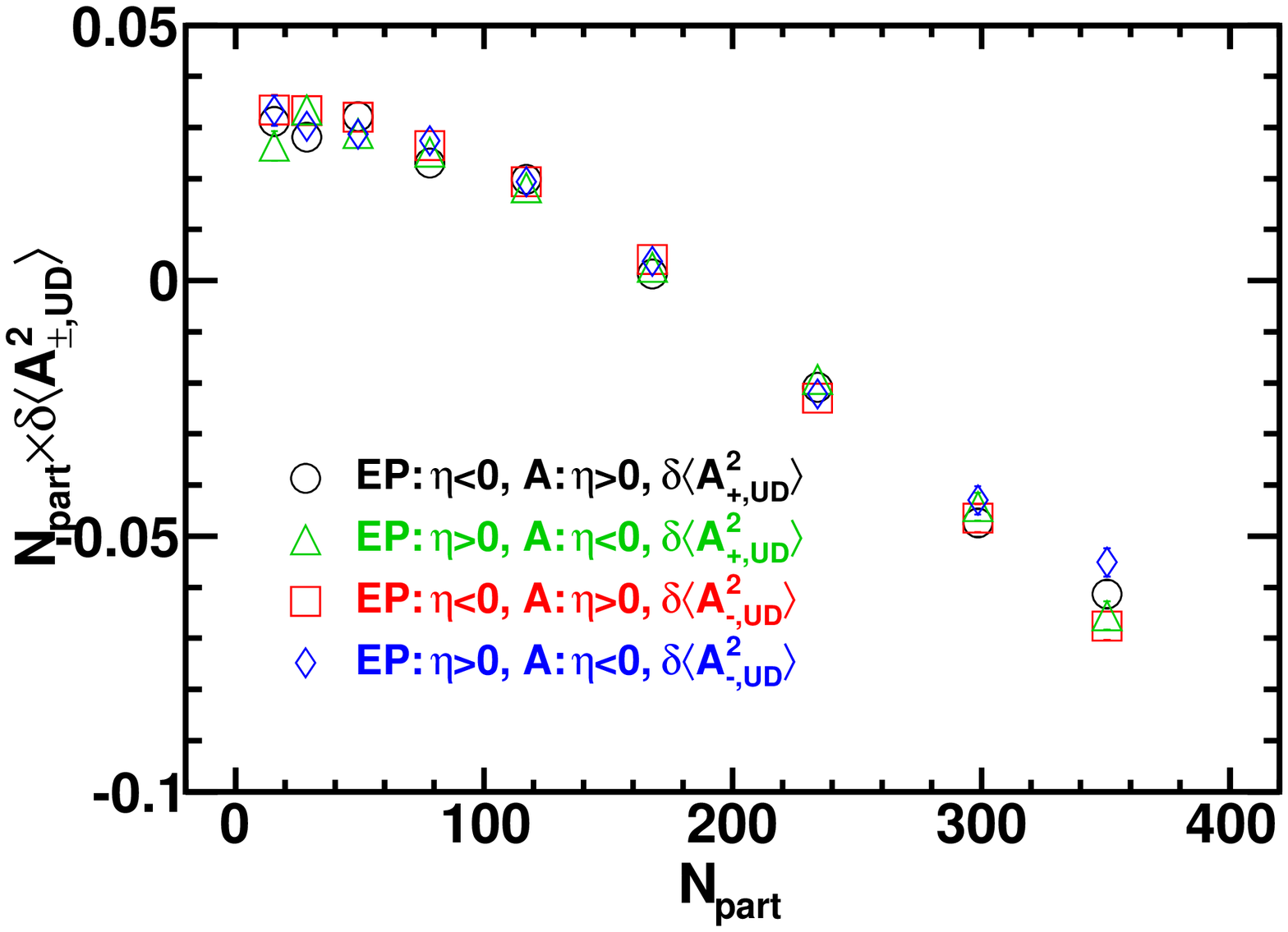}}
		\subfigure[$\delta\langle A^{2}_{LR} \rangle$]{\label{fig:cons-f}\includegraphics[width=0.4\textwidth]{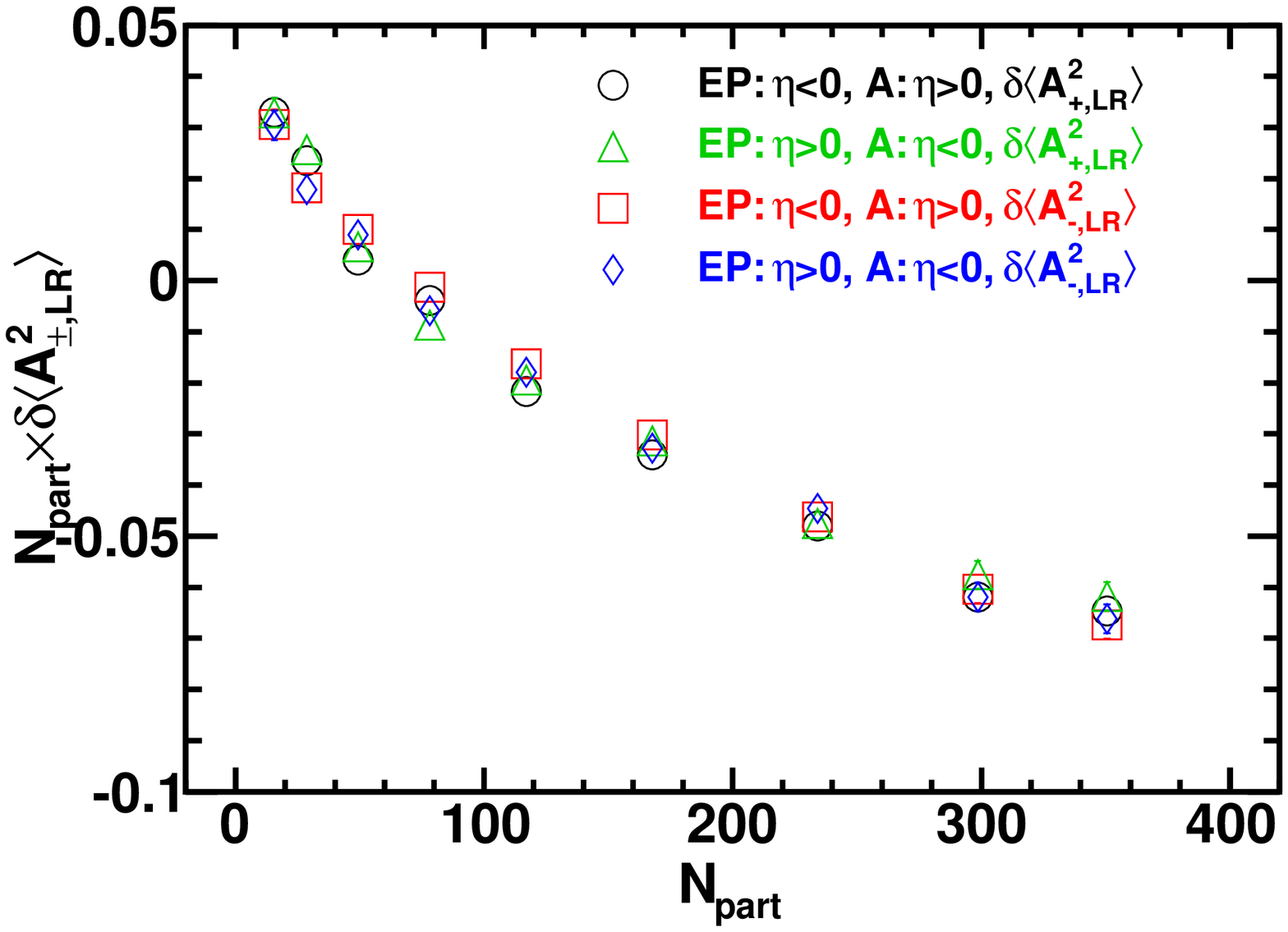}}
	\end{center}
	\caption[Consistency check of the variances]{Consistency check of the variances as a function of centrality for panel (a): $\langle A^{2}_{+,UD} \rangle$, (b): $\langle A^{2}_{+,LR} \rangle$,
	(c): $\langle A^{2}_{-,UD} \rangle$, (d): $\langle A^{2}_{-,LR} \rangle$ and their statistical fluctuation and detector effects.
	And the dynamic variances panel (e): $\delta\langle A^{2}_{UD} \rangle$ and (f): $\delta\langle A^{2}_{LR} \rangle$.
	The data are RUN IV 200 GeV Au+Au collisions with particle $p_T$ range of $0.15<p_T <2$ GeV/$c$.
	All correlations are scaled by $N_{part}$.}
	\label{fig:consA2}
\end{figure}

In the final result, the charge asymmetry correlations are averaged over different pseudo-rapidity regions for variances and covariances,
and also for different charge combinations for variances.
In this section, we check all the individual asymmetry correlations to make sure they are consistent with each other.

Figure \ref{fig:consA2} shows the asymmetry variances with their statistical fluctuation and detector effects superimposed in lines.
We show $\langle A^2_{+,UD} \rangle$ in figure \ref{fig:cons-a}), $\langle A^2_{+,LR} \rangle$ in figure \ref{fig:cons-b}),  $\langle A^2_{-,UD} \rangle$ in figure \ref{fig:cons-c} and $\langle A^2_{-,LR} \rangle$ in figure \ref{fig:cons-d} with $p_T$ range of $0.15 < p_T < 2.0 $ GeV/$c$ as a function of $N_{part}$.
The red data points and lines denote the variances and their statistical fluctuation and detector effects from $\eta<0$ region, and blue ones are from $\eta>0$ region.
The asymmetry variances are scaled by $N_{part}$ to better show the differences.
\red{
From the plots we can see all the asymmetry correlations are with similar magnitude and centrality dependence.
They are largely dominated by the statistical fluctuations.
All variances in $\eta<0$ region (red squares) are slightly larger than the corresponding variances in $\eta>0$ region (blue circles).
This is because of the detector efficiency deficit in $\eta<0$ region.
The estimate of statistical fluctuation plus detector effects using adding-$\pi$ method is also shown in lines in each plot.
The red line shows the $\eta<0$ region, which is also slightly larger than the blue line ($\eta>0$ region) because of the same reason.

After subtracting the statistical fluctuation plus detector effects, we obtain the dynamical variances in $UD$ direction, which are shown in figure \ref{fig:cons-e},
and in $LR$ directions which are shown in figure \ref{fig:cons-f}.
The dynamical variances show good consistency between the different pseudo-rapidity regions and different charges.
Therefore, we take the average over the different combinations to increase the statistics.
}

\begin{figure}[thb]
	\begin{center}
		\psfig{figure=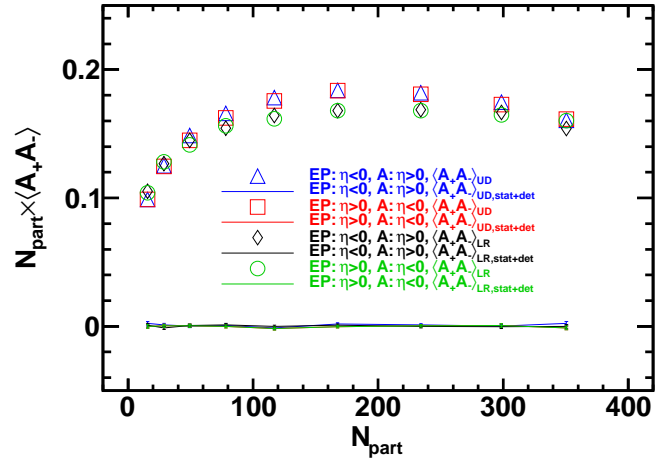,width=0.6\textwidth}
	\end{center}
	\caption[Consistency check of the covariances]{Consistency check of the covariances and their statistical fluctuation and detector effect scaled by $N_{part}$ as a function of $N_{part}$.
	The data are RUN IV 200 GeV Au+Au collisions with particle $p_T$ range of $0.15 < p_T < 2.0$ GeV/$c$.
	}
	\label{fig:consAA}
\end{figure}

\red{
The covariance consistency check is shown in figure \ref{fig:consAA}.
The statistical fluctuation plus detector effects are consistent with zero for all centralities and combinations, so we do not have to subtract them.
Again we take the average of the different combinations.
}

\section{Systematic Uncertainties}
\label{systematic}

\red{
In order to study systematic effects, we have studied various conditions and cuts made in the analysis.
}

\subsection{Sanity checks}

\begin{figure}[thb]
	\begin{center}
		\subfigure[Dynamical asymmetry correlations]{\label{fig:rndep-a} \includegraphics[width=0.45\textwidth]{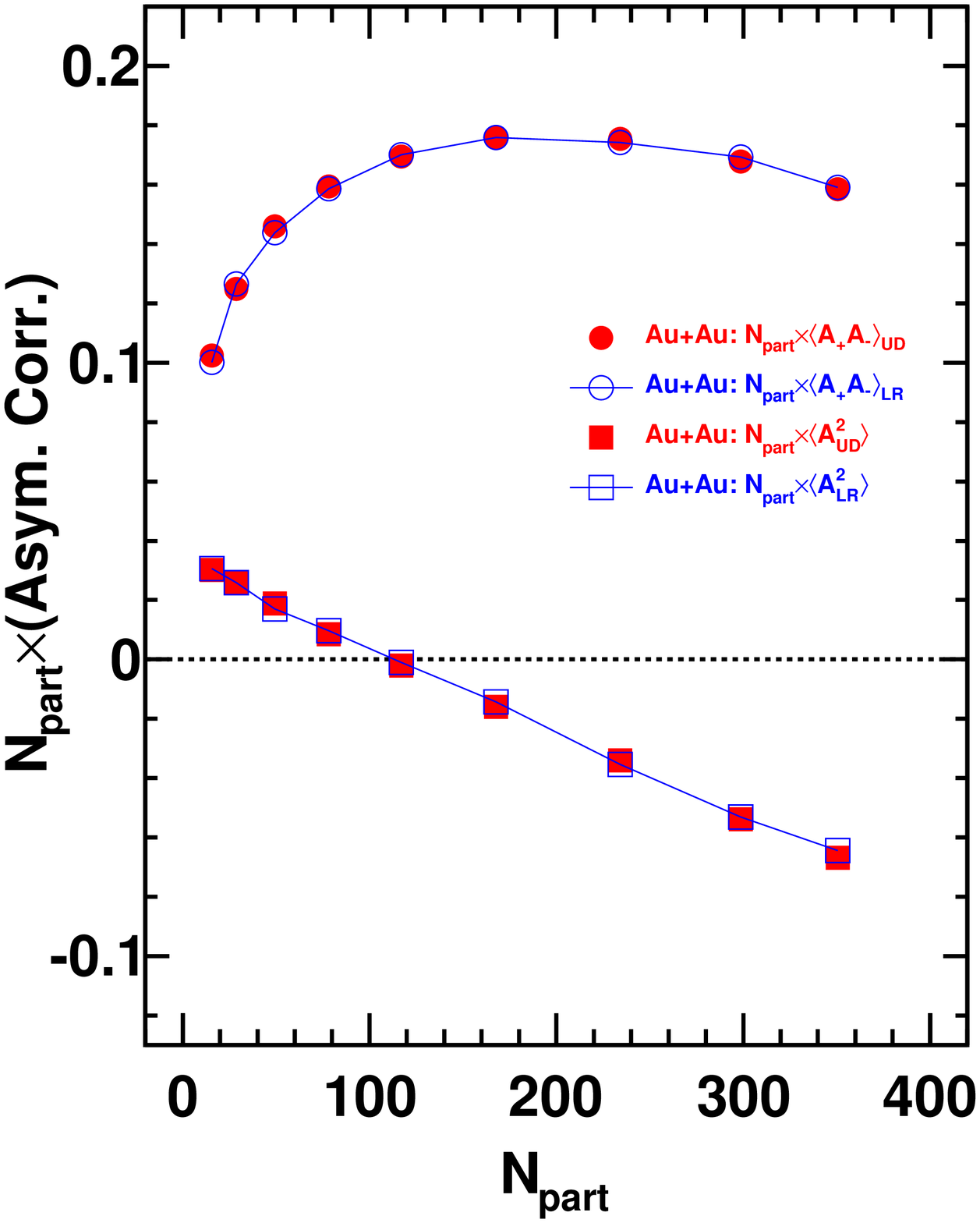}}
		\subfigure[$UD-LR$ asymmetry correlations]{\label{fig:rndep-b} \includegraphics[width=0.45\textwidth]{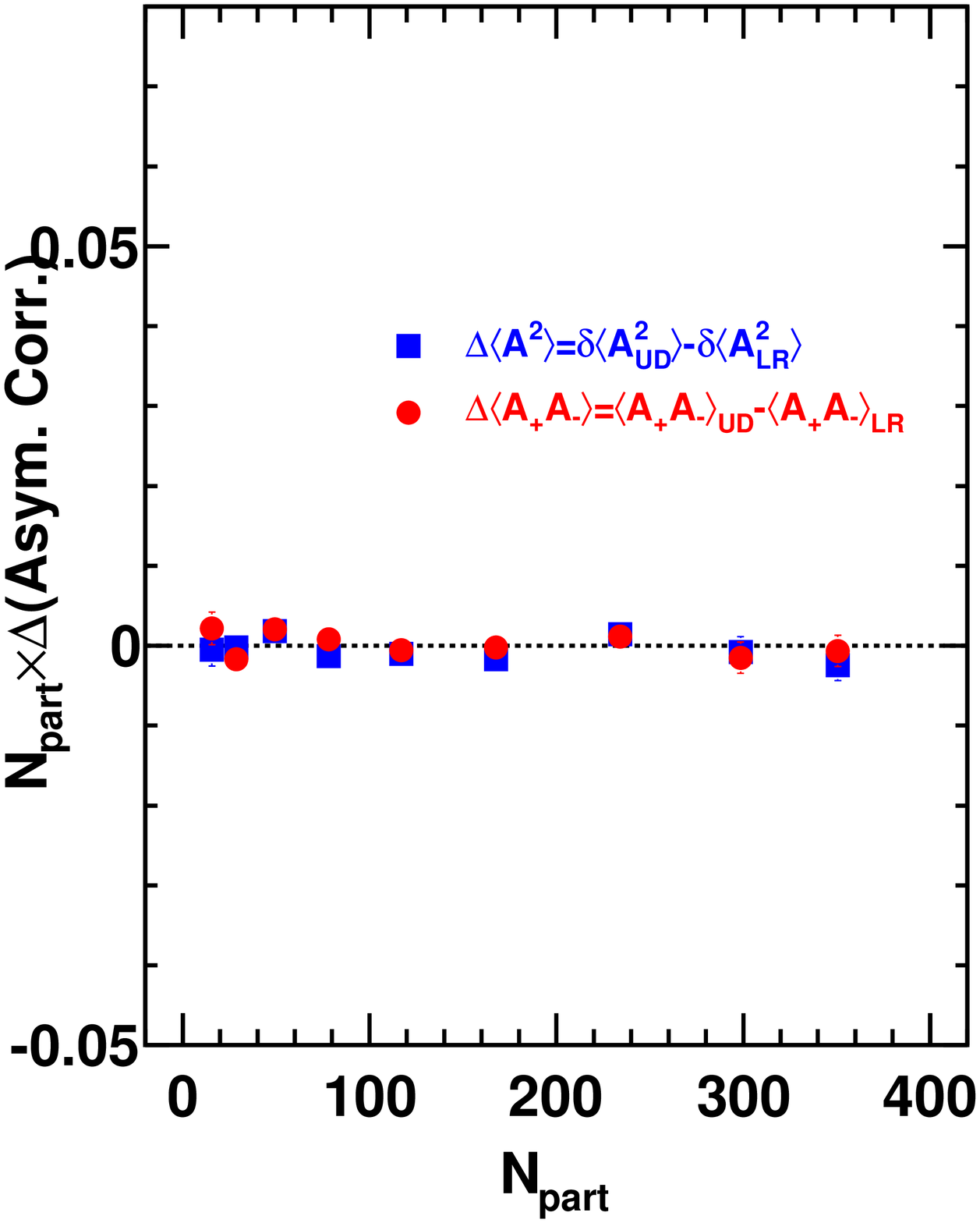}}
	\end{center}
	\caption[Charge asymmetry correlations with random EP]
	{Panel (a): Centrality dependences of the charge multiplicity dynamical correlations $\delta\langle A^{2} \rangle$ and the opposite-sign correlations $\langle A_{+}A_{-} \rangle$ for RUN IV 200 GeV Au+Au collisions with respect to random EP.
	Panel (b): The $UD-LR$ correlations of the same-sign and opposite-sign correlations.
	The asymmetry correlations are scaled by the number of participants $N_{part}$ to better show the magnitude.
	The particle $p_{T}$ range is integrated over $0.15 < p_T < 2.0$ GeV/$c$ for the asymmetry calculation.
	Error bars are statistical errors only.
	}
	\label{fig:rndep}
\end{figure}

\red{
To show the charge asymmetry correlations are indeed caused by the event-plane effect,
we use random event-plane to check the asymmetry correlations.
The analysis procedure is the same, except we generate a uniform distributed number $\phi_{EP}$ in range of $0 < \phi_{EP} < 2\pi$ as the event-plane azimuthal angle.
The asymmetries and their correlations are then calculated with respect to the randomly generated event-plane direction.
The results are shown in figure \ref{fig:rndep}, where the asymmetry correlations are scaled by $N_{part}$ to better show the scale.
Because the event-plane is random, the asymmetries of $UD$ and $LR$ cannot be distinguished.
Thus the asymmetry correlations are then identical for $UD$ and $LR$, which is shown in figure \ref{fig:rndep-a}.
We can see that both the same-sign and opposite-sign asymmetry correlations are consistent between $UD$ and $LR$.
Figure \ref{fig:rndep-b} shows the difference between $UD$ and $LR$ correlations.
They are consistent with zero for the same-sign and opposite-sign correlations,
i.e. $\langle A_+A_- \rangle_{UD} = \langle A_+A_- \rangle_{LR}$ and $\langle A^2_{UD} \rangle = \langle A^2_{LR} \rangle$.
These results give us confidence that the charge asymmetry correlations we present are indeed caused by the event-plane.
The asymmetry correlation differences between the $UD$ and $LR$ are indeed correlated to the event initial geometry configuration.
}

\begin{figure}[thb]
	\begin{center}
		\subfigure[Dynamical asymmetry correlations]{\label{fig:eprotate-a} \includegraphics[width=0.45\textwidth]{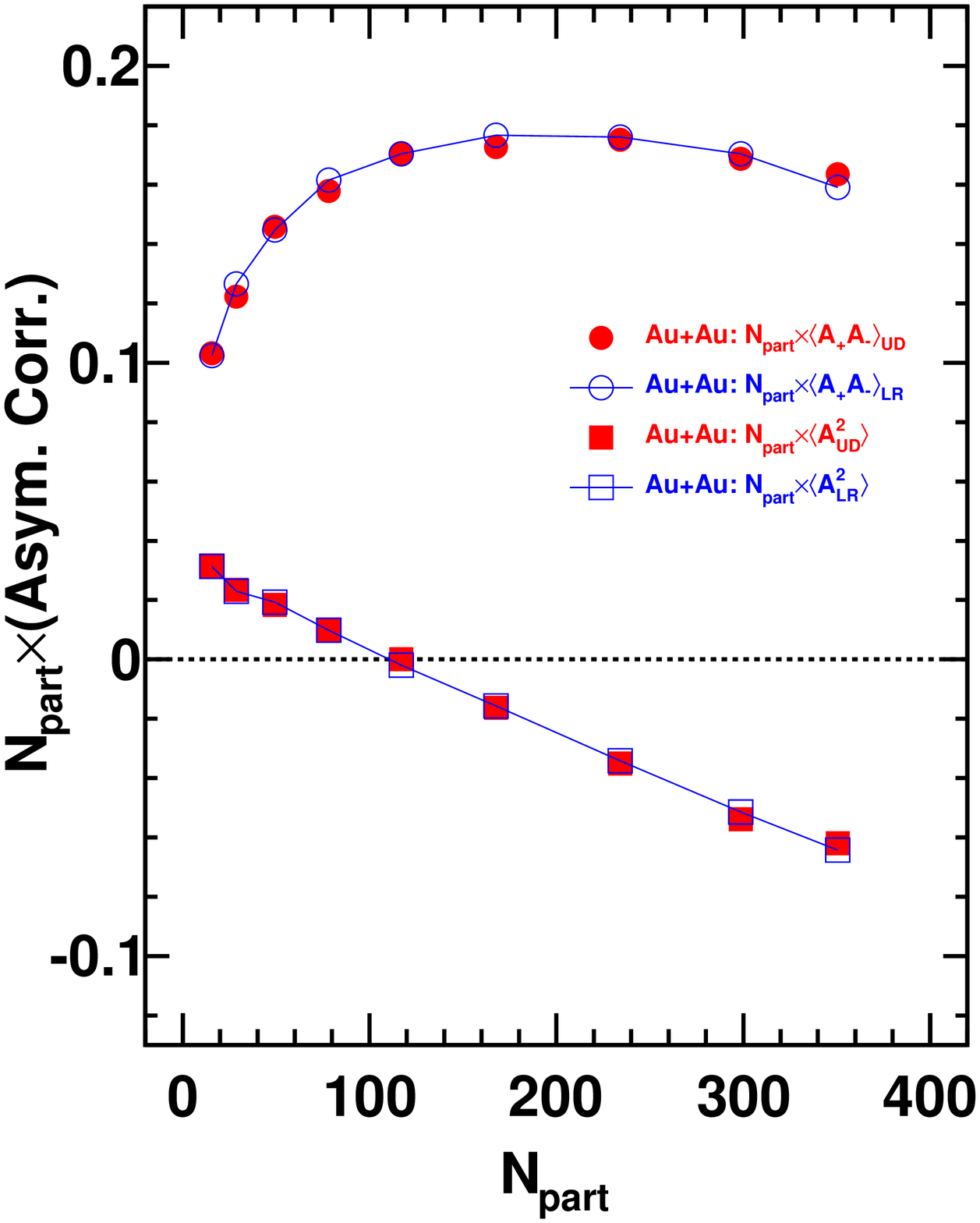}}
		\subfigure[$UD-LR$ asymmetry correlations]{\label{fig:eprotate-b} \includegraphics[width=0.45\textwidth]{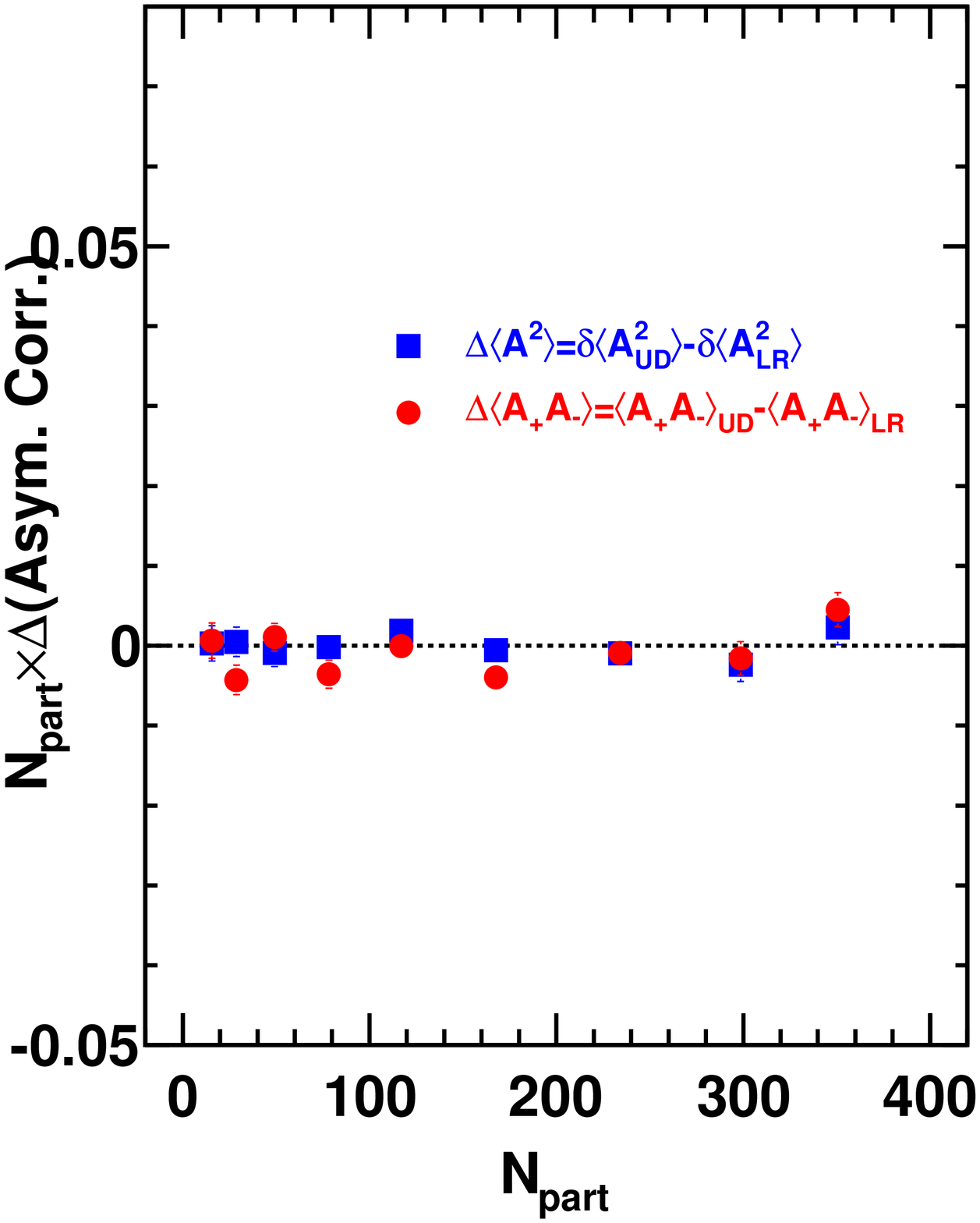}}
	\end{center}
	\caption[Charge asymmetry correlations with rotated EP]
	{Panel (a): Centrality dependences of the charge multiplicity dynamical correlations $\delta\langle A^{2} \rangle$ and the opposite-sign correlations $\langle A_{+}A_{-} \rangle$ for RUN IV 200 GeV Au+Au collisions with respect to EP rotated by $\pi/4$.
	Panel (b): The $UD-LR$ correlations of the same-sign and opposite-sign correlations.
	The asymmetry correlations are scaled by the number of participants $N_{part}$ to better show the magnitude.
	The particle $p_{T}$ range is integrated over $0.15 < p_T < 2.0$ GeV/$c$ for the asymmetry calculation.
	Error bars are statistical errors only.
	}
	\label{fig:eprotate}
\end{figure}

\red{
We also rotate the reconstructed event-plane by $\pi/4$ counterclockwise to study the systematic uncertainty.
The asymmetries and their correlations are then calculated with respect to the rotated event-plane.
The results are shown in figure \ref{fig:eprotate}.
After rotation, the $UD$ and $LR$ are shifted, so that the new $UD$ and $LR$ hemispheres are equally mixed by the original $UD$ and $LR$ hemispheres.
As expected, the asymmetry correlations are destroyed, $UD$ and $LR$ correlations cannot be distinguished as seen in figure \ref{fig:eprotate-a}.
The differences between $UD$ and $LR$ are consistent with zero, and show no centrality dependence in figure \ref{fig:eprotate-b}.
}

\begin{figure}[thb]
	\begin{center}
		\subfigure[Asymmetry correlations within $|\eta|<0.5$]{\label{fig:asymeta2-a} \includegraphics[width=0.45\textwidth]{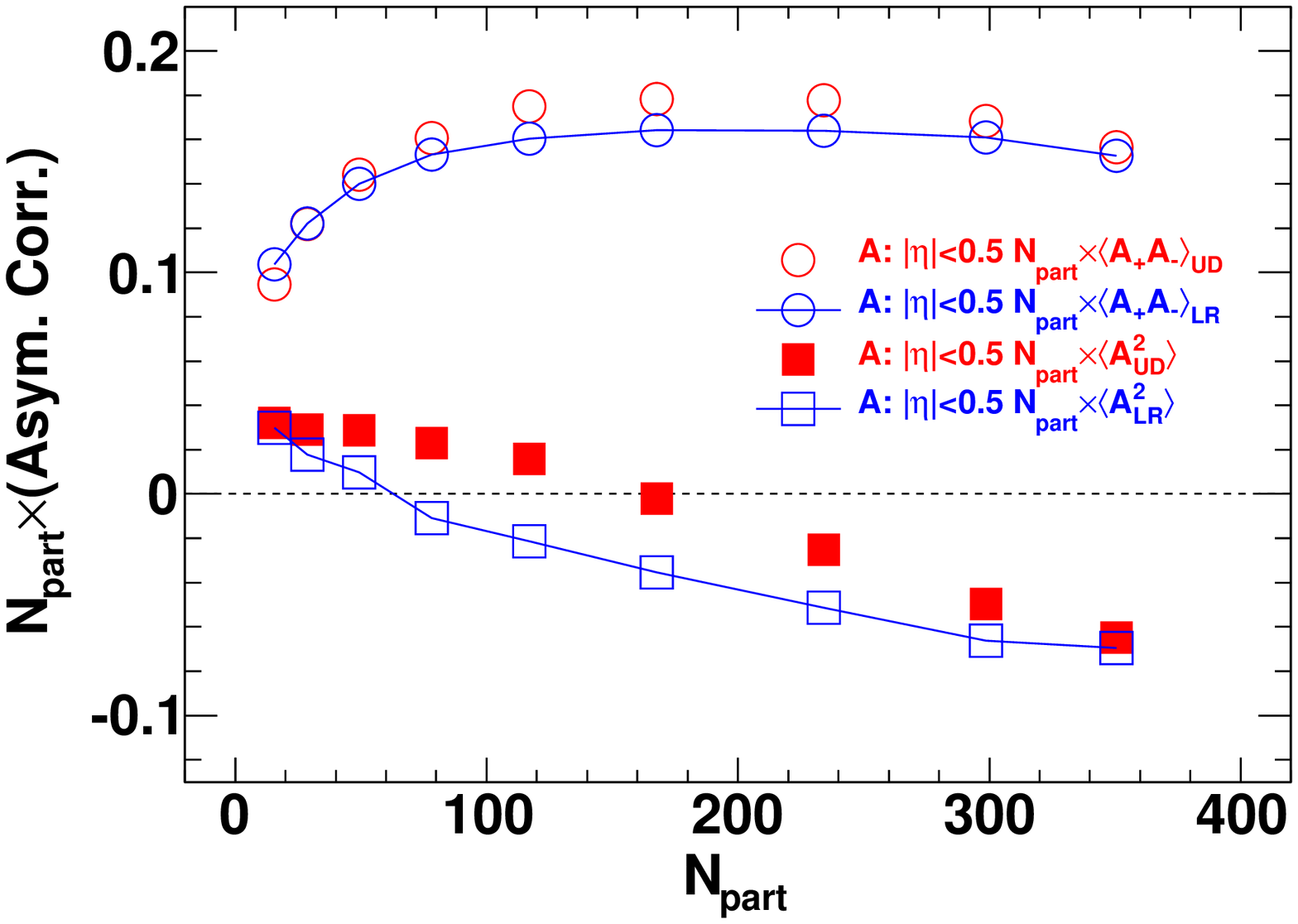}}
		\subfigure[Correlation differences between $|\eta|<0.5$ and half TPC sub-events]{\label{fig:asymeta2-b} \includegraphics[width=0.45\textwidth]{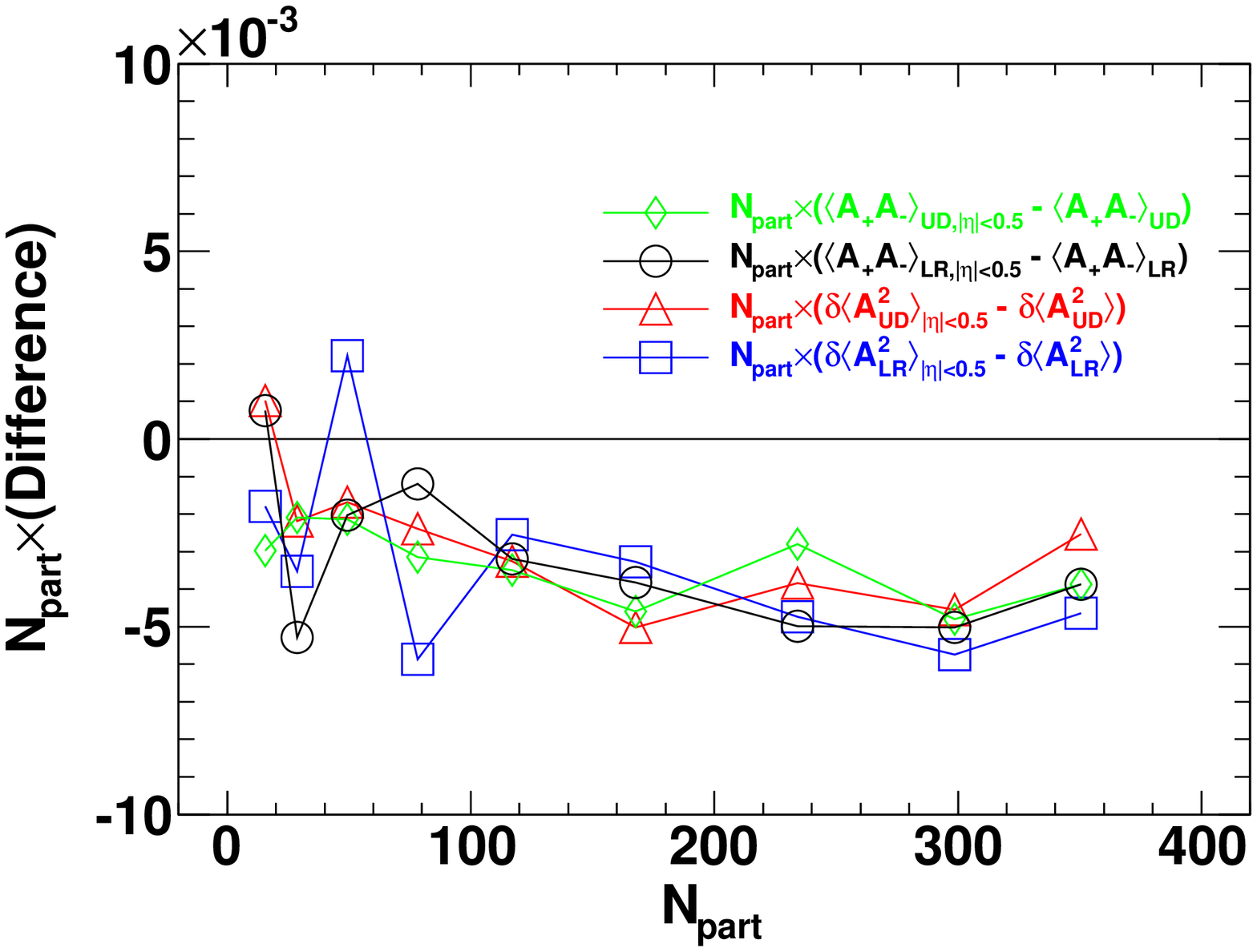}}
		\subfigure[$UD-LR$ correlation differences between $|\eta|<0.5$ and half TPC sub-events]{\label{fig:asymeta2-c} \includegraphics[width=0.45\textwidth]{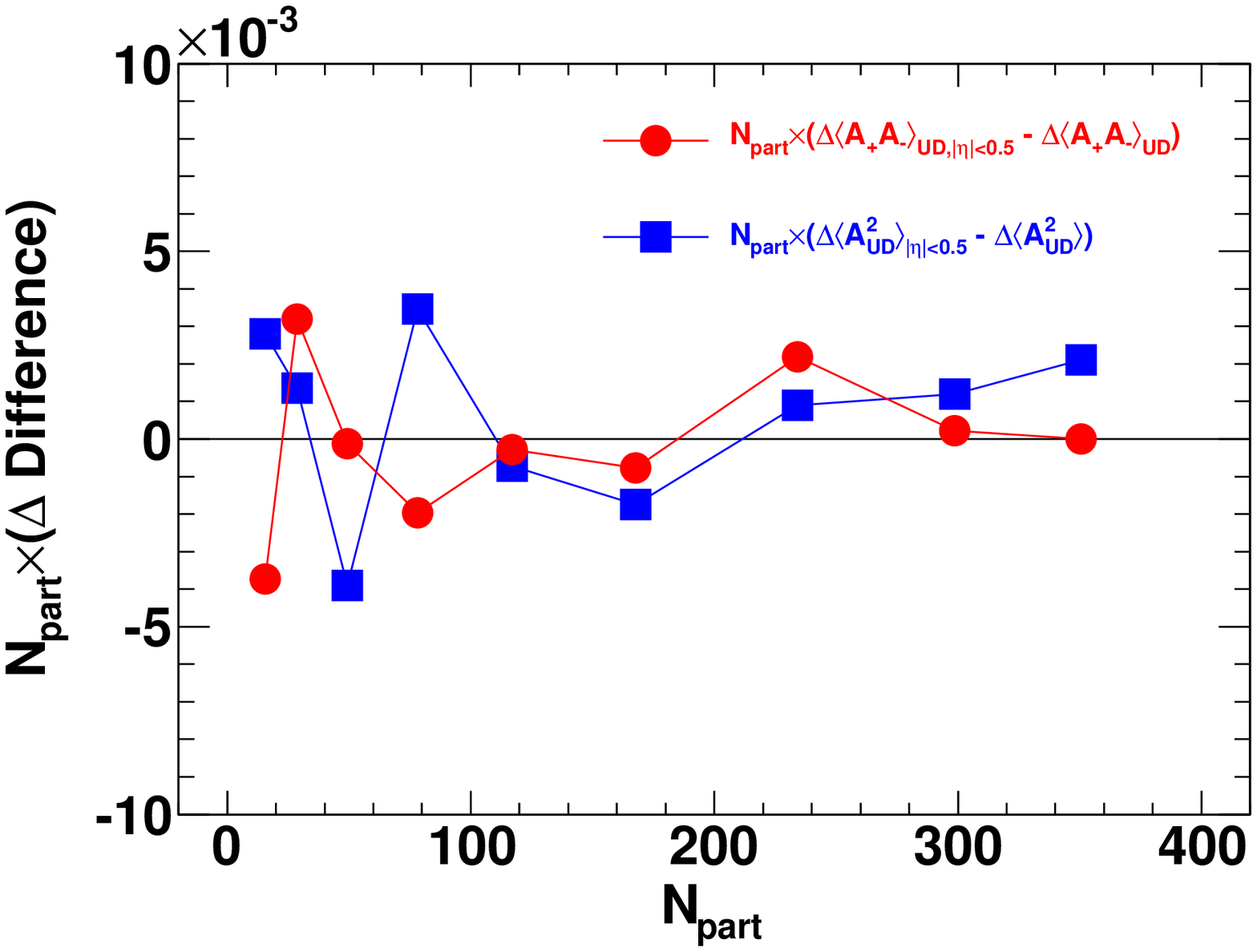}}
	\end{center}
	\caption[Directed flow systematics]{
	Panel (a): Asymmetry correlations scaled by number of participants $N_{part}$ as a function of centrality for particles from $|\eta|<0.5$ region and event-plane reconstructed from particles from $0.5<|\eta|<1.0$ region.
	Panel (b): The differences between the correlations from $|\eta|<0.5$ regions and the half TPC sub-event correlations (figure \ref{fig:asym}).
	Panel (c): The differences between the $UD-LR$ correlations from $|\eta|<0.5$ regions and the half TPC sub-event correlations.
	}
	\label{fig:asymeta2}
\end{figure}

\red{
To estimate the directed flow effects, we separate the event with pseudo-rapidity range of $|\eta|<0.5$ and $0.5<|\eta|<1.0$.
Since the directed flow is an odd function of $\eta$, and close to zero in $|\eta|<0.5$ region, we calculate the asymmetries and their correlations from sub-event in $|\eta|<0.5$ with respect to the event-plane reconstructed from $0.5<|\eta|<1.0$ region.
The results are shown in figure \ref{fig:asymeta2-a}, which are consistent with using $\eta<0$ and $\eta>0$ separated sub-event results in figure \ref{fig:asym}.
For better comparison, we show the difference between the two in figure \ref{fig:asymeta2-b}, and also the differences between the $UD-LR$ correlations in \ref{fig:asymeta2-c}.
The difference is close to zero, and does not depend on centrality.
}

\subsection{Analysis Cut Variations}

\begin{figure}[thb]
	\begin{center}
		\subfigure[Systematics of $\delta\langle A^2 \rangle$]{\label{fig:sysA2} \includegraphics[width=0.45\textwidth]{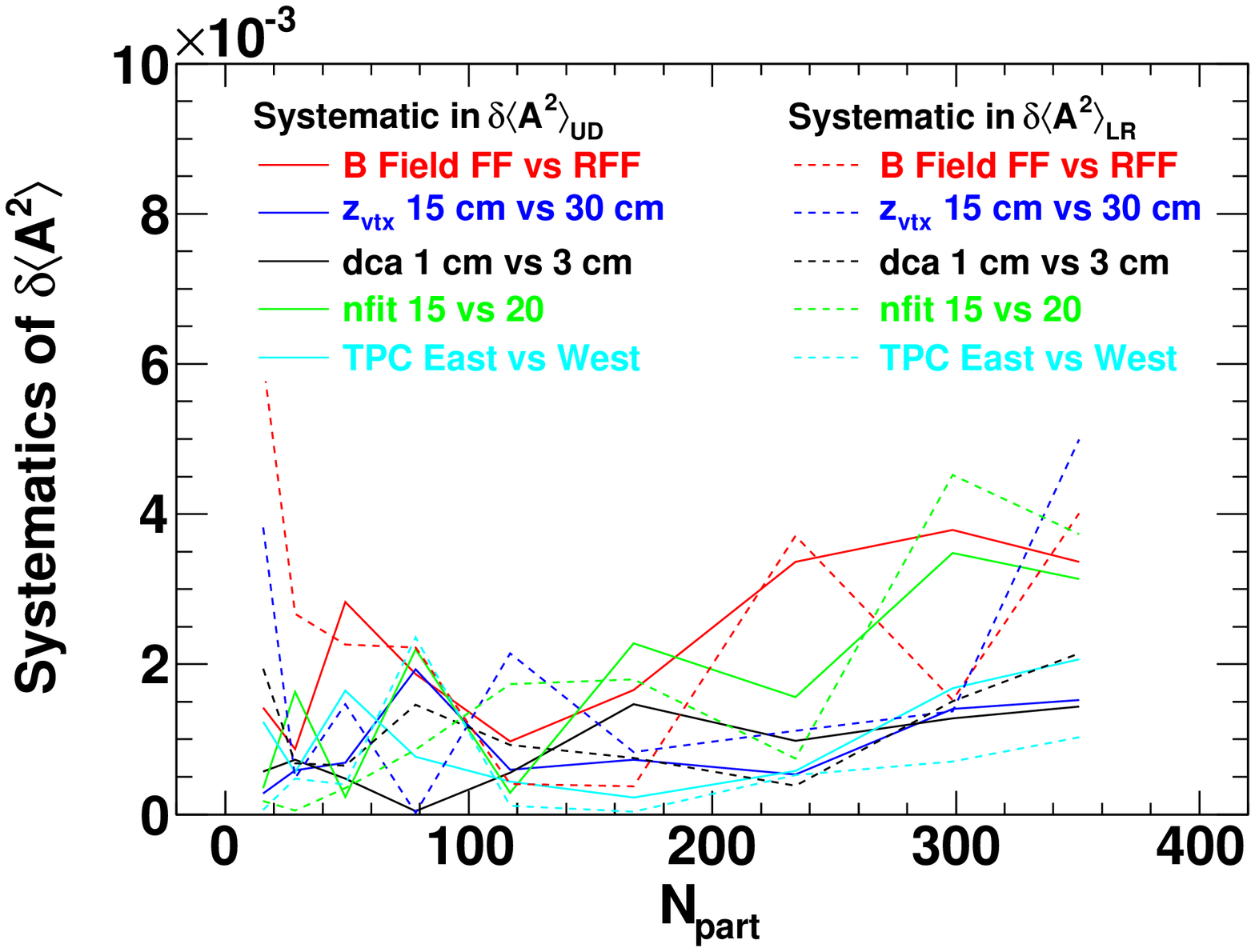}}
		\subfigure[Systematics of $\langle A_+A_- \rangle$]{\label{fig:sysAA} \includegraphics[width=0.45\textwidth]{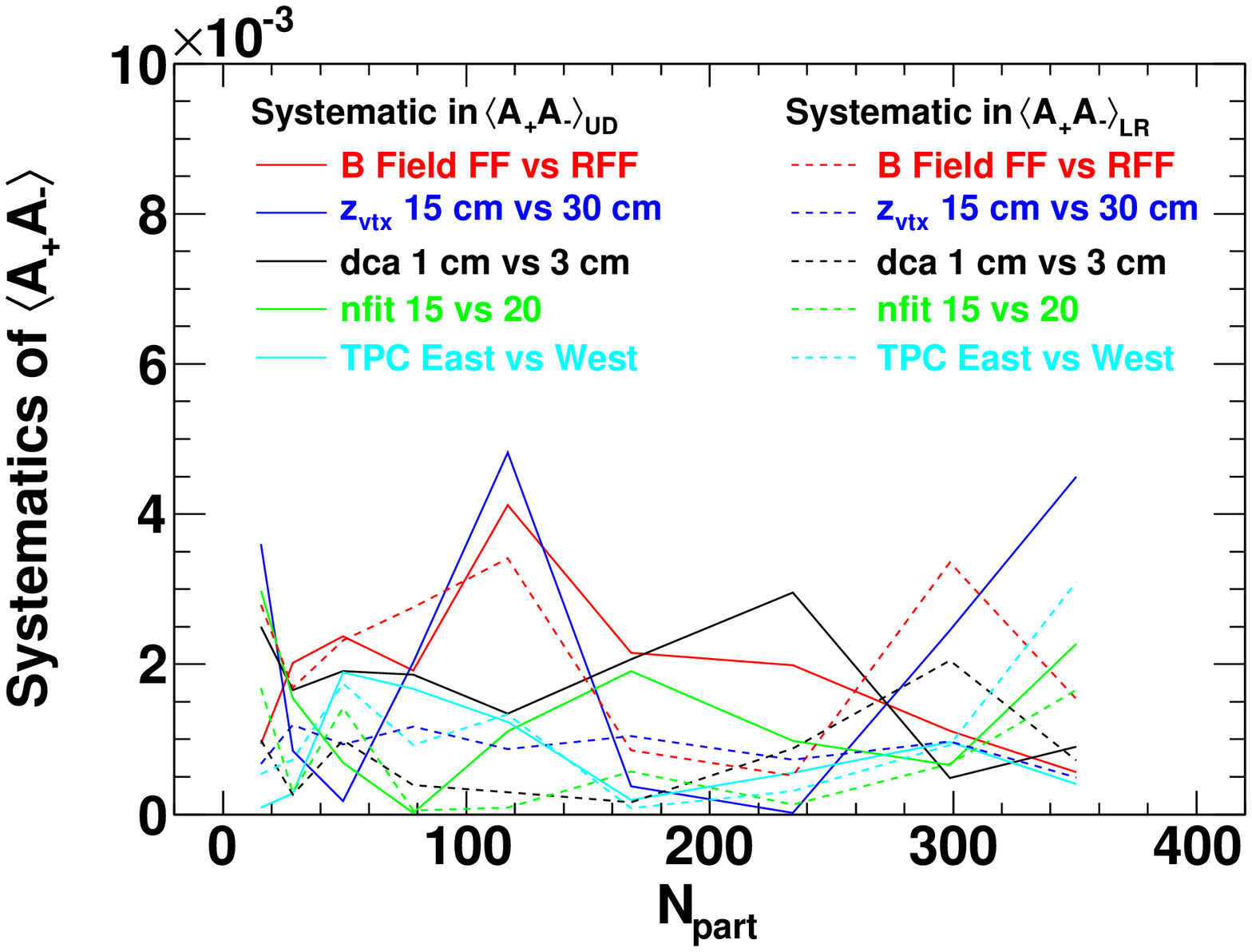}}
		\subfigure[Systematics of $UD-LR$]{\label{fig:sysDelta} \includegraphics[width=0.45\textwidth]{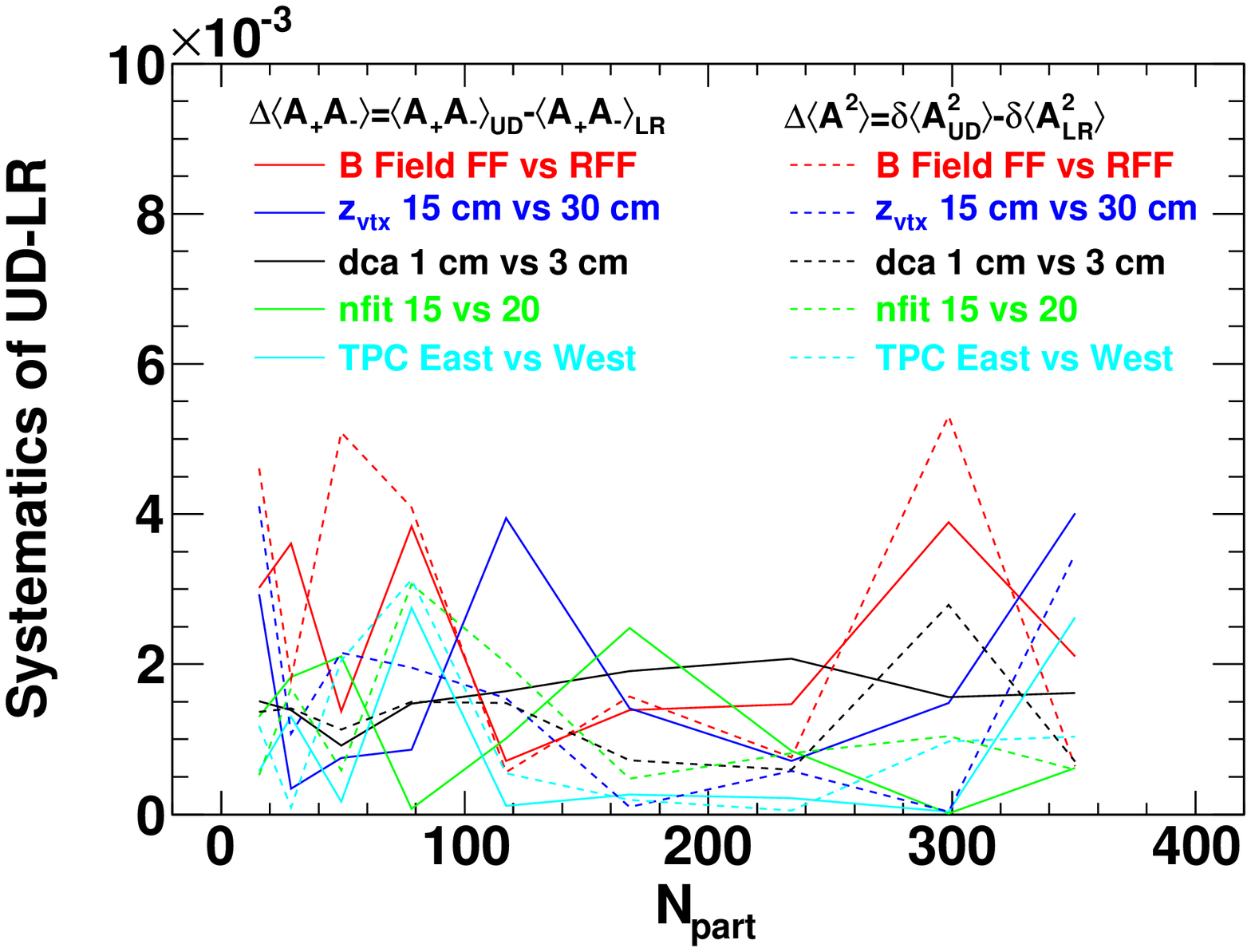}}
	\end{center}
	\caption[Systematic uncertainties]{
	Systematic uncertainties from difference sources as a function centrality in dynamical variances panel (a), variances panel (b) and the differences between $UD$ and $LR$ panel (c).
	Refer the text for details.
	}
	\label{fig:sys}
\end{figure}

\red{
The systematic uncertainties can be estimated by varying various STAR standard cuts.
Firstly, we check the magnetic field polarity uncertainties.
We separate the data according to the magnetic field direction, full magnetic field (FF) and reversed full magnetic field (RFF).
We use $C$ collectively denotes the final correlation values of either variances or covariance with any charge or $UD$/$LR$ directions combinations with standard cuts shown in table \ref{tab:cuts}.
For different magnetic fields, we have the correlations $C_{FF}$ and $C_{RFF}$ separately.
The systematic uncertainty $S_{BF}$ is then estimated as $|C-C_{FF}|$ or $|C-C_{RFF}|$ whichever is larger,
where $C$ is the magnetic field integrated value.
The systematic uncertainties are shown in figure \ref{fig:sys} as red lines.

Secondly, we check the vertex position $v_{z}$ cut.
The standard cut we used in this analysis is $|v_z|<$ 30 cm.
We then vary the cut to $|v_z|<$ 15 cm to study the difference $|C_{|v_z|<15}-C|$ 
which is denoted as the systematic uncertainty $S_{v_z}$.
The results are shown in figure \ref{fig:sys} as blue lines.

Thirdly, we study the standard DCA cut of $DCA <$ 2 cm.
We vary it to $DCA <$ 1 cm and $DCA<$ 3 cm to study the changes of the correlations.
The difference $|C-C_{DCA<1}|$ or $|C-C_{DCA<3}|$ whichever is larger is then treated as the systematics of the DCA cut, and noted as $S_{DCA}$.
The results are shown in figure \ref{fig:sys} as black lines.

Fourthly, the number of fit points can also contribute to the systematics.
We vary the cut to $nfit < 15$ from the standard $nfit<20$ and treat the difference $|C-C_{nfit<15}|$ as another systematic uncertainty $S_{nfit}$.
The results are shown in figure \ref{fig:sys} as green lines.

Lastly, the asymmetries from east- and west-side of the TPC may contribute to the systematics.
We use the asymmetry correlations from east- and west-side of the TPC comparing to the average values to estimate the uncertainties.
The systematics $S_{EW}$ are then defined as $|C-C_{East}|$ or $|C-C_{West}|$ whichever is larger.
The results are shown in figure \ref{fig:sys} as cyan lines.
}

\red{
The total systematic uncertainties are obtained by combining all five sources
\begin{equation}
	S = \sqrt{S_{BF}^2+S_{v_z}^2+S_{DCA}^2+S_{nfit}^2+S_{EW}^2}.
	\label{eq:sys}
\end{equation}
The systematic uncertainty values are detailed in table \ref{tab:sysAA} for covariances, table \ref{tab:sysA2} for dynamical variances and table \ref{tab:sysDelta} for the $UD-LR$ differences.
And they are shown in figure \ref{fig:asym} and figure \ref{fig:asymdiff} with the shaded areas.
}

\begin{sidewaystable}[thb]
	\centering
	\caption[Systematic uncertainties of $\langle A_+A_- \rangle$]{Systematic uncertainties of the opposite-sign correlations $\langle A_+A_- \rangle$. All the numbers are scaled by the corresponding number of participants $N_{part}$.}
	\begin{tabular}{c|c|c|c|c|c|c|c|c|c|c|c|c}
		\hline
		\hline
		\multirow{2}{*}{
		} & \multicolumn{6}{c|}{$\langle A_+A_- \rangle_{UD}$} & \multicolumn{6}{c}{$\langle A_+A_- \rangle_{LR}$} \\
		\cline{2-13}
		\begin{sideways}
			Centrality\,
		\end{sideways}
		 & 
		 \multicolumn{1}{c|}{
		 \begin{sideways}
			B Field FF vs RFF\,
		 \end{sideways}
		 }
		 &
		 \multicolumn{1}{c|}{
		 \begin{sideways}
			$v_z$ 15 cm vs 30 cm\,
		 \end{sideways}
		 }
		 &
		 \multicolumn{1}{c|}{
		 \begin{sideways}
			 dca 1 cm vs 3 cm\,
		 \end{sideways}
		 }
		 &
		 \multicolumn{1}{c|}{
		 \begin{sideways}
			$nfit$ 15 vs 20\,
		 \end{sideways}
		 }
		 &
		 \multicolumn{1}{c|}{
		 \begin{sideways}
			 TPC East vs West\,
		 \end{sideways}
		 }
		 &
		 \multicolumn{1}{c|}{
		 \begin{sideways}
			 Total\,
		 \end{sideways}
		 }
		 &
		 \multicolumn{1}{c|}{
		 \begin{sideways}
			FF vs RFF\,
		 \end{sideways}
		 }
		 &
		 \multicolumn{1}{c|}{
		 \begin{sideways}
			$v_z$ 15 cm vs 30 cm\,
		 \end{sideways}
		 }
		 &
		 \multicolumn{1}{c|}{
		 \begin{sideways}
			 dca 1 cm vs 3 cm\,
		 \end{sideways}
		 }
		 &
		 \multicolumn{1}{c|}{
		 \begin{sideways}
			$nfit$ 15 vs 20\,
		 \end{sideways}
		 }
		 &
		 \multicolumn{1}{c|}{
		 \begin{sideways}
			 TPC East vs West\,
		 \end{sideways}
		 }
		 &
		 \multicolumn{1}{c}{
		 \begin{sideways}
			 Total\,
		 \end{sideways}
		 }
		 \\

		 \hline
0-5\%	&0.0009	&0.0036	&0.0025	&0.0030	&0.0001	&0.0054	&0.0028	&0.0007	&0.0010	&0.0017	&0.0005	&0.0035	\\
5-10\%	&0.0020	&0.0008	&0.0017	&0.0015	&0.0003	&0.0032	&0.0017	&0.0012	&0.0003	&0.0003	&0.0007	&0.0022	\\
10-20\%	&0.0024	&0.0002	&0.0019	&0.0007	&0.0019	&0.0037	&0.0023	&0.0009	&0.0010	&0.0014	&0.0017	&0.0035	\\
20-30\%	&0.0019	&0.0020	&0.0019	&0.0000	&0.0017	&0.0037	&0.0028	&0.0012	&0.0004	&0.0001	&0.0009	&0.0032	\\
30-40\%	&0.0041	&0.0048	&0.0013	&0.0011	&0.0012	&0.0067	&0.0034	&0.0009	&0.0003	&0.0001	&0.0013	&0.0038	\\
40-50\%	&0.0022	&0.0004	&0.0021	&0.0019	&0.0002	&0.0036	&0.0009	&0.0010	&0.0002	&0.0006	&0.0001	&0.0015	\\
50-60\%	&0.0020	&0.0000	&0.0030	&0.0010	&0.0006	&0.0037	&0.0005	&0.0007	&0.0009	&0.0001	&0.0003	&0.0013	\\
60-70\%	&0.0011	&0.0025	&0.0005	&0.0007	&0.0010	&0.0030	&0.0034	&0.0010	&0.0020	&0.0007	&0.0009	&0.0042	\\
70-80\%	&0.0006	&0.0045	&0.0009	&0.0023	&0.0004	&0.0052	&0.0015	&0.0005	&0.0007	&0.0017	&0.0031	&0.0039	\\
		\hline
		\hline
	\end{tabular}
	\label{tab:sysAA}
\end{sidewaystable}

\begin{sidewaystable}[thb]
	\centering
	\caption[Systematic uncertainties of $\delta\langle A^2 \rangle$]{Systematic uncertainties of the same-sign dynamical correlations $\delta\langle A^2 \rangle$. All the numbers are scaled by the corresponding number of participants $N_{part}$.}
	\begin{tabular}{c|c|c|c|c|c|c|c|c|c|c|c|c}
		\hline
		\hline
		\multirow{2}{*}{
		} & \multicolumn{6}{c|}{$\delta\langle A^2_{UD} \rangle$} & \multicolumn{6}{c}{$\delta\langle A^2_{LR} \rangle$} \\
		\cline{2-13}
		\begin{sideways}
			Centrality\,
		\end{sideways}
		 & 
		 \multicolumn{1}{c|}{
		 \begin{sideways}
			B Field FF vs RFF\,
		 \end{sideways}
		 }
		 &
		 \multicolumn{1}{c|}{
		 \begin{sideways}
			$v_z$ 15 cm vs 30 cm\,
		 \end{sideways}
		 }
		 &
		 \multicolumn{1}{c|}{
		 \begin{sideways}
			 dca 1 cm vs 3 cm\,
		 \end{sideways}
		 }
		 &
		 \multicolumn{1}{c|}{
		 \begin{sideways}
			$nfit$ 15 vs 20\,
		 \end{sideways}
		 }
		 &
		 \multicolumn{1}{c|}{
		 \begin{sideways}
			 TPC East vs West\,
		 \end{sideways}
		 }
		 &
		 \multicolumn{1}{c|}{
		 \begin{sideways}
			 Total\,
		 \end{sideways}
		 }
		 &
		 \multicolumn{1}{c|}{
		 \begin{sideways}
			FF vs RFF\,
		 \end{sideways}
		 }
		 &
		 \multicolumn{1}{c|}{
		 \begin{sideways}
			$v_z$ 15 cm vs 30 cm\,
		 \end{sideways}
		 }
		 &
		 \multicolumn{1}{c|}{
		 \begin{sideways}
			 dca 1 cm vs 3 cm\,
		 \end{sideways}
		 }
		 &
		 \multicolumn{1}{c|}{
		 \begin{sideways}
			$nfit$ 15 vs 20\,
		 \end{sideways}
		 }
		 &
		 \multicolumn{1}{c|}{
		 \begin{sideways}
			 TPC East vs West\,
		 \end{sideways}
		 }
		 &
		 \multicolumn{1}{c}{
		 \begin{sideways}
			 Total\,
		 \end{sideways}
		 }
		 \\

		 \hline
0-5\%	&0.0014	&0.0003	&0.0006	&0.0003	&0.0012	&0.0020	&0.0060	&0.0038	&0.0019	&0.0002	&0.0001	&0.0074	\\
5-10\%	&0.0009	&0.0006	&0.0007	&0.0016	&0.0006	&0.0021	&0.0027	&0.0005	&0.0007	&0.0000	&0.0005	&0.0028	\\
10-20\%	&0.0028	&0.0007	&0.0005	&0.0002	&0.0016	&0.0034	&0.0023	&0.0015	&0.0007	&0.0003	&0.0004	&0.0028	\\
20-30\%	&0.0019	&0.0019	&0.0000	&0.0022	&0.0008	&0.0036	&0.0022	&0.0000	&0.0015	&0.0009	&0.0024	&0.0037	\\
30-40\%	&0.0010	&0.0006	&0.0006	&0.0003	&0.0004	&0.0014	&0.0004	&0.0021	&0.0009	&0.0017	&0.0001	&0.0029	\\
40-50\%	&0.0017	&0.0007	&0.0015	&0.0023	&0.0002	&0.0033	&0.0004	&0.0008	&0.0007	&0.0018	&0.0000	&0.0022	\\
50-60\%	&0.0034	&0.0005	&0.0010	&0.0016	&0.0006	&0.0039	&0.0037	&0.0011	&0.0004	&0.0007	&0.0005	&0.0040	\\
60-70\%	&0.0038	&0.0014	&0.0013	&0.0035	&0.0017	&0.0057	&0.0015	&0.0014	&0.0015	&0.0045	&0.0007	&0.0052	\\
70-80\%	&0.0034	&0.0015	&0.0014	&0.0031	&0.0021	&0.0055	&0.0040	&0.0050	&0.0021	&0.0037	&0.0010	&0.0078	\\
		\hline
		\hline
	\end{tabular}
	\label{tab:sysA2}
\end{sidewaystable}

\begin{sidewaystable}[thb]
	\centering
	\caption[Systematic uncertainties of the $UD-LR$]{Systematic uncertainties of the $UD-LR$ correlation differences. All the numbers are scaled by the corresponding number of participants $N_{part}$.}
	\begin{tabular}{c|c|c|c|c|c|c|c|c|c|c|c|c}
		\hline
		\hline
		\multirow{2}{*}{}
		& \multicolumn{6}{c|}{$\Delta\langle A_+A_- \rangle$} & \multicolumn{6}{c}{$\Delta\langle A^2 \rangle$} \\
		\cline{2-13}
		\multicolumn{1}{c|}{
		\begin{sideways}
			Centrality \,
		\end{sideways}
		}
		 & 
		 \multicolumn{1}{c|}{
		 \begin{sideways}
			B Field FF vs RFF\,
		 \end{sideways}
		 }
		 &
		 \multicolumn{1}{c|}{
		 \begin{sideways}
			$v_z$ 15 cm vs 30 cm\,
		 \end{sideways}
		 }
		 &
		 \multicolumn{1}{c|}{
		 \begin{sideways}
			 dca 1 cm vs 3 cm\,
		 \end{sideways}
		 }
		 &
		 \multicolumn{1}{c|}{
		 \begin{sideways}
			$nfit$ 15 vs 20\,
		 \end{sideways}
		 }
		 &
		 \multicolumn{1}{c|}{
		 \begin{sideways}
			 TPC East vs West\,
		 \end{sideways}
		 }
		 &
		 \multicolumn{1}{c|}{
		 \begin{sideways}
			 Total\,
		 \end{sideways}
		 }
		 &
		 \multicolumn{1}{c|}{
		 \begin{sideways}
			FF vs RFF\,
		 \end{sideways}
		 }
		 &
		 \multicolumn{1}{c|}{
		 \begin{sideways}
			$v_z$ 15 cm vs 30 cm\,
		 \end{sideways}
		 }
		 &
		 \multicolumn{1}{c|}{
		 \begin{sideways}
			 dca 1 cm vs 3 cm\,
		 \end{sideways}
		 }
		 &
		 \multicolumn{1}{c|}{
		 \begin{sideways}
			$nfit$ 15 vs 20\,
		 \end{sideways}
		 }
		 &
		 \multicolumn{1}{c|}{
		 \begin{sideways}
			 TPC East vs West\,
		 \end{sideways}
		 }
		 &
		 \multicolumn{1}{c}{
		 \begin{sideways}
			 Total\,
		 \end{sideways}
		 }
		 \\

		 \hline
0-5\%	&0.0030	&0.0029	&0.0015	&0.0013	&0.0006	&0.0047	&0.0046	&0.0041	&0.0014	&0.0005	&0.0012	&0.0064	\\
5-10\%	&0.0036	&0.0003	&0.0014	&0.0018	&0.0013	&0.0045	&0.0018	&0.0011	&0.0014	&0.0017	&0.0001	&0.0030	\\
10-20\%	&0.0014	&0.0008	&0.0009	&0.0021	&0.0002	&0.0028	&0.0051	&0.0022	&0.0011	&0.0006	&0.0020	&0.0060	\\
20-30\%	&0.0038	&0.0009	&0.0015	&0.0001	&0.0028	&0.0050	&0.0041	&0.0020	&0.0015	&0.0031	&0.0031	&0.0065	\\
30-40\%	&0.0007	&0.0039	&0.0016	&0.0010	&0.0001	&0.0045	&0.0006	&0.0015	&0.0015	&0.0020	&0.0005	&0.0030	\\
40-50\%	&0.0014	&0.0014	&0.0019	&0.0025	&0.0003	&0.0037	&0.0016	&0.0001	&0.0007	&0.0005	&0.0002	&0.0018	\\
50-60\%	&0.0015	&0.0007	&0.0021	&0.0008	&0.0002	&0.0028	&0.0008	&0.0006	&0.0006	&0.0008	&0.0001	&0.0014	\\
60-70\%	&0.0039	&0.0015	&0.0016	&0.0000	&0.0000	&0.0044	&0.0053	&0.0000	&0.0028	&0.0010	&0.0010	&0.0062	\\
70-80\%	&0.0021	&0.0040	&0.0016	&0.0006	&0.0026	&0.0055	&0.0006	&0.0035	&0.0007	&0.0006	&0.0010	&0.0038	\\
		\hline
		\hline
	\end{tabular}
	\label{tab:sysDelta}
\end{sidewaystable}

%% file: result.tex
%
%
%

\chapter{RESULTS AND DISCISSIONS}

In this chapter, we report the charge multiplicity asymmetry correlation results for Au+Au collisions at center of mass energy of $\sqrt{s_{NN}} = 200$ GeV.
We also report the asymmetry correlations as function of transverse momentum~($p_T$), event-by-event anisotropies~($v_2^{obs}$), wedge size and wedge axis location.
We discuss the implication of the results and compare with previously published results.

\section{Charge Asymmetry Correlations}
\label{CAC}

\begin{figure}[thb]
	\begin{center}
		\subfigure[$\langle A \rangle_{UD}$]{\label{fig:asymmean-a} \includegraphics[width=0.45\textwidth]{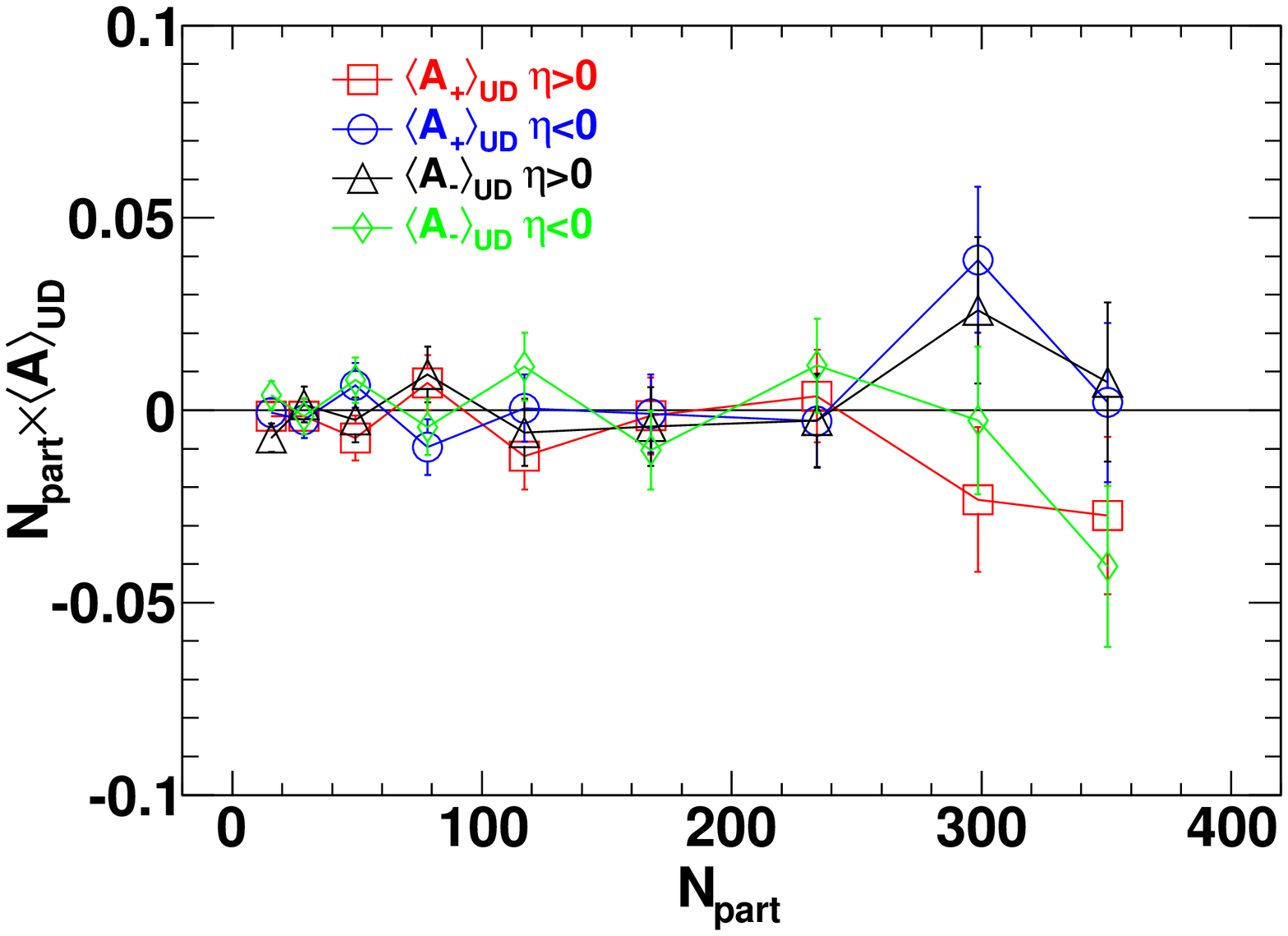}}
		\subfigure[$\langle A \rangle_{LR}$]{\label{fig:asymmean-b} \includegraphics[width=0.45\textwidth]{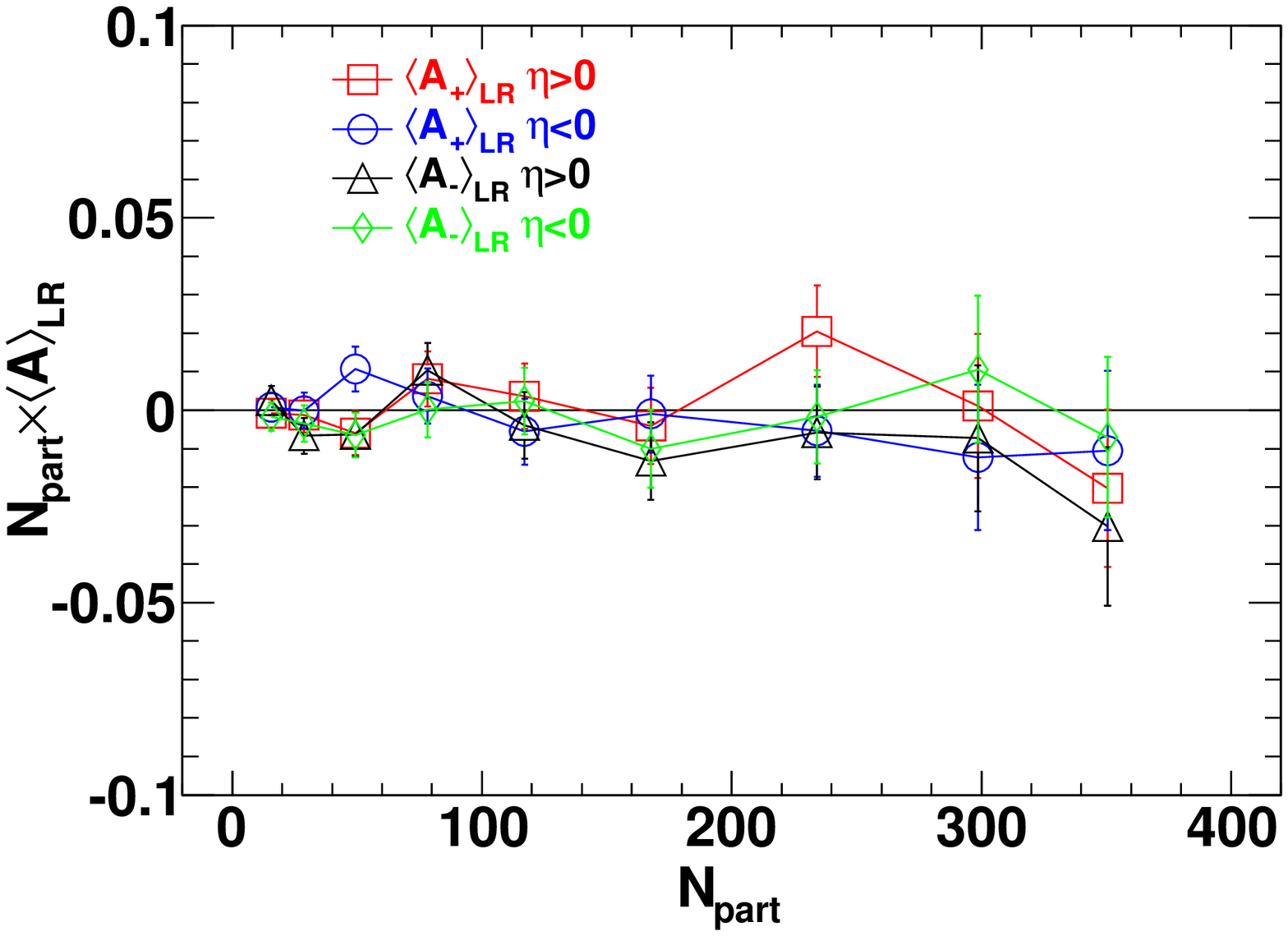}}
	\end{center}
	\caption[Mean value of asymmetries]{Mean value of the charge multiplicity asymmetries of $UD$ direction in panel (a) and $LR$ direction in panel (b).
	The data used are from Au+Au 200 GeV RUN IV with $p_T$ ranges of $0.15<p_T<2.9$ GeV/$c$ for both the asymmetry calculations and event-plane reconstruction.
	Error bars are statistical errors only.
	}
	\label{fig:asymmean}
\end{figure}

It is obvious that the single charge multiplicity asymmetries ($\langle A_{+,UD}\rangle$, $\langle A_{-,UD}\rangle$, $\langle A_{+,LR}\rangle$, $\langle A_{-,LR}\rangle$) are, 
by definition, zero because the directions of up (left) and down (right) hemispheres are all random from event to event.
The up (left) and down (right) directions only have relative meaning, and cannot be measured experimentally.
\red{
The data indeed show that they are centered at zero within statistical errors in figure \ref{fig:asymmean},
where data are from RUN IV 200 GeV Au+Au collisions.
The asymmetries are calculated from one side of the TPC tracks with respect to the event-plane reconstructed from the other side of the TPC tracks with $p_T$ ranges of $0.15<p_T<2.0$ GeV/$c$ for both the asymmetry calculation and event-plane reconstruction.
The average values of the asymmetries as a function of the centrality are shown in figure \ref{fig:asymmean-a} for $UD$ direction and figure \ref{fig:asymmean-b} for $LR$ direction.
They are multiplied by the number of participants $N_{part}$.
}

\begin{figure}[thb]
	\begin{center}
		\psfig{figure=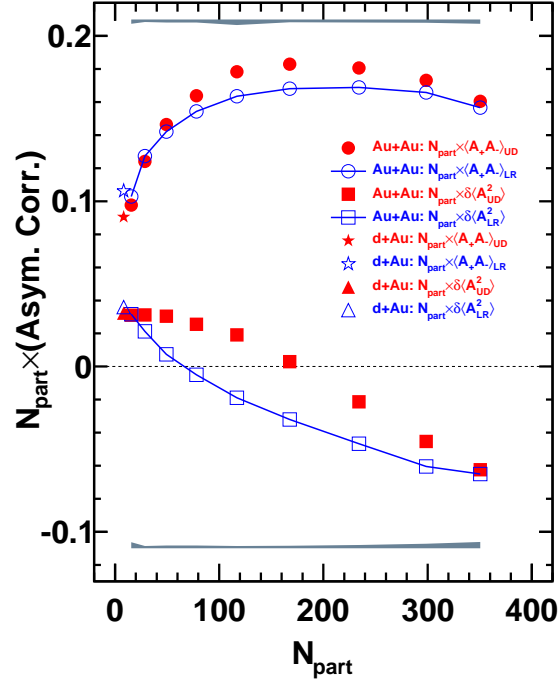,width=0.5\textwidth}
	\end{center}
	\caption[Dynamical charge asymmetry correlations]
	{Centrality dependence of the charge multiplicity dynamical correlations $\delta\langle A^{2} \rangle$ and the opposite-sign correlations $\langle A_{+}A_{-} \rangle$ for RUN IV 200 GeV Au+Au collisions.
	The asymmetries are calculated between hemispheres from a half event divided by event-plane ($UD$) reconstructed from the other half event,
	and the plane perpendicular to the event-plane ($LR$).
	The asymmetry correlations are scaled by the number of participants $N_{part}$ to better show the magnitude.
	The particle $p_{T}$ range is integrated over $0.15 < p_T < 2.0$ GeV/$c$ for both the asymmetry calculation and event-plane reconstruction.
	The upper and lower shaded areas illustrate the half size of the systematic uncertainties on $\langle A_+A_- \rangle$ and $\langle A^2 \rangle$ respectively in $UD$ or $LR$ direction whichever is larger of each centrality.
	The error bars are smaller than the symbols.
	}
	\label{fig:asym}
\end{figure}

We report the variances and covariances of the charge multiplicity asymmetries shown in figure~\ref{fig:asym},
which are obtained from the Au+Au and d+Au collisions with the center of mass energy of $\sqrt{s_{NN}} = 200$ GeV.
The Au+Au data are from RUN IV with centrality cuts and $N_{part}$ shown in table~\ref{tab:cent}.
d+Au data are from RUN III, and Glauber model shows the average number of participants for d+Au collision at $\sqrt{s_{NN}} = 200$ GeV is $8.31$ \cite{:2008ez,Abelev:2007nt}.
All the charge asymmetry correlations are scaled by the number of the participants $N_{part}$.
The dynamics and properties for d+Au collisions are very similar to the very peripheral Au+Au and p+p (proton to proton) collisions, so we present the d+Au result as a reference.
It is believed that the QGP can hardly be formed in p+p, d+Au and very peripheral Au+Au collisions,
hence chiral symmetry cannot be restored in such collisions.
Therefore the correlations in d+Au and the most peripheral Au+Au collisions are the background that is not due to CME/LPV.
We present d+Au data to show the trend in the correlations is consistent when moving from Au+Au peripheral collisions to d+Au collisions.
Note that in d+Au collisions, the reaction-plane is not defined because there is no anisotropy flow in such small multiplicity collisions.
If we apply the same algorithm of the second order event-plane reconstruction as shown in section \ref{TPCEP}, the event-plane reconstructed is mostly dominated by fluctuation and/or di-jets, similar to the peripheral Au+Au collisions.

The dynamical asymmetry variances (same-sign correlations) are shown in figure~\ref{fig:asym} as square symbols.
The solid squares are the dynamical variance in $UD$ direction ($\delta\langle A_{UD}^2\rangle$),
and open squares are the dynamical variance in $LR$ direction ($\delta \langle A_{LR}^2\rangle$).
A positive variance $\delta \langle A^2\rangle$ indicates broadening of the single asymmetry distributions of $A_+$ and $A_-$ due to any dynamical processes beyond statistical fluctuation, such that the same-sign particles are emitted more likely in the same direction.
On the other hand, a negative variance indicates narrowing of such distributions due to symmetric (back-to-back) correlations.
From figure \ref{fig:asym}, in peripheral collisions, both $\delta\langle A_{UD}^2 \rangle$ and $\delta \langle A_{LR}^2 \rangle$ are positive,
which suggests that the same-sign particles are more likely emitted in the same direction (i.e., they have small angle correlation) within one unit of pseudorapidity, $-1<\eta<0$ or $0<\eta<1$.
When moving toward more central collisions, both of the variances become negative.
Same-sign particles are more preferentially emitted back-to-back, in other words, more symmetrical in central collisions than in peripheral.
And the small angle correlation is stronger in $UD$ direction than that in $LR$ direction, i.e. $\delta\langle A_{UD}^2\rangle$ is larger than $\delta\langle A_{LR}^2\rangle$ for all centralities.

Figure~\ref{fig:asym} also shows the covariances $\langle A_+A_- \rangle_{UD}$ in solid circles and $\langle A_+A_- \rangle_{LR}$ in open circles.
The correlations between the positive and negative asymmetries are large and positive for both $UD$ and $LR$ directions.
They are on the order of about $10^{-3}$ before scaled by $N_{part}$, suggesting the individual asymmetries are as large as a few percent, comparable to the elliptic flow ($v_2$) magnitude.
As discussed, positive correlations suggest that the positive and negative charged particles tend to be emitted in the same direction (small angle correlation).
The small angle correlation is stronger in the $UD$ than that in $LR$ direction in most centralities. 
And the correlations are much stronger in the opposite-sign than the same-sign.

We also show the same-sign and opposite-sign correlations in d+Au collisions in the leftmost symbols.
Those correlations follow the peripheral collisions of the Au+Au collision, suggesting a smooth trend.

Figure~\ref{fig:asym}  depicts the following picture.
d+Au collisions and very peripheral Au+Au collisions fall in the same trend.
Particles, within one unit of pseudorapidity ($-1<\eta<0$ and $0<\eta<1$), are preferentially emitted in the same direction (with small angle correlation), no matter same-sign pairs or opposite-sign pairs.
The magnitude of the small angle correlation is, however, stronger in the opposite-sign than that in same-sign pairs,
and is stronger in the out-of-plane ($UD$) than in-plane ($LR$) direction regardless of charge combinations.
From mid-central to central collisions, the same-sign pairs are preferentially back-to-back,
while the opposite-sign pairs are still preferentially aligned in the same direction and more so than in peripheral collisions.
The opposite-sign pair small angle correlation is always stronger in the out-of-plane ($UD$) than in-plane ($LR$) direction.
On the other hand, the back-to-back emission tendency of same-sign pairs is weaker in the out-of-plane ($UD$) that in-plane ($LR$) direction.
It is interesting to see that same-sign pair correlations show different behavior in peripheral collisions and most central collisions,
which implies different mechanisms dominate in each situation.

\red{
The systematic uncertainties are shown in the shaded areas for the covariances (upper band) of $UD$ or $LR$ direction whichever is larger, and for the variances (lower band) of $UD$ and $LR$ whichever is larger.
The systematic uncertainties are negligible compare to the magnitude of the correlations.
}

In order to investigate the possible CME/LPV effect, we study the difference between $UD$ and $LR$ asymmetry correlations of the variance ($\Delta\langle A^2\rangle = \delta\langle A_{UD}^2\rangle - \delta\langle A_{LR}^2\rangle$) and covariance ($\Delta\langle A_+A_- \rangle = \langle A_+A_-\rangle_{UD} - \langle A_+A_- \rangle_{LR}$).
The charge independent background not related to event-plane cancels by taking the difference.
For variances, we take the difference between the dynamical correlations, where statistical fluctuation and detector effects have been subtracted according to different charges and $\eta$ regions respectively.

\begin{figure}[thb]
	\begin{center}
		\psfig{figure=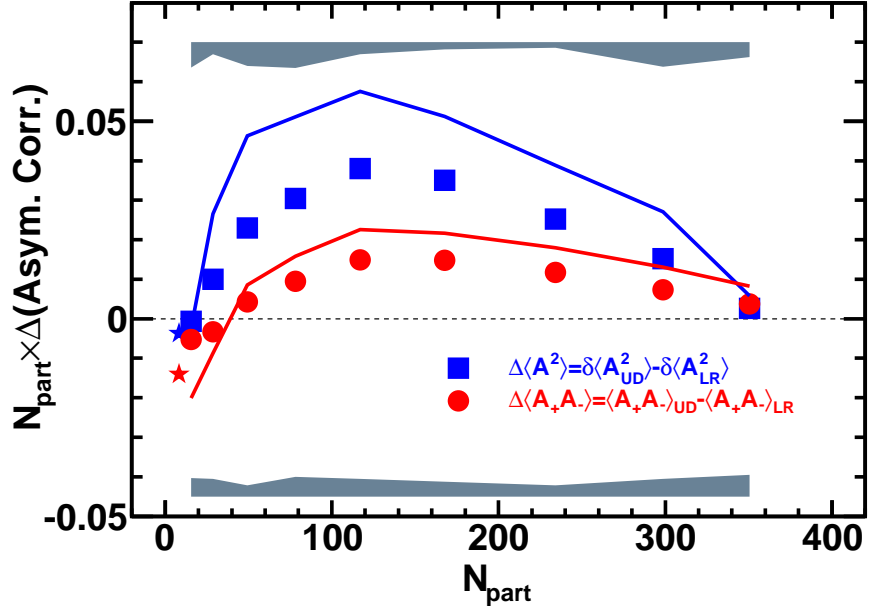,width=0.8\textwidth}
	\end{center}
	\caption[Asymmetry $UD-LR$ correlations]{
	The correlation differences between out-of-plane $UD$ and in-plane $LR$ of the same-sign $\Delta\langle A^{2} \rangle = \langle A^{2}_{UD} \rangle - \langle A^{2}_{LR} \rangle$
	and opposite-sign $\Delta\langle A_+A_- \rangle = \langle A_+A_- \rangle_{UD} - \langle A_+A_- \rangle_{LR}$.
	Both correlations are multiplied by $N_{part}$.
	The particle $p_T$ range of $0.15 < p_T < 2.0$ GeV/$c$ is used for both asymmetry calculation and event-plane reconstruction.
	The leftmost data points are from d+Au data.
	The curves are the linear-extrapolated values of the $\Delta\langle A^{2} \rangle$ (blue) and $\Delta\langle A_+A_- \rangle$ (red) corresponding to the perfect event-plane resolution of unity.
	The shaded bands are the systematic uncertainties of  $\Delta\langle A_+A_- \rangle$ (upper band) and $\Delta\langle A^{2} \rangle$ (lower band).
	Statistical errors are smaller than the symbols.
	}
	\label{fig:asymdiff}
\end{figure}

Figure~\ref{fig:asymdiff} shows the centrality dependence of $\Delta\langle A^2\rangle$ and $\Delta\langle A_+A_- \rangle$ scaled by the number of participants $N_{part}$.
Also shown in curves are the $\Delta\langle A^2\rangle$ and $\Delta\langle A_+A_- \rangle$ values with event-plane resolution correction.
It is done by linear extrapolation to unity of the correlations if higher-order harmonic terms can be ignored.

The shaded areas are the systematic uncertainties of $\Delta\langle A^2\rangle$ in upper band and $\Delta\langle A_+A_- \rangle$ in lower band both multiplied by $N_{part}$.

From figure~\ref{fig:asymdiff}, we can conclude the following.
First, same-sign variance is positive $\Delta\langle A^2\rangle>0$ ($\delta\langle A_{UD}^2\rangle>\delta\langle A_{LR}^2\rangle$) in all centralities.
The $UD$ asymmetry distribution is always broader than the $LR$ distribution.
The dynamical same-sign correlations $\delta\langle A^2\rangle$ are the width of single asymmetry correlation beyond the statistical fluctuation.
More small angle pairs will increase the dynamical correlation, while more back-to-back pairs will decrease it.
Thus, there are more small angle pairs in the out-of-plane ($UD$) direction than that in the in-plane ($LR$) direction, or,
more back-to-back pairs in-plane than out-of-plane.
Or, it is possible both cases are true.

Second, opposite-sign covariance is positive $\Delta\langle A_+A_-\rangle>0$ ($\langle A_+A_-\rangle_{UD} > \langle A_+A_-\rangle_{LR}$) in all centralities except the most peripheral bins.
Similar to the discussion above, more small angle opposite-sign pairs are emitted in out-of-plane ($UD$) direction than that in in-plane ($LR$) direction.

Third, the magnitude of opposite-sign correlation $\Delta\langle A_+A_-\rangle$ is small relative to the correlations ($\langle A_+A_- \rangle_{UD}$ and $\langle A_+A_- \rangle_{LR}$) themselves.
This indicates that the majority of the strong background correlations between opposite-sign pairs are unrelated to the reaction-plane.
The $UD-LR$ difference, which is related to the reaction plane, is a small effect compared to the background.

Lastly, $\Delta\langle A_{UD}^2\rangle$ and $\Delta\langle A_+A_-\rangle$ have similar centrality dependence.
They both increase with centrality from most peripheral collisions, and reach a maximum in medium central collisions.
Then, they decrease with centrality to almost zero in most central collisions.
While the opposite-sign $\Delta\langle A_+A_-\rangle$ starts slightly negative in very peripheral collisions, which is also true for d+Au collisions for both same-sign and opposite-sign correlations.
It is mainly due to non-flow effect, such as di-jets, dominating the EP reconstruction.
The decrease towards most central collisions is most likely due to the lower EP resolution and the dilution effect \red{by the large multiplicities} in the most central collisions.
In most central collisions, the low EP resolution cannot distinguish between up and down, or left and right.
Thus the difference between $UD$ and $LR$ vanishes.

\red{
CME/LPV predicts that charge separation is along the system angular momentum direction \cite{Fukushima:2008xe,Kharzeev:2004ey,Kharzeev:2007jp}. 
In the absence of the medium effect,
the charge asymmetry correlations between light quarks across the event-plane will carry to the final hadronic state.
Thus the final measured charged particle correlations are expected to have the charge separation effect \cite{Kharzeev:2007tn}.
CME/LPV expects that more negatively charged particles going down (up) across the reaction-plane,
if more positively charged particles going up (down) across the reaction-plane.
In other words, positive and negative charges are anti-correlated in $UD$ direction.
On the other hand, CME/LPV effect does not contribute to the $LR$ direction.
Therefore the covariance in $UD$ direction should be smaller than that in $LR$ direction,
$\langle A_+A_- \rangle_{UD} < \langle A_+A_- \rangle_{LR}$, or $\Delta\langle A_+A_- \rangle < 0$.
Since CME/LPV generates additional correlations in $UD$ direction rather than $LR$ direction,
one should expect the variance in $UD$ is larger than that in $LR$ direction.
This can be written in the language of charge asymmetry correlations as 
$\delta\langle A_{\pm,UD}^2 \rangle > \delta\langle A_{\pm,LR}^2 \rangle$, or $\Delta\langle A^2 \rangle > 0$.

CME/LPV effect has to occur in the early time of the collision while the low mass quarks are deconfined, chiral symmetry restored, and the magnetic field is large \cite{Kharzeev:2007jp}.
The correlations due to CME/LPV could be modified by the interactions with the hot dense medium.
Effect of medium interactions has been observed in the jet-like two-particle correlations with strong back-to-back correlation suppression at high $p_{T}$ and enhancement at low $p_T$ \cite{Adler:2002tq,:2009qa,Petersen:2010di}.
One may expect similar suppression effect on the back-to-back charge asymmetry correlations,
which are the opposite-sign correlations across the reaction-plane.

Data shown in figure \ref{fig:asymdiff} seem to suggest that the same-sign correlation is consistent with CME/LPV expectation that additional correlations broaden the asymmetry distribution in $UD$ direction.
However, the dynamical variances of $\delta\langle A^{2}_{UD} \rangle$ and $\delta\langle A^{2}_{LR} \rangle$ in figure \ref{fig:asym} show that they are negative in mid-central to most central collisions.
In other words, the same-sign particles are more likely emitted back-to-back, which is inconsistent with CME/LPV expectation.
The opposite-sign correlations are also inconsistent with the expectation that the opposite-sign pairs are strongly correlated in the same direction, and the effect is stronger in $UD$ than $LR$.
It appears that CME/LPV alone cannot explain the data.
}

\red{
As we discussed in section \ref{CME}, the magnitude of the opposite-sign correlation expected from CME/LPV is negative around $\sim 10^{-5}$ to $\sim 10^{-4}$, and $\sim 10^{-3}$ to $\sim 10^{-2}$ after multiplied by the number of participants \cite{Kharzeev:2004ey,Kharzeev:2007tn,Fukushima:2008xe,Kharzeev:2007jp}.
The covariances in $UD$ and $LR$ directions are significantly larger than the expected value by 1 to 2 orders and they are positive,
which suggests that they are largely dominated by the event-plane unrelated background.
After taking the difference between $UD$ and $LR$, it has the order of $10^{-2}$ for mid-central collisions but still in positive.

It has been suggested that the charge separation effect can be invoked by QCD processes without the need of CME/LPV to create such asymmetry correlations \cite{Asakawa:2010bu}.
However, the effects are estimated to be orders of magnitude smaller than our observed charge asymmetry correlations.
We note that such estimates have large uncertainties due to uncertainty in various effects, such as the effect of medium interaction \cite{Muller:2010jd,Ma:2011uma}.
}

\section{Charge Asymmetry $p_{T}$ Dependence}

\begin{figure}[thb]
	\begin{center}
		\subfigure[Most central 0-20\%]{\label{fig:asympt-b} \includegraphics[width=0.45\textwidth]{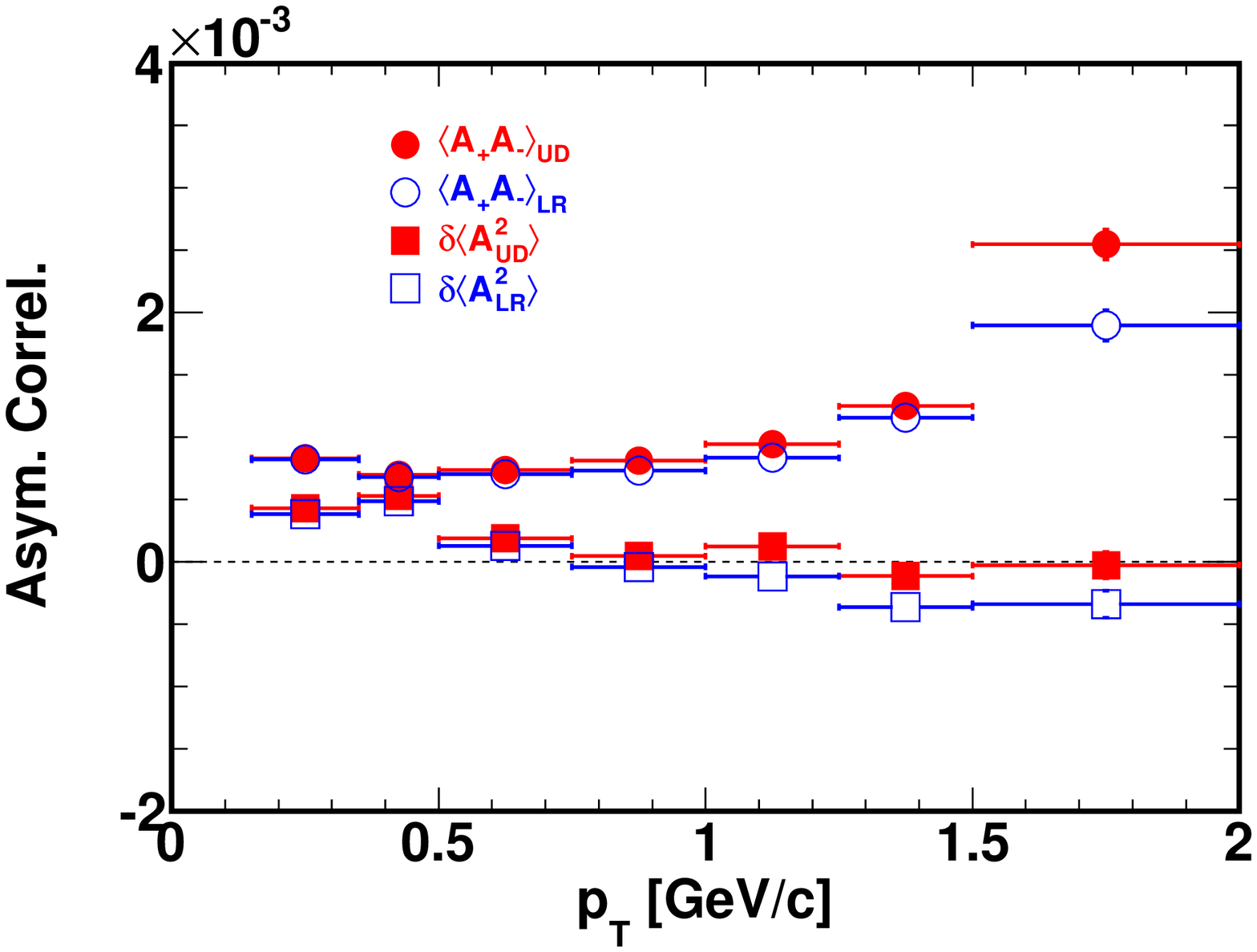}}
		\subfigure[Mid-central 20-40\%]{\label{fig:asympt-a} \includegraphics[width=0.45\textwidth]{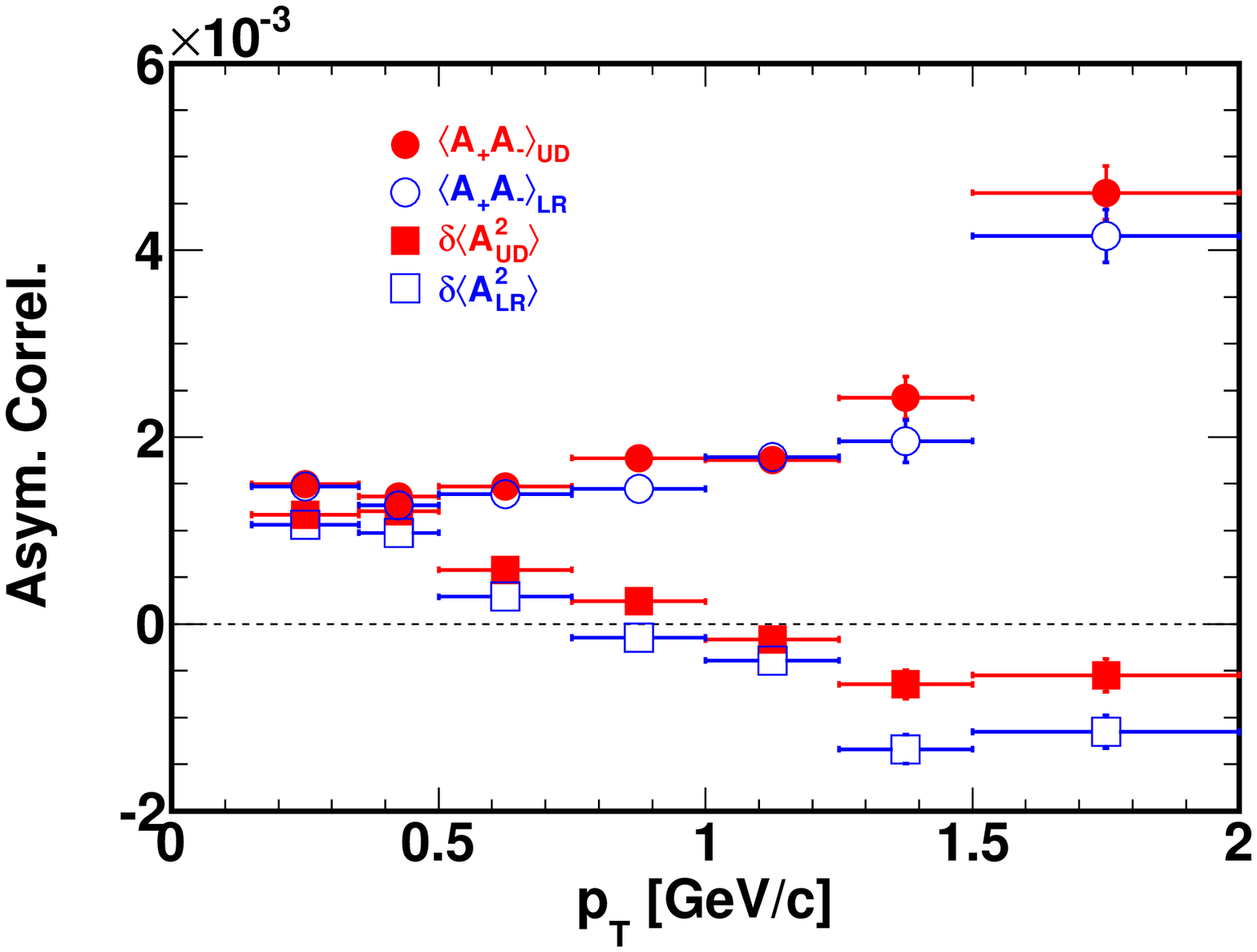}}
		\subfigure[Peripheral 40-80\%]{\label{fig:asympt-c} \includegraphics[width=0.45\textwidth]{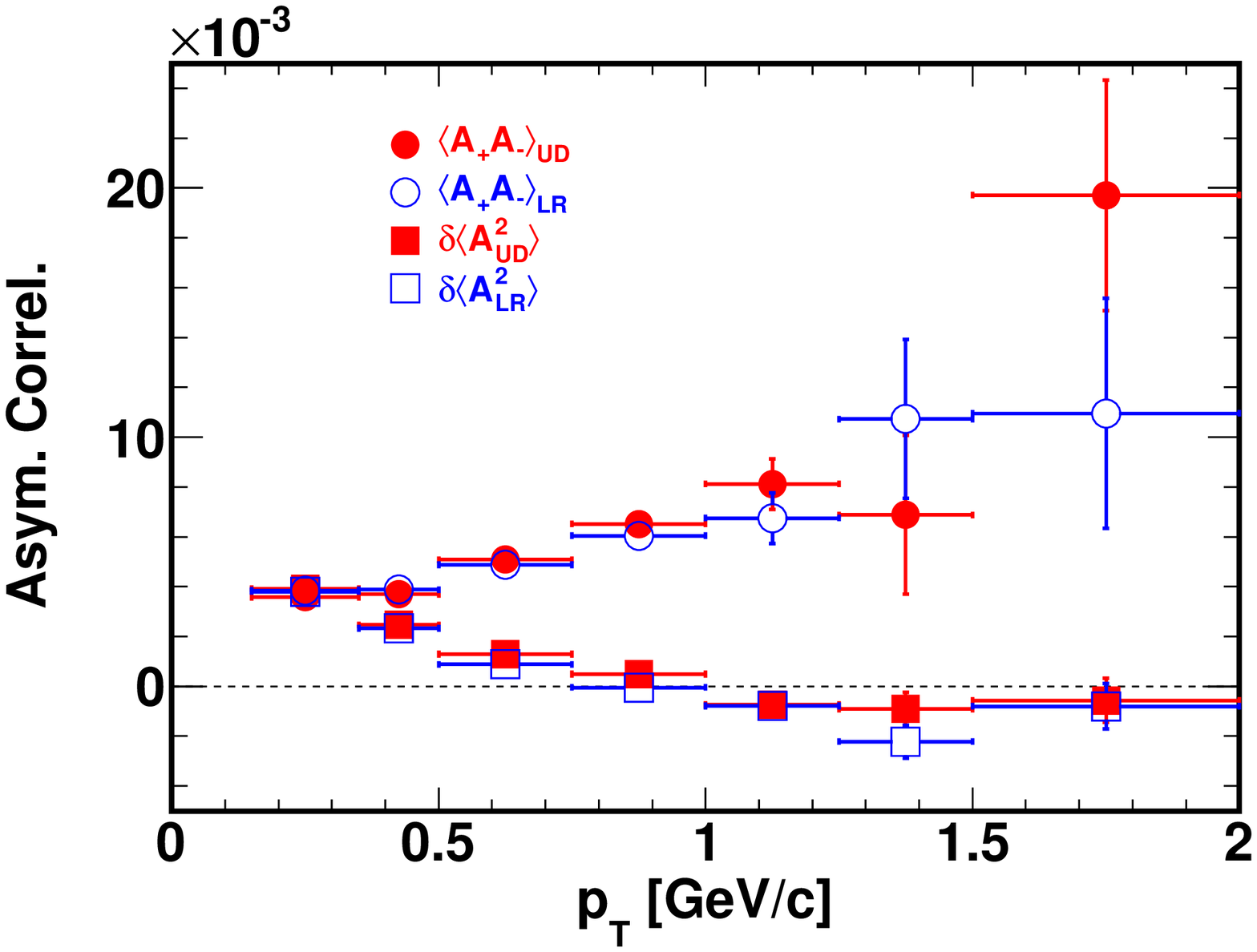}}
	\end{center}
	\caption[The $p_T$ dependence of the asymmetry correlations]{
	The $p_T$ dependence of the charge asymmetry dynamical correlations $\delta\langle A^{2} \rangle$ and the opposite-sign charge asymmetry correlations $\langle A_+A_- \rangle$ for RUN IV Au+Au 200 GeV collisions.
	The most central 0-20\% collisions are shown in panel (a), medium central 20-40\% collisions are shown in panel (b), and the peripheral 40-80\% collisions in panel (c).
	The asymmetries are calculated between hemispheres separated by event-plane ($UD$) and the plane perpendicular to the event-plane ($LR$).
	The particle $p_T$ range of $0.15 < p_T < 2.0$ GeV/$c$ is used for event-plane reconstruction.
	Error bars are statistical only.
	}
	\label{fig:asympt}
\end{figure}

We study the $p_T$ dependence of the asymmetry correlations which has been shown in figure \ref{fig:asympt-b} for 0-20\%the most central collisions, figure \ref{fig:asympt-a} for 20-40\% mid-central collisions,
and figure \ref{fig:asympt-c} for 40-80\% peripheral collisions.
The $p_T$ ranges for particles used in asymmetry calculation are 0.15-0.35, 0.35-0.5, 0.5-0.75, 0.75-1.0, 1.0-1.25, 1.25-1.5, and 1.5-2.0 GeV/$c$.
The event-plane is still reconstructed from particles with $0.15<p_T<2.0$ GeV/$c$.
All dynamical variances and covariances in $UD$ and $LR$ are positive at very low $p_T$ for all centralities.
The dynamical variances $\delta\langle A^2\rangle$ shown as squares drop rapidly to zero with $p_T$ up to 1 GeV/$c$ (soft particles) for all centralities.
The $\delta\langle A_{LR}^2\rangle$ becomes negative at $p_T>$ 1 GeV/$c$,
and the $\delta\langle A_{UD}^2\rangle$ is around zero, slightly negative at $p_T>$ 1 GeV/$c$ for all centralities.
The same-sign correlation $p_T$ dependence results indicate that, small angle pairs dominate at low-$p_T$ ($p_T < 1$ GeV/$c$).
The tendency for back-to-back emission of same-sign pairs increases with increasing $p_T$.

On the other hand, opposite-sign correlations $\langle A_+A_-\rangle$ shown as circles remain relatively constant with $p_T$ up to 1 GeV/$c$, and then rapidly increase with increasing $p_T$.
The results indicate that, opposite-sign pairs are emitted in the same direction at low-$p_T$, and the small angle correlation increases strongly with $p_T>1$ GeV/$c$ for both out-of-plane ($UD$) and in-plane ($LR$) directions.
The correlations are qualitatively similar for all centralities.

\begin{figure}[thb]
	\begin{center}
		\subfigure[Most central 0-20\%]{\label{fig:asymptdiff-b} \includegraphics[width=0.45\textwidth]{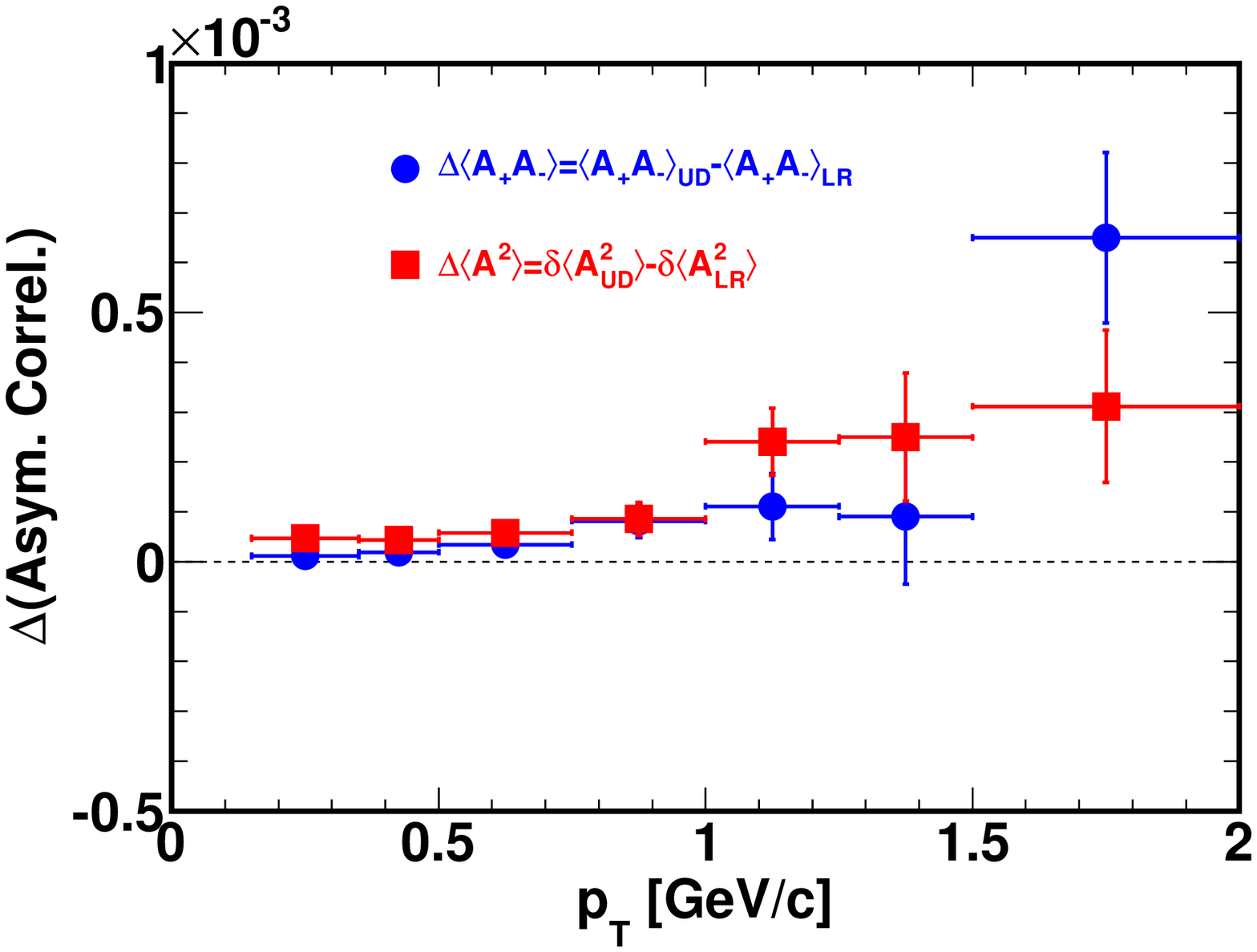}}
		\subfigure[Mid-central 20-40\%]{\label{fig:asymptdiff-a} \includegraphics[width=0.45\textwidth]{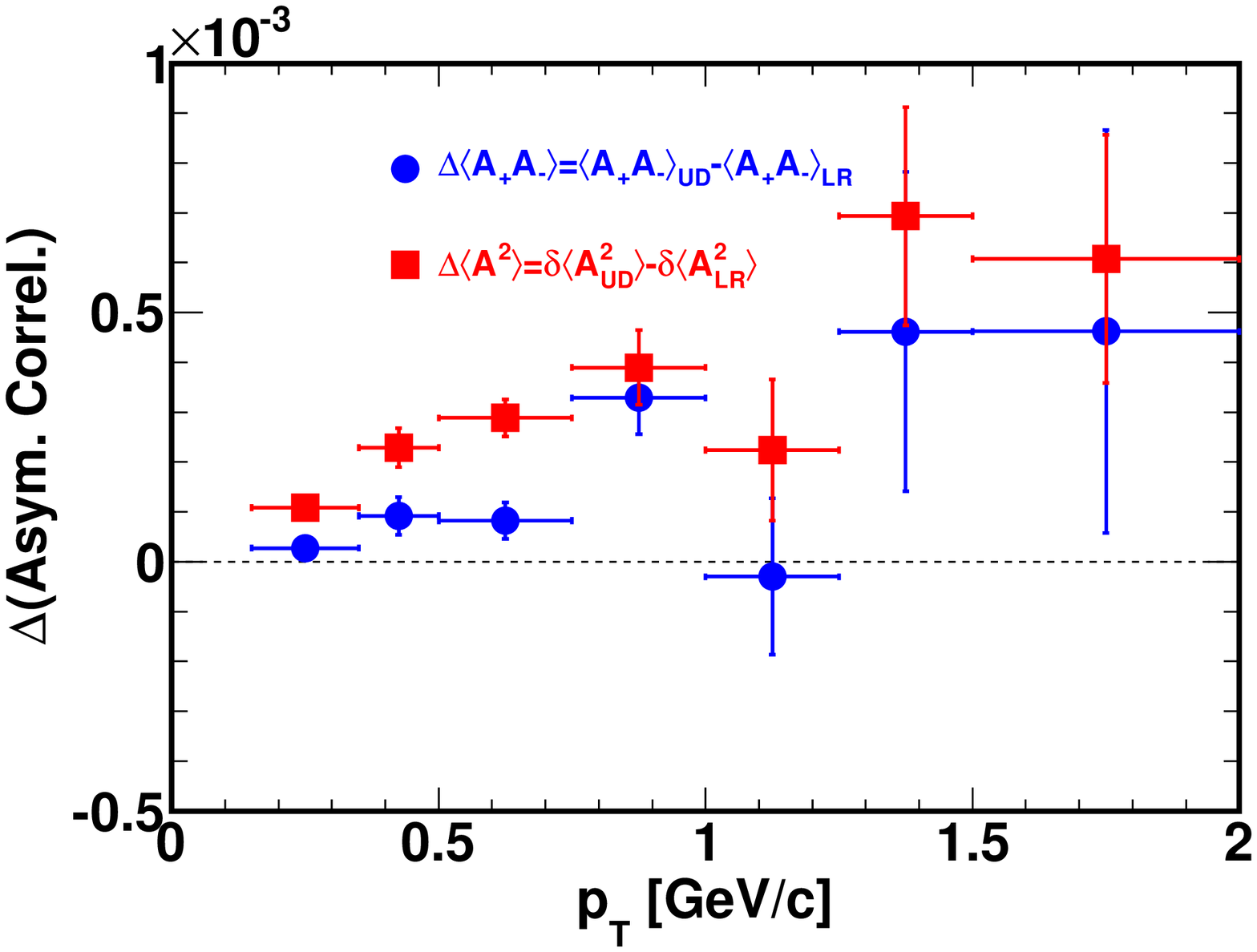}}
		\subfigure[Peripheral 40-80\%]{\label{fig:asymptdiff-c} \includegraphics[width=0.45\textwidth]{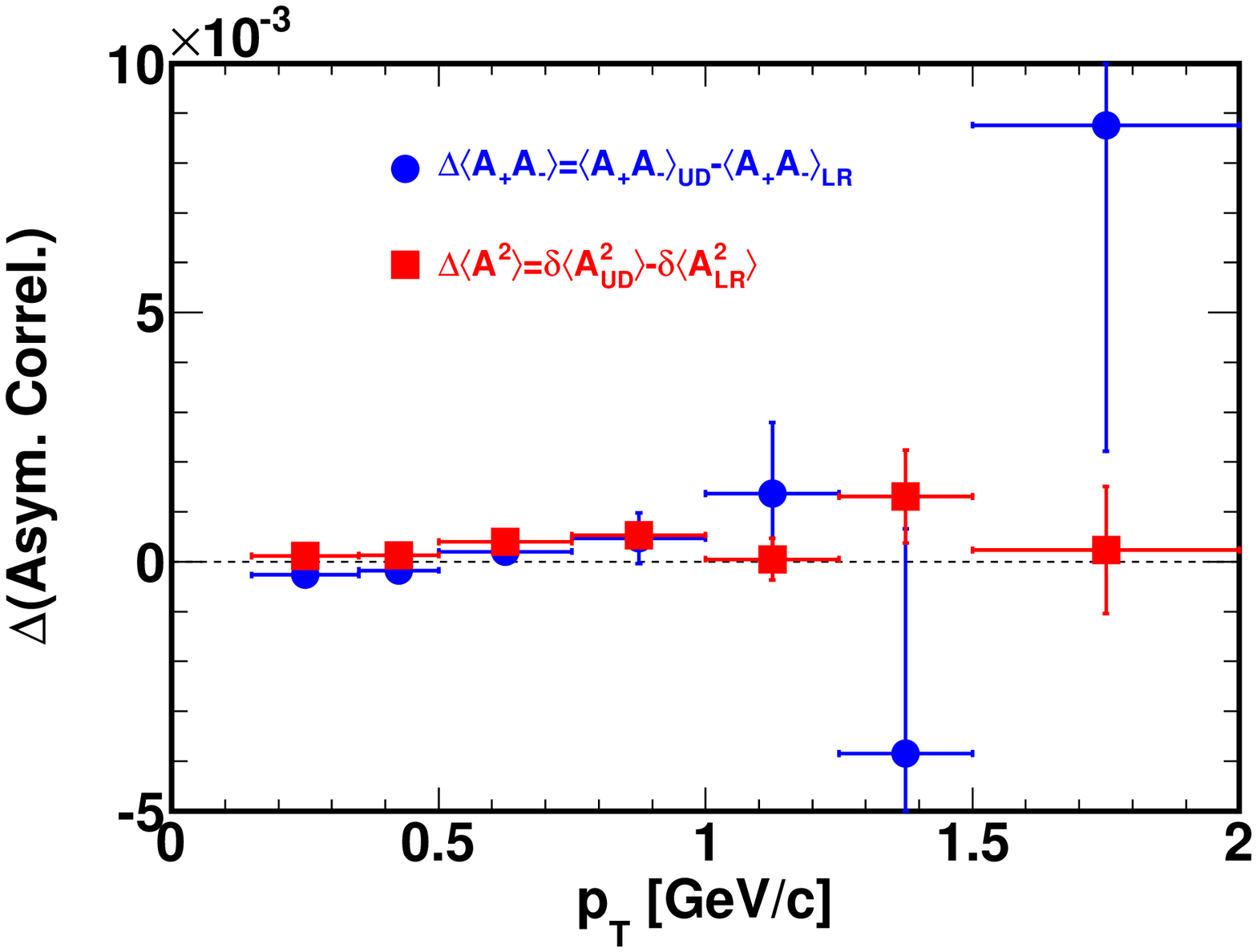}}
	\end{center}
	\caption[The $p_T$ dependence of $UD-LR$ correlations]{
	The $p_T$ dependence of the charge asymmetry correlation differences between $UD$ and $LR$ of same-sign $\Delta\langle A^2 \rangle = \delta\langle A^2_{UD} \rangle - \delta\langle A^2_{LR} \rangle$,
	and opposite-sign $\Delta\langle A_+A_- \rangle = \langle A_+A_- \rangle_{UD} - \langle A_+A_- \rangle_{LR}$.
	The event-plane is constructed from particles $p_T$ range of $0.15 < p_T < 2.0$ GeV/$c$.
	Error bars are statistical.
	}
	\label{fig:asymptdiff}
\end{figure}

Figure~\ref{fig:asymptdiff} shows the $UD-LR$ of same-sign and opposite-sign correlations as a function of $p_T$ for the most central in figure \ref{fig:asymptdiff-b}, mid-central in figure \ref{fig:asymptdiff-a} and peripheral collisions in figure \ref{fig:asymptdiff-c}.
Both same-sign and opposite-sign correlations grow with increasing $p_T$, and show similar $p_T$ dependence.
\red{
The same-sign correlation is stronger than the opposite-sign correlation in mid-central collisions,
while the difference between same-sign and opposite-sign correlations is not obvious in the most central and peripheral collisions.
}

\red{
The CME/LPV effect is non-perturbative and subjected to soft particle production correlation.
The limit of the transverse momentum $p_T$ range for such effect being experimentally tested at center of mass energy $\sqrt{s_{NN}}=200$ GeV is estimated smaller than 1 GeV/$c$ \cite{Kharzeev:2007jp},
which may be subjected to radial flow effect.
Thus, CME/LPV expects charge separation at low-$p_T$ and little separation at high-$p_T$.
However this is not obvious in the data due to limited statistics.
}

It is worthwhile to mention that the charge asymmetry correlations for each $p_T$ range, in this section, are calculated only within the $p_T$ bin.
Meanwhile, the centrality dependence of figure~\ref{fig:asym} and figure~\ref{fig:asymdiff} is calculated from all particles within $0.15<p_T<2.0$ GeV/$c$.
As such, the data shown in the previous section cannot be readily obtained from those in figure~\ref{fig:asympt} of the corresponding centrality bin.

\section{Event-by-Event Anisotropy ($v_2^{obs}$) Dependence}

The physics mechanism for such event-plane dependent charge asymmetry correlations is unclear.
There are alternative models other than CME/LPV effect,
such as, general cluster particle correlations with anisotropies that could generate sizeable difference between in-plane and out-of-plane particle correlations \cite{Wang:2009kd}.
Or, the momentum conservation and local charge conservation together with elliptic flow could yield event-plane dependent correlations that differ between same- and opposite-sign pairs \cite{Pratt:2010gy,Ma:2011uma}.
Path-length dependent jet-quenching effect could be another mechanism \cite{Petersen:2010di,Wang:2009kd,Abelev:2009jv}.
For most of these, one qualitatively expects event anisotropy dependence.

\red{
The similar trends of same-sign and opposite-sign charge asymmetry correlations in figure \ref{fig:asymdiff}, are qualitatively consistent with that of elliptic flow as a function of centrality \cite{:2008ed}.
The asymmetry correlations have maxima at mid-central collisions, and drop in central and peripheral collisions for both same-sign and opposite-sign correlations.
Motivated by these considerations, we examine the event shape dependence of the asymmetry correlations.
}

We investigate the dynamical variances $\delta\langle A^2 \rangle$, covariances $\langle A_+A_- \rangle$, and their difference between out-of-plane ($UD$) and in-plane ($LR$) $\Delta\langle A^2 \rangle$ and $\Delta\langle A_+A_- \rangle$ as a function of event-by-event azimuthal anisotropy ($v_2^{obs}$) of high-$p_T$ ($p_T>2$ GeV/$c$) and low-$p_T$ ($p_T<2$ GeV/$c$) particles respectively.
The high-$p_T$ anisotropy dependence may be sensitive to the jet-quenching effect where the high-$p_T$ particles are suppressed in out-of-plane direction \cite{Adams:2004bi,Petersen:2010di}.
The low-$p_T$ anisotropy dependence may characterize the bulk event shape of the charge combinations.

\red{
The event anisotropy is defined as $v_2^{obs} \equiv \langle \cos 2(\phi-\psi_{EP})\rangle$ with low-$p_T$ and high-$p_T$ particles.
For low-$p_T$ anisotropy $\langle \cos 2(\phi-\psi_{EP}) \rangle$, the average is taken from all particles used in asymmetry calculation~(within one half of the TPC), and the event-plane is reconstructed from the other half of the TPC.
The $v^{obs}_{2,p_T < 2~\text{GeV/}c}$ denotes the low-$p_T$ anisotropy.
For high-$p_T$ anisotropy, in order to increase statistics, particles with $p_T > 2$ GeV/$c$ from the entire event ($\left|\eta\right|<1$) are used for the anisotropy calculation, while the EP is reconstructed from different sides of the TPC.
We use $v_{2,p_T > 2~\text{GeV/}c}^{obs}$ to stand for high-$p_T$ anisotropy.
}

\red{
A positive $v_{2}^{obs}$ indicates an event with more particles of the interest emitted in-plane, i.e. the event is elongated in the event-plane direction.
On the other hand, a negative $v_{2}^{obs}$ indicates more interest particles emitted out-of-plane, i.e. the event is elongated perpendicular to the event-plane.
A zero $v_2^{obs}$ means the event is spherical, i.e. the emitted particles are isotropic in azimuth.
We use term ``spherical'' to refer events with such shape.
}

\begin{figure}[thb]
	\begin{center}
		\subfigure[$\delta\langle A^{2} \rangle$ vs high-$p_T$ $v_{2}^{obs}$]{\label{fig:asymv2h-a} \includegraphics[width=0.45\textwidth]{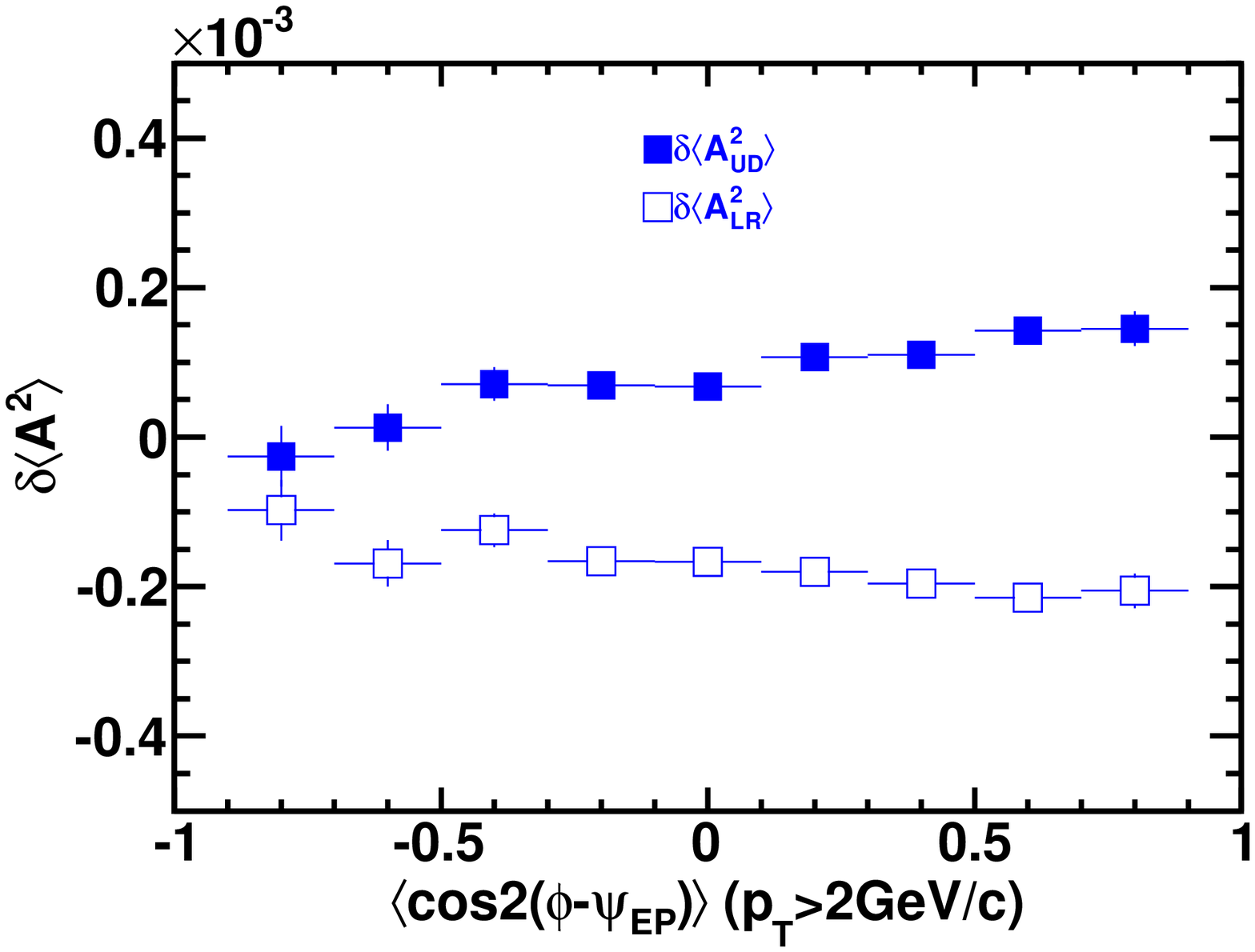}}
		\subfigure[$\delta\langle A_{+}A_{-} \rangle$ vs high-$p_T$ $v_{2}^{obs}$]{\label{fig:asymv2h-b} \includegraphics[width=0.45\textwidth]{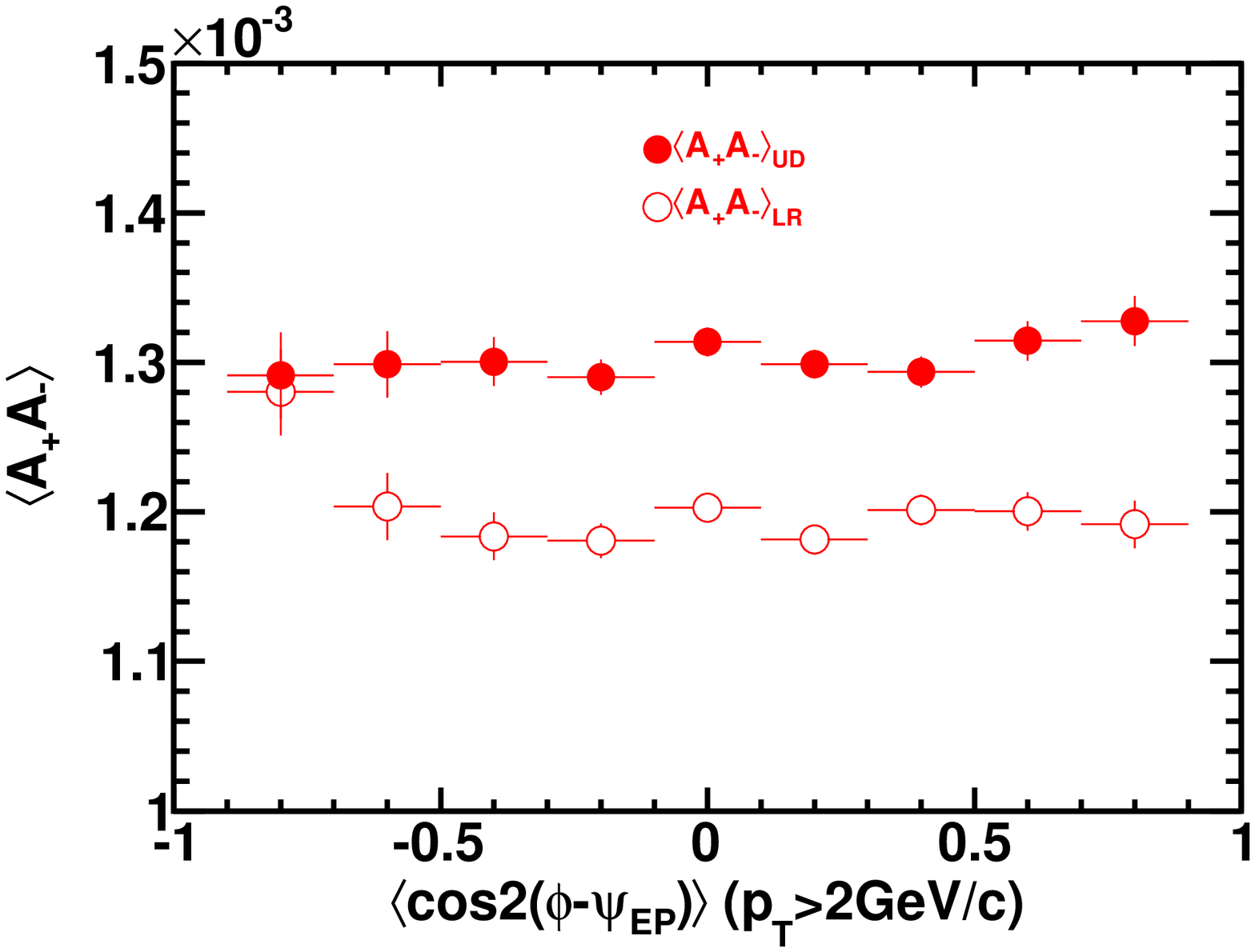}}
		\subfigure[High-$p_T$ $v_{2}^{obs}$ $UD-LR$ correlations]{\label{fig:asymv2h-c} \includegraphics[width=0.45\textwidth]{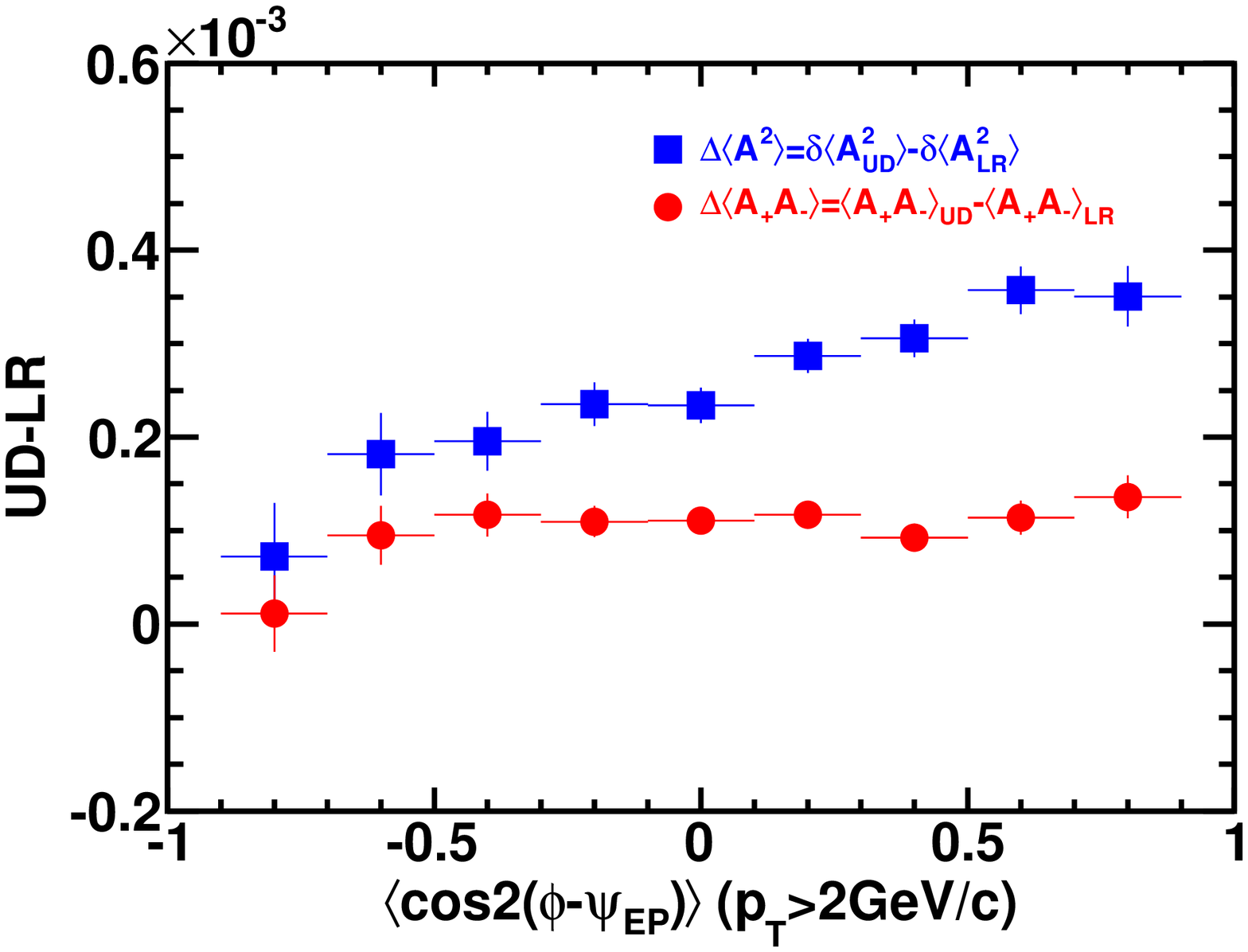}}
	\end{center}
	\caption[Mid-central asymmetry correlations vs high-$p_T$ event-by-event $v_{2}^{obs}$]{
	High-$p_T$ event-by-event anisotropy $v_2^{obs}$ dependence of (a) the dynamical charge asymmetry variances, (b) asymmetry covariances, and (c) their differences between $UD$ and $LR$ hemispheres, in mid-central 20-40\% Au+Au 200 GeV collisions from RUN IV.
	The asymmetries are calculated from all particles from one side of the TPC tracks with respect to the event-plane reconstructed from the other side of the TPC tracks.
	Both the particles used in asymmetry calculation and event-plane reconstruction are with $p_T$ range of $0.15 < p_T < 2.0$ GeV/$c$.
	The asymmetries are correlated to the high-$p_T$ event-by-event anisotropy $v_{2}^{obs}$, which is calculated from particles with $p_T > 2.0$ GeV/$c$ in the entire TPC to increase statistics.
	Error bars are statistical only.
	}
	\label{fig:asymv2h}
\end{figure}

Figure~\ref{fig:asymv2h} shows the asymmetry correlation results in mid-central 20-40\% centrality Au+Au 200 GeV RUN IV collisions as a function of event-by-event anisotropy for high-$p_T$ $v_{2,p_T>2~\text{GeV/}c}^{obs}$.
The high-$p_T$ anisotropy~($v_{2,p_T>2~\text{GeV/}c}^{obs}$) dependence of same-sign dynamical variances $\delta\langle A^2\rangle$ is shown in figure \ref{fig:asymv2h-a}, and the opposite-sign covariances $\langle A_+A_- \rangle$ is shown in figure \ref{fig:asymv2h-b}.
No significant high-$p_T$ event-by-event anisotropy dependence is observed of same- and opposite-sign correlations in both in-plane and out-of-plane directions, which may suggest that the path-length dependence jet-quenching has little effect on the charge asymmetry correlations.
The differences between the $UD$ and $LR$ of the variance and covariance are shown in figure~\ref{fig:asymv2h-c}.
Both $\Delta\langle A^{2} \rangle$ and $\Delta\langle A_+A_- \rangle$ are positive and flat over a large range of the high-$p_T$ anisotropy.
This is consistent with the finding of medium modified jets and initial state fluctuations in \cite{Petersen:2010di}.

\begin{figure}[thb]
	\begin{center}
		\subfigure[$\delta\langle A^{2} \rangle$ vs low-$p_T$ $v_{2}^{obs}$]{\label{fig:asymv2l-a} \includegraphics[width=0.45\textwidth]{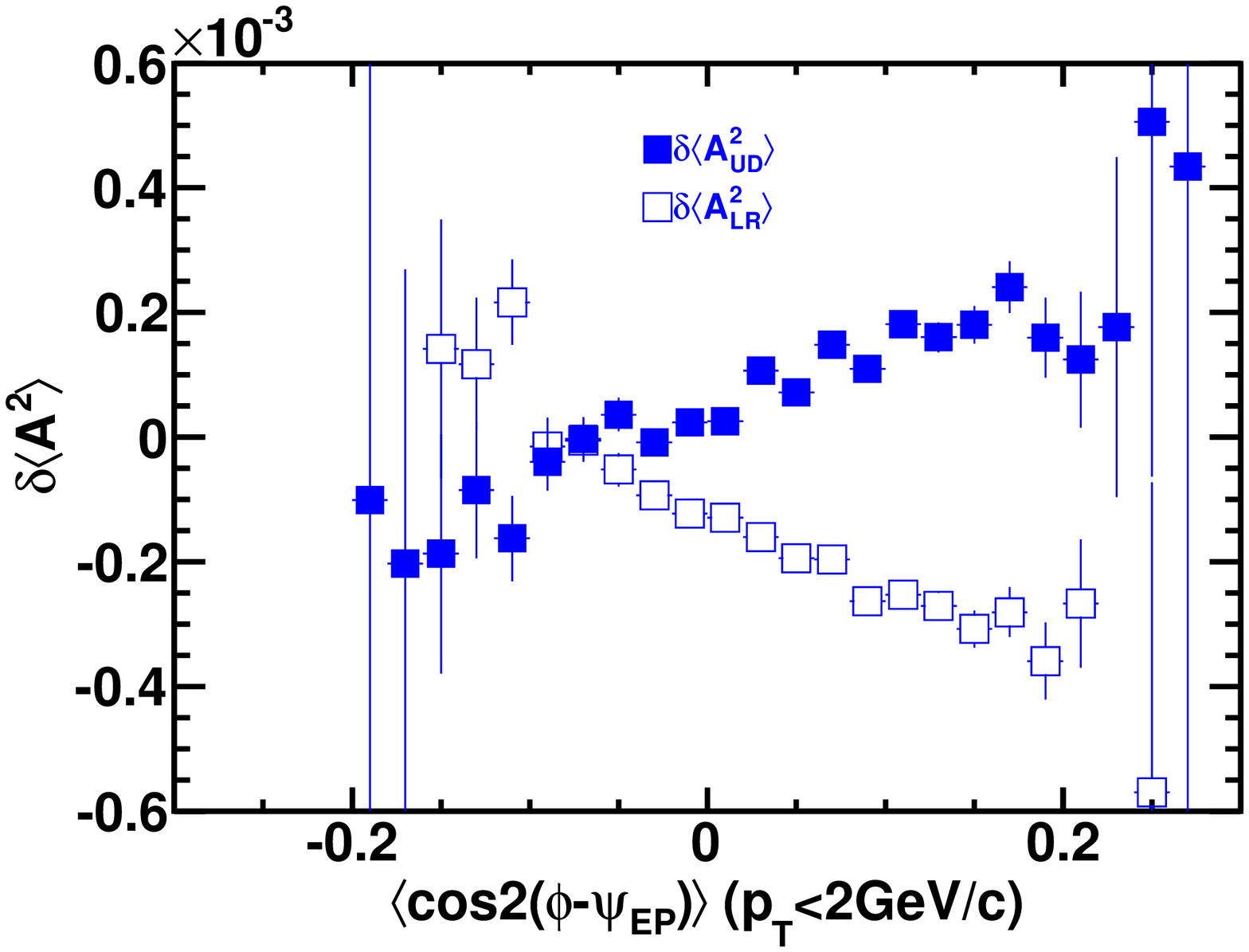}}
		\subfigure[$\delta\langle A_{+}A_{-} \rangle$ vs low-$p_T$ $v_{2}^{obs}$]{\label{fig:asymv2l-b} \includegraphics[width=0.45\textwidth]{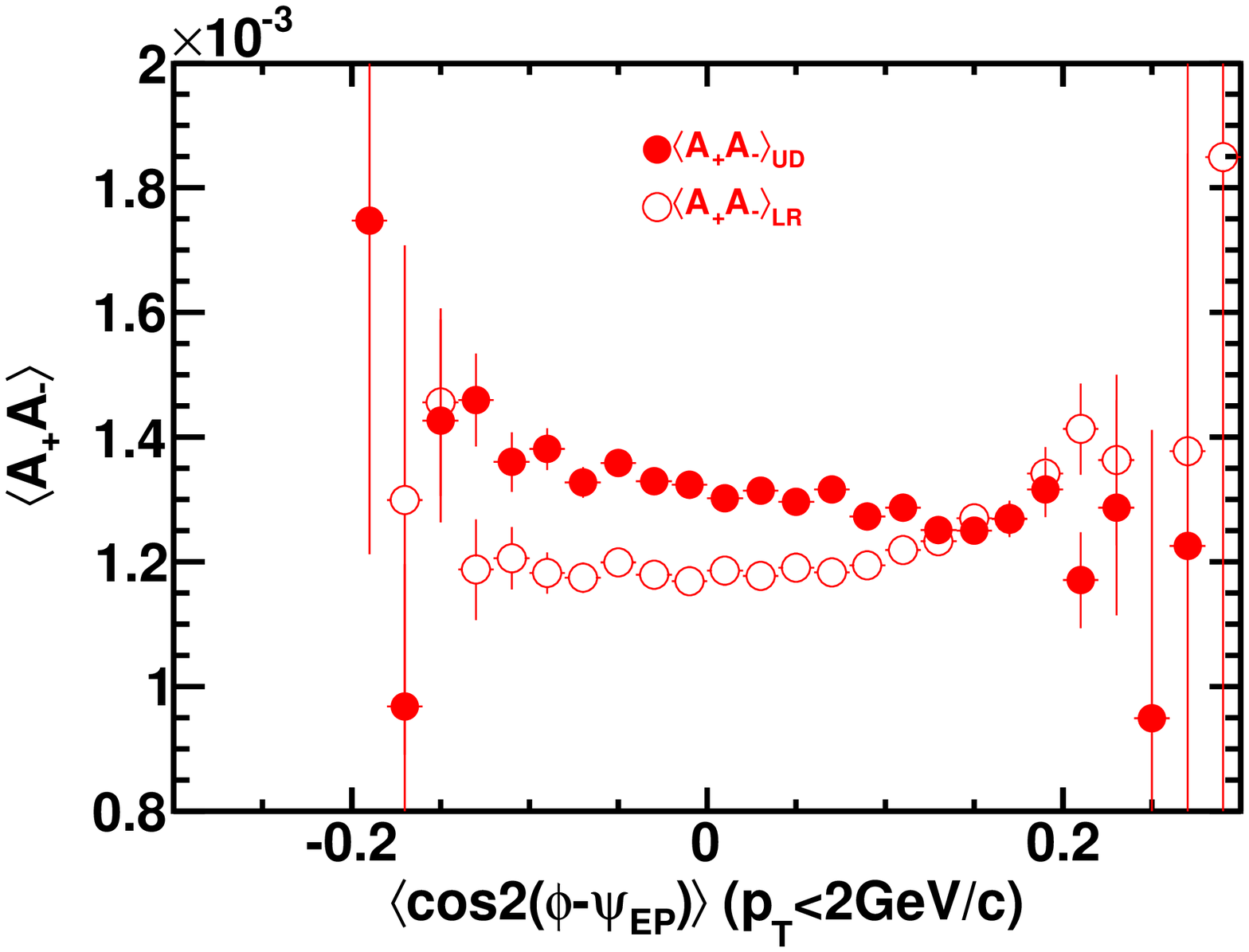}}
		\subfigure[Low-$p_T$ $v_{2}^{obs}$ $UD-LR$ correlations]{\label{fig:asymv2l-c} \includegraphics[width=0.45\textwidth]{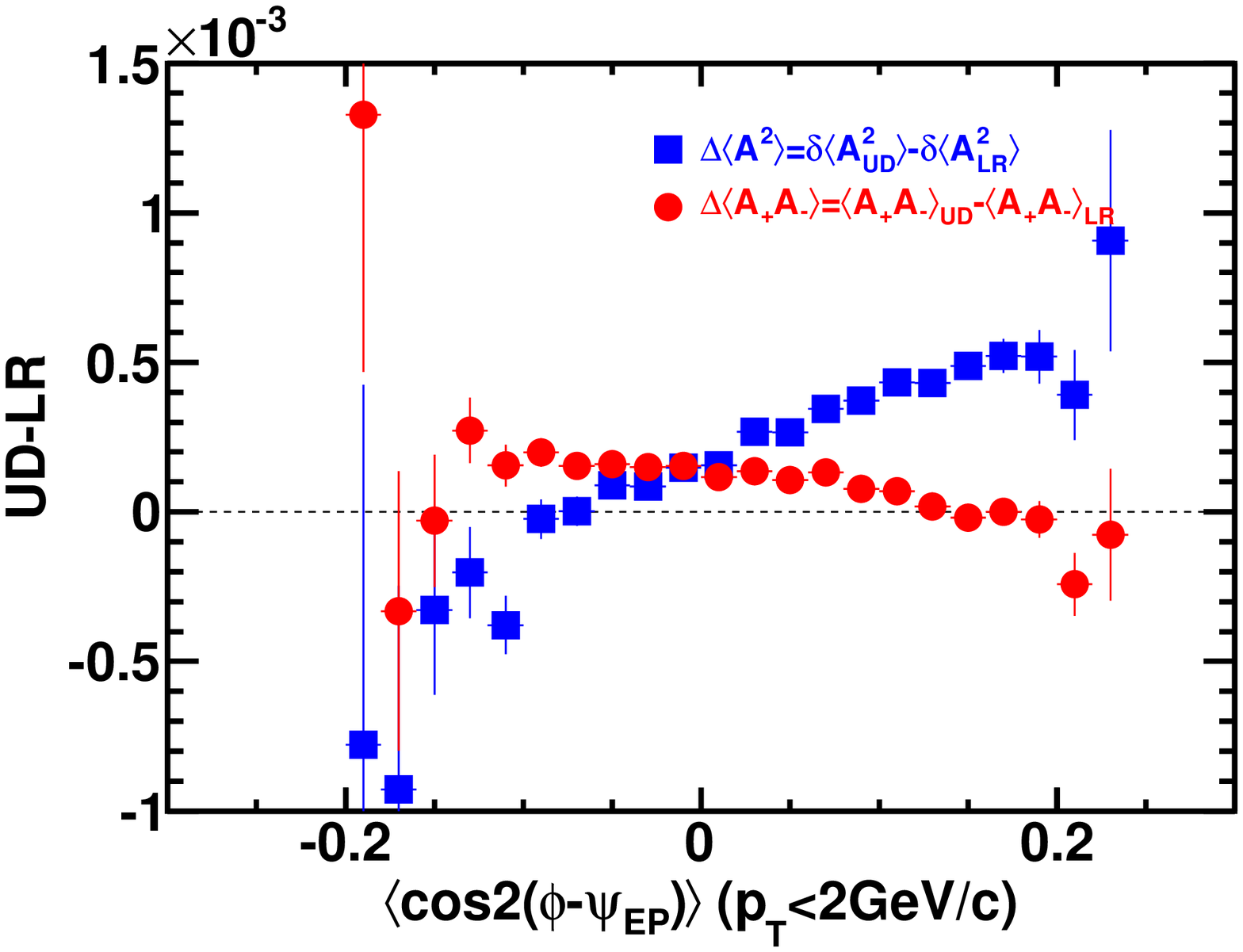}}
	\end{center}
	\caption[Mid-central asymmetry correlations vs low-$p_T$ event-by-event $v_{2}^{obs}$]{
	Low-$p_T$ event-by-event anisotropy $v_2^{obs}$ dependence of (a) the dynamical charge asymmetry variances, (b) asymmetry covariances, and (c) their differences between $UD$ and $LR$ hemispheres, in mid-central 20-40\% Au+Au 200 GeV collisions from RUN IV.
	The asymmetries are calculated from all particles from one side of the TPC tracks with respect to the event-plane reconstructed from the other side of the TPC tracks.
	Both the particles used in asymmetry calculation and event-plane reconstruction are with $p_T$ range of $0.15 < p_T < 2.0$ GeV/$c$.
	The asymmetries are correlated to the low-$p_T$ event-by-event anisotropy $v_{2}^{obs}$, which is calculated from the same particles used for the asymmetries.
	Error bars are statistical only.
	}
	\label{fig:asymv2l}
\end{figure}

Figure~\ref{fig:asymv2l} shows the low-$p_T$ anisotropy dependence of the charge asymmetry correlations.
Different from high-$p_T$ anisotropy, significant $v_{2}^{obs}$ dependence is observed in the dynamical variances $\delta\langle A^2\rangle$ in figure \ref{fig:asymv2l-a}.
The out-of-plane variance $\delta\langle A^2_{UD}\rangle$ increases with low-$p_T$ $v_2^{obs}$, while the in-plane variance $\delta\langle A^2_{LR}\rangle$ decreases with $v_2^{obs}$.
The different trend results in a strong $v_2^{obs}$ dependence of the difference between $UD$ and $LR$ ($\Delta\langle A^2\rangle$), which is shown in figure~\ref{fig:asymv2l-c} in squares.
Some low-$p_T$ $v_2^{obs}$ dependence is observed in the covariances $\langle A_+A- \rangle$, but is significantly weaker than the variances.
However, we can still see opposite trend in the covariances, that a slightly decreasing trend for low-$p_T$ $v_2^{obs}$ in $UD$ and increasing trend in $LR$ as shown in figure \ref{fig:asymv2l-b}.
The difference of the covariance $\Delta\langle A_+A_- \rangle$ is shown in figure~\ref{fig:asymv2l-c} in circles,
which decreases with increasing $v_2^{obs}$.
It is interesting to see that the same-sign and opposite-sign $UD-LR$ correlations have different trend with low-$p_T$ $v_2^{obs}$.

It is important to point out in figure \ref{fig:asymv2l-c} that, $\Delta\langle A^2\rangle$ and $\Delta\langle A_+A_- \rangle$ cross at $v_2^{obs} \approx 0$, and the crossing value is positive.
For events with $v_2^{obs} \approx 0$, no significant charge difference is observed between same-sign and opposite-sign pair correlations, at which the particle distribution is isotropic.
We know that the average $v_2^{obs}$ over the entire event sample is positive due to the elliptic flow, table \ref{tab:cent}.
\red{
The different trend of the variance and covariance as a function of $v_2^{obs}$  leads to the integrated values of variance and covariance to diverge.
In other words, after integration, the $UD-LR$ variance is larger than the covariance, i.e. $\Delta\langle A^2 \rangle > \Delta\langle A_+A_- \rangle$ for all centralities at $\langle v_{2,p_T<2\text{GeV/}c}^{obs} \rangle > 0$.
}

\red{
There is centrality dependence of the magnitudes of the asymmetry correlations as a function of $v_{2}^{obs}$.
We show the central 0-20\% Au+Au 200 GeV $v_2^{obs}$ asymmetry correlation dependence in figure \ref{fig:appasymv220} and the peripheral 40-80\% collisions in figure \ref{fig:appasymv280} in the appendix.
The features are qualitatively similar as the mid-central collisions.
For high-$p_T$ $v_2^{obs}$, the dynamical variances and covariances are consistent with constant for all centralities.
The difference between $UD$ and $LR$ does not depend on high-$p_T$ $v_2^{obs}$.
However, the dynamical variances and covariances strongly depend on low-$p_T$ $v_2^{obs}$,
and the difference between $UD$ and $LR$ shows opposite trend of the variance and covariance in all centralities.
The magnitudes are different though.
}

\begin{figure}[thb]
	\begin{center}
		\subfigure{\label{fig:asymv2etagap-a} \includegraphics[width=0.45\textwidth]{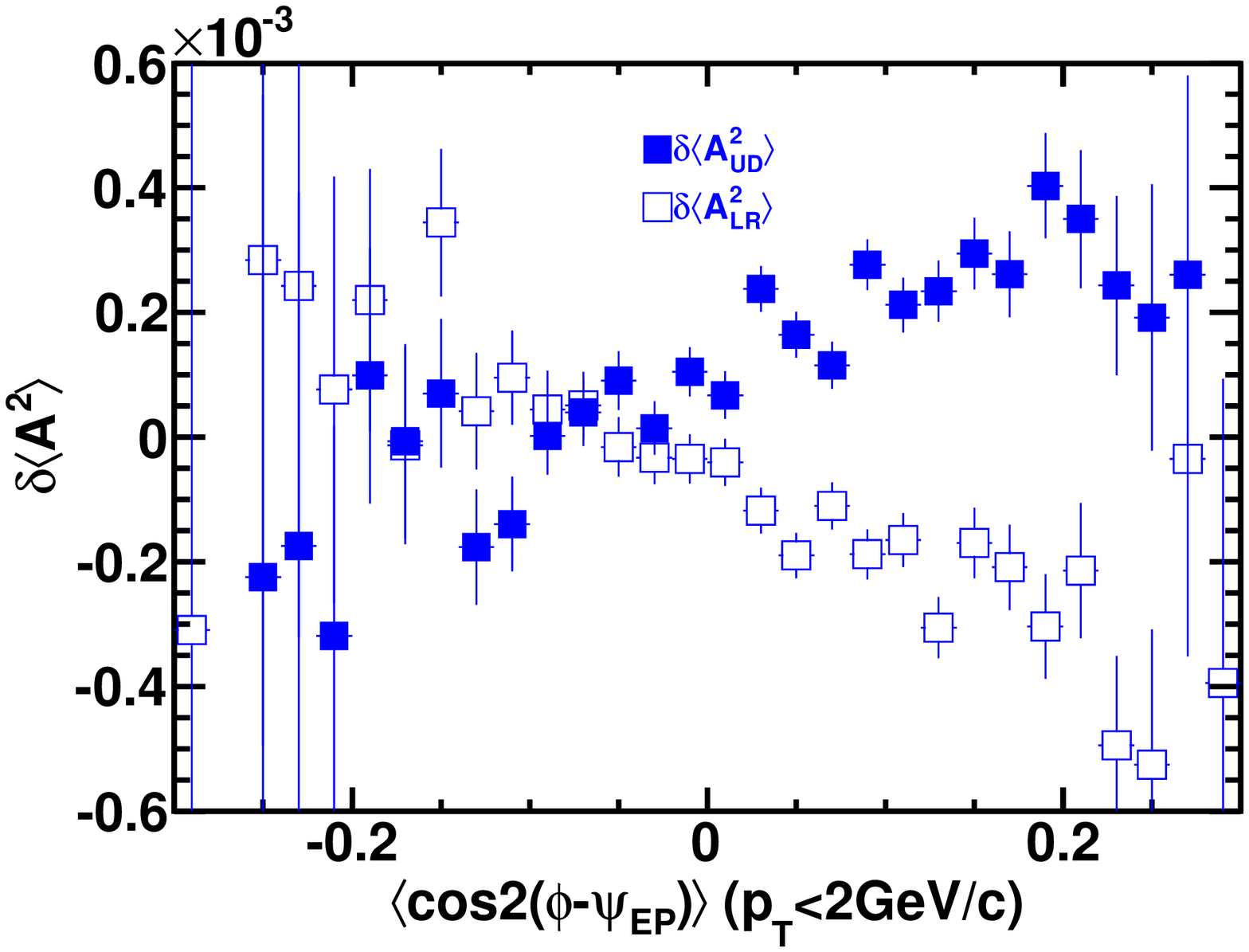}}
		\subfigure{\label{fig:asymv2etagap-b} \includegraphics[width=0.45\textwidth]{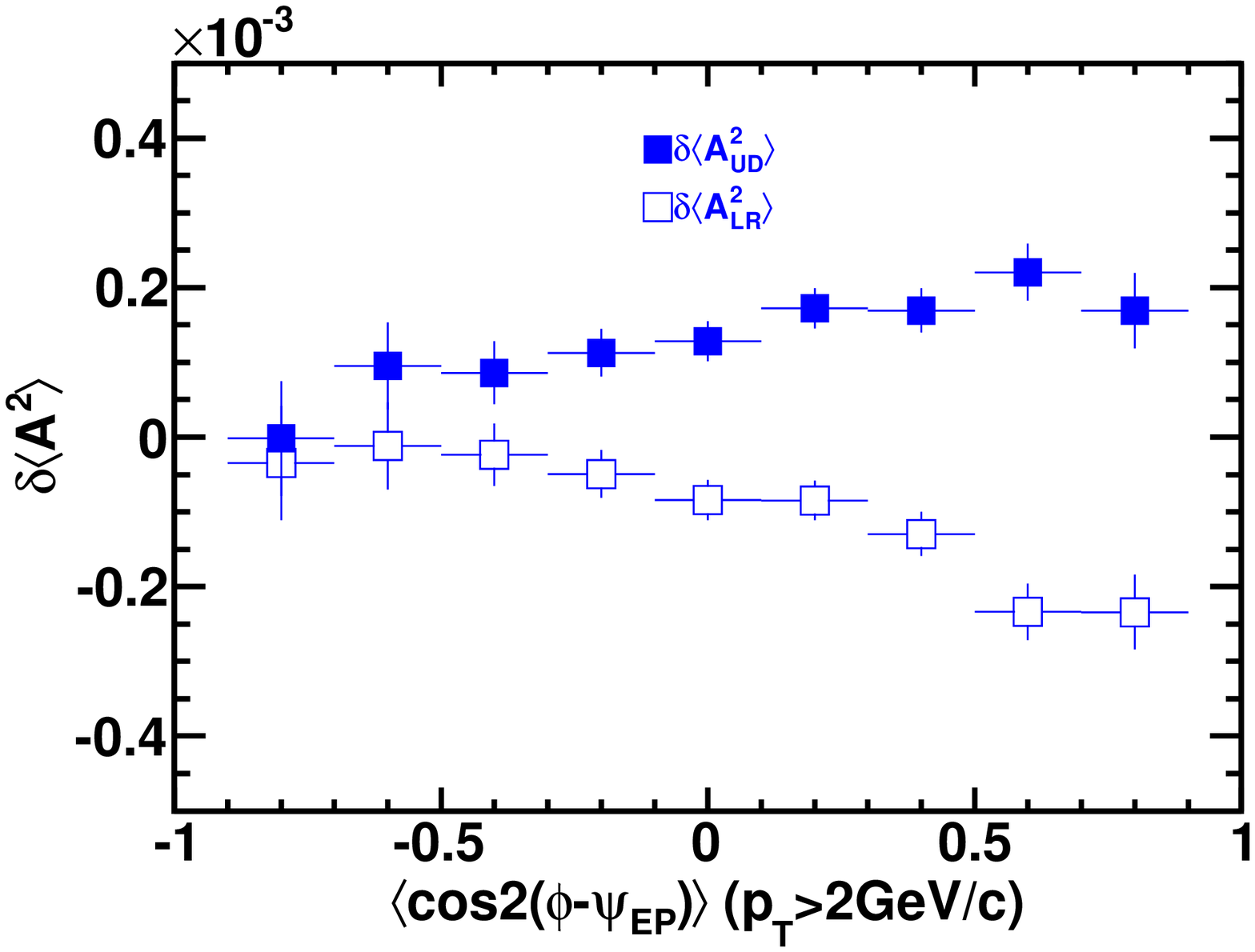}}
		\subfigure{\label{fig:asymv2etagap-c} \includegraphics[width=0.45\textwidth]{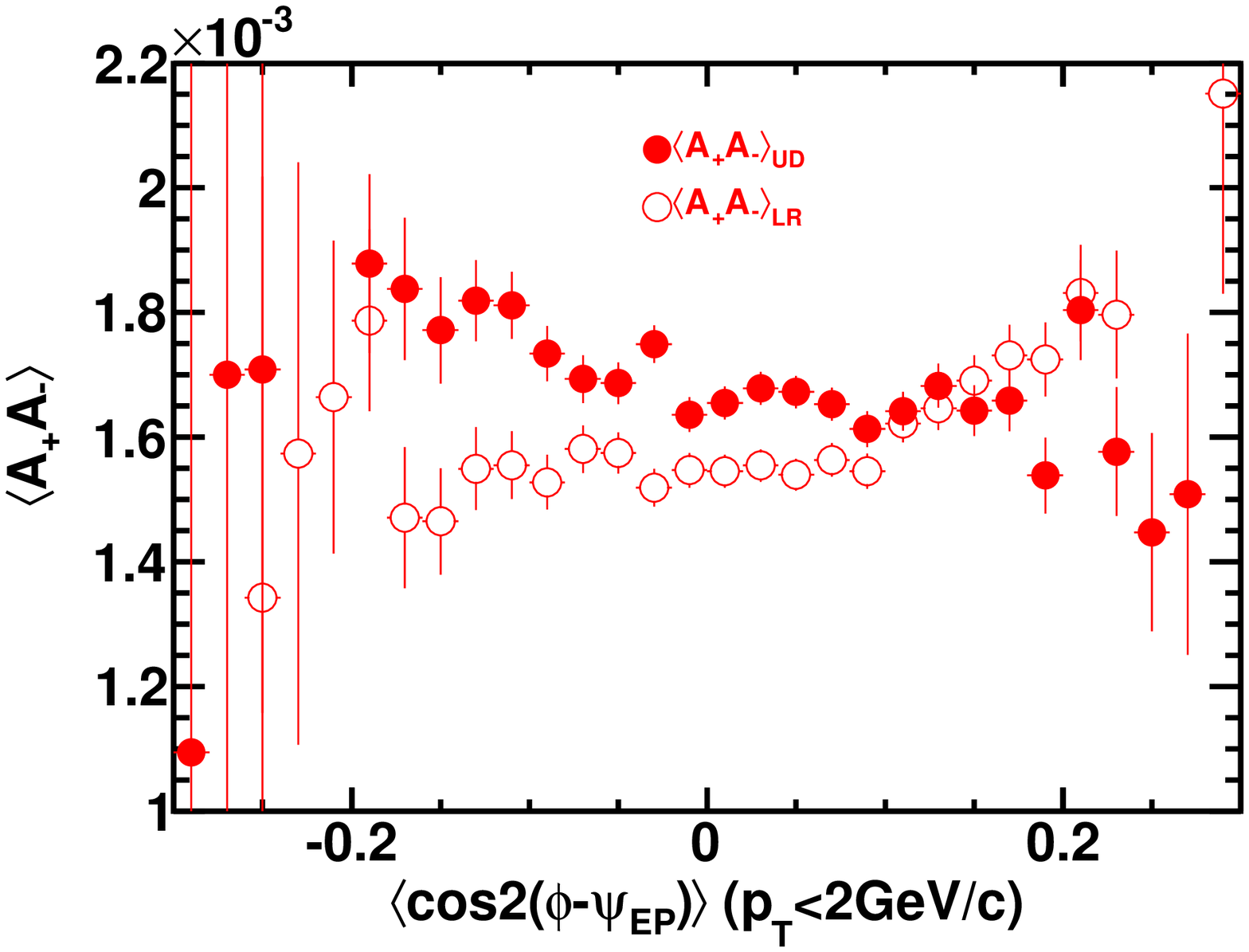}}
		\subfigure{\label{fig:asymv2etagap-d} \includegraphics[width=0.45\textwidth]{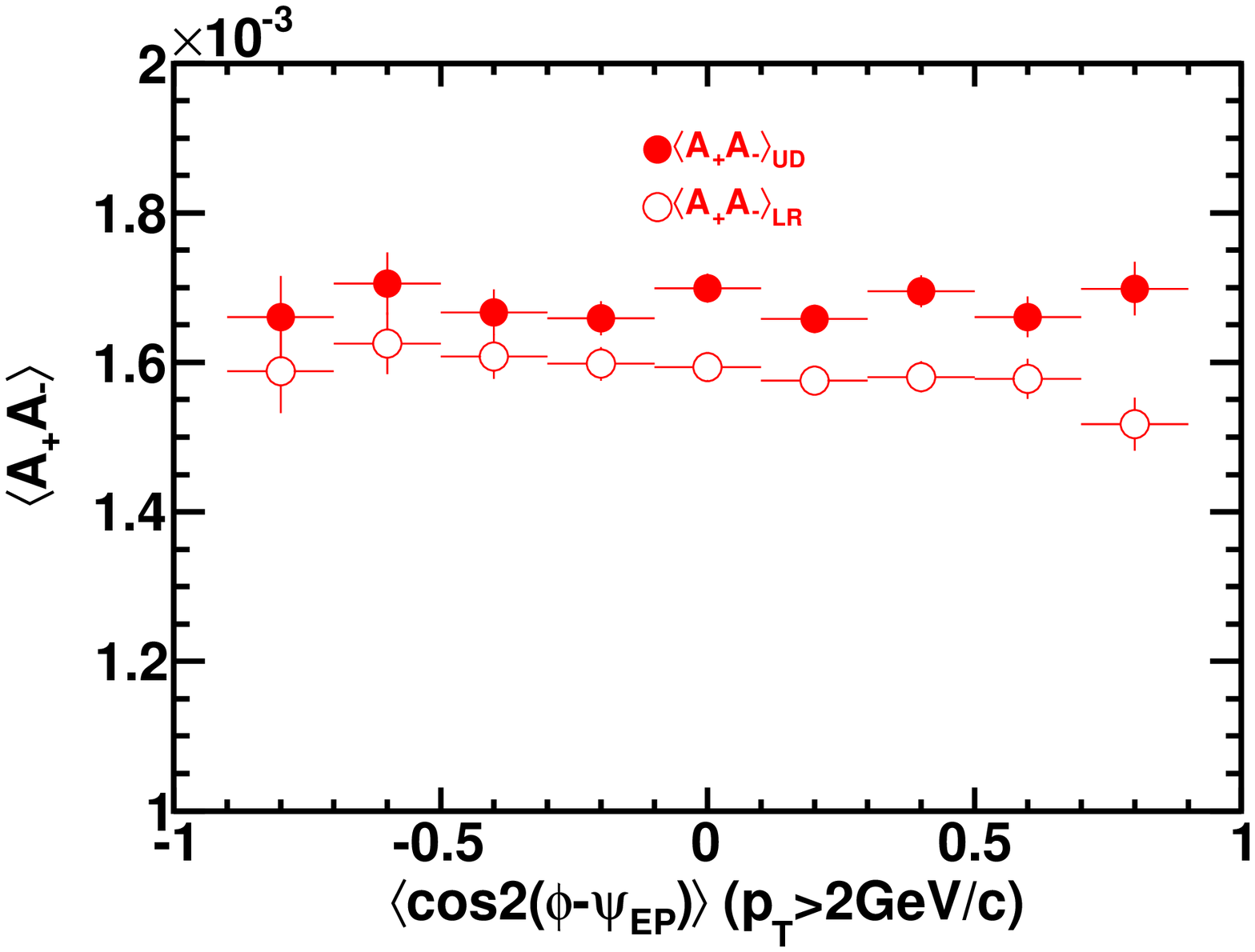}}
		\subfigure{\label{fig:asymv2etagap-e} \includegraphics[width=0.45\textwidth]{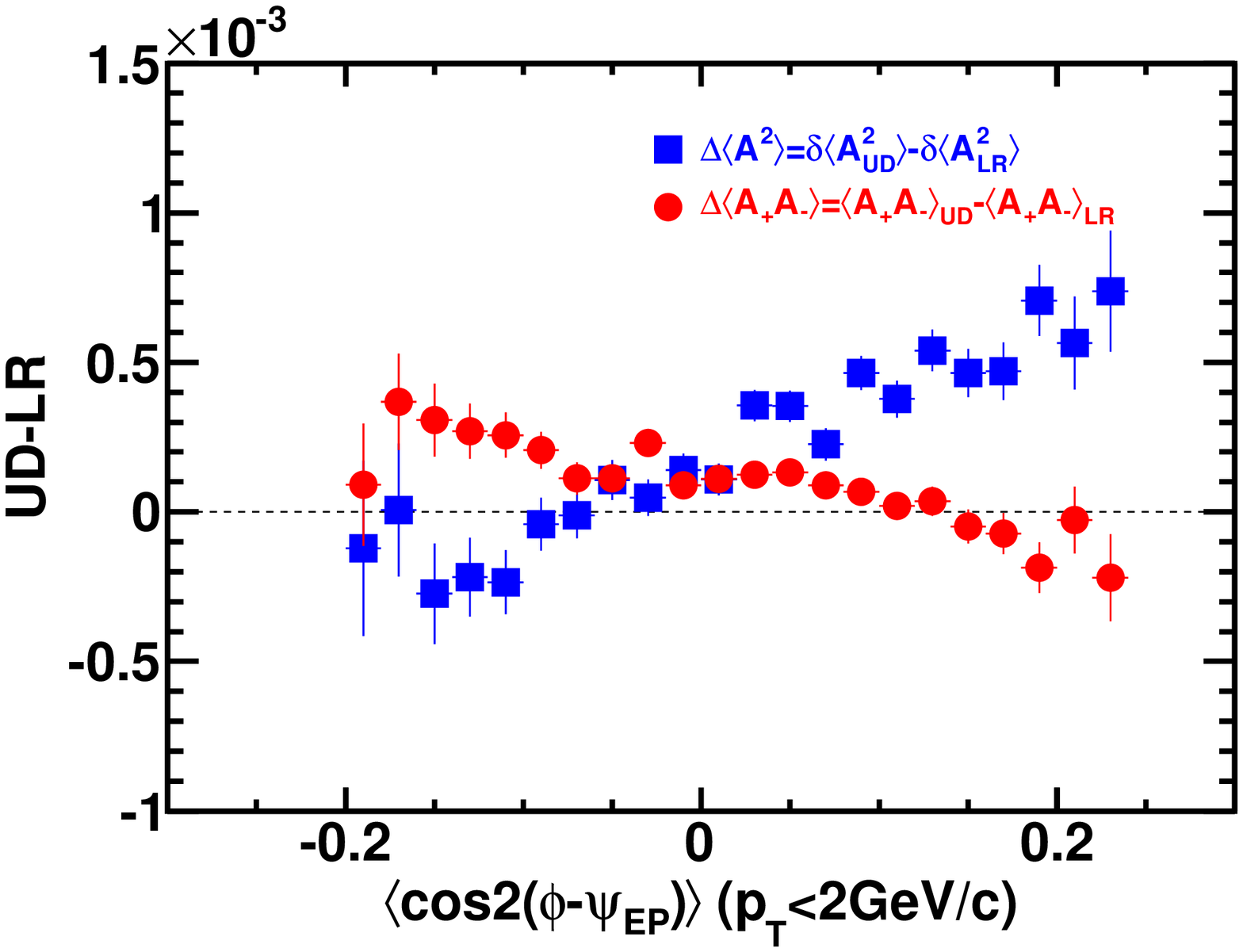}}
		\subfigure{\label{fig:asymv2etagap-f} \includegraphics[width=0.45\textwidth]{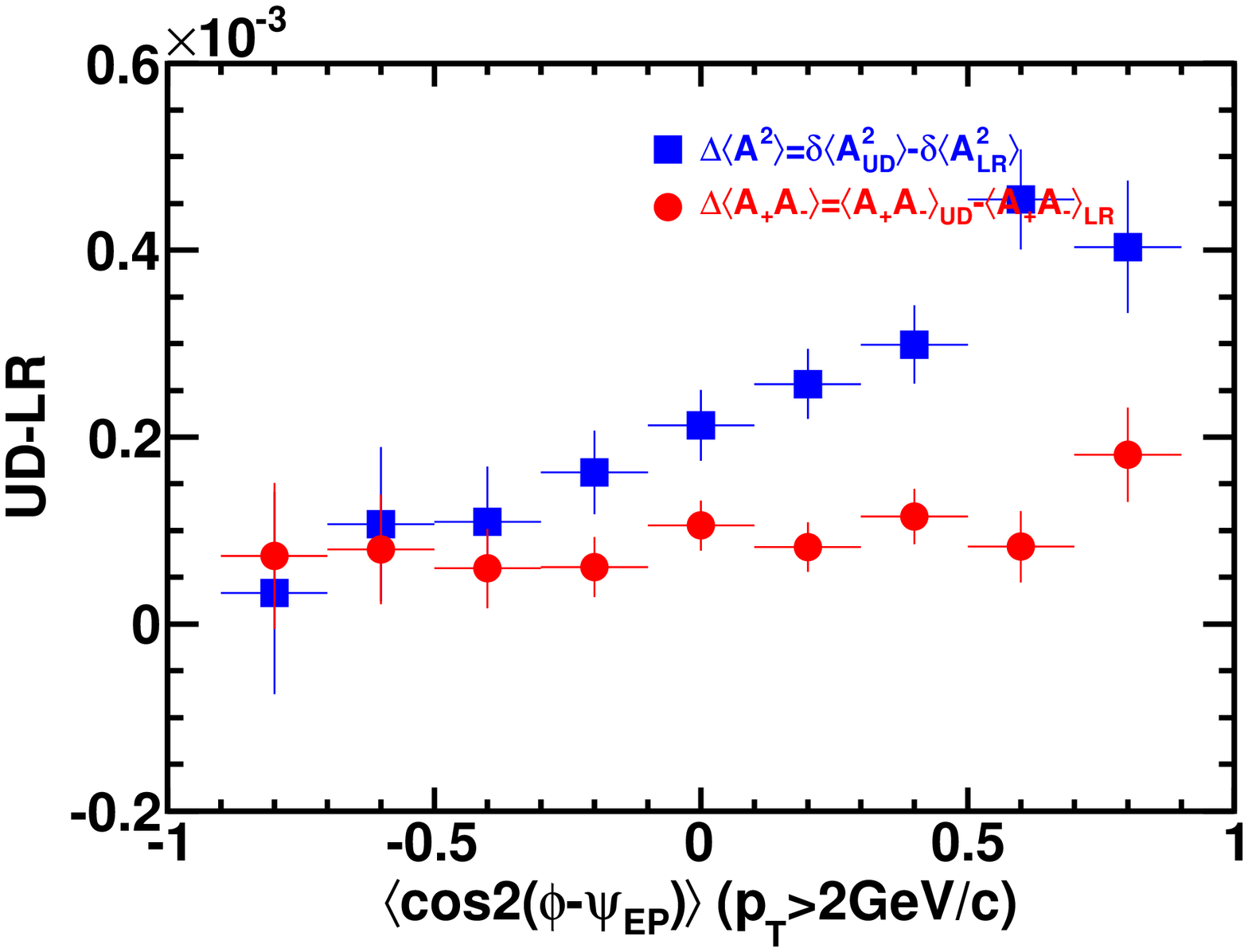}}
	\end{center}
	\caption[Mid-central asymmetry correlations vs $v_{2}^{obs}$ with $eta$ gap]{
	Event-by-event $v_2^{obs}$ dependence of the dynamical variances (top panel), covariances (middle panel) and $UD-LR$ (bottom panel) of RUN IV 20-40\% 200 GeV Au+Au data.
	Particles used for asymmetries and EP reconstruction are separated by one unit in $\eta$.
	Asymmetries are correlated to low-$p_T$ (left column) and high-$p_T$ (right column) $v_{2}^{obs}$.
	Error bars are statistical only.
	}
	\label{fig:asymv2etagap}
\end{figure}

\red{
To further remove possible short range correlation between the particles used in asymmetry calculation and the event-plane reconstruction, 
we choose particles with a large pseudo-rapidity gap between the two particle sets for event-plane reconstruction and the charge multiplicity asymmetry correlation.
Instead of separating event by $-1<\eta<0$ and $0<\eta<1$, 
we separate the event with $-1.0<\eta<-0.5$ and $0.5<\eta<1.0$, such that the two sub events have a one unit pseudo-rapidity gap, which is the usual soft particle bulk correlation span in pseudo-rapidity space.
}

\red{
We show similar $v_2^{obs}$ dependence results of mid-central Au+Au 200 GeV data in figure \ref{fig:asymv2etagap}.
Low-$p_T$ $v_{2}^{obs}$ ($p_T<2.0$ GeV/$c$) is calculated from the same particles used for the asymmetries.
High-$p_T$ $v_{2}^{obs}$ is calculated from all particles with $p_T>2.0$ GeV/$c$ to increase statistics.
The corresponding $UD-LR$ correlation with $\eta$ gap is shown in the bottom panel in figure \ref{fig:asymv2etagap}.
The result is consistent with what we observed without $\eta$ gap but with larger error bars because of the limited statistics.
The central and peripheral collision results are shown in figure \ref{fig:appasymv2etagap20} and figure \ref{fig:appasymv2etagap80}.
They are also qualitatively consistent with the mid-central result.
}

\begin{figure}[thb]
	\begin{center}
		\subfigure{\label{fig:asymv2zdc-a} \includegraphics[width=0.45\textwidth]{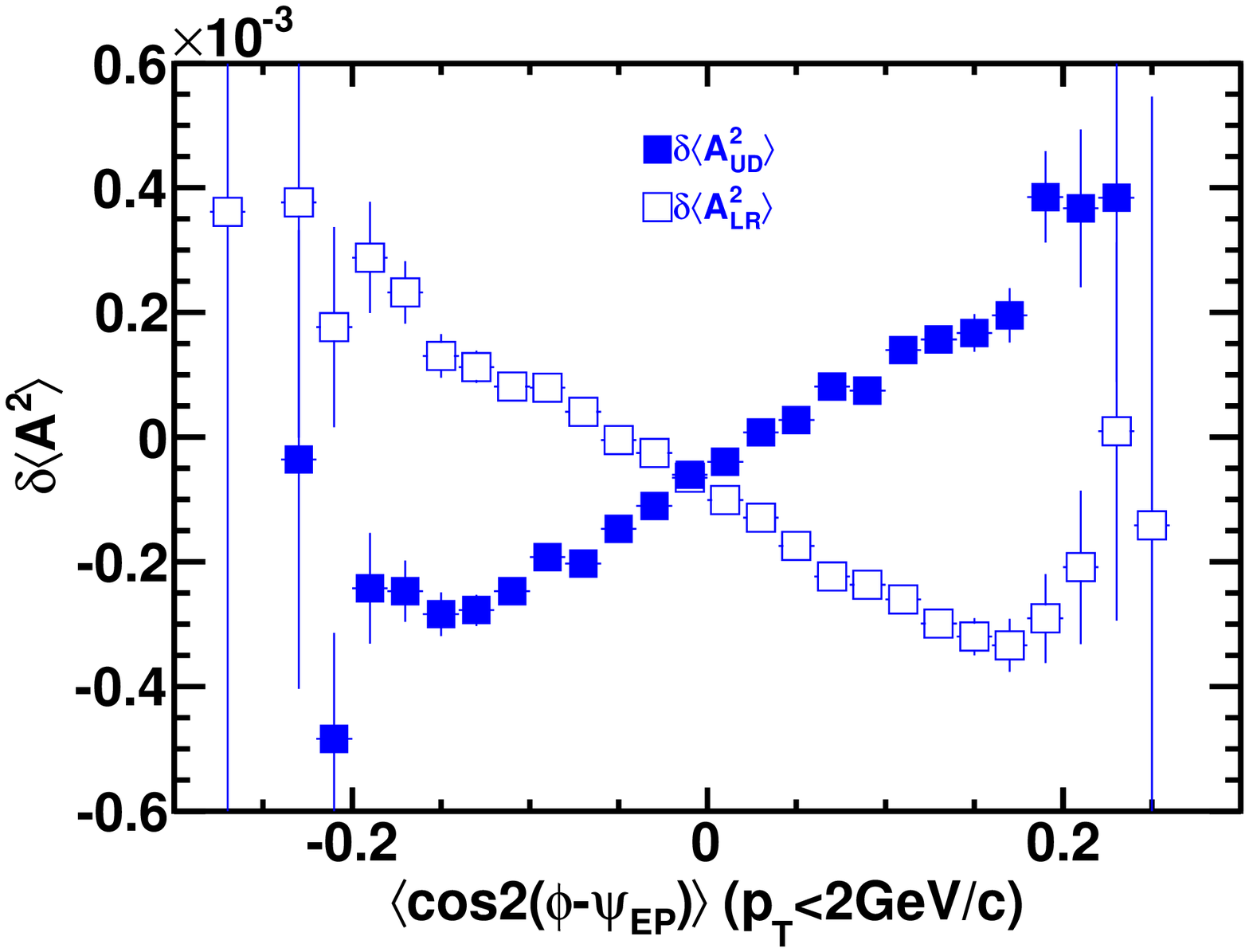}}
		\subfigure{\label{fig:asymv2zdc-b} \includegraphics[width=0.45\textwidth]{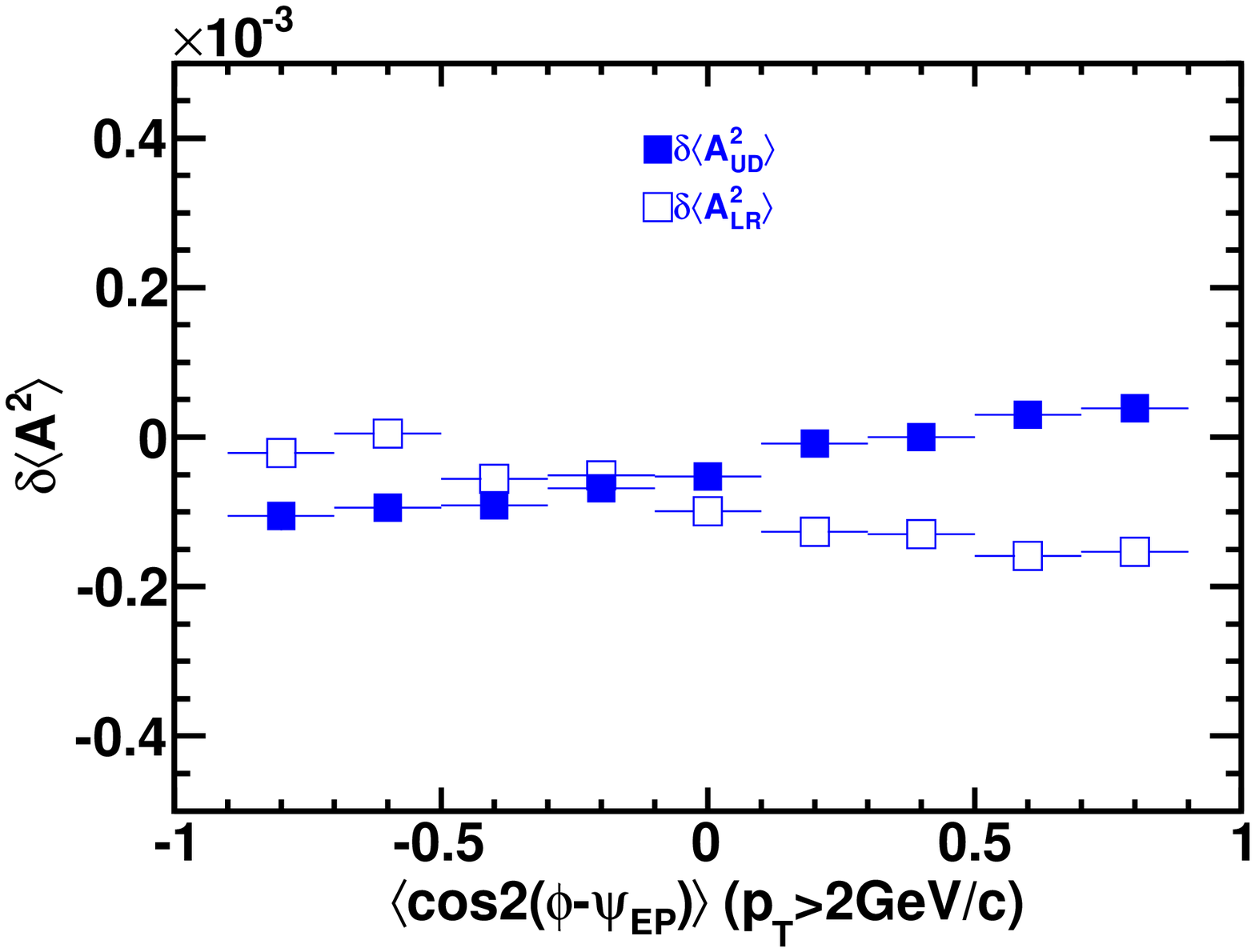}}
		\subfigure{\label{fig:asymv2zdc-c} \includegraphics[width=0.45\textwidth]{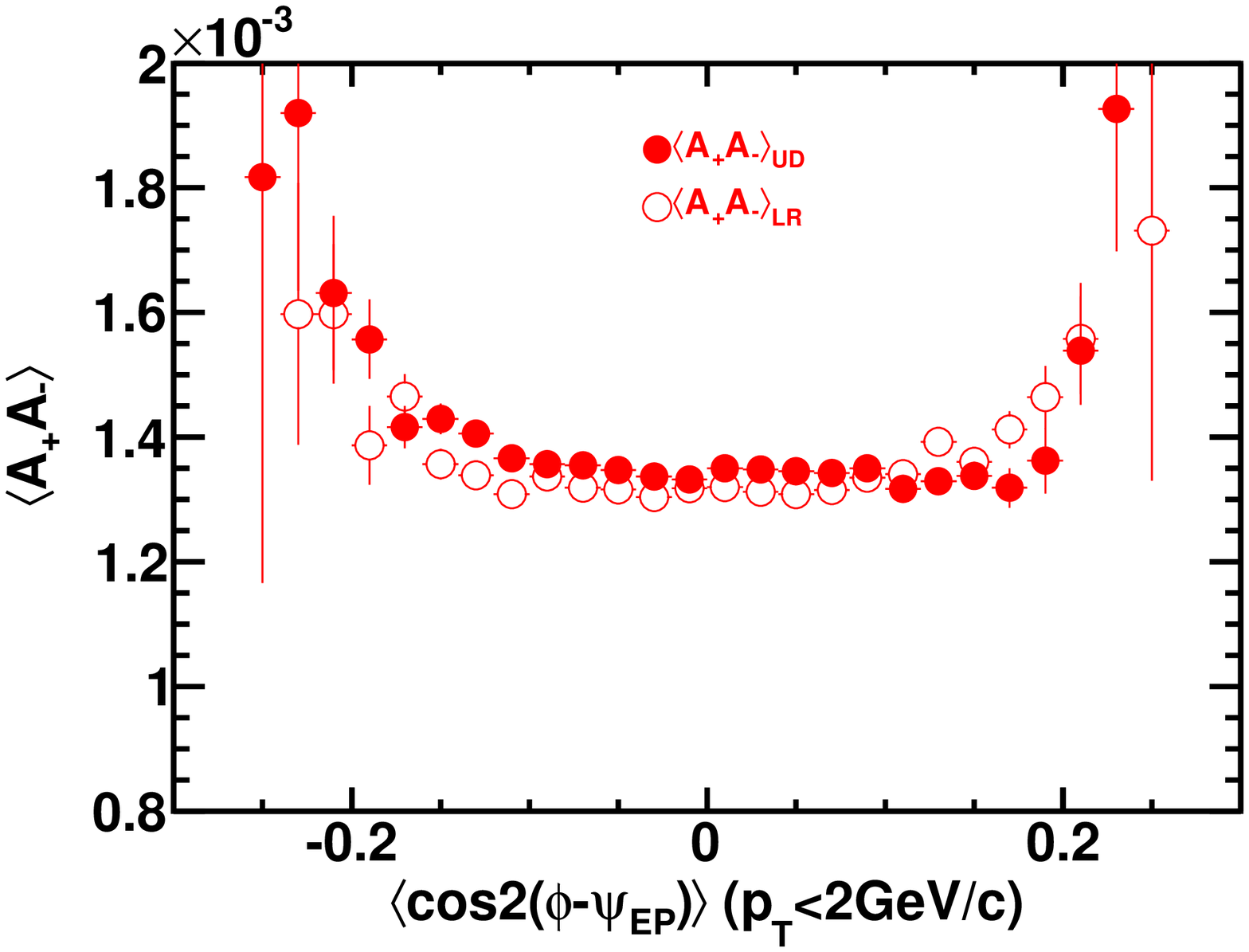}}
		\subfigure{\label{fig:asymv2zdc-d} \includegraphics[width=0.45\textwidth]{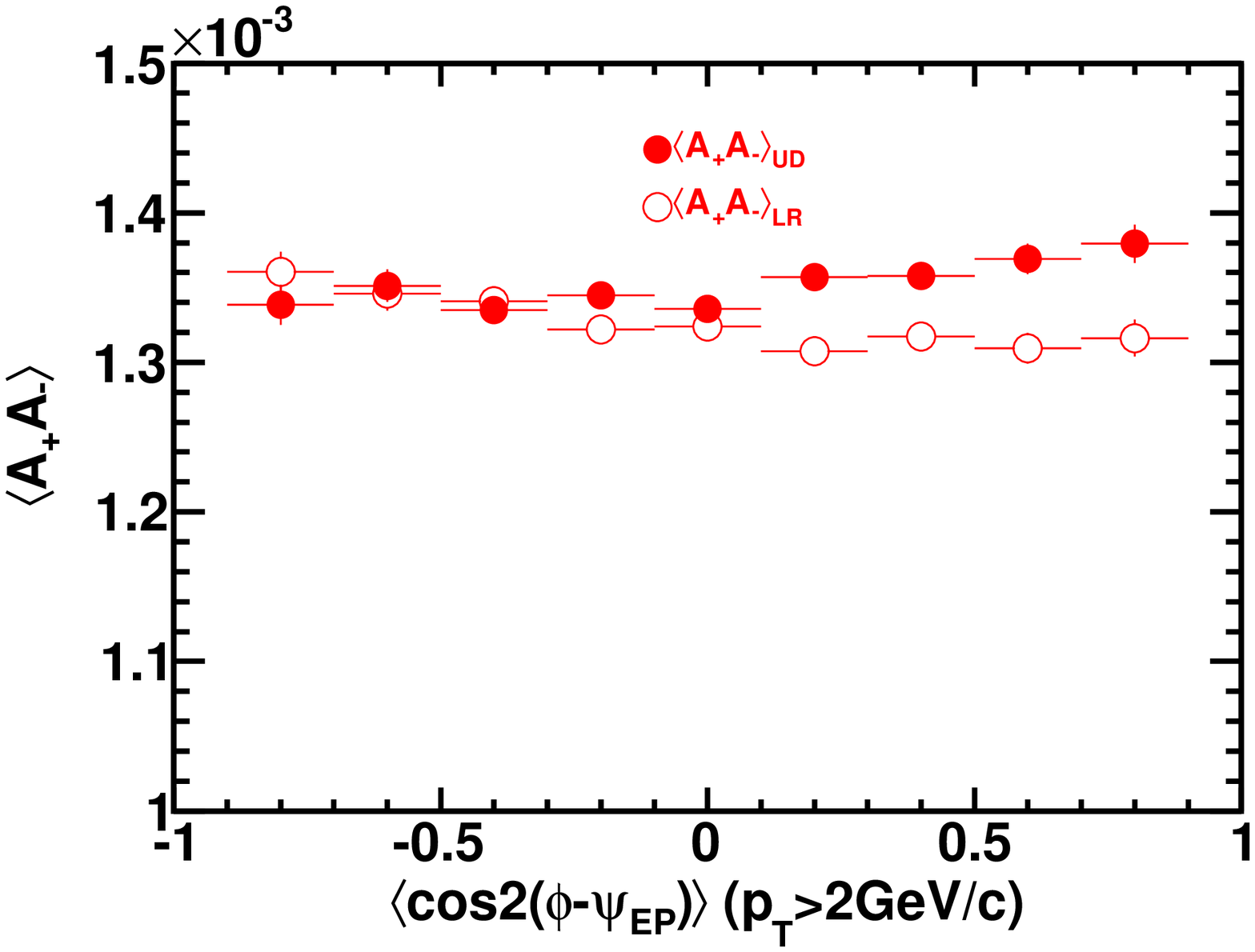}}
		\subfigure{\label{fig:asymv2zdcd-a} \includegraphics[width=0.45\textwidth]{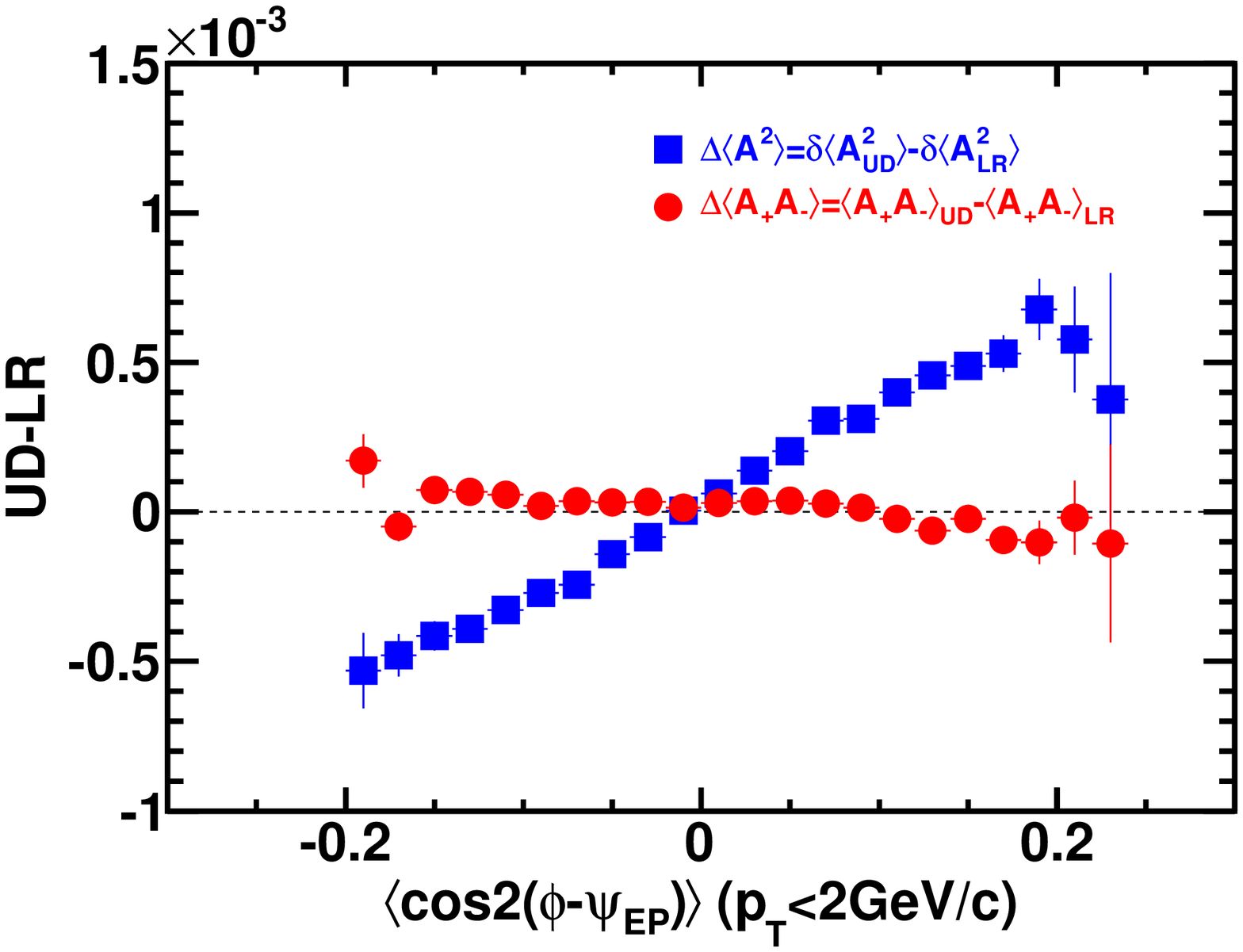}}
		\subfigure{\label{fig:asymv2zdcd-b} \includegraphics[width=0.45\textwidth]{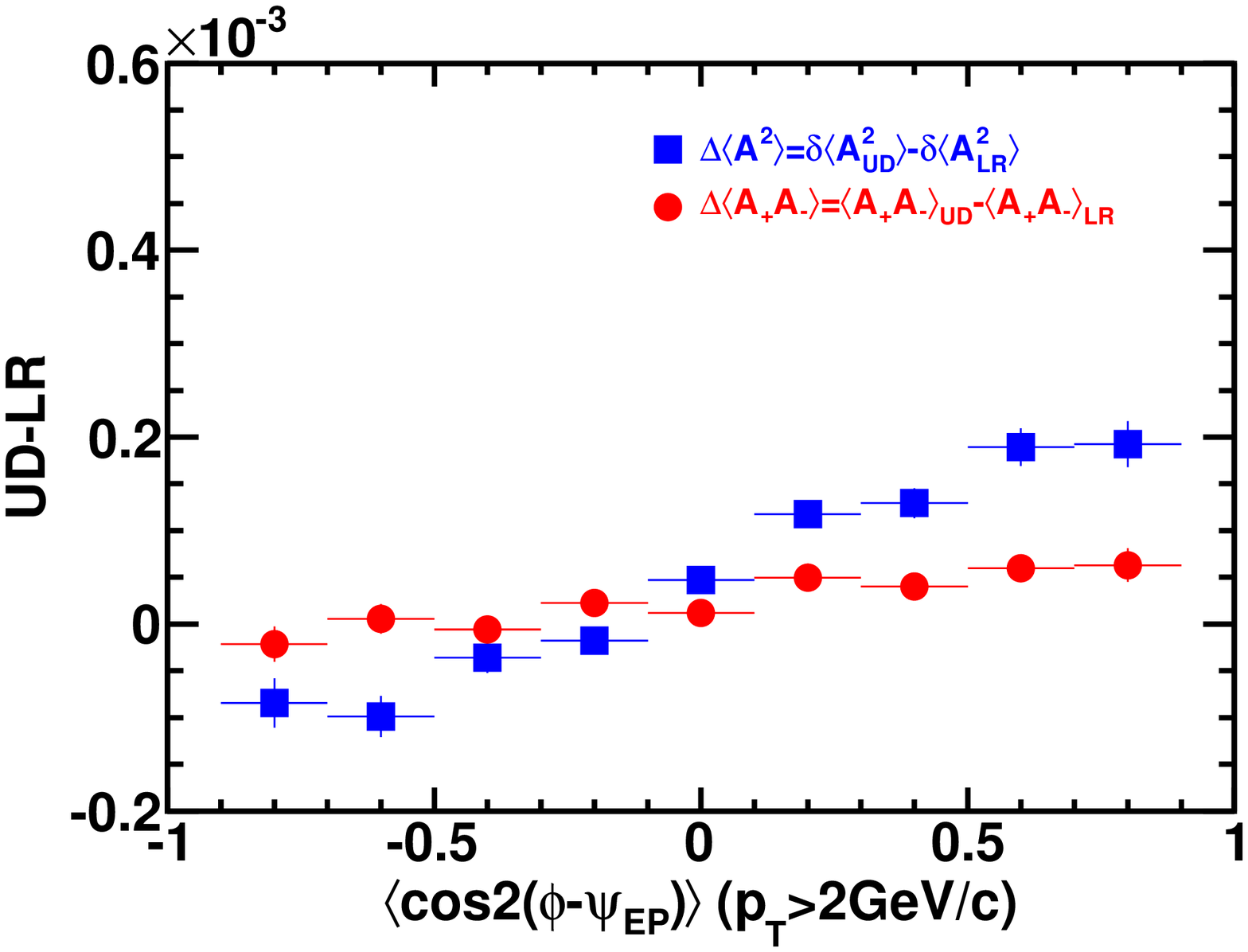}}
	\end{center}
	\caption[Mid-central asymmetry correlations vs $v_{2}^{obs}$ with ZDC-SMD EP]{
	Asymmetry correlations vs event-by-event $v_{2}^{obs}$ of RUN VII 200 GeV Au+Au mid-central 20-40\% collisions.
	The particles used for asymmetry calculation are from half side of the TPC with respect to the first order event-plane reconstructed from ZDC-SMD.
	Error bars are statistical only.
	}
	\label{fig:asymv2zdc}
\end{figure}

\red{
To even further remove the short range correlation, we use the first order event-plane reconstructed with the ZDC-SMD detector, which is $|\eta|>6$ away from the main TPC in pseudo-rapidity direction.
The large distance in pseudo-rapidity space removes short range correlation almost completely.
We calculate the charge asymmetry correlations and the event-by-event $v_2^{obs}$ with respect to the first order ZDC-SMD event-plane using RUN VII Au+Au 200 GeV data.
We show the result in figure \ref{fig:asymv2zdc} which has very good consistency with the second order event-plane result.
The central and peripheral asymmetry correlations vs $v_2^{obs}$ results are shown in figure \ref{fig:appasymv2zdc20} and figure \ref{fig:appasymv2zdc80}, where similar dependence is observed.
}

\begin{figure}[thb]
	\begin{center}
		\subfigure{\label{fig:asymv2top2-a} \includegraphics[width=0.45\textwidth]{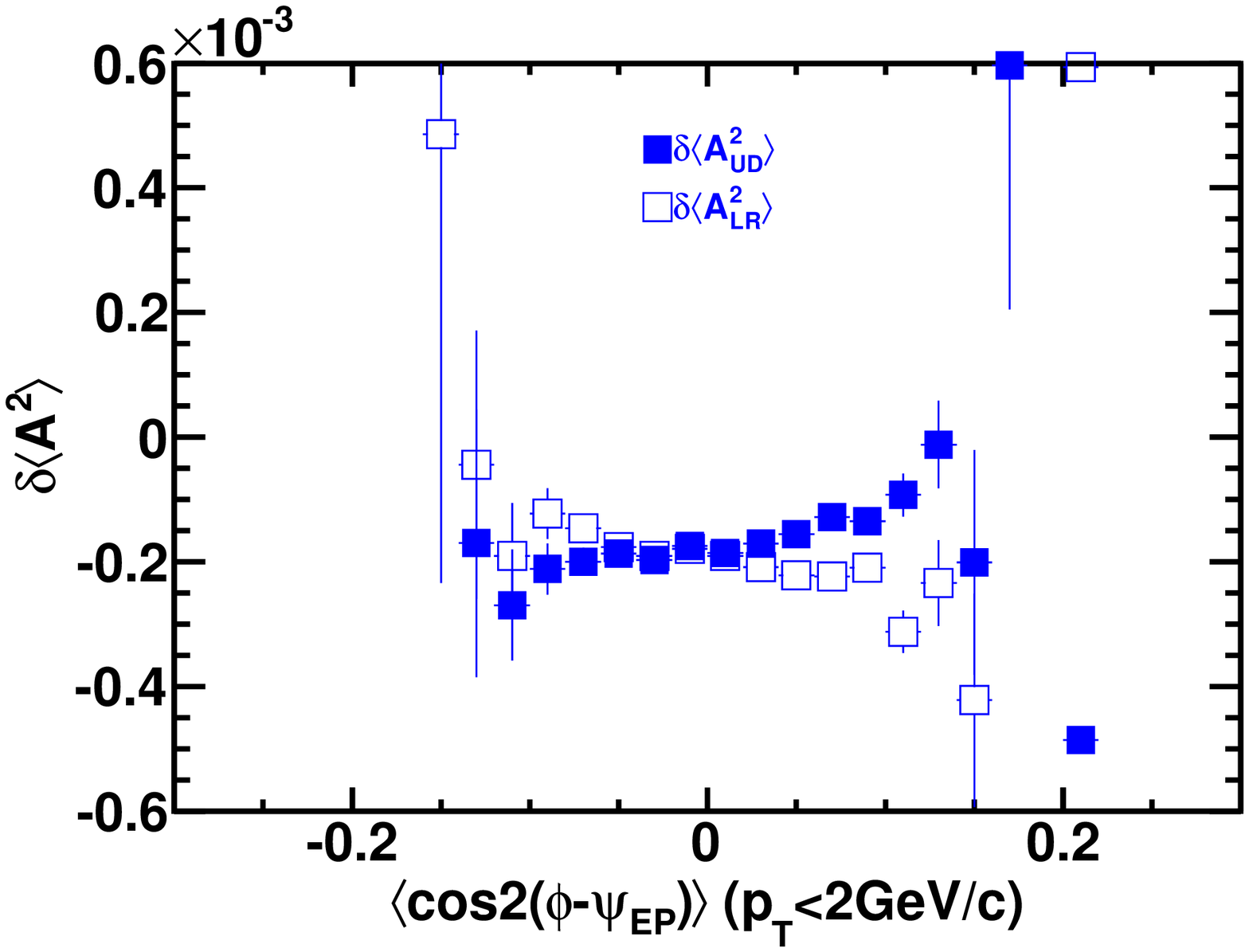}}
		\subfigure{\label{fig:asymv2top2-b} \includegraphics[width=0.45\textwidth]{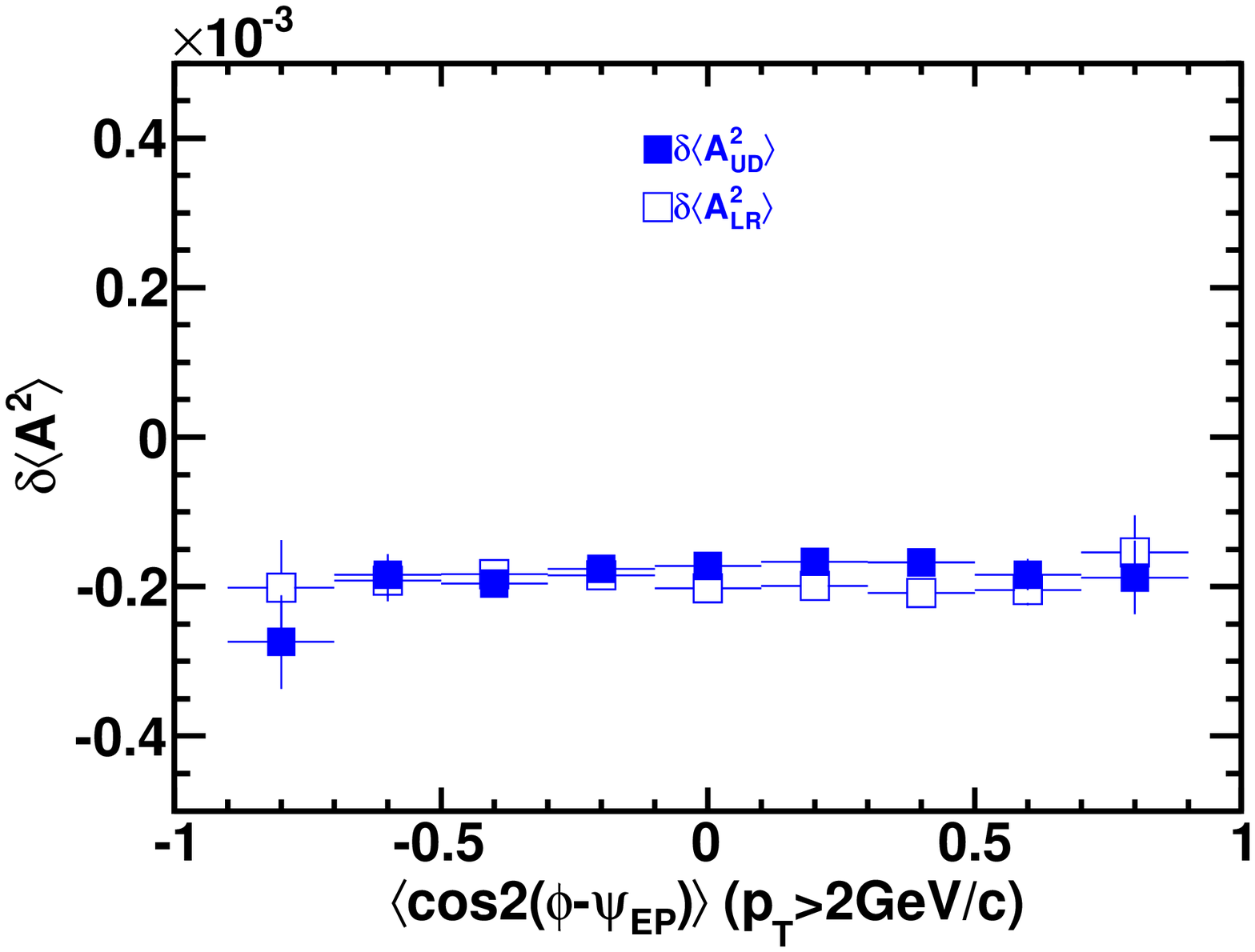}}
		\subfigure{\label{fig:asymv2top2-c} \includegraphics[width=0.45\textwidth]{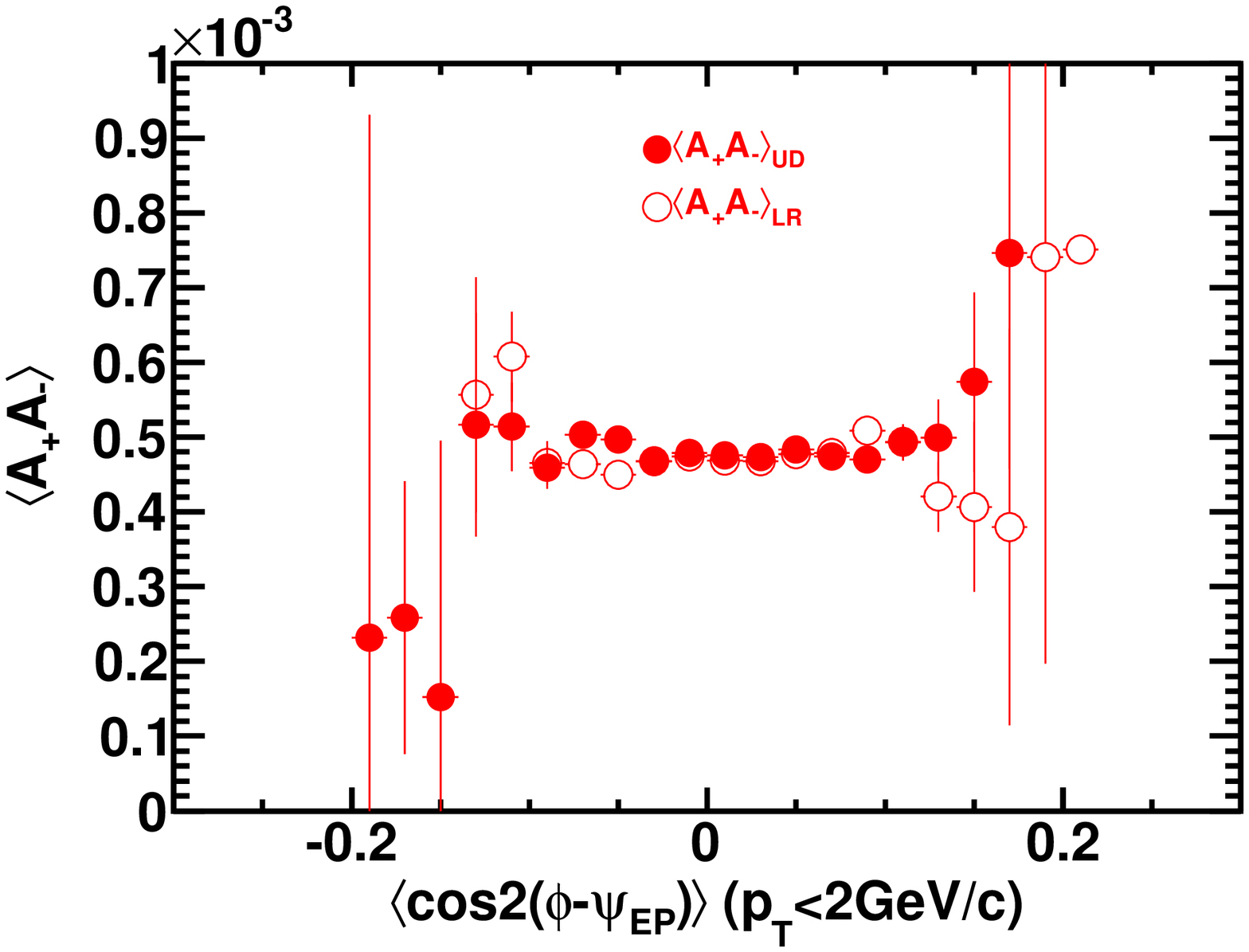}}
		\subfigure{\label{fig:asymv2top2-d} \includegraphics[width=0.45\textwidth]{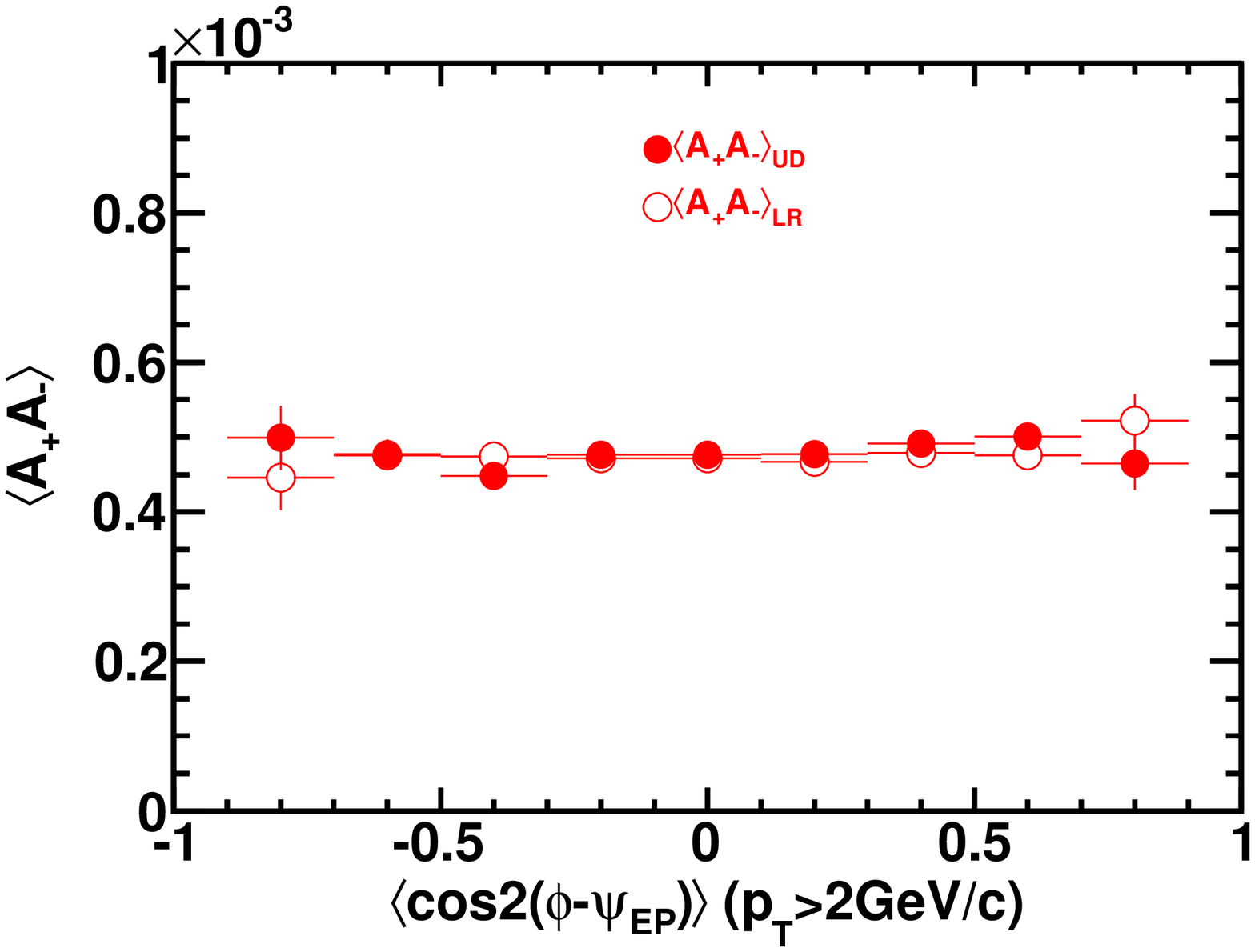}}
		\subfigure{\label{fig:asymv2top2d-a}\includegraphics[width=0.45\textwidth]{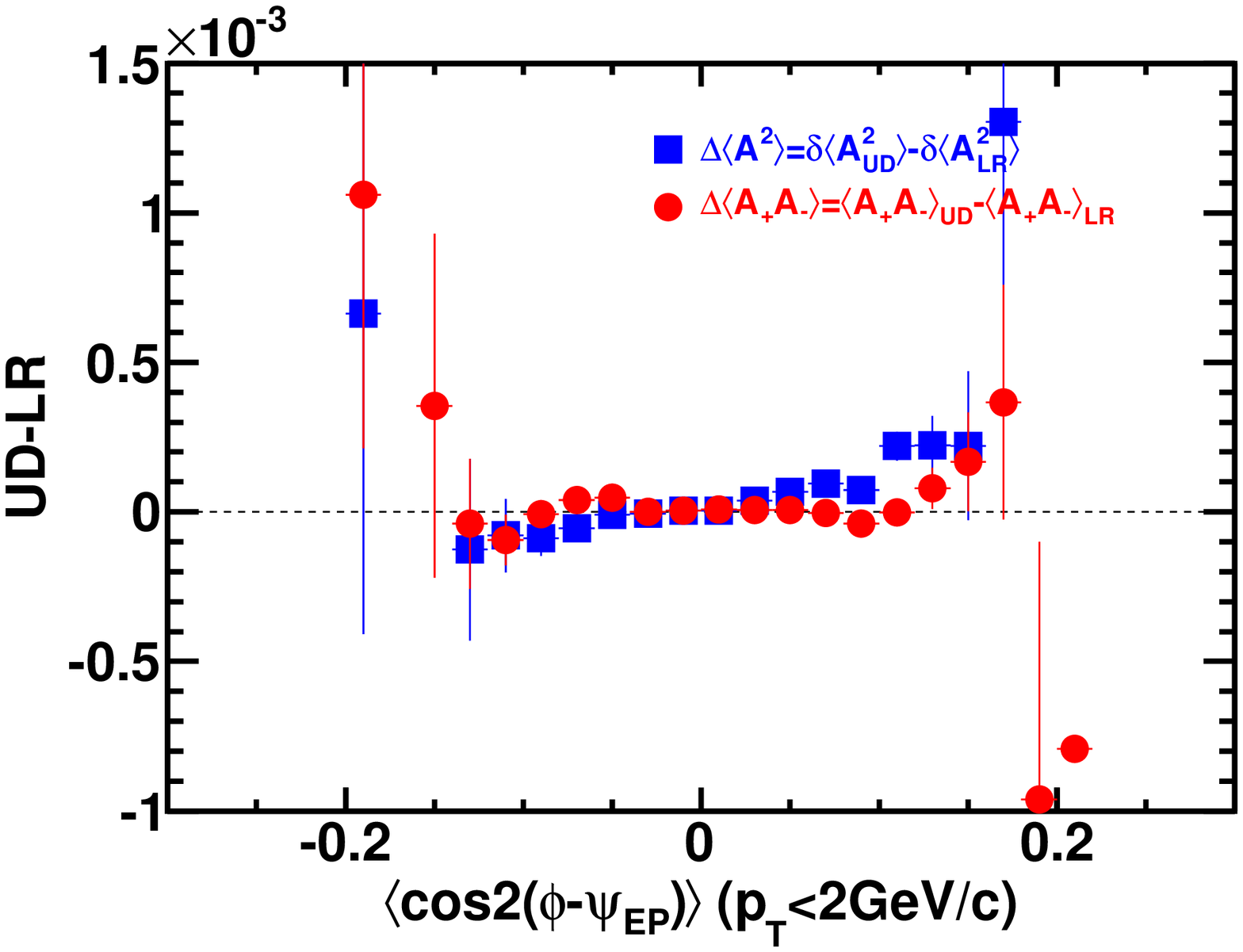}}
		\subfigure{\label{fig:asymv2top2d-b}\includegraphics[width=0.45\textwidth]{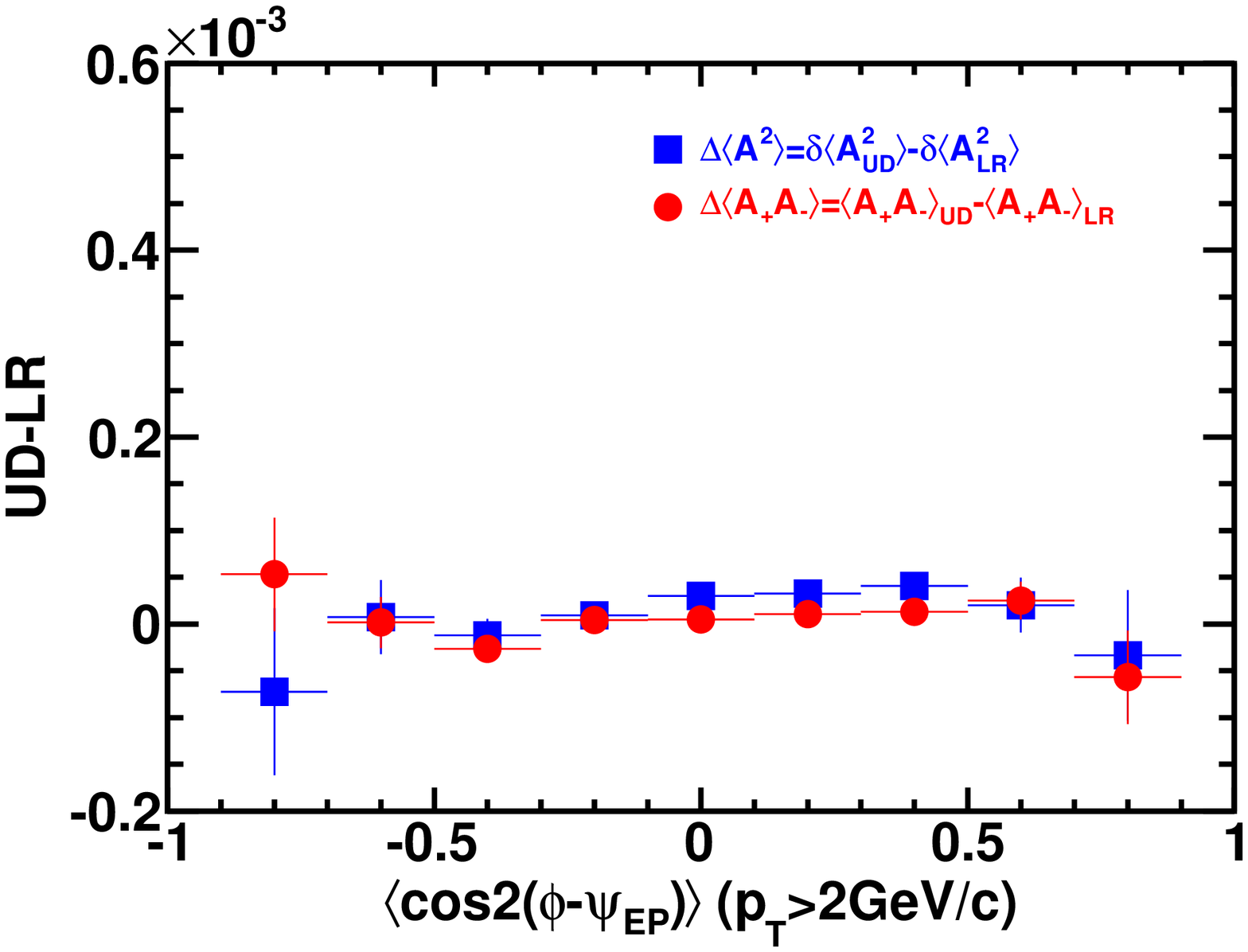}}
	\end{center}
	\caption[Top 2\% most central asymmetry correlations vs $v_{2}^{obs}$]{
	Asymmetry correlations vs event-by-event $v_{2}^{obs}$ of RUN IV 200 GeV Au+Au ZDC triggered top 2\% most central collisions.
	The particles used for asymmetry calculation are from half side of the TPC with respect to the event-plane reconstructed from another side of the TPC.
	Error bars are statistical only.
	}
	\label{fig:asymv2top2}
\end{figure}

\red{
One other suggestion is to evaluate charge separation effect in the top 2\% most central collisions \cite{Voloshin:2004vk}.
In such events, the magnetic field generated by the wounded nuclei is moderate with less fluctuation over time \cite{Voronyuk:2011jd}.
The fluctuation of event-by-event $v_2^{obs}$ is also small in the very most central collisions.
Thus CME/LPV should be relatively constant, and not depend on the event shape $v_2^{obs}$.
To select such top 2\% central events, we use the ZDC triggered RUN IV 200 GeV Au+Au events, which triggers on the top 12\% most central collisions.
In addition, we require the sum of the ADC signal from ZDC-SMD detectors to be less than 78, which will roughly select the top 2\% top central collisions.
We apply the standard STAR quality cuts as we do in the minimum-bias triggered data, and the corresponding acceptance correction.
There are in total 5.5 million events of the top 2\% central collisions in the final data sample.
Our results are shown on figure \ref{fig:asymv2top2}.
Similar features of the event-by-event anisotropy dependence are seen in the top 2\% most central collisions as for 0-20\% centrality (figure \ref{fig:appasymv220}), but the error bars are large due to limited EP resolution.
}

\red{
To summarize, in this section we presented the asymmetry correlations in Au+Au 200 GeV 20-40\% mid-central centrality as a function of event-by-event $v_2^{obs}$ defined using the second order event-plane with and without $\eta$ gap, first order event-plane from ZDC-SMD detector, and we also checked the results against those for the top 2\% most central collisions.
All the results are consistent with each other, which gives us confidence that the $v_2^{obs}$ dependence is indeed the main cause of the observed final state charge asymmetry correlations.
}

\section{Wedge Size and Location Dependence}

All the results we have presented above are charge multiplicity asymmetries obtained from hemispheres divided by event-plane ($UD$) and the plane perpendicular to the event-plane ($LR$).
The results show the asymmetry correlations, either $UD$ or $LR$, which are calculated from the same set of particles but divided into different sub groups.
In this way, statistical correlation and detector effect are mostly canceled between $UD$ and $LR$.
However, using the same set of particles will not be sensitive enough to distinguish the fine angular charge asymmetry structure.
For example, the hemisphere study cannot tell us whether the charge separation happens in-plane or out-of-plane, 
and how the charge asymmetry correlations evolve from in-plane~($LR$) to out-of-plane~($UD$) direction.
As we discussed in previous chapter, one of the advantages of these observables is that we can vary the azimuthal opening angle $\Delta\phi_\text{w}$ in which the analysis particles are included.
By restricting the azimuthal range, we can study the charge separation as a function of the wedge size and location.
The wedge size and location study will give us further insight into the fine structure of the charge separation.

\begin{figure}[thb]
	\begin{center}
		\subfigure[Asymmetry correlations vs wedge size]{\label{fig:asymwedgesize-a} \includegraphics[width=0.6\textwidth]{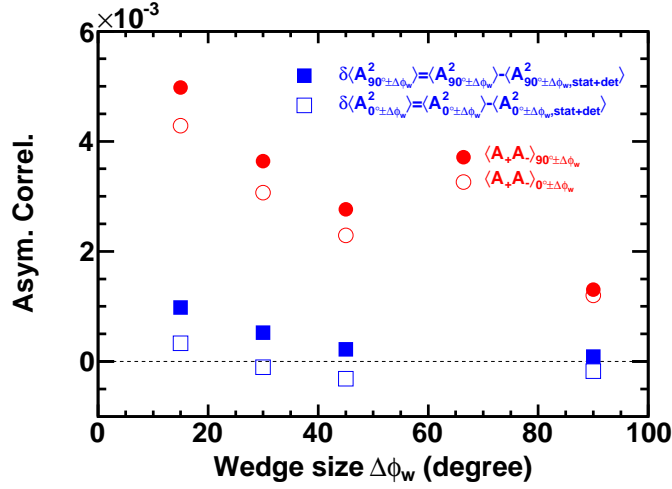}}
		\subfigure[$UD-LR$ correlations vs wedge size]{\label{fig:asymwedgesize-b} \includegraphics[width=0.6\textwidth]{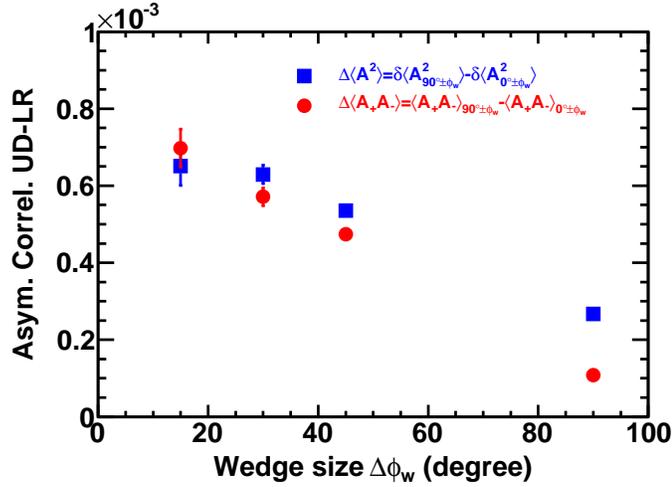}}
	\end{center}
	\caption[Mid-central wedge size dependence of charge asymmetry correlations]{
	The wedge size dependence of charge asymmetry correlations in panel (a) and their differences between out-of-plane ($UD$) and in-plane ($LR$) correlations
	$\Delta\langle A^2_{\Delta\phi_{\text{w}}} \rangle$ and $\Delta\langle A_+A_- \rangle_{\Delta\phi_{\text{w}}}$ in panel (b).
	Data are from 20-40\% centrality RUN IV 200 GeV Au+Au collisions.
	The particle $p_T$ range of $0.15 < p_T < 2.0$ GeV/$c$ is used for both asymmetry calculation and event-plane reconstruction.
	Error bars are statistical.
	}
	\label{fig:asymwedgesize}
\end{figure}

We present the wedge size dependence of the mid-central 20-40\% 200 GeV Au+Au collision data in figure~\ref{fig:asymwedgesize} for $\Delta\phi_{\text{w}} = 15\degree, 30\degree, 45\degree, 90\degree$, where $\Delta\phi_\text{w}=90\degree$ is identical to the hemisphere analysis.
Figure~\ref{fig:asymwedgesize-a} shows the charge dynamical asymmetry variances $\delta\langle A^2_{90\degree,\pm\Delta\phi_\text{w}}\rangle$, $\delta\langle A^2_{0\degree,\pm\Delta\phi_\text{w}}\rangle$, 
and covariances $\langle A_+A_-\rangle_{90\degree\pm\Delta\phi_\text{w}}$, $\langle A_+A_-\rangle_{0\degree\pm\Delta\phi_\text{w}}$ as a function of the wedge size $\Delta\phi_\text{w}$.
Note that the statistical fluctuation and detector effects are subtracted for each charge and wedge separately in the variances.
Both covariance correlations increase with decreasing wedge size $\Delta\phi_\text{w}$.
The dynamical variances seem to show similar increasing trend with decreasing $\Delta\phi_\text{w}$.
The smaller the wedge size, the stronger the correlations are.
This could possibly suggest that the charge asymmetry correlations are local, and the major effect is in the opposite-sign correlations.

Figure~\ref{fig:asymwedgesize-b} shows the difference of the asymmetry correlations between out-of-plane and in-plane directions, for variance 
\begin{equation}
	\Delta\langle A^2_{\Delta\phi_\text{w}}\rangle = \delta\langle A^2_{90\degree\pm\Delta\phi_\text{w}}\rangle - \delta\langle A^2_{0\degree\pm\Delta\phi_\text{w}}\rangle
\end{equation}
and covariance  
\begin{equation}
	\Delta\langle A_+A_- \rangle_{\Delta\phi_\text{w}} = \langle A_+A_- \rangle_{90\degree\pm\delta\phi_\text{w}} - \langle A_+A_- \rangle_{0\degree\pm\delta\phi_\text{w}}.
\end{equation}
Both of them increase with decreasing wedges size, and they show similar trend.
The difference between these two seems to disappear when $\Delta\phi_\text{w}$ decreases.

\begin{figure}[thb]
	\begin{center}
		\psfig{figure=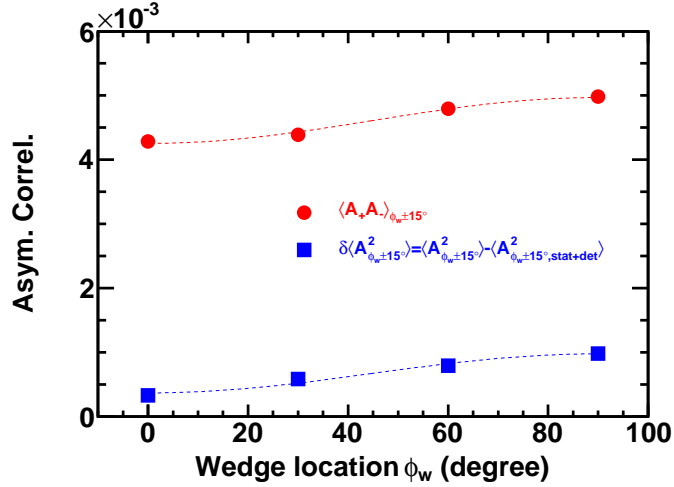,width=0.6\textwidth}
	\end{center}
	\caption[Mid-central wedge location dependence of charge asymmetry correlations]{
	Charge multiplicity correlations as a function of the wedge location $\phi_\text{w}$ of 20-40\% centrality 200 GeV Au+Au collisions.
	The wedge opening angle is $30\degree$ ($\Delta\phi_\text{w} = 15\degree$), and the back-to-back wedges are located at $0\degree$, $30\degree$, $60\degree$ and $90\degree$ relatively to the event-plane.
	The curves are the cosine modulation $a\times(1+2b\cos(2\phi_\text{w}))$ to guide the eye.
	The particle $p_T$ range of $0.15 < p_T < 2.0$ GeV/$c$ is used for both the asymmetry calculation and event-plane reconstruction.
	}
	\label{fig:asymwedgeloc}
\end{figure}

So far, we show the charge asymmetry correlations focusing only on in-plane and out-of-plane directions and the difference between them.
It is important to understand how the correlations evolve from in-plane to out-of-plane.
To do so, we study the correlation for back-to-back wedges with fixed wedge opening angle of $30\degree$ ($\Delta\phi_\text{w} = 15\degree$), and vary the wedge axis location from in-plane ($\phi_\text{w} = 0\degree$) to out-of-plane ($\phi_\text{w} = 90\degree$).
Figure~\ref{fig:asymwedgeloc} shows the dynamical variance and covariance of the $30\degree$-size back-to-back wedges in different locations relative to the event-plane ($\phi_\text{w} = 0\degree, 30\degree, 60\degree, 90\degree$).
The data are from mid-central 20-40\% centrality 200 GeV Au+Au collisions, same as figure \ref{fig:asymwedgesize}.
Both the dynamical variance and covariance increase from in-plane to out-of-plane, which is expected from figure~\ref{fig:asymwedgesize-a},
and both are modulated by a cosine function ($\sim \cos(2\phi_{\text{w}})$) as shown in the dashed lines.

\red{
We also show the wedge size and location study of the central collisions in figure \ref{fig:appasymwedgesize20} and peripheral collisions in figure \ref{fig:appasymwedgesize80}.
Although the magnitudes are different, they are qualitatively similar.
}

\section{Discussion}
\label{discuss}

In this section, we first make connections between our observables and the three-particle correlators used in previous STAR analysis.
Then we discuss the implication of our charge asymmetry correlation results.
We show the additional information from our results of charge separation directions and the event-by-event anisotropy analysis.
Through the study of event-by-event anisotropy, we may have a further understanding of the charge separation signal and/or background.

\subsection{Connection to Three-Particle Correlators}
\label{3P}

Our study of the charge multiplicity correlations is motivated by the CME/LPV.
To observe the effect, chiral symmetry restoration and large magnetic field are required \cite{Kharzeev:1998kz}.
The hot QCD matter is a particle environment to test the effect.
CME/LPV predicts that light-quark electric charge separates along the direction of the magnetic field,
which is the same as the system's orbital angular momentum direction. 
The effect yields charge separation in the final state hadrons, such as pions \cite{Kharzeev:2004ey}.

Previous STAR published result on three-particle correlator $\langle \cos (\alpha+\beta-2c) \rangle$ \cite{Voloshin:2004vk,:2009txa,:2009uh}, we have measured a negative same-sign correlator, and a close to zero, even slightly negative opposite-sign correlator.
The positive same-sign correlator is qualitatively consistent with the local parity violation expectation.
However, the close to zero opposite-sign correlator seems inconsistent with the naive expectation from CME/LPV.
Whether the charge separation can survive the hydrodynamic evolution to the final state is still an open question.
A recent study shows that a large percentage of charge separation in the initial state is needed for charge separation to be detected in the final state \cite{Ma:2011uma}, which suggests the in-medium interaction can strongly modify the charge separation correlations.
As a result, the back-to-back correlated pairs are quenched, because at least one of the pair quarks would be affected by the in-medium interactions.
Such in-medium effects could qualitatively explain the close to zero opposite-sign correlator observed at STAR \cite{Kharzeev:2007jp,:2009uh,:2009txa}.

STAR has reported the first measurement of three-particle azimuthal correlator to search for CME/LPV in relativistic heavy ion collisions \cite{:2009uh,:2009txa}.
The correlators were introduced in \cite{Voloshin:2004vk}, and can be measured as
\begin{align}
	\langle \cos(\phi_{\alpha}+\phi_{\beta}-2\psi_{RP})\rangle &\approx \langle \cos (\phi_{\alpha}+\phi_{\beta}-2\phi_{c})\rangle / v_{2,c}, \nonumber \\
	&= \left \langle {1 \over N_{\alpha}(N_{\beta}-\delta_{\alpha\beta})} \sum_{\alpha,\beta=0,\alpha\ne \beta}^{N_{\alpha},N_{\beta}} \cos (\phi_{\alpha}+\phi_{\beta}-2\psi_{RP})\right\rangle
	\label{eq:3part}
\end{align}
where the average in the last step is taken over event sample, and $\alpha$ and $\beta$ represent positive or negative charges.
$v_{2,c}$ is the elliptic flow of particle $c$ which serves to measure the reaction plane.
$\psi_{RP}$ stands for the reaction plane angle, $\delta_{\alpha\beta}$ is the Kronecker delta function.
In equation~\ref{eq:3part}, it is assumed that the event-plane dependent background is negligible, which is not necessarily true.

Our $UD-LR$ asymmetry correlations $\Delta\langle A_{\pm}^2 \rangle = \delta\langle A_{\pm,UD}^2 \rangle - \delta\langle A_{\pm,LR}^2\rangle$ and $\Delta\langle A_+A_- \rangle = \langle A_+A_- \rangle_{UD} - \langle A_+A_- \rangle_{LR}$ are related to the above three-particle azimuthal correlators, but with significant differences.
The charge multiplicity asymmetries can be expanded into Fourier series of step-function in $\phi-\psi_{EP}$ (particle azimuthal angle relative to the event-plane) as 
\begin{align}
	A_{\pm,UD} &= {4 \over \pi N_{\pm}} \sum_{i=1}^{N_{\pm}} \sum_{n=0}^{\infty} {\sin [(2n+1)(\phi_{\pm,i}-\psi_{EP})] \over 2n+1},\nonumber\\
	A_{\pm,LR} &= {4 \over \pi N_{\pm}} \sum_{i=1}^{N_{\pm}} \sum_{n=0}^{\infty} {\cos [(2n+1)(\phi_{\pm,i}-\psi_{EP})] \over 2n+1}.
	\label{eq:asymexpand}
\end{align}
The difference between $UD$ and $LR$ is then
\begin{multline}
	\langle A_{\alpha}A_{\beta} \rangle_{LR} - \langle A_{\alpha}A_{\beta} \rangle_{UD} = \left( {4 \over \pi}\right)^{2} \\
	\times \left\langle {1 \over N_{\alpha}N_{\beta}} \sum_{i,j=0}^{N_{\alpha},N_{\beta}} \sum_{n,m=0}^{\infty} {\cos \left[ (2n+1)(\phi_{\alpha,i}-\psi_{EP}) + (2m+1)(\phi_{\beta,j}-\psi_{EP})\right] \over (2n+1)(2m+1)}\right\rangle,
	\label{eq:asymdiffexpand}
\end{multline}
where $\alpha$ and $\beta$ represent positive or negative charges.
The asymmetry correlation differences contain all possible harmonic terms, including cross terms.
While the three-particle correlators in equation~\ref{eq:3part} contain only the first order terms.
The correlators are measured in terms of azimuthal angle with respect to reaction-plane angle $\psi_{RP}$,
while the asymmetry correlations are measured using azimuthal angle relative to event-plane angle $\psi_{EP}$.
Thereby, our asymmetry observables are affected by event-plane resolution.
Hence, the asymmetry correlation observables are related to three-particle correlators but are essentially different.

\begin{figure}[ht]
	\begin{center}
		\psfig{figure=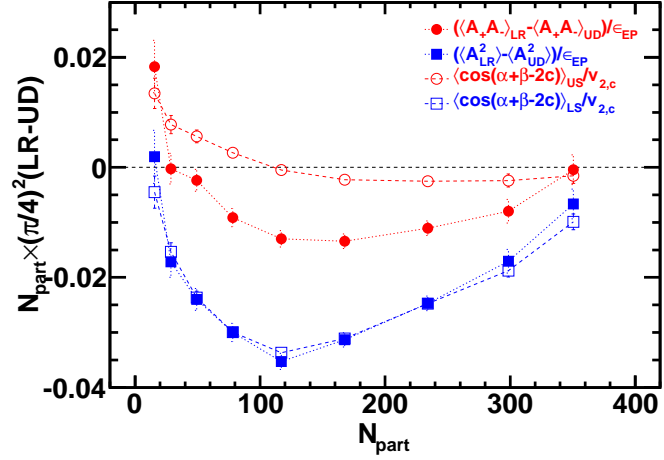,width=0.6\textwidth}
	\end{center}
	\caption[Asymmetry correlations compare to three-particle correlators]{
	The solid data points are asymmetry correlation $LR-UD$ differences of variances $\Delta\langle A^{2} \rangle = \delta\langle A^{2}_{LR} \rangle - \delta\langle A^{2}_{UD} \rangle$ (blue) and variances $\Delta\langle A_+A_- \rangle = \langle A_+A_- \rangle_{LR} - \langle A_+A_- \rangle_{UD}$ (red), 
	scaled by the number of participants $N_{part}$ times the scaling constant $(\pi/4)^2$, and divided by event-plane resolution $1/\epsilon_{EP}$ .
	Also shown are the three-particle correlators in open symbols, $\langle \cos(\phi_{\alpha}+\phi_{\beta}-2\phi_{c}) \rangle/v_{2,c}\approx \langle \cos (\phi_{\alpha}+\phi_{\beta}-2\psi_{RP}) \rangle$ of the like-sign (blue) and unlike-sign particle pairs (red). 
	The asymmetry correlations and the correlator particle $\alpha$ and $\beta$ are calculated with particles from one side of the TPC $\eta<0$ ($\eta>0$).
	The event-plane for the asymmetry correlations and particle $c$ of the three-particle correlators are from the other side of the TPC $\eta>0$ ($\eta<0$).
	Particle $p_T$ range of $0.15 < p_T < 2.0$ GeV/$c$ is used for asymmetry calculation, correlator calculation and event-plane reconstruction.
	Error bars are statistical only.
	}
	\label{fig:asym3part-a}
\end{figure}

To gain more insight into the relationship and the differences, we compare the two different observables in figure~\ref{fig:asym3part-a}.
In the figure, the charge asymmetry correlation $LR-UD$ and the three-particle correlators $\langle \cos (\phi_{a}+\phi_{\beta}-2\psi_{RP})\rangle$ are shown as a function of centrality ($N_{part}$).
The charge asymmetry correlations are shown in solid symbols, and the three-particle correlators are shown in open symbols, where ``LS'' stands for like-sign (same-charge), and ``US'' stands for unlike-sign (opposite-charge).
The three-particle correlators and $LR-UD$ correlations are scaled by $N_{part}$.
The $LR-UD$ correlations are also scaled by a constant $(\pi/4)^2$ according to equation~\ref{eq:asymdiffexpand}, and divided by the event-plane resolution in order to make direct comparison.
To be consistent, the correlators are calculated using the same set of particles as the asymmetry correlations for $\alpha$ and $\beta$, i.e. in the same $\eta$ region.
Particle $c$ is used from the same set of particles to reconstruct the event-plane for asymmetry correlations, which is from the other side of the $\eta$ region to avoid self-correlation.
The three-particle correlator results shown in figure~\ref{fig:asym3part-a} are not identical to those in published paper \cite{:2009uh}, where particle $\alpha$, $\beta$ and $c$ are from the TPC with pseudo-rapidity range of $-1<\eta<1$.

From equations \ref{eq:3part} and \ref{eq:asymexpand}, we know that the three-particle correlators contain only one out of the infinitely many of harmonic terms in the difference of $UD$ and $LR$ asymmetry correlations.
The same-sign correlation $\delta\langle A^2_{LR}\rangle - \delta\langle A^2_{UD}\rangle$ in figure \ref{fig:asym3part-a} is comparable to the three-particle like-sign correlator $\langle \cos (\phi_{\alpha} + \phi_{\beta} - 2 \phi_{c})\rangle$.
This indicates higher order terms plus the cross terms in same-sign correlation $\Delta\langle A^2\rangle$ contribute very little and can be neglected.
However, our opposite-sign correlation $\Delta\langle A_+A_- \rangle$ differs from the opposite-sign correlator significantly.
The $\Delta\langle A_+A_- \rangle$ is significantly negative, while unlike-sign correlator is close to zero.
This suggests that the higher order terms plus the cross terms are important.
Note that the event-plane resolution correction for higher order terms may not be correct in this case.
But the event-plane resolution effect can only smear out the correlations.
Thus, the real magnitude of the $\Delta\langle A_+A_- \rangle$ should be more negative with respect to the true reaction plane.
See the discussion in section \ref{TPCEP}.

\begin{figure}[ht]
	\begin{center}
		\psfig{figure=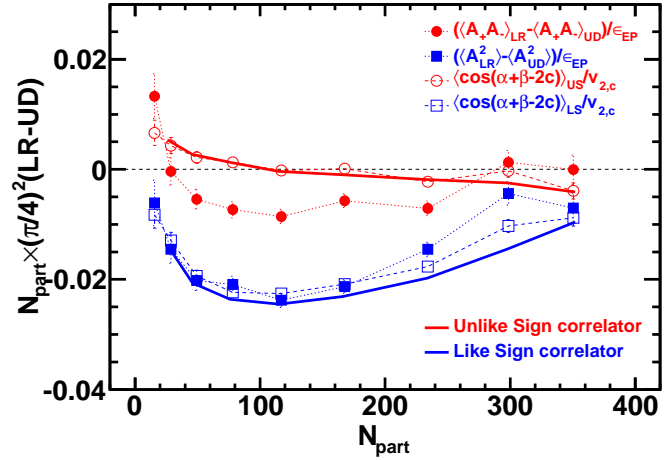,width=0.6\textwidth}
	\end{center}
	\caption[Asymmetry correlation compare to published three-particle correlator]{
	This figure shows the same comparison as figure \ref{fig:asym3part-a} but with the results obtained by randomly dividing the event into two equal sub groups of tracks within the full $\eta$ range of the TPC ($-1<\eta<1$).
	The asymmetries and correlators are then calculated from one sub group with respect to the event-plane reconstructed from the other sub group.
	The solid curves are the published three-particle correlator values \cite{:2009uh} superimposed upon our results.
	Error bars are statistical only.
	}
	\label{fig:asym3part-b}
\end{figure}

In order to make direct comparison to the published result, we analyze the charge asymmetry correlations within the entire TPC region.
To avoid self-correlation, we randomly divide the event into two halves (sub-events) regardless of the $\eta$ region.
We use one sub-event to calculate the asymmetries and their correlations relative to the event-plane reconstructed from the other sub-event.
We also calculate the three-particle correlators using particles $\alpha$ and $\beta$ from the sub-event for asymmetry calculation, and particle $c$ from the sub-event for the event-plane reconstruction in the asymmetry calculation regardless of the charges.
The results are shown in figure \ref{fig:asym3part-b}, where the solid points are our $LR-UD$ asymmetry correlations; the open points are our calculated the three-particle correlations; and the solid lines are the published data from \cite{:2009uh}.
Our calculated three-particle correlators match the published data very well, so we have confidence in the correctness of our analysis code. 
We also observe the qualitatively similar results to what we show in figure \ref{fig:asym3part-a}, namely that the same-sign correlation is consistent with the like-sign three-particle correlator, and opposite-sign correlation is significantly more negative than the unlike-sign three-particle correlator.
So we conclude that the high order terms and the cross terms in the $UD-LR$ Fourier expansion are negligible for same-sign correlations, however, they are significant for opposite-sign correlations.

\subsection{Interpretation of Charge Asymmetry Correlations}

Our charge asymmetry correlations provide an opportunity to study the event-plane dependence of the charge separation effect.
The published three-particle correlators assume that the higher order terms and cross terms are negligible.
However, we showed in the previous section that the higher order terms and the cross terms contribute significantly to the opposite-sign correlations.
Furthermore, our study can provide more information on the charge asymmetry correlations in the in-plane and out-of-plane regions separately.
In this section, we compare the CME/LPV expectations to our results, and try to interpret what data tells us.

First, we check the CME/LPV charge combination expectation. 
The CME/LPV would produce additional same-sign pairs in the up and down hemispheres.
Such additional correlation will result in wider distributions (larger variances) of both the positive and negative charge multiplicities in the same hemisphere.
Hence we expect larger dynamical correlation $\delta\langle A^2 \rangle$ in $UD$ than that in $LR$ direction.
On the other hand, the CME/LPV will produce back-to-back opposite-sign pairs in $UD$ direction,
which gives anti-correlation between the positive charge and negative charge particle pairs in the $UD$ direction.
The result is a smaller covariance $\langle A_+A_-\rangle_{UD}$ than $\langle A_+A_- \rangle_{LR}$.

Our data show that the same-sign dynamic correlation satisfy $\delta\langle A_{UD}^2 \rangle > \delta\langle A_{LR}^2\rangle$ of all centralities in Au+Au 200 GeV collisions (figure \ref{fig:asym}).
This is qualitatively consistent with CME/LPV expectation because the multiplicity distribution is wider in $UD$ direction than $LR$ direction.
However, the same-sign correlations are negative in mid-central to central collisions regardless of whether they are measured in-plane or out-of-plane.
Negative variances suggest that the same-sign charged pairs are preferentially emitted back-to-back, in another words more symmetrically distributed between hemispheres.
This is contrary to the expectation from the CME/LPV that same-sign charged pairs tend to be emitted in the same direction.

The opposite-sign correlations show $\langle A_+A_- \rangle_{UD} > \langle A_+A- \rangle_{LR}$ in all centralities except for the most peripheral collisions,
which suggests the opposite-sign particle pairs are likely emitted in the same hemisphere, and more strongly in $UD$ than $LR$ direction.
This contradicts the naive expectation from CME alone that opposite-sign charged pairs are emitted back-to-back.
The very peripheral collisions of Au+Au collision data are consistent with CME/LPV, however the same effect is also observed in d+Au collisions,
where CME/LPV is not expected.
The peripheral result can be explained by non-flow effect in low multiplicity events.
When we divided an event into two sub-events, the low multiplicity collisions are more sensitive to non-flow correlation, such as di-jet.
The non-flow correlation will become important and dominate the reconstructed event-plane orientation with limited multiplicity.
The event-plane is then preferentially sitting in the di-jet direction, 
and separating the event into two more equal halves in up and down hemispheres,
which makes the di-jets lie in the $LR$ hemispheres separately.
This leads to a large fluctuation in the $LR$ multiplicity, yielding large $LR$ asymmetry.
But for the $UD$ hemispheres, they are divided more or less symmetrically by the event-plane axis, which has smaller asymmetries of the opposite-sign.
Thus, the $\langle A_+A_- \rangle_{UD} < \langle A_+A- \rangle_{LR}$ for peripheral collisions is reasonable.

Secondly, we look at the magnitude of the charge asymmetry correlation.
There is little theoretical guidance to the quantitative magnitude of the charge separation itself and the charge correlations.
However, there are model estimates suggesting a few percent charge asymmetry, 
which implies for our charge multiplicity asymmetry correlation magnitudes of the order of $10^{-4}$ to $10^{-3}$.
Considering the in-medium interaction and the multiplicity dilution effect, the estimated charge asymmetries could be reduced by one order in magnitude because the interaction with medium conserves parity and destroys only the correlations.
This has been shown in \cite{Ma:2011uma}.
Or, it could be even lower than $10^{-6}$ as shown in \cite{Muller:2010jd}.
Our results do indicate much larger correlations than these estimates, see figure \ref{fig:asymdiff} (divide the data points with $N_{part}$ from \ref{tab:cent}).
Both the variance and covariance in mid-central collisions have the magnitudes around $\sim 10^{-4}$.
There are also models arguing that the charge asymmetries and the magnitude can be explained by known QCD processes without invoking CME.
Nevertheless, the estimated charge asymmetry correlations are, at least, a few orders of magnitude smaller than the measurements.
Our measurements show strong correlations between same-sign and opposite-sign correlation in both $UD$ and $LR$ directions.

Thirdly, we look at the $p_T$ dependence of the charge asymmetry correlations.
Figure~\ref{fig:asymptdiff} shows that the charge asymmetry $UD-LR$ correlations grow with $p_T$ for both the same-sign and opposite-sign.
While CME/LPV expects the charge separation to take place in low-$p_T$ region with $p_T < 1$ GeV/$c$ \cite{Kharzeev:2007jp},
the data are inconsistent with that naive expectation.

The wedge location dependence will be discussed in the following section \ref{secwedge}.

Despite that both same-sign and opposite-sign correlations in figure~\ref{fig:asymdiff} are all positive except the most peripheral collisions,
the same-sign correlation $\Delta\langle A^2 \rangle$ is larger than the opposite-sign correlation $\Delta\langle A_+A_- \rangle$.
A yet unknown underlying background which is coupled with flow but exhibits no charge dependence could produce similar correlation.
In that case, the correlation from underlying background could possibly lie in between our same-sign and opposite-sign correlation results.
Thus, after subtracting the common background, the same-sign and opposite-sign correlations will have different sign, which is then consistent with CME/LPV expectation, assuming medium effect does not change correlation signs and the common background does not depend on charge combinations.
Taking this idea, we define the charge separation observable ($\Delta$) across the event-plane as
\begin{equation}
	\Delta \equiv \Delta\langle A^2 \rangle - \Delta \langle A_+A_- \rangle.
	\label{eq:chargesep}
\end{equation}
It is the correlation difference between the same-sign and opposite-sign $UD-LR$ correlations.
Equation \ref{eq:chargesep} would quantify the charge separation effect if the reaction-plane dependent backgrounds were the same for the same-sign and opposite-sign correlations.
$\Delta$ should be consistent with 0 if there is no charge separation.
$\Delta > 0$ would suggest a charge separation effect with same-sign charged particles emitted in the same direction (small angle correlation) and/or opposite-sign charged particles emitted in the opposite direction.
$\Delta < 0$ would suggest the contrary situation with same-sign charge particles emitted in back-to-back direction and/or opposite-sign charged particles emitted in the same direction.
We further test $\Delta$ in the following sections.

\subsection{In-Plane or Out-of-Plane?}
\label{secwedge}

\begin{figure}[thb]
	\begin{center}
		\subfigure[Charge separation out-of-plane]{\label{fig:inout-a} \includegraphics[width=0.4\textwidth]{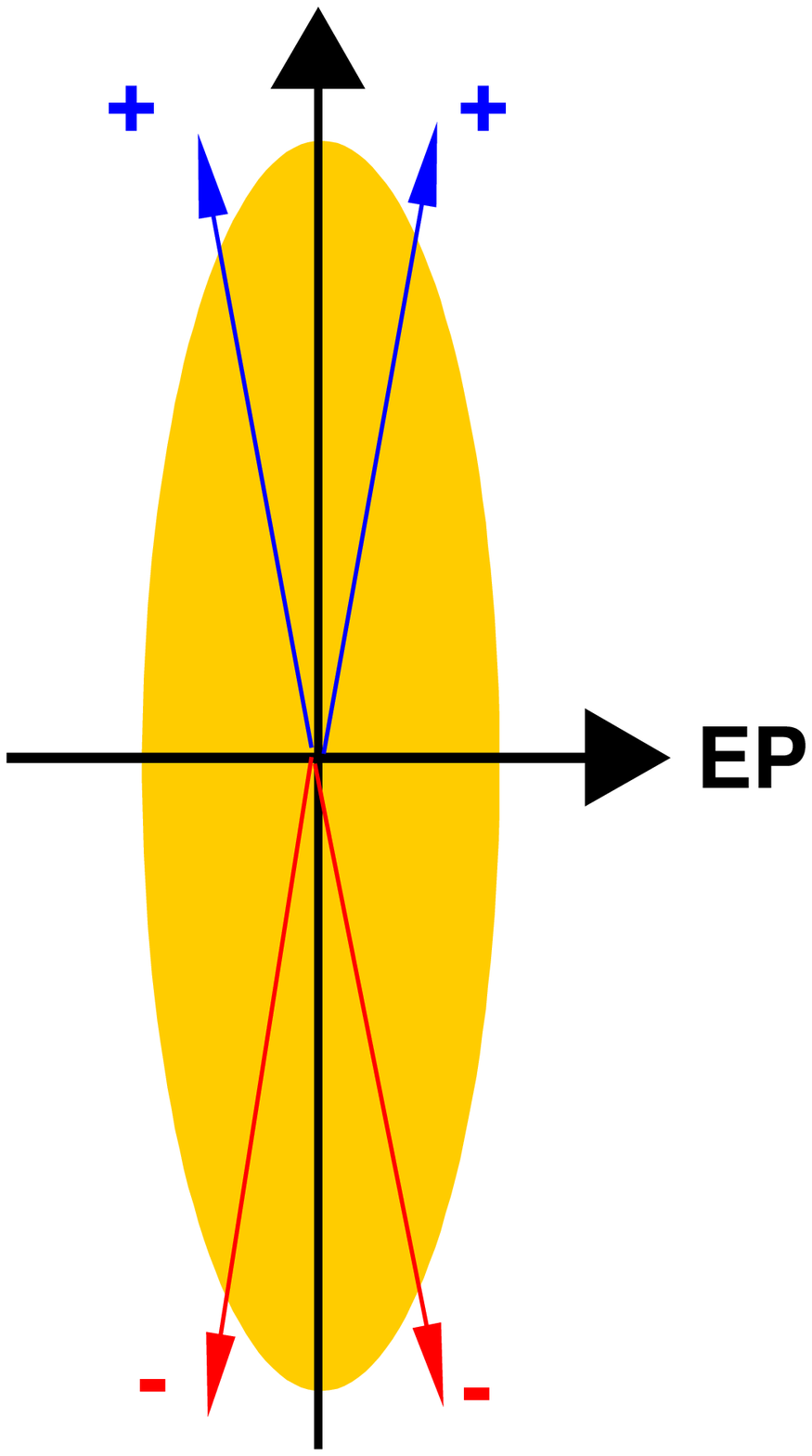}}
		\subfigure[Charge separation in-plane]{\label{fig:inout-b} \includegraphics[width=0.4\textwidth]{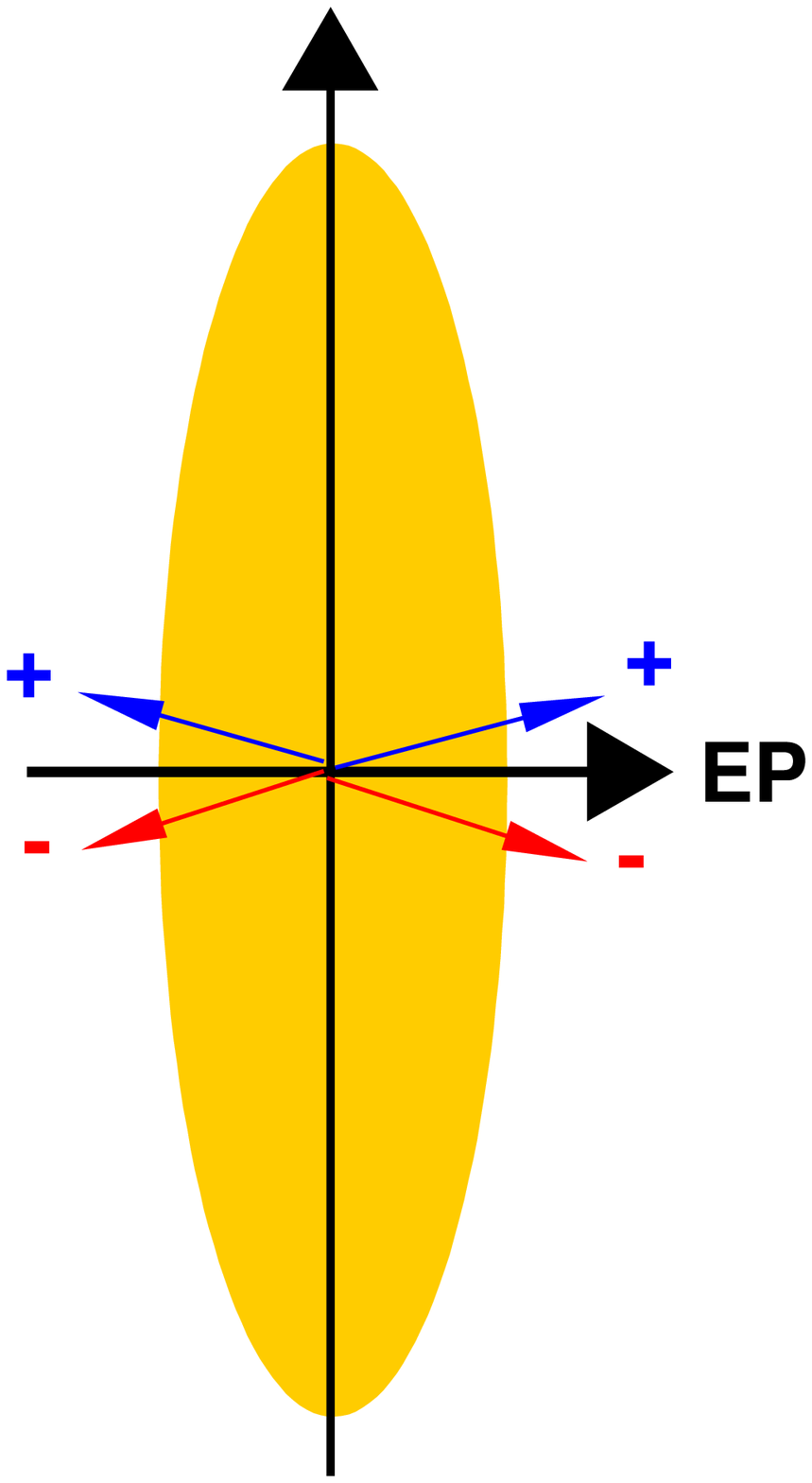}}
	\end{center}
	\caption[Charge separation direction]{
	Schematic plot depicting the possible charge separation direction.
	The wedge size dependence of the difference between same-sign and opposite-sign $UD-LR$ correlations ($\Delta(\Delta\phi_{\text{w}})$)	will increase with the decreasing of the wedge size, 
	if the charge separation is out-of-plane as shown in panel (a).
	Otherwise, $\Delta(\Delta\phi_{\text{w}})$ decreases with the decreasing of the wedge size,
	if the charge separation is in-plane as shown in panel (b).
	Our results favor in figure (b).
	}
	\label{fig:inout}
\end{figure}

If CME/LPV effect in the same-sign correlation could survive through the hydrodynamic evolution to the final state,
and its preferred direction is still along the system orbital angular momentum direction (magnetic field direction),
then the charge separation $\Delta$ from equation~\ref{eq:chargesep} should depend on the wedge size and is possible to measure, as illustrated in figure \ref{fig:inout-a}.
The smaller the wedge size, the stronger charge separation effect we should expect, for
\begin{equation}
	\Delta(\Delta\phi_\text{w}) = \Delta\langle A^2_{\Delta\phi_\text{w}} \rangle - \Delta\langle A_+A_- \rangle_{\Delta\phi_\text{w}}.
	\label{eq:chargesepwedge}
\end{equation}
It is also possible that in-medium interaction and final state interaction will modify the preferred direction of the same-sign asymmetry correlation,
so that $\Delta(\Delta\phi_\text{w})$ may not necessarily increase with decreasing the wedge angle $\Delta\phi_\text{w}$.
For example, the elliptic flow pushes the correlated pairs into the in-plane direction, as illustrated in figure \ref{fig:inout-b}.

\begin{figure}[thb]
	\begin{center}
		\psfig{figure=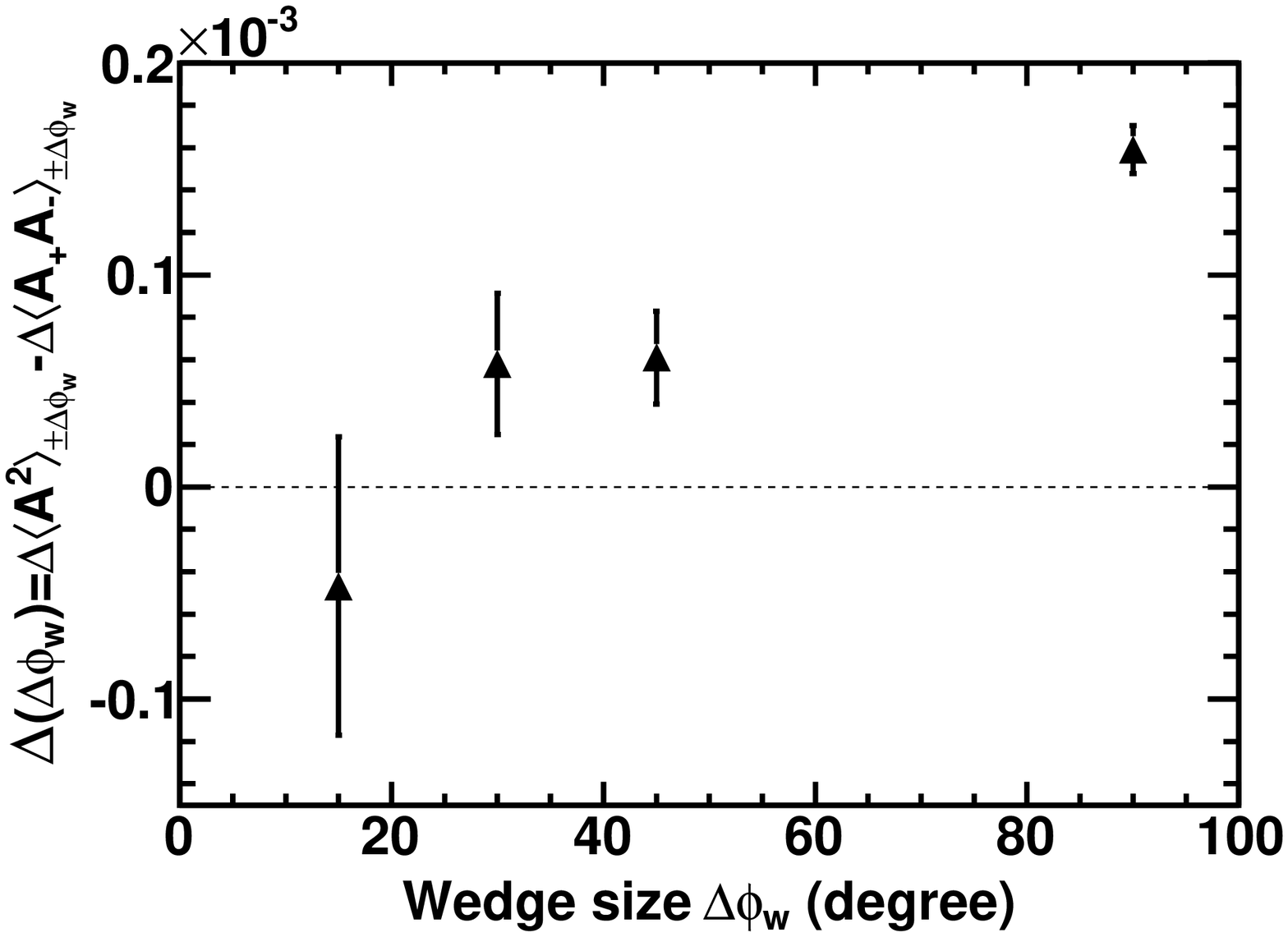,width=0.6\textwidth}
	\end{center}
	\caption[Wedge size dependence of charge separation]{
	The wedge size dependence of the charge separation $\Delta(\Delta\phi_\text{w})$ of Mid-central 20-40\% Au+Au collisions.
	Error bars are statistical.
	}
	\label{fig:chargesepwedge}
\end{figure}

The $\Delta(\Delta\phi_{\text{w}})$ result is shown in figure~\ref{fig:chargesepwedge} of 20-40\% Au+Au 200 GeV collisions.
Charge separation decreases with decreasing the wedge size $\Delta\phi_\text{w}$.
The charge separation seems to disappear at very small wedge open angle in mid-central collisions.
The same-sign and opposite-sign pair correlations have the maximum difference in the hemispheres.
This may suggest that the effect of charge separation across the event-plane happens in the vicinity of the in-plane direction rather than the out-of-plane direction in mid-central collisions.
\red{
Figure \ref{fig:appasymwedgesize20-d} shows the central 0-20\% centrality charge separation wedge size dependence $\Delta(\Delta\phi_{\text{w}})$,
and figure \ref{fig:appasymwedgesize80-d} shows the peripheral 40-80\% centrality $\Delta(\Delta\phi_{\text{w}})$.
Charge separation decreases with decreasing wedge size for all centralities.
It even goes negative in central and peripheral collisions.
}

\subsection{Signal or Background?}

As discussed above, the charge separation along the system angular momentum direction~($UD$) is possibly due to CME,
if the event-plane dependent common background lies between the same-sign and opposite-sign correlation in figure~\ref{fig:asymdiff}.
However, we find inconsistencies with the naive CME/LPV expectation.
Firstly, the observed same-sign correlation shows more back-to-back symmetry in both $UD$ and $LR$ direction; it is stronger in $UD$ than $LR$.
Secondly, the observed opposite-sign correlation shows that the opposite charged pairs are emitted in the same direction, and more strongly emitted in the $UD$ than $LR$ direction.
Thirdly, the charge separation seems to happen in the vicinity of the event-plane instead of out-of-plane.
Fourthly, the asymmetry correlations increase with transverse momentum.
Except the $p_{T}$ dependence, the results may be explained by CME/LPV if an unsubtracted background in these observables is charge independent and falls in between the same- and opposite-charge measurements.

\begin{figure}[thb]
	\begin{center}
		\psfig{figure=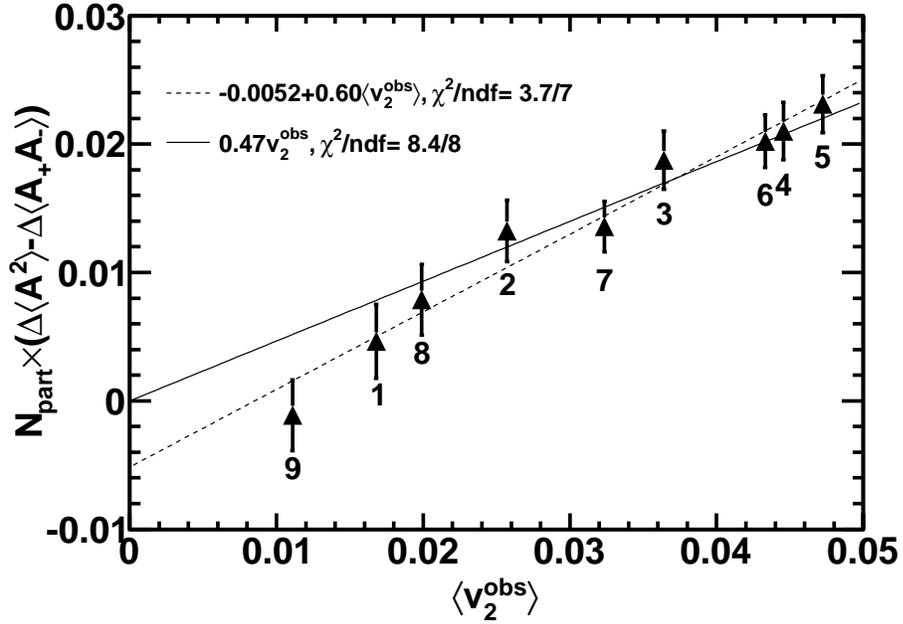,width=0.8\textwidth}
	\end{center}
	\caption[Charge separation vs $\langle v_2^{obs} \rangle$]{
	Charge separation $\Delta$ scaled by $N_{part}$ as a function of the measured average anisotropy $\langle v_2^{obs} \rangle$ in RUN IV 200 GeV Au+Au collisions.
	The centrality bin is labeled below each data point (see table \ref{tab:cent}).
	The particle $p_T$ range of $0.15 < p_T < 2.0$ GeV/$c$ is used for both asymmetry calculation and event-plane reconstruction.
	Error bars are statistical.
	}
	\label{fig:chargesepv2}
\end{figure}

On the other hand, the assumption that the event-plane dependent background is charge independent may not be true.
It is possible that the background for the same-sign and opposite-sign correlations is quite different due to different physics mechanisms.
As argued in \cite{Pratt:2010gy} by Pratt, charge conservation combined with collective flow naturally creates charge separation, with a difference between same-sign and opposite-sign correlations.
It is already shown in figure~\ref{fig:asymdiff} that the correlation $UD-LR$ differences as a function of centrality are qualitatively similar to the elliptic flow $v_2$ centrality dependence \cite{:2008ed}.
The correlations are peaked at medium central collisions, and drop close to zero in peripheral and most central collisions.
Thus, we plot the charge separation $\Delta$ (defined in equation \ref{eq:chargesep}) against the average event-by-event anisotropy $\langle v_2^{obs} \rangle$.
The $\langle v_2^{obs} \rangle$ value of each centrality is listed in table \ref{tab:cent}.
The result is shown in figure \ref{fig:chargesepv2}.
The charge separation is scaled by number of participants $N_{part}$.
We show the centrality bin numbers under the data points according to table \ref{tab:cent}.
The dashed line is a linear fit of the data points, which shows a very good agreement of the event shape dependence of the charge separation.
The solid line is also a linear fit but with fixed intercept at zero.
The result is qualitatively consistent with Pratt's suggestion that the charge separation is proportional to the elliptic flow in \cite{Pratt:2010gy}, which is also suggested in \cite{Voloshin:2004vk} and consistent with the finding in STAR previous measurement in \cite{:2009txa}.
The linear dependence of the charge separation effect on $v_{2}^{obs}$ suggests that there may be a $v_{2}^{obs}$-dependent background in our measured charge separation.

Recall the correlation dependence of $v_2^{obs}$ in figure~\ref{fig:asymv2l}.
The event-by-event anisotropy charge asymmetry correlations also suggest that the different behavior of the same-sign and opposite-sign correlations may be subject to different physics mechanism.
The background may be due to a significant bulk correlation related to the event structure,
which causes the background to be indeed different for same-sign and opposite-sign correlations.
For proper interpretation of figure~\ref{fig:asymv2l}, it is important to understand the charge separation background and signal, and the cause of the possible signal.

For the low-$p_T$ $v_2^{obs}$ dependent same-sign correlation, 
the increasing trend in $UD$ direction suggests more and more back-to-back same-sign charged particle pairs are emitted in-plane while the event is more elongated in the in-plane direction, i.e. increasing $v_2^{obs}$.
Equivalently, there might be fewer particles emitted out-of-plane with the increase of event-by-event anisotropy $v_2^{obs}$.
It is also possible both are true.
Thus the overall effect would be fewer back-to-back same-sign pairs emitted in out-of-plane direction than in-plane,
causing the dynamic correlation in $UD$ to increase with increasing $v_2^{obs}$.
The decreasing trend in $LR$ direction would be similar to that in $UD$ direction.
There are more particles emitted in the in-plane direction than out-of-plane direction with larger $v_2^{obs}$, thus more abundant back-to-back same-sign particle pairs in-plane than out-of-plane,
which reduces the dynamical correlation in $LR$.

For the opposite-sign correlations, the trend seems opposite to the same-sign correlations, but with much weaker effect.
This may suggest different origins of the same-sign and opposite-sign particle pairs.
The behaviors of $\delta\langle A^2 \rangle$ and $\langle A_+A_- \rangle$ as a function of low-$p_T$ $v_2^{obs}$ could also be a possible effect of cluster or resonance decay correlations overlaid with the elliptic flow \cite{Wang:2009kd}.

\begin{figure}[ht]
	\begin{center}
		\subfigure[Charge separation vs $v_2^{obs}$]{\label{fig:chargesepv2stdy4-a} \includegraphics[width=0.6\textwidth]{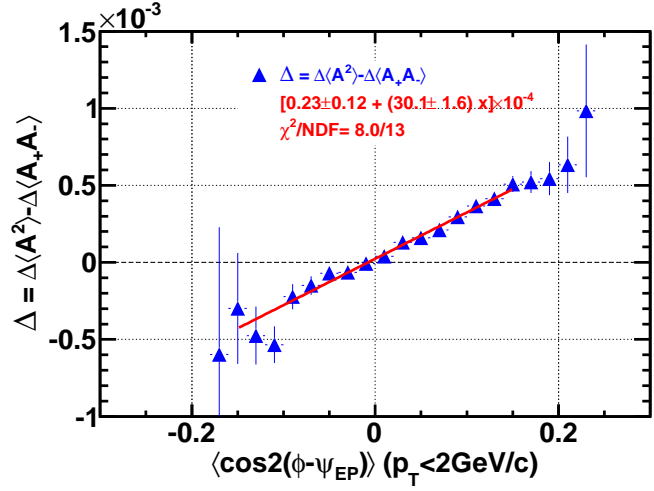}}
		\subfigure[EP resolution vs $v_2^{obs}$]{\label{fig:chargesepv2stdy4-b} \includegraphics[width=0.6\textwidth]{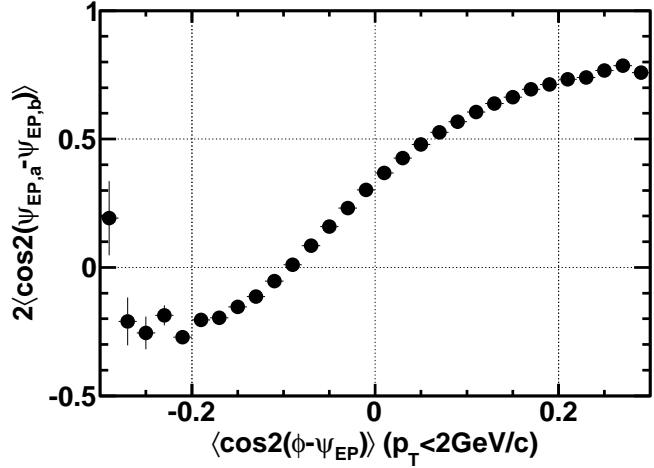}}
	\end{center}
	\caption[Mid-centrality charge separation vs $v_2^{obs}$ from 2nd order EP]{
	Panel (a): RUN IV Au+Au 200 GeV 20-40\% centrality charge separation $\Delta$ as a function of low-$p_T$ event-by-event anisotropy $v_2^{obs}$.
	Panel (b): Event-plane resolution squared as a function of $v_2^{obs}$.
	The asymmetries and $v_2^{obs}$ are calculated relative to the second order event-plane reconstructed from the other side of the TPC tracks.
	The particle $p_T$ range of $0.15 < p_T < 2.0$ GeV/$c$ is used for asymmetry, $v_2^{obs}$ and event-plane reconstruction.
	Error bars are statistical.
	}
	\label{fig:chargesepv2stdy4}
\end{figure}

We study the charge separation $\Delta$ as a function of the low-$p_T$ event-by-event anisotropy $v_2^{obs}$ to show the event shape dependence.
In figure~\ref{fig:chargesepv2stdy4-a}, we plot the charge separation between the same-sign and opposite-sign correlations $\Delta = \Delta \langle A^2 \rangle - \Delta \langle A_+A_- \rangle$ as a function of the low-$p_T$ event-by-event elliptic anisotropy $v_2^{obs}$.
It is simply the difference between the same-sign and opposite-sign correlations shown in figure~\ref{fig:asymv2l-c}.
Here $v_2^{obs}$ is calculated on the event-by-event basis from particles used in asymmetry calculation relative to the event-plane reconstructed from the other half side of TPC tracks.
Data are from RUN IV 20-40\% centrality Au+Au 200 GeV collisions.
For those events with large positive $v_2^{obs}$, the charge separation is large and positive, $\Delta \langle A^2 \rangle > \Delta \langle A_+A_- \rangle$, which is consistent with CME/LPV.
However the charge separation flips sign with more negative $v_2^{obs}$, which appears inconsistent with CME/LPV expectation.

One can argue that, at very negative $v_2^{obs}$, event-plane has very poor resolution, thus, the reconstructed EP is more orthogonal to the real reaction-plane rather than align with it.
The $UD$ and $LR$ direction are then flipped if it is the real situation.
So, the charge separation for those events is actually positive.
The event-plane resolution does vary with the particle elliptic anisotropy $v_2^{obs}$.
To show the dependence of event-plane resolution on $v_2^{obs}$,
the square of the EP resolution from the half event is shown in figure~\ref{fig:chargesepv2stdy4-b}.
We obtain the resolution by randomly dividing the half event into two quarter-events noted by subscript $a$ and $b$.
Then we can reconstruct the event-plane from the quarter events to get $\psi_{EP,a}$ and $\psi_{EP,b}$.
Similarly, the half event-plane resolution can be assessed by $\epsilon_{EP}^{2} = 2 \langle \cos 2(\psi_{EP,a}-\psi_{EP,b}) \rangle$.
Although the particles for event-plane reconstruction and event anisotropy calculation are from different phase space,
we can still see the correlation between them in the figure.
Note that at significant negative $v_2^{obs}$, the $\langle \cos 2(\psi_{EP,a}-\psi_{EP,b}) \rangle$ turns negative,
which suggests that the reconstructed event-plane does not reflect the true reaction-plane.
It might be more likely orthogonal to the reaction-plane rather than aligned with it.
This could mean that the $UD$ and $LR$ are indeed flipped.
Therefore, the results at very negative $v_{2}^{obs}$ are also consistent with CME.

As shown, the resolution squared at $v_2^{obs} \sim 0$ is sizably positive, which means the event-plane reconstructed with an isotropic half event on the other side is considerably good.
The charge separation $\Delta$ vanishes or slightly positive at $v_2^{obs} \sim 0$.
For events with modest negative $v_2^{obs}$ ($v_2^{obs} > -0.1$ for this particular centrality), the event-plane resolution from two sub-events method is well defined.
Within $-0.1 < v_2^{obs} \lesssim 0$, the charge separation is negative with reasonable EP resolution.
And the charge separation has a good linear dependence on the event-by-event anisotropy $v_2^{obs}$, which is not expected by CME/LPV.
We know that if we integrate over all events within 20-40\% centrality bins, the average $\langle v_2^{obs} \rangle$ is positive with the magnitude of a few percent, see table \ref{tab:cent}.
The charge separation of the integrated result will be positive due to the linear dependence,
but it doesn't necessarily mean the charge separation is caused by CME/LPV.
The $v_2^{obs}$ dependent charge separation result suggests that CME/LPV may not be the reason for the charge separation observed in the study.

\red{
We show the charge separation vs $v_2^{obs}$ results for central collision 0-20\% centrality in figure \ref{fig:appchargesepv2stdy420}, and for peripheral collisions 40-80\% centrality in figure \ref{fig:appchargesepv2stdy480}.
The results are qualitatively similar to the mid-central results.
}

\begin{figure}[ht]
	\begin{center}
		\subfigure[Charge separation of isotropic events]{\label{fig:Deltav2Y4-a} \includegraphics[width=0.6\textwidth]{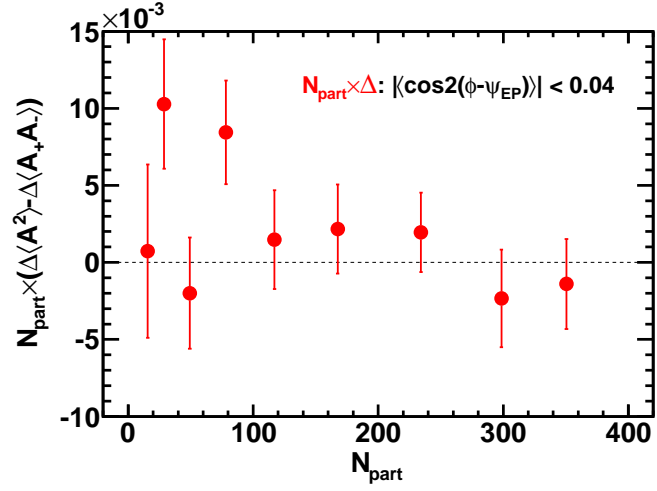}}
		\subfigure[Event-plane resolution of isotropic events]{\label{fig:Deltav2Y4-b} \includegraphics[width=0.6\textwidth]{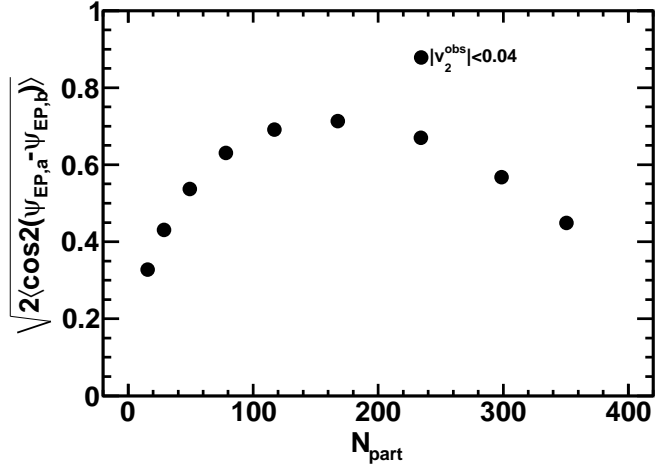}}
	\end{center}
	\caption[Charge separation ($|v_2^{obs}|<0.04$) from 2nd order event-plane]{
	Panel (a): The charge separation $\Delta$ scaled by $N_{part}$ as a function of centrality requiring the event isotropic condition $|v_2^{obs}|<0.04$.
	Panel (b): The event-plane resolution for those isotropic events.
	Data are from RUN IV 200 GeV Au+Au collisions, and particle $p_T$ range of $0.15<p_T<2.0$ GeV/$c$ is used for asymmetry calculation, $v_2^{obs}$ calculation and event-plane reconstruction.
	Error bars are statistical.
	}
	\label{fig:Deltav2Y4}
\end{figure}

It has been pointed out by several authors that there could be charge dependent physics background proportional to the event-by-event $v_2^{obs}$ due to the net effect of particle intrinsic correlation and production anisotropy \cite{:2008ed,:2009txa,Pratt:2010gy}.
The charge dependent background will cause the final state charge separation.
We thus fit the charge separation $\Delta(v_2^{obs})$ to a linear polynomial in $v_2^{obs}$ with the fitting range of $-0.15<v_2^{obs}<0.15$.
The result is shown as red line in figure~\ref{fig:chargesepv2stdy4-a}.
We have $\Delta(v_2^{obs}) = (0.23\pm0.12)\times10^{-4}+(3.0\pm0.2)\times10^{-3} v_2^{obs}$.
The slope could be a measurement of the particle intrinsic correlation strength.
And the intercept could then be a more sensitive measurement of the charge separation.
In other words, the measurement of CME/LPV has to take place in the phase space where the particle multiplicity distribution is more isotropic, so that the measured charge separation could be more sensitive to CME/LPV.
To do so, we apply a cut requiring the event-by-event $|v_2^{obs}|<0.04$ for the asymmetry calculation,
and then plot the charge separation $\Delta$ as a function of centrality.
The result is shown in figure~\ref{fig:Deltav2Y4-a}, with the charge separation $\Delta$ scaled by $N_{part}$.
The charge separation is consistent with zero within our present statistical precision,
which suggests no substantial charge separation is observed in those approximately isotropic events of the measured particles.
By taking the intercept value of the linear fit fuction of charge separation $\Delta=0.23\times10^{-4}$ at $v_2^{obs}=0.$, with twice of the uncertainty $\sigma\Delta = 0.12 \times10^{-4}$ away from the intercept, we get the upper limit of the charge separation for the isotropical events in mid-central collisions, which is $\Delta = 4.7 \times 10^{-5}$ with 98\% CL.

The event-plane resolution for those nearly isotropic events is shown in figure~\ref{fig:Deltav2Y4-b},
which shows that, with one spheric shape event on one side of the TPC, the event-plane reconstructed from the other side still has a good resolution.

\begin{figure}[ht]
	\begin{center}
		\subfigure[Charge separation vs $v_2^{obs}$ with $\eta$ gap]{\label{fig:etagap-a} \includegraphics[width=0.55\textwidth]{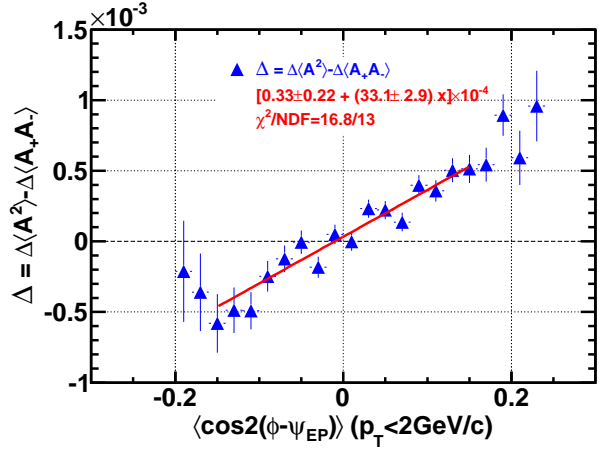}}
		\subfigure[EP resolution of $\eta$ gap]{\label{fig:etagap-b} \includegraphics[width=0.55\textwidth]{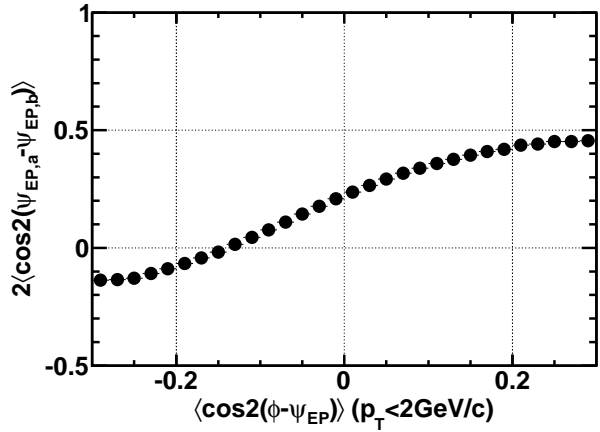}}
	\end{center}
	\caption[Mid-central charge separation with $\eta$ gap vs $v_2^{obs}$ from 2nd order EP]{
	Panel (a): The charge separation $\Delta$ scaled by $N_{part}$ as a function of event-by-event anisotropy $v_2^{obs}$.
	The asymmetries and event-plane reconstruction are taken place in sub events with one unit pseudo-rapidity separation, $-1.0<\eta<-0.5$ and $0.5<\eta<1.0$.
	The charge separation is fitted to a linear polynomial as shown in red line.
	Panel (b): The event-plane resolution squared as a function of $v_2^{obs}$.
	Data are from RUN IV 200 GeV Au+Au collisions in 20-40\% centrality, and the particle $p_T$ range of $0.15<p_T<2.0$ GeV/$c$ is used for asymmetry calculation, $v_2^{obs}$ calculation and event-plane reconstruction.
	Error bars are statistical.
	}
	\label{fig:etagap}
\end{figure}

For consistency, we check the charge separation vs $v_2^{obs}$ with one unit $\eta$ gap result.
It is shown in figure \ref{fig:etagap-a} and the corresponding event-plane resolution squared is shown in figure~\ref{fig:etagap-b}.
With the large pseudo-rapidity gap, we further remove the short range bulk correlations from the soft particles.
The charge separation fitted to a linear polynomial as a function of $v_2^{obs}$ gives $(0.33\pm0.22)\times10^{-4}+(3.3\pm0.3)\times10^{-3}v_2^{obs}$ with the fitting range of $-0.15<v_2^{obs}<0.15$.
The linear dependence of the charge separation with event-by-event $v_2^{obs}$ is still seen after removing partially the short range correlations.
Note for those isotropic events with $v_2^{obs} \sim 0$, the charge separation is close to zero with large errors due to the statistics.
And resolution squared of the event-plane reconstructed from the half unit of pseudo-rapidity is reasonable positive at $v_2^{obs} \sim 0$ as shown in figure \ref{fig:etagap-b}.

The central (0-20\% centrality) and peripheral (40-80\% centrality) collision results are shown in figure \ref{fig:appetagap20} and figure \ref{fig:appetagap80}, which also show the charge separation is linearly dependent on the event-by-event $v_2^{obs}$.

\begin{figure}[ht]
	\begin{center}
		\subfigure[Charge separation vs $v_2^{obs}$ of ZDC-SMD EP]{\label{fig:chargesepv2zdc-a} \includegraphics[width=0.55\textwidth]{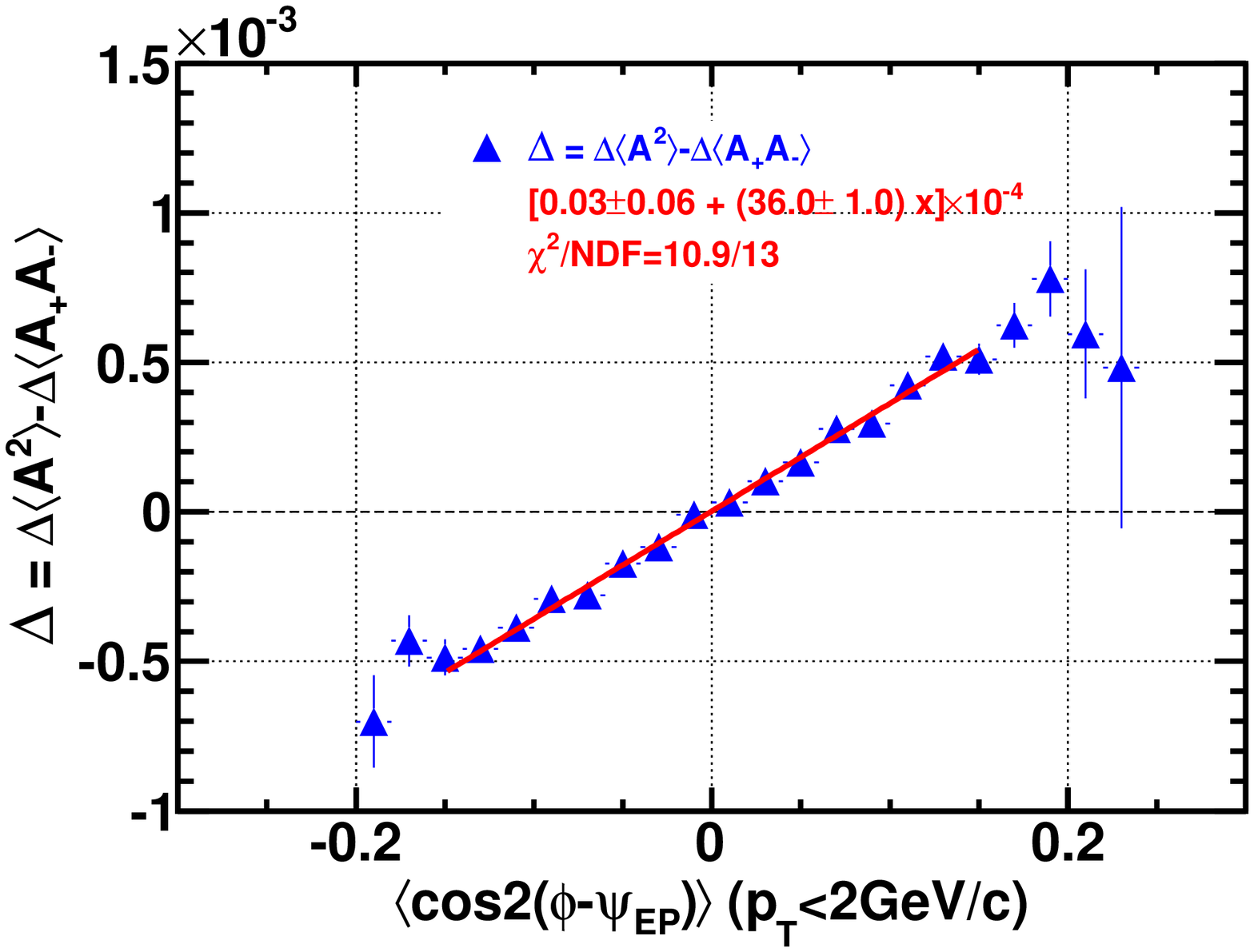}}
		\subfigure[ZDC-SMD EP resolution]{\label{fig:chargesepv2zdc-b} \includegraphics[width=0.55\textwidth]{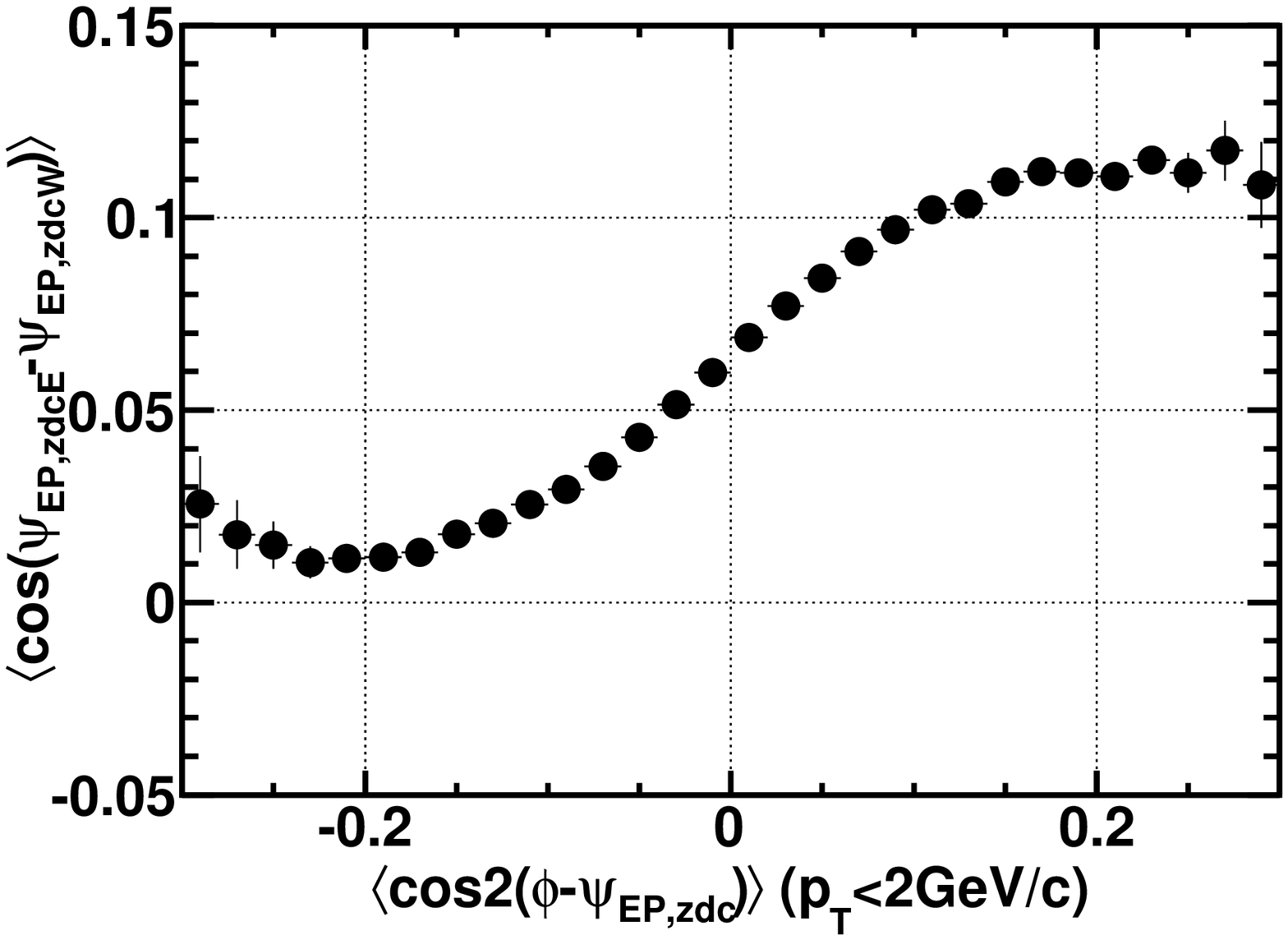}}
	\end{center}
	\caption[Mid-central charge separation from 1st order EP]{
	Panel (a): The charge separation $\Delta$ scaled by $N_{part}$ as a function of event-by-event anisotropy $v_2^{obs}$.
	The asymmetries and $v_2^{obs}$ are calculated from half TPC tracks of an event, with respect to the first order event-plane reconstructed from ZDC-SMD detectors.
	The charge separation is fitted to a linear polynomial as shown in red line.
	Panel (b): The first order event-plane resolution squared as a function of $v_2^{obs}$.
	Data are from RUN VII 200 GeV Au+Au collisions in 20-40\% centrality, and the particle $p_T$ range of $0.15<p_T<2.0$ GeV/$c$ is used for asymmetry calculation and $v_2^{obs}$ calculation.
	Error bars are statistical.
	}
	\label{fig:chargesepv2zdc}
\end{figure}

We can also study the charge separation using the first order event-plane reconstructed from the ZDC-SMD detector, which is not correlated to the TPC tracks.
The charge separation with respect to the first order event-plane result is shown in figure \ref{fig:chargesepv2zdc-a} along with the linear fit of the charge separation against $v_2^{obs}$ also with respect to the first order event-plane within the range of $-0.15<v_2^{obs}<0.15$.
We observe linear dependence of charge separation $(0.03\pm0.06)\times10^{-4}+(3.6\pm0.1)\times10^{-3} v_2^{obs}$, which suggests the charge separation is indeed correlated with the event-by-event shape.

At $v_2^{obs} \sim 0$, the charge separation is consistent with zero, indicating no charge separation effect for those spherical shaped events.
Figure~\ref{fig:chargesepv2zdc-b} shows the first order event-plane resolution squared as a function of $v_2^{obs}$.
The large statistics of RUN VII data helps to limit the uncertainty while the first order event-plane resolution is relatively low compared to the second order event-plane.

We also show the first order event-plane central (0-20\% centrality) and peripheral (40-80\% centrality) collision results in figure \ref{fig:appchargesep20v2zdc} and figure \ref{fig:appchargesep80v2zdc}.
They are qualitatively consistent with the linear dependence and the intercepts are consistent with zero.

\begin{figure}[ht]
	\begin{center}
		\subfigure[Charge separation vs $v_2^{obs}$ of top 2\% centrality]{\label{fig:chargesepv2top2-a} \includegraphics[width=0.55\textwidth]{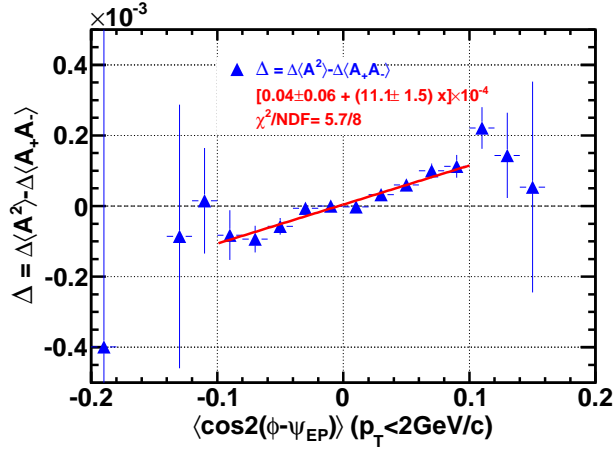}}
		\subfigure[EP resolution of top 2\% centrality]{\label{fig:chargesepv2top2-b} \includegraphics[width=0.55\textwidth]{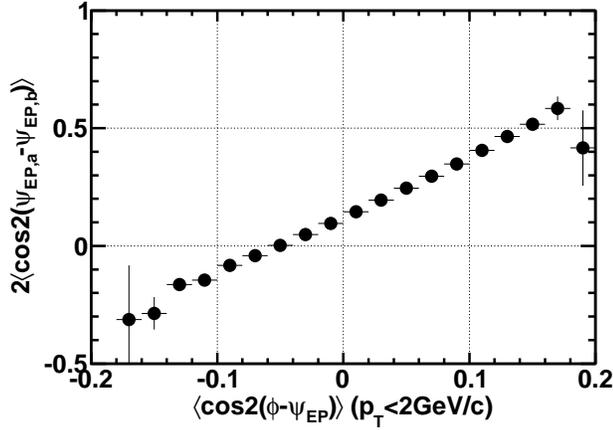}}
	\end{center}
	\caption[Charge separation vs $v_2^{obs}$ of top 2\% central events]{
	Panel (a): The charge separation $\Delta$ scaled by $N_{part}$ as a function of event-by-event anisotropy $v_2^{obs}$ of RUN IV ZDC triggered top 2\% centrality 200 GeV Au+Au collisions.
	The asymmetries and $v_2^{obs}$ are calculated from half TPC tracks of an event, with respect to the event-plane reconstructed from the other half TPC tracks.
	The charge separation is fitted to a linear polynomial as shown in red line.
	Panel (b): The top 2\% centrality event event-plane resolution squared as a function of $v_2^{obs}$.
	The particle $p_T$ range of $0.15<p_T<2.0$ GeV/$c$ is used for asymmetry calculation and $v_2^{obs}$ calculation.
	Error bars are statistical.
	}
	\label{fig:chargesepv2top2}
\end{figure}

Results for the top 2\% most central collisions are shown in figure \ref{fig:chargesepv2top2},
exhibiting charge separation that linearly depends on the event-by-event $v_2^{obs}$ within the range of $-0.1<v_2^{obs}<0.1$, though the slope is smaller than the other three measurements above in figure \ref{fig:chargesepv2top2-a}.
The event-plane resolution of the top 2\% central collision is shown in \ref{fig:chargesepv2top2-b}.
The linear dependence is qualitatively similar to the results presented earlier.

To summarize, we have studied four cases of charge separation ($\Delta$) as a function of event-by-event anisotropy ($v_2^{obs}$).
The results agree well with a linear dependence on the $v_2^{obs}$.
The charge separation is larger with larger event elliptic shape $v_2^{obs}$, meaning the more elongated shape in in-plane direction, the more same-sign pairs going in the same direction and/or more opposite-sign pairs going back-to-back across the event-plane.
The charge separation turns negative when $v_2^{obs}$ goes negative, meaning the more elongated shape in out-of-plane direction, the more same-sign pairs going back-to-back and more opposite-sign pairs going in the same direction across the event-plane.
Such linear effect is not expected from CME/LPV, thus it cannot be explained by CME/LPV alone.

%% file: summary.tex
%
%

\chapter{SUMMARY}

Motivated by the Chiral Magnetic Effect (CME), a possible signature of local parity violation (LPV) in heavy ion collisions, we have measured the charge multiplicity asymmetries and their correlations for both positively and negatively charged particles in the directions across the event-plane ($UD$) and the plane perpendicular to the event-plane ($LR$).
Each event is divided into two equally halves (sub-events) according to the pseudo-rapidity of the tracks within TPC ($-1 < \eta < 0$ and $0 < \eta < 1$).
The event-plane is reconstructed from one sub-event, and the asymmetries are calculated in the other sub-event with respect to the event-plane in order to reduce self-correlation.
The dynamical variances ($\delta\langle A^2_{\pm} \rangle$) and covariances ($\delta\langle A_+A_- \rangle$) of positive and negative charges are presented in $UD$ and $LR$ directions with Au+Au and d+Au at 200 GeV center of mass energy collisions.
We use asymmetry correlation from $LR$ direction as a null-reference to study the CME/LPV which is an effect supposedly along the system angular momentum direction, i.e. $UD$ direction.

As shown in figure~\ref{fig:asym}, the same-sign dynamical variances $\delta\langle A^2 \rangle$ are positive in d+Au collisions and peripheral Au+Au collisions both out-of-plane ($UD$) and in-plane ($LR$), and turn negative in mid-central and central collisions.
The positive variances in d+Au and peripheral Au+Au collisions indicate a broadening in the multiplicity asymmetry distributions,
which suggests the same-sign charged particle pairs are positively correlated, or in other words, preferentially emitted in the same direction (small angle pairs).
On the other hand, in mid-central and central collisions, the variances are turning negative, which indicates narrowing in the multiplicity asymmetry distributions,
suggesting the same-sign charged pairs are negatively correlated, and preferentially emitted in the back-to-back direction.
The dynamical asymmetry variance $\delta\langle A^2_{UD} \rangle$ is larger than $\delta\langle A^2_{LR} \rangle$ except the most peripheral bin.
This indicates that there are more same-sign small angle pairs emitted in out-of-plane than in-plane direction, 
equivalently, more same-sign back-to-back pairs in in-plane than out-of-plane.

The opposite-sign covariances $\langle A_+A_- \rangle$ are largely positive in d+Au and all Au+Au collisions both in-plane and out-of-plane.
It is an indication of strong positive correlation of opposite-sign charged particle pairs, 
which suggests the positively charged and negatively charged particles are preferentially emitted in the same direction regardless of the reaction plane direction.
There are more significant positive correlation in out-of-plane than in-plane direction, $\langle A_+A_- \rangle_{UD} > \langle A_+A_- \rangle_{LR}$, 
for most centralities except the most peripheral two bins, which is consistent with d+Au collisions.
The additional positive asymmetry correlation indicates opposite-sign charged pairs in out-of-plane are more preferentially emitted in small angle than that in in-plane direction.

By taking the difference between the correlations out-of-plane ($UD$) and in-plane ($LR$), we obtain asymmetry correlations $UD-LR$ by subtracting the charge dependent correlation background which is not related to the reaction-plate as shown in figure~\ref{fig:asymdiff}.
The variance difference $\Delta\langle A^2 \rangle = \delta\langle A^2_{UD} \rangle - \delta\langle A^2 \rangle_{LR}$ and covariance difference $\Delta\langle A_+A_- \rangle = \langle A_+A_- \rangle_{UD} - \langle A_+A_- \rangle_{LR}$ show similar centrality dependence, and are all positive except most peripheral covariance bins.
The $UD-LR$ result shows inconsistence with CME/LPV for the opposite-sign correlation, which is naively expected to be negative due to CME/LPV charge separation.
The same-sign variance is consistent with the CME/LPV expectation that there is additional broadening of charge multiplicity asymmetry in out-of-plane direction compared to the in-plane direction.
However, we find (section \ref{CAC}) that the variances $\delta\langle A_+A_- \rangle_{UD}$ and $\delta\langle A_+A_- \rangle_{LR}$ are negative in mid-central to central collisions, 
suggesting additional back-to-back same-sign pairs both in-plane and out-of-plane, which is inconsistent with CME/LPV expectation.

The transverse momentum $p_T$ dependence figure \ref{fig:asympt} shows that, the charge asymmetry dynamical variances $\delta\langle A^2 \rangle_{UD}$ and $\delta\langle A^2 \rangle_{LR}$ decrease with $p_T$.
They are all positive at low $p_T$, then decrease below zero for dynamical variances and covariances above $1$ GeV/$c$.
Data indicates the small angle same-sign correlations are stronger in low $p_T$, suggesting a bulk effect.
The charge asymmetry covariances $\langle A_+A_- \rangle_{UD}$ and $\langle A_+A_- \rangle_{LR}$ are all positively correlated and increase with $p_T$,
and $\langle A_+A_- \rangle_{UD}$ grows faster than $\langle A_+A_- \rangle_{LR}$ with $p_T$.
This suggests, for the entire $p_T$ range, opposite-sign particle pairs are always positively correlated, i.e. positive and negative charges are preferentially emitted in same direction, 
and with stronger effect in higher $p_T$.

STAR has previous measurements of three-particle correlators which were consistent with CME/LPV expectation together with medium interaction.
However, the possible physics backgrounds which are related to reaction-plane have not been addressed thoroughly.
We investigated the connection between our charge asymmetry correlation measurements and the three-particle correlator measurements, 
and showed that the differences are due to higher orders and cross terms in the Fourier expansion of asymmetry correlations as a function of azimuth,
especially in the opposite-sign covariances, shown in figure~\ref{fig:asym3part-a} and figure \ref{fig:asym3part-b}.

To investigate charge separation, 
we study the wedge size and wedge location dependence of the charge asymmetry correlations.
The charge asymmetry correlations as a function of the wedge size is shown in figure~\ref{fig:asymwedgesize}.
Both the dynamical variances and covariances increase with decreasing wedge size,
which suggests the reaction-plane dependent charge asymmetry correlations are more likely local in azimuth.
The $UD-LR$ measurements show the difference both $\Delta\langle A_+A_- \rangle$ and $\Delta\langle A^2 \rangle$ increase with decrease wedge size.
However, the difference between same-sign and opposite-sign correlations vanish for small wedge size.

A common unknown background for the same-sign and opposite-sign correlations could possibly lie in between the correlations.
Then the difference between variance and covariance $\Delta \equiv \Delta\langle A^2 \rangle - \Delta\langle A_+A_- \rangle$, might be sensitive to the charge separation induced by CME/LPV.
We find that the charge separation $\Delta_{\Delta\phi_{\text{w}}}$ decreases with decreasing of wedge size, figure~\ref{fig:chargesepwedge}, which suggests the charge separation across the event-plane happens in the vicinity of the in-plane rather than out-of-plane direction.

We have also studied the charge asymmetry correlations as a function of the event-by-event anisotropy $v_2^{obs}$ of the measured asymmetry particles in figure~\ref{fig:asymv2h} and figure \ref{fig:asymv2l}.
The $UD-LR$ correlations show little dependence on high-$p_T$ $v_2^{obs}$.
However, the $UD-LR$ of the variances $\Delta\langle A^2 \rangle$ increases strongly with low-$p_T$ $v_2^{obs}$,
while the covariance $\Delta\langle A_+A_- \rangle$ decreases slowly with low-$p_T$ $v_2^{obs}$.
Also the variance and covariance intercept at the same positive value at $v_2^{obs} \approx 0$.

We reported the charge separation $\Delta$ as a function of $v_2^{obs}$ with different event-plane reconstruction methods and event selection cuts. 
The results show a robust linear dependence on $v_2^{obs}$.
For all the instances, the charge separation across the event-plane $\Delta$ is qualitatively consistent with zero for those events with least event-by-event anisotropy ($v_2^{obs} \approx 0$).
For the most isotropic events ($|v_2^{obs}|<0.04$) where such backgrounds may vanish,
we find the charge separation effect is consistent with zero as shown in figure \ref{fig:Deltav2Y4}.

We show the charge separation $\Delta$ as a function of the average $v_{2}^{obs}$, $\langle v_{2}^{obs} \rangle$, in figure \ref{fig:chargesepv2}.
A very good linear dependence is observed over the event shape, which suggests an event anisotropy dependent charge separation.

It is also possible the physics backgrounds are different for same-sign and opposite-sign $UD-LR$ correlations.
The linear dependence of charge separation on event-by-event anisotropy $v_2^{obs}$ suggests that the intrinsic particle correlation and event shape can play an important role in the charge asymmetry correlations.
Also we suggest that alternative contributions can naturally create such effect as well.
By taking the events with minimum anisotropy ($v_2^{obs} \approx 0$), we found the charge separation magnitude to be $\Delta = (2.3\pm1.2)\times 10^{-5}$ for mid-central collisions (20-40\% centrality), from figure \ref{fig:chargesepv2stdy4}.
Then, we conclude that CME/LPV can't explain the charge separation alone, and we provide the upper limit with current measurements with charge separation $\Delta = 4.7\times 10^{-5}$ with 98\% CL.

%% file: bibliography.tex
%
%
%

\bibliography{all}


%
%
%
%

%% file: moreplots.tex
\chapter{APPENDIX}

\begin{sidewaysfigure}[htb]
	\begin{center}
		\psfig{figure=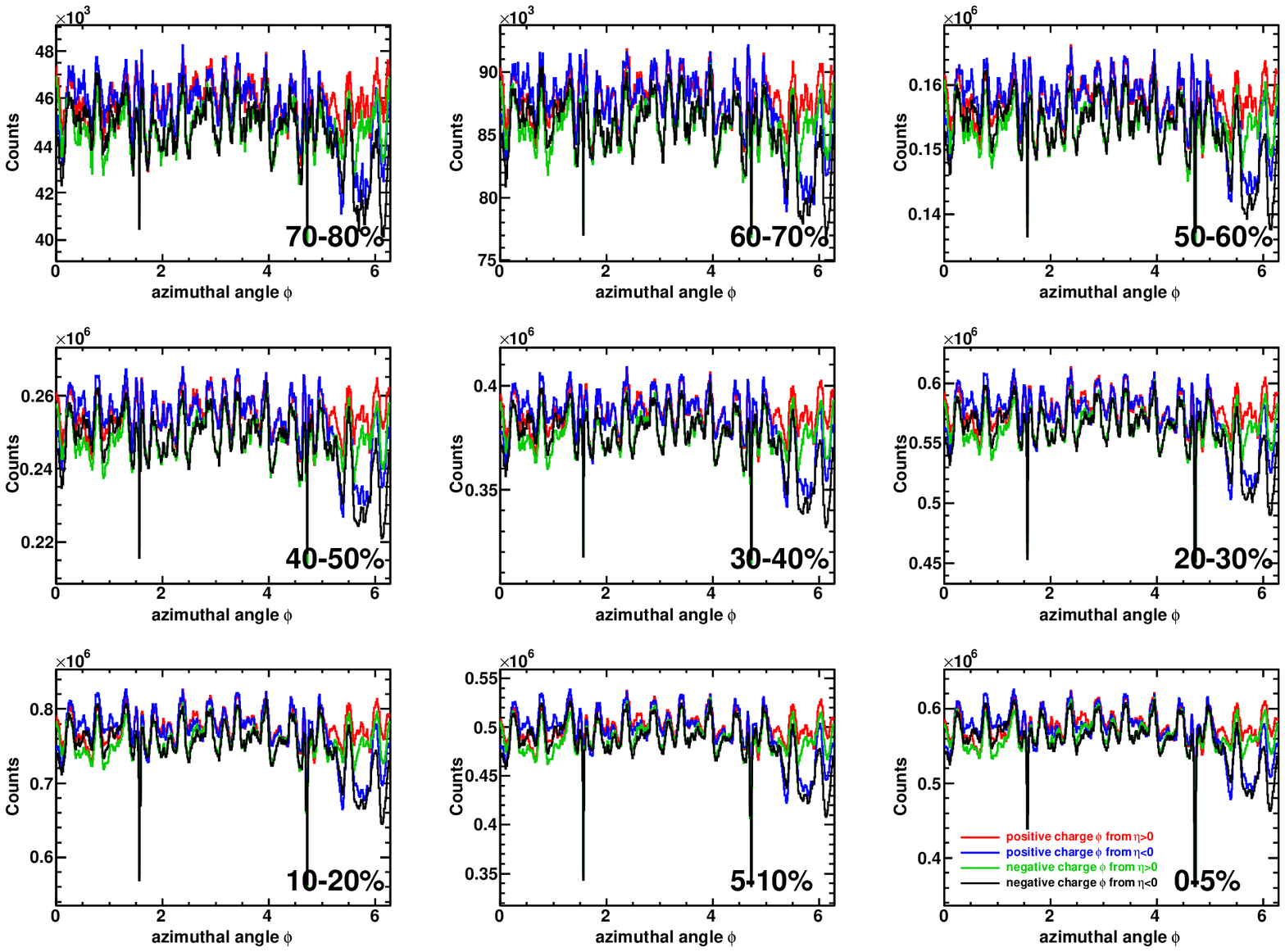,width=0.8\textwidth}
	\end{center}
	\caption[Uncorrected $\phi$ distribution for all centralities]{Single track azimuthal angle distributions for all centralities before any acceptance corrections. Red and blue lines are for positive charged particles from $\eta>0$ and $\eta<0$ regions, and green and black lines are negative charged particles from $\eta>0$ and $\eta<0$ regions respectively.}
	\label{fig:appacc1}
\end{sidewaysfigure}

\begin{sidewaysfigure}[htb]
	\begin{center}
		\psfig{figure=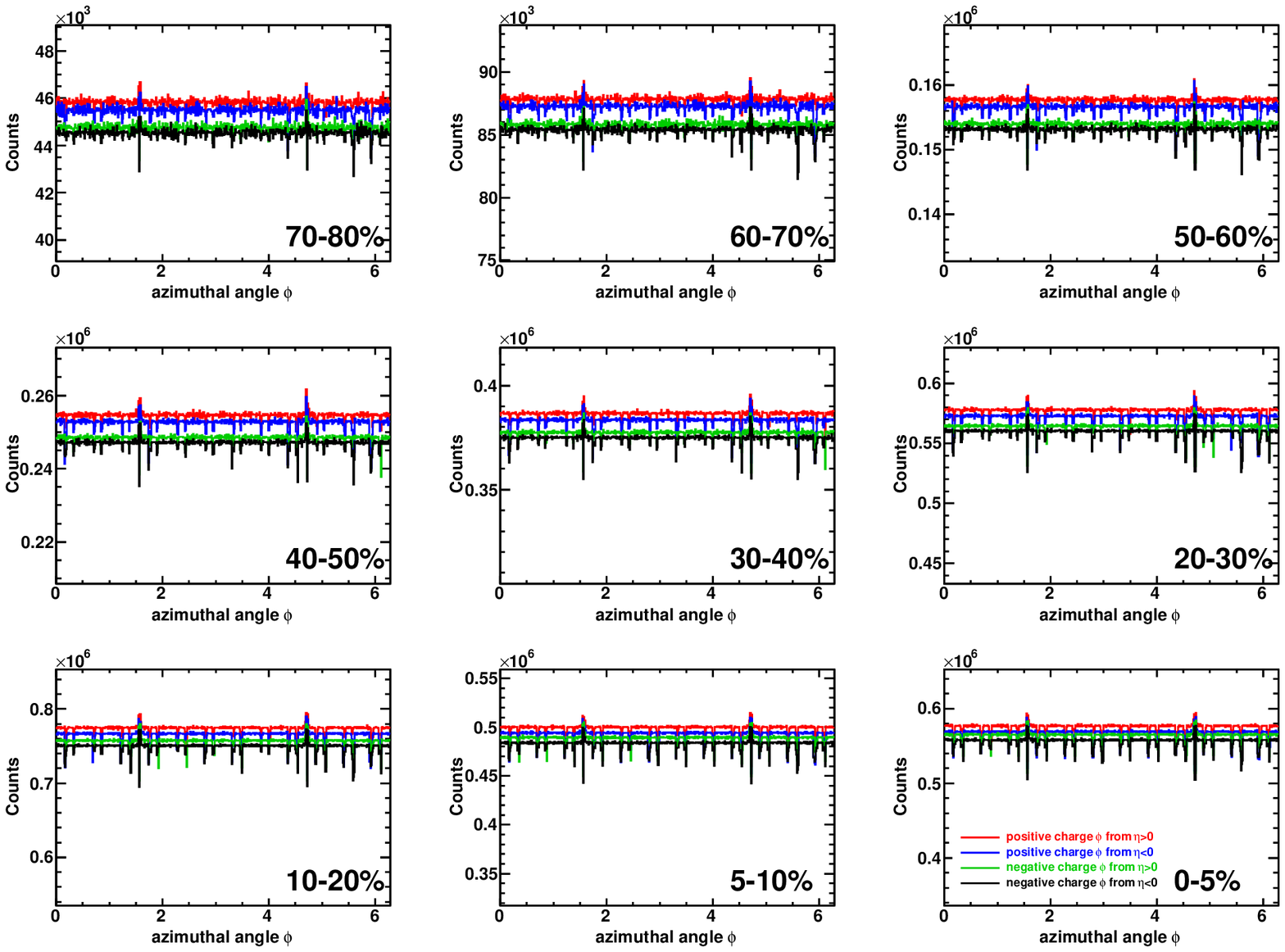,width=0.8\textwidth}
	\end{center}
	\caption[Corrected $\phi$ distribution for all centralities]{Single track azimuthal angle distributions for all centralities after the acceptance corrections. Red and blue lines are for positive charged particles from $\eta>0$ and $\eta<0$ regions, and green and black lines are negative charged particles from $\eta>0$ and $\eta<0$ regions respectively.}
	\label{fig:appacc2}
\end{sidewaysfigure}

\begin{sidewaysfigure}[htb]
	\begin{center}
		\psfig{figure=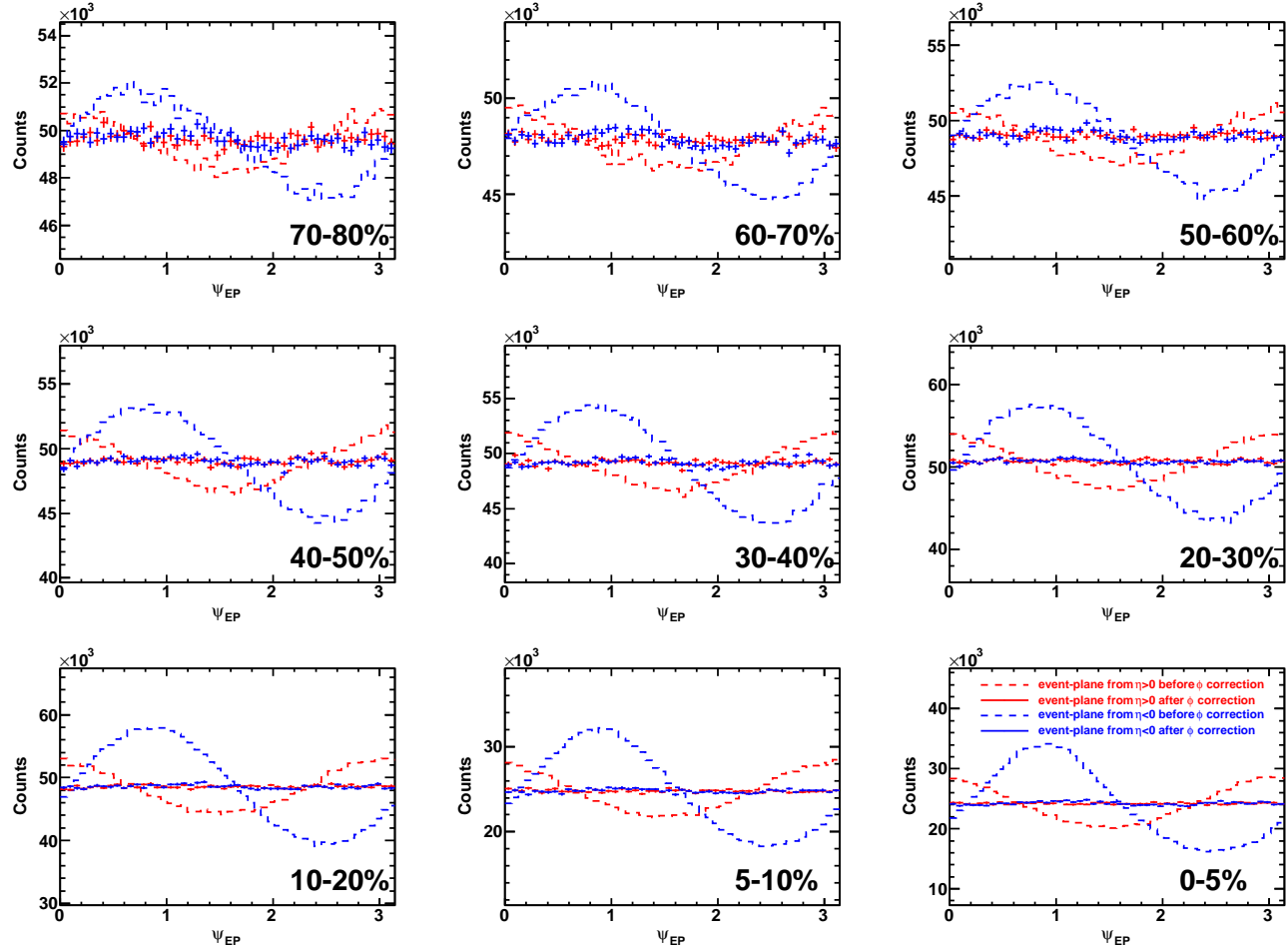,width=0.8\textwidth}
	\end{center}
	\caption[TPC second order EP distribution all centralities]{Reconstructed second order event-plane azimuthal distributions for RUN IV $Au+Au$ 200 GeV collisions in all centralities.
	The event-plane is reconstructed from charged particles within $0.15 < p_T < 2~GeV/c$ from $\eta<0$ (red) and $\eta>0$ (blue) separately.
	Error bars are statistical only.}
	\label{fig:appEP}
\end{sidewaysfigure}

\begin{figure}[htb]
	\begin{center}
		\subfigure[$\langle A_{+,UD}^2\rangle_{\eta>0}$]{\label{fig:appaccasymUD-a}\includegraphics[width=0.3\textwidth]{img/accasym-A2posUD0.eps}}
		\subfigure[$\langle A_{+,UD}^2\rangle_{\eta<0}$]{\label{fig:appaccasymUD-b}\includegraphics[width=0.3\textwidth]{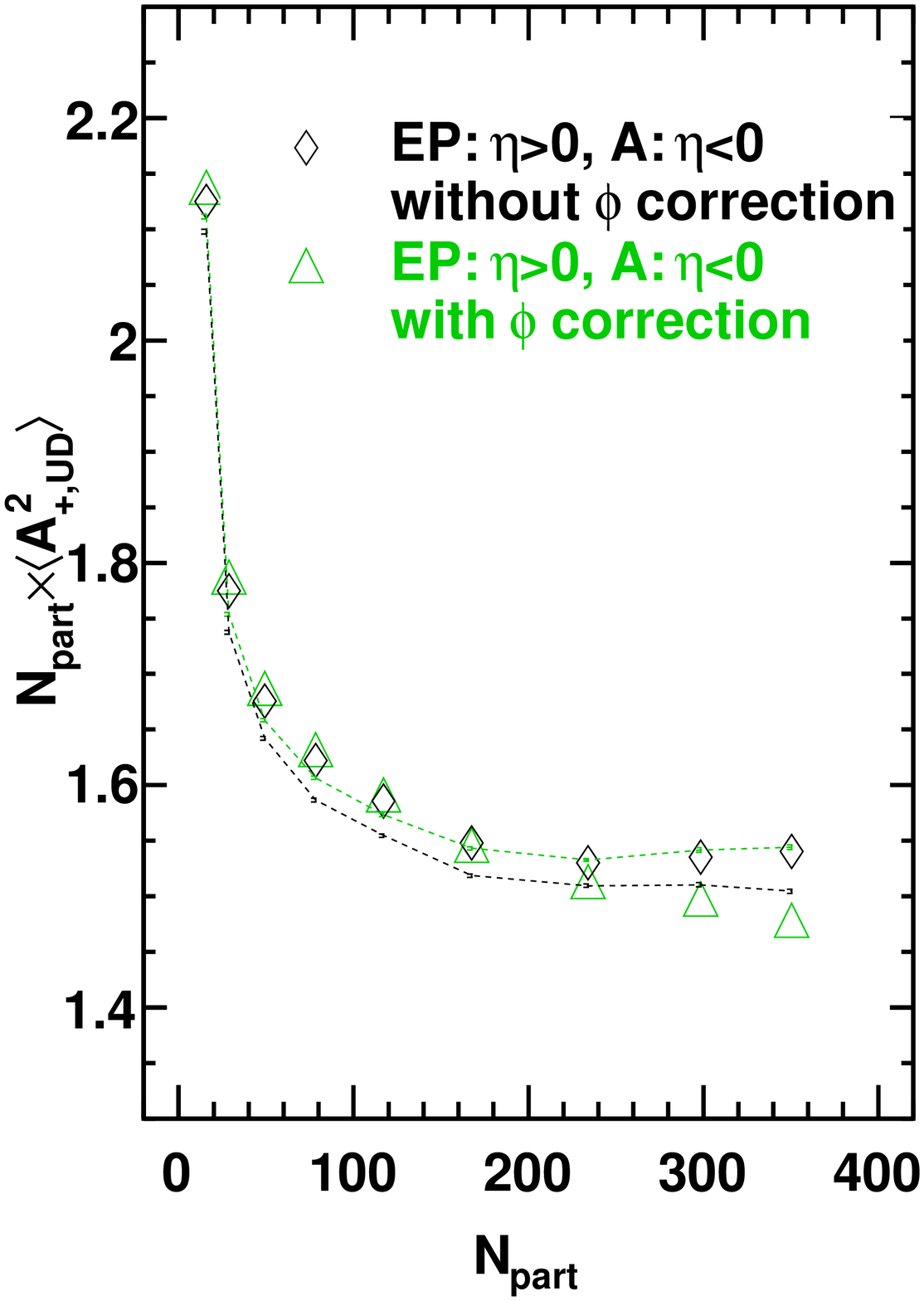}}
		\subfigure[$\langle A_+ A_- \rangle_{UD}$]{\label{fig:appaccasymUD-c}\includegraphics[width=0.3\textwidth]{img/accasym-AAUD.eps}}
		\subfigure[$\langle A_{-,UD}^2\rangle_{\eta>0}$]{\label{fig:appaccasymUD-d}\includegraphics[width=0.3\textwidth]{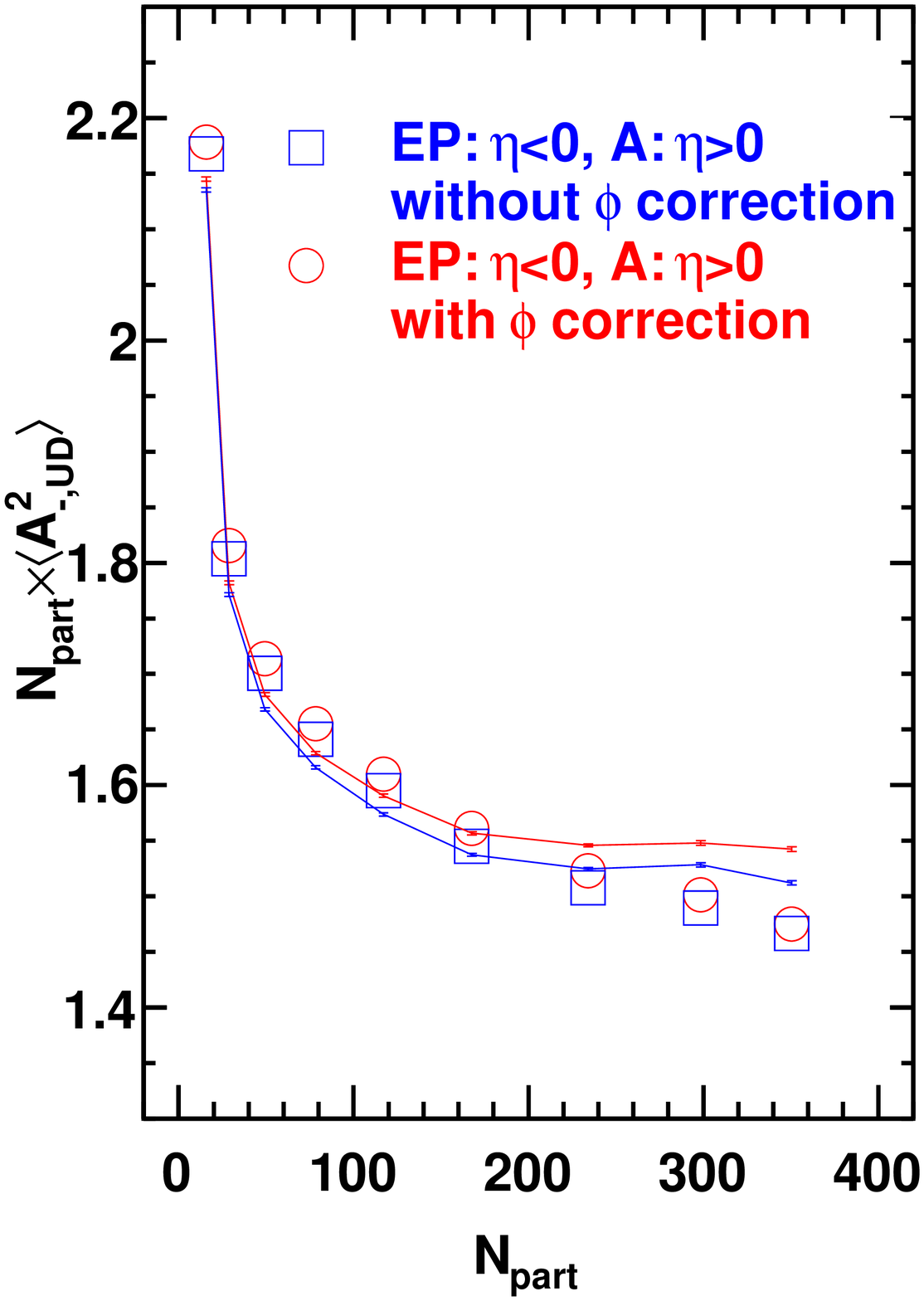}}
		\subfigure[$\langle A_{-,UD}^2\rangle_{\eta<0}$]{\label{fig:appaccasymUD-e}\includegraphics[width=0.3\textwidth]{img/accasym-A2negUDZ.eps}}
	\end{center}
	\caption[Charge asymmetry correlations efficiency correction in $UD$]{
	Asymmetry correlations: panel (a) $\langle A_{+,UD}^2\rangle_{\eta>0}$, panel (b) $\langle A_{+,UD}^2\rangle_{\eta<0}$,  panel (c) $\langle A_+ A_- \rangle_{UD}$,
	panel (d) $\langle A_+ A_- \rangle_{UD}$, panel (e) $\langle A_{-,UD}^2\rangle_{\eta<0}$
	(scaled by number of participants, $N_{part}$) before and after single particle corrections for the $\phi$ dependent acceptance $\times$ efficiency.
	The EP is reconstructed by charged particles with $p_T$ range of $0.15 < p_T < 2.0$~GeV/$c$ from one side of the TPC, and the asymmetry correlations are calculate in the same $p_T$ range but from the other side of the TPC.
	}
	\label{fig:appaccasymUD}
\end{figure}

\begin{figure}[htb]
	\begin{center}
		\subfigure[$\langle A_{+,LR}^2\rangle_{\eta>0}$]{\label{fig:appaccasymLR-a}\includegraphics[width=0.3\textwidth]{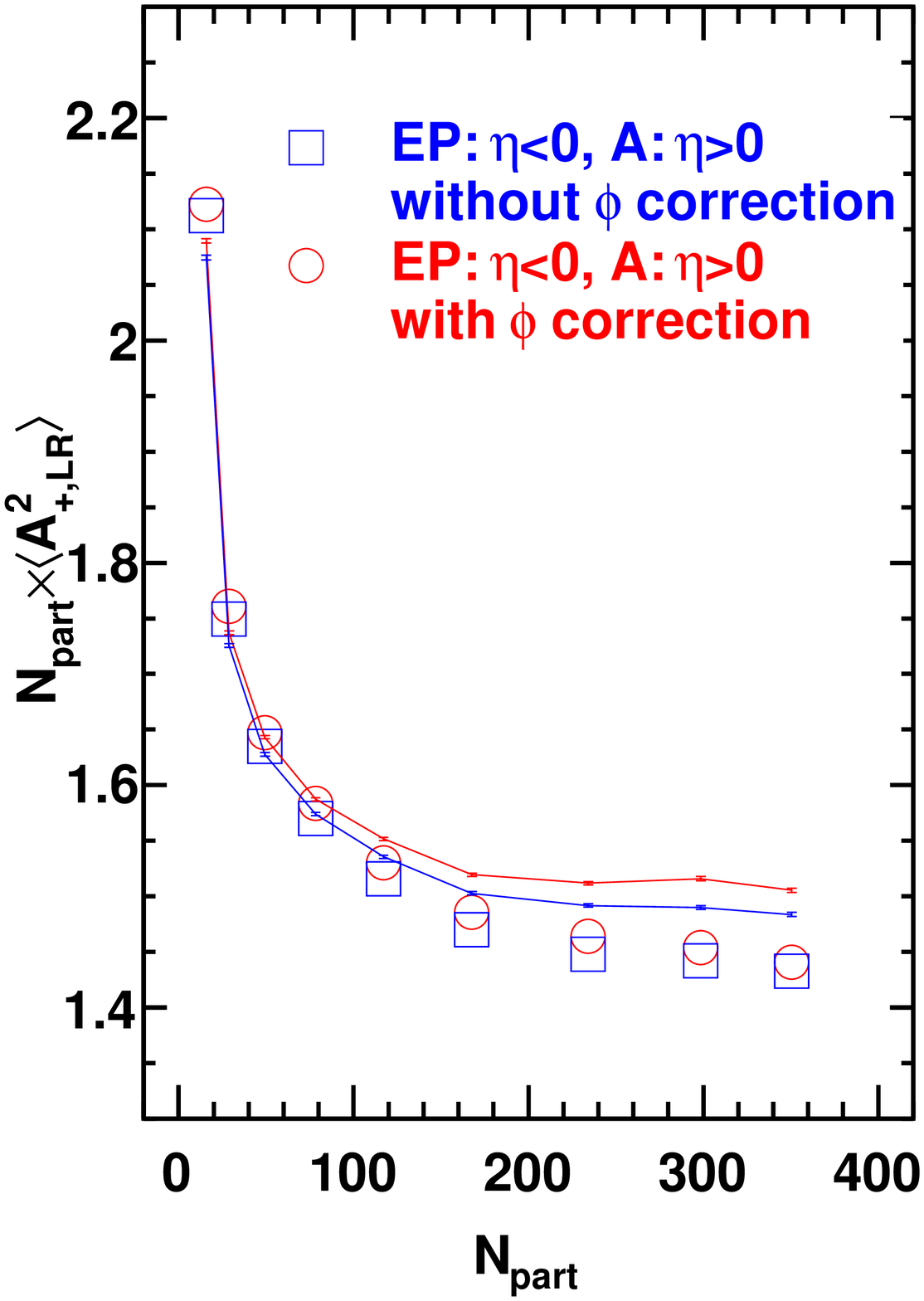}}
		\subfigure[$\langle A_{+,LR}^2\rangle_{\eta<0}$]{\label{fig:appaccasymLR-b}\includegraphics[width=0.3\textwidth]{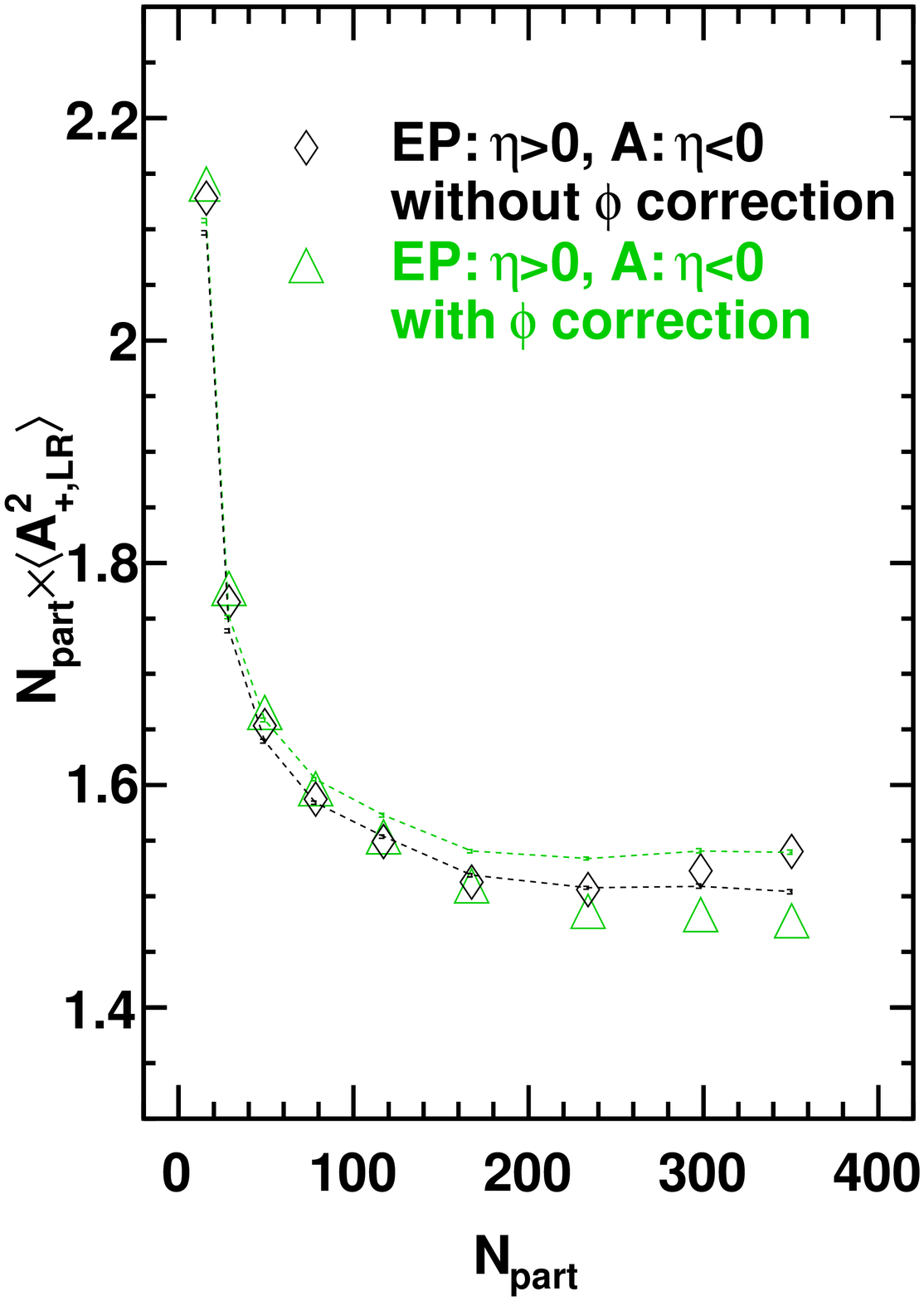}}
		\subfigure[$\langle A_+ A_- \rangle_{LR}$]{\label{fig:appaccasymLR-c}\includegraphics[width=0.3\textwidth]{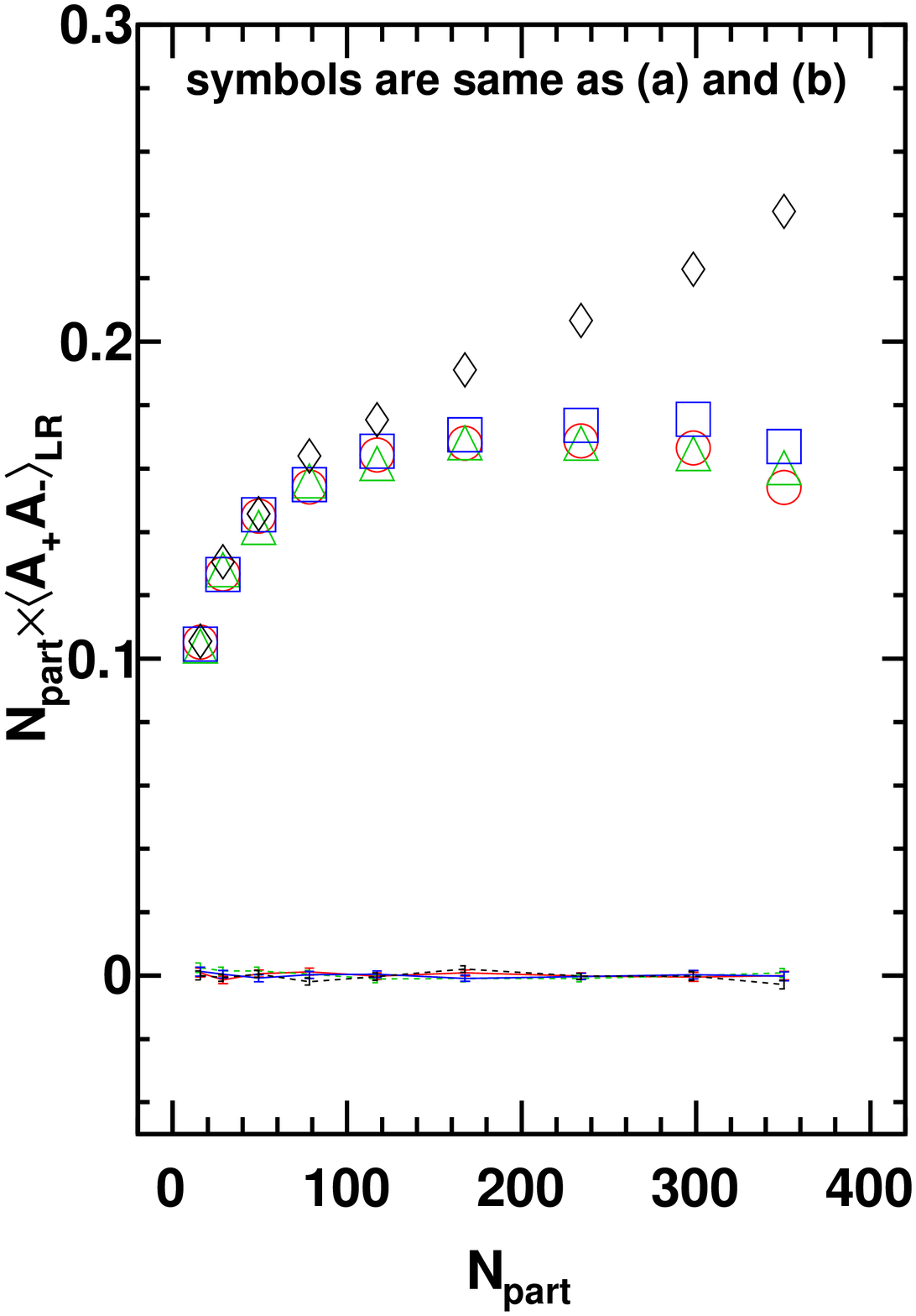}}
		\subfigure[$\langle A_{-,LR}^2\rangle_{\eta>0}$]{\label{fig:appaccasymLR-d}\includegraphics[width=0.3\textwidth]{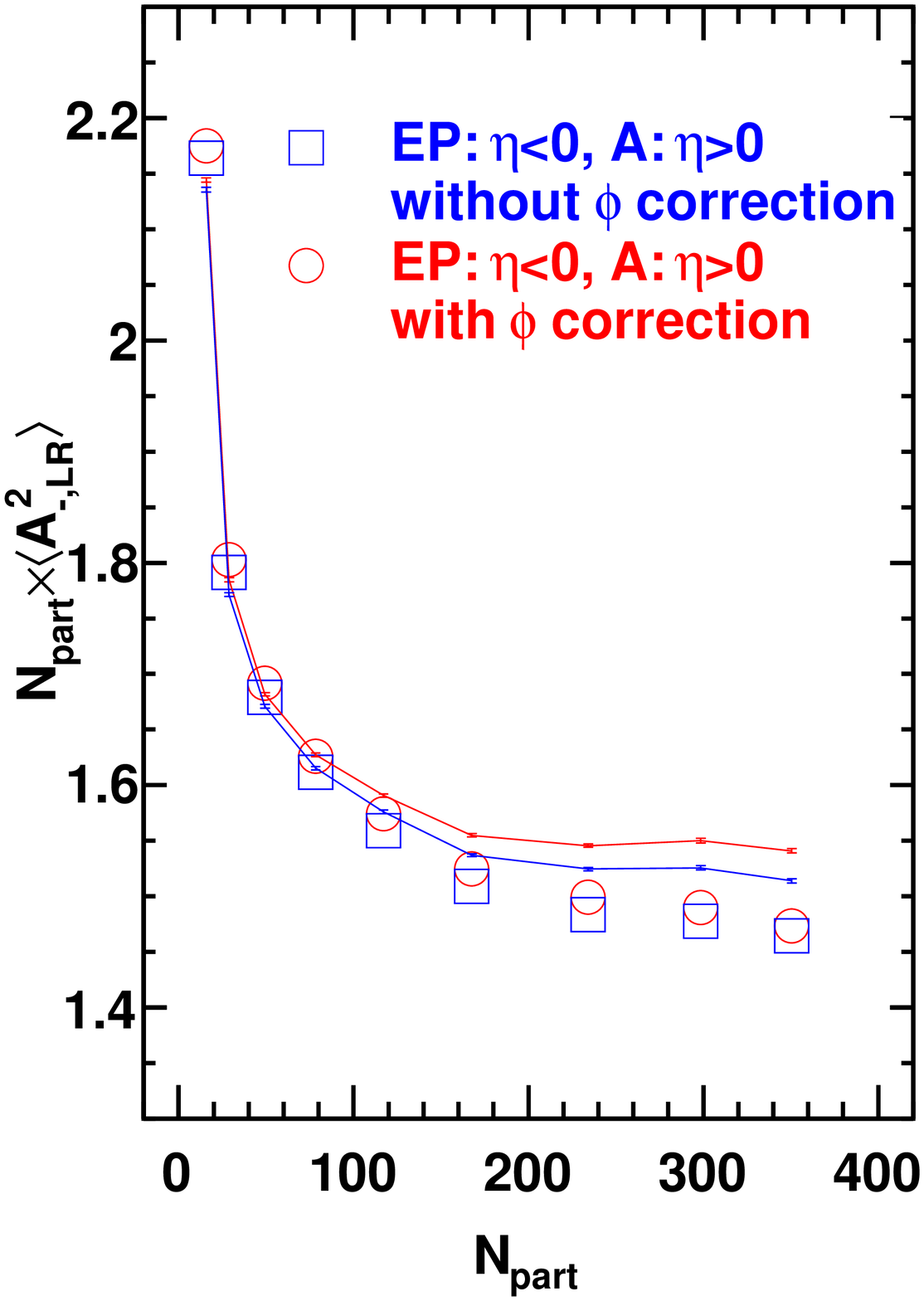}}
		\subfigure[$\langle A_{-,LR}^2\rangle_{\eta<0}$]{\label{fig:appaccasymLR-e}\includegraphics[width=0.3\textwidth]{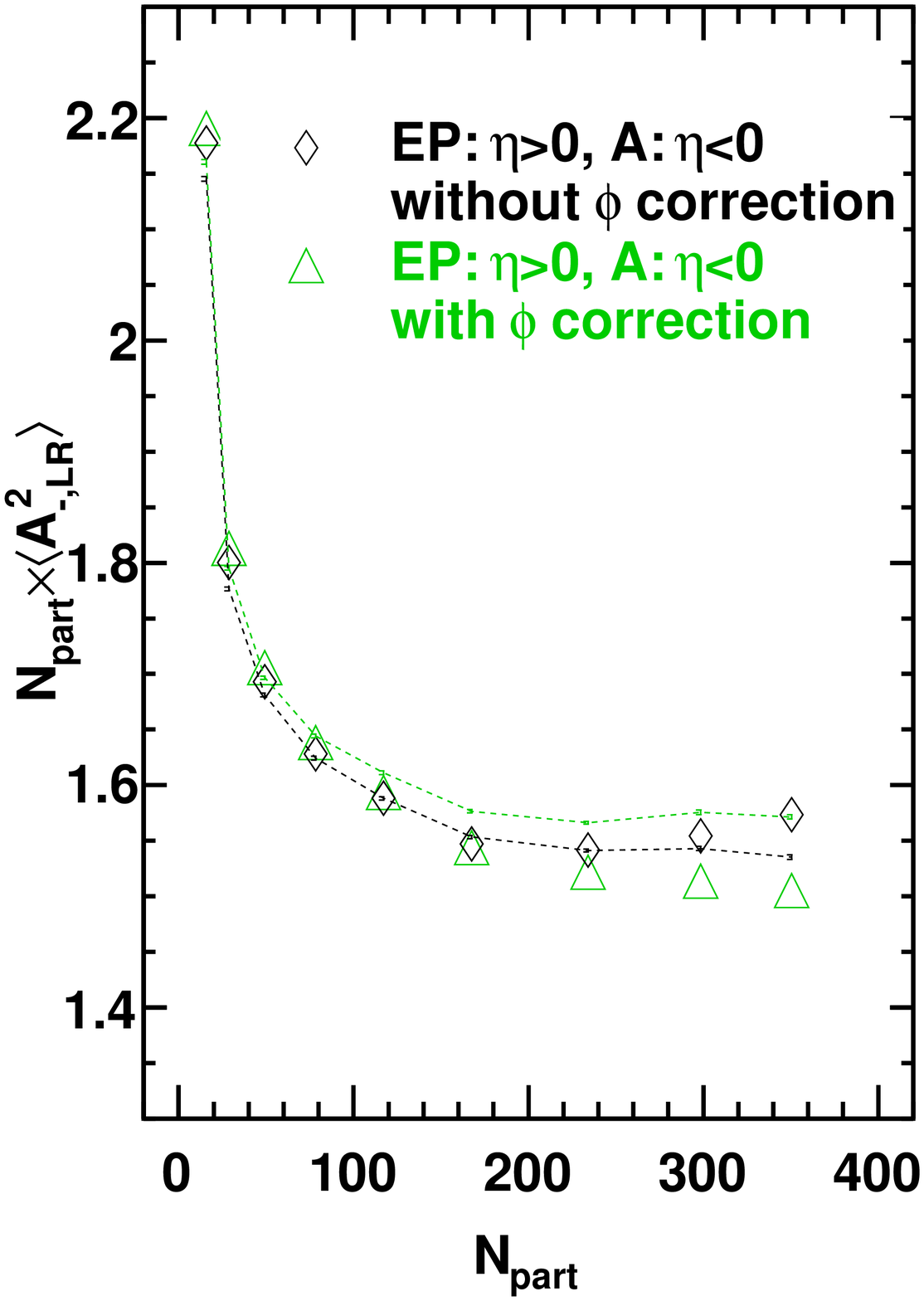}}
	\end{center}
	\caption[Charge asymmetry correlations efficiency correction in $LR$]{
	Asymmetry correlations: panel (a) $\langle A_{+,LR}^2\rangle_{\eta>0}$, panel (b) $\langle A_{+,LR}^2\rangle_{\eta<0}$,  panel (c) $\langle A_+ A_- \rangle_{LR}$,
	panel (d) $\langle A_+ A_- \rangle_{LR}$, panel (e) $\langle A_{-,LR}^2\rangle_{\eta<0}$
	(scaled by number of participants, $N_{part}$) before and after single particle corrections for the $\phi$ dependent acceptance $\times$ efficiency.
	The EP is reconstructed by charged particles with $p_T$ range of $0.15 < p_T < 2.0$~GeV/$c$ from one side of the TPC, and the asymmetry correlations are calculate in the same $p_T$ range but from the other side of the TPC.
	}
	\label{fig:appaccasymLR}
\end{figure}
\clearpage

\begin{figure}[htb]
	\begin{center}
		\subfigure[$\langle A^{2} \rangle$ vs EP resolution (most central)]{\label{fig:appEPres-a}\includegraphics[width=0.45\textwidth]{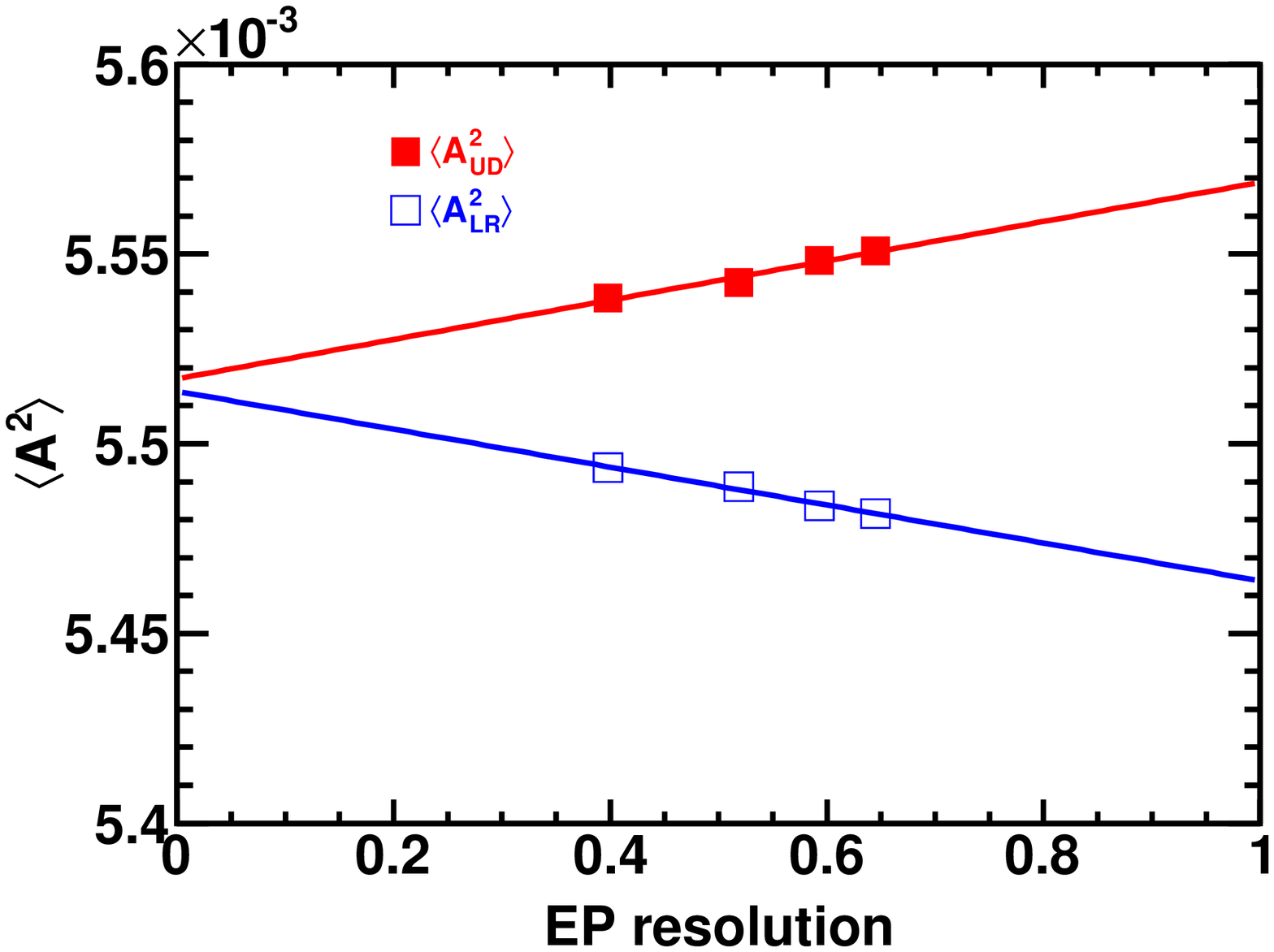}}
		\subfigure[$\langle A^{2} \rangle$ vs EP resolution (peripheral)]{\label{fig:appEPres-b}\includegraphics[width=0.45\textwidth]{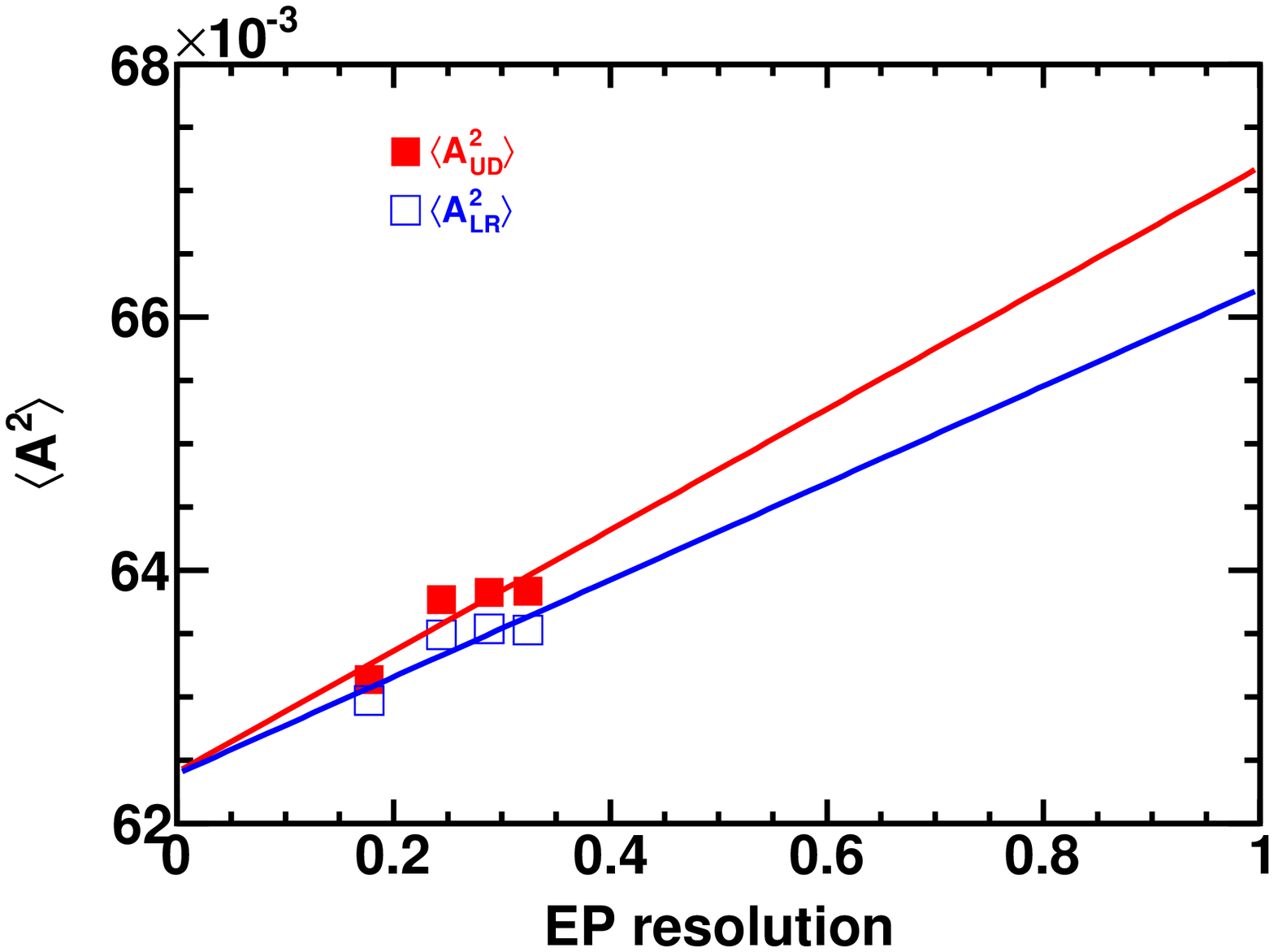}}
		\subfigure[$\langle A_+A_- \rangle$ vs EP resolution (most central)]{\label{fig:appEPres-c}\includegraphics[width=0.45\textwidth]{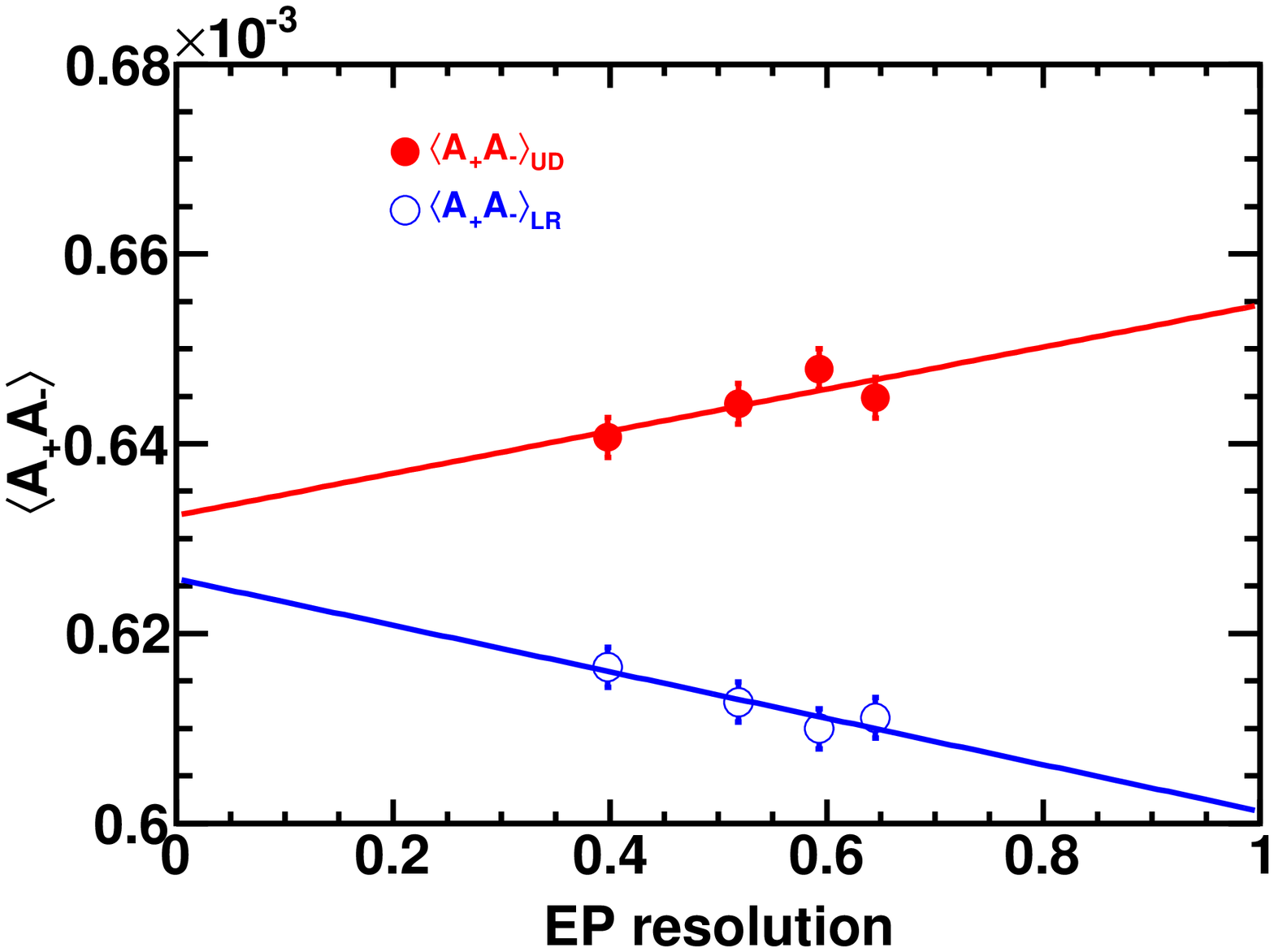}}
		\subfigure[$\langle A_+A_- \rangle$ vs EP resolution (peripheral)]{\label{fig:appEPres-d}\includegraphics[width=0.45\textwidth]{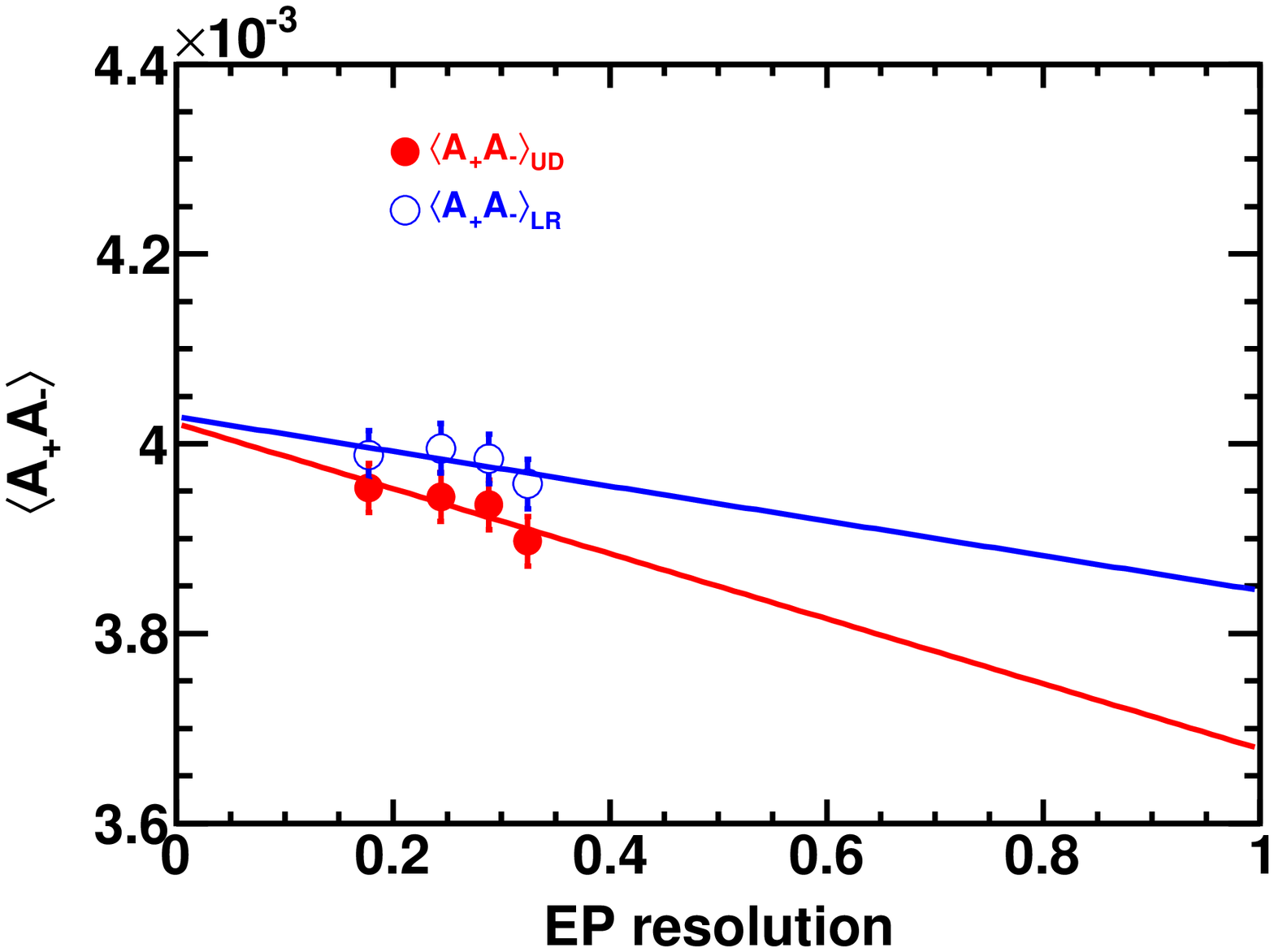}}
		\subfigure[$UD-LR$ vs EP resolution (most central)]{\label{fig:appEPres-e}\includegraphics[width=0.45\textwidth]{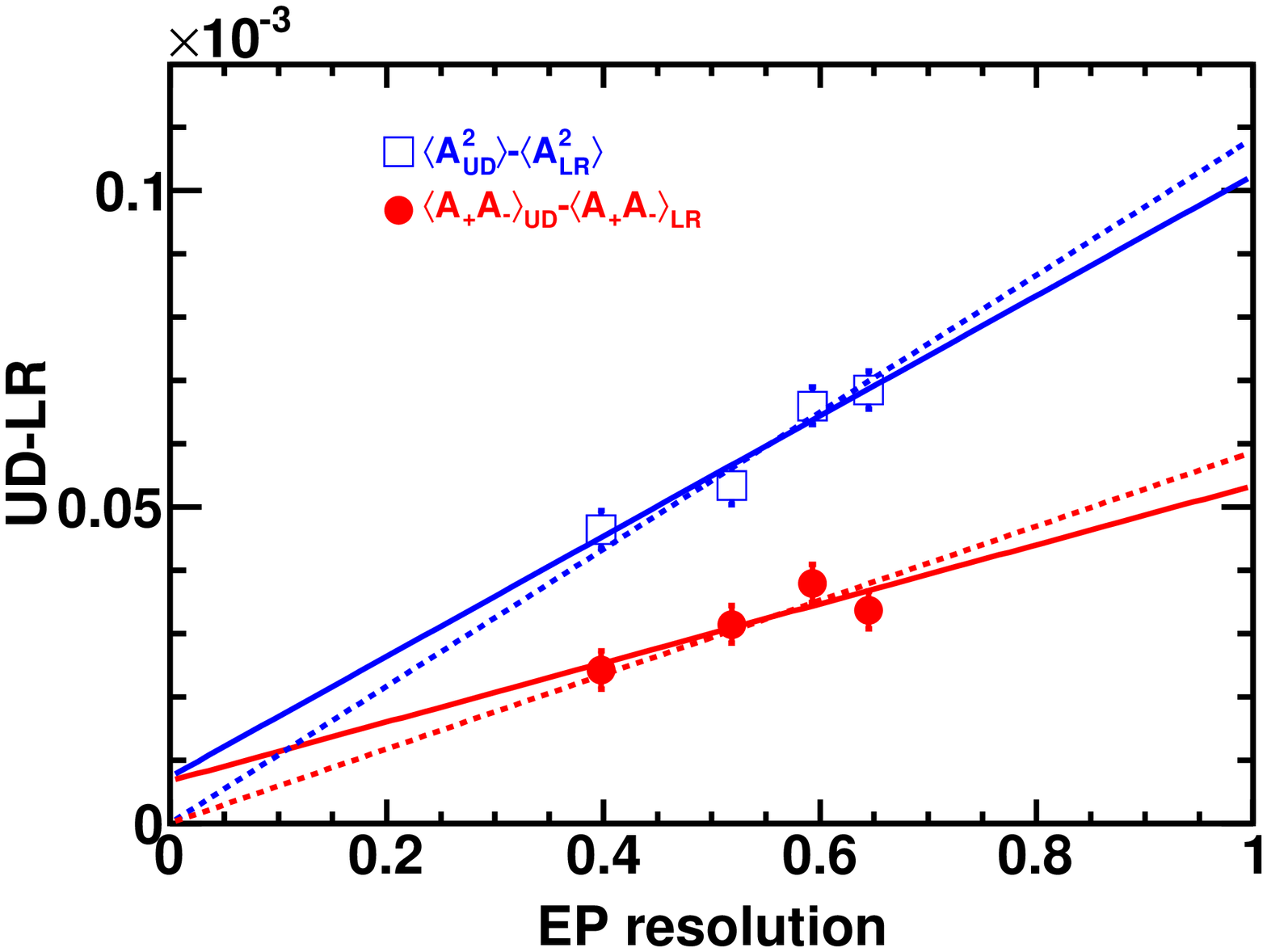}}
		\subfigure[$UD-LR$ vs EP resolution (peripheral)]{\label{fig:appEPres-f}\includegraphics[width=0.45\textwidth]{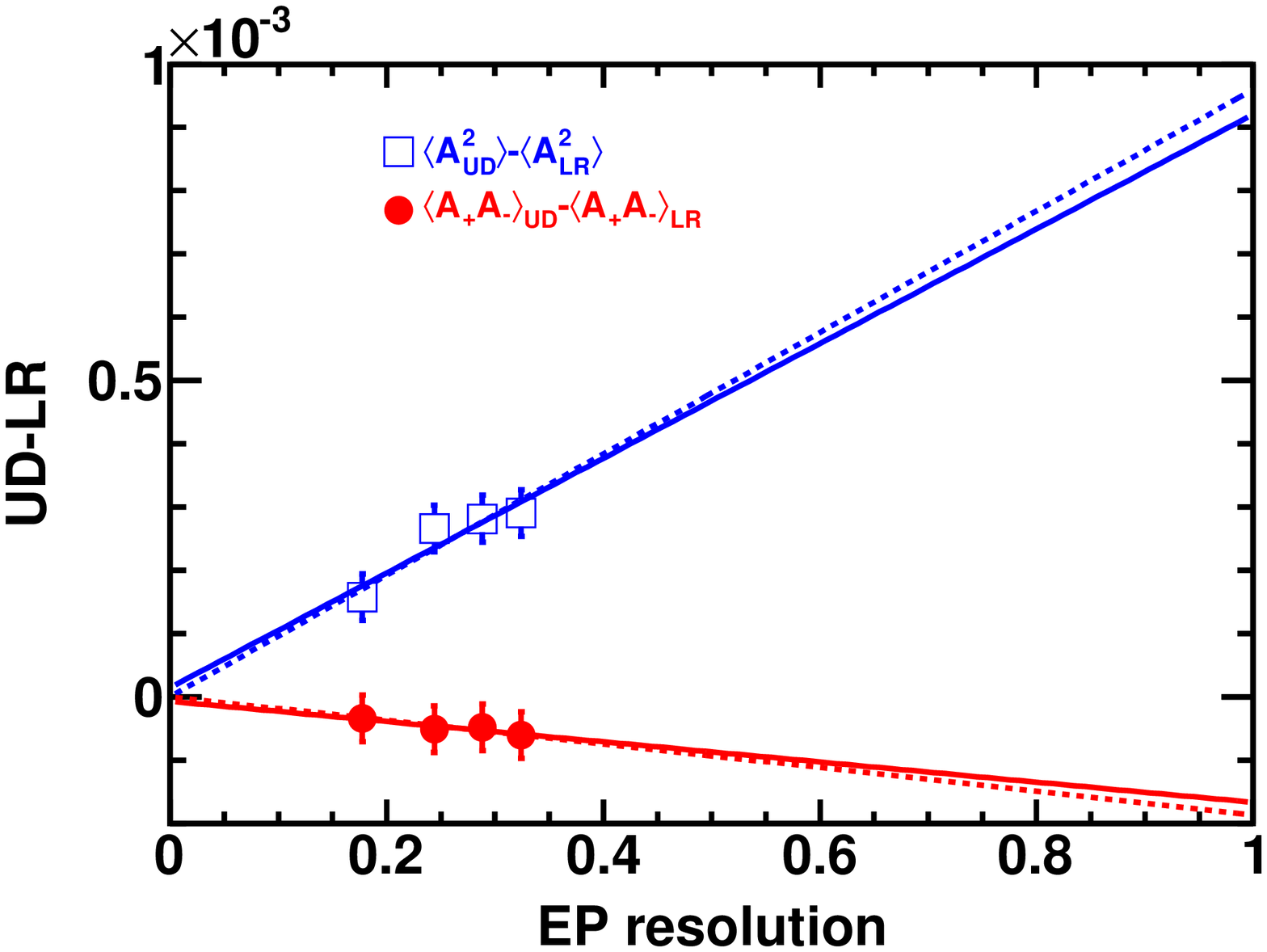}}
	\end{center}
	\caption[EP resolution dependences]{Charge multiplicity asymmetry correlations as a function of the EP resolution.
	Top and middle panels show the variances $\langle A^{2} \rangle$ and covariances $\langle A_+A_- \rangle$ EP resolution dependences.
	Bottom panel shows the $UD-LR$ differences as a function of the EP resolution.
	Left column shows 20-40\% the most central collisions, and the right column shows 40-80\% the most peripheral collisions.
	}
	\label{fig:appEPres}
\end{figure}

\begin{sidewaysfigure}[htb]
	\begin{center}
		\psfig{figure=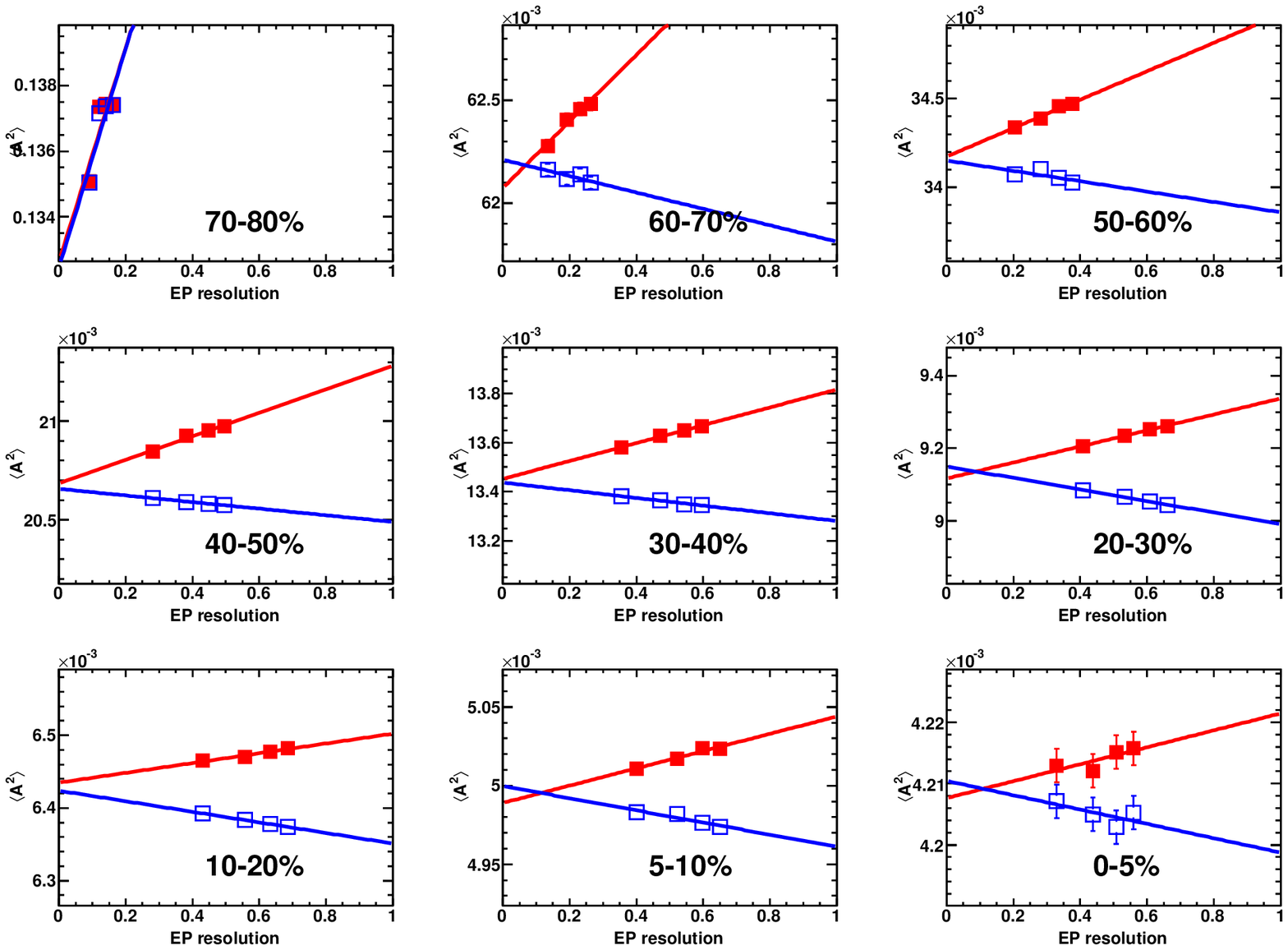,width=0.8\textwidth}
	\end{center}
	\caption[All centrality $\langle A^{2} \rangle$ vs EP resolution.]{
	Charge multiplicity asymmetry correlations $\langle A^{2} \rangle$ as a function of the event-plane resolution $\epsilon_{EP}$ in all centralities.
	The solid lines are linear fits to the data.
	Error bars are statistical.
	}
	\label{fig:appEPresA2}
\end{sidewaysfigure}

\begin{sidewaysfigure}[htb]
	\begin{center}
		\psfig{figure=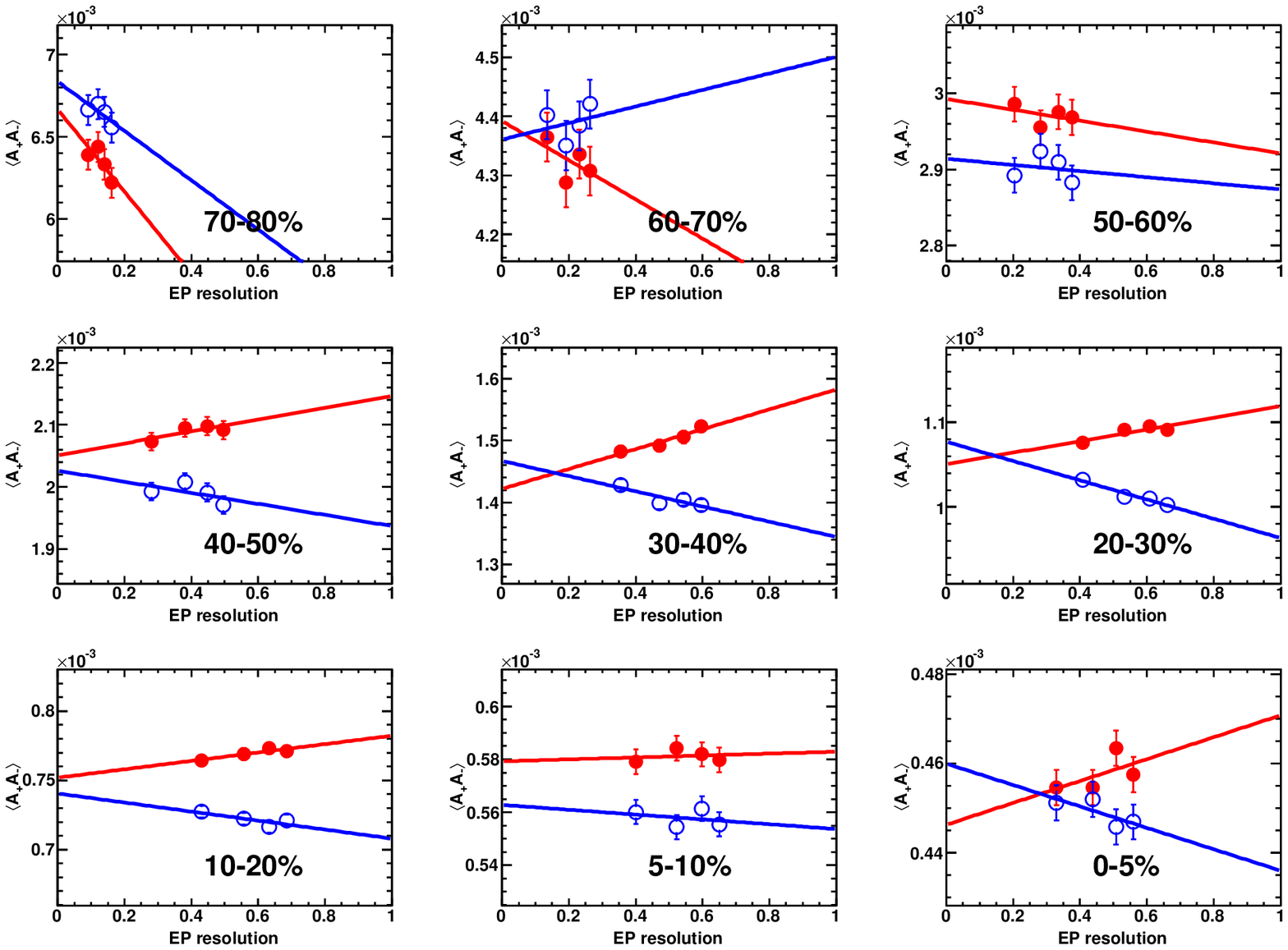,width=0.8\textwidth}
	\end{center}
	\caption[All centrality $\langle A_+A_- \rangle$ vs EP resolution.]{
	Charge multiplicity asymmetry correlations $\langle A_+A_- \rangle$ as a function of the event-plane resolution $\epsilon_{EP}$ in all centralities.
	The solid lines are linear fits to the data.
	Error bars are statistical.
	}
	\label{fig:appEPresAA}
\end{sidewaysfigure}

\begin{sidewaysfigure}[htb]
	\begin{center}
		\psfig{figure=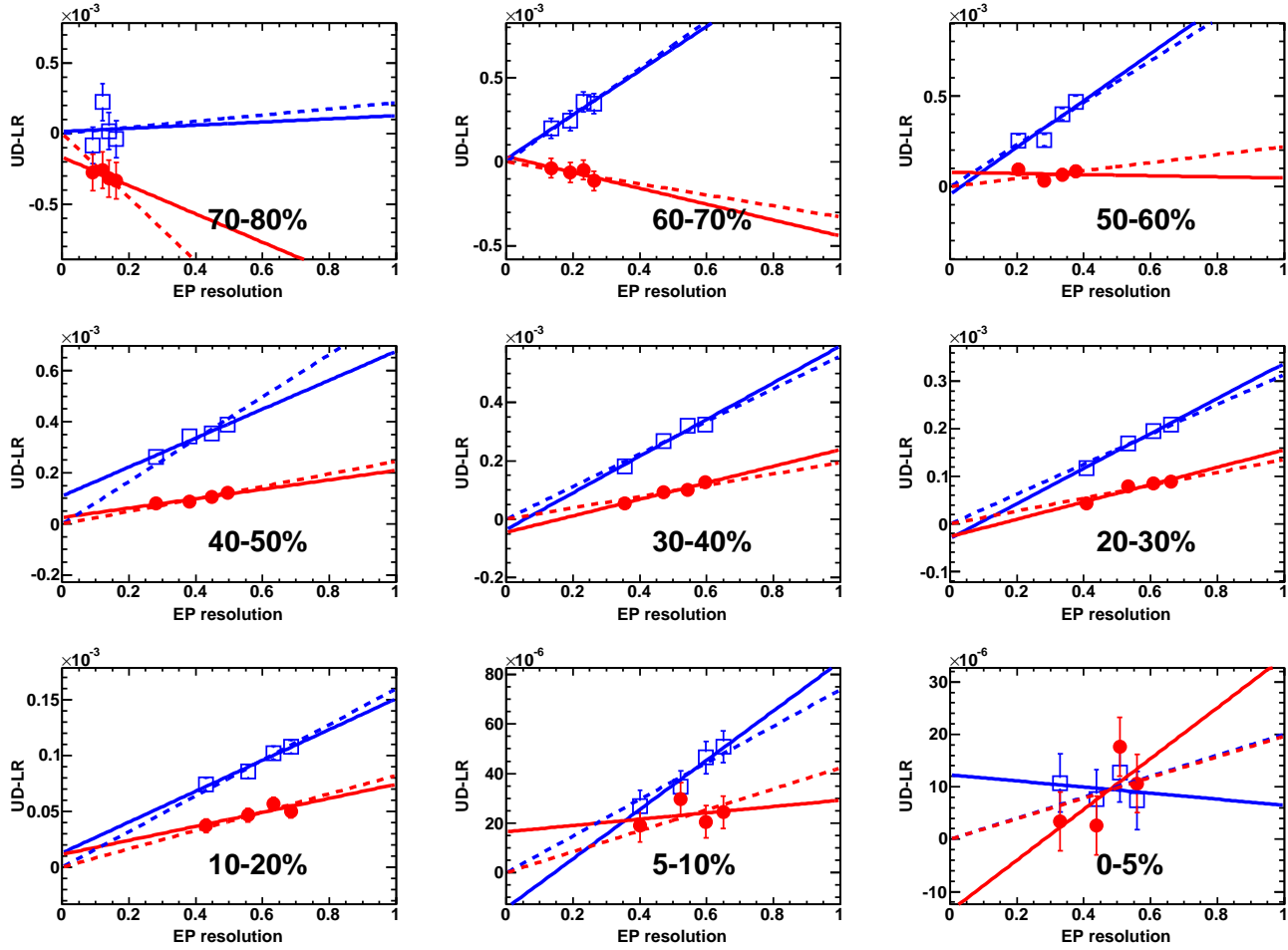,width=0.8\textwidth}
	\end{center}
	\caption[All centrality $UD-LR$ differences vs EP resolution.]{
	Charge multiplicity asymmetry correlations differences between $UD$ and $LR$ as a function of the event-plane resolution $\epsilon_{EP}$ in all centralities.
	The solid lines are linear fits to the data.
	The dashed lines are linear fits with fixed zero intercept at $\epsilon_{EP} = 0$.
	Error bars are statistical.
	}
	\label{fig:appEPresd}
\end{sidewaysfigure}

\begin{sidewaysfigure}[htb]
	\begin{center}
		\psfig{figure=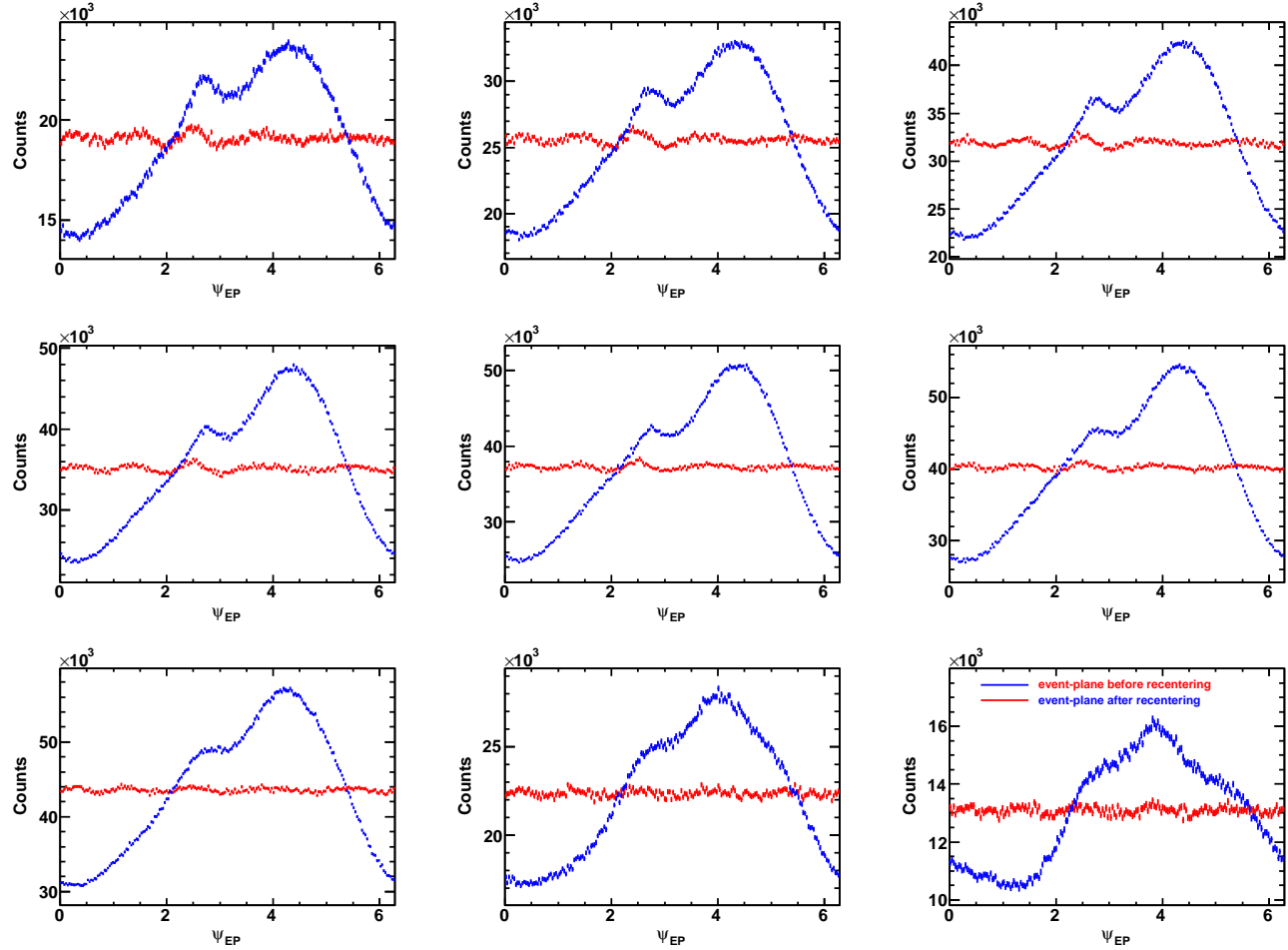,width=0.8\textwidth}
	\end{center}
	\caption[ZDC-SMD first order EP distribution all centralities]{Reconstructed first order event-plane azimuthal distributions for RUN VII Au+Au 200 GeV collisions in all centralities.
	The raw event-plane distributions are shown in blue data points, and the corrected event-plane distributions (recentering method) are shown in red data points.
	}
	\label{fig:appZDCEP}
\end{sidewaysfigure}


\begin{figure}[htb]
	\begin{center}
		\subfigure[$stat+det$ for $\langle A_{+}^2 \rangle_{LR}$ in $\eta<0$]
		{\label{fig:appstat-a}\includegraphics[width=0.45\textwidth]{img/stat-a.eps}}
		\subfigure[$stat+det$ for $\langle A_{+}^2 \rangle_{LR}$ in $\eta>0$]
		{\label{fig:appstat-a0}\includegraphics[width=0.45\textwidth]{img/stat-a0.eps}}
		\subfigure[$stat+det$ for $\langle A_{+}^2 \rangle_{UD}$ in $\eta<0$]
		{\label{fig:appstat-a1}\includegraphics[width=0.45\textwidth]{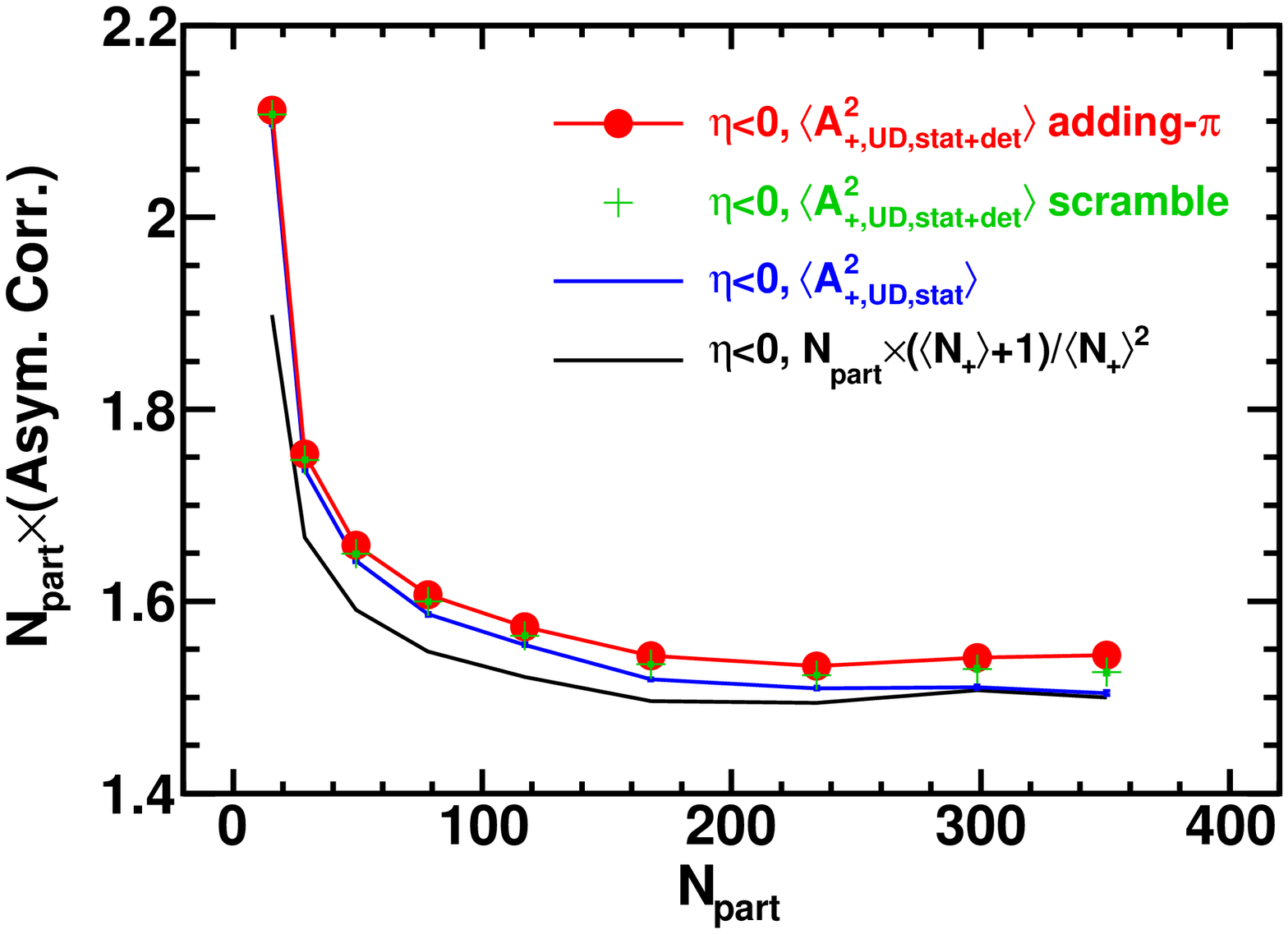}}
		\subfigure[$stat+det$ for $\langle A_{+}^2 \rangle_{UD}$ in $\eta>0$]
		{\label{fig:appstat-a2}\includegraphics[width=0.45\textwidth]{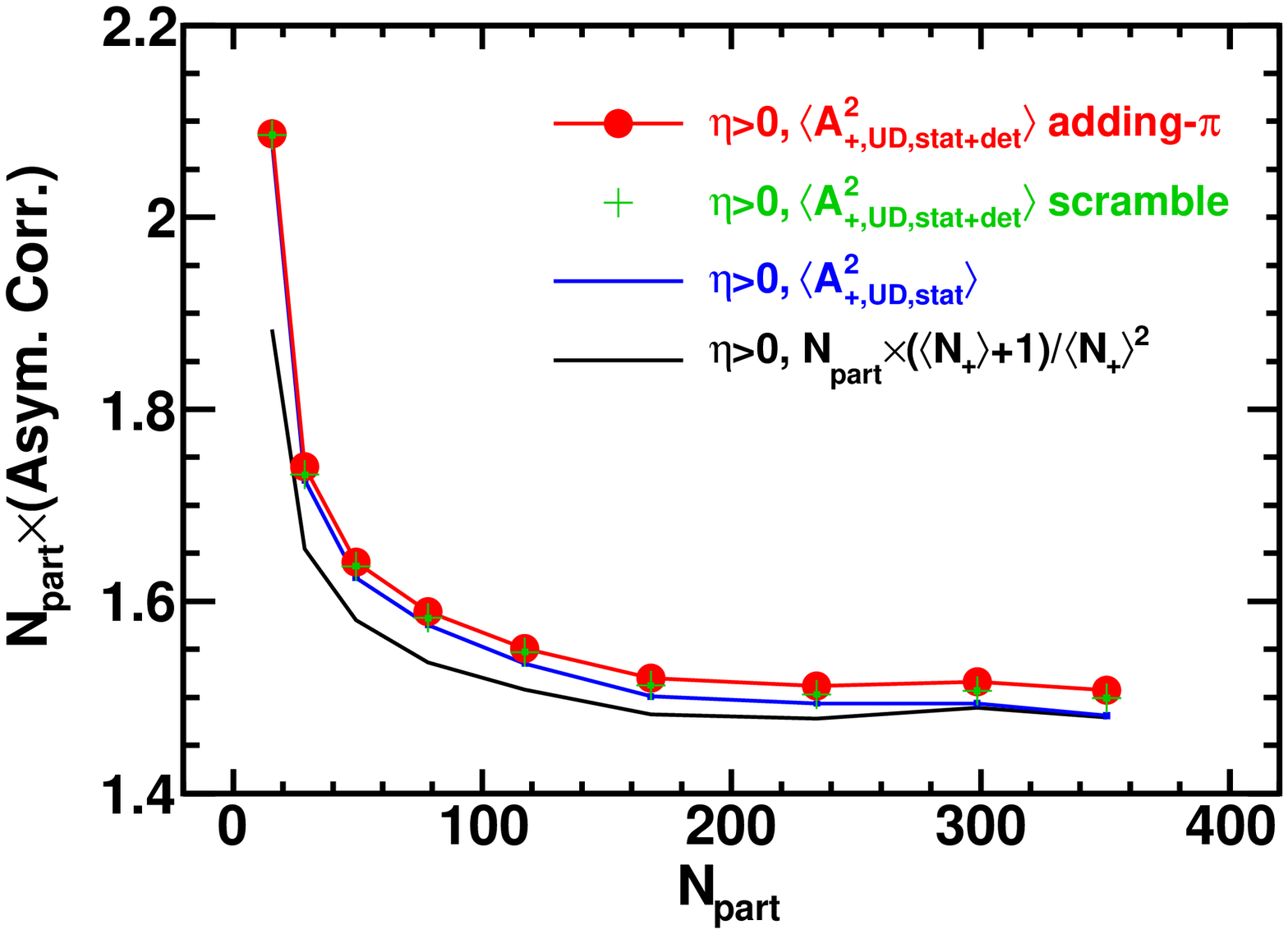}}
	\end{center}
	\caption[Statistical fluctuation and detector effect positive charge]{
	Panel (a): Statistical fluctuation and detector effects in charge asymmetry variances scaled by the number of participants $N_{part}$ from east-side of the TPC, $\eta<0$ region, 
	with respect to the EP reconstructed from west-side of the TPC, $\eta>0$ region.
	The black curve shows the ``$1/N$'' approximation by equation~\ref{eq:stat1}.
	The blue curve shows the pure statistical fluctuation $\langle A^2_{+,LR,stat} \rangle$ with ``50-50'' method.
	The statistical fluctuation plus detector effects $\langle A^2_{+,LR,stat+det} \rangle$ are shown in green crosses
	with scramble method, and red circles with flipping-$\pi$ method.
	Panel (b): Same as (a) but for $\eta>0$ region with EP reconstructed from $\eta<0$ region.
	Panel (c): Same as panel (a) for $UD$.
	Panel (d): Same as panel (b) for $UD$.
	Data are from RUN IV Au+Au 200 GeV collisions. The particle $p_T$ range is integrated over $0.15 < p_T < 2.0$ GeV/$c$.
	}
	\label{fig:appstat}
\end{figure}

\begin{figure}[htb]
	\begin{center}
		\subfigure[$stat+det$ for $\langle A_{-}^2 \rangle_{LR}$ in $\eta<0$]
		{\label{fig:appstat-n}\includegraphics[width=0.45\textwidth]{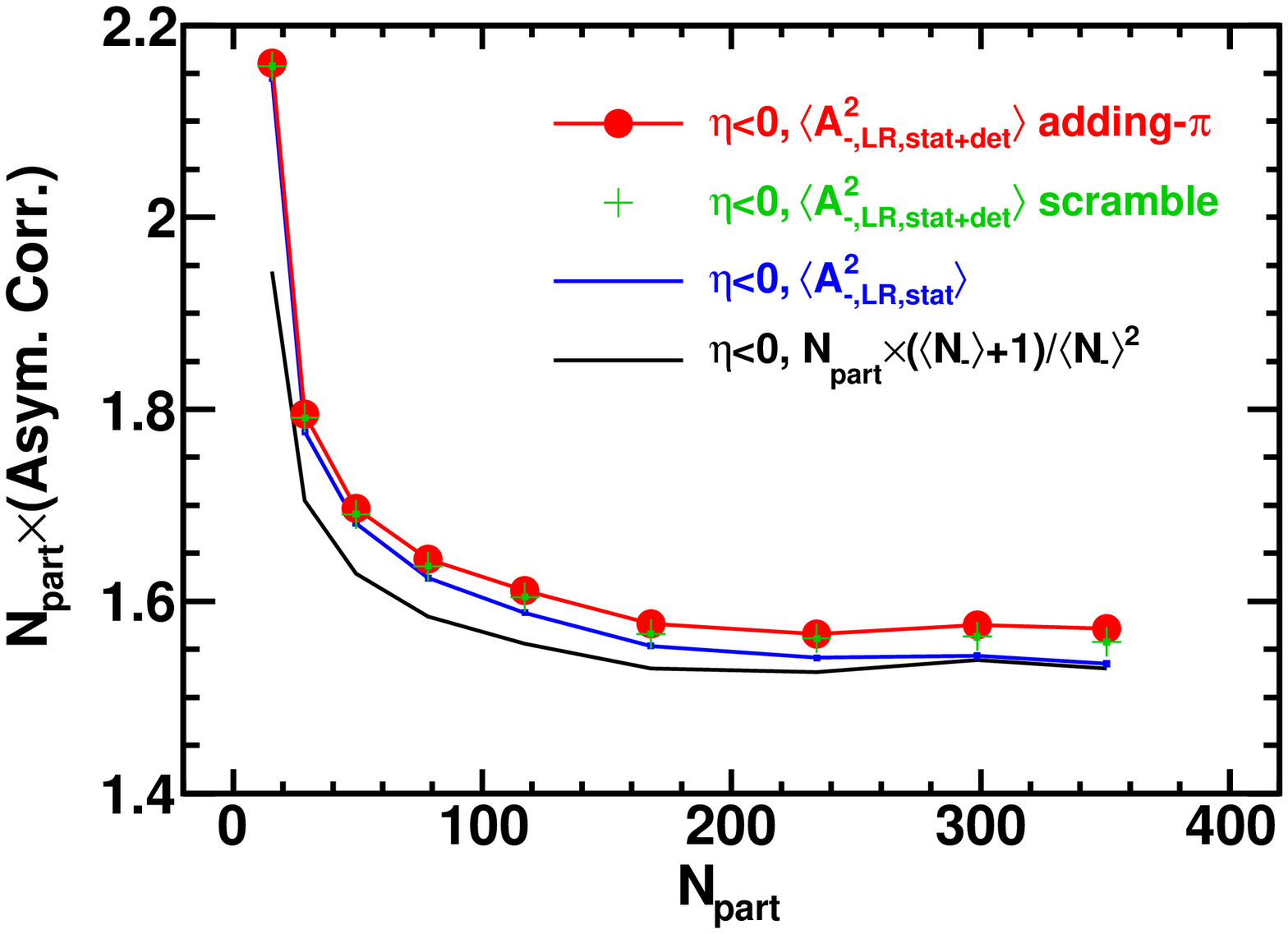}}
		\subfigure[$stat+det$ for $\langle A_{-}^2 \rangle_{LR}$ in $\eta>0$]
		{\label{fig:appstat-n0}\includegraphics[width=0.45\textwidth]{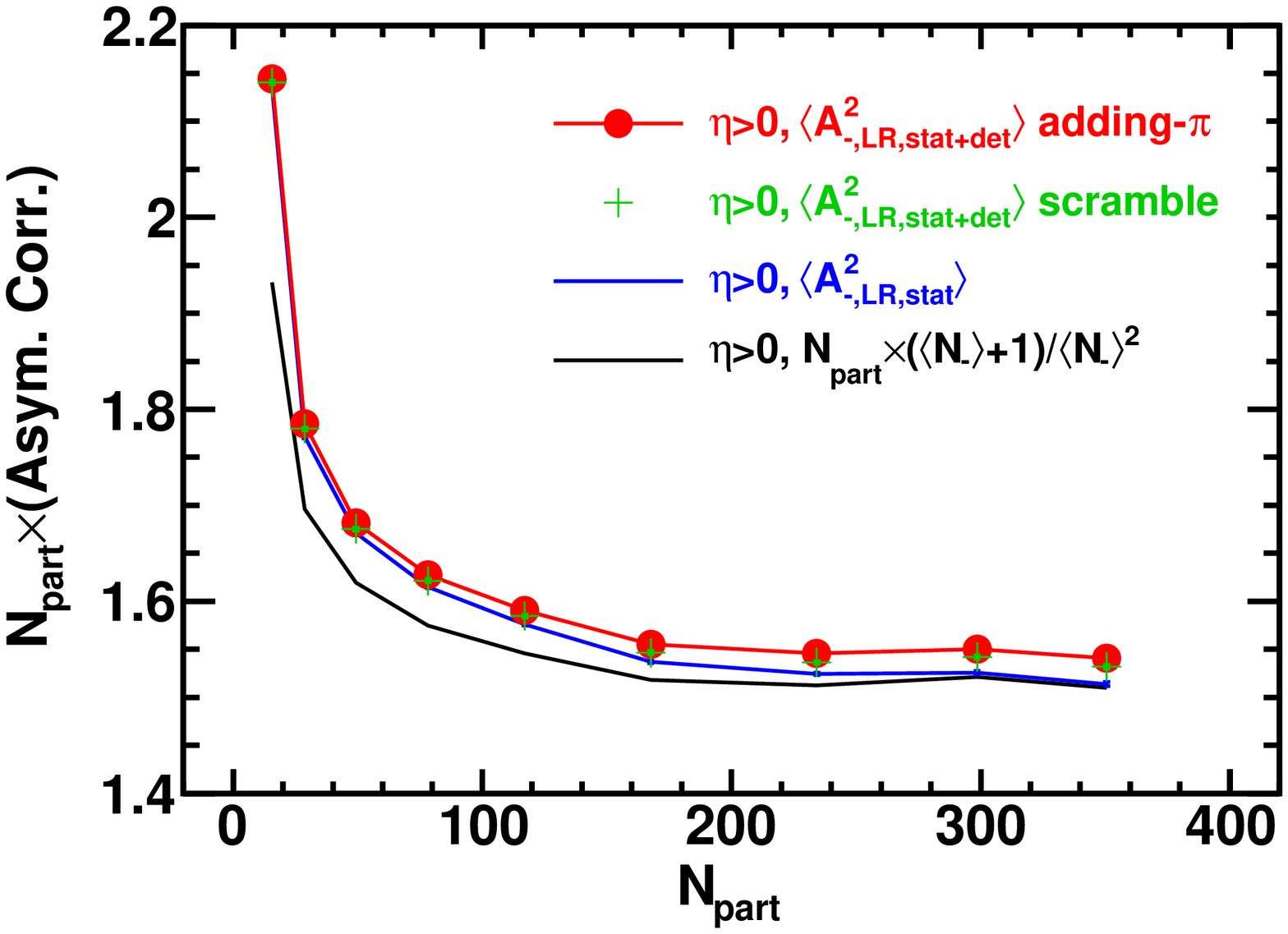}}
		\subfigure[$stat+det$ for $\langle A_{-}^2 \rangle_{UD}$ in $\eta<0$]
		{\label{fig:appstat-n1}\includegraphics[width=0.45\textwidth]{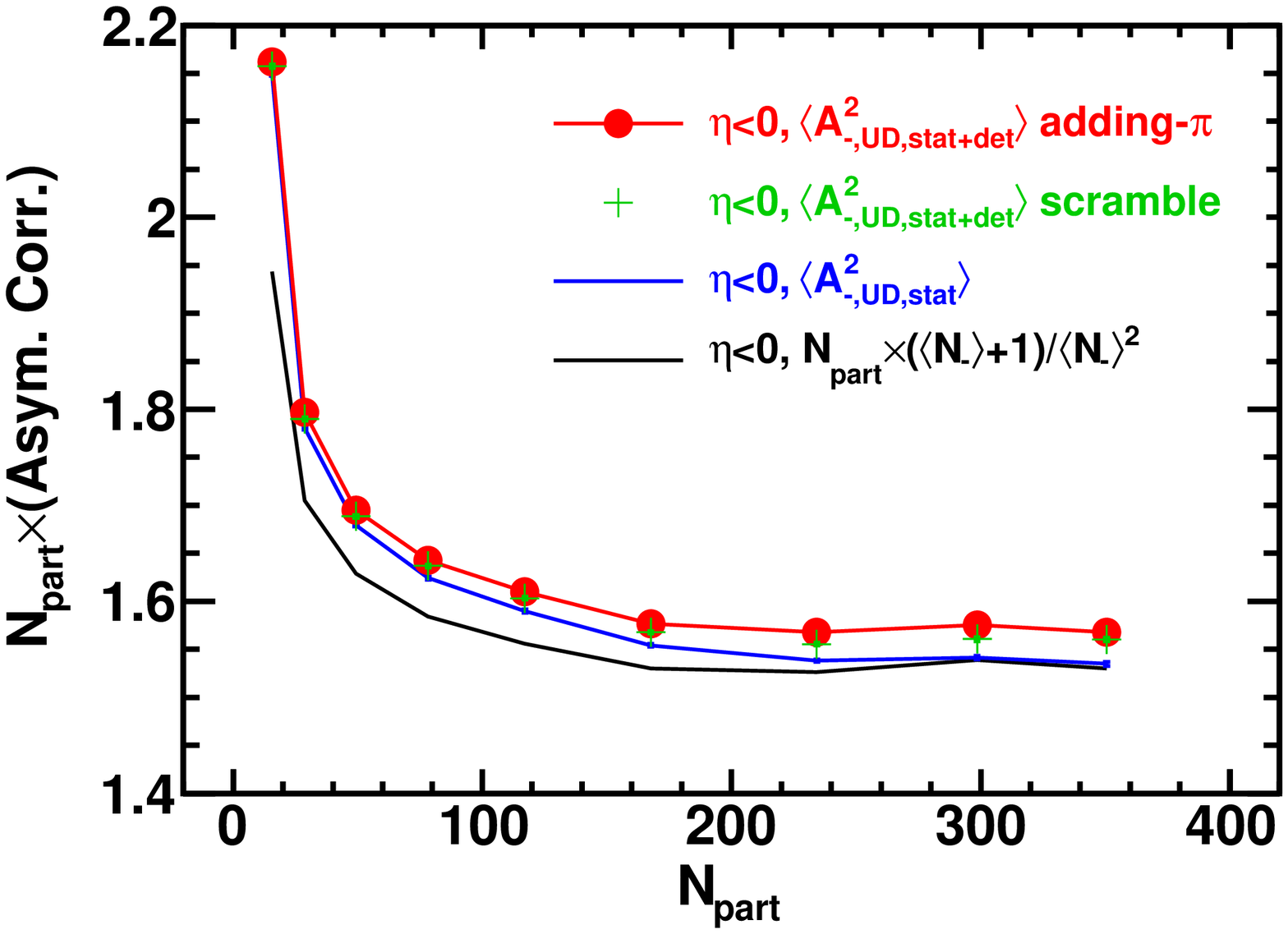}}
		\subfigure[$stat+det$ for $\langle A_{-}^2 \rangle_{UD}$ in $\eta>0$]
		{\label{fig:appstat-n2}\includegraphics[width=0.45\textwidth]{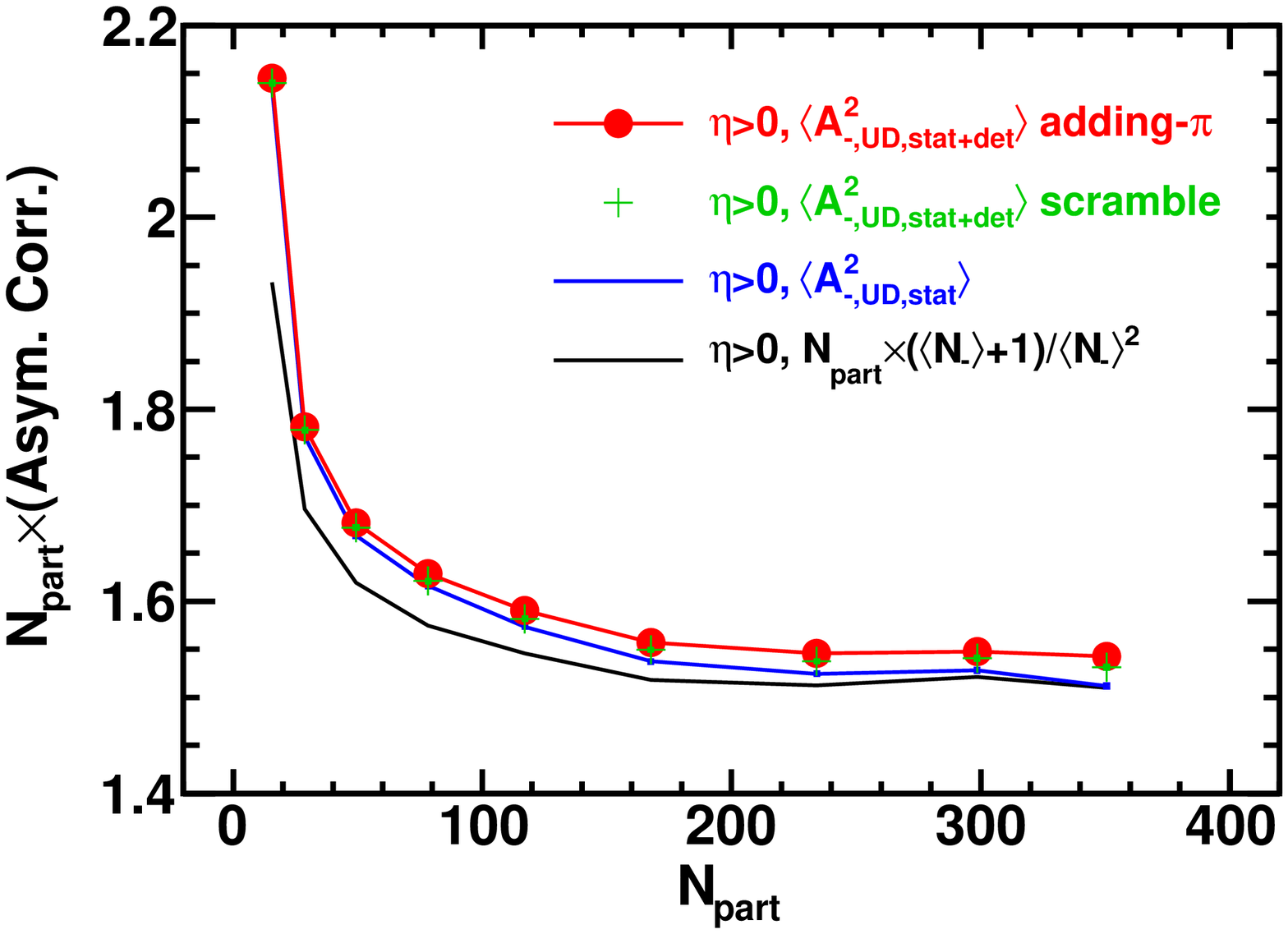}}
	\end{center}
	\caption[Statistical fluctuation and detector effect negative charge]{
	Panel (a): Statistical fluctuation and detector effects in charge asymmetry variances scaled by the number of participants $N_{part}$ from east-side of the TPC, $\eta<0$ region, 
	with respect to the EP reconstructed from west-side of the TPC, $\eta>0$ region.
	The black curve shows the ``$1/N$'' approximation by equation~\ref{eq:stat1}.
	The blue curve shows the pure statistical fluctuation $\langle A^2_{-,LR,stat} \rangle$ with ``50-50'' method.
	The statistical fluctuation plus detector effects $\langle A^2_{-,LR,stat+det} \rangle$ are shown in green crosses
	with scramble method, and red circles with flipping-$\pi$ method.
	Panel (b): Same as panel (a) but for $\eta>0$ region with EP reconstructed from $\eta<0$ region.
	Panel (c): Same as panel (a) for $UD$.
	Panel (d): Same as panel (b) for $UD$.
	Data are from RUN IV Au+Au 200 GeV collisions. The particle $p_T$ range is integrated over $0.15 < p_T < 2.0$ GeV/$c$.
	}
	\label{fig:appstatn}
\end{figure}

\begin{figure}[htb]
	\begin{center}
		\subfigure{\label{fig:appasymv220-a} \includegraphics[width=0.45\textwidth]{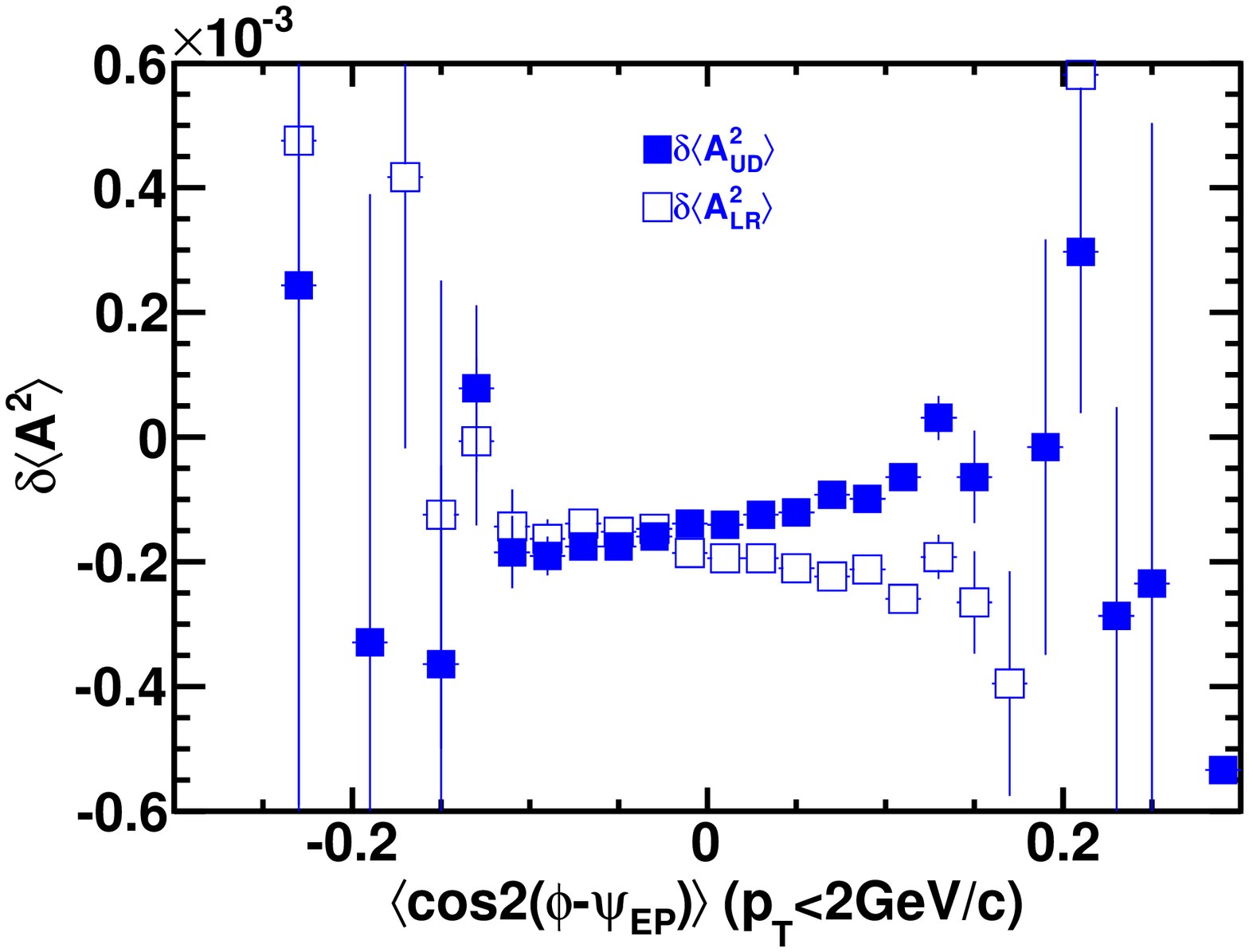}}
		\subfigure{\label{fig:appasymv220-b} \includegraphics[width=0.45\textwidth]{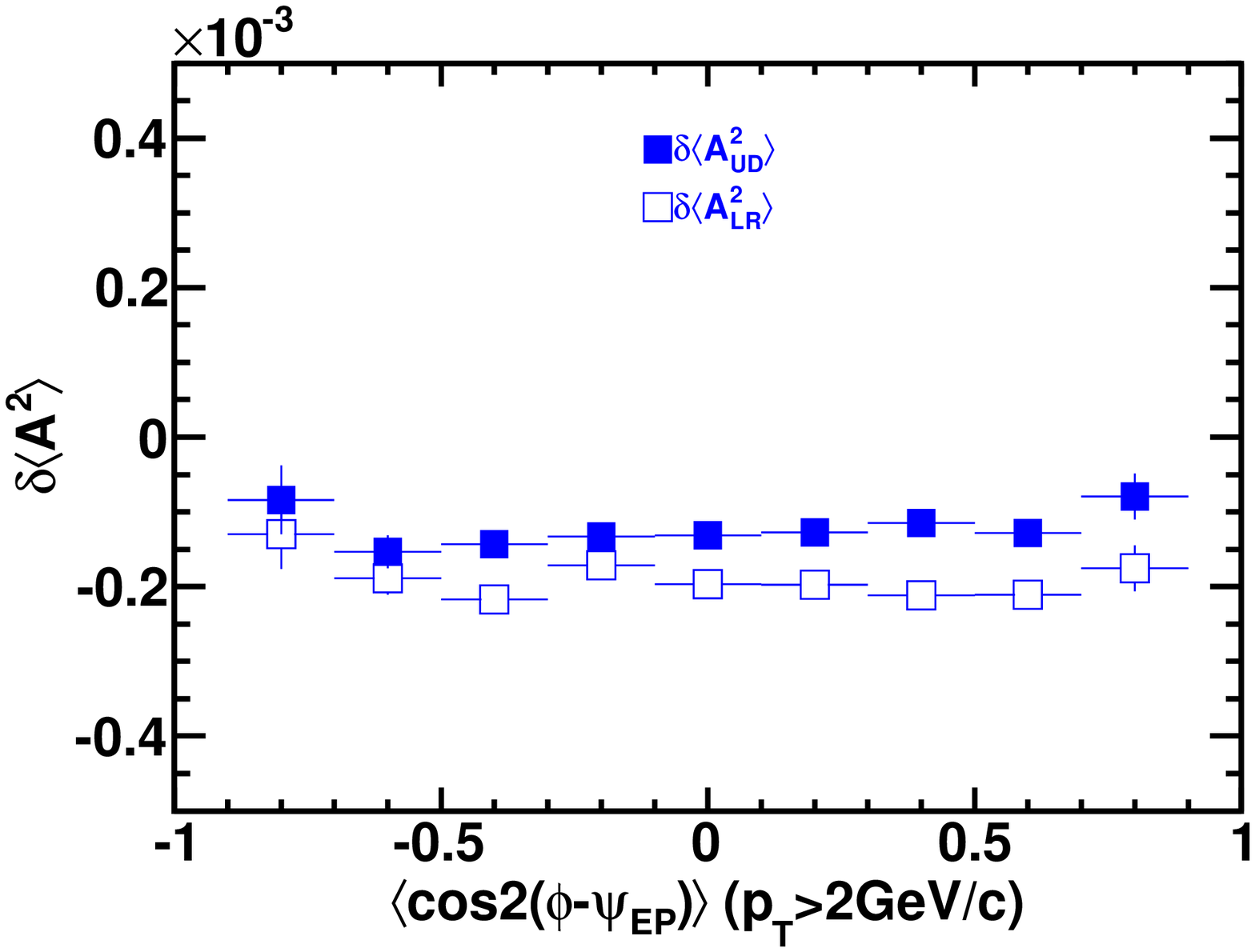}}
		\subfigure{\label{fig:appasymv220-c} \includegraphics[width=0.45\textwidth]{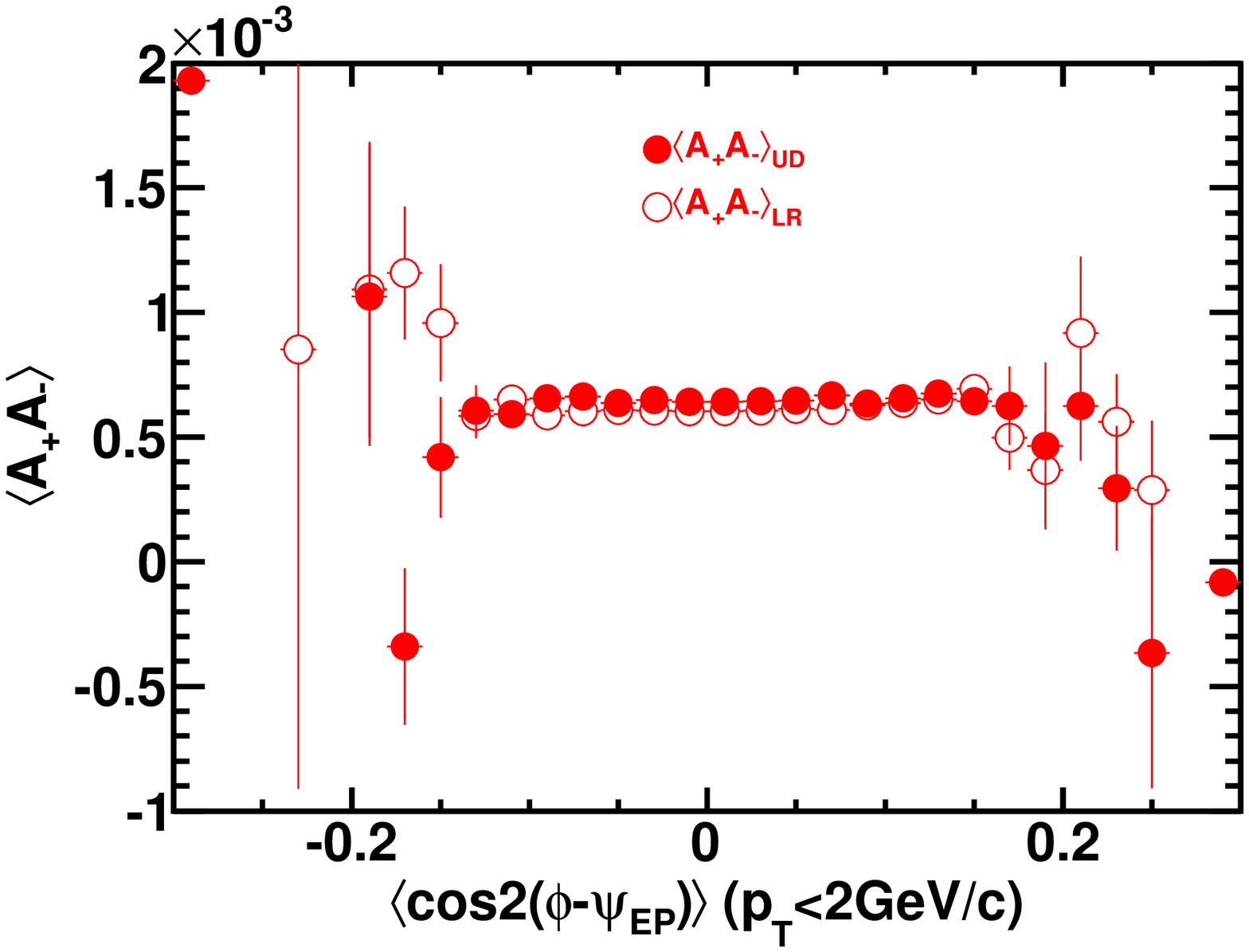}}
		\subfigure{\label{fig:appasymv220-d} \includegraphics[width=0.45\textwidth]{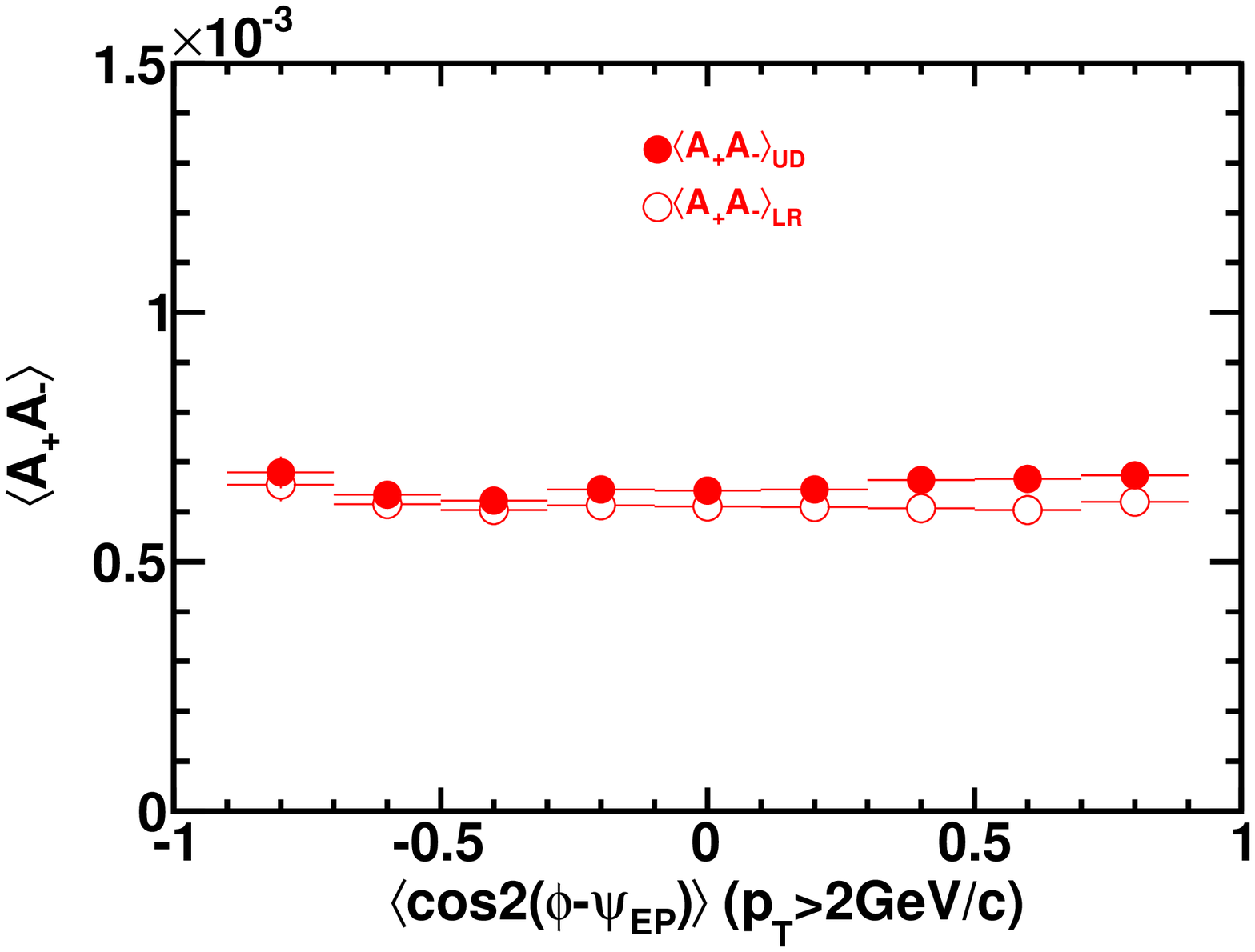}}
		\subfigure{\label{fig:appasymv2d20-a} \includegraphics[width=0.45\textwidth]{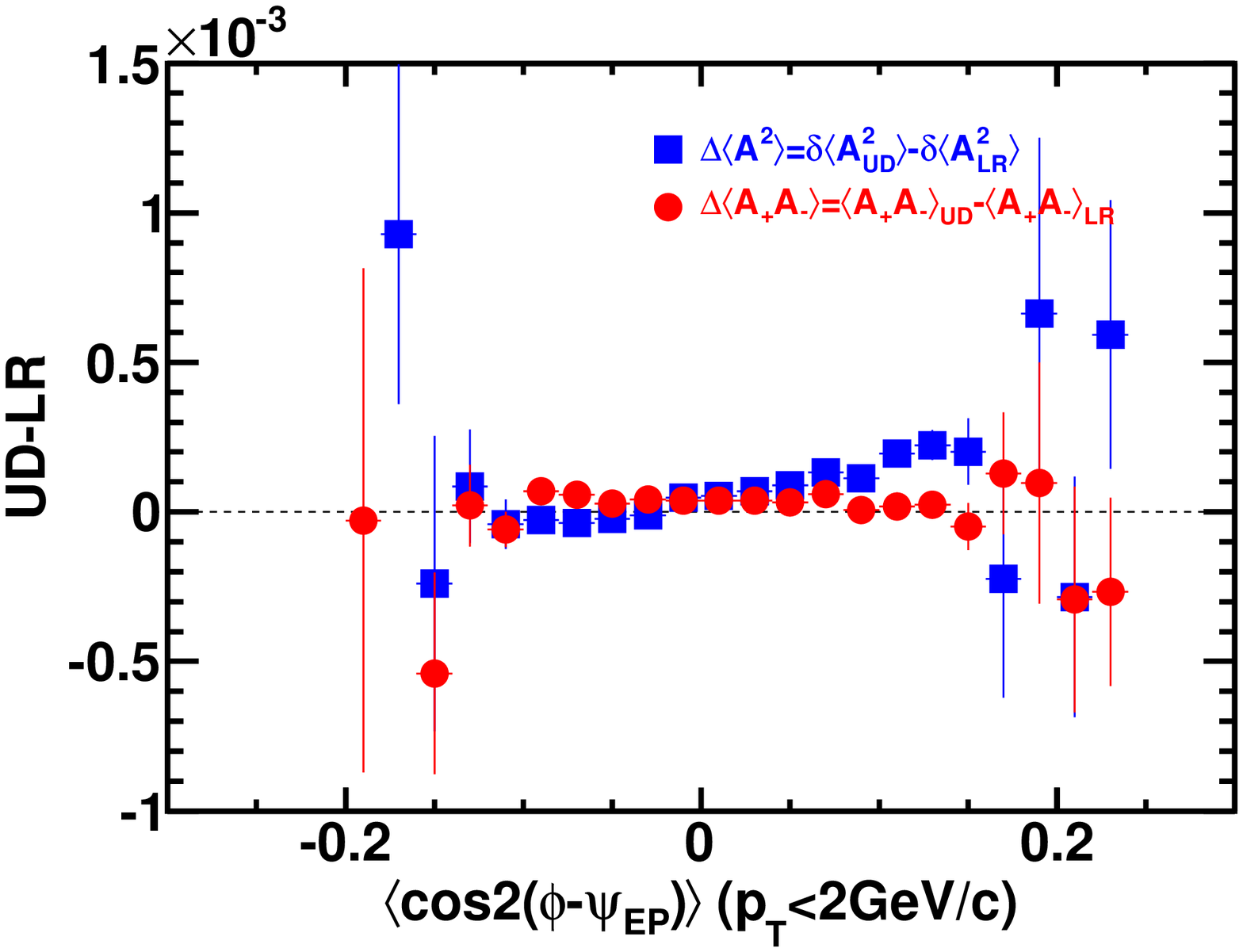}}
		\subfigure{\label{fig:appasymv2d20-b} \includegraphics[width=0.45\textwidth]{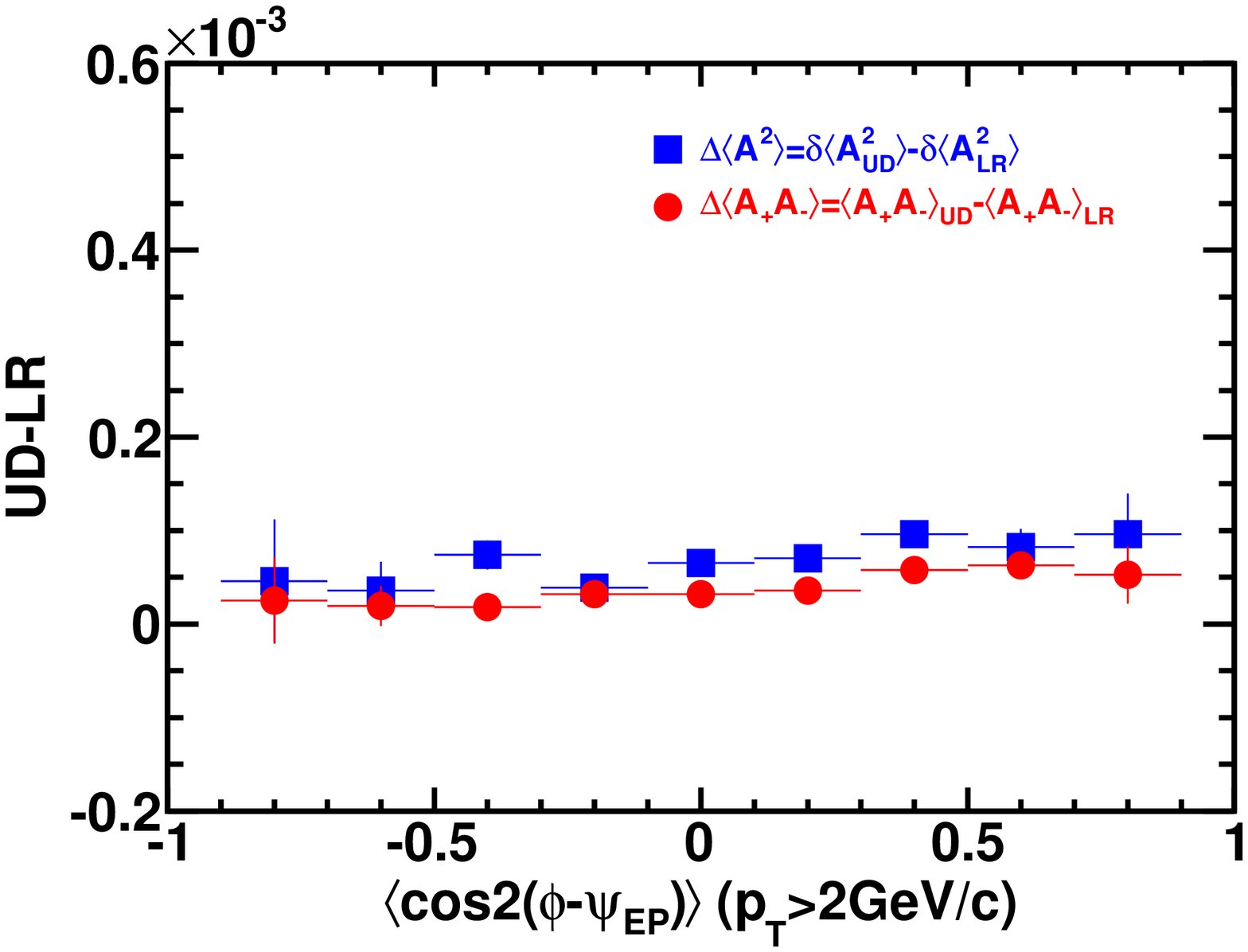}}
	\end{center}
	\caption[Central asymmetry correlations vs event-by-event $v_{2}^{obs}$]{
	Asymmetry correlations vs event-by-event $v_{2}^{obs}$ of RUN IV 200 GeV Au+Au central 0-20\% collisions.
	Error bars are statistical only.
	}
	\label{fig:appasymv220}
\end{figure}

\begin{figure}[htb]
	\begin{center}
		\subfigure{\label{fig:appasymv280-a} \includegraphics[width=0.45\textwidth]{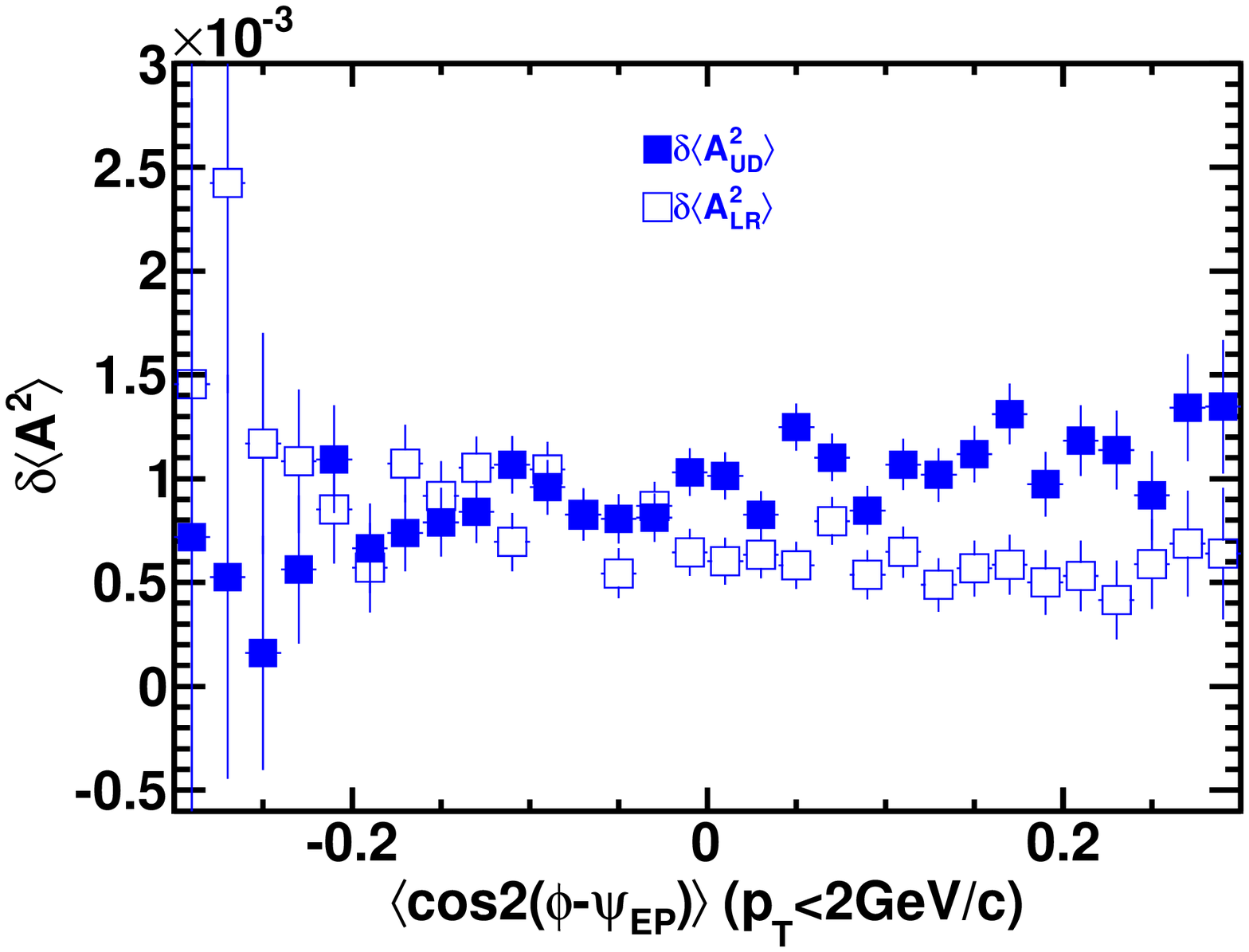}}
		\subfigure{\label{fig:appasymv280-b} \includegraphics[width=0.45\textwidth]{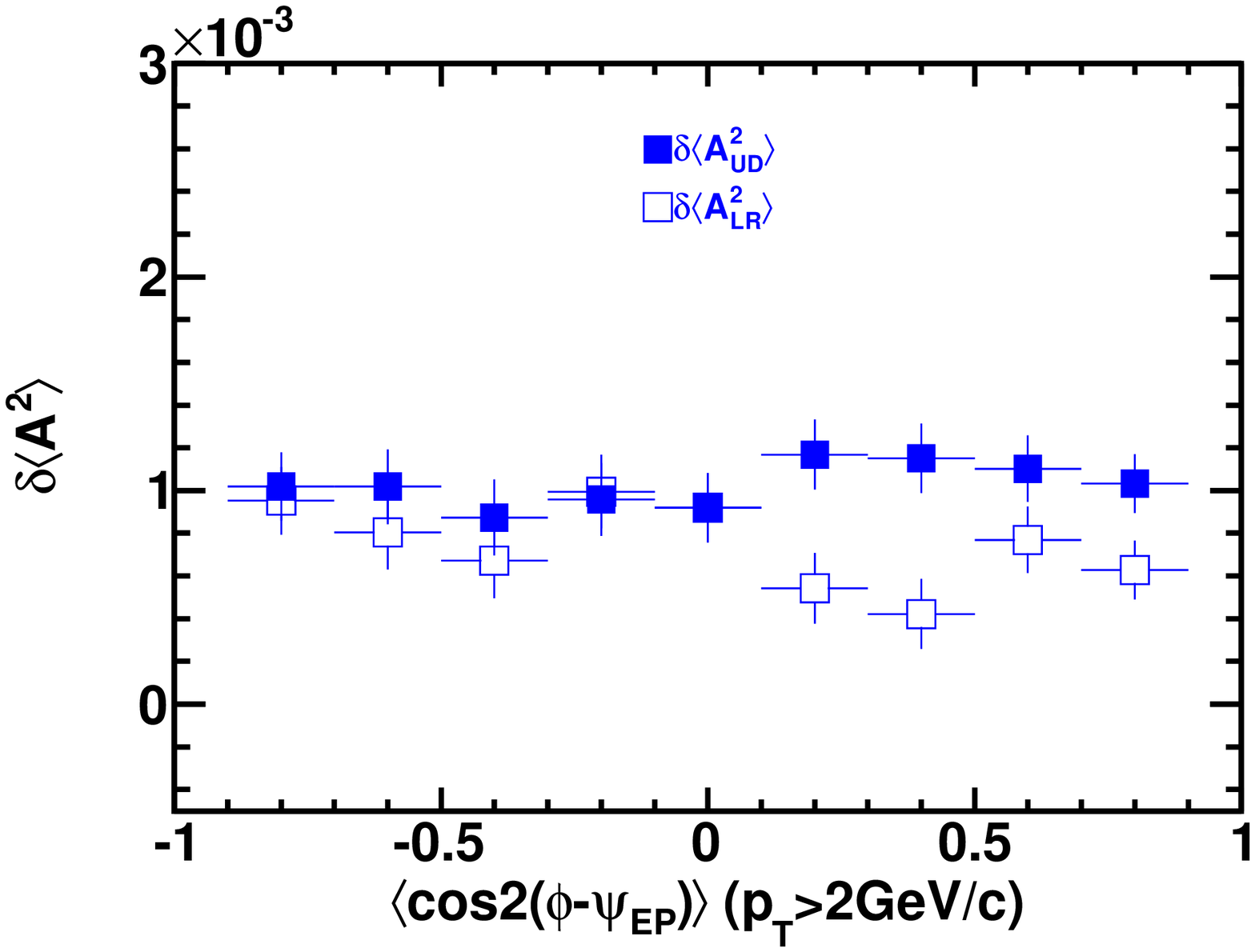}}
		\subfigure{\label{fig:appasymv280-c} \includegraphics[width=0.45\textwidth]{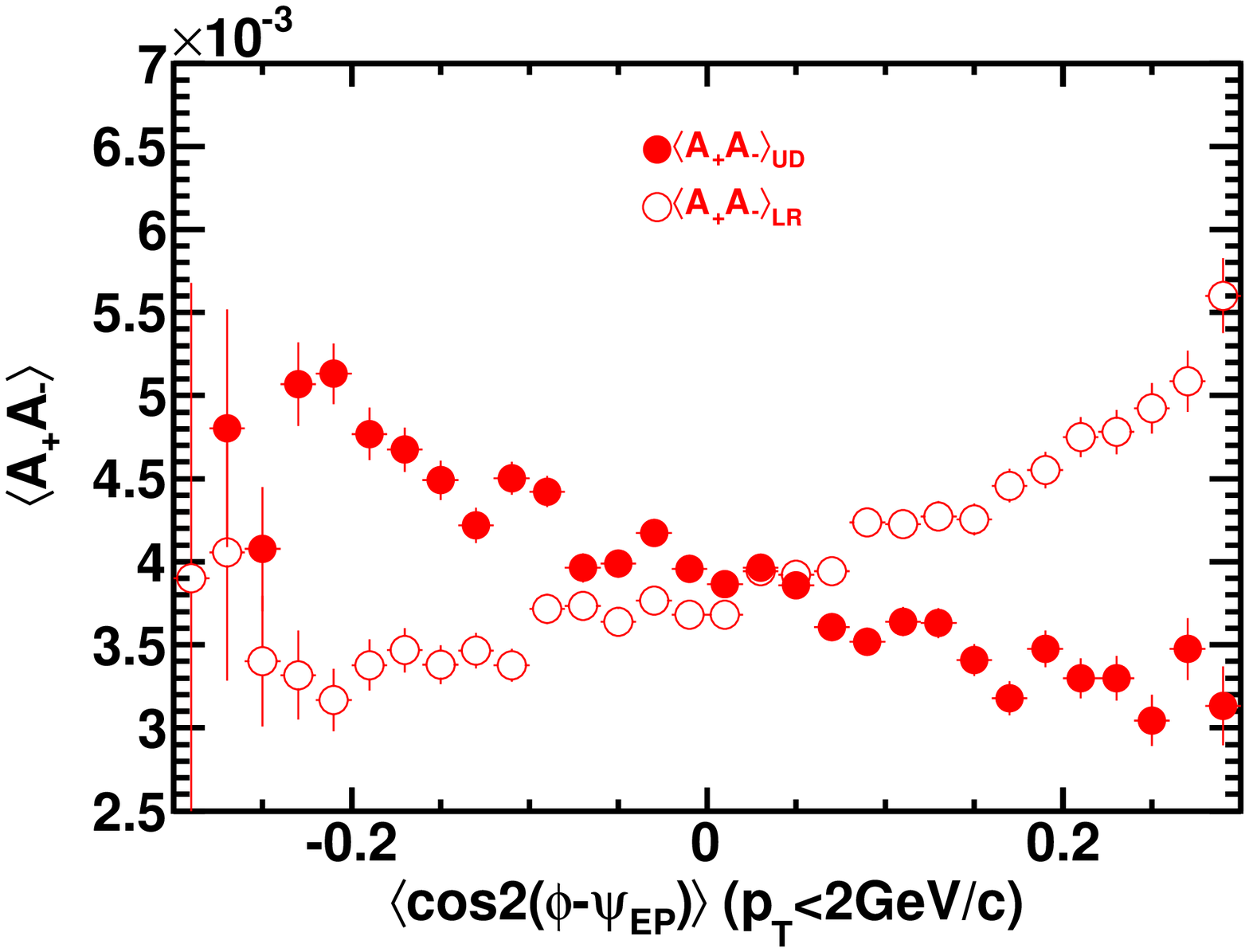}}
		\subfigure{\label{fig:appasymv280-d} \includegraphics[width=0.45\textwidth]{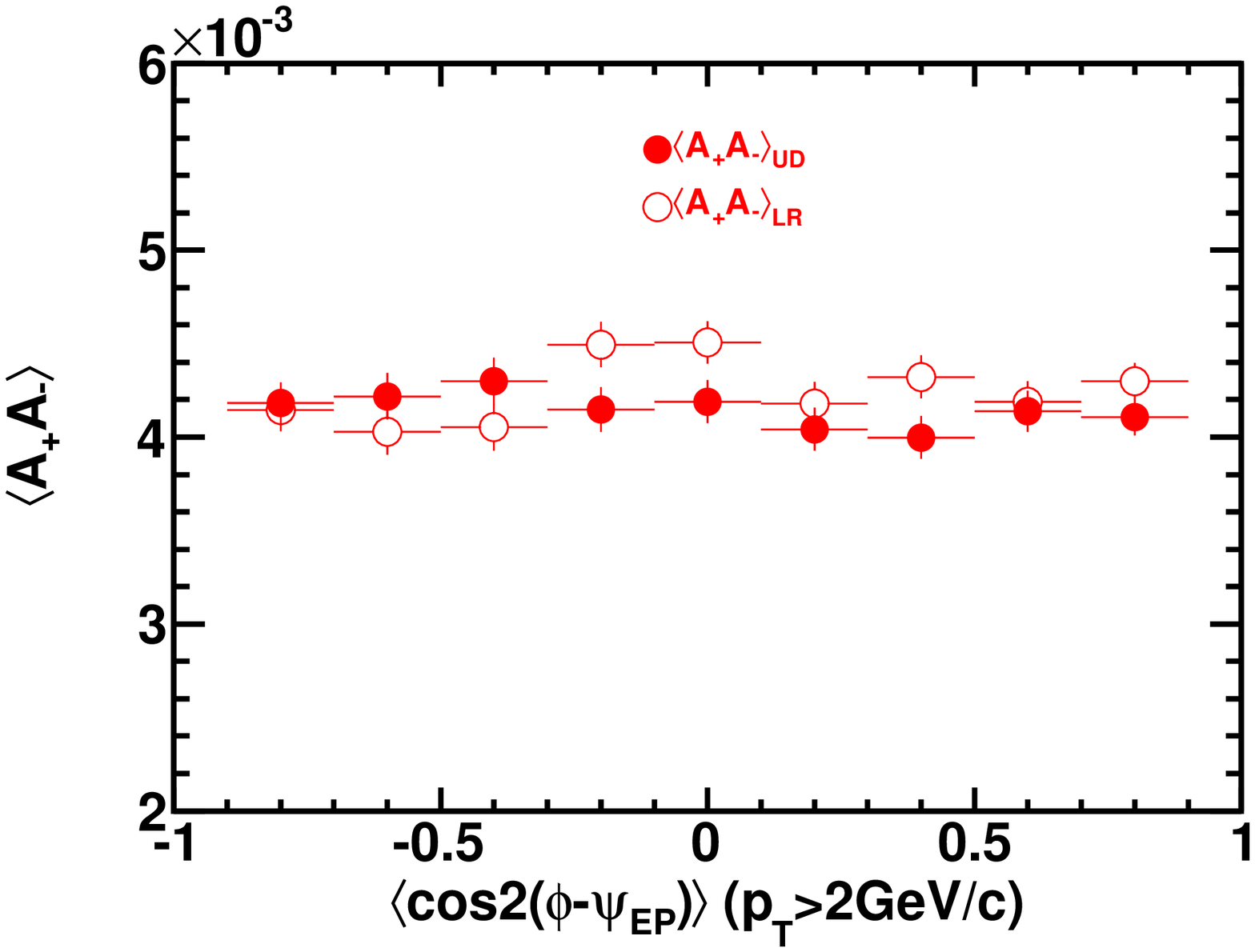}}
		\subfigure{\label{fig:appasymv2d80-a} \includegraphics[width=0.45\textwidth]{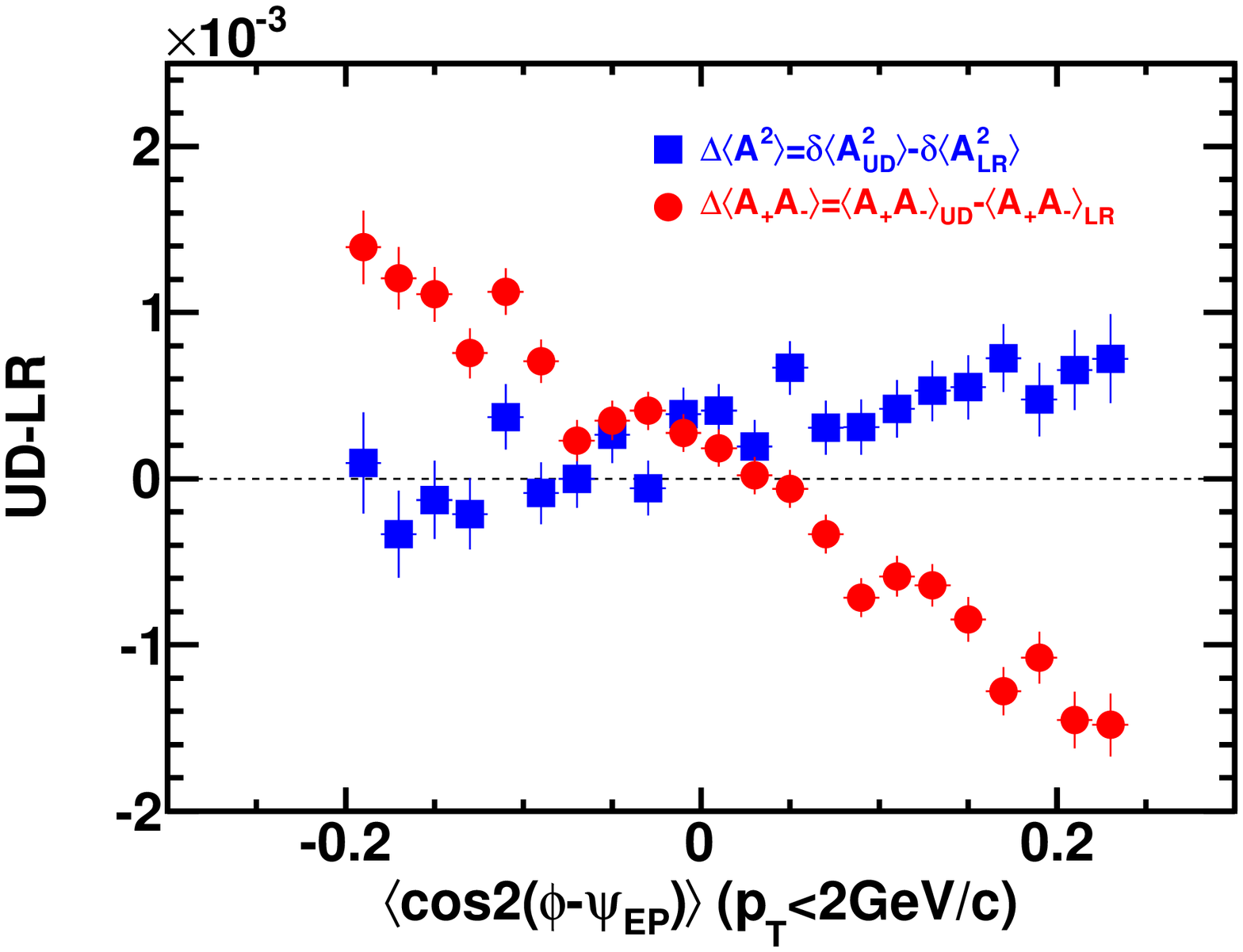}}
		\subfigure{\label{fig:appasymv2d80-b} \includegraphics[width=0.45\textwidth]{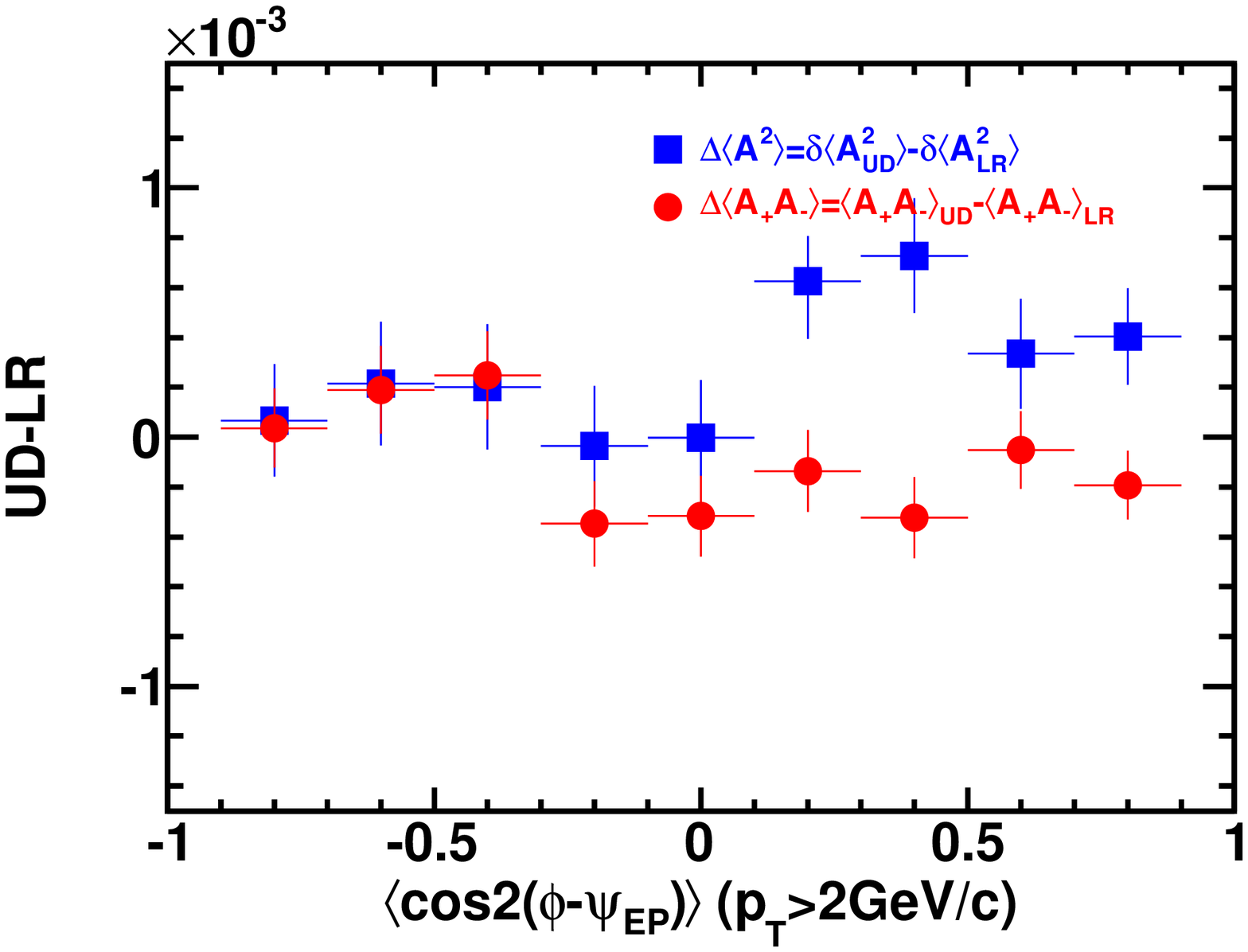}}
	\end{center}
	\caption[Peripheral asymmetry correlations vs event-by-event $v_{2}^{obs}$]{
	Asymmetry correlations vs event-by-event $v_{2}^{obs}$ of RUN IV 200 GeV Au+Au peripheral 40-80\% collisions.
	Error bars are statistical only.
	}
	\label{fig:appasymv280}
\end{figure}

\begin{figure}[htb]
	\begin{center}
		\subfigure{\label{fig:appasymv2etagap20-a} \includegraphics[width=0.45\textwidth]{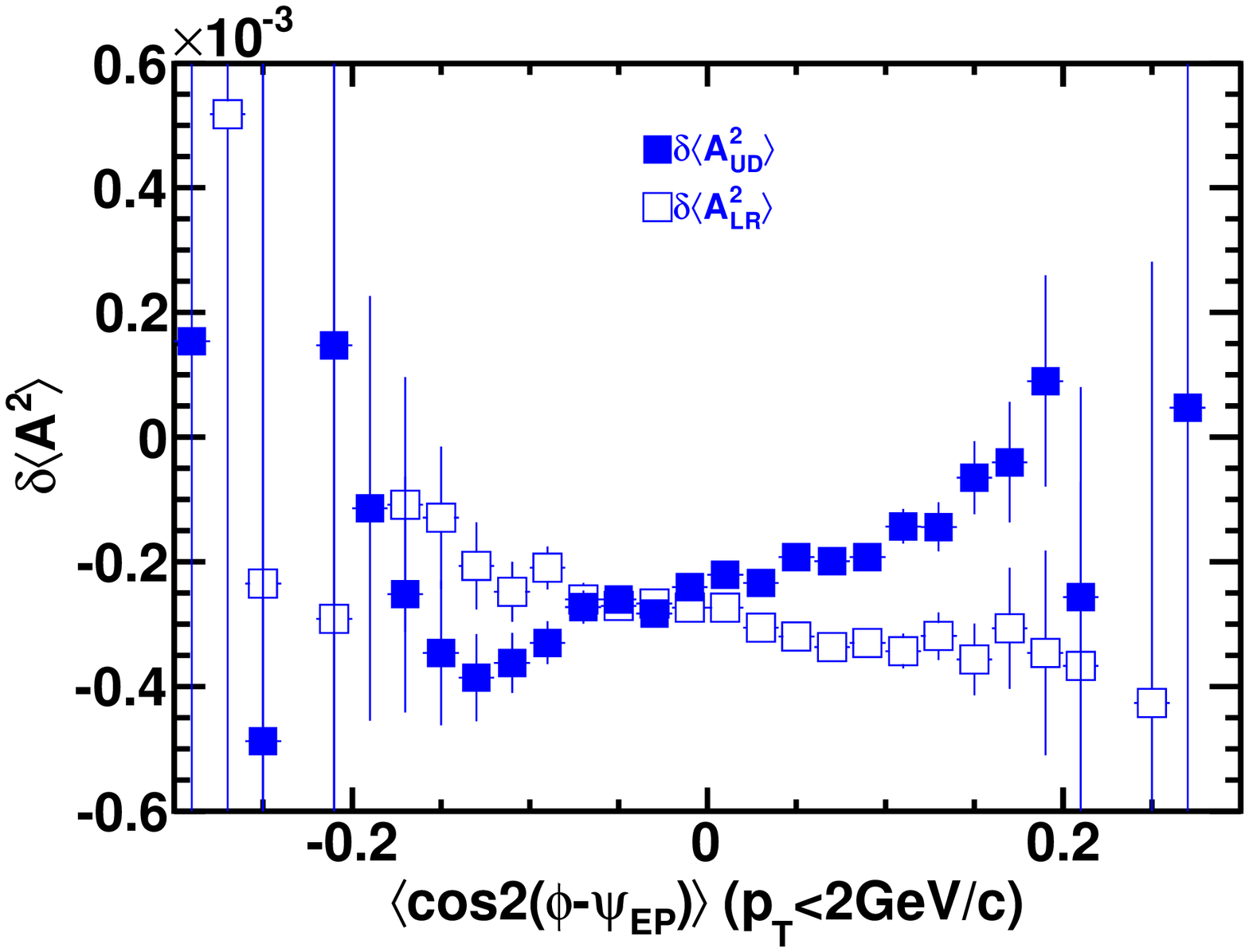}}
		\subfigure{\label{fig:appasymv2etagap20-b} \includegraphics[width=0.45\textwidth]{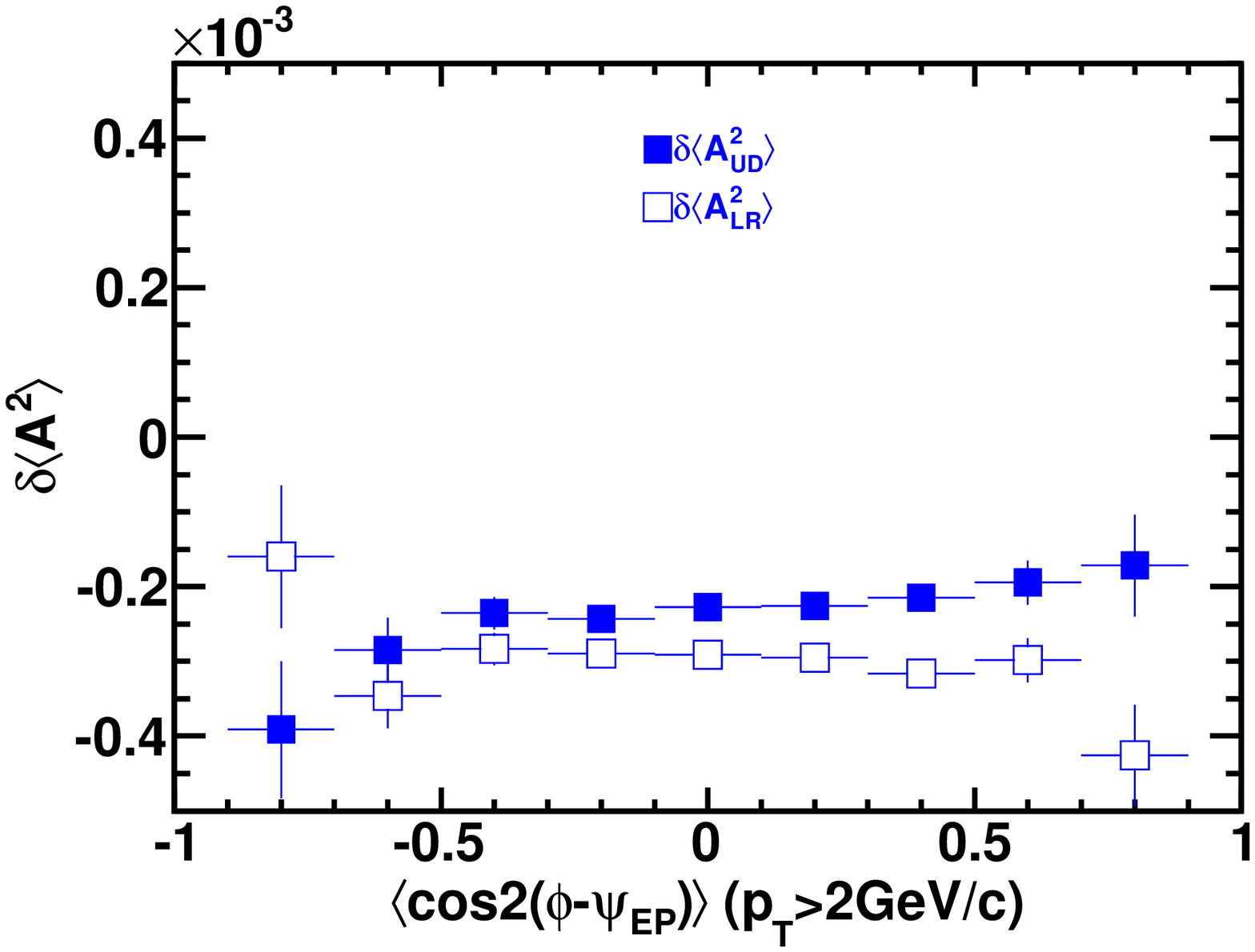}}
		\subfigure{\label{fig:appasymv2etagap20-c} \includegraphics[width=0.45\textwidth]{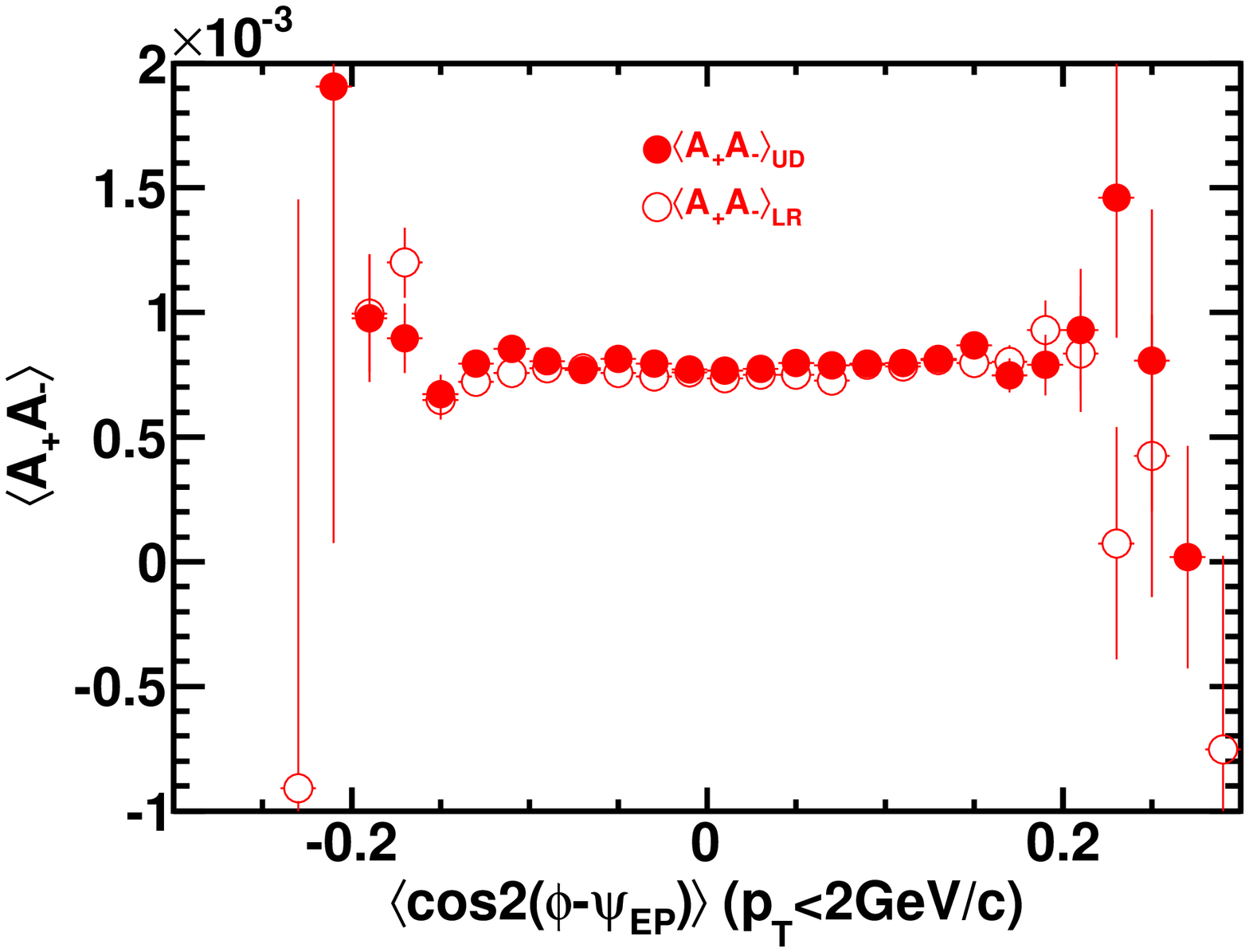}}
		\subfigure{\label{fig:appasymv2etagap20-d} \includegraphics[width=0.45\textwidth]{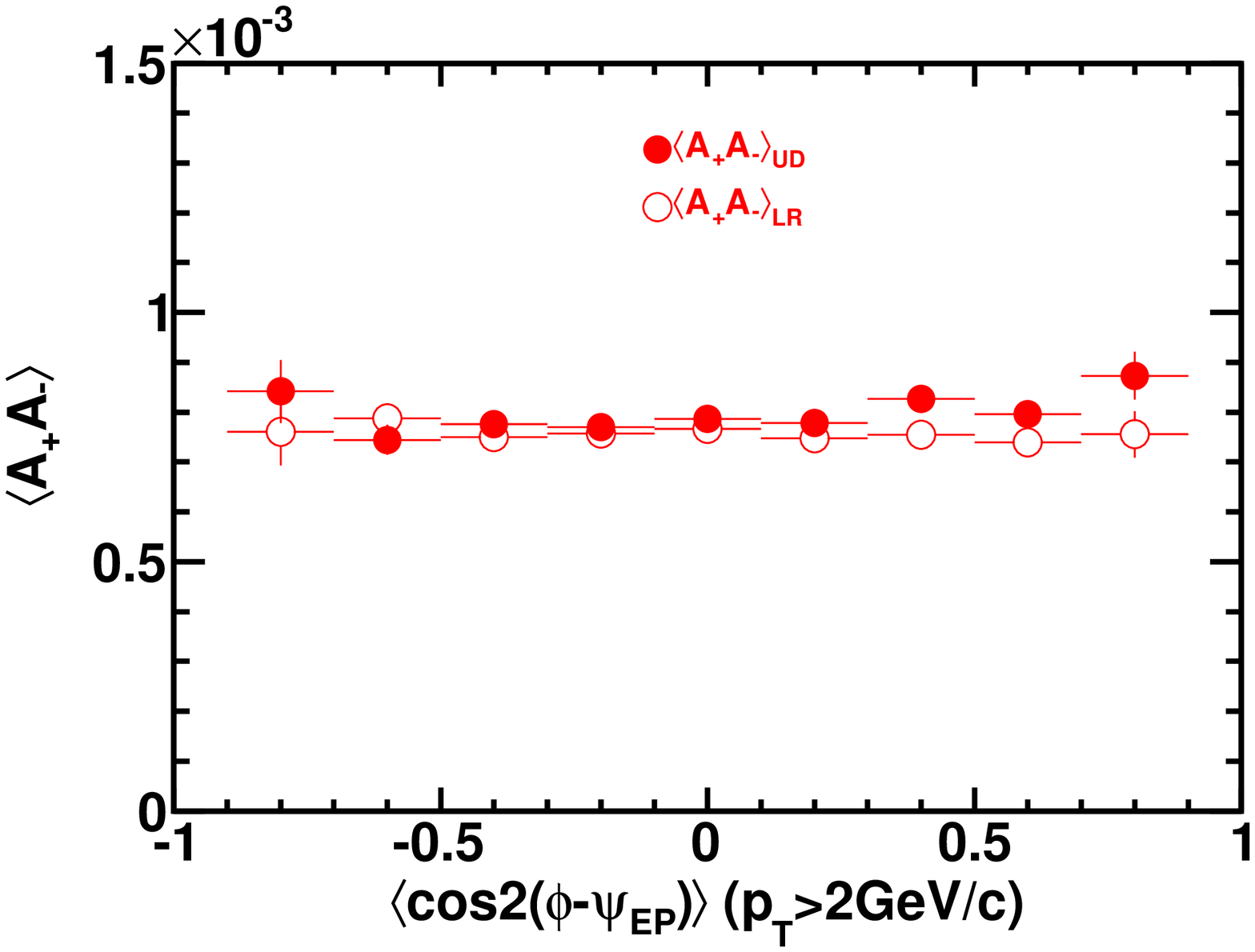}}
		\subfigure{\label{fig:appasymv2etagapd20-a} \includegraphics[width=0.45\textwidth]{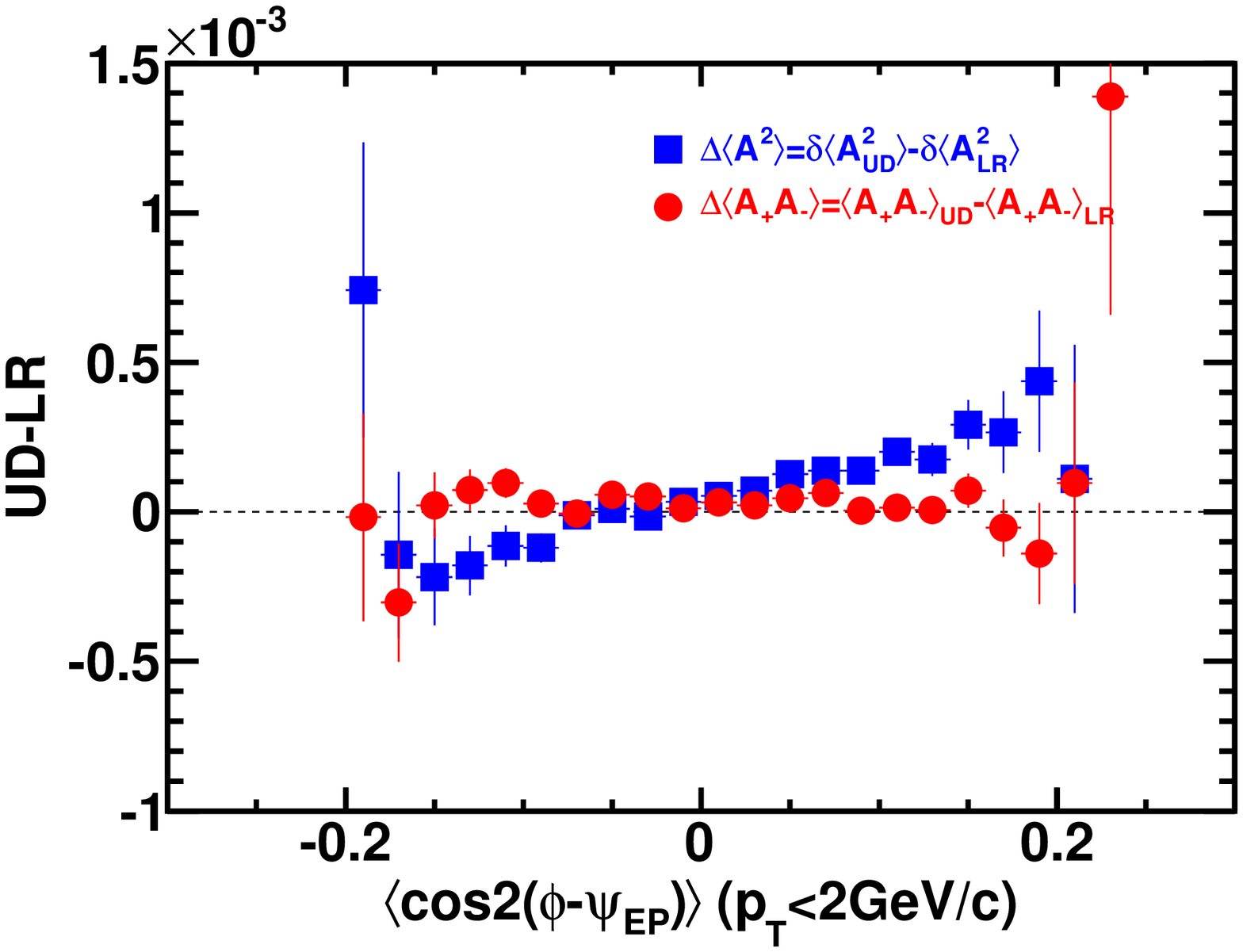}}
		\subfigure{\label{fig:appasymv2etagapd20-b} \includegraphics[width=0.45\textwidth]{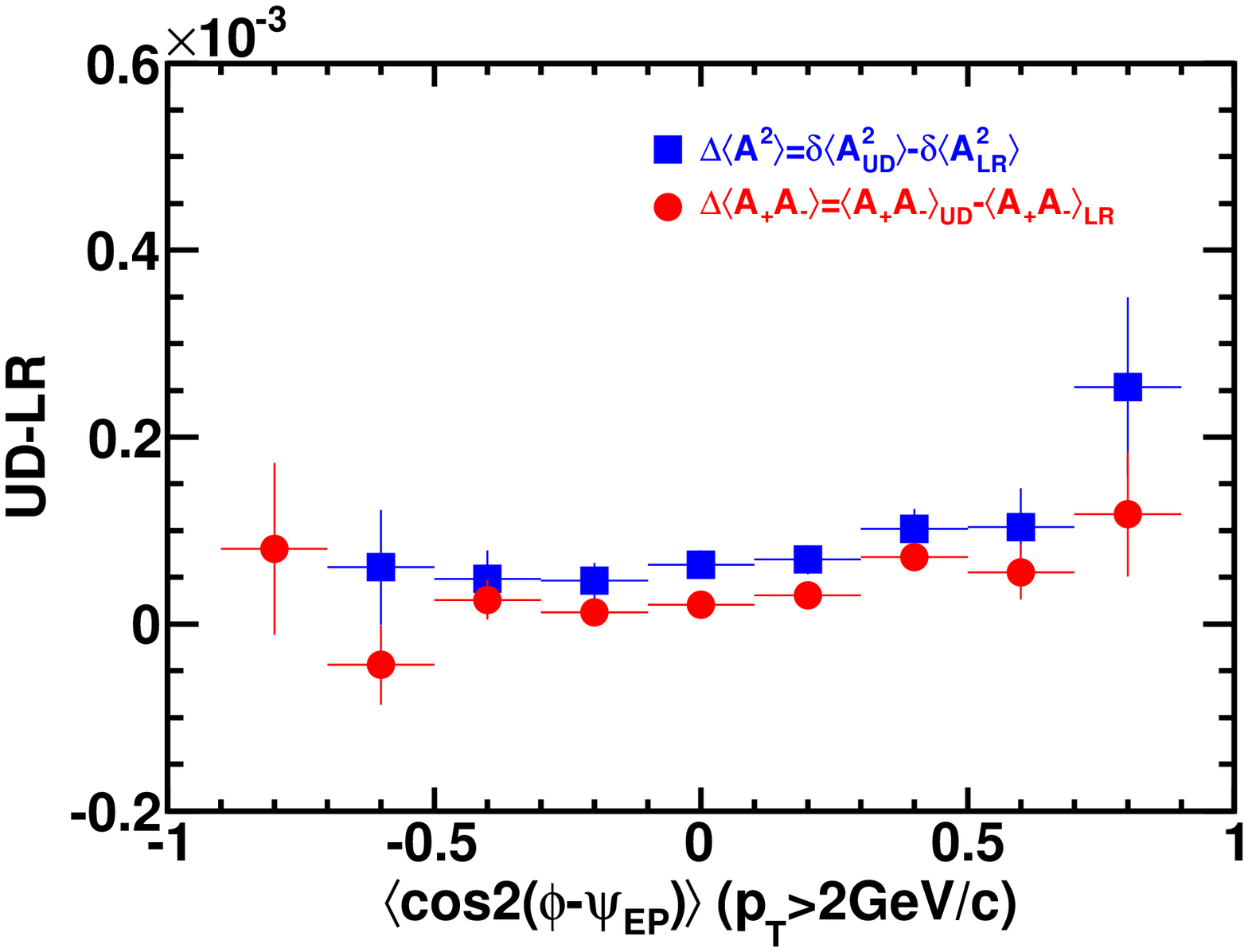}}
	\end{center}
	\caption[Central asymmetry correlations with $\eta$ gap vs $v_{2}^{obs}$]{
	Asymmetry correlations vs event-by-event $v_{2}^{obs}$ of RUN IV 200 GeV Au+Au central 0-20\% collisions.
	The particles used for asymmetry calculation and event-plane reconstruction are divided by pseudo-rapidity $-1.0<\eta<-0.5$ and $0.5<\eta<1.0$.
	Error bars are statistical only.
	}
	\label{fig:appasymv2etagap20}
\end{figure}

\begin{figure}[htb]
	\begin{center}
		\subfigure{\label{fig:appasymv2etagap80-a} \includegraphics[width=0.45\textwidth]{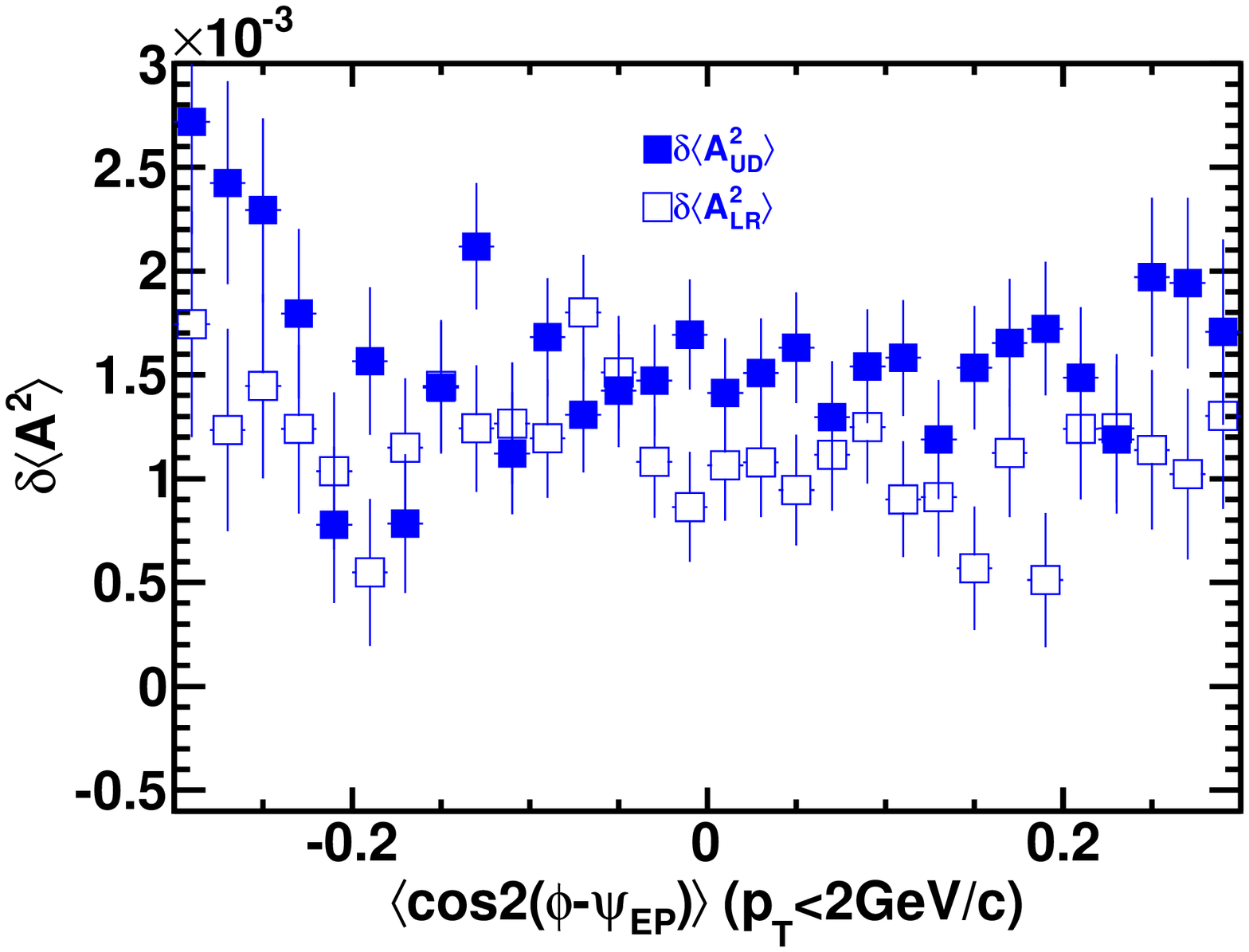}}
		\subfigure{\label{fig:appasymv2etagap80-b} \includegraphics[width=0.45\textwidth]{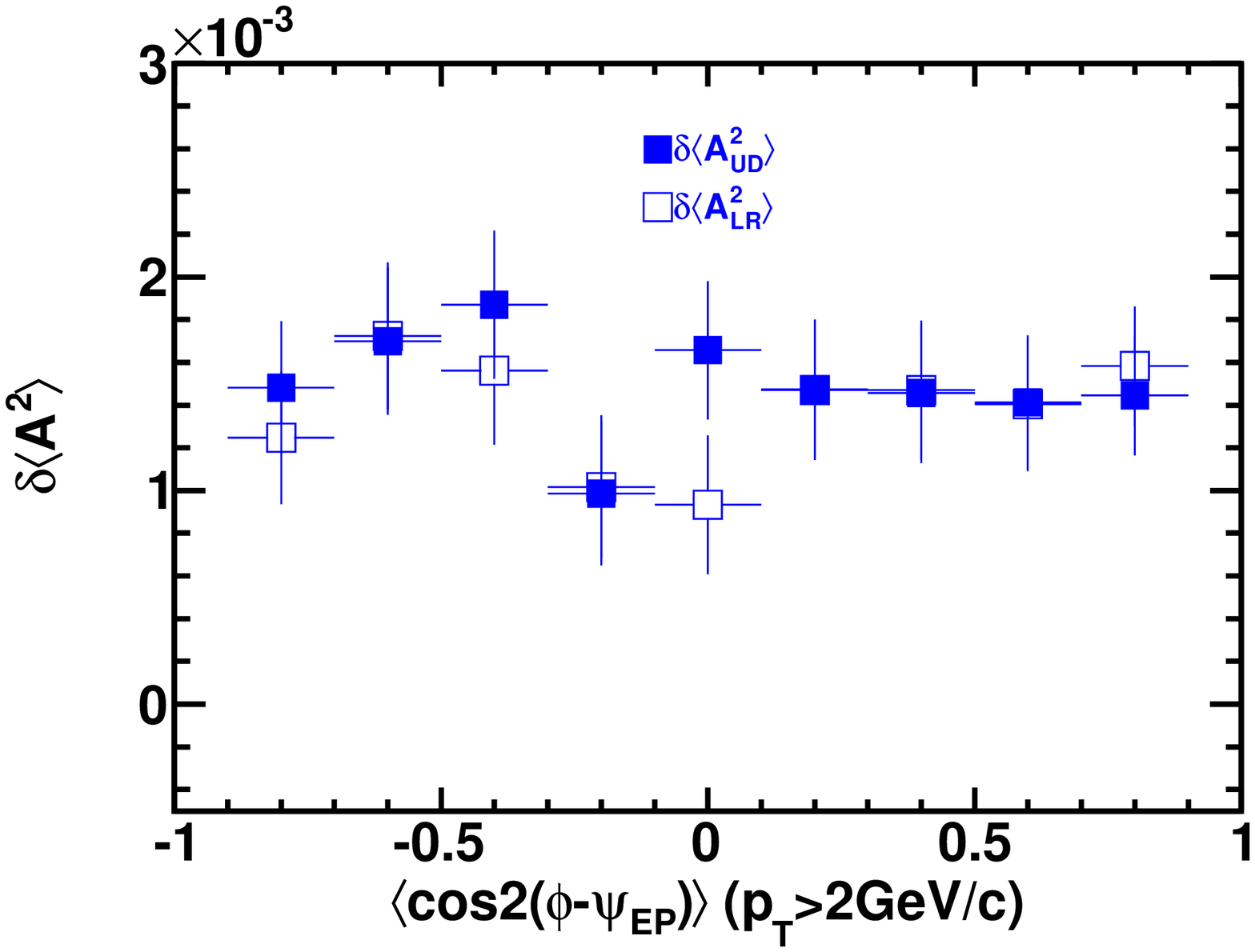}}
		\subfigure{\label{fig:appasymv2etagap80-c} \includegraphics[width=0.45\textwidth]{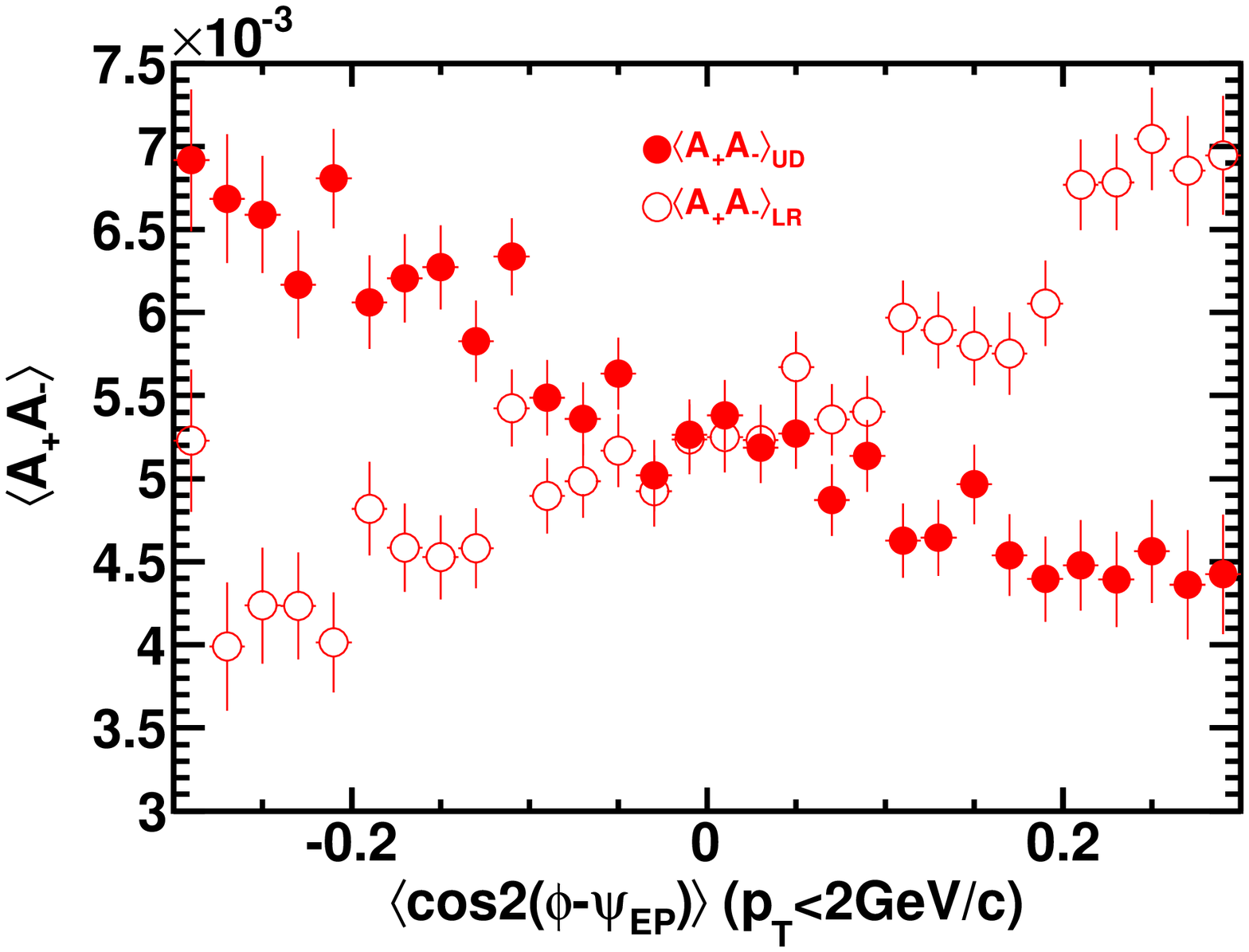}}
		\subfigure{\label{fig:appasymv2etagap80-d} \includegraphics[width=0.45\textwidth]{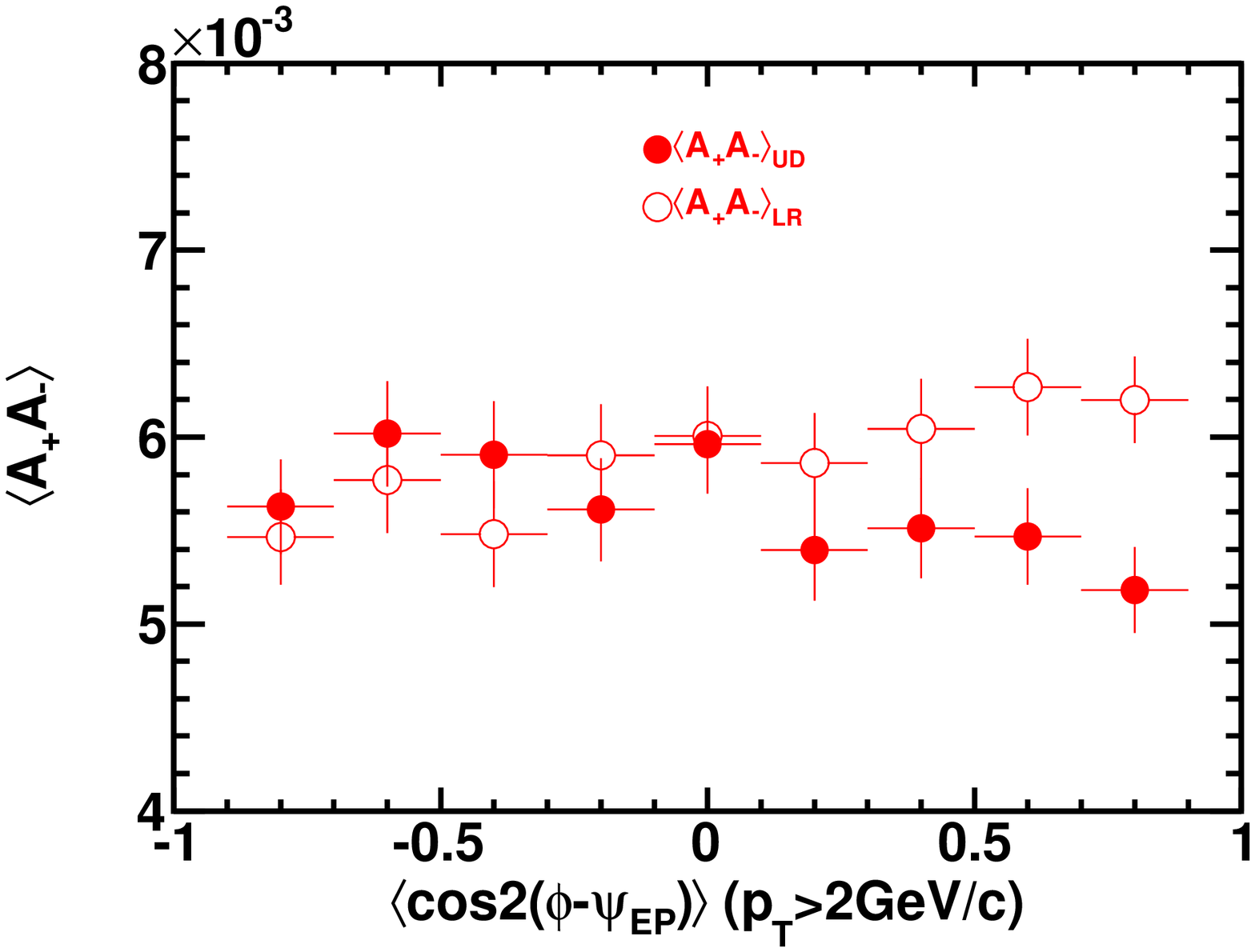}}
		\subfigure{\label{fig:appasymv2etagapd80-a} \includegraphics[width=0.45\textwidth]{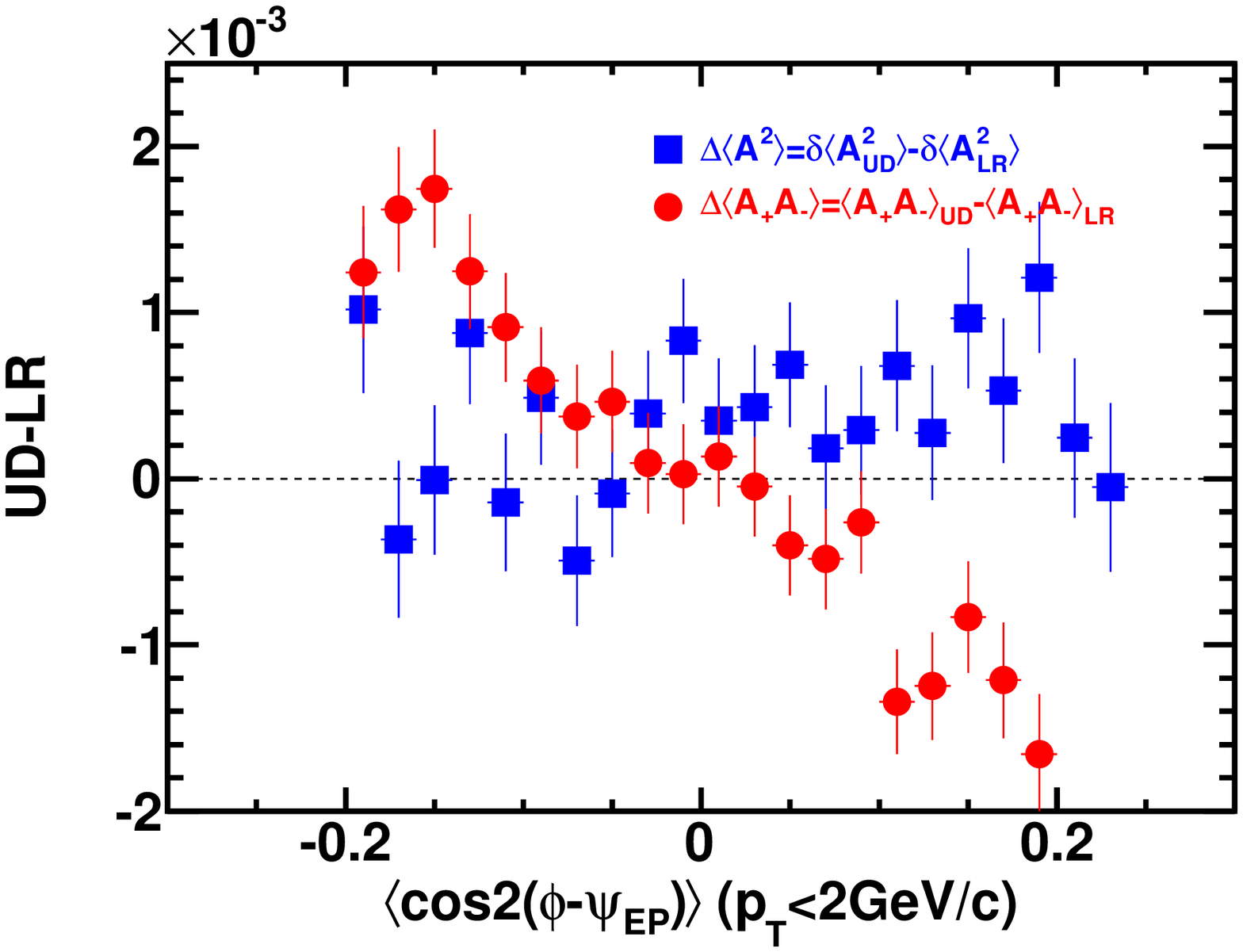}}
		\subfigure{\label{fig:appasymv2etagapd80-b} \includegraphics[width=0.45\textwidth]{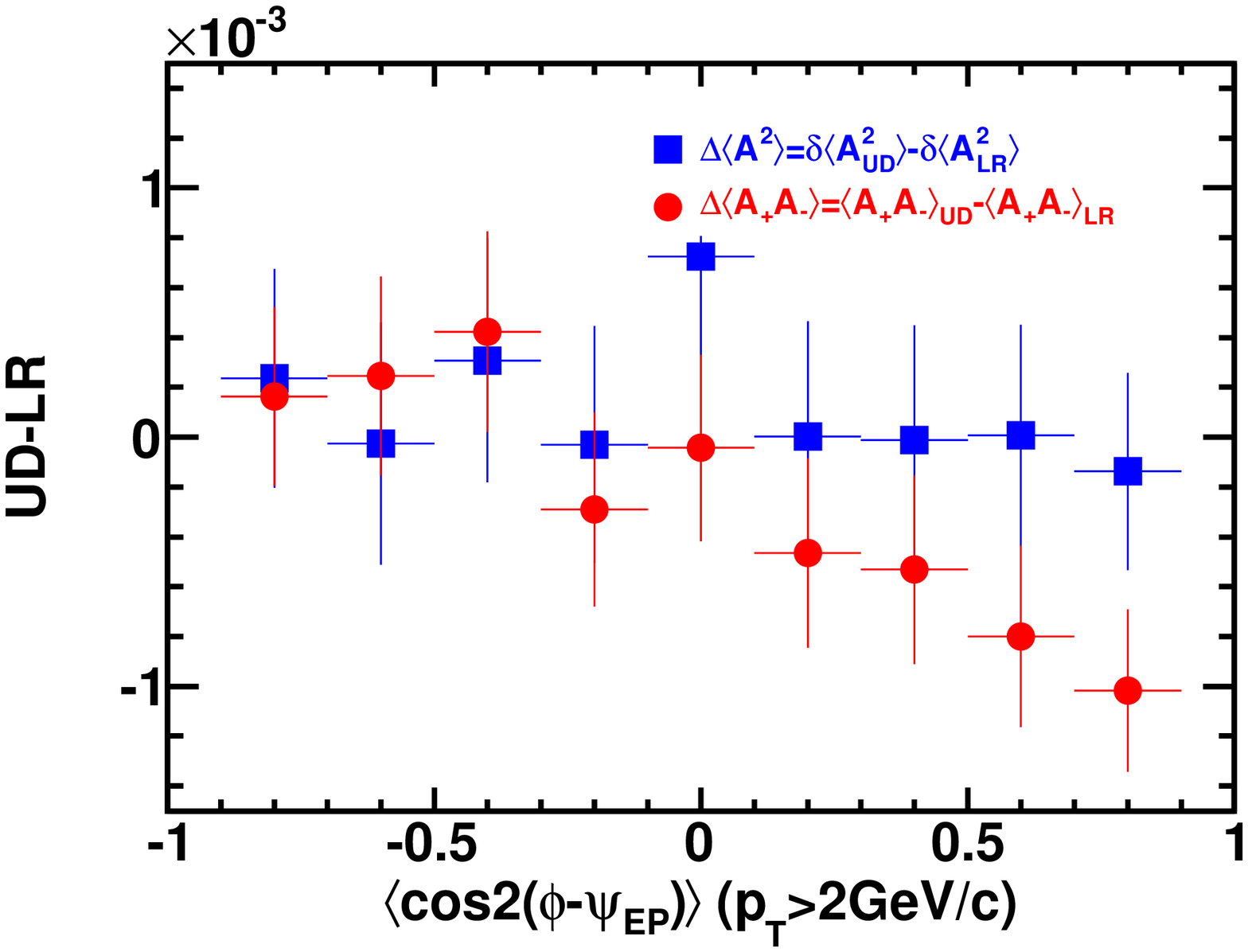}}
	\end{center}
	\caption[Peripheral asymmetry correlations with $\eta$ gap vs $v_{2}^{obs}$]{
	Asymmetry correlations vs event-by-event $v_{2}^{obs}$ of RUN IV 200 GeV Au+Au peripheral 40-80\% collisions.
	The particles used for asymmetry calculation and event-plane reconstruction are divided by pseudo-rapidity $-1.0<\eta<-0.5$ and $0.5<\eta<1.0$.
	Error bars are statistical only.
	}
	\label{fig:appasymv2etagap80}
\end{figure}

\begin{figure}[htb]
	\begin{center}
		\subfigure{\label{fig:appasymv2zdc20-a} \includegraphics[width=0.45\textwidth]{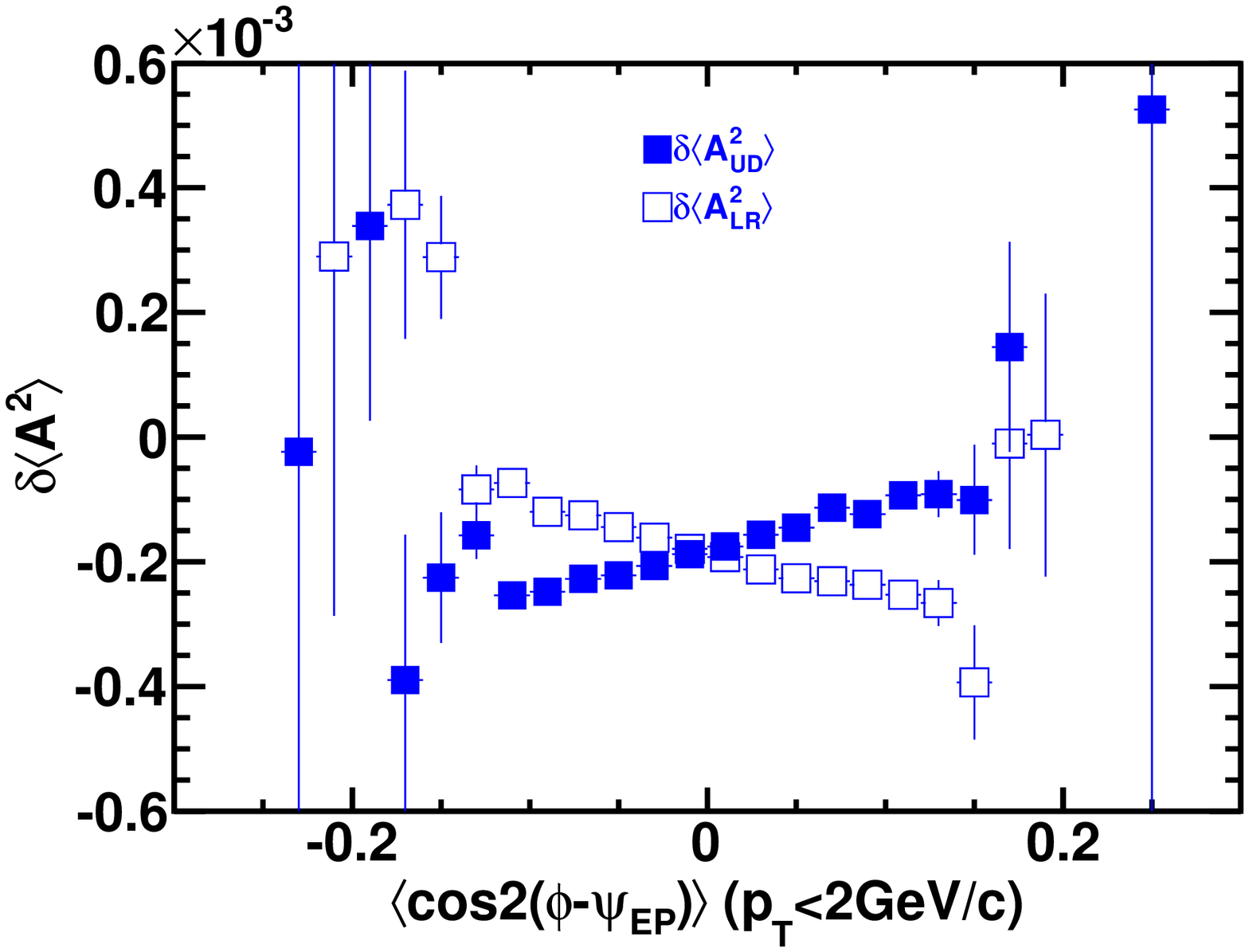}}
		\subfigure{\label{fig:appasymv2zdc20-b} \includegraphics[width=0.45\textwidth]{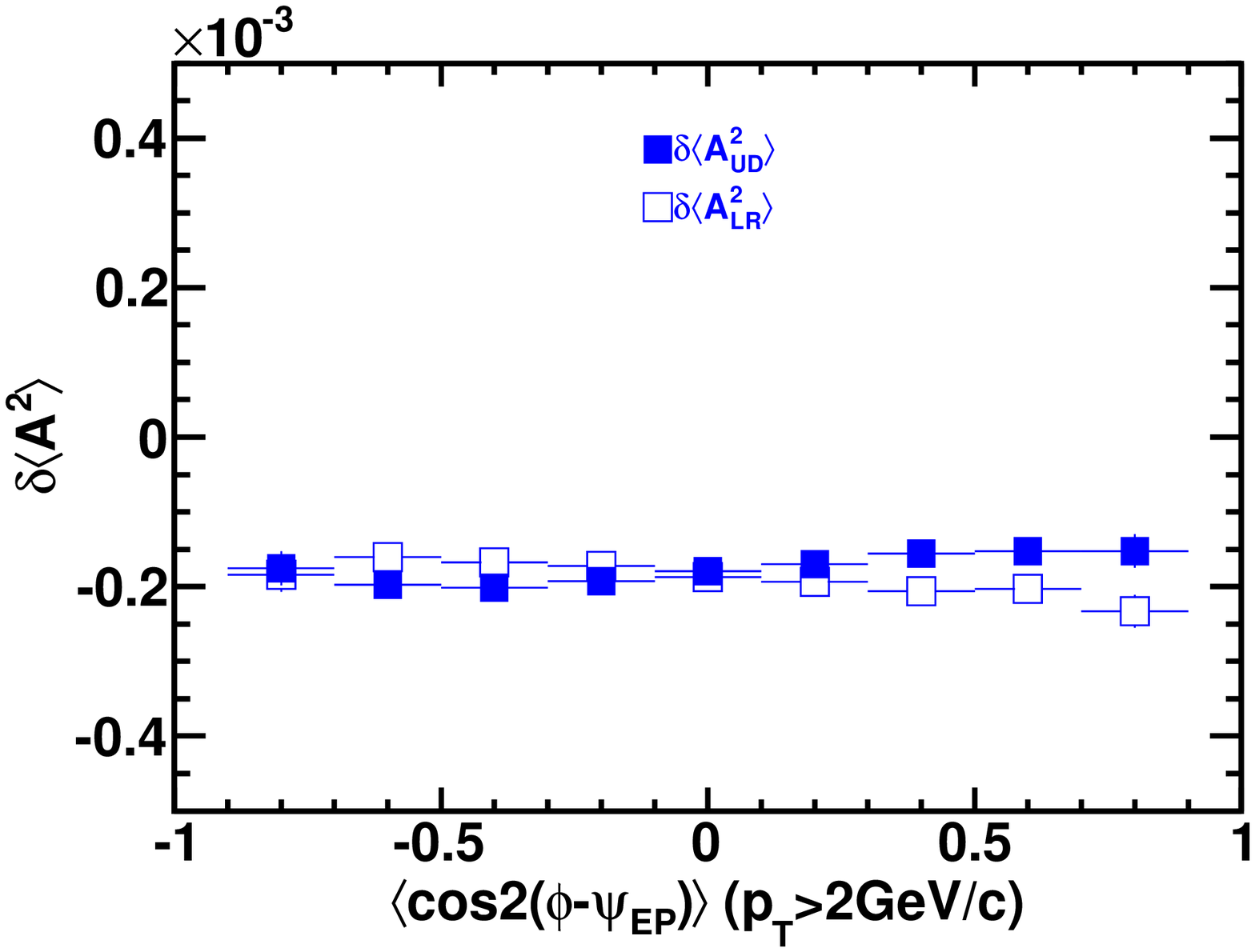}}
		\subfigure{\label{fig:appasymv2zdc20-c} \includegraphics[width=0.45\textwidth]{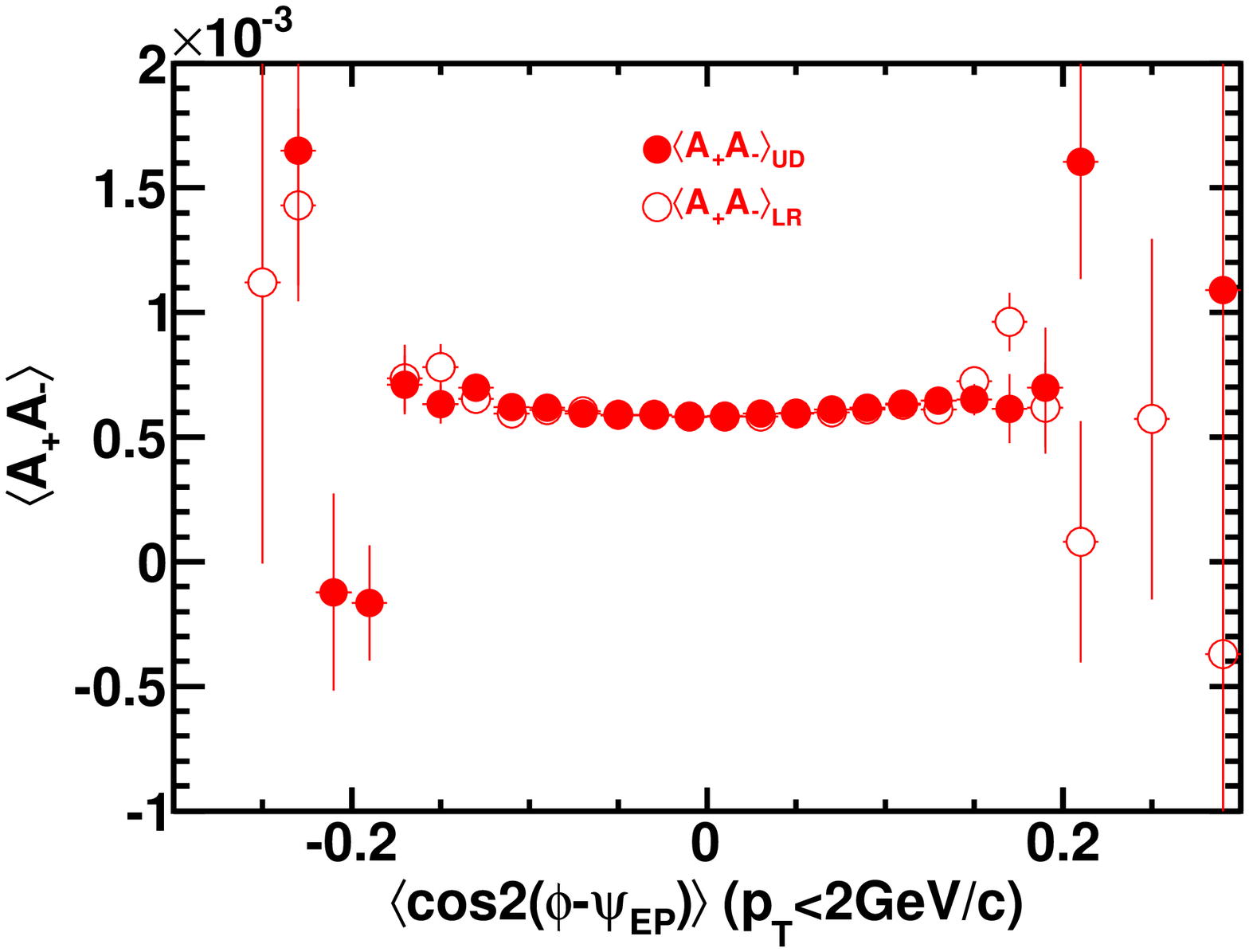}}
		\subfigure{\label{fig:appasymv2zdc20-d} \includegraphics[width=0.45\textwidth]{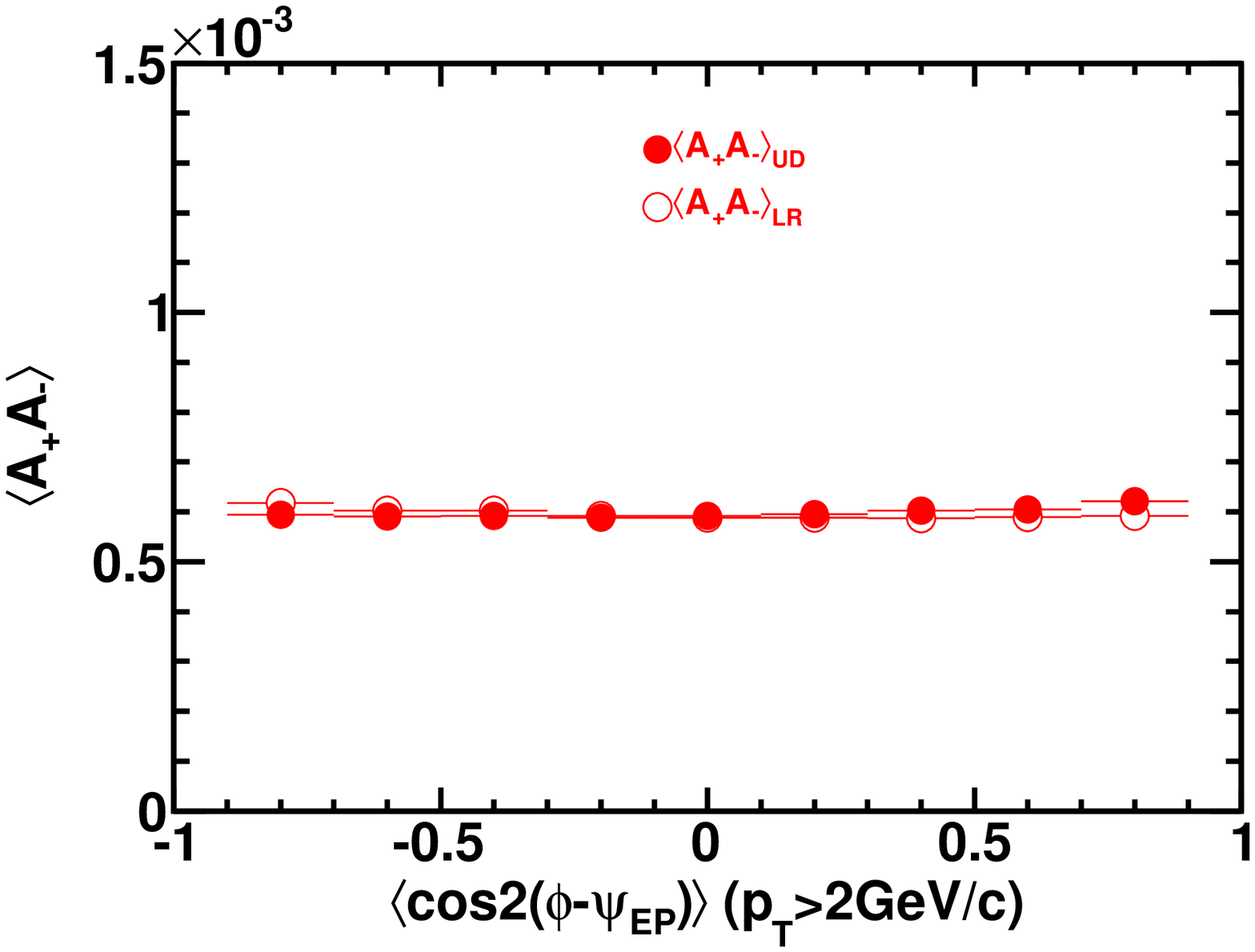}}
		\subfigure{\label{fig:appasymv2zdcd20-a} \includegraphics[width=0.45\textwidth]{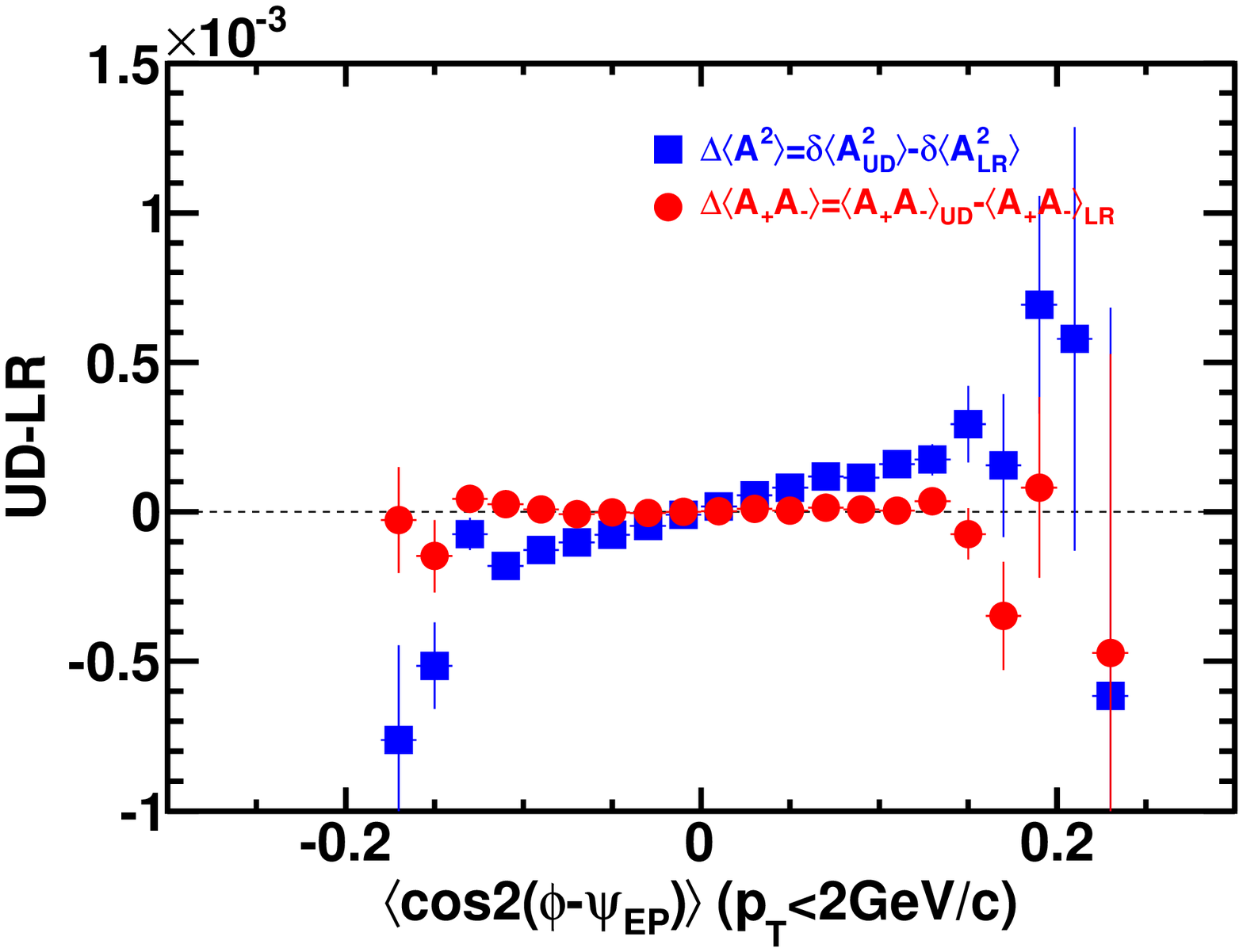}}
		\subfigure{\label{fig:appasymv2zdcd20-b} \includegraphics[width=0.45\textwidth]{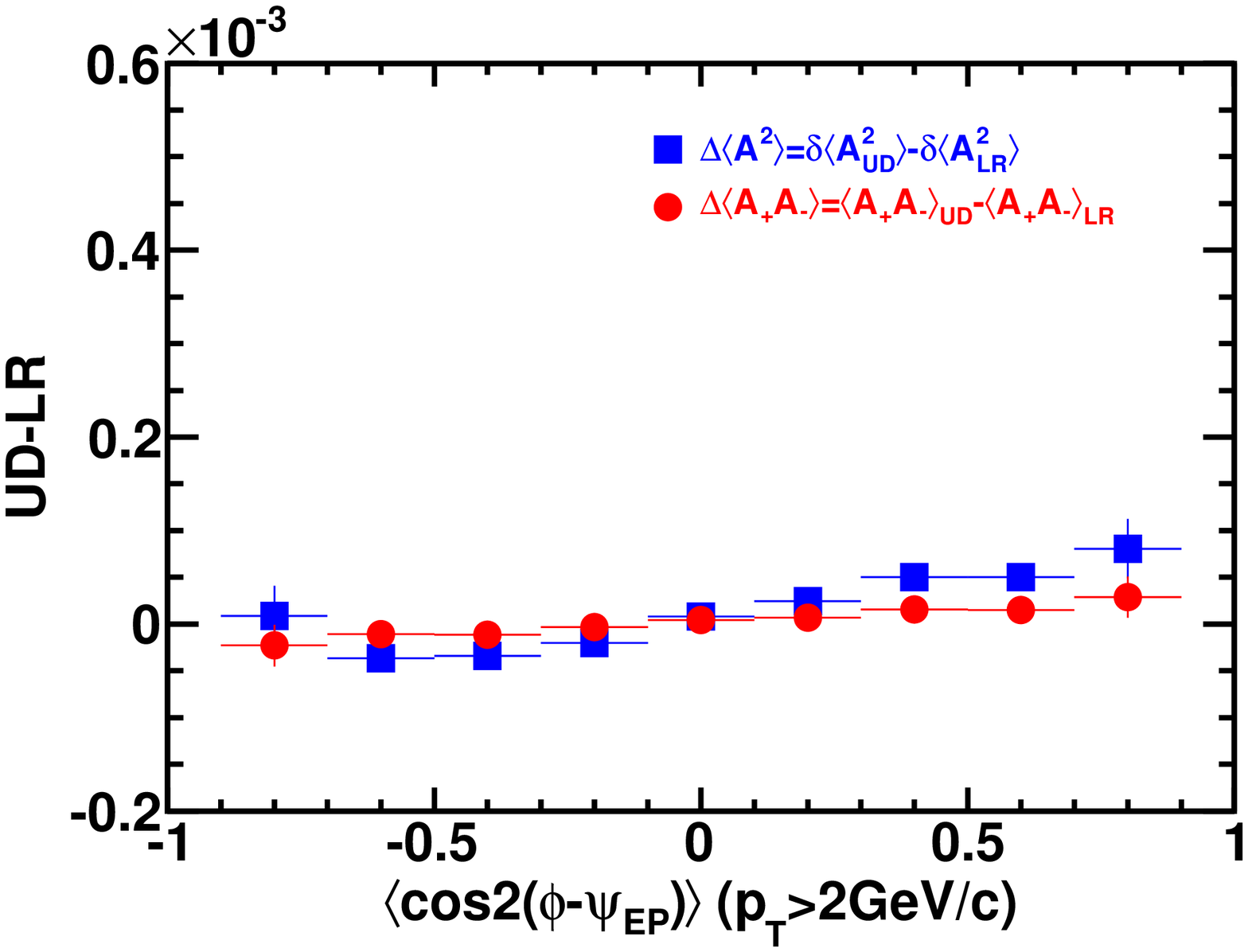}}
	\end{center}
	\caption[Central asymmetry correlations vs $v_{2}^{obs}$ with ZDC-SMD EP]{
	Asymmetry correlations vs event-by-event $v_{2}^{obs}$ of RUN VII 200 GeV Au+Au central 0-20\% collisions.
	The particles used for asymmetry calculation are from half side of the TPC with respect to the first order event-plane reconstructed from ZDC-SMD.
	Error bars are statistical only.
	}
	\label{fig:appasymv2zdc20}
\end{figure}

\begin{figure}[htb]
	\begin{center}
		\subfigure{\label{fig:appasymv2zdc80-a} \includegraphics[width=0.45\textwidth]{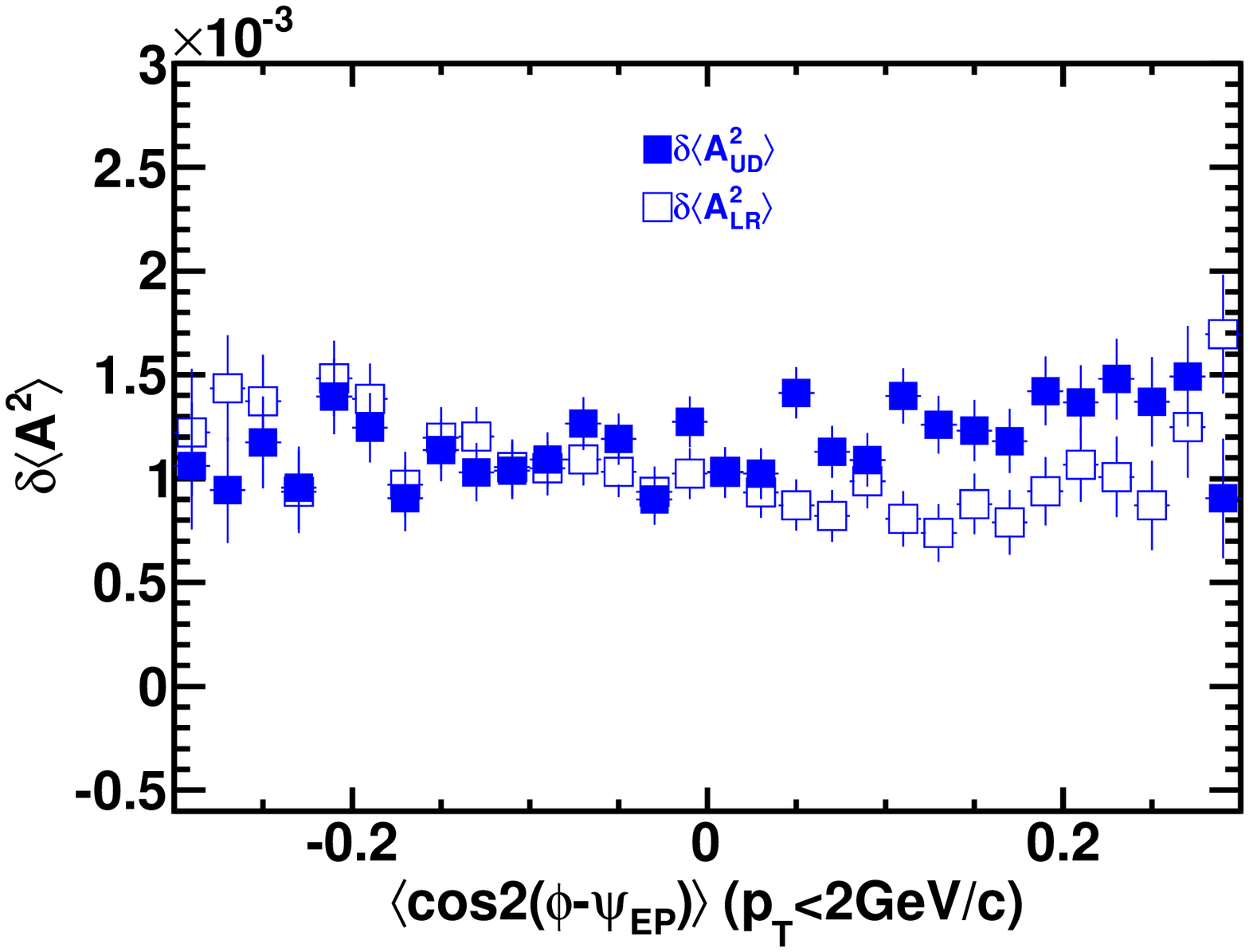}}
		\subfigure{\label{fig:appasymv2zdc80-b} \includegraphics[width=0.45\textwidth]{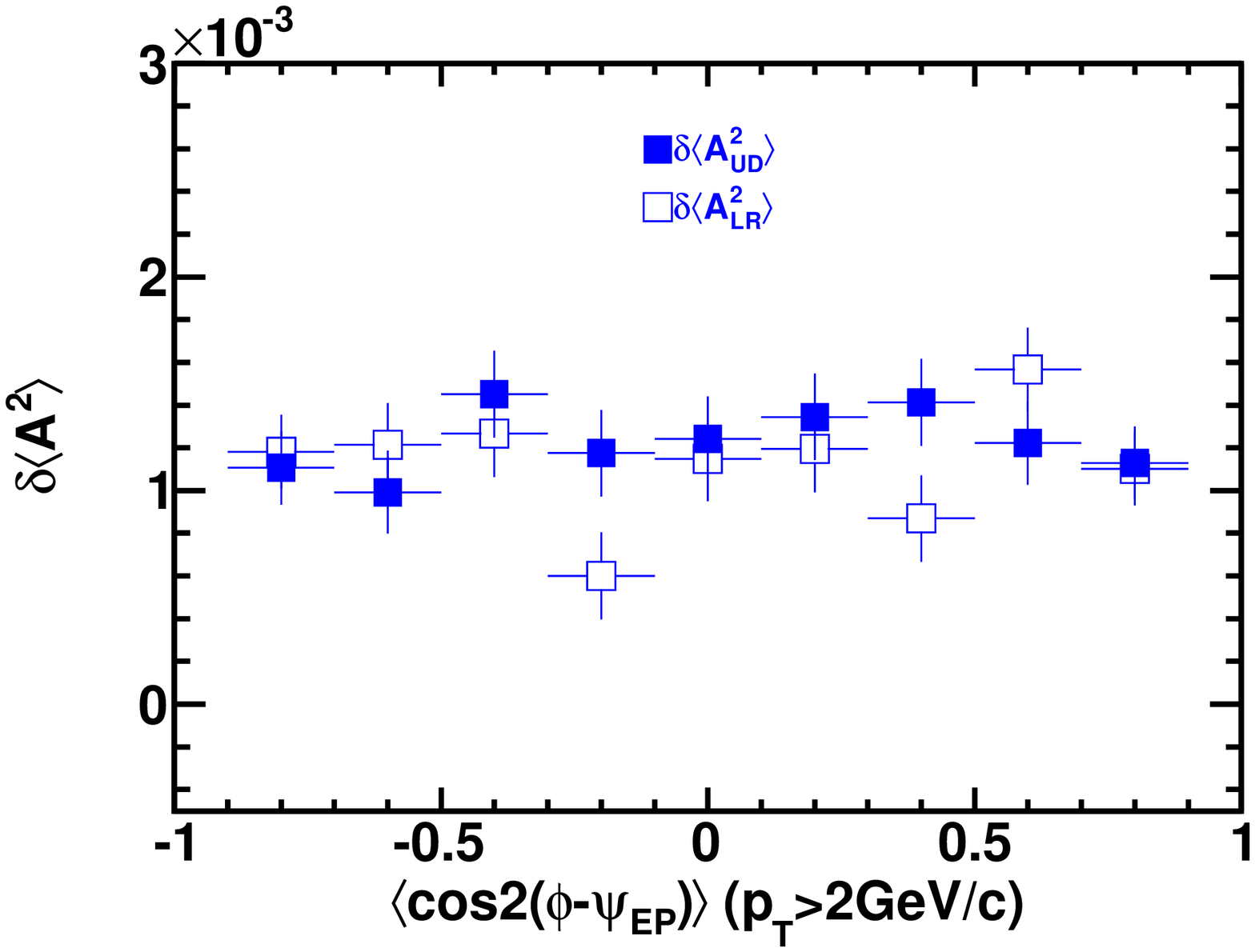}}
		\subfigure{\label{fig:appasymv2zdc80-c} \includegraphics[width=0.45\textwidth]{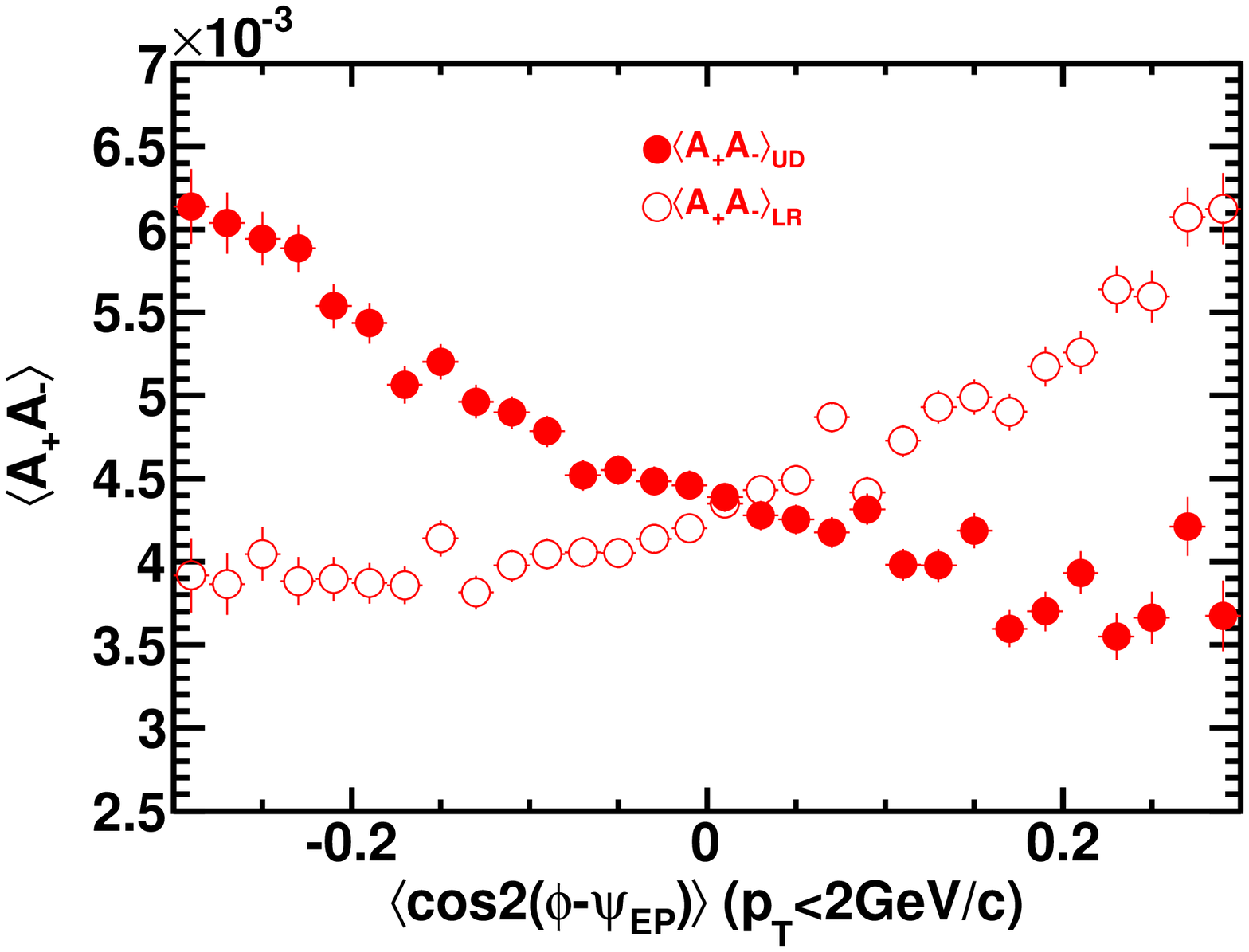}}
		\subfigure{\label{fig:appasymv2zdc80-d} \includegraphics[width=0.45\textwidth]{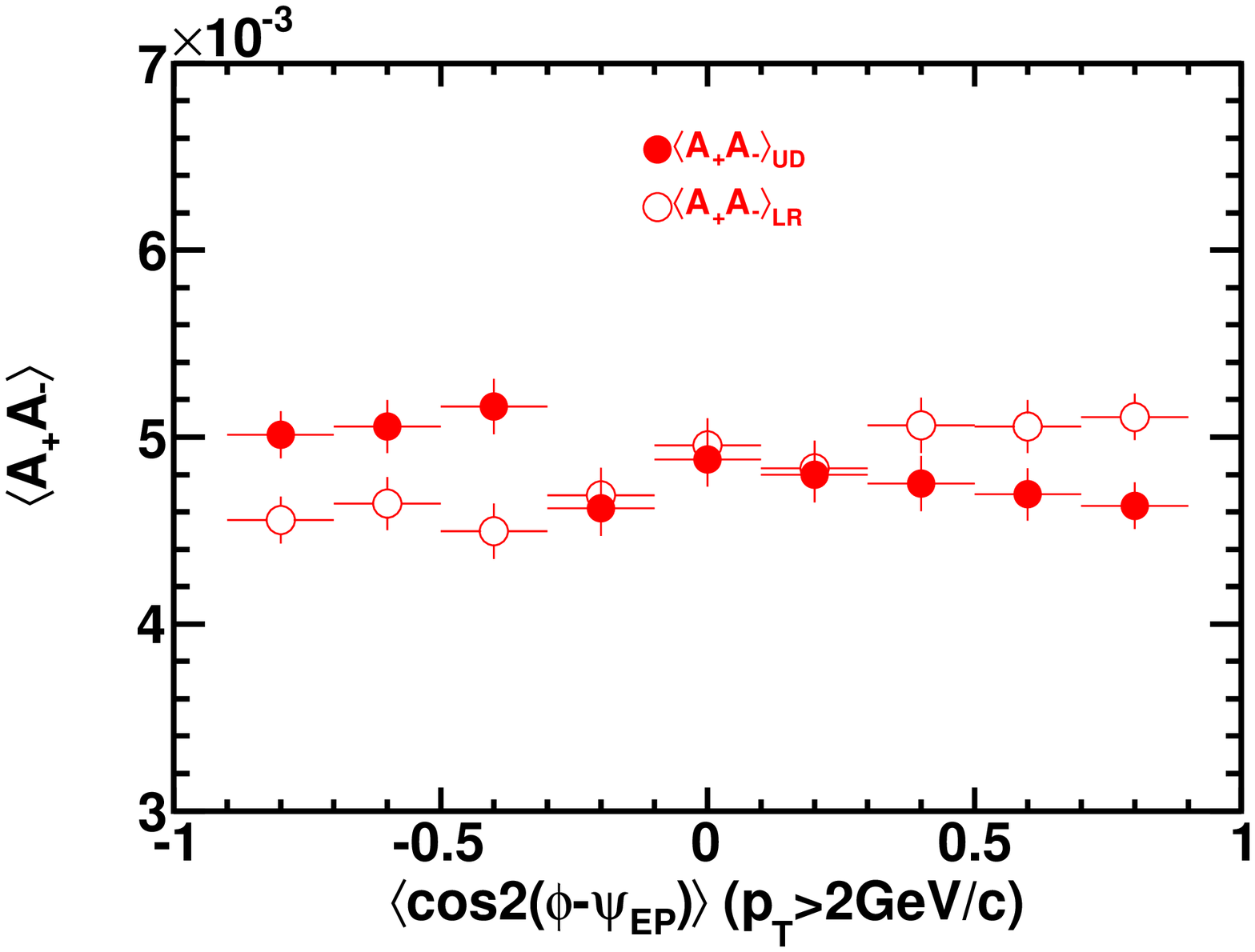}}
		\subfigure[Low-$p_T$ $v_{2}^{obs}$ $UD-LR$ correlations]{\label{fig:appasymv2zdcd80-a} \includegraphics[width=0.45\textwidth]{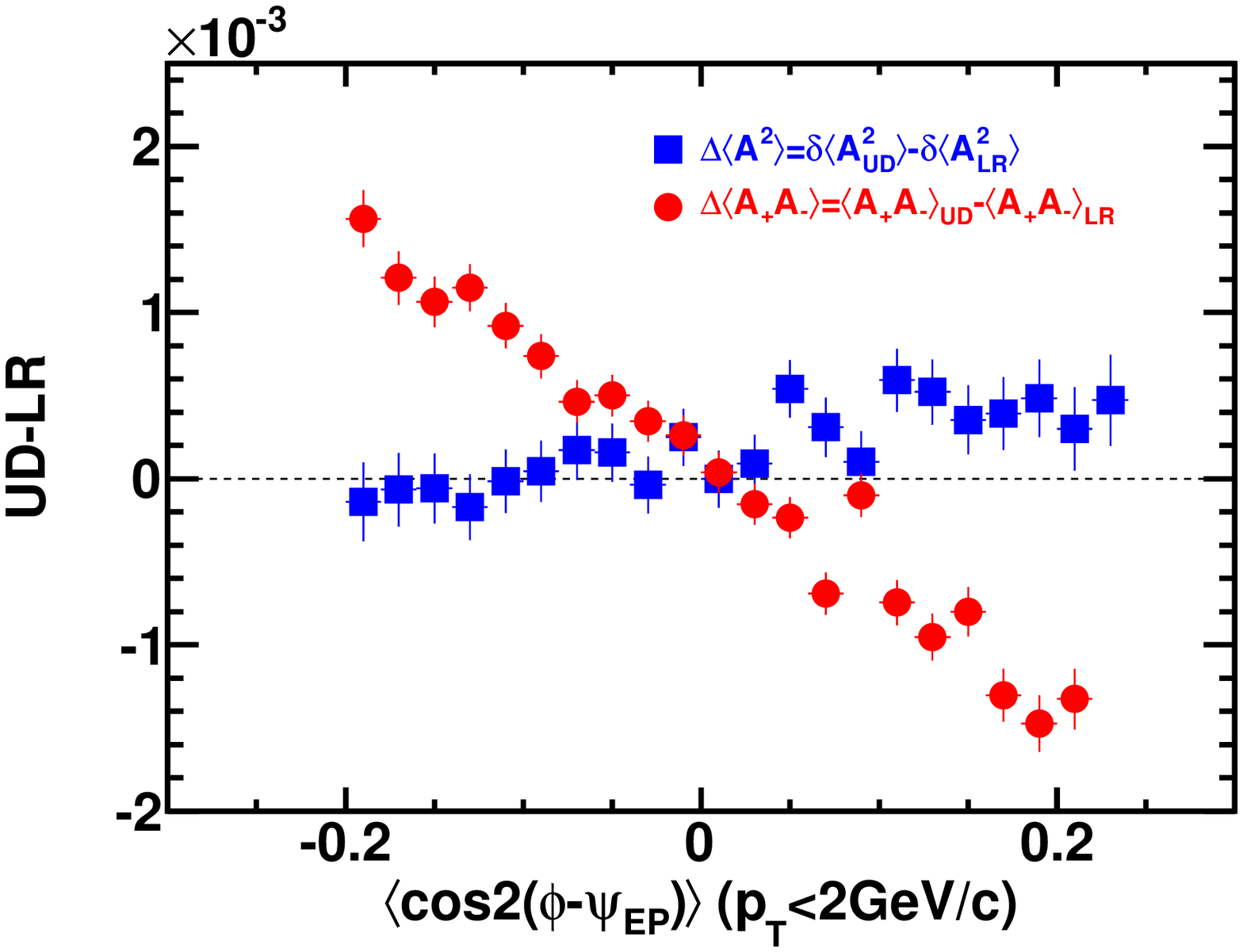}}
		\subfigure[High-$p_T$ $v_{2}^{obs}$ $UD-LR$ correlations]{\label{fig:appasymv2zdcd80-b} \includegraphics[width=0.45\textwidth]{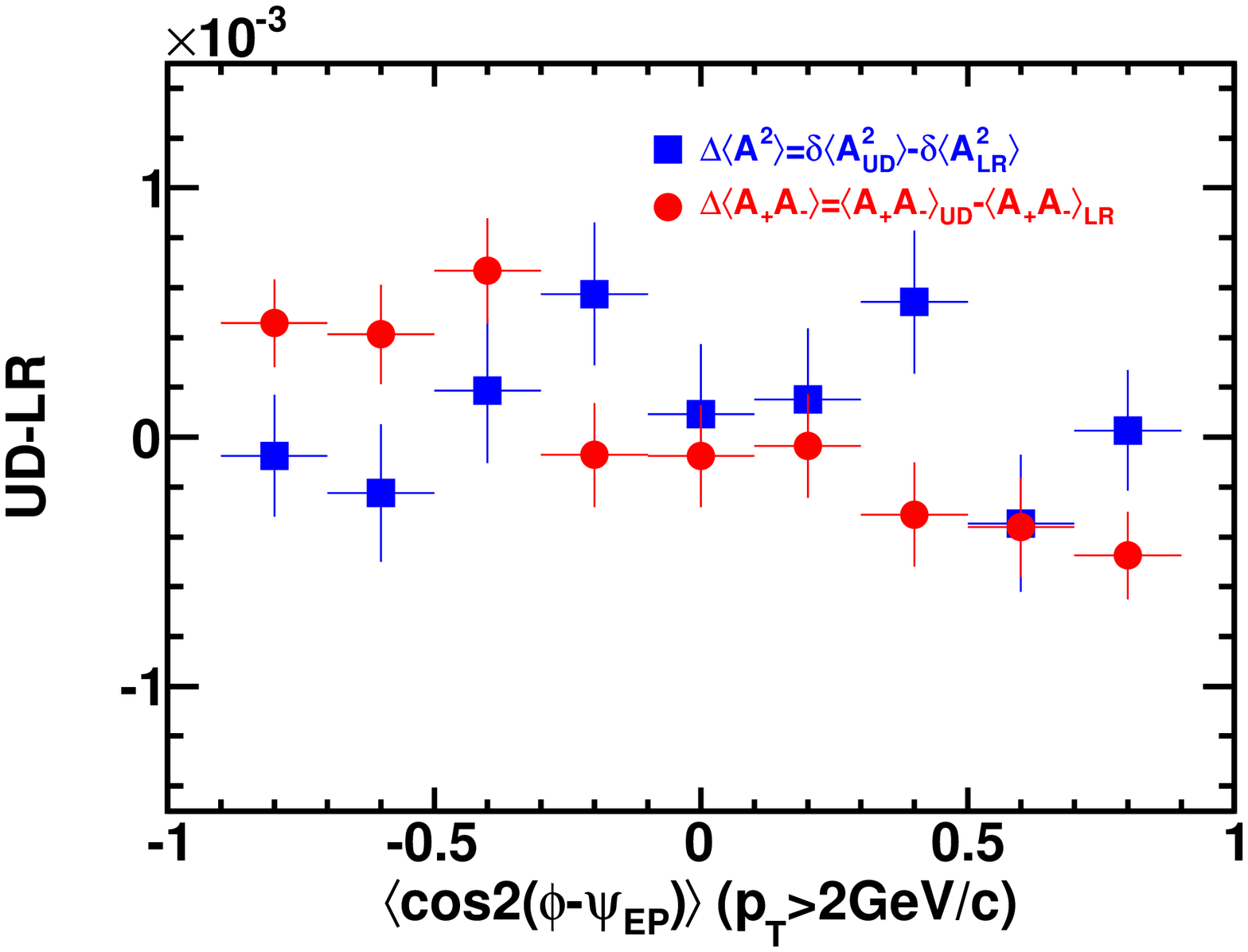}}
	\end{center}
	\caption[Peripheral asymmetry correlations vs $v_{2}^{obs}$ with ZDC-SMD EP]{
	Asymmetry correlations vs event-by-event $v_{2}^{obs}$ of RUN VII 200 GeV Au+Au peripheral 40-80\% collisions.
	The particles used for asymmetry calculation are from half side of the TPC with respect to the first order event-plane reconstructed from ZDC-SMD.
	Error bars are statistical only.
	}
	\label{fig:appasymv2zdc80}
\end{figure}

\begin{figure}[htb]
	\begin{center}
		\subfigure[Asymmetry correlations vs wedge size]{\label{fig:appasymwedgesize20-a} \includegraphics[width=0.45\textwidth]{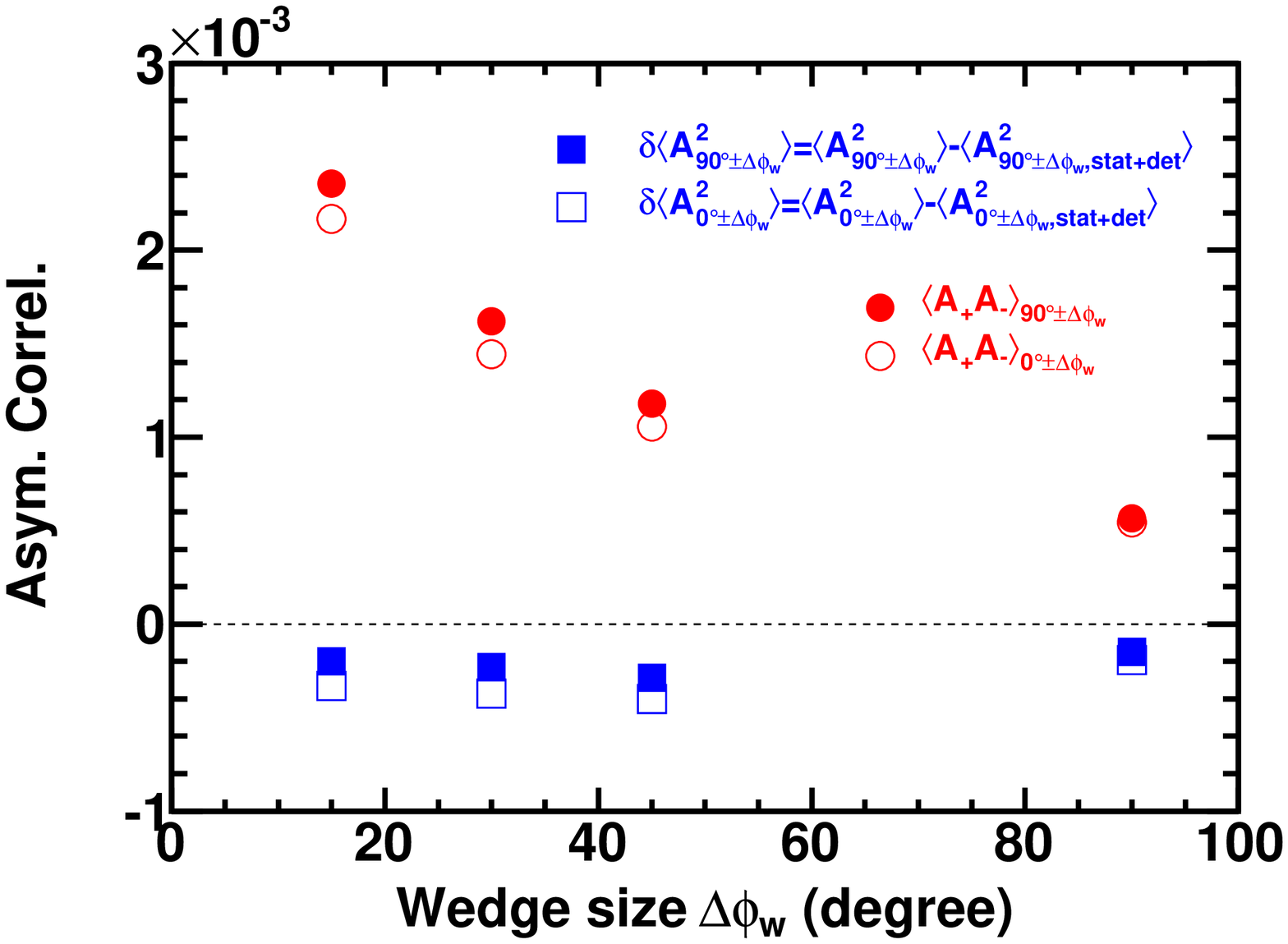}}
		\subfigure[$UD-LR$ correlations vs wedge size]{\label{fig:appasymwedgesize20-b} \includegraphics[width=0.45\textwidth]{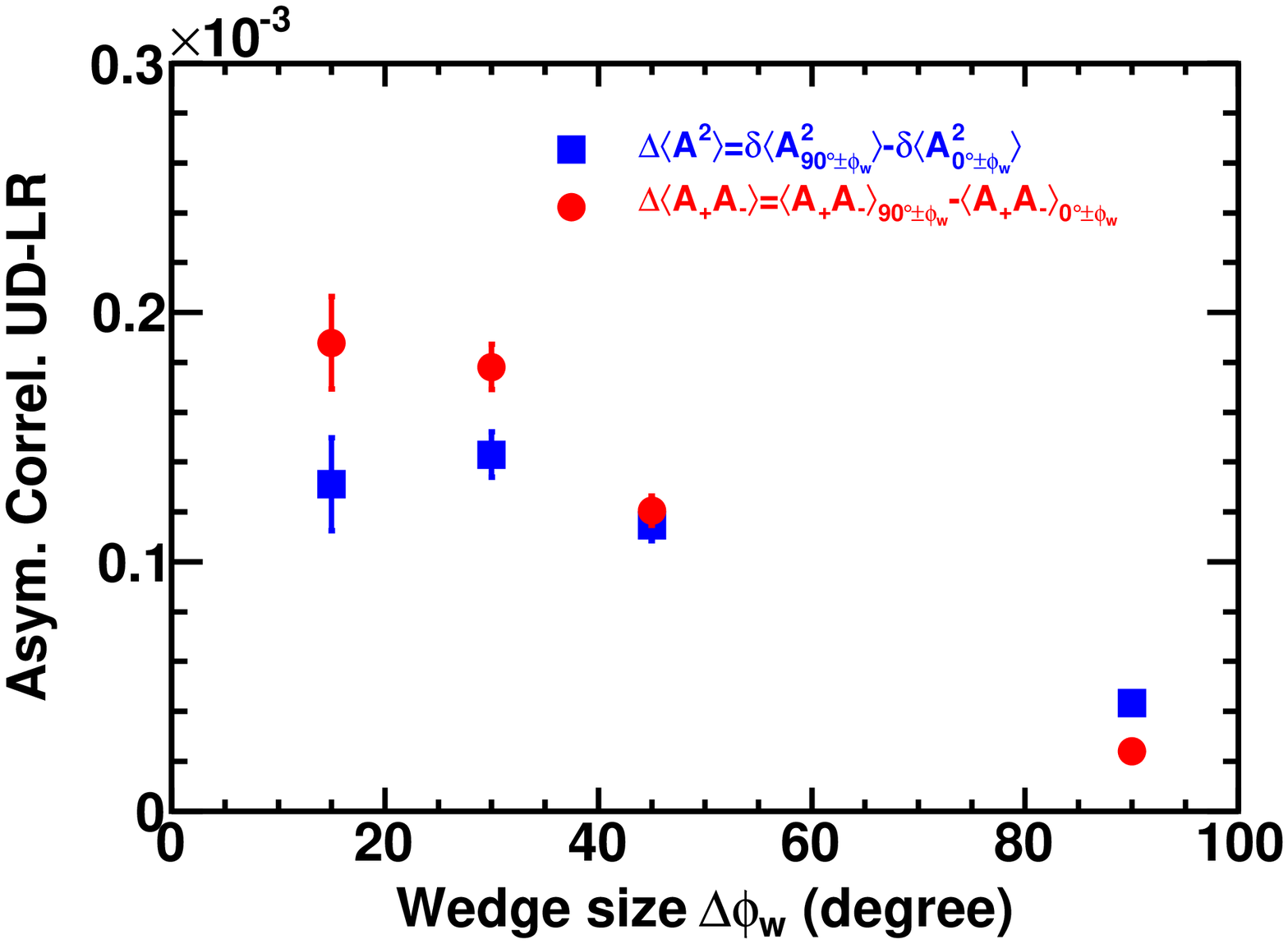}}
		\subfigure[Asymmetry correlations vs wedge location]{\label{fig:appasymwedgesize20-c} \includegraphics[width=0.45\textwidth]{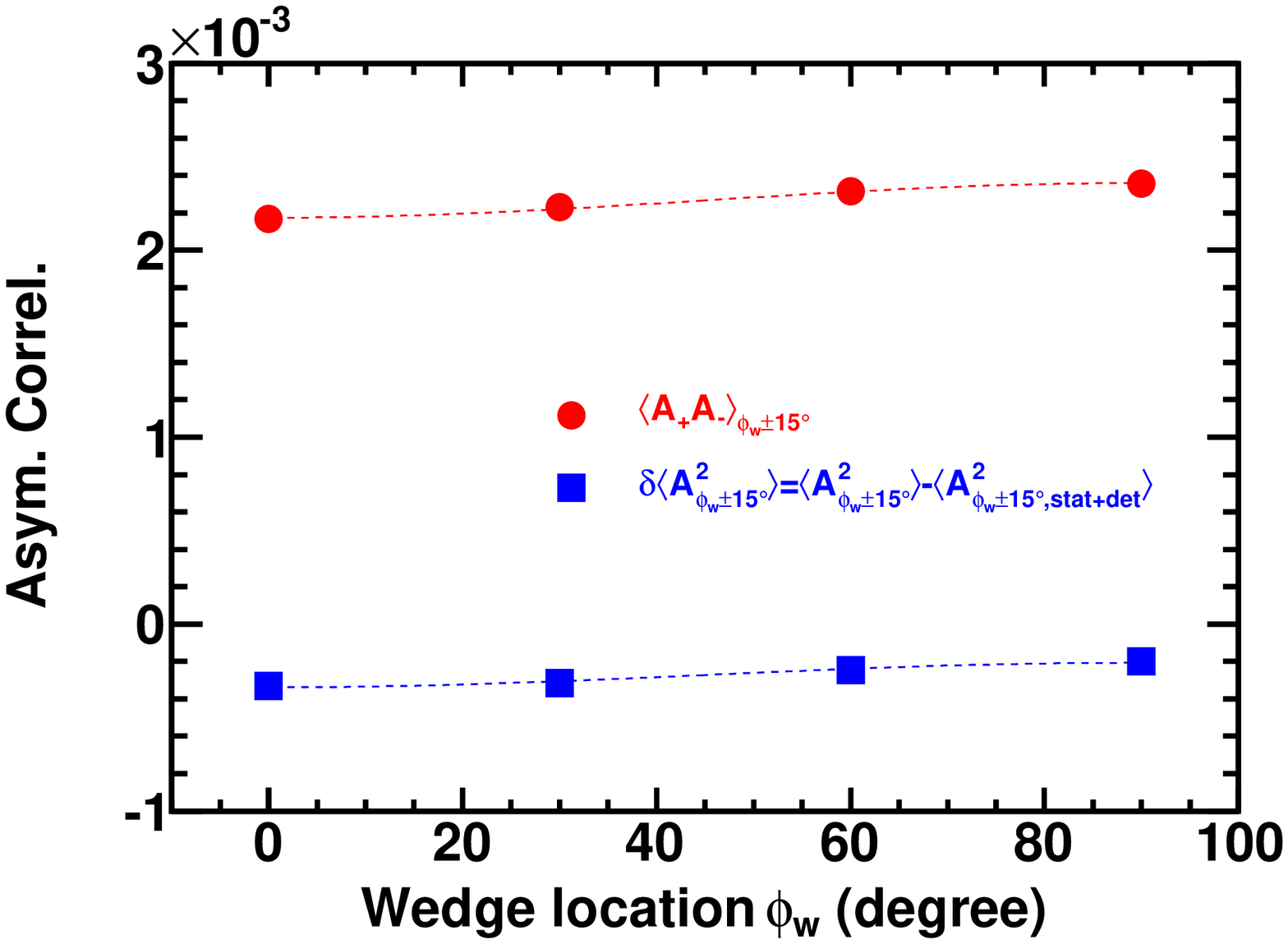}}
		\subfigure[Charge separation $\Delta(\Delta\phi_\text{w})$]{\label{fig:appasymwedgesize20-d} \includegraphics[width=0.45\textwidth]{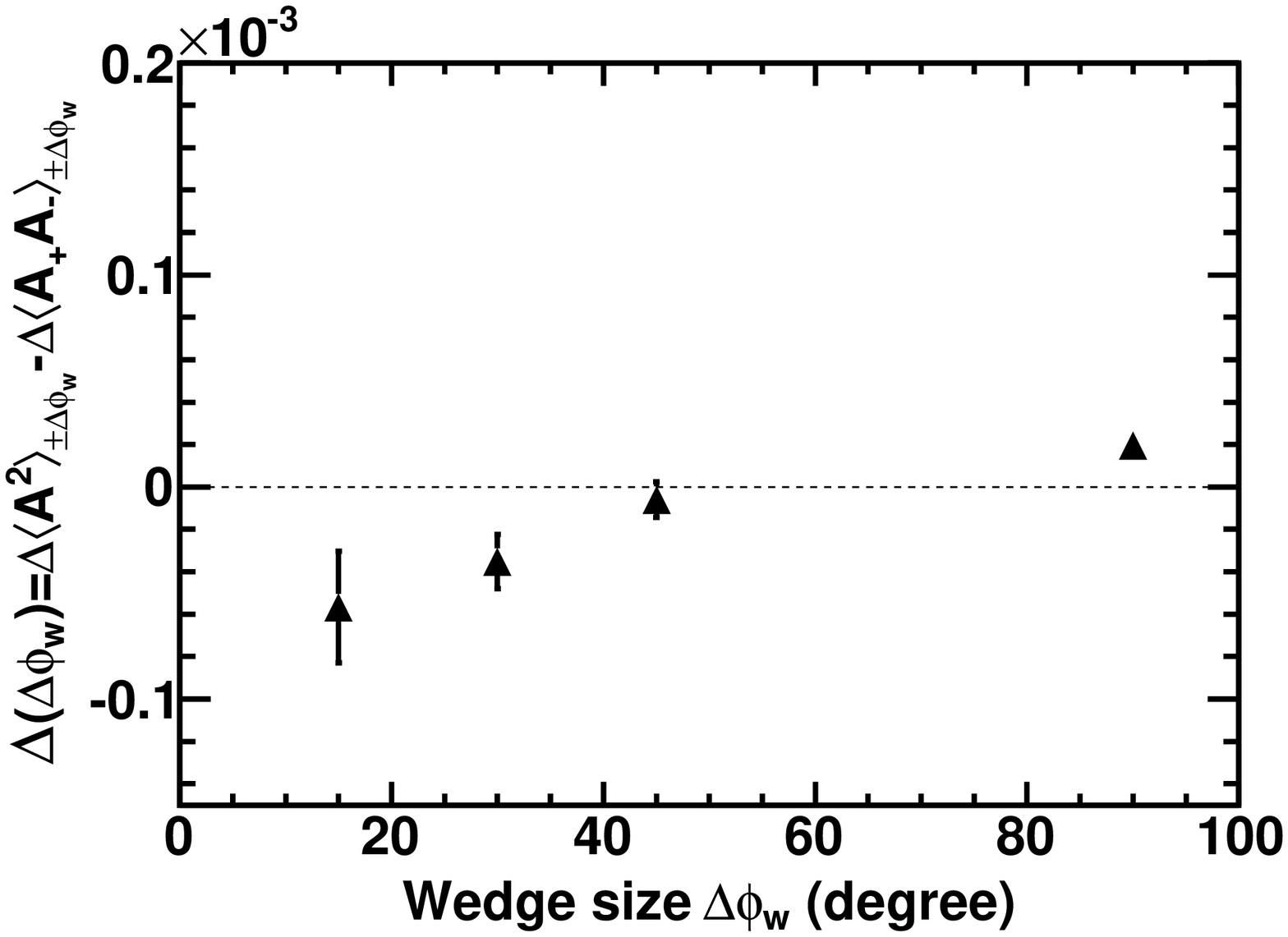}}
	\end{center}
	\caption[Central wedge size and location dependence of asymmetry correlations]{
	The wedge size dependence of asymmetry correlations in panel (a) and their differences between $UD$ and $LR$ correlations
	$\Delta\langle A^2_{\Delta\phi_{\text{w}}} \rangle$ and $\Delta\langle A_+A_- \rangle_{\Delta\phi_{\text{w}}}$ in panel (b).
	Wedge location dependence is shown in panel (c) with opening angle of $30\degree$ ($\Delta\phi_\text{w} = 15\degree$).
	Charge separation $\Delta(\Delta\phi_\text{w})$ as a function of the wedge size panel (d).
	Data are from 0-20\% centrality RUN IV 200 GeV Au+Au collisions.
	Error bars are statistical.
	}
	\label{fig:appasymwedgesize20}
\end{figure}

\begin{figure}[htb]
	\begin{center}
		\subfigure[Asymmetry correlations vs wedge size]{\label{fig:appasymwedgesize80-a} \includegraphics[width=0.45\textwidth]{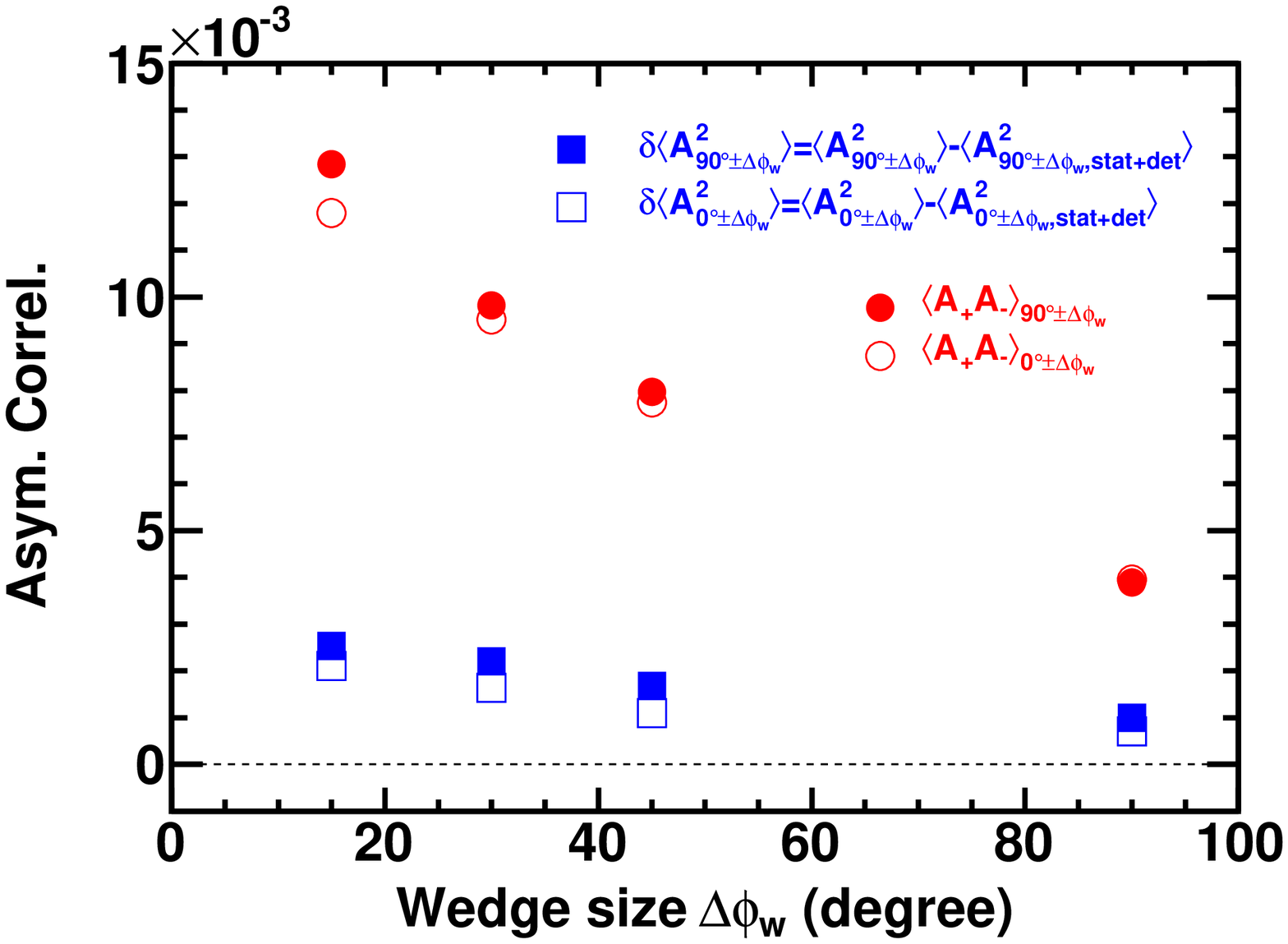}}
		\subfigure[$UD-LR$ correlations vs wedge size]{\label{fig:appasymwedgesize80-b} \includegraphics[width=0.45\textwidth]{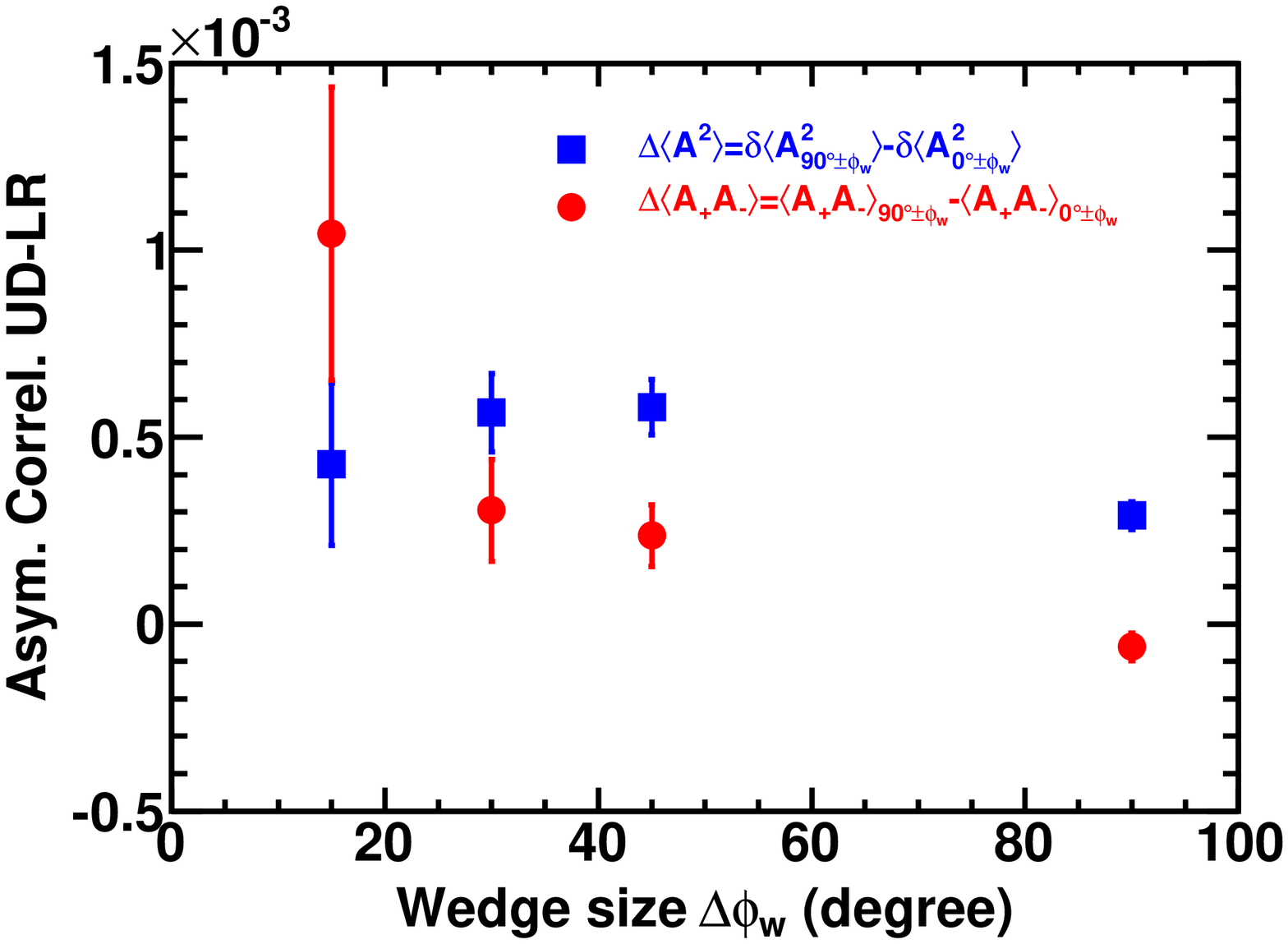}}
		\subfigure[Asymmetry correlations vs wedge location]{\label{fig:appasymwedgesize80-c} \includegraphics[width=0.45\textwidth]{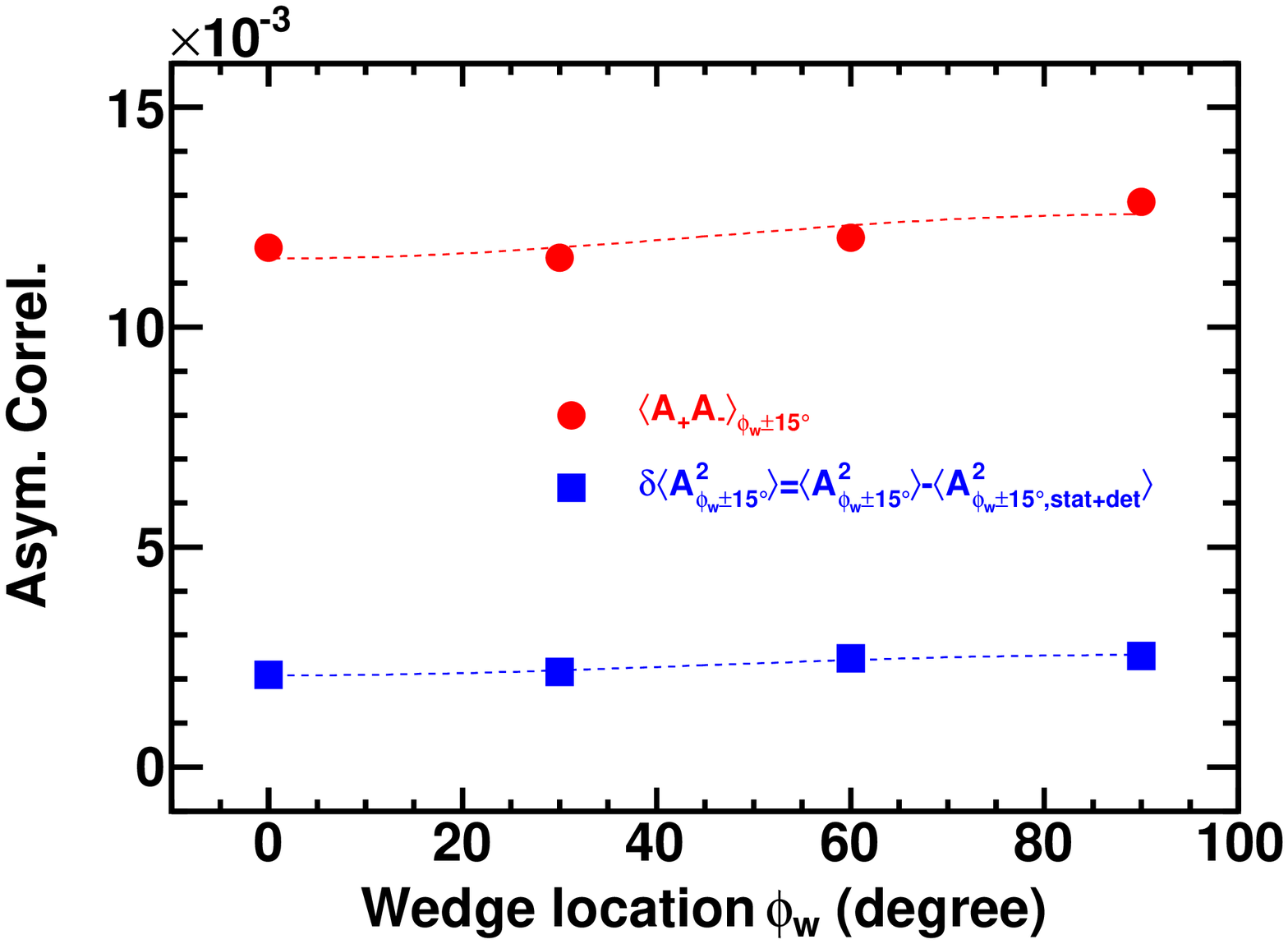}}
		\subfigure[Charge separation $\Delta(\Delta\phi_\text{w})$]{\label{fig:appasymwedgesize80-d} \includegraphics[width=0.45\textwidth]{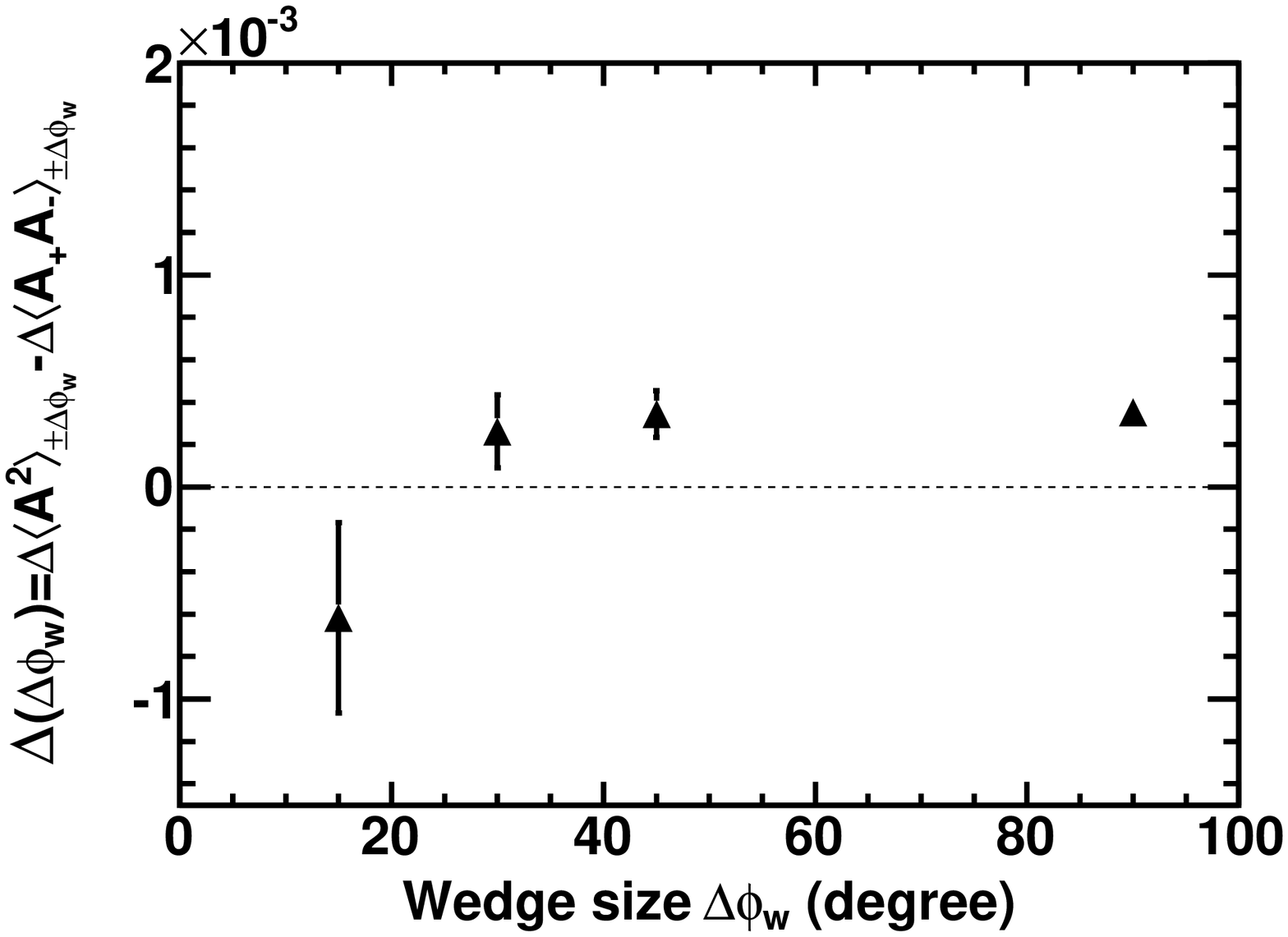}}
	\end{center}
	\caption[Peripheral wedge result]{
	The wedge size dependence of asymmetry correlations in panel (a) and their differences between $UD$ and $LR$ correlations
	$\Delta\langle A^2_{\Delta\phi_{\text{w}}} \rangle$ and $\Delta\langle A_+A_- \rangle_{\Delta\phi_{\text{w}}}$ in panel (b).
	Wedge location dependence is shown in panel (c) with opening angle of $30\degree$ ($\Delta\phi_\text{w} = 15\degree$).
	Charge separation $\Delta(\Delta\phi_\text{w})$ as a function of the wedge size panel (d).
	Data are from 40-80\% centrality RUN IV 200 GeV Au+Au collisions.
	Error bars are statistical.
	}
	\label{fig:appasymwedgesize80}
\end{figure}

\clearpage

\begin{figure}[ht]
	\begin{center}
		\subfigure[Charge separation vs $v_2^{obs}$]{\label{fig:appchargesepv2stdy420-a} \includegraphics[width=0.6\textwidth]{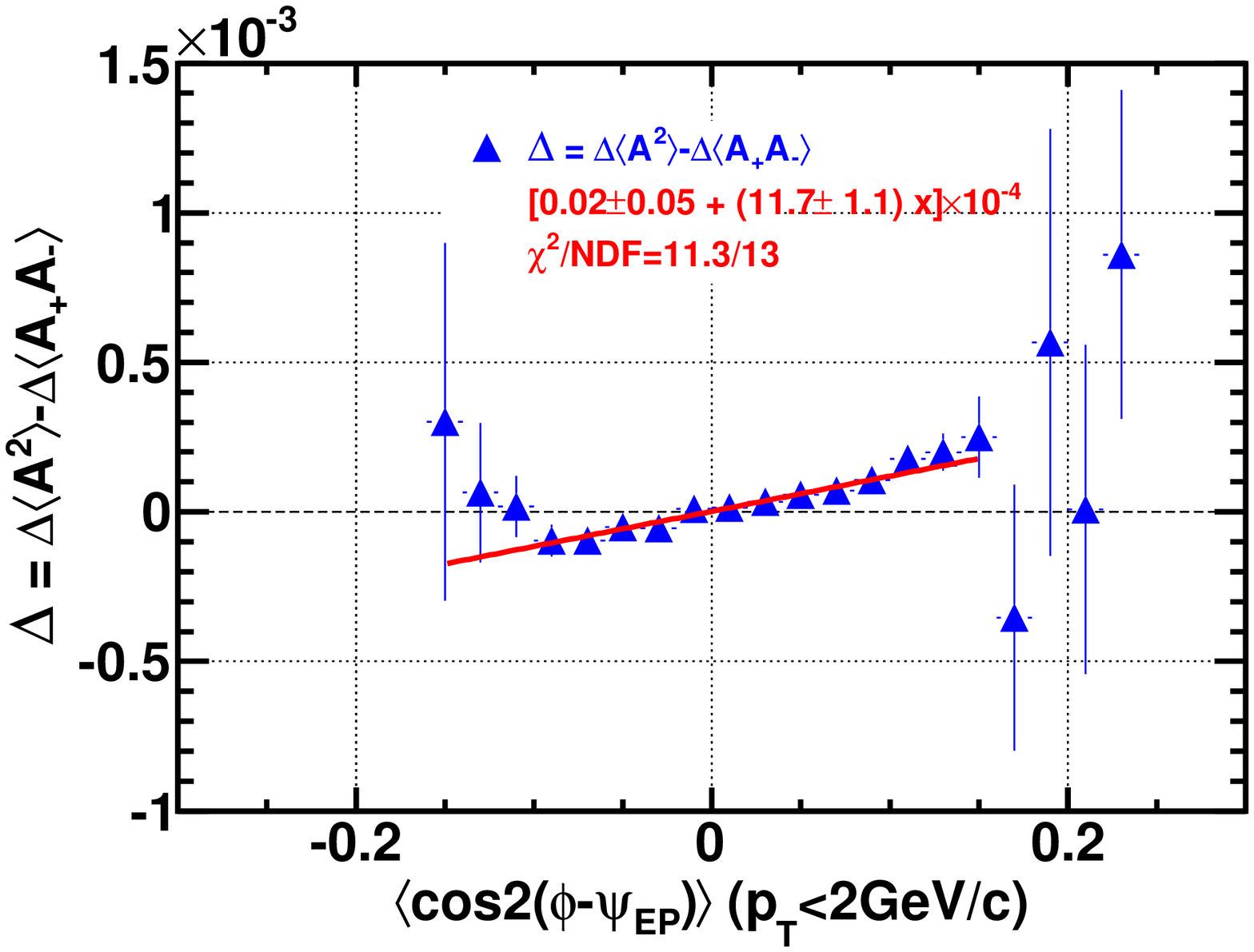}}
		\subfigure[Event-plane resolution vs $v_2^{obs}$]{\label{fig:appchargesepv2stdy420-b} \includegraphics[width=0.6\textwidth]{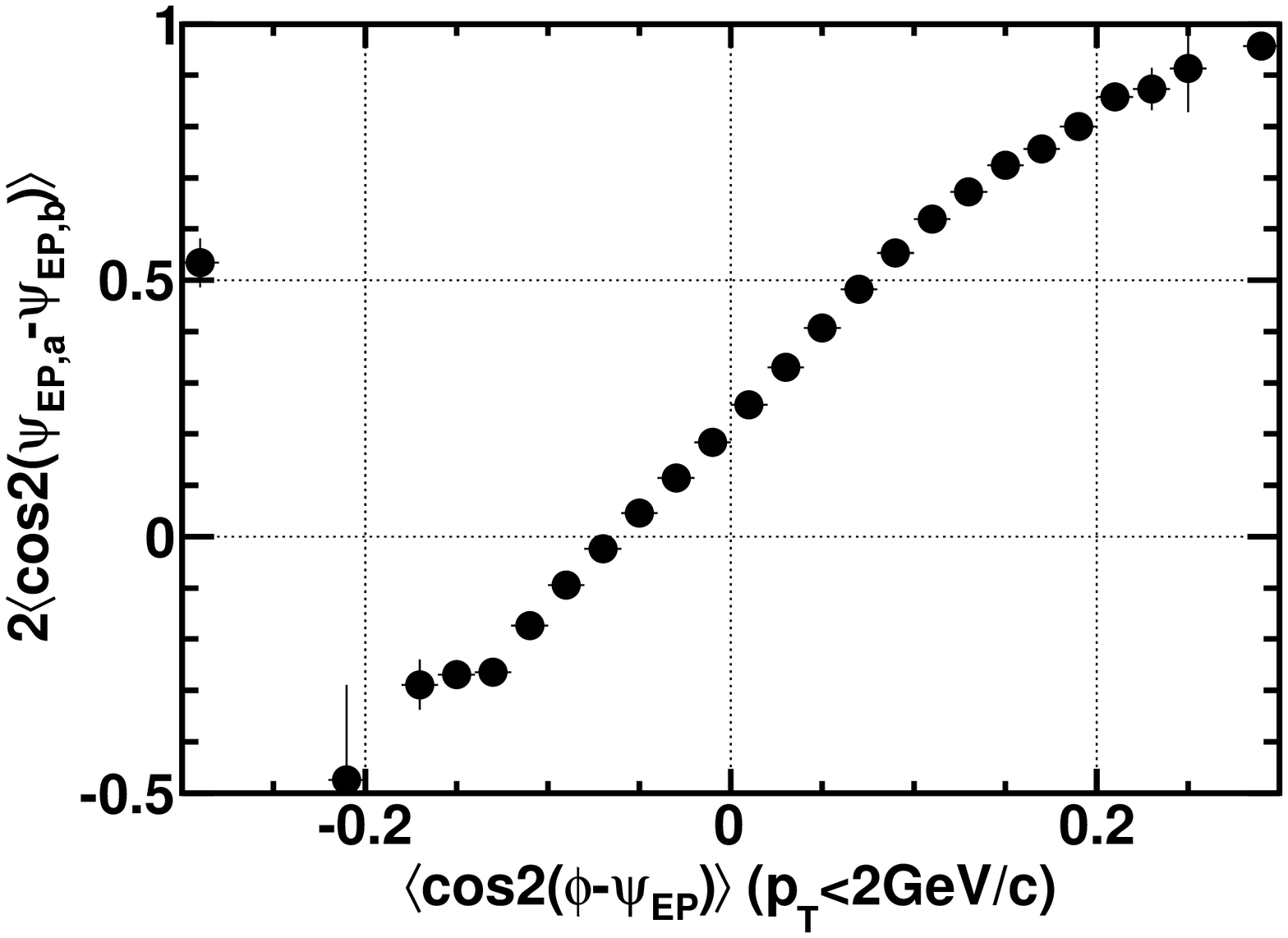}}
	\end{center}
	\caption[Central charge separation vs $v_2^{obs}$ from 2nd order EP]{
	Panel (a): RUN IV Au+Au 200 GeV 0-20\% centrality charge separation $\Delta$ as a function of low-$p_T$ event-by-event anisotropy $v_2^{obs}$.
	Panel (b): Event-plane resolution squared $\epsilon^{2}$ as a function of $v_2^{obs}$.
	The asymmetries and $v_2^{obs}$ are calculated relative to the second order event-plane reconstructed from the other side of the TPC tracks.
	The particle $p_T$ range of $0.15 < p_T < 2.0$ GeV/$c$ is used for asymmetry, $v_2^{obs}$ and event-plane reconstruction.
	Error bars are statistical.
	}
	\label{fig:appchargesepv2stdy420}
\end{figure}

\begin{figure}[ht]
	\begin{center}
		\subfigure[Charge separation vs $v_2^{obs}$]{\label{fig:appchargesepv2stdy480-a} \includegraphics[width=0.6\textwidth]{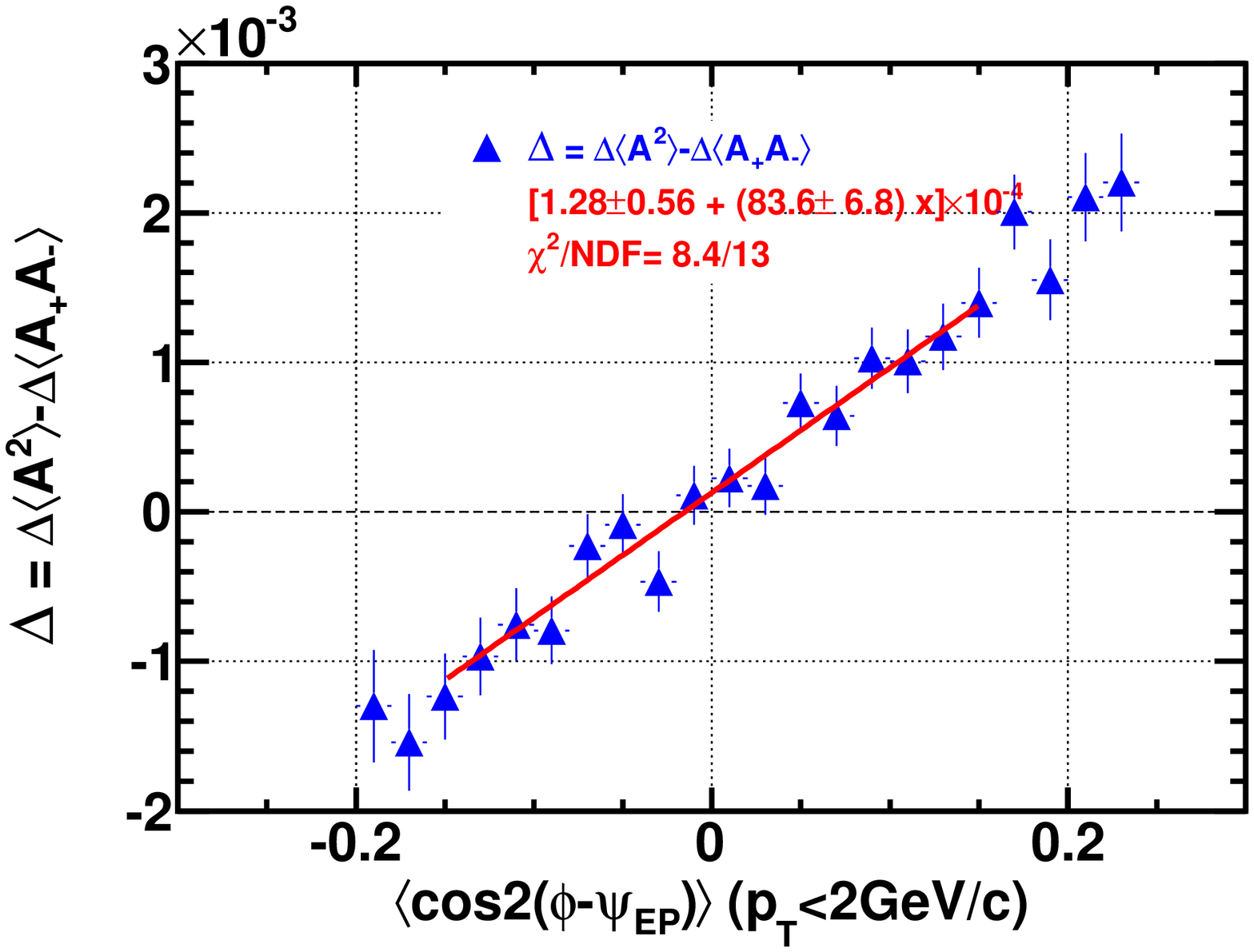}}
		\subfigure[Event-plane resolution vs $v_2^{obs}$]{\label{fig:appchargesepv2stdy480-b} \includegraphics[width=0.6\textwidth]{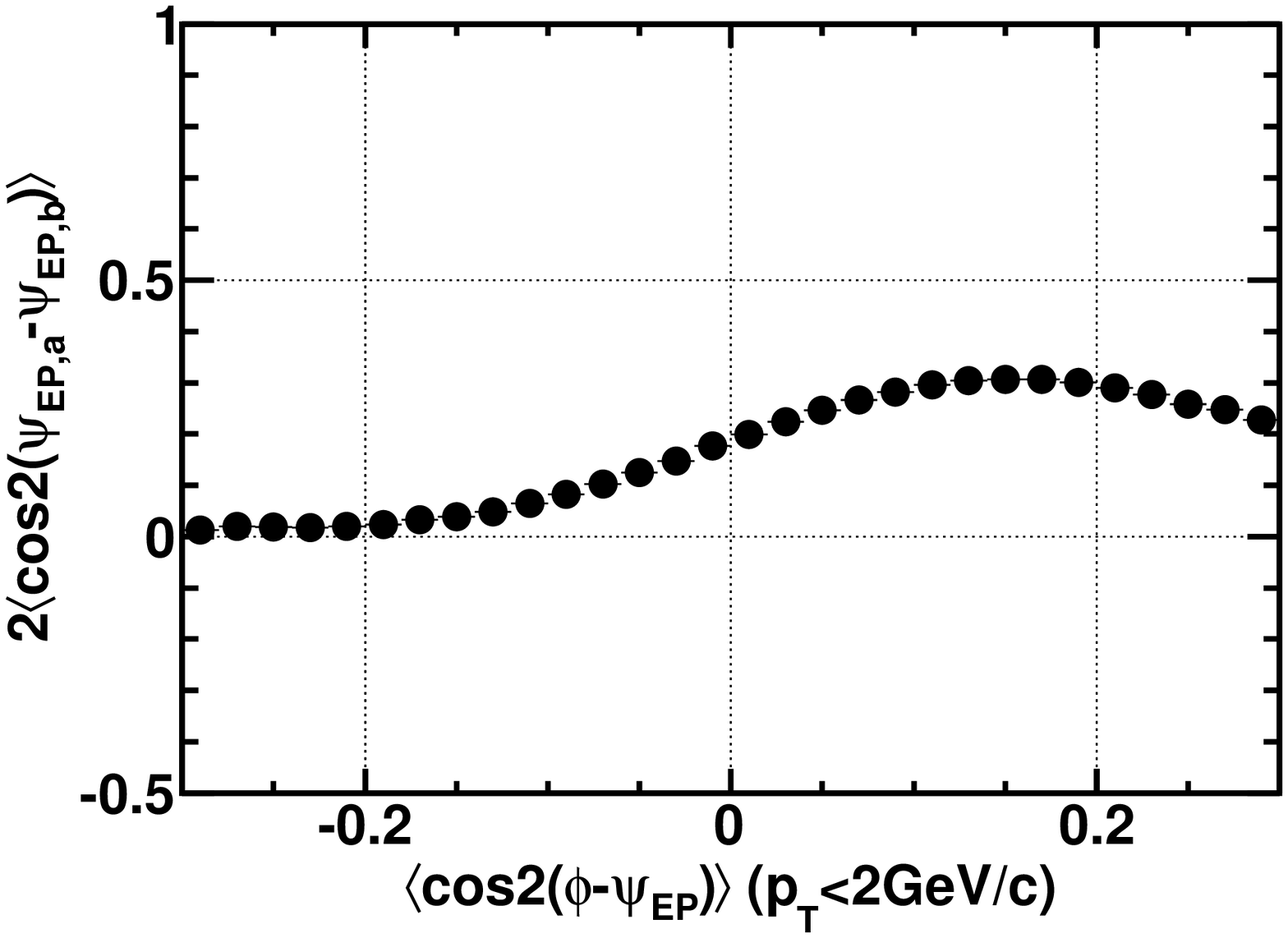}}
	\end{center}
	\caption[Peripheral charge separation vs $v_2^{obs}$ from 2nd order EP]{
	Panel (a): RUN IV Au+Au 200 GeV 40-80\% centrality charge separation $\Delta$ as a function of low-$p_T$ event-by-event anisotropy $v_2^{obs}$.
	Panel (b): Event-plane resolution squared $\epsilon^{2}$ as a function of $v_2^{obs}$.
	The asymmetries and $v_2^{obs}$ are calculated relative to the second order event-plane reconstructed from the other side of the TPC tracks.
	The particle $p_T$ range of $0.15 < p_T < 2.0$ GeV/$c$ is used for asymmetry, $v_2^{obs}$ and event-plane reconstruction.
	Error bars are statistical.
	}
	\label{fig:appchargesepv2stdy480}
\end{figure}

\begin{figure}[ht]
	\begin{center}
		\subfigure[Charge separation vs $v_2^{obs}$ with $\eta$ gap]{\label{fig:appetagap20-a} \includegraphics[width=0.55\textwidth]{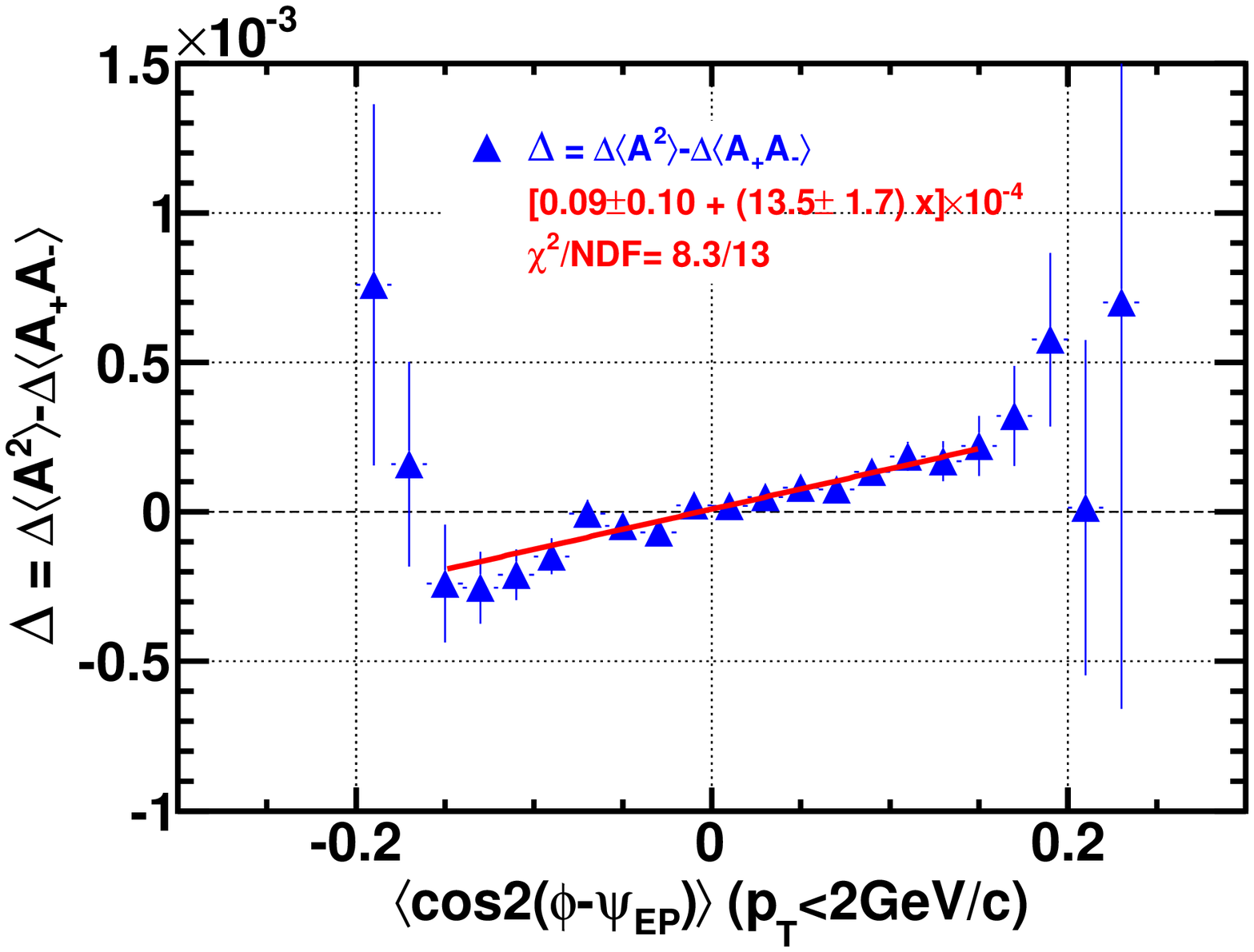}}
		\subfigure[EP resolution of $\eta$ gap]{\label{fig:appetagap20-b} \includegraphics[width=0.55\textwidth]{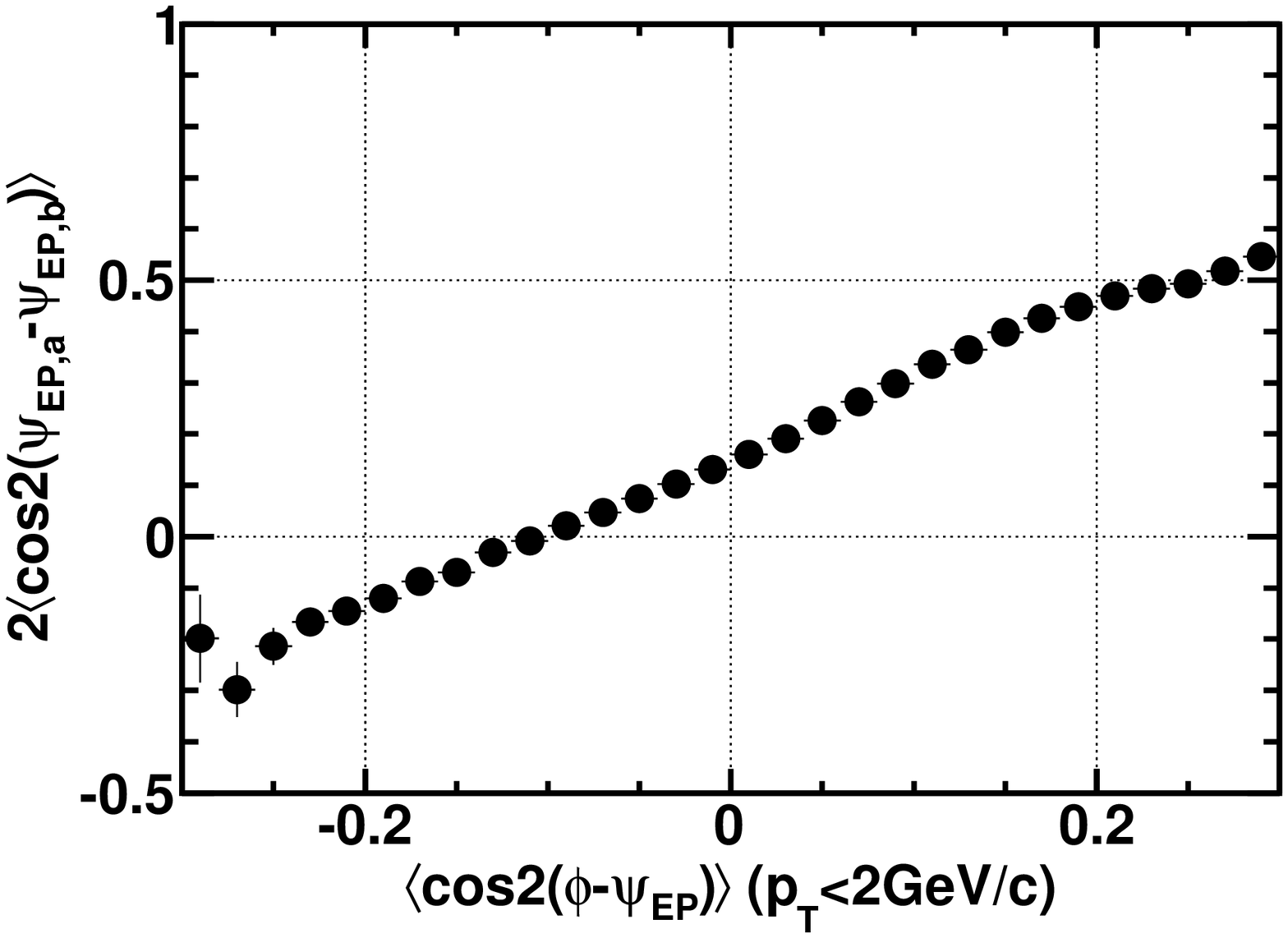}}
	\end{center}
	\caption[Central charge separation with $\eta$ gap vs $v_2^{obs}$ from 2nd order EP]{
	Panel (a): The charge separation $\Delta$ scaled by $N_{part}$ as a function of event-by-event anisotropy $v_2^{obs}$.
	The asymmetries and event-plane reconstruction are taken place in sub events with one unit pseudo-rapidity separation, $-1.0<\eta<-0.5$ and $0.5<\eta<1.0$.
	The charge separation is fitted to a linear polynomial as shown in red line.
	Panel (b): The event-plane resolution squared as a function of $v_2^{obs}$.
	Data are from RUN IV 200 GeV Au+Au collisions in 0-20\% centrality, and the particle $p_T$ range of $0.15<p_T<2.0$ GeV/$c$ is used for asymmetry calculation, $v_2^{obs}$ calculation and event-plane reconstruction.
	Error bars are statistical.
	}
	\label{fig:appetagap20}
\end{figure}

\begin{figure}[ht]
	\begin{center}
		\subfigure[Charge separation vs $v_2^{obs}$ with $\eta$ gap]{\label{fig:appetagap80-a} \includegraphics[width=0.55\textwidth]{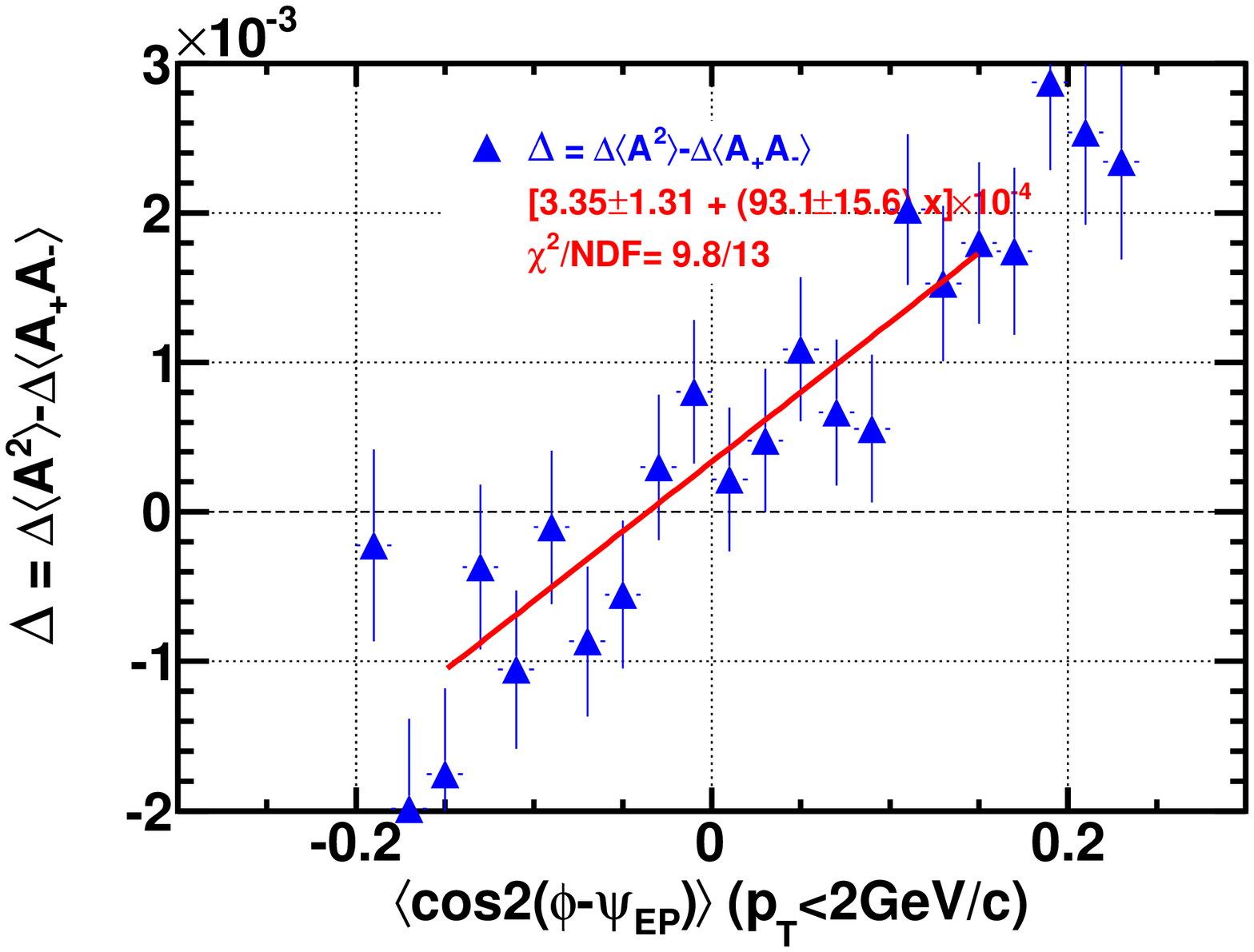}}
		\subfigure[EP resolution of $\eta$ gap]{\label{fig:appetagap80-b} \includegraphics[width=0.55\textwidth]{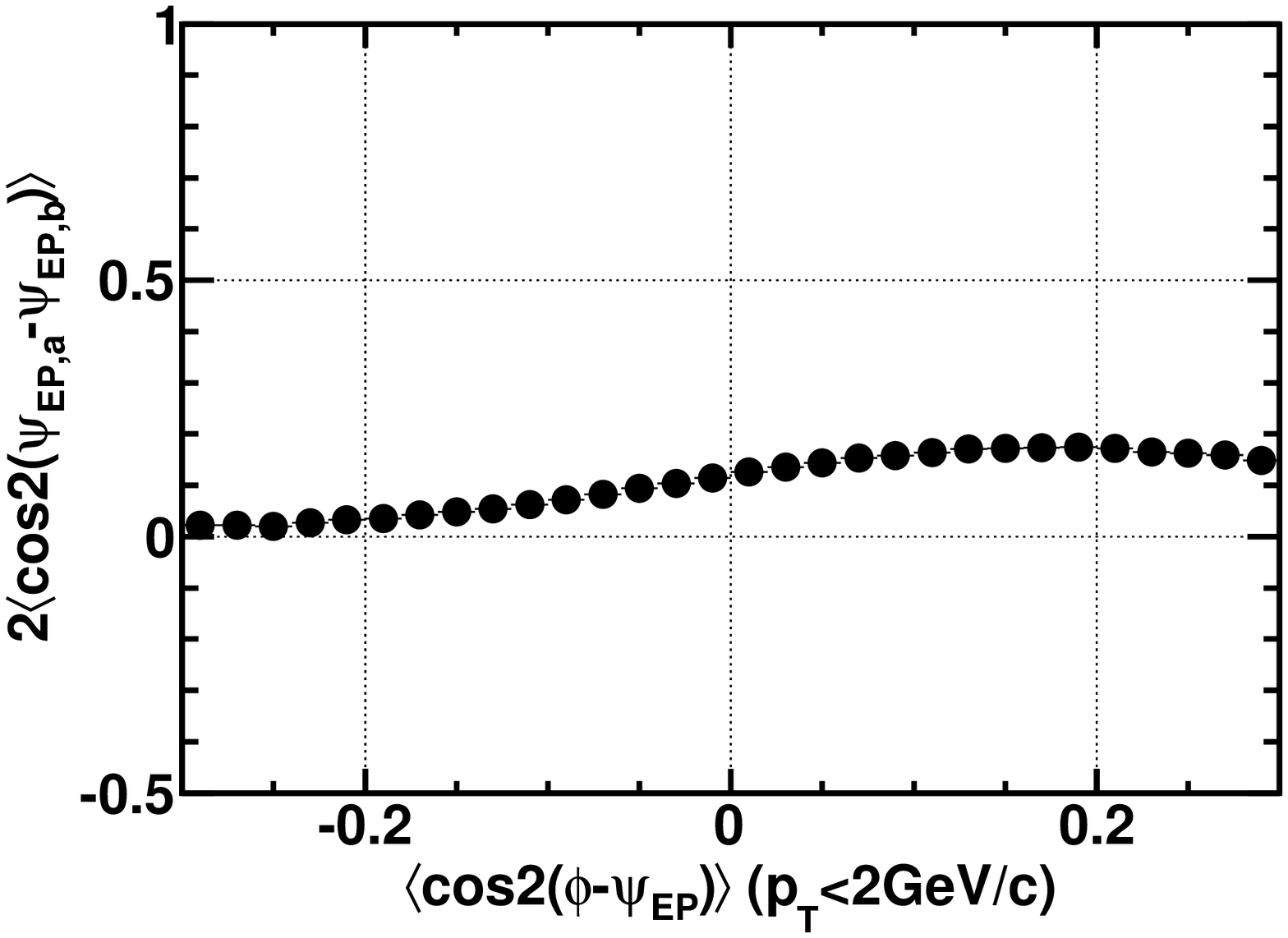}}
	\end{center}
	\caption[Peripheral charge separation with $\eta$ gap vs $v_2^{obs}$ from 2nd order EP]{
	Panel (a): The charge separation $\Delta$ scaled by $N_{part}$ as a function of event-by-event anisotropy $v_2^{obs}$.
	The asymmetries and event-plane reconstruction are taken place in sub events with one unit pseudo-rapidity separation, $-1.0<\eta<-0.5$ and $0.5<\eta<1.0$.
	The charge separation is fitted to a linear polynomial as shown in red line.
	Panel (b): The event-plane resolution squared as a function of $v_2^{obs}$.
	Data are from RUN IV 200 GeV Au+Au collisions in 40-80\% centrality, and the particle $p_T$ range of $0.15<p_T<2.0$ GeV/$c$ is used for asymmetry calculation, $v_2^{obs}$ calculation and event-plane reconstruction.
	Error bars are statistical.
	}
	\label{fig:appetagap80}
\end{figure}

\begin{figure}[ht]
	\begin{center}
		\subfigure[Charge separation vs $v_2^{obs}$ of ZDC-SMD EP]{\label{fig:appchargesep20v2zdc-a} \includegraphics[width=0.55\textwidth]{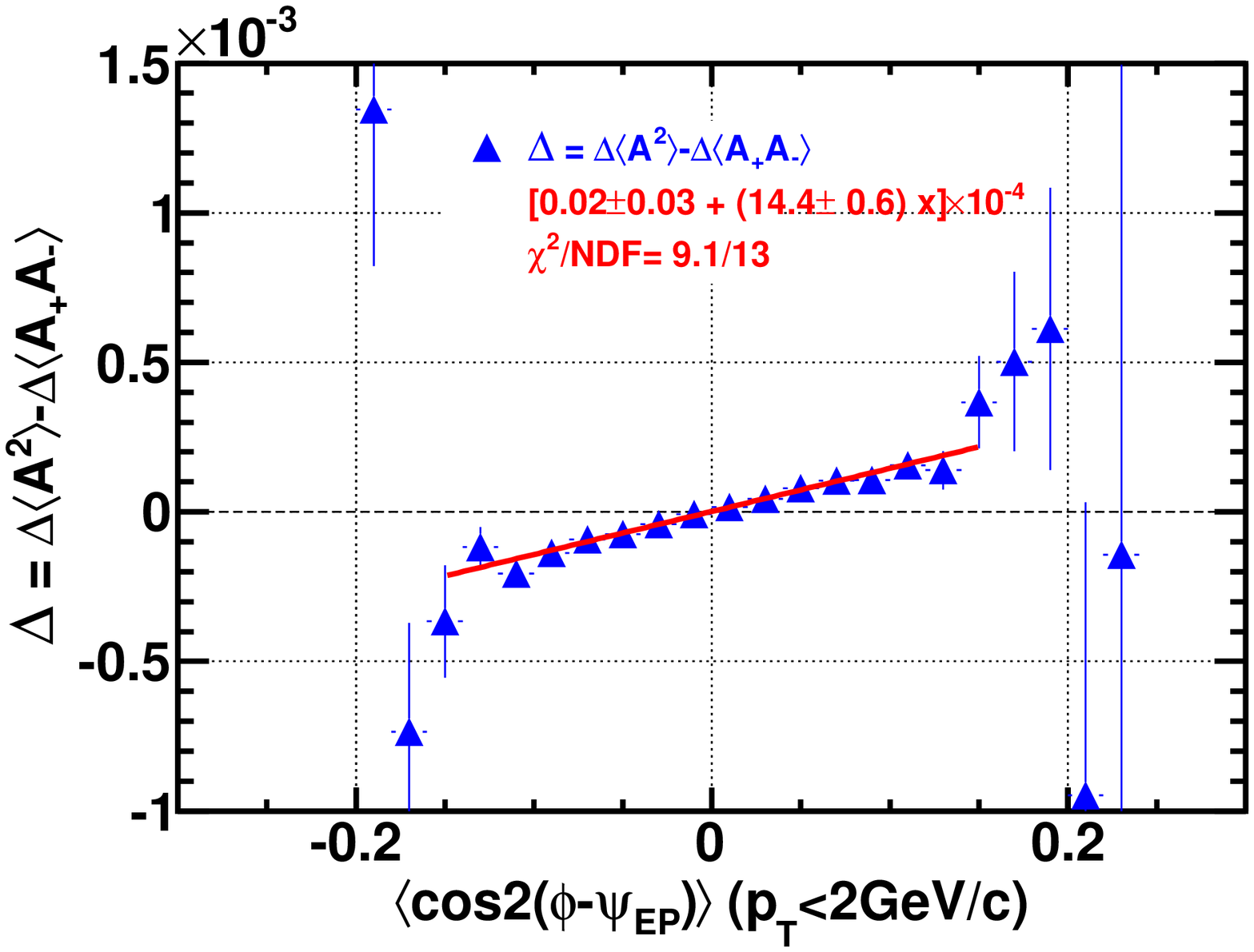}}
		\subfigure[ZDC-SMD EP resolution]{\label{fig:appchargesep20v2zdc-b} \includegraphics[width=0.55\textwidth]{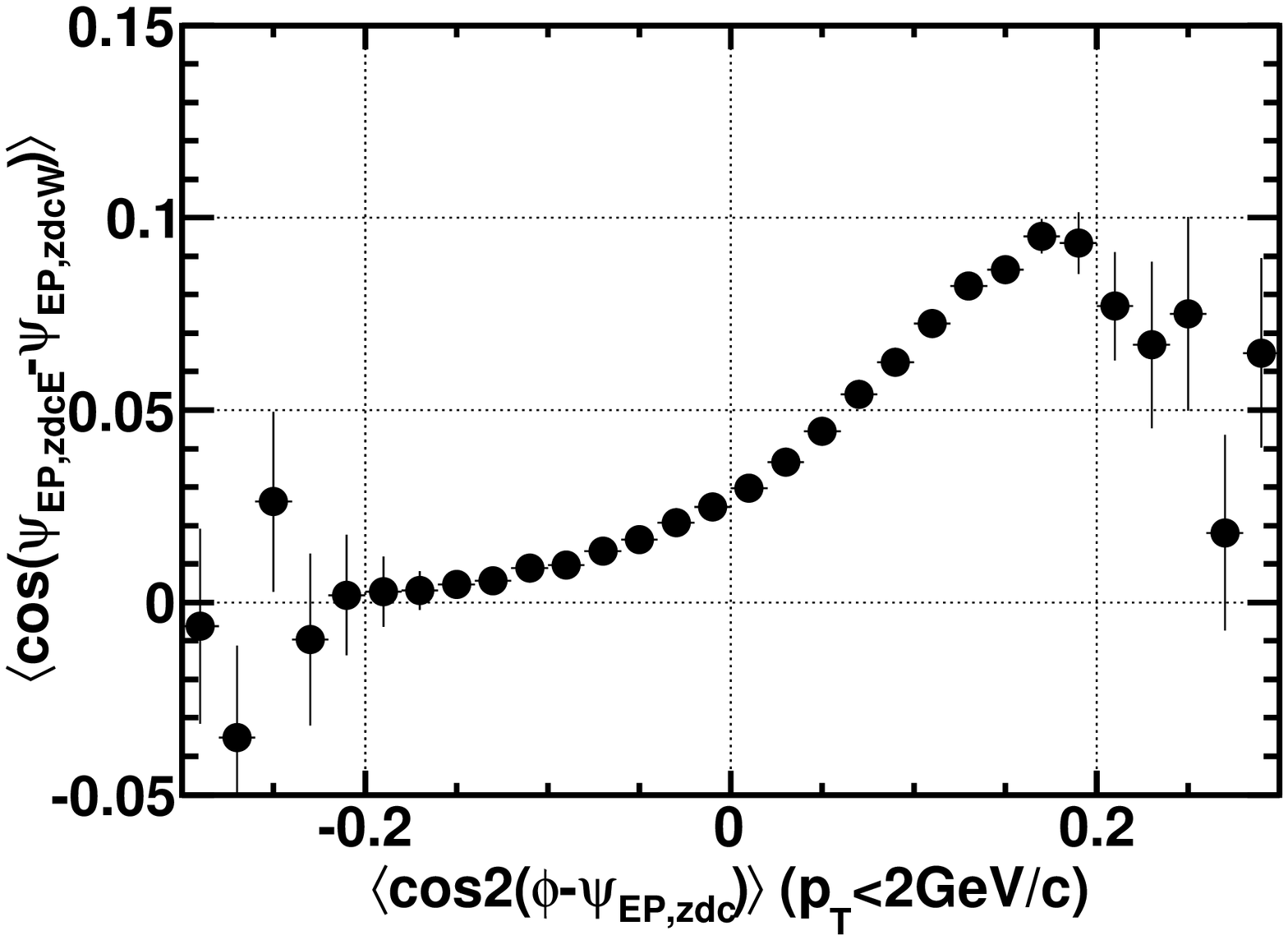}}
	\end{center}
	\caption[Central charge separation from 1st order EP]{
	Panel (a): The charge separation $\Delta$ scaled by $N_{part}$ as a function of event-by-event anisotropy $v_2^{obs}$.
	The asymmetries and $v_2^{obs}$ are calculated from half TPC tracks of an event, with respect to the first order event-plane reconstructed from ZDC-SMD detectors.
	The charge separation is fitted to a linear polynomial as shown in red line.
	Panel (b): The first order event-plane resolution squared as a function of $v_2^{obs}$.
	Data are from RUN VII 200 GeV Au+Au collisions in 0-20\% centrality, and the particle $p_T$ range of $0.15<p_T<2.0$ GeV/$c$ is used for asymmetry calculation and $v_2^{obs}$ calculation.
	Error bars are statistical.
	}
	\label{fig:appchargesep20v2zdc}
\end{figure}

\begin{figure}[ht]
	\begin{center}
		\subfigure[Charge separation vs $v_2^{obs}$ of ZDC-SMD EP]{\label{fig:appchargesep80v2zdc-a} \includegraphics[width=0.55\textwidth]{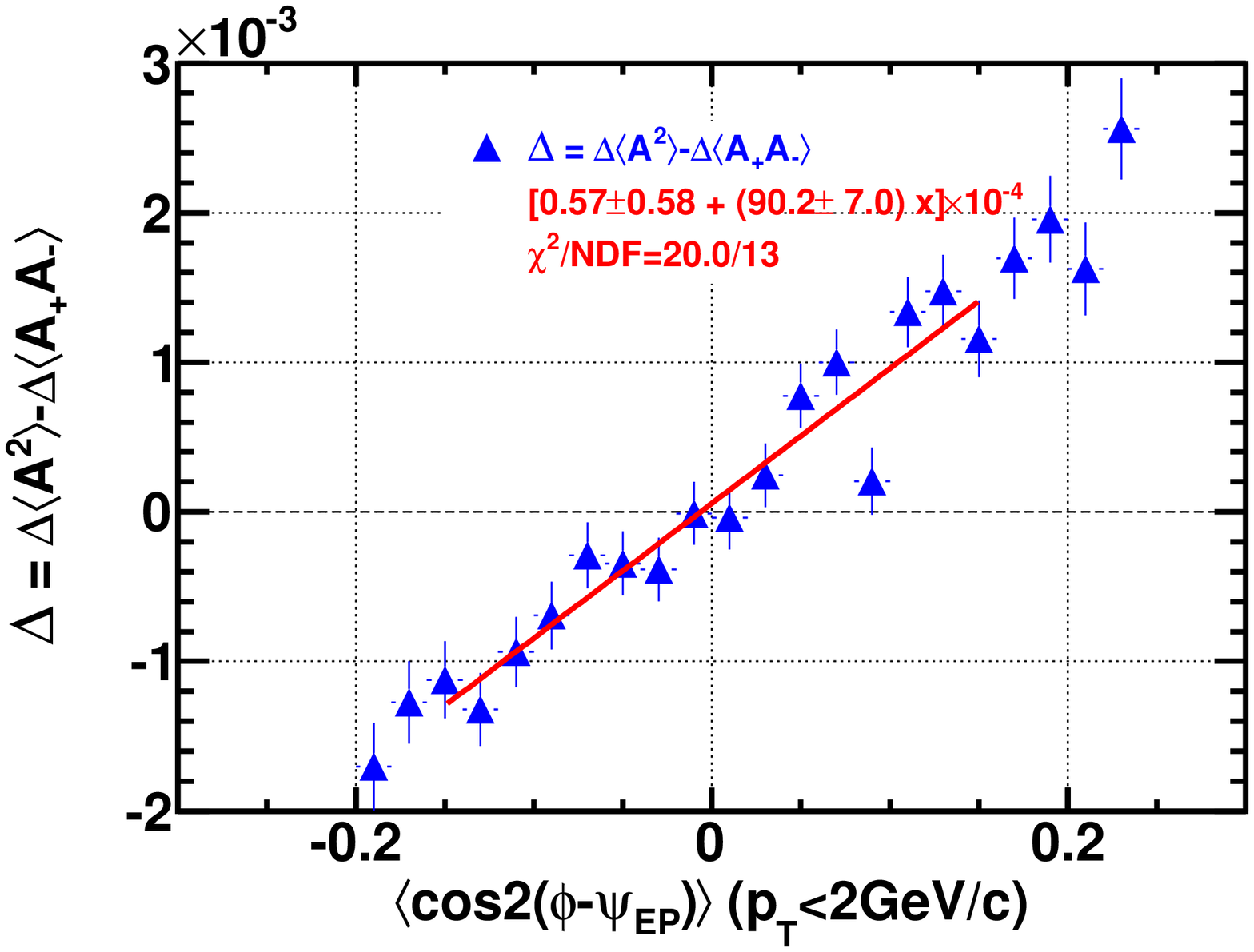}}
		\subfigure[ZDC-SMD EP resolution]{\label{fig:appchargesep80v2zdc-b} \includegraphics[width=0.55\textwidth]{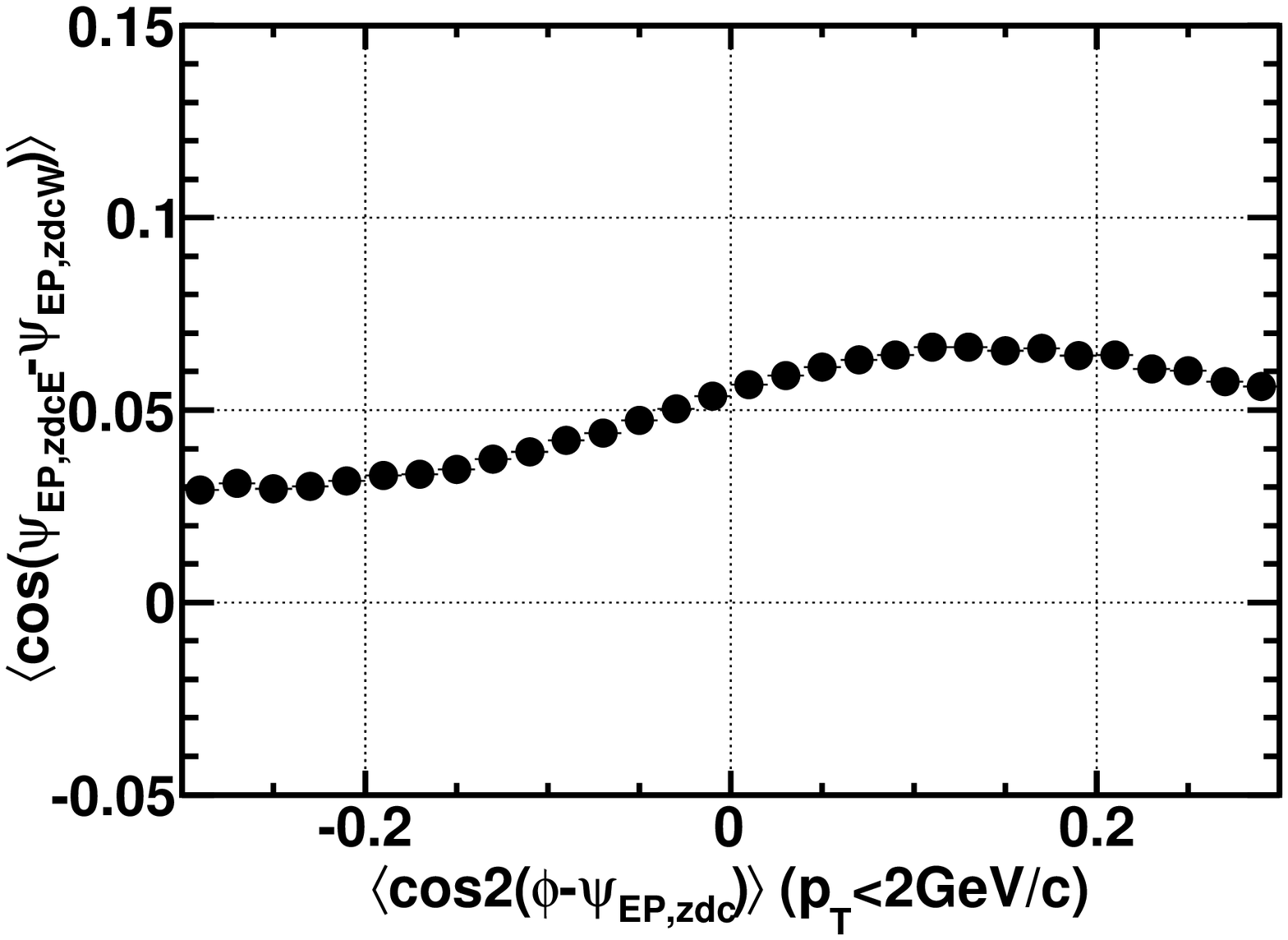}}
	\end{center}
	\caption[Central charge separation from 1st order EP]{
	Panel (a): The charge separation $\Delta$ scaled by $N_{part}$ as a function of event-by-event anisotropy $v_2^{obs}$.
	The asymmetries and $v_2^{obs}$ are calculated from half TPC tracks of an event, with respect to the first order event-plane reconstructed from ZDC-SMD detectors.
	The charge separation is fitted to a linear polynomial as shown in red line.
	Panel (b): The first order event-plane resolution squared as a function of $v_2^{obs}$.
	Data are from RUN VII 200 GeV Au+Au collisions in 40-80\% centrality, and the particle $p_T$ range of $0.15<p_T<2.0$ GeV/$c$ is used for asymmetry calculation and $v_2^{obs}$ calculation.
	Error bars are statistical.
	}
	\label{fig:appchargesep80v2zdc}
\end{figure}

%% file: vita.tex
%
%
%

\begin{vita}

	Quan Wang was born in Shanhaiguan, China on Feburary 22nd, 1979.
	He received his Bachelar of Science Degree in Physics in May 2002 from University of Science and Technology of China~(USTC). 
	He received his Master of Science Degree from USTC in experimental particle physics in Physics from USTC in December 2005.
	The theis title is ``Magnetic Spectrometer for Thermal Particles in Sub-picosecond Pulse''.
	He served as graduate student teaching assistent and research assistent at USTC during 2002-2005.
	He received his Ph.D. in experimental high energy nuclear physics from Purdue University in May 2012 and hired as graduate student teaching assistent in 2005-2007, and reasearch assistent in 2007-2012.
	In the thesis, he used correlation method to give further insight of the charge separation effect in relativistic heavy ion collisions, which suggests a flow related background.
	The thesis title is ``Charge Multiplicity Asymmetry Correlation Study Searching for Local Parity Violation At RHIC For STAR Collaboration''.

%
\end{vita}

%% file: thesis.bbl
\begin{thebibliography}{10}

\bibitem{Adams:2005dq}
John Adams et~al.
\newblock Experimental and theoretical challenges in the search for the quark
  gluon plasma: {T}he {STAR} collaboration's critical assessment of the
  evidence from {RHIC} collisions.
\newblock {\em Nucl. Phys.}, A757:102--183, 2005.

\bibitem{Adcox:2004mh}
K.~Adcox et~al.
\newblock Formation of dense partonic matter in relativistic nucleus nucleus
  collisions at {RHIC}: {E}xperimental evaluation by the {PHENIX}
  collaboration.
\newblock {\em Nucl. Phys.}, A757:184--283, 2005.

\bibitem{Arsene:2004fa}
I.~Arsene et~al.
\newblock Quark {G}luon {P}lasma and {C}olor {G}lass {C}ondensate at {RHIC}?
  {T}he perspective from the {BRAHMS} experiment.
\newblock {\em Nucl. Phys.}, A757:1--27, 2005.

\bibitem{Back:2004je}
B.B. Back et~al.
\newblock The {PHOBOS} perspective on discoveries at {RHIC}.
\newblock {\em Nucl.Phys.}, A757:28--101, 2005.

\bibitem{Morley:1983wr}
P.D. Morley and I.A. Schmidt.
\newblock {Strong P, CP, T Violations In Heavy Ion Collisions}.
\newblock {\em Z.Phys.}, C26:627, 1985.

\bibitem{Kharzeev:1998kz}
Dmitri Kharzeev, R.D. Pisarski, and Michel~H.G. Tytgat.
\newblock {Possibility of spontaneous parity violation in hot QCD}.
\newblock {\em Phys.Rev.Lett.}, 81:512--515, 1998.

\bibitem{Kharzeev:2004ey}
Dmitri Kharzeev.
\newblock {Parity violation in hot QCD: Why it can happen, and how to look for
  it}.
\newblock {\em Phys.Lett.}, B633:260--264, 2006.

\bibitem{Kharzeev:2007tn}
D.~Kharzeev and A.~Zhitnitsky.
\newblock Charge separation induced by {P}-odd bubbles in {QCD} matter.
\newblock {\em Nucl.Phys.}, A797:67--79, 2007.

\bibitem{Fukushima:2008xe}
Kenji Fukushima, Dmitri~E. Kharzeev, and Harmen~J. Warringa.
\newblock {The Chiral Magnetic Effect}.
\newblock {\em Phys.Rev.}, D78:074033, 2008.

\bibitem{Kharzeev:2007jp}
Dmitri~E. Kharzeev, Larry~D. McLerran, and Harmen~J. Warringa.
\newblock {The Effects of topological charge change in heavy ion collisions:
  `Event by event P and CP violation'}.
\newblock {\em Nucl.Phys.}, A803:227--253, 2008.

\bibitem{Stephanov:2004wx}
Mikhail~A. Stephanov.
\newblock {QCD phase diagram and the critical point}.
\newblock {\em Prog.Theor.Phys.Suppl.}, 153:139--156, 2004.

\bibitem{Aggarwal:2010cw}
M.M. Aggarwal et~al.
\newblock {An Experimental Exploration of the QCD Phase Diagram: The Search for
  the Critical Point and the Onset of De-confinement}.
\newblock 2010.

\bibitem{Adams:2003xp}
J.~Adams et~al.
\newblock {Identified particle distributions in pp and Au+Au collisions at
  $\sqrt{s_{NN}}$ = 200 GeV}.
\newblock {\em Phys.Rev.Lett.}, 92:112301, 2004.

\bibitem{Ma:2011uma}
Guo-Liang Ma and Bin Zhang.
\newblock {Effects of final state interactions on charge separation in
  relativistic heavy ion collisions}.
\newblock {\em Phys.Lett.}, B700:39--43, 2011.

\bibitem{Muller:2010jd}
Berndt Muller and Andreas Schafer.
\newblock {Charge Fluctuations from the Chiral Magnetic Effect in Nuclear
  Collisions}.
\newblock {\em Phys.Rev.}, C82:057902, 2010.

\bibitem{:2009txa}
B.I. Abelev et~al.
\newblock {Observation of charge-dependent azimuthal correlations and possible
  local strong parity violation in heavy ion collisions}.
\newblock {\em Phys.Rev.}, C81:054908, 2010.

\bibitem{:2011eg}
Adamczyk et~al.
\newblock {Directed Flow of Identified Particles in Au + Au Collisions at
  $\sqrt{s_{NN}} = 200$ GeV at RHIC}.
\newblock 2011.

\bibitem{:2008ed}
B.I. Abelev et~al.
\newblock {Centrality dependence of charged hadron and strange hadron elliptic
  flow from $\sqrt{s_{NN}} = 200$ GeV Au + Au collisions}.
\newblock {\em Phys.Rev.}, C77:054901, 2008.

\bibitem{Voloshin:2004vk}
Sergei~A. Voloshin.
\newblock {Parity violation in hot QCD: How to detect it}.
\newblock {\em Phys.Rev.}, C70:057901, 2004.

\bibitem{:2009uh}
B.I. Abelev et~al.
\newblock {Azimuthal Charged-Particle Correlations and Possible Local Strong
  Parity Violation}.
\newblock {\em Phys.Rev.Lett.}, 103:251601, 2009.

\bibitem{Ackermann:2002ad}
K.H. Ackermann et~al.
\newblock {STAR detector overview}.
\newblock {\em Nucl.Instrum.Meth.}, A499:624--632, 2003.

\bibitem{Bergsma:2002ac}
F.~Bergsma et~al.
\newblock {The STAR detector magnet subsystem}.
\newblock {\em Nucl.Instrum.Meth.}, A499:633--639, 2003.

\bibitem{Ackermann:1999kc}
K.H. Ackermann et~al.
\newblock {The STAR time projection chamber}.
\newblock {\em Nucl.Phys.}, A661:681--685, 1999.

\bibitem{Anderson:2003ur}
M.~Anderson, J.~Berkovitz, W.~Betts, R.~Bossingham, F.~Bieser, et~al.
\newblock {The STAR time projection chamber: A Unique tool for studying high
  multiplicity events at RHIC}.
\newblock {\em Nucl.Instrum.Meth.}, A499:659--678, 2003.

\bibitem{Adler:2003sp}
C.~Adler, A.~Denisov, E.~Garcia, M.~Murray, H.~Strobele, et~al.
\newblock {The RHIC zero-degree calorimeters}.
\newblock {\em Nucl.Instrum.Meth.}, A499:433--436, 2003.

\bibitem{Bieser:2002ah}
F.S. Bieser, H.J. Crawford, J.~Engelage, G.~Eppley, L.C. Greiner, et~al.
\newblock {The STAR trigger}.
\newblock {\em Nucl.Instrum.Meth.}, A499:766--777, 2003.

\bibitem{Miller:2007ri}
Michael~L. Miller, Klaus Reygers, Stephen~J. Sanders, and Peter Steinberg.
\newblock Glauber modeling in high energy nuclear collisions.
\newblock {\em Ann.Rev.Nucl.Part.Sci.}, 57:205--243, 2007.

\bibitem{:2008ez}
B.I. Abelev et~al.
\newblock {Systematic Measurements of Identified Particle Spectra in $p p$, d +
  Au and Au + Au Collisions from STAR}.
\newblock {\em Phys.Rev.}, C79:034909, 2009.

\bibitem{Adams:2004cb}
J.~Adams et~al.
\newblock {Measurements of transverse energy distributions in Au + Au
  collisions at $\sqrt{s_{NN}}$ = 200 GeV}.
\newblock {\em Phys.Rev.}, C70:054907, 2004.

\bibitem{Poskanzer:1998yz}
Arthur~M. Poskanzer and S.A. Voloshin.
\newblock {Methods for analyzing anisotropic flow in relativistic nuclear
  collisions}.
\newblock {\em Phys.Rev.}, C58:1671--1678, 1998.

\bibitem{Adams:2004bi}
J.~Adams et~al.
\newblock {Azimuthal anisotropy in Au+Au collisions at s(NN)**(1/2) = 200-GeV}.
\newblock {\em Phys.Rev.}, C72:014904, 2005.

\bibitem{JiayunChen:phd}
Jiayun Chen.
\newblock {Directed Flow at STAR}.
\newblock Ph.D. thesis.

\bibitem{GangWang:phd}
Gang Wang.
\newblock {Correlations Relative to the Reaction Plane at the Relativistic
  Heavy Ion Collider Based on Transverse Deflection of Spectator Neutrons }.
\newblock Ph.D. thesis.

\bibitem{Abelev:2007nt}
B.I. Abelev et~al.
\newblock {Charged particle distributions and nuclear modification at high
  rapidities in d + Au collisions at $\sqrt{s_{NN}}$ = 200 GeV}.
\newblock {\em Phys.Lett.B}, 2007.

\bibitem{Adler:2002tq}
C.~Adler et~al.
\newblock {Disappearance of back-to-back high $p_{T}$ hadron correlations in
  central Au+Au collisions at $\sqrt{s_{NN}}$ = 200 GeV}.
\newblock {\em Phys.Rev.Lett.}, 90:082302, 2003.

\bibitem{:2009qa}
B.I. Abelev et~al.
\newblock {Long range rapidity correlations and jet production in high energy
  nuclear collisions}.
\newblock {\em Phys.Rev.}, C80:064912, 2009.

\bibitem{Petersen:2010di}
Hannah Petersen, Thorsten Renk, and Steffen~A. Bass.
\newblock {Medium-modified Jets and Initial State Fluctuations as Sources of
  Charge Correlations Measured at RHIC}.
\newblock {\em Phys.Rev.}, C83:014916, 2011.

\bibitem{Asakawa:2010bu}
Masayuki Asakawa, Abhijit Majumder, and Berndt Muller.
\newblock {Electric Charge Separation in Strong Transient Magnetic Fields}.
\newblock {\em Phys.Rev.}, C81:064912, 2010.

\bibitem{Wang:2009kd}
Fuqiang Wang.
\newblock {Effects of Cluster Particle Correlations on Local Parity Violation
  Observables}.
\newblock {\em Phys.Rev.}, C81:064902, 2010.

\bibitem{Pratt:2010gy}
Scott Pratt.
\newblock {Alternative Contributions to the Angular Correlations Observed at
  RHIC Associated with Parity Fluctuations}.
\newblock 2010.
\newblock arXiv:1002.1758.

\bibitem{Abelev:2009jv}
B.I. Abelev et~al.
\newblock {Three-particle coincidence of the long range pseudorapidity
  correlation in high energy nucleus-nucleus collisions}.
\newblock {\em Phys.Rev.Lett.}, 105:022301, 2010.

\bibitem{Voronyuk:2011jd}
V.~Voronyuk, V.D. Toneev, W.~Cassing, E.L. Bratkovskaya, V.P. Konchakovski,
  et~al.
\newblock {(Electro-)Magnetic field evolution in relativistic heavy-ion
  collisions}.
\newblock {\em Phys.Rev.}, C83:054911, 2011.

\end{thebibliography}
